\documentclass[b5paper,10pt,twoside]{book}
\RequirePackage[papersize={17cm,24cm},bindingoffset=3mm,vcentering,dvips=true,twoside]{geometry}
\usepackage{graphicx}
\usepackage{amsmath,amssymb}             % AMS Math
\usepackage{rotating}                    % Sideways of figures & tables
\usepackage{lscape}                      %landscape
\usepackage{appendix}                    % needed for chapter appendices
\usepackage{fancyhdr}                    % Fancy Header and Footer
\usepackage{setspace}                    % Line spacing
\usepackage{txfonts}                     % Public Times New Roman text & math font
\usepackage[rightcaption]{sidecap}       % TK:for captions on the side of figures.

\usepackage{longtable}
\usepackage{natbib}                      % Cross-reference package (Natural BiB)
\bibpunct{(}{)}{;}{a}{}{,}               % to follow the A&A style

\newcommand{\captionfonts}{\small}
\makeatletter  % Allow the use of @ in command names
\long\def\@makecaption#1#2{%
  \vskip\abovecaptionskip
  \sbox\@tempboxa{{\captionfonts #1: #2}}%
  \ifdim \wd\@tempboxa >\hsize
    {\captionfonts #1: #2\par}
  \else
    \hbox to\hsize{\hfil\box\@tempboxa\hfil}%
  \fi
  \vskip\belowcaptionskip}
\makeatother   % Cancel the effect of \makeatletter

    \makeatletter
    \def\cleardoublepage{\clearpage\if@twoside \ifodd\c@page\else%
        \hbox{}%
        \thispagestyle{empty}%              % Empty header styles
        \newpage
        \if@twocolumn\hbox{}\newpage\fi\fi\fi}
    \makeatother

\let\la=\lesssim            % For Springer A&A compliance... 
\let\ga=\gtrsim

\def\apjl{ApJ}%
\def\apjs{ApJS}%
\def\ao{Appl.~Opt.}%
\def\aap{A\&A}%
\renewcommand\maketitle{\begin{titlepage}
  \let\footnotesize\small
  \let\footnoterule\relax
  \let \footnote \thanks
  \begin{center}
    {\LARGE The primordial binary population\\in the association Sco~OB2 \par}
  \end{center}
   \newpage
   \noindent Cover illustration: a large fraction of the stars is known to be in binary or multiple systems. It is therefore likely that many exoplanets have more than one Sun in the sky. The cover illustration represents an imaginary landscape on a planet in such a system. The original photograph shows the surface of the planet Mars, obtained with the exploration rover {\em Spirit}; courtesy of NASA/JPL/Cornell.
   \clearpage
   \begin{center}
    {\LARGE  The primordial binary population\\in the association Sco~OB2 \par}
    \vskip 3em
    {\large De oorspronkelijke dubbelsterpopulatie\\in de associatie Sco~OB2 \par}
    \vskip 1em
    {\normalsize 
    \par}
    \vskip 2em
    {\large Academisch Proefschrift \par}
    \vskip 2em

     {\normalsize ter verkrijging van de graad van doctor \\
     aan de Universiteit van Amsterdam, \\
     op gezag van de Rector Magnificus prof. mr. P.~F. van der Heijden,\\
     ten overstaan van een door het college voor promoties ingestelde commissie, \\
     in het openbaar te verdedigen in de Aula der Universiteit op
     \par}

    \vskip 2em
   {\large  donderdag 28 september 2006, te 12:00 uur \par}
    \vskip 5em
     {\normalsize door \par}
      \vskip 2em
    {\large Mattheus Bartholomeus Nicolaas Kouwenhoven \par}
    \vskip 5em
     {\normalsize geboren te Kwintsheul\par}
    \vfill
    {\normalsize 
 \par}
  \end{center}
  \newpage
      \begin{tabular}[t]{rl}
  \textsc{Promotiecommissie} &  \\ 
                             &  \\ 
        \textsc{Promotores}  & prof. dr. L. Kaper  \\
	                     & prof. dr. E.P.J. van den Heuvel  \\
	\textsc{Co-promotores}  & dr. A.G.A. Brown        \\
			     & dr. S.F. Portegies Zwart          \\
      \textsc{Overige leden} & prof. dr. H.F. Henrichs    \\ 
                             & dr. A. de Koter    \\ 
                             & prof. dr. L.B.F.M. Waters   \\ 
                             & prof. dr. R.A.M.J. Wijers   \\ 
			     & prof. dr. P.T. de Zeeuw \\
                             & dr. H. Zinnecker    \\ 
                            
      \end{tabular}
      \vskip 5em
      \noindent Faculteit der Natuurwetenschappen, Wiskunde en Informatica \\

      \vfill

      \begin{center}
      \includegraphics[width=0.4\textwidth,height=!]{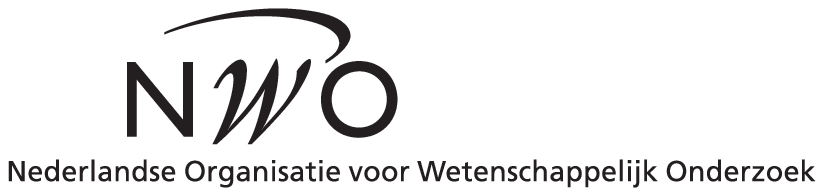}
      \end{center}

      \noindent Het hier beschreven onderzoek werd mede mogelijk gemaakt door steun van NWO.      

      \vskip 0.7cm

      \noindent ISBN--10: 90--6464--033--5 \\
      \noindent  ISBN--13: 978--90--6464--033--9\\
      \\
      Printed by Ponsen \& Looijen, Wageningen. Copyright 2006 \copyright \ Thijs Kouwenhoven.

  \end{titlepage}
}

\begin{document}

\pagestyle{empty}

\maketitle

\pagestyle{fancy}                       % Sets fancy header and footer

    %%%%%%%%%%%%%%%%%%%%%%%%%%%%%%%%%%%%%%%%%%%%%%%%%%%%%%%%%%%%%%%%%%%%%%%%%%%%%%%
    %   Title Page                                                                %
    %%%%%%%%%%%%%%%%%%%%%%%%%%%%%%%%%%%%%%%%%%%%%%%%%%%%%%%%%%%%%%%%%%%%%%%%%%%%%%%
    %\begin{titlepage}
    %\begin{center}
    %The Primordial Binary Population
    %\end{center}
    %\end{titlepage}
    %%%%%%%%%%%%%%%%%%%%%%%%%%%%%%%%%%%%%%%%%%%%%%%%%%%%%%%%%%%%%%%%%%%%%%%%%%%%%%%
    %   Dedication Page                                                           %
    %%%%%%%%%%%%%%%%%%%%%%%%%%%%%%%%%%%%%%%%%%%%%%%%%%%%%%%%%%%%%%%%%%%%%%%%%%%%%%%
    %\thispagestyle{empty}
    \newenvironment{dedication}
      {\cleardoublepage \thispagestyle{empty} \vspace*{\stretch{1}} \begin{center} \em}
      {\end{center} \vspace*{\stretch{3}} \clearpage}
    \begin{dedication}
    %Your dedication goes here ...
    \end{dedication}
    % \cleardoublepage generates one blank page for the next 'Abstract'
    % to be on an odd page.
    \thispagestyle{empty} \cleardoublepage

\renewcommand{\thepage}{\roman{page}}
\setcounter{page}{1}

\tableofcontents

\cleardoublepage

%\listoftables
%\listoffigures

%\setcounter{page}{12}
%\setcounter{chapter}{0}

\renewcommand{\thepage}{\arabic{page}}
\setcounter{page}{1}

\chapter{Introduction} \label{chapter: intro}

Observations have shown that a large fraction of the stars are in binary or multiple systems. The
term ``double star'' was first used in Ptolemy's ``Almagest'' in the
description of $\nu$~Sagittarii, which consists of two stars of
fifth magnitude ($\nu^1$~Sgr and $\nu^2$~Sgr) separated by
$0.23^\circ$. In 1767, the reverend John Mitchell had already pointed out 
that, seen through a telescope, there are many more stars with close companions 
than one would have expected if the stars had been randomly distributed over the 
sky. He therefore concluded that many double stars must be physical pairs, 
gravitationally bound to each other \citep[see][]{heintz1978}.
The term ``binary star'' was introduced by William
Herschel, who defines a binary as ``{\em a real double star --- the
union of two stars, that are formed together in one system, by the
laws of attraction}'' \citep{herschel1802}. 
He was also the first to discover in 1804 the orbital motion of a binary 
star \citep{herschel1804}, for the system of Castor ($\alpha$~Geminorum). 
Herschel realized that a
large fraction of the candidate binary stars are likely the result of
projection effects; these are called ``optical doubles''.
A few decades later, Friedrich Bessel obtained precise positional 
measurements of Sirius and found that it was slowly moving,
as if it were being pulled by another star. 
In 1844 he announced that Sirius must have an unseen companion \citep{bessel1844}.
Since that time, a large number of binarity surveys have been performed,
and new observational techniques have become available; for a historical overview
we refer to the works of \cite{aitken1918}, \cite{batten1973}, and
\cite{heintz1978}. 

Over the last couple of decades it has become clear that indeed many
stars have physical companions. It is an observational challenge to
detect a companion, due to the often small angular separation, large
brightness contrast, long orbital period, and small radial-velocity
variations induced on the primary. For these reasons, binary searches are inevitably biased
towards relatively bright and nearby stars (e.g., solar neighbourhood
or OB-type stars) and towards relatively massive companions. Therefore, when
determining the intrinsic stellar multiplicity fraction, one has to account
for the companions that have remained undetected due to observational
limitations. When taking these observational biases into account, it
was soon realized that the true multiplicity fraction may be large, 
possibly even up to 100\% \citep{poveda1982,abt1983}. 

Multiplicity studies often focus on a particular selection of
primaries.  \cite{garmany1980} report a binary fraction of $36 \pm
7$\% based on a radial-velocity study among a sample of O stars
brighter than $V = 7$~mag. \cite{gies1987} find a binary fraction of
56\% for O stars in clusters and associations, versus 26\% for field
O stars. \cite{abt1990} report a binary fraction of 74\% for B2--B5
primaries. A recent radial-velocity survey of 141 early-type stars in
the Cygnus OB2 association indicates that the intrinsic binary fraction
must be in the range 70--100\% \citep{kobulnicky2006}.

Among solar-type F3-G2 stars \cite{abtlevy1976} report a multiplicity
fraction of 55\%. The CORAVEL spectroscopic survey of 167 F7 to G9
primaries by \cite{duquennoy1991} shows that only about one third
of the G-dwarf primaries may be single stars, i.e. having no companion
above 0.01~M$_{\odot}$. Thus, the Sun (and its planetary system) would
be atypical as a single star \citep[although a distant companion named
``Nemesis'' has been proposed; see][]{davis1984,whitmire1984}.
Also pre-main sequence stars (T~Tauri stars) show a high multiplicity:
the binary star frequency in the projected separation range $16-252$~AU
is $60\pm 17\%$, four times higher than that of solar-type main-sequence
stars \citep{ghez1993,ghez1997}. 
However, \cite{lada2006} points out that the binary frequency declines
for systems with a primary mass later than type~G: the binary
fraction is around 30\% for M~stars, and as low as 15\% for L and T
dwarfs, objects near and below the hydrogen burning limit ({\em brown
dwarfs}). Given a decline in binary frequency as a function of primary
mass, Lada argues that most stellar systems in the Galaxy consist of
single rather than binary or multiple stars.

This issue touches upon the existence of a so-called {\em brown-dwarf
desert}. To date, over 1500 F, G, K, and M stars have been observed
using radial-velocity techniques with sufficient sensitivity to detect
brown dwarf companions orbiting at a distance smaller than 5~AU
\citep{mccarthy2004}. The results of these searches are clear
and striking: there is a deficit of brown dwarf companions, both
relative to the stellar companion fraction
\citep{duquennoy1991,fischer1992}, and relative to the observed
frequency of extrasolar planets
\citep{marcy2000,marcy2003}. Also at larger separations
(75--1200~AU) the frequency of brown-dwarf companions is a few percent
at most \citep{mccarthy2004}. Obviously, one would like to
know whether the existence of a brown-dwarf desert is restricted to
solar-type primaries, or perhaps extends into the regime of the more
massive primaries. As one would not expect that the star formation
process is ``aware'' of the hydrogen burning limit, the observation of a
brown-dwarf desert poses an important constraint on the theory of star
formation and the early evolution of star clusters.

One of the main motivations to measure the binary fraction of stars is
to better understand the process of star formation. If many stars are
formed in binary or multiple systems, multiplicity is a fundamental
aspect of star formation. Also, the period, mass-ratio and
eccentricity distribution of the binary population contain
important information on the star formation process itself 
\citep[see, e.g.,][]{mathieu1994,tohline2002}. Naively
one may expect stars to form in multiple systems so that the excessive
amount of angular momentum, that would otherwise prevent the formation
of the star, is stored in orbital motion \citep[even though the angular momentum 
is partially stored in stellar rotation; e.g.,][]{mokiem2006}.

The binary population resulting from the star formation process is
the {\em primordial binary population}, which is defined as {\em the population of
binaries as established just after the gas has been removed from the
forming system, i.e., when the stars can no longer accrete gas from
their surroundings} \citep{brown2001,kouwenhoven2005}. This definition is chosen
such as to provide a natural end point for calculations of the actual
formation of binaries \citep[e.g., ][]{bate2003}, as well as a
starting point for understanding the origins of the evolved cluster
and field binary population \citep[e.g., ][]{ecology4}. In reality,
the processes of accretion and stellar dynamics will overlap, meaning
that this idealized primordial population does not really exist and can only
be established approximately.

Characterizing the primordial binary population is of key importance
to stellar population and cluster evolution studies. Binaries are
invoked to explain a wide variety of astrophysical phenomena, from
short and long gamma-ray bursts \citep{fryer1999}, type Ib/c and
blue-type supernovae \citep{podsialowski1992}, to the fastest OB
runaway stars \citep{blaauw1961,hoogerwerf2001} and the entire
menagerie of binary systems with compact remnants: X-ray binaries,
millisecond pulsars, double neutron-star systems
\citep{vandenheuvel1994,fryer1997} and the progenitors to supernovae type Ia
\citep{yungelson1998,hillebrandt2000}, the standard candles used in
modern cosmology. Binaries (and higher-order multiples) also play a
fundamental role in determining the dynamical evolution of dense
stellar clusters \citep{hut1992,ecology4}.

The most likely place to find a binary population best approximating
the primordial state would be very young, low density stellar groupings
containing a wide spectrum of stellar mass. The youth
implies that the binary parameters of only a handful of the most
massive systems have changed due to stellar evolution, and their low stellar density guarantees
that little dynamical evolution has taken place. 
This naturally leads to
the study of the local ensemble of OB associations.
OB~associations are young ($5-50$~Myr)
and low-density ($< 0.1~\mbox{M}_\odot\,{\rm pc}^{-3}$) stellar
groupings -- such that they are unlikely to be gravitationally bound -- containing a
significant population of B stars. Their projected dimensions range
from $\sim 10$ to $\sim 100$~pc and their mass spectra reach all the
way down to brown dwarf masses \citep{brown1999}. Thanks to
the {\em Hipparcos} satellite the stellar content of the nearby OB
associations has been established with unprecedented accuracy
\citep{dezeeuw1999} to a completeness limit of $V \sim 7-8$~mag, or
about 2.5~M$_{\odot}$ for the nearest associations, resulting in a well
defined sample to start from. Lately, the population of low-mass
pre-main-sequence stars has
also been extensively studied in, e.g., the Sco~OB2 association
\citep{preibisch1999} and Ori~OB1 \citep{briceno2005}.

\begin{figure}[btp]
  \centering
  \includegraphics[width=1\textwidth,height=!]{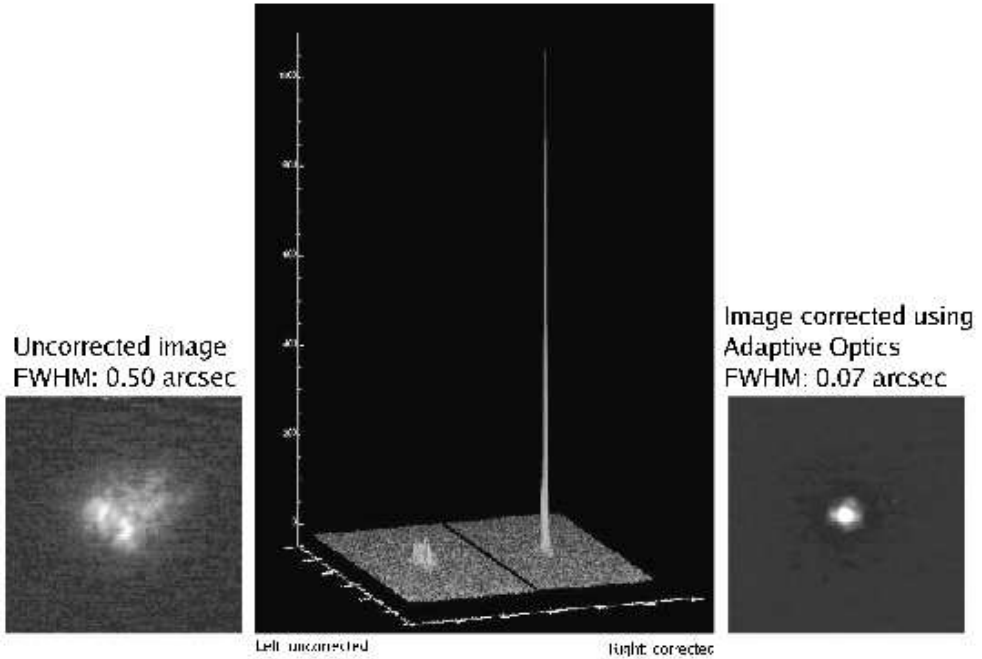}
  \caption{A near-infrared $K_S$ band image of a star ($V=8$~mag) obtained before ({\em left}) and after ({\em right)} the adaptive optics was switched on. The corresponding three-dimensional intensity profiles of both images are shown in the middle panel, demonstrating the angular resolution and central intensity gain obtained with the adaptive optics technique. These images are obtained with NAOS/CONICA on the {\em Very Large Telescope} at Paranal, Chile. Courtesy of the European Southern Observatory (ESO). \label{figure: naco_ao} }
\end{figure}

\begin{figure}[btp]
  \centering
  \includegraphics[width=1\textwidth,height=!]{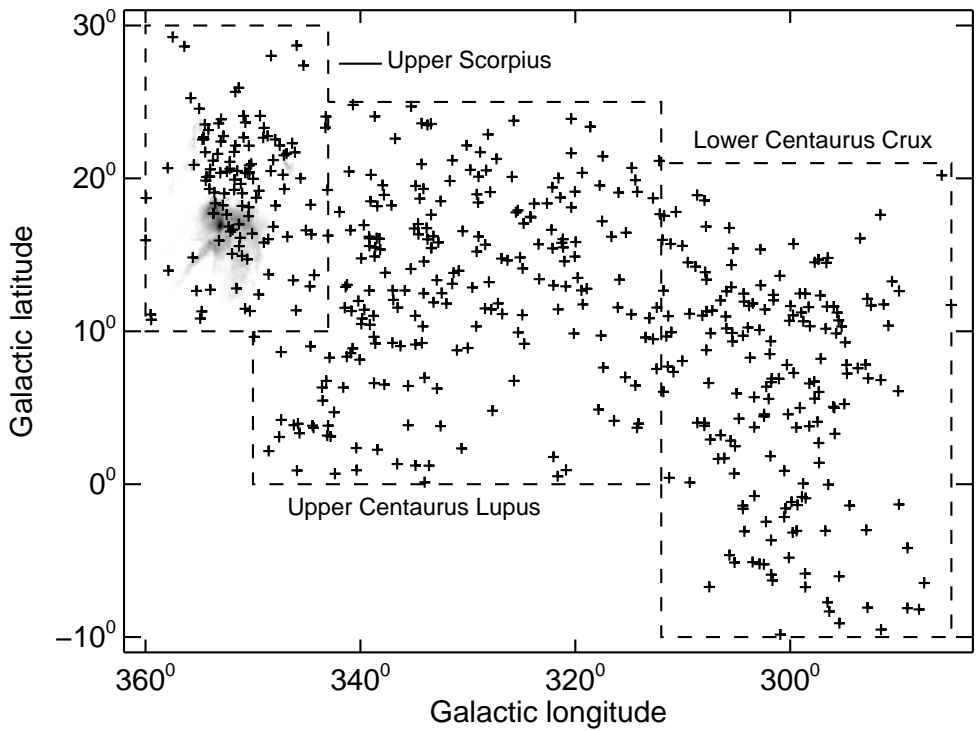}
  \caption{The three subgroups of Sco~OB2: Upper Scorpius (US), Upper Centaurus Lupus (UCL), and Lower Centaurus Crux (LCC). The locations of the association members according to the study of \cite{dezeeuw1999} are indicated with the plusses. The $\rho$~Ophiuchus star forming region is visible in the upper-left corner of the figure in gray scale (IRAS 100~$\mu$m). The structure of Sco~OB2 is likely the result of sequential star formation. The subgroups UCL and LCC, aged 20~Myr, triggered star formation in Sco~OB2. In turn, the 5~Myr aged subgroup US triggered star formation in the $\rho$~Ophiuchus region. \label{figure: scocen_subgs} }
\end{figure}

\begin{figure}[btp]
  \centering
  \includegraphics[width=1\textwidth,height=!]{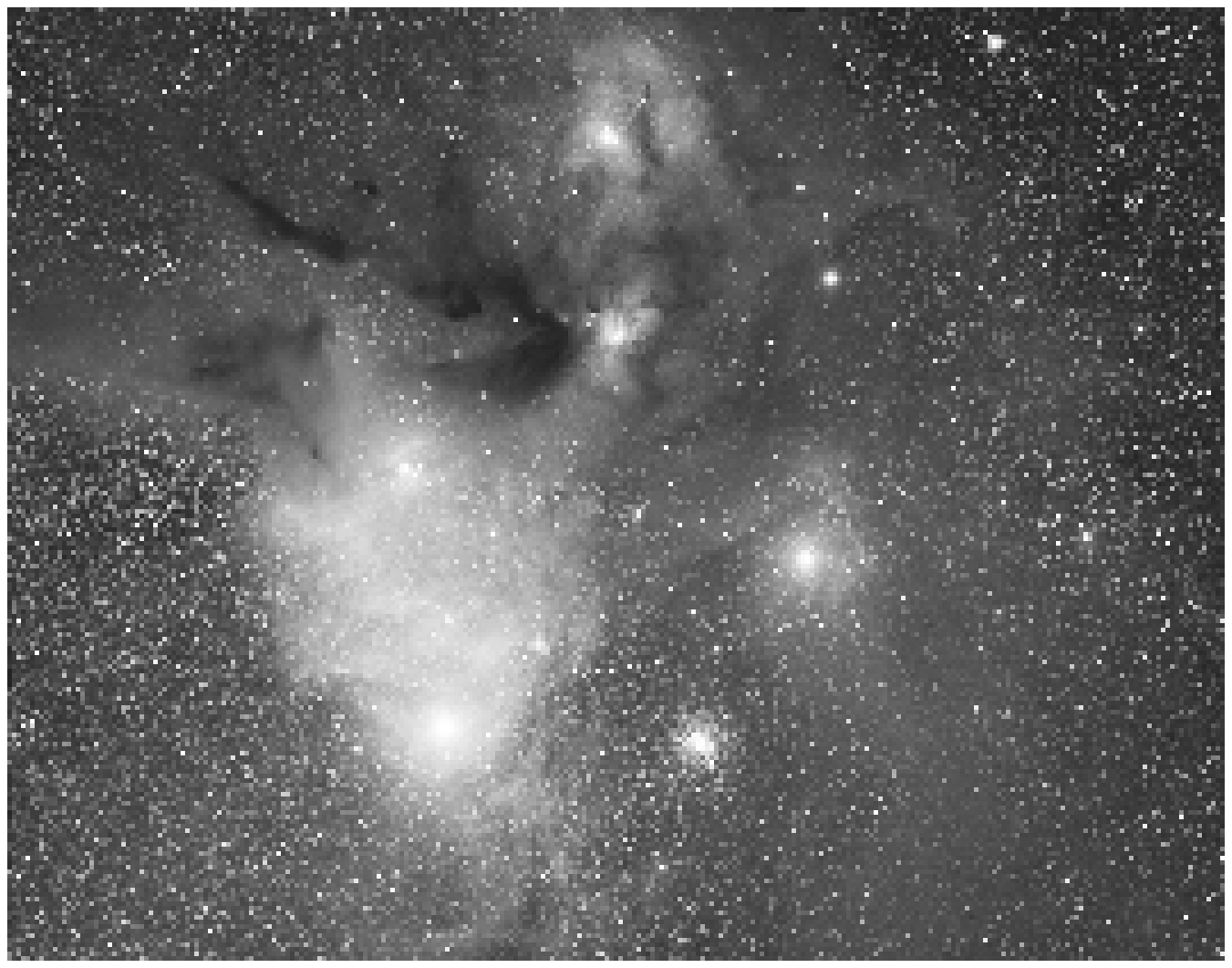}
  \caption{The $\rho$~Ophiuchus star forming region near the Sco~OB2 association (cf. Figure~\ref{figure: scocen_subgs}). The dust clouds near the North (top) are illuminated by the young star $\rho$~Ophiuchus. The bright star at the bottom is the cool supergiant Antares ($\alpha$~Scorpii). The globular cluster M4 is visible in the background, to the West (right) of Antares. The image size is approximately six degrees. Courtesy of NASA. 
  \label{figure: scocen} }
\end{figure}

\section{Observational challenges}

Due to observational limitations only a small, heavily biased subset
of the binary population is detectable as such. As a result, we are
observing only the ``tip of the iceberg'' of the binary parameter
space. 
In the last century many efforts have been undertaken to determine
how many companion stars have remained undetected, and what their
properties are. Here we refer to the work of Gerard Kuiper
\citep{kuiper1935a,kuiper1935b}, Jean-Louis Halbwachs
\citep{halbwachs1981,halbwachs1983,halbwachs1986,halbwachs2003}, and
Sake Hogeveen \citep{hogeveen1990,hogeveen1992a,hogeveen1992b}, among
several others. Binary systems are studied using different techniques, each associated with its specific selection effects. Observationally, binary systems can be roughly classified in the following categories:
\begin{itemize}\addtolength{\itemsep}{-0.5\baselineskip}
\item[--] {\bf Visual binaries.} These are binary systems for which both components are resolved with imaging techniques; also including Speckle imaging and adaptive optics. Orbital motion is often undetected due to their long period of visual binaries. Occasionally, two stars have a small separation due to chance projection, while they are physically unbound; these are so-called optical doubles. It is often difficult to discriminate between visual binaries and optical doubles.
\item[--] {\bf Spectroscopic binaries.} For these binaries the presence of a companion star is inferred from the spectrum of the target. For a single-lined spectroscopic binary (SB1), the projected orbital motion (of the primary) is inferred from the movement of the spectral lines due to the Doppler effect. For a double-lined spectroscopic binary (SB2), the spectrum is a superposition of the spectra of the two individual stars. Most exoplanets have been found using the Doppler technique.
\item[--] {\bf Astrometric binaries.} Astrometric binaries are similar to visual binaries, for which one of the components is not detected. For these binaries, the presence of a companion star is inferred from the orbital motion of the visible star in the plane of the sky. Observations of astrometric binaries can provide much information on the orbit; however, these binaries are difficult to detect.
\item[--] {\bf Other binaries.} These include 
eclipsing binaries \citep[e.g.][]{todd2005}, 
common proper motion binaries \citep[e.g.][]{greaves2004}, 
lunar occultation binaries \citep[e.g.][]{simon1995}, 
differential-photometry binaries \citep[e.g.][]{tenbrummelaar2000}, 
variability-induced movers \citep[e.g.][]{pourbaix2003}, 
and photometric/astrometric lensing binaries \citep[e.g.,][]{han2006}.
\end{itemize}

Clearly, the investigation of stellar and substellar companions has
greatly benefited from the recent advance in astronomical
instrumentation. The development of large telescopes such as the 8.2~meter ESO
{\em Very Large Telescope} and the 10~meter Keck telescope, the
technical improvement of infrared detectors, and the availability of
adaptive optics instrumentation has enormously improved the capacity
to detect faint and close companions. 

With adaptive optics a deformable mirror is used, that is able to
correct the observed stellar image that is blurred due to atmospheric
turbulence. Given the rapid (fraction of a second) time-variable
nature of atmospheric turbulence the image has to be corrected in real
time using a reference star. The reference object is a bright
point source in the field of view; as its instrumental profile (the
so-called point spread function) is known in advance, the measured
deviations in the point spread function can be used to determine on-line 
how the atmospheric turbulence has affected the stellar
image. Bright stars are not very common nor homogeneously spread over
the sky, so developments are taking place to produce artificial
reference stars using the backscattering of the light of a laser in the upper
atmospheric layers (so-called laser guide stars). When one is looking
for companions of bright stars, the bright primary itself can be used
to apply the corrections to the deformable mirror.  

Figure~\ref{figure: naco_ao} demonstrates the power of adaptive optics using
the instrument NAOS/CONICA at the ESO
{\em Very Large Telescope}. The figure shows the improvement one
obtains in the shape of the point-spread function, almost reaching the
diffraction limit of an 8.2m telescope in the 2.2~$\mu$m $K_S$ band. The
improvement in the quality of the obtained image is often expressed in
terms of the Strehl ratio, i.e. the ratio of the peak intensity
of the observed and ideal (diffraction-limited) point source image.

The big advantage that adaptive optics instrumentation offers to the
detection of binary companions is that it provides the opportunity to
bridge the gap in orbital separation between the wide visual binaries and
close spectroscopic binaries, potentially unveiling a large fraction
of the binary population.

\section{Recovering the true binary population in Sco OB2}

The immediate aim of this thesis is to characterize the binary
population in the nearby OB association Scorpius~OB2. It is the
nearest young OB~association and a prime candidate for studying the
primordial binary population. The proximity of Sco~OB2 (118--145~pc) facilitates
the detection of faint and close companions. The young
age ($5-20$~Myr) provides some confidence that stellar and dynamical evolution have not
significantly altered the binary population since the moment of gas
removal. Sco~OB2 consists of three subgroups: Upper Scorpius (US),
Upper Centaurus Lupus (UCL), and Lower Centaurus Crux (LCC); 
see Figure~\ref{figure: scocen_subgs}. The three
subgroups are located at a distance of 145~pc, 140~pc, and 118~pc, respectively, and
have an age of 5~Myr, $15-22$~Myr, and $17-23$~Myr, respectively. The
membership and stellar content of the association was established by
\cite{dezeeuw1999} using {\em Hipparcos} parallaxes and proper
motions. The structure of the Sco~OB2 complex is likely the result of
sequential star formation. The LCC and UCL subgroups are the oldest,
and may have triggered star formation in US, which in turn may have
triggered star formation in the star forming region $\rho$~Oph
\citep{degeus1992,preibisch1999}; the latter region is shown in Figure~\ref{figure: scocen}. 
By studying the properties of the
binary population in the three different subgroups in Sco~OB2, one can
in principle establish whether the binary population has evolved as a
function of time.

\begin{figure}[btp]
  \centering
  \includegraphics[width=0.7\textwidth,height=!]{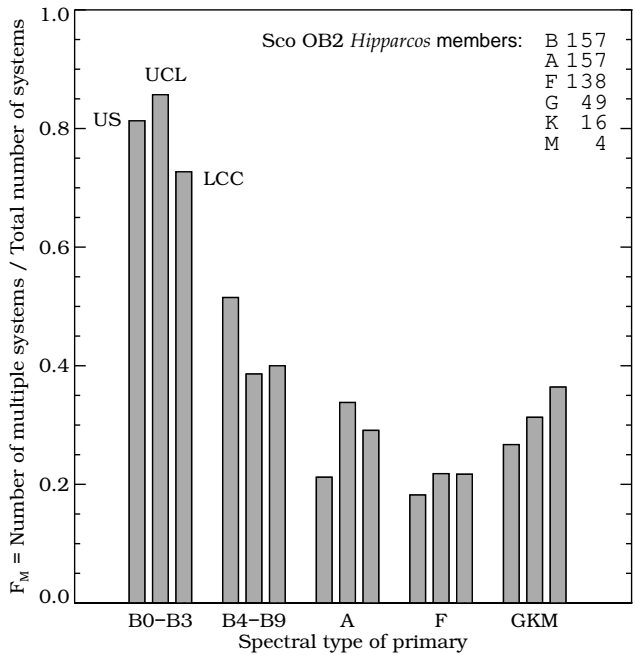}
  \caption{The (observed) fraction of stellar systems which is multiple versus the spectral type of the primary, for the three subgroups of Sco~OB2 \citep{brown2001}, at the time before the adaptive optics observations presented in this thesis. The spectral types of the companions (not included in this plot) are always later than those of the primary stars. This figure suggests that multiplicity is a function of spectral type, but in this thesis we show that this conclusion is premature, as selection effects should be taken into account. Recent adaptive optics imaging surveys \citep[Chapters~2 and~3;][]{shatsky2002} have significantly increased the known multiplicity fraction among intermediate mass stars in Sco~OB2; cf. Figure~\ref{figure: mulspt}. In Chapter~5 we study the selection effects in detail, and find that the intrinsic binary fraction among intermediate-mass stars is close to 100\%.
  \label{figure: intro_mulspt} }
\end{figure}

Many studies regarding the multiplicity of Sco~OB2 members are
available in the literature, although the data is quite scattered. The
latter has become less of a problem thanks to the availability of
on-line catalogues such as provided by the SIMBAD and Vizier services
of the Centre de Donn\'{e}es astronomiques de Strasbourg. For each
{\em Hipparcos} member of Sco~OB2 all available information has been collected
and summarized in terms of visual, astrometric and spectroscopic
binaries. This literature study \citep{brown2001,kouwenhoven2005}
shows that the fraction of multiple systems with B0--B3 primary is as
high as 80\%, declining to less than 40\% for the A star
primaries; see Figure~\ref{figure: intro_mulspt}. 
The drastic decline in binary fraction going from
early-B to A stars is striking, but one has to bear in mind the
observational biases. Apart from the incompleteness of the {\em Hipparcos} sample beyond spectral type B9--A0 there is also a bias
which may lead to artificially low binary fractions among A and F
stars. The latter are in between the very bright, and therefore often
observed B stars, and the faint later-type stars, which are very
interesting because of the presence of e.g. T-Tauri stars among
them. Hence the A and F stars are the least surveyed for
multiplicity in this association. 

To at least partly overcome this situation, we set out to perform an
adaptive optics survey targeted at the A star {\em Hipparcos} members
of Sco~OB2. This survey should allow for the detection of stellar and
substellar visual companions down to an angular separation of
0.1~arcsec in the $K_S$ band. A similar study has recently been performed
around the B star primaries of Sco~OB2 \citep{shatsky2002}.  

In parallel to improving on the statistics of the multiplicity in
Sco~OB2 and to obtaining a better determination of the distributions
of the binary parameters (primary mass, mass ratio, orbital
separation, eccentricity, etc.), we construct models of OB~associations with different properties.
The outcome of these model
calculations can be converted into simulated observations taking into account
the selection effects. Comparison with the real observations then provides insight
into the effects of the observational biases and allows for the
reconstruction of the true binary population in Sco~OB2.

\subsection{Outline of this thesis}

Our ultimate goal is to derive the primordial binary population in
Sco~OB2. Our strategy is to collect data on binarity in Sco~OB2 using
observations and literature data (Chapters~2
and~3), study the selection effects introduced by
the different observing techniques (Chapter~4), and
recover the current and primordial binary population in Sco~OB2
(Chapter~5).
Astronomers have attempted to recover the primordial binary population for
over a century. Two technological developments in the 1990s have opened
the possibility to accurately recover the primordial binary population.
The first is the availability of the adaptive optics technique (see above), 
making it possible to bridge the gap between visual
and spectroscopic binaries. The second development is the rapid
increase in computer power. 
With respect to the past, there is less of a need for simplifications, 
such as very basic models for selection effects, random pairing between the binary components, circular orbits, and the absence of background stars. Realistic simulations of evolving star
clusters, including stellar and binary evolution, can now be performed on
a personal computer. 
These developments play an important role in the recovery of the primordial binary population, and are therefore an integral part of the work presented in this thesis, which is organized as follows.
\vskip0.3cm
\noindent
{\bf Chapter~2}. In order to find the primordial binary population, the first step is to characterize the {\em current} binary population in Sco~OB2, by performing a literature study on binarity, and increasing the available dataset with new observations. I and my co-authors present in Chapter~2 the results of our
near-infrared adaptive optics binarity survey among 199~A and late-B
members of Sco~OB2. We performed our observations with the ADONIS system,
mounted on the ESO 3.6\,meter telescope on La~Silla, Chile. Each target is
observed in the $K_S$ band; several in the $J$ and $H$ bands as well.
The observations were obtained in the near-infrared, as in this wavelength regime the brightness contrast between a massive primary and a low-mass companion is less than in the optical.
We detect 151~stellar components (other than the target stars), of which
74~are companion stars, and 77~are background stars. 
The companion stars and background stars are separated using a simple magnitude 
criterion ($K_S=12$~mag).
With our survey we
report 41~previously undocumented companion stars, which significantly
increases the number of known companions among intermediate-mass stars in
Sco~OB2. For example, with our study we have doubled the number of known binaries among A~stars in Sco~OB2. The observed mass ratio distribution is consistent with the form
$f_q(q) \propto q^{-0.33}$, and excludes random
pairing of the binary components.
\vskip0.3cm
\noindent
{\bf Chapter~3}. In this chapter we present the results of a $JHK_S$ near-infrared
follow-up survey of 22~A and late-B stars in Sco~OB2, with the
near-infrared adaptive optics system NAOS/CONICA, which is mounted on UT4
of the ESO {\em Very Large Telescope} at Paranal, Chile. 
The follow-up study was performed to (1) study the validity of the
criterion used in the ADONIS survey to separate companion stars and background 
stars, and (2) investigate the scarcity of close brown dwarf
companions among intermediate-mass stars in Sco~OB2. 
Using multi-color observations of each secondary, we compare its location in the color-magnitude diagram with that of the isochrone. Based on this comparison, each secondary is classified as a companion star or a background star. 
Our analysis indicates that the classification criterion adopted for the ADONIS survey
correctly classifies most companion stars and background stars. Our study
confirms the very small brown dwarf companion fraction, and the small
substellar-to-stellar companion ratio. These properties, often referred
to as the brown dwarf desert, are a natural result of the mass
ratio distribution $f_q(q) \propto q^{-0.33}$ as derived in Chapter~2. 
If star formation results in a mass
ratio distribution of the above form, the embryo ejection scenario is not
necessary to explain the small number of brown dwarf companions.
Our results suggest that brown dwarfs form like stars, and that the brown dwarf 
desert can be ascribed to an excess of planetary companions, rather than by a 
lack of brown dwarf companions.
%When even further extrapolated, this mass ratio distribution would predict
%a very small number of planetary companions, contrary to the observations of solar-type stars in the solar neighbourhood, supporting the idea that stars and planets form in different ways.
\vskip0.3cm
\noindent
{\bf Chapter~4}. In this chapter we address the issue of
deriving the binary population from an observational dataset that is 
hampered by selection effects. We show that, in order
to do so, a comparison between {\em simulated observations} and the
observations is necessary, rather than naively correcting the
observations for the selection effects. We study in detail the selection
effects associated with imaging surveys, radial velocity surveys, and
astrometric binarity surveys. Furthermore, we identify five 
methods of pairing the components of a binary system (i.e., pairing
functions), and discuss their differences. In this chapter we demonstrate the 
power of the
method of simulating observations, by recovering the true binary population
from an artificial dataset.
\vskip0.3cm
\noindent
{\bf Chapter~5}. In the last chapter we discuss the
recovery of the binary population in Sco~OB2 from observations. We use the
technique of simulating observations that was extensively discussed in
Chapter~4, and find that the binary fraction in Sco~OB2 is close to $100\%$ (larger than $70\%$ with $3\sigma$ confidence). We exclude random pairing and primary-constrained random pairing between the binary components. The mass ratio distribution is of the form $f_q(q) \propto q^{\gamma_q}$, with $\gamma_q = -0.4 \pm 0.2$. The semi-major axis distribution has the form $f_a(a) \propto a^{\gamma_a}$ with $\gamma_a=-1.0 \pm 0.15$, corresponding to \"{O}pik's law. The log-normal period distribution of \cite{duquennoy1991} is inconsistent with the observations. The observed eccentricity distribution, although poorly constrained by observations, is consistent with a flat distribution. Due to the youth and low stellar density, one expects that stellar and dynamical evolution have only mildly affected the binary population since the moment of gas removal. The current binary population of Sco~OB2 is thus very similar to its primordial binary population. The major result presented in this Chapter, i.e. that practically all intermediate mass stars have formed in a binary system, provides fundamental information to our understanding of the star forming process.
\vskip0.3cm

\subsection{Future work}

In this thesis we have identified the current binary population in Sco~OB2, which is likely a good representation of the primordial binary population.  Throughout our analysis we have adopted simplified models and we have made several assumptions that need to be confirmed observationally. 

In each chapter of this thesis we discuss the difficulty of distinguishing between background stars and physical companion stars. A follow-up photometric (or spectroscopic) survey is necessary to identify the true nature of each candidate companion. Future models should therefore include an accurate description of the background star population. Our models for the {\em Hipparcos} selection effects are simplifications of reality, making it difficult to compare the simulated observations directly with {\em Hipparcos} astrometric binaries. For spectroscopic binaries we can predict which binary system would have been detected due to its radial velocity variations, but we have not modeled the derivation of the orbital elements for single- and double-lined spectroscopic binaries, which is necessary for more detailed comparison with the observations, e.g., in order to accurately recover the eccentricity distribution. 

Another observational complication is that not only binaries, but also variable stars may show radial velocity variations, and sometimes mimic spectroscopic binaries. 
Although it is known from the observations in Chapters~2 and~3 that triple and higher-order systems are present in Sco~OB2, we neglect these in the analysis of Chapters~4 and~5. Our simulation software allows the inclusion of these higher-order systems, but we decided not to do so because of the non-trivial comparison with the observations, the complicated selection effects, and the small number of known higher-order multiples. In Chapter~4 and~5 we have also assumed the independence of binary parameters; it is of great importance to study whether this assumption holds. The comparison between observations and simulations needs further attention, particularly on the issue of the relative weight that is given to the comparison of individual properties, when combining the results. For example, what relative weight is assigned to the spectroscopic binary fraction, and to the observed mass ratio distribution for visual binaries, respectively?

In Chapter~5 we include the results of six major binarity surveys among Sco~OB2 members. We have not included the smaller surveys and individual discoveries, as each of these has its associated, often not well-documented selection effects. Inclusion of these will provide a more accurate description of the binary population in Sco~OB2. Moreover, we have not included the common proper motion pairs in our analysis, which is necessary for a good understanding of the wide binary population in Sco~OB2, which can provide information on the past dynamical evolution of the binary population Sco~OB2.

The properties of the brown dwarf companions need to be studied in more detail. In particular, the relation between the pairing function and the existence of a brown dwarf desert (Chapter~2) needs observational confirmation. Our analysis in Chapter~3 has shown that the mass ratio distribution for binary systems with a brown dwarf companion is an extrapolation of that for stellar companions, and that the brown dwarf desert could be a natural outcome of the star forming process. A follow-up binarity survey among the low-mass members of Sco~OB2 will provide further information on the differences between the formation of brown dwarf companions and planetary companions.

\begin{figure}[t]
  \centering
  \includegraphics[width=1\textwidth,height=!]{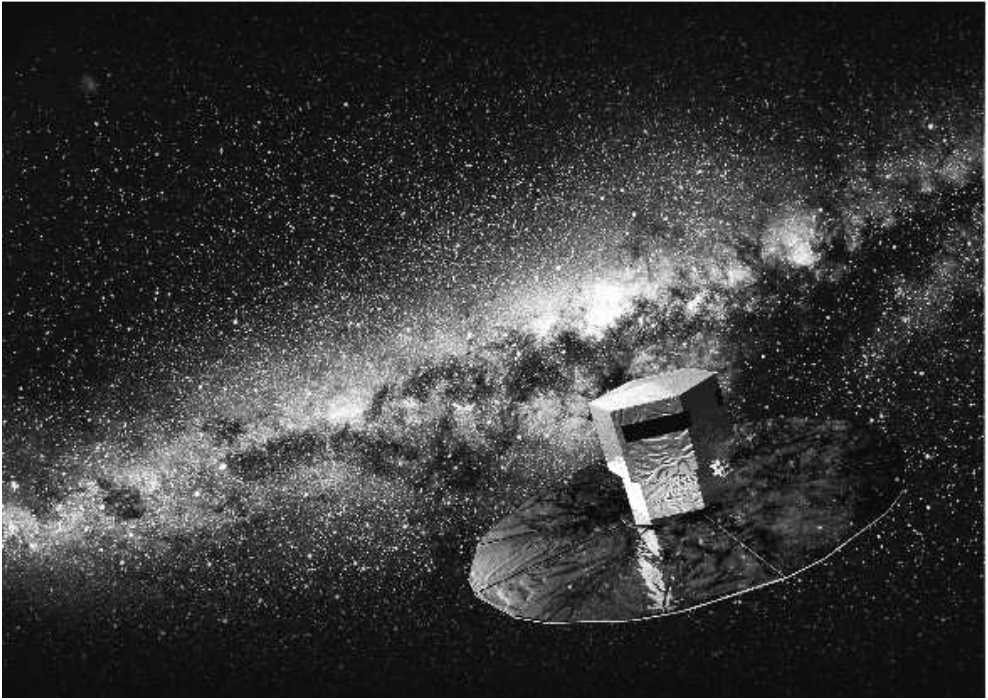}
  \caption{The {\em Gaia} satellite is expected to be launched in 2011 by the {\em European Space Agency}, and will measure the position, distance, space motion, and other physical properties of over a billion stars with unprecedented accuracy. {\em Gaia} will find millions of new visual, spectroscopic, astrometric, and eclipsing binary systems, as well as thousands of exoplanets and brown dwarf binaries. The resulting {\em Gaia} dataset can be used to determine the membership, stellar content, and binarity in the nearby associations down into the brown dwarf regime, and will thus provide crucial information for our understanding of the star forming process. Courtesy of the {\em European Space Agency}. 
  \label{figure: intro_gaia} }
\end{figure}

The current binary population in Sco~OB2 may be a fossil record of the primordial binary population. The young age and low stellar density of Sco~OB2 suggest that stellar evolution
has affected only a handful of the most massive binaries, and that dynamical
evolution of the binary population has been modest. Although the
former statement is correct, the second statement needs to be
verified using numerical simulations of OB~associations. Whether the effect of dynamical
evolution has been negligible over the lifetime of Sco~OB2, depends on
the initial conditions. If Sco~OB2 was born as a low-density
association, similar to its present state, this is likely. On the other
hand, if the association has expanded significantly over the last
5--20~Myr, dynamical evolution may have been more prominent.
The definition of the primordial binary population as given above refers to the ideal situation where the gas (out of which the stars have formed) is rapidly removed, and accretion halts. In practice, however, the process of gas removal by stellar winds takes time, and several stars may retain their accretion disk. Detailed simulations including both stars and gas are necessary to further investigate this issue.
Another issue is the variation of the primordial binary population for different stellar groupings. Due to variations in density, turbulence, temperature, and metallicity in different star forming regions, each of these may produce a characteristic binary population. For example, \cite{koehler2006} study multiplicity among
members of the Orion Nebula Cluster, and find a binary fraction among M~stars
that is less than that of solar-type main sequence stars, and 3.5--5 times
lower than that of the Sco~OB2 and Taurus-Auriga groupings. Two possible explanations
for these differences are that the Orion Nebula Cluster was originally
much denser, or that its primordial binary fraction may have been much lower.
The observational techniques used in this paper, as well as the method of simulating observations, are in principle applicable to any stellar population. As long as the observations provide a reasonable binarity dataset, and the selection effects are well modeled, an estimate for the primordial binary population can be obtained for e.g., $\rho$~Ophiuchus, Taurus-Auriga, the field star population, and the population near the Galactic center.

Due to selection effects a relatively small number of visual, spectroscopic, and astrometric binaries is known among the low-mass members of Sco~OB2 (see Figure~\ref{figure: intro_mulspt}). As a result of this scarcity it is difficult to derive the properties of the low-mass stellar population. Our knowledge on the low-mass binary population in OB~associations has to be improved significantly for a complete understanding of star formation.
The {\em Gaia} space mission \citep{perryman2001,turon2005} may provide the answer. The {\em Gaia} satellite is expected to be launched in 2011 by the {\em European Space Agency} and will survey over a billion stars in our Galaxy and the Local Group.
It will provide an enormous dataset of visual, eclipsing, spectroscopic, and astrometric binaries \citep{soderhjelm2005}. 
{\em Gaia} will resolve all binaries with $\rho \ga 20$~mas up to several kiloparsecs, as long as the magnitude difference between the components is moderate. A large fraction of the binaries with a period between 0.03--30 year will be resolved astrometrically, and the brightest of these (up to $V \approx 15$~mag) also spectroscopically. {\em Gaia} is also expected to find millions of eclipsing binaries with an orbital period less than a week. Furthermore, thousands of exoplanets and brown dwarf binaries are expected to be discovered. 
With the results of {\em Gaia} it will be possible to accurately determine the membership and stellar content of nearby OB~associations, down into the brown dwarf regime. The {\em Gaia} dataset will be homogeneous, and its selection effects can therefore be modeled in detail. The available resulting dataset of binaries will be larger and more complete than any other binarity survey in OB~associations, star clusters, and the field star population thus far.

\chapter[Close visual companions to A star members of Scorpius OB2]{A near-infrared adaptive optics search for close visual companions to A star members of Scorpius OB2} \label{chapter: adonis}

\begin{center}
M.B.N. Kouwenhoven, A.G.A. Brown, H. Zinnecker, L. Kaper, \& S.F. Portegies Zwart

\vspace{0.2cm}
{\it Astronomy \& Astrophysics}, 430, 137 (2005)
\end{center}

% ====================================================================
% ====================================================================
% ====================================================================
% ==ABSTRACT==========================================================
% ====================================================================
% ====================================================================
% ====================================================================

\section*{Abstract}

We present the results of a near-infrared adaptive optics survey
with the aim to detect close companions to {\it Hipparcos} members in the
three subgroups of the nearby OB~association Sco~OB2: Upper Scorpius (US),
Upper Centaurus Lupus (UCL) and Lower Centaurus Crux (LCC). We have targeted
199 A-type and late B-type stars in the $K_S$ band, and a subset also in the
$J$ and $H$ band. We find 151 stellar components other than the target
stars. A brightness criterion is used to separate these components into 77
background stars and 74 candidate physical companion stars. 
Out of these 74 candidate companions, 41 have not
been reported before (14 in US; 13 in UCL; 14 in LCC). 
The angular separation between primaries and observed companion stars 
ranges from $0.22''$ to $12.4''$. At the mean distance of Sco~OB2 (130~pc) 
this corresponds to a projected separation of $28.6$~AU to $1612$~AU. 
Absolute magnitudes are derived for all primaries and observed companions
using the parallax and interstellar extinction for each star individually.
For each object we derive the mass from $K_S$, assuming an age of 5~Myr for the
US subgroup,
and 20~Myr for the UCL and LCC subgroups.
Companion star masses range 
from $0.10~{\rm M}_\odot$ to $3.0~{\rm M}_\odot$.
The mass ratio distribution follows $f_q(q) = q^{-\Gamma}$ with $\Gamma=0.33$, 
which excludes random pairing.
No close ($\rho \leq 3.75''$) companion stars or background stars 
are found in the magnitude range $12~{\rm mag}\leq K_S \leq 14~{\rm mag}$.
The lack of stars with these properties
cannot be explained by low-number statistics, and may imply a lower
limit on the companion mass of $\sim 0.1~{\rm M}_\odot$. 
Close stellar components with $K_S>14~{\rm mag}$ are observed. If these
components are very low-mass companion stars, a gap
in the companion mass distribution might be present.
The small number of close low-mass companion stars
could support the embryo-ejection formation scenario for brown dwarfs.
Our findings are compared
with and complementary to visual, spectroscopic, and astrometric data on
binarity in Sco~OB2. 
We find an overall companion star fraction of 0.52 in this association. 
This is a lower limit since the data from the observations and
from literature are
hampered by observational biases and selection effects.
This paper is the first step toward our goal 
to derive the primordial binary population in Sco~OB2.

% ====================================================================
% ====================================================================
% ====================================================================
% ==INTRODUCTION======================================================
% ====================================================================
% ====================================================================
% ====================================================================

\section{Introduction}

Duplicity and multiplicity 
properties of newly born stars are among the most important clues
to understanding the process of star formation \citep{blaauw1991}.
Observations of star forming regions over the past two decades have revealed
two important facts: (1) practically all (70--90\%) stars form in clusters 
\citep[e.g.,][]{LL2003} and, (2) within these clusters most stars are 
formed in binaries \citep{mathieu1994}.
Consequently, the star formation community has
shifted its attention toward understanding the formation of multiple systems
--- from binaries to star clusters --- by means of both observations and
theory.

The observational progress in studies of very young embedded as well as
exposed star clusters is extensively summarized in the review by
\cite{LL2003}. On the theoretical side the numerical simulations of cluster
formation have become increasingly sophisticated, covering the very earliest
phases of the development of massive dense cores in giant molecular clouds
\citep[e.g.,][]{klessen2000}, 
the subsequent clustered formation of
stars and binaries \citep[e.g.,][]{bate2003}, 
as well as the early evolution of the binary population
during the phase of gas expulsion from the
newly formed cluster \citep{kroupaaarseth2001}.
At the same time numerical
simulations of older exposed clusters have become more realistic
by incorporating detailed stellar and binary evolution effects in N-body
simulations \citep[e.g.,][]{ecology4}. This has led to the creation of a number
of research networks that aim at synthesizing the modeling and observing
efforts into a single framework which covers all the stages from the formation
of a star cluster to its eventual dissolution into the Galactic field, an
example of which is the MODEST collaboration \citep{modest1,modest2}.

Such detailed models require stringent observational constraints in the form
of a precise characterization of the stellar content of young
clusters. Investigations of the stellar population in young clusters have
mostly focused on single stars. However, as pointed out by \cite{larson2001},
single stars only retain their mass from the time of formation whereas
binaries retain three additional parameters, their mass ratio, angular
momentum and eccentricity. Thus, the properties of the binary population can
place much stronger constraints on the physical mechanisms underlying the star
and cluster formation process.

Ideally, one would like an accurate description of the `primordial' binary
population. This population was defined by \cite{brown2001} as ``the
population of binaries as established just after the formation process itself
has finished, i.e., after the stars have stopped accreting gas from their
surroundings, but before stellar or dynamical evolution have had a chance to
alter the distribution of binary parameters''. This definition is not entirely
satisfactory as stellar and dynamical evolution will take place already during
the gas-accretion phase.

Here we revise this definition to: 
{\em``the population of binaries as established
just after the gas has been removed from the forming system, i.e., when the
stars can no longer accrete gas from their surroundings''}.
This refers to the
same point in time, but the interpretation of the `primordial binary
population' is somewhat different. The term now refers to the point in time
beyond which the freshly formed binary population is affected by {\em
stellar/binary evolution and stellar dynamical effects only}. Interactions with a surrounding gaseous medium no longer take place.

From the point of view of theoretical/numerical models of star cluster
formation and evolution the primordial binary population takes on the
following meaning. It is the final population predicted by simulations of the
formation of binaries and star clusters, and it is the initial population for
simulations that follow the evolution of star clusters and take into account
the details of stellar dynamics and star and binary evolution. 
The primordial
binary population as defined in this paper 
can be identified with the initial binary population,
defined by \cite{kroupa1995b} as the binary population at the 
instant in time when the pre-main-sequence eigenevolution
has ceased, 
and when dynamical evolution of the stellar cluster becomes effective.
\cite{kroupa1995a,kroupa1995c} infers the
initial binary population by the so-called inverse dynamical population
synthesis technique.
This method involves the evolution of simulated stellar
clusters forward in time for different initial binary populations, where the
simulations are repeated until a satisfactory fit with the present day binary
population is found.

Our aim is to obtain a detailed observational characterization of the
primordial binary population as a function of stellar mass, binary parameters,
and (star forming) environment. The most likely sites where this population
can be found are very young (i.e., freshly exposed), 
low density stellar groupings containing a
wide spectrum of stellar masses. 
The youth of such a stellar grouping implies that stellar
evolution will have affected the binary parameters of only a handful of the
most massive systems.
The low stellar densities guarantee that
little dynamical evolution has taken place
after the gas has been removed from the forming system.
These constraints naturally
lead to the study of the local ensemble of OB associations. 
Star clusters are older and have a higher density than OB associations
and are therefore less favorable.
For example, in the Hyades and Pleiades,
the binary population has significantly changed 
due to dynamical and stellar evolution \citep{kroupa1995c}.
Note that OB associations may start out as dense clusters 
\citep{kroupaaarseth2001,kroupaboily2002}.
However, they rapidly expel their gas and evolve into low-density systems. 
This halts any further dynamical evolution of the binary population.
\cite{brown1999} define OB associations as 
``young ($\la$\,50~Myr) stellar groupings
of low density ($\lesssim 0.1 {\rm M}_\odot {\rm pc}^{-3}$) 
---~such that they are likely to be unbound~--- 
containing a significant population of B stars.''
Their projected dimensions range from
$\sim$\,10 to $\sim$\,100~pc and their mass spectra cover the mass range from
O~stars all the way down to brown dwarfs. 
For reviews on OB
associations we refer to \cite{blaauw1991} and \cite{brown1999}.  Thanks to
the {\it Hipparcos} Catalogue \citep{esa1997} the stellar content of the nearby OB
associations has been established with
unprecedented accuracy to a completeness limit of $V\,\sim\,10.5~{\rm mag}$, or about
1~M$_\odot$ for the stars in the nearest associations 
\citep[][]{dezeeuw1999, hoogerwerf2000}. Beyond this limit the population of
low-mass pre-main-sequence stars has been intensively studied in, e.g., the
Sco~OB2 association \citep[][]{preibisch2002,mamajek2002}.

The latter is also the closest and best studied of the OB associations in the
solar vicinity and has been the most extensively surveyed for binaries. 
The association consists of three
subgroups: Upper Scorpius (US, near the Ophiuchus star forming region, at a
distance of 145\,pc), Upper Centaurus Lupus (UCL, 140\,pc) and Lower Centaurus
Crux (LCC, 118\,pc). The ages of the subgroups range from 5 to $\sim20$\,Myr
and their stellar content has been established from OB stars down to brown
dwarfs
\citep[for details see][]{degeus1989, dezeeuw1999, debruijne1999,
hoogerwerf2000, mamajek2002, preibisch2002}. 
Surveys targeting the binary
population of Sco~OB2 include the radial-velocity study by \cite{levato1987},
the speckle interferometry study by \cite{koehler2000}, and 
the adaptive optics study by 
\cite{shatsky2002}. Because of their brightness many of the B-star members of
Sco OB2 have been included in numerous binary star surveys, in which a
variety of techniques have been employed. The literature data on the binary
population in Sco~OB2 is discussed by \cite{brown2001} and reveals that
between 30 and 40 per cent of the {\it Hipparcos} members of Sco~OB2 are known to be
binary or multiple systems. However, these data are incomplete and suffer from
severe selection effects, which, if not properly understood, will prevent a
meaningful interpretation of the multiplicity data for this association in
terms of the primordial binary population. The first problem can be addressed
by additional multiplicity surveys of Sco~OB2. In this paper we report on
our adaptive optics survey of Sco~OB2 which was aimed at surveying all the
{\it Hipparcos} members of spectral type A and late B, 
using the ADONIS instrument on the
3.6m telescope at ESO, La Silla.

We begin by describing in Sect.~\ref{sec:theaosurvey} our observations, the
data reduction procedures and how stellar components other than the target
stars were detected in our images. These components have to be separated into
background stars and candidate physical companions. We describe how this was
done in Sect.~\ref{sec: backgroundstars}. The properties of the physical
companions are described in Sect.~\ref{sec: properties}. In Sect.~\ref{sec:
literaturedata} we discuss which of the physical companions are new by
comparing our observations to data in the literature and we provide updated
statistics of the binary population in Sco~OB2. We summarize this work in
Sect.~\ref{sec:conclusions} and outline the next steps of this study which are
aimed at addressing in detail the problem of selection biases associated with
multiplicity surveys and subsequently characterizing the primordial binary
population.

% Sco OB2 details

% ====================================================================
% ====================================================================
% ====================================================================
% ==OBSERVATIONS AND DATA REDUCTION===================================
% ====================================================================
% ====================================================================
% ====================================================================

\section{Observations and data reduction}
\label{sec:theaosurvey}

\subsection{Definition of the sample} \label{sec: sample}

A census of the stellar content of the three subgroups of Sco~OB2 based on
positions, proper motions, and parallaxes was presented by \cite{dezeeuw1999}.
Our sample is extracted from their list of \textit{Hipparcos} member stars and
consists of A and B stars. In order to avoid saturating the detector
during the observations, we were restricted to observing stars fainter than
$V\sim6~{\rm mag}$. The sample therefore consists mainly of late-B and A stars which at
the mean distance of Sco~OB2 (130 pc) corresponds to stars with $6~{\rm mag}\lesssim
V\lesssim 9~{\rm mag}$, which translates to very similar limits in $K_S$. We observed
199 stars (listed in the appendix) from the resulting \textit{Hipparcos} member sample. 
Not all late-B and A member stars could be observed due to time limitations.
The distribution of target stars over spectral type is: 83~(157)~B, 113~(157)~A,
2~(138)~F, and 1~(48)~G, where the numbers in brackets denote the total
number of Sco~OB2 {\it Hipparcos} members of the corresponding spectral type.

\subsection{Observations} \label{sec: observations}

The observations were performed with the ADONIS/SHARPII+ system 
\citep{beuzit1997} on the ESO
3.6~meter telescope at La Silla, Chile. This is an adaptive optics system
coupled to an infrared camera with a NICMOS3 detector array. The field of
view of the $256 \times 256$ pixel detector 
is $12.76'' \times 12.76''$. The plate scale
is $0.0495'' \pm 0.0003''$ per pixel, and the orientation of the
field is $-0^\circ.20 \pm 0^\circ.29$ (measured from North to
East)\footnote{The astrometric calibrations are based on observations of the
  astrometric reference field ($\theta$~Ori) for ADONIS/SHARPII+/NICMOS3 (see
  \texttt{http://www.imcce.fr/priam/adonis}). Mean values for the period
  9/1999 to 12/2001 are used. Plate scale and position angle differences
  between the two observing runs fall within the error bars.}. 
Wavefront sensing was performed directly on the target stars.

\begin{SCfigure}[][btp]
  \centering
  \includegraphics[width=0.6\textwidth,height=!]{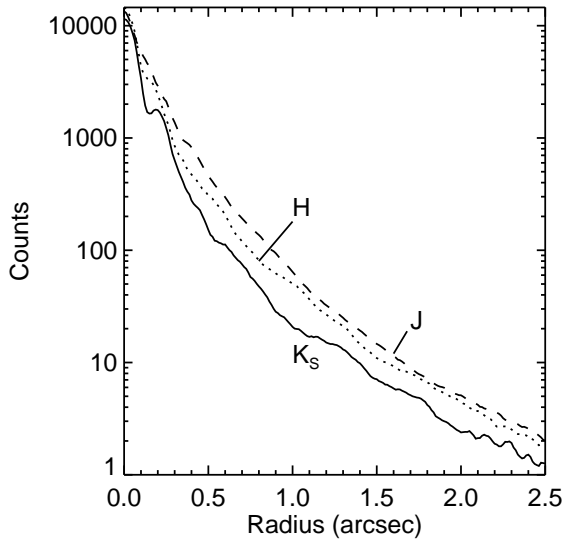}
  \caption{The radial profile of the PSF for the target star
    HIP53701. The observations are obtained using the 
    ADONIS/SHARPII+ system on June 7, 2001 in $J$, $H$, and $K_S$.
    The corresponding Strehl ratios for these observations
    are 18\% in $J$, 26\% in $H$
    and 31\% in $K_S$.
    \label{figure: psfprofile_adonis} }
\end{SCfigure}

Our observing campaign was carried out in two periods: a period from May 31 to
June 4, 2000, and one from June 5 to 9, 2001. Out of the eight observing
nights about 1.5 were lost due to a combination of bad weather and technical
problems. Each target star is observed in the $K_S$ ($2.154~\mu$m) band. A
subset is additionally observed in the $J$ ($1.253~\mu$m) and $H$
($1.643~\mu$m) band. In our search for companion stars, near-infrared
observations offer an advantage over observations in the optical. In the
near-infrared the luminosity contrast between the primary star and its (often
later type) companion(s) is lower, which facilitates the detection of faint
companions. The performance of the AO system is measured by the Strehl ratio, 
which is the ratio between the maximum of the ideal point spread
function (PSF) of the system and the measured maximum of the PSF. Typical
values for the Strehl ratio in our observations were 5--15\% in $J$, 15--20\% in $H$ and
20--35\% in $K_S$ (Figure~\ref{figure: psfprofile_adonis}).

\begin{SCfigure}[][btp]
  \centering
  \includegraphics[width=0.5\textwidth,height=!]{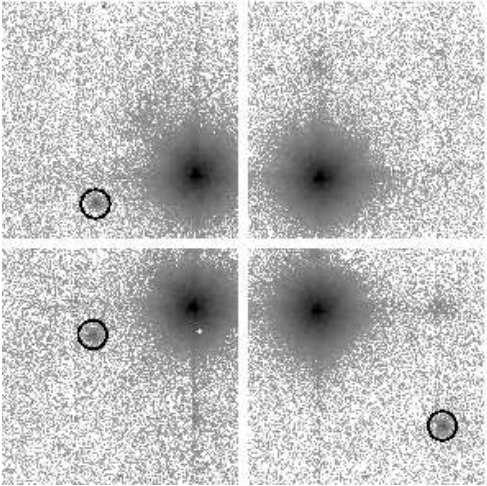}
  \caption{An example illustrating the observing strategy.
  Observations are obtained with ADONIS/SHARPII+ system in $J$, $H$, 
  and $K_S$.
  Four observations of each star are made, each observed with a
  different offset, in order to maximize the field of view. The frame size
  of the individual quadrants is $12.76'' \times 12.76''$. Combination of the quadrants
  results in an effective field of view of $19.1'' \times 19.1''$.
  This figure shows $K_S$ band
  observations of the A0~V star HIP75957. Two stellar components at
  angular separations of $5.6''$ and $9.2''$ 
  (indicated with the black circles) 
  are detected.
  These components are visible to the left (East)
  and bottom-right (South-West) of HIP75957 and are probably background 
  stars. All other features are artifacts.
  \label{figure: mosaic}}
\end{SCfigure}

Each star was observed at four complementary pointings in order to enhance the
sensitivity of the search for close companions and to maximize the available
field of view (Figure~\ref{figure: mosaic}). The two components of the known
and relatively wide binaries HIP77315/HIP77317 and HIP80324 
were observed individually
and combined afterwards. For all targets each observation consists of 4 sets of
30 frames (i.e. 4 data cubes). The integration time for each frame was
$200-2000$ ms, depending on the brightness of the source. After each
observation, thirty sky frames with integration times equal to those of the
corresponding target star were taken in order to measure background
emission. The sky frames were taken 10 arcmin away from the target star and
care was taken that no bright star was present in the sky frame. Nevertheless,
several sky frames contain faint background stars. The error on the target
star flux determination due to these background stars is less than $0.06\%$ of
the target star flux and therefore negligible for our purposes. Dark frames
and flatfield exposures (using the internal lamp) were taken and standard
stars were regularly observed.

We did not use the coronograph in our setup. The coronograph obscures most of
the target star's flux, which complicates flux calibration of the target stars
and their companions. Furthermore, the use of the coronograph prevents
detection of close companions with angular separations
less than $1''$. The advantage of the
coronograph is that one can detect fainter objects, which would otherwise
remain undetected due to saturation of the target stars. However, the large
majority of these faint components are likely background stars (see
\S\ref{sec: backgroundstars}).

\subsection{Data reduction procedures} \label{sec: datareduction}

The primary data reduction was performed with the ECLIPSE package
\citep{devillard1997}. Information about the sensitivity of each pixel is derived
from the (dark current subtracted) flatfield images with different exposure
times. The number of counts as a function of integration time is linearly
fitted for each pixel. The results are linear gain maps (pixel sensitivity
maps) for each filter and observing night. Bad pixel maps for each observing
run and filter are derived from the linear gain maps: each pixel with a
deviation larger than $3\sigma$ from the median is flagged as a bad pixel. The
average sky map is subtracted from the corresponding standard star and target
star data cubes. The resulting data cubes are then corrected using the dark
current maps, the linear gain maps, and bad pixel maps.

Image selection and image combination was done with software that was written
specifically for this purpose. 
Before combination, frames with low Strehl ratios with respect to the
median were removed.  Saturated frames (i.e. where the stellar flux
exceeds the SHARPII+ linearity limit) and frames with severe wavefront
correction errors were also removed. 
The other frames were combined by taking the median.
Typically, about 2 to 3 out of 30 frames were rejected
before combination.

\subsection{Component detection} \label{sec: componentdetection}

We used the STARFINDER package \citep{diolaiti2000} to determine the position
and instrumental flux of all objects in the images. For each component other
than the target star we additionally measured the correlation between the PSF
of the target star and that of the component. Components with peak fluxes less
than 2 to 3 times the noise in the data, or with correlation coefficients less
than $\sim 0.7$, were considered as spurious detections.

Since the PSF of the target star is generally not smooth, it was sometimes
difficult to decide whether a speckle in the PSF halo of the star is a stellar
component, or merely a part of the PSF structure. In this case four
diagnostics were used to discriminate between stellar components and PSF
artifacts. First, a comparison between the images with the target star located
in four different quadrants was made. Objects that do not appear in all
quadrants where they {\it could} be detected were considered
artifacts. Second, a similar comparison was done with the individual
uncombined raw data frames. Third, a radial profile was fit to the PSF and
subtracted from the image to increase the contrast, so that the stellar
component becomes more obvious. And finally, a comparison between the PSF of
the target star and the PSF of the star previous or next in the program with a
similar position on the sky was made by blinking the two images. These two
PSFs are expected to be similar and therefore faint close companion stars can
be detected.

\begin{SCfigure}[][btp]
  \includegraphics[width=0.5\textwidth,height=!]{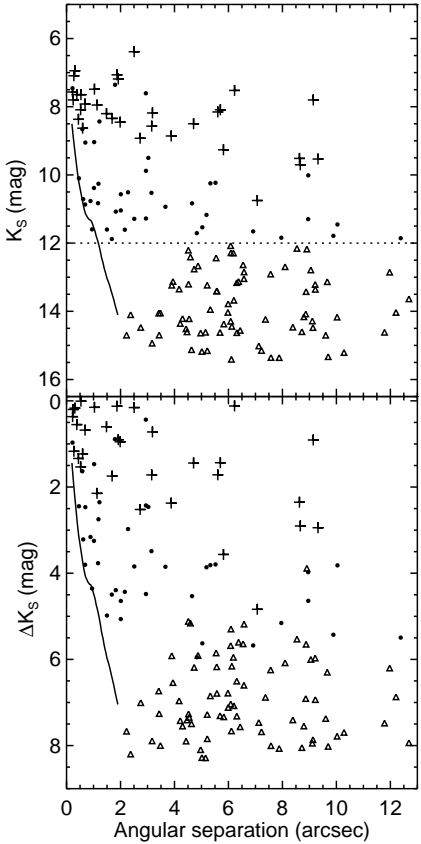}   
  \caption{Angular separation between the target stars and observed components as a
  function of the component's $K_S$ magnitude ({\it top}) and component's
  $K_S$ magnitude relative to that of the target star ({\it bottom}). 
  The figures show 
  the 41 new (dots) and 33 previously known (plusses) 
  companion stars ($K_S < 12~{\rm mag}$),
  respectively.
  The triangles are the 77 background stars 
  ($K_S > 12~{\rm mag}$; \S~\ref{sec: backgroundstars}). The solid curves show the
  estimated detection limit as a function of angular separation. 
  These
  figures clearly demonstrate
  that faint close companions are not detected because of
  the extended PSF halo of the target star. 
  The effective field of view is $19.1'' \times 19.1''$.
  Companion stars with an angular separation larger than
  $\frac{1}{2}\sqrt{2} \cdot 19.1'' = 13.5''$ cannot be detected.
  For small
  angular separation ($\rho < 3.5''$) no faint ($K_S > 12~{\rm mag}$) components
  are found. This could indicate a gap in the companion mass function, or a
  lower mass limit for companion stars at small separation 
  (see \S~\ref{sec: masses}). 
  \label{figure: detectionlimits_adonis}}
\end{SCfigure}

In total we detect 151 stellar components other than the target stars.
A significant fraction of these 151 components are background
stars (see \S~\ref{sec: backgroundstars}). All other components
are companion stars.
Figure~\ref{figure: detectionlimits_adonis} shows the $K_S$ magnitude of
the companion and background stars as
a function of angular separation $\rho$ between the target star
and the companion or background stars. The same plot also shows
$\Delta K_S = K_{S,{\rm component}}-K_{S,{\rm primary}}$ as a
function of $\rho$. Previously unknown companion stars
(see \S~\ref{sec: newcompanions}) and known companions 
are indicated with the dots and 
the plusses, respectively. The background stars
(see \S~\ref{sec: backgroundstars}) are represented with
triangles. This figure shows that the detection probability of
companion stars increases with increasing angular separation and decreasing
magnitude difference, as expected. The lower magnitude limit for component
detection is determined by the background noise; the lower limit for the
angular separation is set by the PSF delivered by the AO system. Note that not
all detected components are necessarily companion stars. A simple criterion is
used to separate the companion stars ($K_S < 12~{\rm mag}$) and the background stars
($K_S > 12~{\rm mag}$). See \S~\ref{sec: backgroundstars} for the motivation of this
choice.

The estimated detection limit as a function of angular separation and
companion star flux is represented in Figure~\ref{figure: detectionlimits_adonis} with a
solid curve. The detection limit is based on $K_S$ band measurements of
simulated companion stars around the B9~V target star HIP65178. In the
image we artificially added a second component, by scaling and shifting a copy
of the original PSF, thus varying the desired flux and angular separation of
the simulated companion. We sampled 30 values for the companion flux linearly
in the range $9.35~{\rm mag} \leq K_S \leq 15.11~{\rm mag}$, 
and 40 values for the angular
separation between primary and companion linearly in the range $0.05''
\leq \rho \leq 2''$. For each value of the angular separation we determined
the minimum flux for which the companion star is detectable (i.e. with a peak
flux of $2-3$ times the noise in the data). In order to minimize biases, we
repeated this procedure for four different position angles and averaged the
results. The standard deviation of the detection limit is $0.07''$ at a
given magnitude and $0.35~{\rm mag}$ at a given angular separation.  Similar
simulations with other target stars (e.g., HIP58859, Strehl 
ratio = 8\%; HIP76048, Strehl ratio =
27\%) show that the detection limit for the observations with different Strehl ratios
is comparable to that of HIP65178 (Strehl ratio = 40\%). All target stars are of
spectral type A or late-B and have comparable distances. The detection limit
shown in Figure~\ref{figure: detectionlimits_adonis} is therefore representative for
our sample.

\begin{SCfigure}[][btp]
  \centering
  \includegraphics[width=0.5\textwidth,height=!]{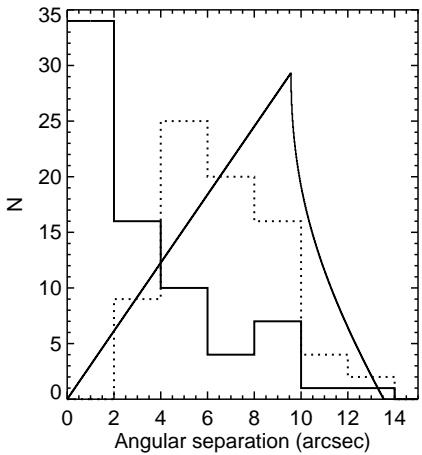} 
  \caption{The number of objects per angular separation bin versus 
  angular separation (bin size $2''$). The
  distribution of presumed background stars (dotted histogram) is clearly
  different from that of the companion stars (solid histogram). The expected
  angular separation distribution 
  for background stars (in units of stars per bin size) is represented 
  with the solid curve. The curve is normalized such that the
  number of stars corresponds to the expected number of background stars
  with $12~{\rm mag}<K_S<16~{\rm mag}$ 
  (cf. Figure~\ref{figure: expectedbackground}).
  The excess of close ($\rho \lesssim 6''$) companions
  could be caused by faint companion stars that are misclassified
  as background stars. 
  Objects with $\rho < 0.1''$ are too deep in the PSF halo of the
  primary to be detectable. Objects with $\rho > 15''$ cannot be
  measured due to the limited field of view of the ADONIS/SHARPII+ system. 
  \label{figure: separationhisto} }
\end{SCfigure}

\subsection{Photometry}

Near-infrared magnitudes for the observed standard stars are taken from
\cite{vanderbliek1996}, \cite{carter1995}, and the 2MASS catalog. All magnitudes are
converted to the 2MASS system using the transformation formula's of
\cite{carpenter2001}. The standard stars HD101452, HD190285 and HD96654 were
found to be double. The fluxes for the primary and secondary of the standard
stars HD101452 and HD96654 are added for calibration, since the secondaries
are unresolved in \cite{vanderbliek1996}. For HD190285, only the flux of the primary
is used for calibration since the secondary is resolved in the 2MASS catalog.

The zeropoint magnitudes are calculated for each observing night and each
filter individually. There are non-negligible photometric zeropoint differences
between the four detector quadrants. Application of the reduction procedure to simulated
observations (but using the ADONIS/SHARPII+ dark frames and flatfield exposures) 
also shows this zeropoint offset, unless the flatfielding was
omitted. Hence, the zeropoint offset between the four quadrants is caused by
large-scale variations in the flatfield exposures. The four detector quadrants
were consequently calibrated independently. 

Since the distribution of observed standard stars over airmass is not
sufficient to solve for the extinction coefficient $k$, we use the mean
extinction coefficients for La~Silla\footnote{see
\texttt{http://www.ls.eso.org/lasilla/sciops/ntt/sofi}}: 
$k_J=0.081~{\rm mag/airmass}$, $k_H=0.058~{\rm mag/airmass}$ and
$k_{K_S}=0.113~{\rm mag/airmass}$. The standard stars
have similar spectral type as the target stars, which allows us to neglect the
color term in the photometric solution.

% ====================================================================
% ====================================================================
% ====================================================================
% ==BACKGROUND=STARS==================================================
% ====================================================================
% ====================================================================
% ====================================================================

\section{Background stars} \label{sec: backgroundstars}

\begin{figure}[btp]
  \centering
  \includegraphics[width=0.6\textwidth,height=!]{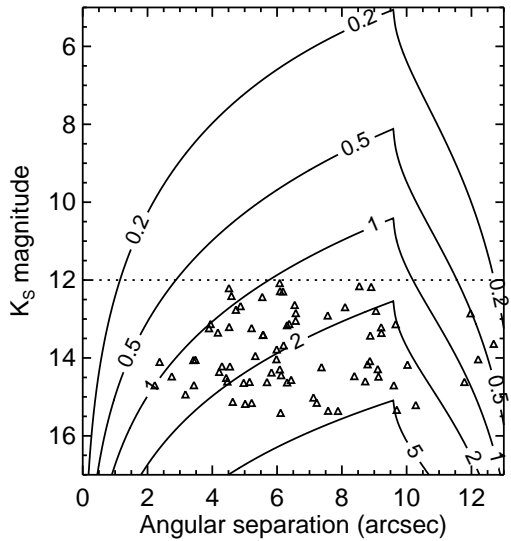}
  \caption{The expected background star distribution 
    as a function of angular separation
    and the $K_S$ magnitude distribution
    of background stars as observed by \cite{shatsky2002}. 
    The distribution of
    background stars over the field of view is assumed to be homogeneous.
    The triangles represent the observed stellar objects that we
    classify as background stars. 
    The contours indicate the background star density in units of
    arcsec$^{-1}$mag$^{-1}$. The distribution is normalized, so that for $14~{\rm mag}
    \leq K_S \leq 15~{\rm mag}$ and $5'' \leq \rho \leq 13''$ the number of
    background stars equals 19, i.e. the observed number of background
    stars with these properties. Observational biases are not considered. 
    The dashed line represents our criterion to separate
    companion stars ($K_S \leq 12~{\rm mag}$) and background stars 
    ($K_S>12~{\rm mag}$). 
    The excess of background stars at small angular separations 
    could be caused by faint companion stars that are misclassified
    as background stars.
    \label{figure: expectedbackground}}
\end{figure}

\begin{SCfigure}[][btp]
  \includegraphics[width=0.6\textwidth,height=!]{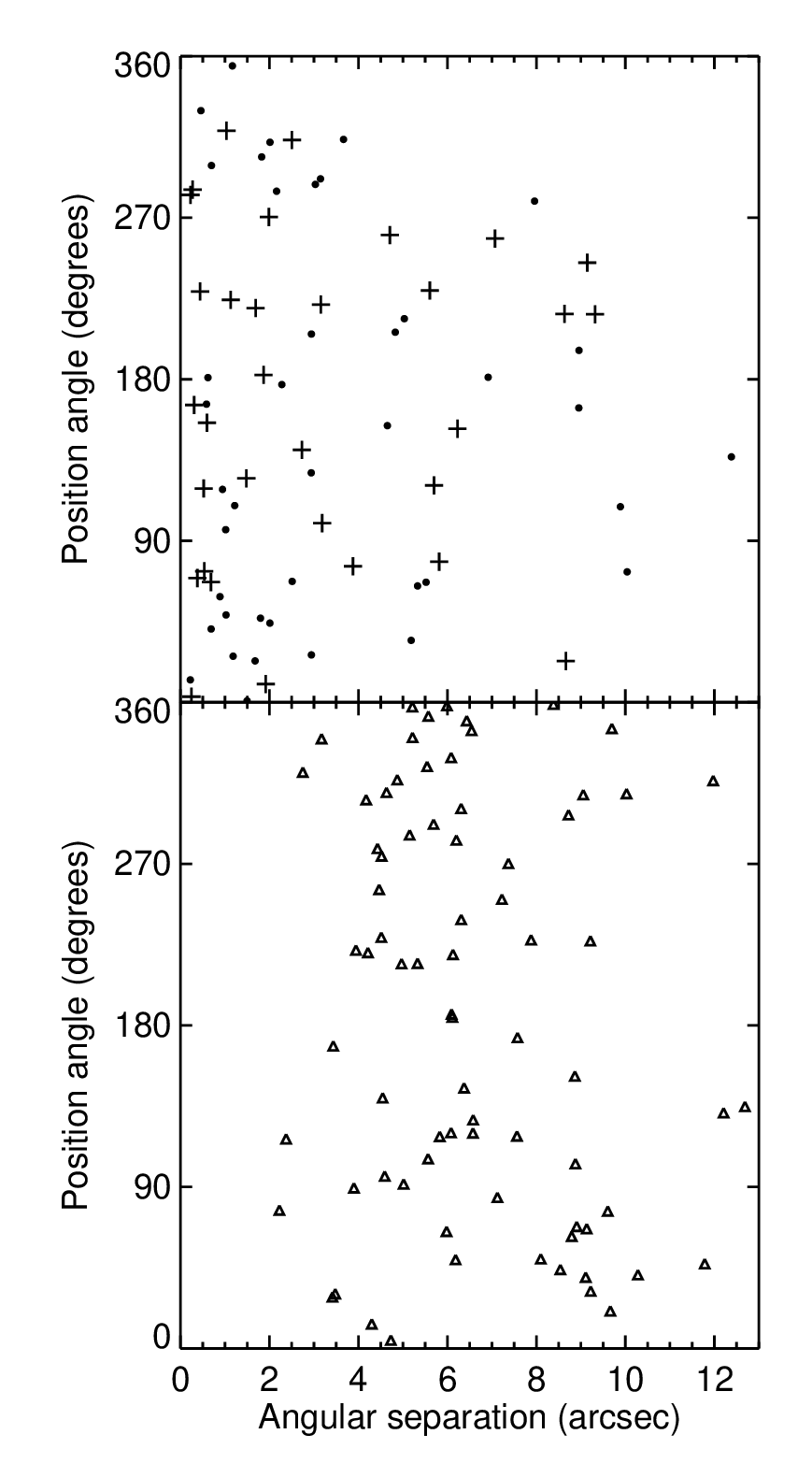} 
  \caption{ {\it Top:} the distribution of presumed companion stars 
    over angular separation and position angle.
    Previously known companions and new companions are indicated with the plusses and 
    dots, respectively.
    {\it Bottom:} the distribution of background stars 
    over angular separation and position angle.
    The distribution of companion stars and background stars over position angle
    is random for angular separation smaller than $9.6''$ (see text).
    The companion stars are more 
    centrally concentrated than the background stars.
    \label{figure: compsfield} }
\end{SCfigure}

Not all of the 151 stellar components that we find around the target
stars are companion stars. 
The chance that a foreground or background star is present in the field is not
negligible. There are several techniques that can be applied to determine
whether a detected component is a background star or a probable companion
star.

Separation of the objects based on the position of the component in the color-magnitude
diagram (CMD) is often applied to discriminate background stars from companion
stars \citep[see, e.g., ][]{shatsky2002}. Given a certain age of the
population, a companion star should be found close to the isochrone that
corresponds to that age. When age and distance spread are properly taken into
account, this method provides an accurate estimate of the status of the
object. We are unable to apply this technique to our data since
only a small fraction of our observations is multi-color (see \S~\ref{sec:
hrdiagram} and Figure~\ref{figure: 3hr}).

\begin{figure}[btp]
  \includegraphics[width=\textwidth,height=!]{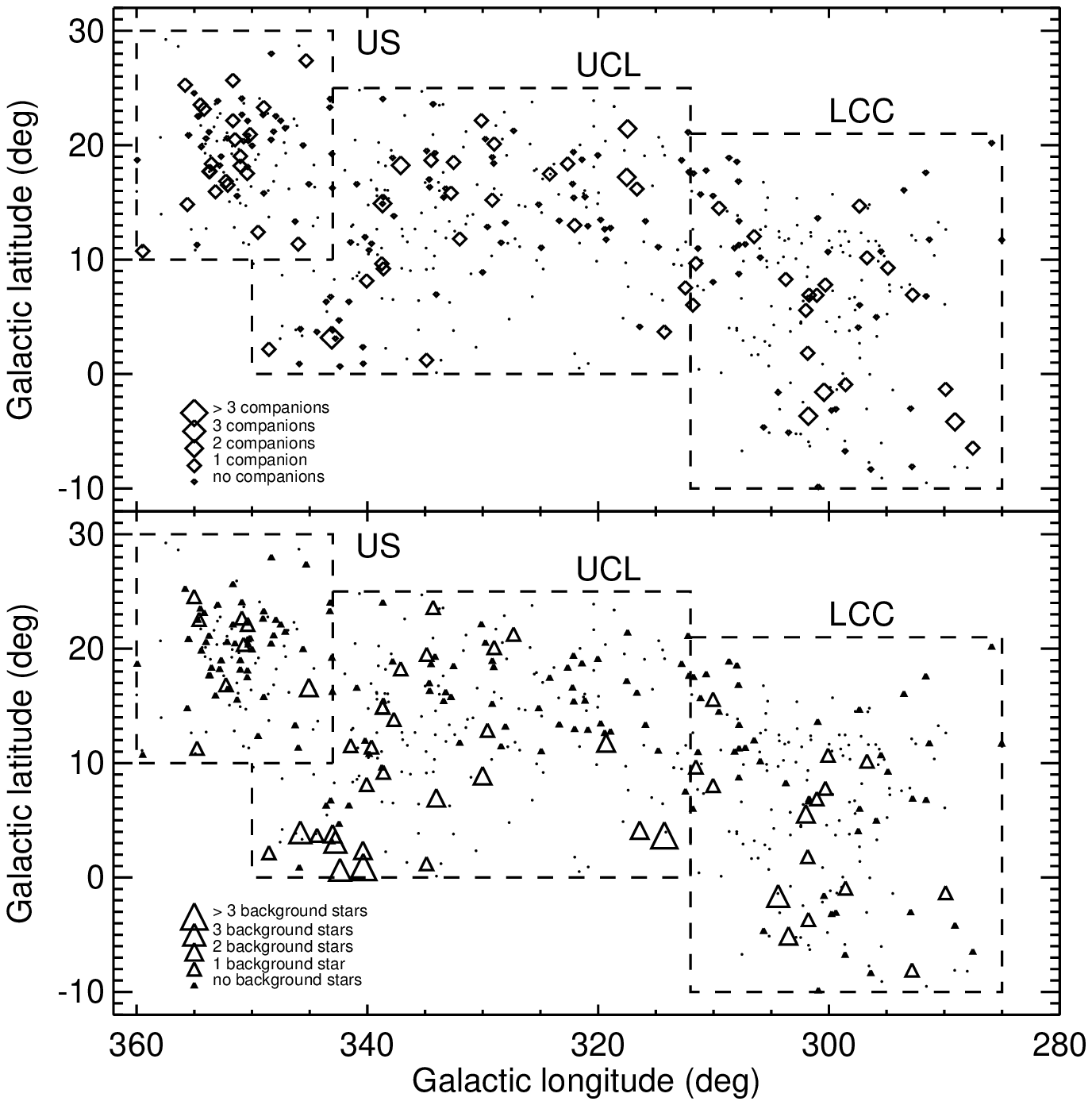}  
  \caption{The distribution of 74 detected candidate companions ({\it top}) and
  the 77 presumed background stars ({\it bottom}) over
  Galactic coordinates. The number of companions and background stars found
  per target star is indicated by the size of the diamonds and triangles,
  respectively. {\it Hipparcos} member stars that were not observed are shown
  as dots. The borders of the three subgroups of Sco~OB2 (Upper Scorpius,
  Upper Centaurus Lupus, and Lower Centaurus Crux) are indicated by the dashed
  lines. The background stars are concentrated towards the location of 
  the Galactic plane (see also Figure~\ref{figure: galacticplanehisto}).
  \label{figure: galacticplane}}
\end{figure}

Following \cite{shatsky2002} we use a simple brightness criterion
to separate the companion stars ($K_S \leq 12~{\rm mag}$) and the background stars ($K_S >
12~{\rm mag}$). At the distance of Sco~OB2, an M5 main sequence star
has approximately $K_S=12~{\rm mag}$.
The three subgroups of Sco~OB2, US, UCL, and LCC, are located at distances of 145,
140, and 118~pc, respectively. We assume an age of 5~Myr for US
\citep{degeus1989, preibisch1999}, 
and 20~Myr for UCL and LCC \citep{mamajek2002}.
Using the isochrones described in \S~\ref{sec: masses}, we find
that stars with $K_S=12~{\rm mag}$ correspond to stars with masses of 
0.08~$~{\rm M}_\odot$, 0.18~$~{\rm M}_\odot$, and
0.13~$~{\rm M}_\odot$ for US, UCL, and LCC, respectively.
Using the $K_S=12~{\rm mag}$ criterion 
we conclude that 77 out of 151 components are likely
background stars. The remaining 74 components are thus candidate companion stars.

Although this $K_S=12~{\rm mag}$ criterion works well for stars,
it is not very accurate below the hydrogen burning limit.
\cite{martin2004} found 28 candidate brown dwarf members of the US
subgroup using $I$, $J$, and $K$ photometry from the DENIS database. All
members have $10.9~{\rm mag} \leq K \leq 13.7~{\rm mag}$ 
and their spectral types are in the range
M5.5--M9. According to the $K_S=12~{\rm mag}$ criterion, 
19 of these brown dwarfs would be classified
as companion stars and 8 as background stars 
(one brown dwarf does not have a $K$ magnitude entry).
A significant fraction of possible brown dwarf companion stars
to our observed target stars
would be classified as background stars.

In this study we cannot determine with absolute certainty 
whether a detected component is a companion star or a 
background star. For example, a foreground star with 
$K_S > 12$~mag will not be excluded by the criterion described 
above. 
However, the differences between the companion 
star distribution and the background star distribution can be used to 
find out if the results are plausible or not.
For this purpose, we derive
the expected distribution of background stars per unit of angular 
separation $P(\rho)$, position angle $Q(\varphi)$,
and $K_S$ band magnitude $R(K_S)$ for the background stars.

% P theory
Background stars are expected to be uniformly distributed over the image field with a surface density of $\Sigma$ stars per unit of solid angle.  The number of observed background stars $P(\rho)$ per unit of angular separation $\rho$ is given by
\begin{equation} \label{eq: theobackgroundgeneral}
P(\rho) = \frac{{\rm d}}{{\rm d}\rho} \int_{\rm \Omega_\rho} \Sigma \  d\Omega_\rho \ ,    
\end{equation}
where $\Omega_\rho$ is the part of the field of view contained within a circle with radius $\rho$ centered on the reference object (the target star). In calculating the expected $P(\rho)$ for our observations, we assume that the reference object is located in the center of the field of view. For a square field of view with dimension $L$, Equation~\ref{eq: theobackgroundgeneral} then becomes
\begin{equation} \label{eq: theobackground}
  P(\rho) \propto \left\{ 
  \begin{array}{llcccccc}	
    2 \pi \rho                               & {\rm for} & 0   & < & \rho &<& L/2&\\
    8\rho \left(\frac{\pi}{4} - \arccos \frac{L}{2\rho}\right)  
    & {\rm for} & L/2 & < & \rho &<& L/\!\sqrt{2}&.\\
    0                                        & {\rm for} &     &   & \rho &>& L/\!\sqrt{2}&\\
  \end{array} \right.
\end{equation}
For our observing strategy (Figure~\ref{figure: mosaic})
we have $L=\frac{3}{2} \cdot 12.76'' = 19.14''$.
For $\rho \leq L/2 = 9.6''$ the background star density for
constant $K_S$ is proportional to $\rho$, after which it decreases down to
$0$ at $\rho = L/\!\sqrt{2}  = 13.6''$. 
Here we have made the assumption that 
each target is exactly in the center of each quadrant.

% Q theory
We can also calculate $Q(\varphi)$, the expected 
distribution of background stars 
over position angle $\varphi$.
For $\rho \leq L/2$,  $Q(\varphi)$ is expected
to be random. For $L/2 < \rho \leq L/\!\sqrt{2}$ one expects
$Q(\varphi)=L^2/8 \cos^2\varphi$ for 
$0^\circ \leq \varphi < 45^\circ$~\footnote{This formula is 
also valid for $90^\circ \leq \varphi < 135^\circ$,
$180^\circ \leq \varphi < 125^\circ$, and $270^\circ \leq \varphi < 315^\circ$. 
$Q(\varphi)=L^2/8 \sin^2\varphi$ for all other position angles between $0^\circ$ 
and $360^\circ$.}. 
$Q(\varphi)$ is undefined for $\rho>L/\!\sqrt{2}$.
The above calculations apply to an ideal situation of our observing strategy. 
In practice, the target star is not always exactly centered in one detector 
quadrant, which influences $P(\rho)$ and $Q(\varphi)$ for $\rho \ga L/\!\sqrt{2}$. 

% R theory
\cite{shatsky2002} analyzed the background stars in their sky frames taken
next to their science targets in Sco~OB2. 
We use their data to derive $R(K_S)$, the expected
$K_S$ magnitude distribution for background stars.
We fit a second order polynomial to the
logarithm of the cumulative $K_S$ as a function of $K_S$ of the background
stars \citep[Figure~4 in][]{shatsky2002}. 
In our fit we only include background stars with 
$12~{\rm mag} < K_S < 16~{\rm mag}$ since
that is the $K_S$ magnitude range of the presumed background stars.
The $K_S$ magnitude distribution
$f_{K_S}(K_S)$ is then given by the derivative of the cumulative $K_S$
distribution. 

% P KS
Figure~\ref{figure: separationhisto} shows the
companion star and background star distribution 
as a function of angular separation.
The two distributions are clearly different. The
companion stars are more centrally concentrated than the
background stars.
The solid curve represents $P(\rho)$, normalized in such a way that 
the number of stars corresponds to the expected number of background stars
with $12~{\rm mag}<K_S<16~{\rm mag}$
(see below and Figure~\ref{figure: expectedbackground}).
More background stars are present at angular separation $4''<\rho<6''$
relative to what is expected. This is confirmed with a
Kolmogorov-Smirnov (KS) test.
The KS test is usually 
quantified with the ``KS significance'', which ranges from 0 to 1 and corresponds to the 
probability that the two datasets are drawn from the same underlying distribution. 
For this KS comparison we only consider those background stars
with $\rho \leq L/\!\sqrt{2}= 9.6''$. For background stars with $\rho > 9.6''$
we expect the predicted distribution to differ slightly from 
the observed distribution (see above).
We find a KS significance level of $1.8 \times 10^{-4}$,
which implies that the excess of background stars with  
$4''<\rho<6''$ is real, assuming that the simple model
for the angular separation distribution is correct.
The excess of close background stars might be
caused by faint ($K_S > 12~{\rm mag}$) companion stars which
are misclassified as background stars.

% Q KS
The position angle distribution of companion stars
and background stars is shown in Figure~\ref{figure: compsfield}.
The objects are expected to be randomly distributed over
position angle for $\rho \leq L/2 = 9.6''$, 
which is illustrated in Figure~\ref{figure: compsfield}.
For objects with  $\rho \leq L/2 = 9.6''$ 
we find that, as expected,  the position angle distribution is random, 
for both the companion stars (KS significance level 0.86)
and the background stars (KS significance level 0.84). For
angular separations larger than $9.6''$ the objects are primarily found
in the 'corners' of the image 
($\varphi = 45^\circ, 135^\circ, 225^\circ, 315^\circ$).

% P/R KS
The expected number of background stars per unit of angular
separation and unit of magnitude, $R(K_S)P(\rho)$ is plotted in Figure~\ref{figure:
expectedbackground}. We observe 19 background stars in the region $14~{\rm mag} \leq K_S
\leq 15~{\rm mag}$ and $5'' \leq \rho \leq 13''$ and normalize the theoretical
background star density accordingly. The brightness of the
PSF halo can explain the smaller number of background stars than expected for
$\rho \la 2.5''$ and $K_S \ga 15~{\rm mag}$. The predicted strong
decline in background stars for $\rho \ga 10''$ is present in
the observations. Few background objects with $12~{\rm mag}<K_S<14~{\rm mag}$ 
are expected in the angular separation range of $2''-4''$. 

\begin{SCfigure}[][btp]
  \centering
  \includegraphics[width=0.6\textwidth,height=!]{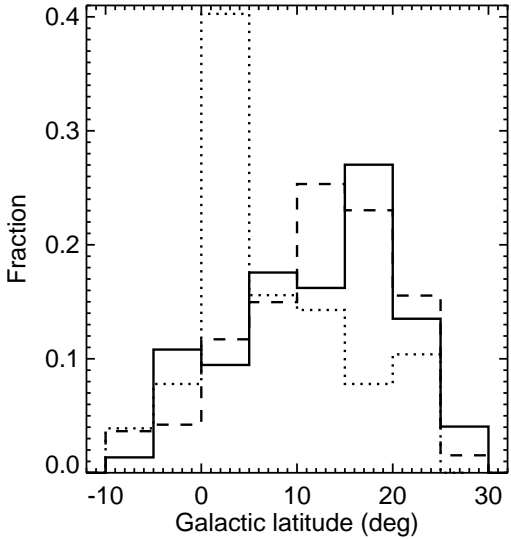} 
  \caption{The distribution of the 74 detected candidate companion stars (solid
  histogram) and 77 presumed background stars (dotted histogram) 
  as a function of Galactic latitude.
  The dashed histogram represents the distribution
  of {\it Hipparcos} member stars over Galactic latitude.
  The bin size is $5^\circ$ for the three histograms.
  The distributions are clearly different: the companion star and
  {\it Hipparcos} member star distribution peak at about the central Sco~OB2
  latitude. As expected, the background star distribution peaks at the 
  location of the Galactic plane.
  \label{figure: galacticplanehisto}}
\end{SCfigure}

The distribution of the detected candidate companions and background stars on
the sky is shown in Figure~\ref{figure: galacticplane}. 
The covered fields close to
the Galactic plane include a larger number of background stars, 
as expected.
This is even better visible in Figure~\ref{figure: galacticplanehisto},
which shows the distribution of observed objects over Galactic latitude.
The distribution of Sco~OB2 {\it
Hipparcos} member stars over Galactic latitude is similar to that of the
companion stars.

In principle, it is also possible that another (undocumented) member star of
Sco~OB2 projects close to the target star. The status of these stars 
(companion star or ``background star'') cannot be determined with
the CMD or the $K_S=12$~mag criterion.
\cite{preibisch2002} estimate the
number of member stars in the US subgroup in the range $0.1-20~{\rm M}_\odot$ to be
about 2500. This corresponds roughly to a stellar surface density of $8\times
10^{-7}$ star\,arcsec$^{-2}$. The chance of finding another US member star
(i.e. not a binary companion) in a randomly pointed observation of US is of
the order $0.03\%$. The detection probabilities for the other two subgroups of
Sco~OB2 are similarly small and negligible for our purpose.

% ====================================================================
% ====================================================================
% ====================================================================
% ==PROPERTIES=OF=COMPANION=STARS=====================================
% ====================================================================
% ====================================================================
% ====================================================================

\section{Properties of the companion stars} \label{sec: properties}

\subsection{General properties} \label{sec: generalproperties}

The angular separation distribution of the companion stars is 
centrally concentrated (Figures~\ref{figure: separationhisto} 
and~\ref{figure: compsfield}). 
The error in
the angular separation is typically $0.0015''$. 
Faint ($K_S \ga 11~{\rm mag}$) objects with angular separation less than $\rho \approx
0.8''$ cannot be detected in the presence of the PSF halo of the bright
primary star, while the upper limit of $13.5''$ is determined by
the field of view. The distribution over position
angle is random for $\rho \leq L/2 = 9.6''$, as expected 
(see \S~\ref{sec: backgroundstars}). 
The median error in position angle is
$0.03^\circ$, increasing with decreasing angular separation.

\begin{figure}[btp]
  \centering
  \begin{tabular}{cc}
    \includegraphics[width=0.46\textwidth,height=!]{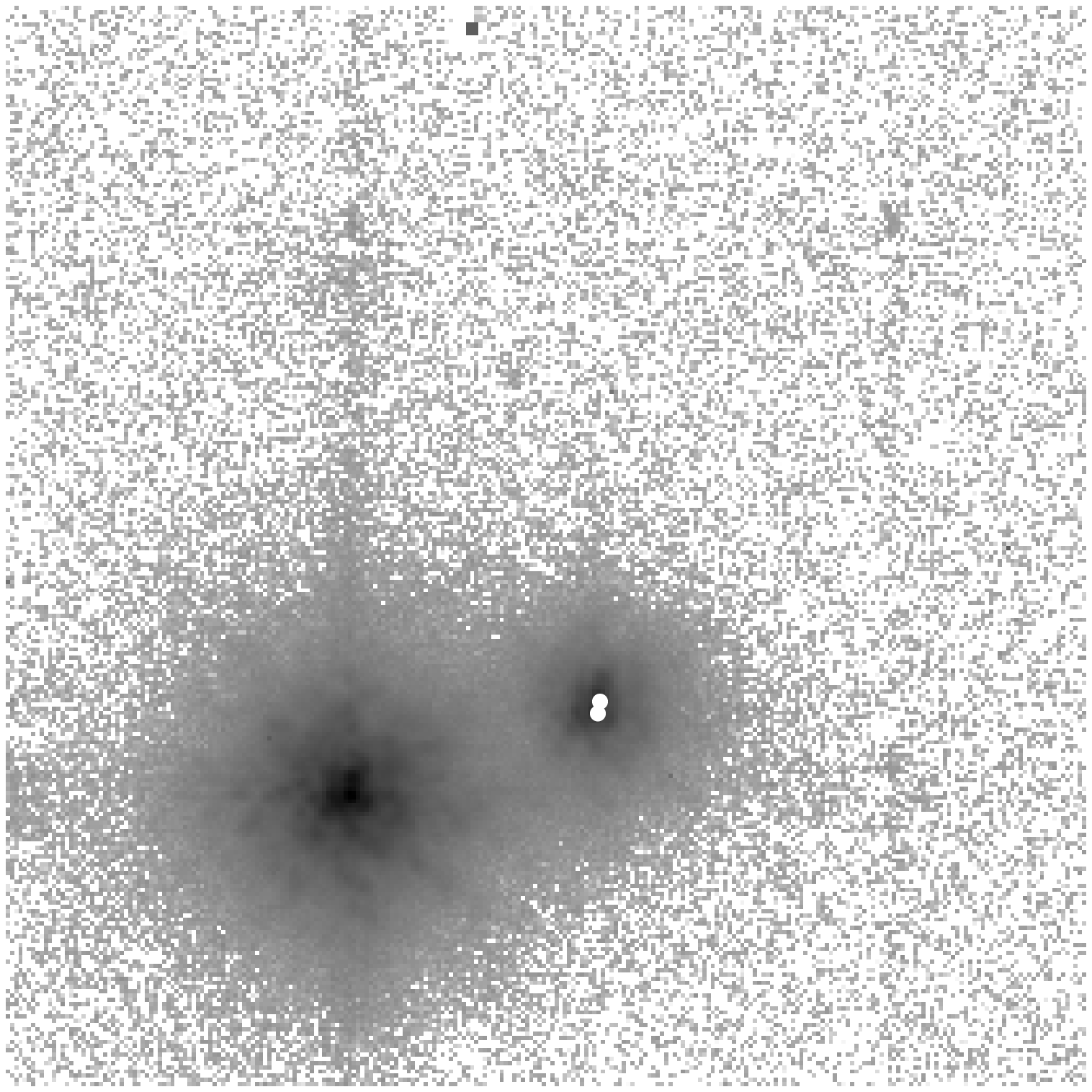} &
    \includegraphics[width=0.46\textwidth,height=!]{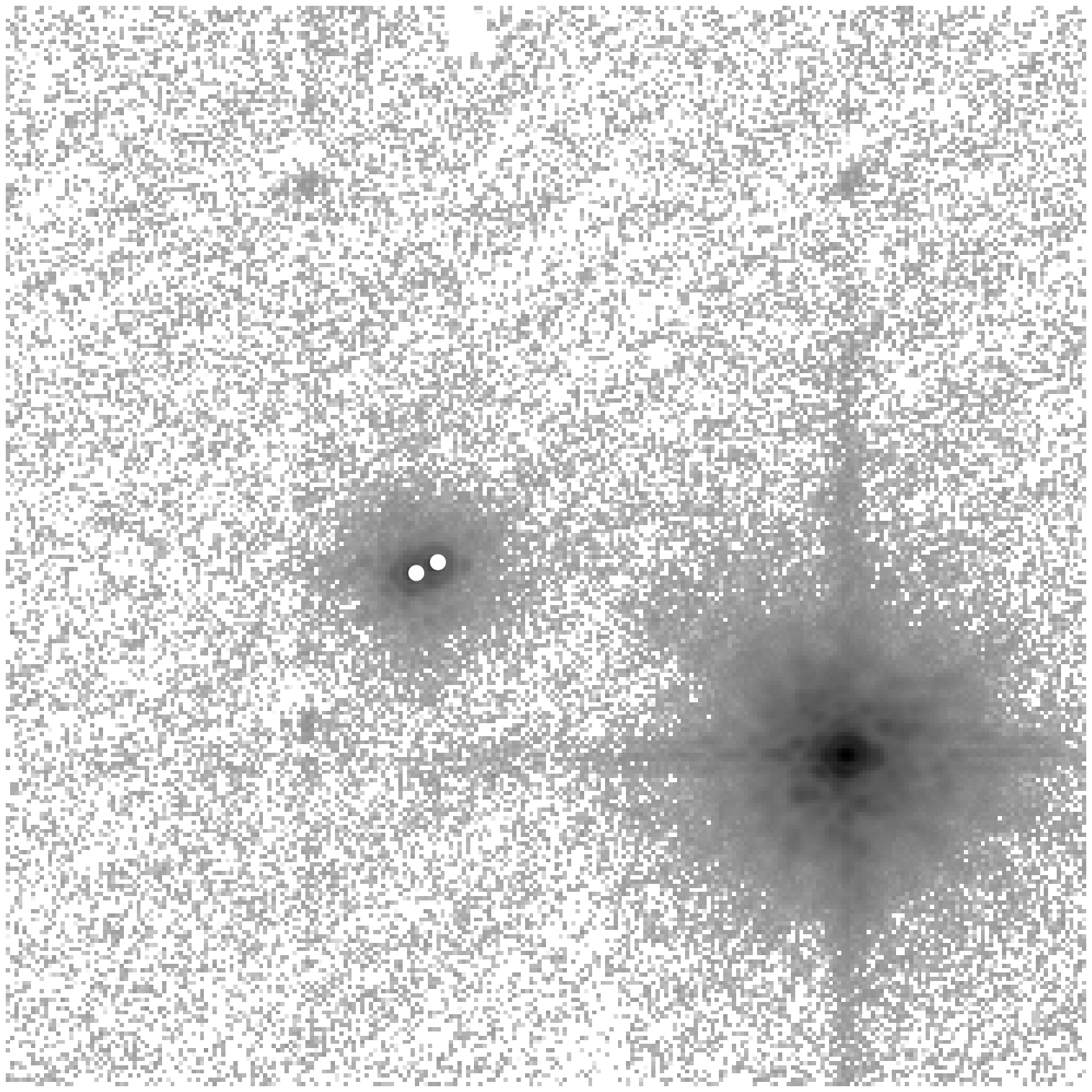} \\
  \end{tabular}
  \caption{
    {\it Left:} The hierarchical triple HIP68532. Both companion stars
    (indicated with the white dots) were previously undocumented.
    {\it Right:} HIP69113, another multiple system, where
    both close companion stars (white dots) 
    were previously undocumented.
    A third known companion star with angular separation of 
    $28''$ is not detected in our survey because of
    the large angular separation.
    \label{figure: triplesadonis}}     
\end{figure}

Two observed systems are clearly hierarchical triples where
a double 'secondary' orbits the primary star. These are HIP68532, where
the double 'secondary' has  $\rho=3.1'' $ and $\varphi=290^\circ$, and 
HIP69113, with $\varphi=65^\circ$ and $\rho=5.4''$.
HIP69113 has a third companion star 
with $\rho=28.6'' $ and $\varphi=35^\circ$ \cite{duflot1995}.
This star is not observed in our survey because of the large distance
from the primary star. 
The nature of the HIP69113 system is extensively discussed
by \cite{huelamo2001}.
Figure~\ref{figure: triplesadonis} shows
HIP68532 and HIP69113 with companion stars.
The target stars HIP52357, HIP61796, and HIP81972 also
have two companion stars. The two companion stars of each of
these systems have approximately the same angular separation and
position angle, but it is unclear
whether these systems are hierarchical in the way described above.

The 199 observed primaries have a $K_S$ magnitude range of $5.2-8.3~{\rm mag}$. The $K_S$
magnitudes of the companion stars (Figure~\ref{figure: detectionlimits_adonis}) range
from $K_S = 6.4~{\rm mag}$ to $12.0~{\rm mag}$, 
the upper limit resulting from the criterion that
was used to remove the background stars from the sample. The
primaries in our sample all have similar spectral type, distance, and
interstellar extinction. The magnitude of the companions relative to that of their
primaries, $\Delta K_S$, spans the range from $0.0~{\rm mag}$ to $6.0~{\rm mag}$. 
The median error is $0.05~{\rm mag}$ in $J$, $0.04~{\rm mag}$ in $H$,
$0.07~{\rm mag}$ in $K_S$, and $0.075~{\rm mag}$ in $J-K_S$.

Properties of the 199 target stars, 74 candidate companion stars, and 77
presumed background stars are listed in Appendix~A. The name of
the star is followed by the $J$, $H$, and $K_S$ magnitude. If the magnitude
is derived from measurements done under non-photometric conditions, a remark
is placed in the last column. The spectral type of each primary is taken from
the \textit{Hipparcos} catalogue. The angular separation and position angle
(measured from North to East) are derived from the combination of all
available observations for a particular star. For each star we also list the
status (p = primary star, 
c = candidate companion star, nc = new candidate
companion star, b = background star) and the subgroup membership.

\subsection{Color-magnitude diagram and isochrones} \label{sec: hrdiagram}

\begin{figure}[btp]
  \centering
  \includegraphics[width=\textwidth,height=!]{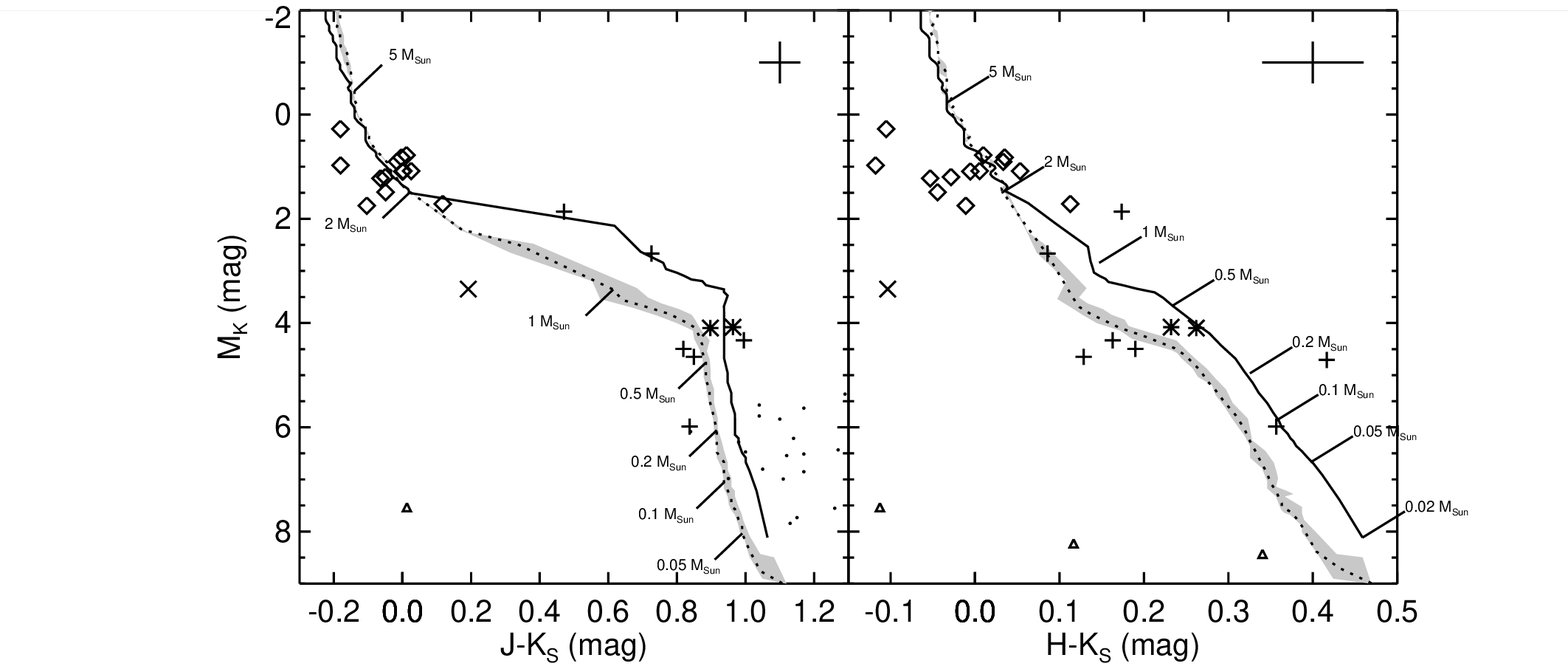}
  \caption{The color-magnitude diagram of the primaries,
    companion stars and background stars for which we obtained
    ADONIS/SHARPII+ multi-color AO observations.
    The $M_{K_S}$ magnitudes are corrected for distance and
    extinction. Companion stars in the three different subgroups are indicated
    with plusses (US), asterisks (UCL) and crosses (LCC). Target stars and
    background stars are indicated with diamonds and triangles, respectively
    (see \S~\ref{sec: backgroundstars}). The curves represent isochrones of
    5~Myr (solid curve) and 20~Myr (dotted curve). 
    The 15~Myr and 23~Myr isochrones enclose the gray-shaded area
    and represent the uncertainty in the age of the
    UCL and LCC subgroups.
    The mass scale is indicated for the 20~Myr isochrone ({\it left})
    and the 5~Myr isochrone ({\it right}).
    The brown dwarf candidates
    identified by \cite{martin2004} are indicated as dots. 
    The median formal errors are indicated 
    as error bars in the top-right corner of each plot.
    The locations of the target stars and companion stars are 
    consistent with the isochrones (within the error bars).
    The presumed background stars are located far away from the 
    isochrones, implying that our criterion to separate
    companion stars and background stars is accurate.
    \label{figure: 3hr}}
\end{figure}

\begin{SCfigure}[][btp]
    \includegraphics[width=0.6\textwidth,height=!]{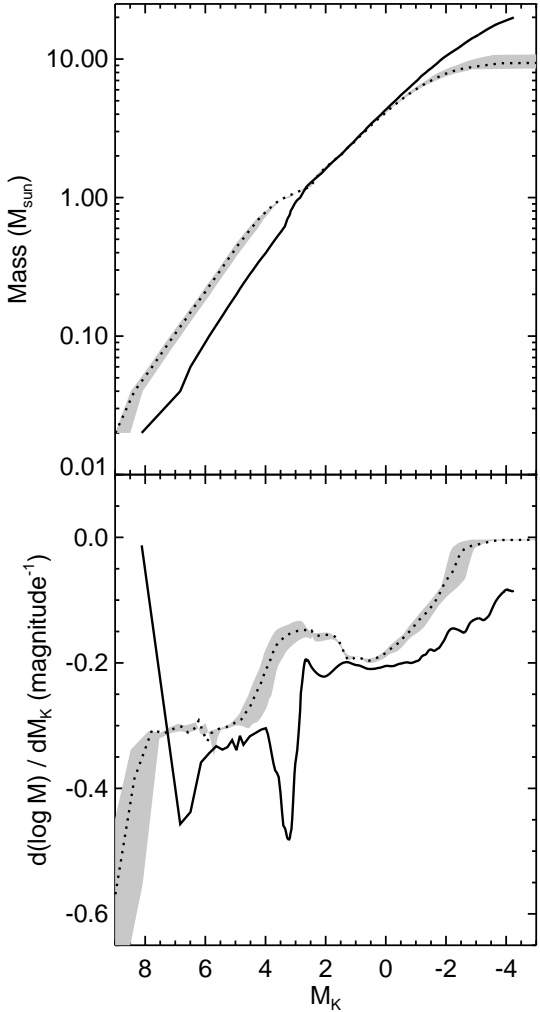}
    \caption{{\it Top:} The relation between mass and $M_{K_S}$ magnitude. The
    5~Myr isochrone (solid curve) 
    is used to find masses of the companion
    stars in US, while the 20~Myr isochrone (dotted curve) 
    is used for the UCL and LCC subgroups. 
    The gray-shaded area is enclosed by the 15~Myr and 23~Myr
    isochrones, and represents the uncertainty in the age of the
    UCL and LCC subgroups.
    For objects at the mean Sco~OB2 distance of 130~pc,
    $M_{K_S}=6$ corresponds to $K_S=11.6$. {\it Bottom:} The derivative of the
    mass-magnitude relation. The conversion from $M_{K_S}$ to mass is most
    accurate where the absolute value of the derivative 
    ${\rm d}M/ {\rm d}M_{K_S}$ is small. \label{figure:
    massmagnituderelation}}
\end{SCfigure}

In \S~\ref{sec: backgroundstars} we mentioned that it is
possible to separate background stars and companion stars
using their position in the CMD. In this section we 
derive the absolute magnitude $M_{K_S}$ for all observed objects.
We additionally derive colors for objects with multi-color 
observations. Only measurements obtained under photometric 
conditions are used. 
We construct a CMD  and
determine whether the observed components are background stars
or companion stars
by comparing their position in the CMD to that of the isochrones.
In \S~\ref{sec: masses} we will 
use the absolute magnitude $M_{K_S}$ and the
isochrones to derive the masses of the primary and
companion stars.

We calculate the absolute magnitude $M_{K_S}$ 
using the distance $D$ and extinction
$A_{K_S}$ for each target star individually. 
In \cite{debruijne1999} the distance to the Sco~OB2
member stars is derived from the 
secular parallax $\pi_{\rm sec}$. 
The secular parallax is calculated from the observed positions and
proper motions of member stars, where the fact that 
all stars in an association 
share the same space motion is exploited. The secular parallax 
can be up to more than two times as precise compared to the
{\it Hipparcos} parallax. 
Secular parallaxes with $g=9$ are used when
available, and with $g=\infty$ otherwise 
\citep[$g$ measures the model-observation discrepancy; see][for
details]{debruijne1999}. The interstellar extinction is also taken from
\cite{debruijne1999}. For the stars that do not have an extinction entry in
\cite{debruijne1999}, an estimate of the extinction is calculated using:
\begin{equation}
A_{K_S} = R_V \times E(B-V) / 9.3,
\end{equation}
where $R_V=3.2$ is the standard ratio of total-to-selective extinction
\citep[see, e.g.,][]{savage1979}. The value 9.3 is the ratio between $V$ band
and $K_S$ band extinction \citep{mathis1990}, and $E(B-V) = (B-V)-(B-V)_0$ is
the color excess. $(B-V)$ is the color listed in \cite{debruijne1999}. The
theoretical color $(B-V)_0$ is derived from the spectral type using the
broadband data for main sequence stars in \cite{kenyon1995}. 
The extinction for the UCL member HIP68958
(spectral type Ap...) is not listed in \cite{debruijne1999} and could not be
derived from \cite{kenyon1995}. We therefore take 
$A_{V,{\rm HIP68958}} = 0.063~{\rm mag}$. This is
the median $A_V$ value for those UCL member
stars which have full photometric data in \cite{debruijne1999}.

Only for a subset of observed targets observations in
more than one filter are available; 
the corresponding CMDs are plotted in Figure~\ref{figure: 3hr}. 
The observed primary stars are all of similar spectral type but show 
scatter in $M_{K_S}$, $J-K_S$, and $H-K_S$.  This
can be explained by the errors in near-infrared photometry, which are
typically $0.07~{\rm mag}$ for our observations. 
The uncertainties in the parallax
and reddening introduce an additional error when 
calculating the absolute magnitudes. Median
errors in $M_J$ and $M_{K_S}$ are consequently about $0.36~{\rm mag}$. The colors
$J-K_S$ and $H-K_S$ have median errors of $0.02~{\rm mag}$ and $0.06~{\rm mag}$,
respectively.

Given the age of the three Sco~OB2 subgroups, 
not all objects are expected to be positioned on the
main sequence. For the young age of Sco~OB2, stars of spectral type~G or later
have not reached the main sequence yet, while the O~stars have already
evolved away from the main sequence.
The age differences between the Sco~OB2 subgroups are relatively
well-determined. Different values for the absolute ages are
derived from the kinematics of the
subgroups \citep{blaauw1964A,blaauw1978} and stellar evolution
\citep{dezeeuw1985,degeus1989,preibisch2002,mamajek2002,sartori2003}. 
The age of the US subgroup is 5~Myr, without a
significant age spread 
\citep{dezeeuw1985,degeus1989,preibisch2002}.
\cite{mamajek2002} derive an age between 
15~and 22~Myr for members of UCL
and 17~and 23~Myr for members of LCC. 
In this paper we adopt an age 
of 20~Myr for both UCL and LCC.

We construct isochrones which are very similar to
those described in \cite{preibisch2002}.
For the low-mass stars ($M < 0.7\, {\rm M}_\odot$) 
we use exactly the same isochrone as \cite{preibisch2002}:
we use the models from \cite{chabrier2000}, which are
based on those from \cite{baraffe1998}. 
For $0.02 \leq M/{\rm M}_\odot < 0.7$ we take the models 
with $[\rm{M/H}]=0$ and $L_{\rm mix} = H_P$, 
where $L_{\rm mix}$ is the mixing length and $H_P$ the pressure scale height.
For $0.7 \leq M/{\rm M}_\odot < 1$ we use the models with
$[\rm{M/H}]=0$ and $L_{\rm mix} = 1.9 H_P$.
We use the models described in \cite{palla1999} for 
$1 \leq M/{\rm M}_\odot < 2$. Near-infrared magnitude tracks,
derived using the procedure described in \cite{testi1998},
were kindly provided by F.~Palla.
Finally, for $M/{\rm M}_\odot > 2$ we use the models from 
\cite{girardi2002} \citep[based on][]{bertelli1994} 
with $Y = 0.352$ and $Z = 0.05$, where
$Y$ and $Z$ are the helium and metal abundance, respectively.

The resulting 5~Myr isochrone (for US) 
and 20~Myr isochrone (for UCL and LCC) 
are plotted in Figure~\ref{figure: 3hr}. 
Both isochrones are continuous.
Figure~\ref{figure: massmagnituderelation} shows
the relationship between stellar mass and $K_S$-band
absolute magnitude for 5~Myr and 20~Myr
populations.
The gray-shaded area in Figures~\ref{figure: 3hr} 
and~\ref{figure: massmagnituderelation} is enclosed
by the 15~Myr and 23~Myr isochrones, and shows the 
effect of the age uncertainty in the UCL and LCC
subgroups. 

The positions of the primary and companion stars in the CMD 
are consistent with the
isochrones (except the companion of the LCC member star HIP53701). 
No obvious correlation of the position of these primaries in the
CMD with spectral type is present due to the errors in the colors and magnitudes.
The presumed background stars are located far from 
the isochrones in the CMD, 
supporting our hypothesis that they are indeed background stars.

In Figure~\ref{figure: 3hr} we added the 
candidate brown dwarf members in US from \cite{martin2004}.
We estimate the absolute magnitude of these brown dwarfs
using the mean US distance of 145~pc and the mean
extinction $A_K = 0.050~{\rm mag}$ (see below).
The colors and magnitudes of these objects are consistent with
the 5~Myr isochrone (solid curve), as is expected for US member stars.

\subsection{Masses and mass ratios} \label{sec: masses}

\begin{SCfigure}[][btp]
    \includegraphics[width=0.6\textwidth,height=!]{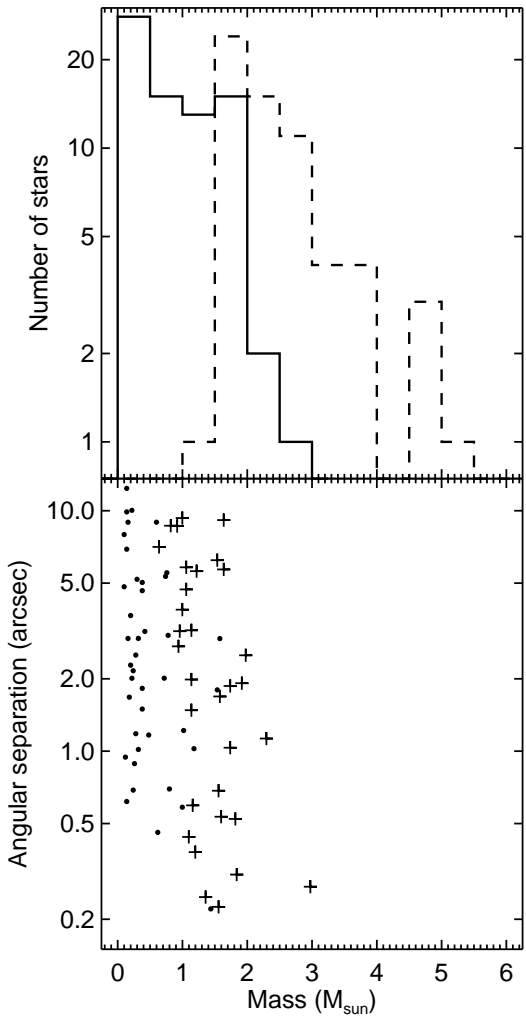}
    \caption{{\it Top:} The mass distribution of 
      the 74 companion stars (solid
      histogram) and corresponding primaries (dashed histogram)
      observed in our AO survey.
      The bin size is $0.5~{\rm M}_\odot$.
      Masses are derived from their $K_S$ magnitude 
      and age (see \S~\ref{sec: masses}). 
      The median errors in the mass are $0.4~{\rm M}_\odot$ for the primaries
      and $0.1~{\rm M}_\odot$ for the companions. 
      {\it Bottom:} Companion star mass versus angular separation. 
      New and previously known companion stars are indicated as
      dots and plusses, respectively.
      Low-mass close
      companion stars are not detected due to the PSF wings of the corresponding
      primary star. \label{figure: massdistributions_ado} }
\end{SCfigure}

We derive the mass of the primary and companion stars 
from their $M_{K_S}$ magnitude
using the isochrones described in \S~\ref{sec: hrdiagram}
(Figures~\ref{figure: 3hr} and~\ref{figure: massmagnituderelation}). 
We use the 5~Myr
isochrone for US and the 20~Myr isochrone for UCL and LCC. The conversion from
$M_{K_S}$ to mass is most accurate for B- and A-type stars,
where the absolute value of the derivative 
${\rm d}M/ {\rm d}M_{K_S}$ is small. 
We find masses between $1.4~{\rm M}_\odot$ and $7.7~{\rm M}_\odot$ for the primaries and masses
between $0.1~{\rm M}_\odot$ and $3.0~{\rm M}_\odot$ for the companions. 
Given the error $\Delta M_{K_S}$ on $M_{K_S}$, we calculate
for each star the masses $M_-$ and $M_+$ 
that correspond to  $M_{K_S}-\Delta M_{K_S}$
and $M_{K_S}+\Delta M_{K_S}$. We define the error on the mass
to be $\frac{1}{2} (M_+-M_-)$.
The median of the error in the mass, which is a lower limit for the
real error, is $0.4~{\rm M}_\odot$ for the primaries 
and $0.1~{\rm M}_\odot$ for the companions. 

Figure~\ref{figure: massdistributions_ado} (top) shows the mass distribution of
the companion stars and of the primary stars 
to which they belong. The mass distribution
of the observed target stars without companions is similar to the latter.
The plot showing companion star mass as
a function of angular separation (Figure~\ref{figure: massdistributions_ado},
bottom) closely
resembles Figure~\ref{figure: detectionlimits_adonis}.
The reason for this is that the companion star mass is
derived from the $K_S$ magnitude.

\begin{SCfigure}[][btp]
    \includegraphics[width=0.6\textwidth,height=!]{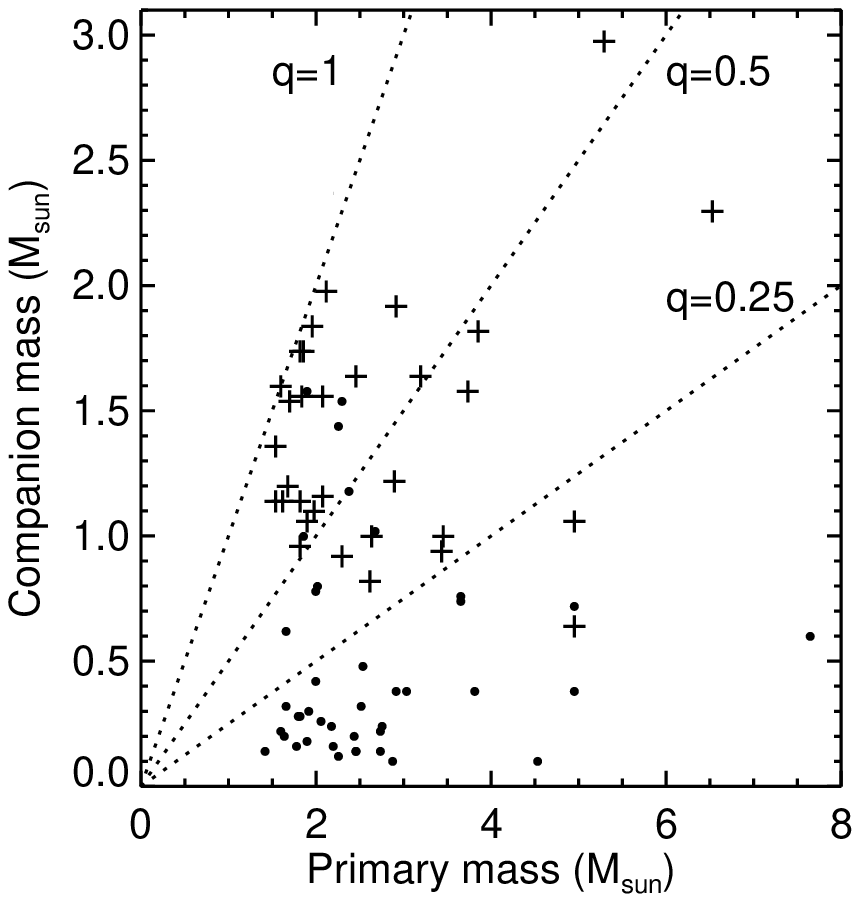} 
    \caption{Companion star mass versus primary star mass. The dashed lines
    represents the $q=0.25$, $q=0.5$, and $q=1$ binaries
    and are shown to guide the eye. 
    The 41 new companion stars
    and the 33 previously known companion stars are indicated with
    dots and plusses, respectively.
    The observed companion stars have masses lower than their associated
    primaries. Most systems with 
    previously undocumented companion stars
    have $q < 0.25$.
    \label{figure: masses}}
\end{SCfigure}

The mass ratio $q$ and corresponding error are calculated for each primary-companion pair. 
The median formal error on the mass ratio is 0.07 and is to first order 
proportional to the error on the companion star mass. The systematic errors
introduced by photometric uncertainties and errors in distance and extinction
are equal for both stars in a system,
so that the magnitude difference is unaffected by these \citep[see, e.g.,][]{shatsky2002}. 
The formal error on $q$ may therefore be slightly overestimated

\begin{SCfigure}[][btp]
  \includegraphics[width=0.5\textwidth,height=!]{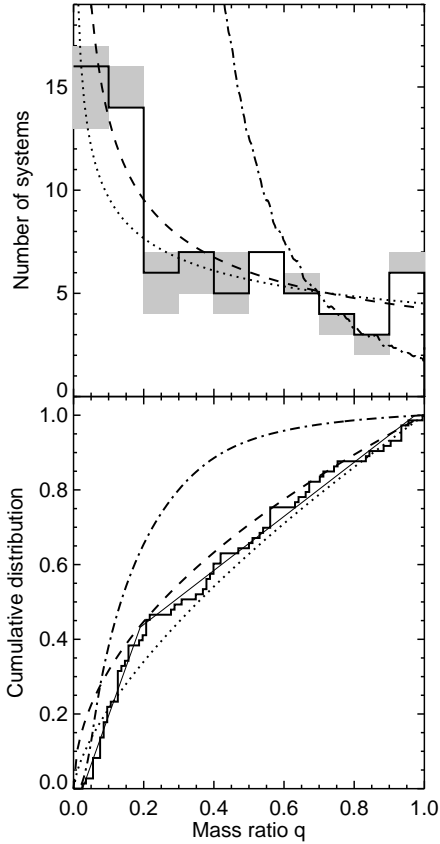}\hspace{0cm}
  \caption{{\it Top:} The mass ratio distribution for the 
    74 systems for which we observe companion stars
    in our near-infrared AO survey (histogram).
    The mass ratio is defined as $q = M_c/M_p$ where
    $M_c$ is the companion star mass and $M_p$ the mass of the 
    corresponding primary star. All primaries are 
    {\it Hipparcos} member stars with A and late-B spectral type.
    The $q$ distribution does not change significantly
    if ages of 15~Myr or 23~Myr for UCL and LCC are assumed 
    (gray-shaded areas).
    The mass ratio distributions $f_q(q) = q^{-\Gamma}$ are represented by the
    curves. 
    For our observations we find $\Gamma = 0.33$ (dotted
    curve). \cite{shatsky2002} find $\Gamma = 0.5$ (dashed curve). 
    The distribution which follows from random pairing
    (dash-dotted curve) is clearly excluded. 
    The curves are normalized using the observed
    mass ratio distribution for the 
    companion stars in the range $0.5 < q < 1.0$.
    {\it Bottom:} The observed cumulative mass
    ratio distribution (histogram). 
    The dotted and
    dashed curve are cumulative distribution functions corresponding to
    $\Gamma = 0.33$ and $\Gamma = 0.5$, respectively. 
    The fitted function described in Equation~\ref{eq: twolinefit} is 
    shown as the thin solid curve. The dash-dotted
    curve represents the cumulative $q$ distribution for random
    pairing. 
    \label{figure: massratiohisto}}
\end{SCfigure}

The mass ratio distribution is shown in Figures~\ref{figure: masses} and~\ref{figure:
massratiohisto}. The smallest mass ratios are found for the companions of
HIP80474 ($q=0.022 \pm 0.008$) and HIP77911 ($q=0.035 \pm 0.012$). 
This corresponds well with the
expected minimum value for $q$: a typical primary in our sample has spectral
type A0~V ($2~{\rm M}_\odot$), while the least massive companions considered have
$K_S=12~{\rm mag}$ ($\sim 0.1~{\rm M}_\odot$), giving $q=0.05$. 
The systems with large mass ratio 
($q>0.9$) are HIP80324 ($q=0.91 \pm 0.23$), HIP50520
($q=0.93 \pm 0.19$), HIP80238 ($q=0.94 \pm 0.29$), HIP64515 ($q=0.94 \pm 0.15$), 
HIP61639 ($q=0.95 \pm 0.15$), and HIP52357 ($q=1.00 \pm 0.34$). 
Note that for these systems the possibility that $q>1$ is included. 
Due to the error in the mass it is not possible to say which star 
in these systems is more massive, and hence, which is the primary star.
Although the uncertainty in the age of UCL and LCC is not negligible,
the effect on the mass ratio distribution is small 
(gray-shaded areas in Figure~\ref{figure: massratiohisto}).

Following \cite{shatsky2002} we fit a power-law of the form $f_q(q) = q^{-\Gamma}$
to the mass ratio distribution. We find that models
with $\Gamma = 0.33$ fit our observations best (KS significance
level 0.33). 
The error in $\Gamma$ due to the age uncertainties in 
the UCL and LCC subgroups is 0.02.
This is in good agreement with \cite{shatsky2002},
who surveyed 115 B-type stars in Sco~OB2 for binarity
and observed 37 physical companions around these stars. 
They find $\Gamma = 0.3 - 0.5$ for their mass ratio distribution.
We observe an excess of systems with $q \sim 0.1$ with
respect to what is expected for models with $\Gamma=0.33$.
This could partially be explained by bright background stars that are
misclassified as companion stars, but this is unlikely to be an important
effect (see \S~\ref{sec: backgroundstars}). 
The observational biases (e.g. the detection limit) are not 
taken into account here. For a more detailed description of the
effect of observational biases
on the mass ratio distribution 
we refer to \cite{hogeveen1990} and \cite{tout1991}.

The cumulative mass ratio distribution $F_q(q)$ can also
be described as a curve consisting of two line segments.
We fit and find that the 
mass ratio distribution
closely follows the function
\begin{equation} \label{eq: twolinefit}
  F_q(q) = \left\{
  \begin{array}{lll}
    2.63 q - 0.067 & {\rm \quad for \quad } & 0.03 < q \leq 0.19 \\
    0.72 q + 0.294 & {\rm \quad for \quad } & 0.19 < q \leq 0.97 \,, \\
  \end{array}
  \right.
\end{equation}
with a root-mean-square residual of $0.019$ 
(see Figure~\ref{figure: massratiohisto}).

We investigate whether the observed mass ratio distribution 
could be the result of random pairing between primary stars
and companion stars, such as observed for solar-type stars
in the solar neighbourhood \citep{duquennoy1991}.
To this end, we use Monte Carlo simulations to calculate the mass ratio distribution
that is expected for random pairing. 

The current knowledge about the brown dwarf
population in Sco~OB2 is incomplete \citep[e.g., Table 2 in][]{preibisch2003}.
Therefore we make a comparison with the 
mass ratio distribution resulting from two different 
initial mass functions (IMFs)
(IMF$_{-0.3}$ and IMF$_{2.5}$),
which differ in slope ${\rm d}N/{\rm d}M$
for substellar masses.
\begin{equation} \label{equation: imf1}
  {\rm IMF_\alpha:~~} \frac{{\rm d}N}{{\rm d}M} \propto \left\{
  \begin{array}{llll}
    M^{\alpha}  & {\rm for \quad } 0.02 & \leq M/{\rm M}_\odot & < 0.08 \\
    M^{-0.9}  & {\rm for \quad } 0.08 & \leq M/{\rm M}_\odot & < 0.6 \\
    M^{-2.8}  & {\rm for \quad } 0.6  & \leq M/{\rm M}_\odot & < 2   \\
    M^{-2.6}  & {\rm for \quad } 2    & \leq M/{\rm M}_\odot & < 20 \ , \\
  \end{array}
  \right.
\end{equation}
For $M \geq 0.10~{\rm M}_\odot$ IMF$_\alpha$ is equal to the IMF in US 
that was derived by \cite{preibisch2002}. The \cite{preibisch2002}
IMF is extrapolated down to $M = 0.08~{\rm M}_\odot$.
For $0.02 \leq M/{\rm M}_\odot < 0.08$ we consider 
$\alpha=-0.3$ \citep{kroupa2002} and $\alpha=2.5$ 
\citep[][Fit~I]{preibisch2003}. Most other models described in
\cite{preibisch2003} have $-0.3<\alpha<2.5$ 
for $0.02 \leq M/{\rm M}_\odot < 0.08$.

We draw the primary and secondary stars from
IMF$_\alpha$. 
Primary and secondary masses are independently drawn from the mass 
range $[0.02,20]~{\rm M}_\odot$ and are swapped
if necessary, so that the primary star is the most massive. Only
those systems with primary mass larger than  $1.4~{\rm M}_\odot$ 
and smaller than $7.7~{\rm M}_\odot$ are considered.
We calculate the resulting mass ratio distribution.
A KS comparison between the mass ratio distribution
resulting from random pairing and the observed distribution
shows that the random-pairing hypothesis can be rejected
for both IMF$_{-0.3}$ and IMF$_{2.5}$.
This is clearly visible in Figure~\ref{figure: massratiohisto}.

Random pairing requires more systems with low mass ratio than we observe.
The error in the mass ratio (ranging between 0.008 and 0.34) 
cannot compensate for the lack of faint companions. 
Even if all close ($\rho \leq 4$) background stars are treated as
low-mass companions, random pairing is excluded. 
Hence, there is a deficit of low-mass companions around
late-B and A type stars compared to what is expected from random pairing. 

Several observed primary stars are actually unresolved binary systems.
Since we derive the mass for this 'primary' using the total $K_S$ flux of such
an unresolved system, we introduce an error in the mass and hence the mass ratio.
Out of the 199 observed primary stars 18 are known 
spectroscopic binaries (14 in US, 3 in UCL, and 1 in LCC). 
These binary systems are too tight to be resolved in our AO survey.
We observe companion stars around three of these primaries:
HIP77911, HIP77939, and HIP79739 (all in US). For these systems
we expect the mass ratio, which we define here as the mass of the
companion star divided by the total mass of the inner part
of the system, to deviate significantly from the value 
derived from our near-infrared observations.

% The discussion about the gap starts here

A closer inspection of Figure~\ref{figure: detectionlimits_adonis} reveals a lack of
stellar components with $\rho \leq 3.75''$ and $12~{\rm mag} \la K_S \la 14~{\rm mag}$. 
If companion stars or background stars
with these properties exist, we should have seen them. 
In \S~\ref{sec: backgroundstars} we showed that few background stars are 
expected in this region. For
example, the brown dwarfs (M5.5--M9) detected by \cite{martin2004} in US have
$11~{\rm mag} \la K \la 14~{\rm mag}$. If similar objects within $\rho \leq 3.75''$ were
present, they would have been detected.
This `gap' is not the result of observational biases: the detection limit shows
that objects with these properties are detectable. 

In order to find out if this `gap' is expected, we generate a stellar population
using a Monte Carlo process and two-dimensional two-sample KS tests.
For comparison with the simulated data we only consider the 59 observed 
companion and background stars with $\rho \leq 4''$. 
Objects with $\rho > 4''$ are not relevant for this test and might 
introduce biases due to observational selection effects.
We randomly draw primary and secondary masses
in the mass range $[0.02,20]~{\rm M}_\odot$.
We select only those binaries for which the primary
mass is in the range $[1.4,7.7]~{\rm M}_\odot$. 
Sets of 59 binary systems are created for 
IMF$_{-0.3}$ and IMF$_{2.5}$ (Equation~\ref{equation: imf1}).
The masses are converted into $K_S$ magnitudes 
using a distance of 145~pc and the 5~Myr
isochrone described in \S~\ref{sec: hrdiagram}. The angular separations for
the secondaries are drawn from a uniform distribution in $[0'',4'']$. We
create $10^3$ realizations which are compared to the observations.
For IMF$_{-0.3}$ we find a mean KS significance 
level of $3.5 \times 10^{-3}$, while for
IMF$_{2.5}$ we find $3.2 \times 10^{-3}$. We repeated the procedure
described above with another angular separation distribution:
$\partial N / \partial \rho \propto \rho$ for $\rho \in [0'',4'']$
(\"{O}pik's law). We find mean KS significance levels of
$8.2 \times 10^{-5}$ and $6.5 \times 10^{-5}$ for IMF$_{-0.3}$ and IMF$_{2.5}$, 
respectively. 
It is therefore unlikely that the observed
distribution (including the `gap') is drawn from the IMFs and angular
separation distributions as described above.

Several faint close components with $\rho \leq 3.75''$ below the `gap'
($K_S \ga 14~{\rm mag}$) are detected. These objects are found next to
the target stars 
HIP61265, HIP67260, HIP73937, HIP78968, HIP79098, HIP79410, and HIP81949.
The Strehl ratios in $K_S$ of the corresponding observations
($\sim 20\%$) are typical. The detection of these close, faint components therefore cannot be
explained by a better performance of the AO system during the observation of
these objects. 
The target stars corresponding to these objects are
not particularly faint; the luminosity contrast between the target star
and the faint close components is typical.
These objects are background stars according to our selection
criterion. However, there is a possibility that these objects are close brown dwarfs
with masses $\la 0.05~{\rm M}_\odot$, 
where the conversion from $K_S$ to mass is
strongly dependent on the age.
This result implies that A and late-B stars do not have close companions with
a mass less than about 0.1~M$_\odot$, unless the assumed background stars {\it
are} physical companions (and thus brown dwarfs). If so, a gap would be
present in the companion mass distribution.

\section{Comparison with literature data} \label{sec: literaturedata}

\subsection{New companions} \label{sec: newcompanions}

\begin{SCtable}[][btp] 
  \begin{tabular}{ll}
    \hline 
    Reference & Detection method \\
    \hline
    \cite{alencar2003} & Spectroscopic \\
    \cite{balega1994} & Visual \\
    \cite{barbier1994} & Spectroscopic\\
    \cite{batten1997} & Spectroscopic \\ 
    \cite{couteau1995} & Combination\\
    The Double Star Library & Combination\\
    \cite{duflot1995} &Spectroscopic\\
    \cite{hartkopf2001} &Visual\\ 
    \cite{jordi1997} &Eclipsing\\
    The {\it Hipparcos} and {\it TYCHO} Catalogues &Astrometric\\ 
    \cite{kraicheva1989} &Spectroscopic\\ 
    \cite{malkov1993} &Combination \\
    \cite{mason1995} &Visual \\
    \cite{mcalister1993} &Visual\\
    Miscellaneous, e.g. SIMBAD &Combination\\
    \cite{miura1992} &Visual\\
    \cite{pedoussaut1996} &Spectroscopic\\
    \cite{shatsky2002} &Visual\\
    \cite{sowell1993} &Visual\\
    \cite{svechnikov1984} &Combination\\
    \cite{tokovininmsc} &Combination\\
    \cite{wds1997} &Combination\\
    \hline 
  \end{tabular}
  \caption{References to literature data with spectroscopic, 
    astrometric, eclipsing, and visual binaries in Sco~OB2. 
    Using these data we
    find that 33 out of the 74 candidate companion stars in our dataset were
    already documented in literature, while 41 were previously unknown.
    \label{table: litlist}}
\end{SCtable}

We compiled a list of known companion stars to all {\it Hipparcos} members of
Sco~OB2 using literature data on binarity and multiplicity. This includes
spectroscopic, astrometric, eclipsing, and visual binaries. The references
used are listed in Table~\ref{table: litlist}.

Since the angular separation and position angle of companion stars change as a
function of time, one has to take care that a `newly' detected companion is
not a displaced known companion star. It is therefore interesting to estimate
how fast the angular separation and position angle change for the observed
companion star.
We obtained the primary mass $M_p$ and companion star mass $M_s$ 
in (\S~\ref{sec: masses}) thus we can estimate the orbital
period of the companion stars using Kepler's third law. If the orbit is
circular and the system is seen face-on, the
orbital period is given by $P = \sqrt{ 4 \pi^2 (D \rho)^3 / G (M_p+M_s) }$,
where $D$ is the distance to the system, $\rho$ the angular separation
between primary and companion star, and $G$ is the gravitational
constant. 
In general, orbits are eccentric and inclined.
However, the effect of nonzero eccentricity 
and inclination on the period is most likely
only of the order of a few per cent \citep{leinert1993}.

In our survey we are sensitive to orbits with a period
between approximately 50~yr and 50,000~yr.
For most of the systems in our sample we find an orbital period of the
order of a few
thousand years. The literature data that we used are several decades old at
most. The angular separation and position angle are not expected to have
changed significantly over this time. If a star with significantly different
values for $\rho$ and/or $\varphi$ is discovered with respect to the known
companion star, this star is considered to be new.
Several close binary systems are observed.
The shortest orbital periods that we find 
(assuming circular and face-on orbits) 
are those for
the companions of HIP62026 (57~yr), HIP62179 (68~yr), HIP76001 (84~yr),
HIP64515 (87~yr), HIP80461 (106~yr), HIP62002 (142~yr), and HIP67260
(213~yr). Differences in angular separation $\rho$ and position angle
$\varphi$ are therefore expected
with respect to previous measurements. 
For example, the observed star HIP76001 has two companions
with angular separations $\rho_1$ and $\rho_2$, and 
position angles $\varphi_1$ and $\varphi_2$, respectively.
In our AO survey we measure $(\rho_1,\varphi_1) = (0.25'',3.2^\circ)$ and
$(\rho_2,\varphi_2) = (1.48'',124.8^\circ)$, 
while \cite{tokovininmsc} quotes 
$(\rho_1,\varphi_1) = (0.094'',6^\circ)$ in 1993 and
$(\rho_2,\varphi_2) = (1.54'',130^\circ)$ in 1991.
Taking the changes in $\rho$ and $\varphi$ into account, we determine
whether an observed companion star was documented before or 
whether it is new.

\begin{SCtable}[][btp]
  \begin{tabular}{l ccccc}
    \hline \hline Type & B1-B3 & B4-B9 & A & F & GKM \\ 
    \hline \multicolumn{6}{l}{Upper Scorpius} \\ 
    \hline Single stars & 3 & 13 & 20 & 17 & 10 \\ 
    Visual & 15 & 9 & 12 & 3 & 2 \\ 
    Astrometric & 0 & 3& 1& 2& 3 \\
    Spectroscopic & 11& 12& 0 & 0& 0 \\ 
    \hline \multicolumn{6}{l}{Upper Centaurus Lupus} \\ 
    \hline 
    Single stars & 3 & 22 & 41 & 43 & 22 \\ 
    Visual  & 15 & 23 & 29 & 7 & 8 \\ 
    Astrometric & 2 & 1 & 3 & 5 & 2 \\ 
    Spectroscopic & 11& 6 & 1& 1& 0 \\ 
    \hline \multicolumn{6}{l}{Lower Centaurus Crux} \\
    \hline 
    Single stars & 3 & 16 & 31 & 47 & 14 \\ 
    Visual & 7 &13 & 24 & 10 & 4 \\ 
    Astrometric & 1 & 3 & 3 & 4 & 4 \\ 
    Spectroscopic & 3 & 3& 1 & 0 & 0 \\ 
    \hline \multicolumn{6}{l}{Scorpius OB2}\\ 
    \hline 
    Single stars & 9 & 51 & 92 & 107 & 46 \\ 
    Visual & 37 & 45 & 65 & 20 &14 \\ 
    Astrometric & 3 & 7 & 7 & 11 & 9 \\ 
    Spectroscopic & 25 & 21 & 2 & 1 & 0 \\ 
    \hline \hline
  \end{tabular}
  \caption{Binarity data for Sco~OB2, showing the number of companion stars
  found with the different techniques. Literature data and the new companion
  stars described in this paper are included. 
  The terms 'visual', 'astrometric', and 'spectroscopic' pertain only to 
  available observations and not to intrinsic properties of the systems.
  Note that 
  each companion of a multiple system is included individually.
  \label{table: summary}}
\end{SCtable}

Table~\ref{table: summary} summarizes the current status on binarity for
Sco~OB2. The number of presently known companion stars is listed as a function
of primary spectral type and detection method. 

We define all companions that
have only been detected with radial velocity studies as spectroscopic. All
companions of which the presence of the companion has only been derived from
astrometric studies of the primary are classified as astrometric. All other
companions, i.e. those that have been detected with (AO) imaging,
Speckle techniques, interferometry, etc., are classified as visual. Note that
our classification refers only to available observations and {\it not} to
intrinsic properties of the binary or multiple systems.

All candidate companions that are found with our AO survey are by
definition visual. 
The overlap between the observed companions in our sample and the 
spectroscopic and astrometric
binaries in literature is zero, although HIP78809 is flagged as 
``suspected non-single'' in the {\it Hipparcos} catalogue.
From our comparison with literature we find
41 new companions (14 in US; 13 in UCL; 14 in UCL), 
while 33 of the candidate companion stars were already
documented.

Several of the stars currently known as 'single' could be unresolved
systems. For a subset of the member stars, multiple companions have been
detected using different techniques. Most systems with early spectral type
primaries are found with spectroscopic methods. Amongst the A and F type
stars, most {\it Hipparcos} member stars are single, although a fraction of
these could be unresolved systems (see \S~\ref{sec: binarystatistics}).

Of the Sco~OB2 {\it Hipparcos} member stars 37 are now known to be
triple. Sixteen of these consist of a primary with two visual
companions. Fifteen consist of a primary, a visual companion, and a
spectroscopic companion. These are only found in UCL and LCC. Six consist of a
primary, a spectroscopic companion and an astrometric companion. Five
quadruple systems are known: two systems consisting of a primary and three
visual companions (HIP69113 and HIP81972) and two consisting of a primary, two
visual companions and a spectroscopic companion (HIP80112 and
HIP78384). HIP77820 is the only quintuple system, consisting of a primary,
three visual companions and one spectroscopic companion. 
The largest known system in Sco~OB2 is
HIP78374, which contains a primary, four visual companions and two
spectroscopic companions.

\subsection{Binary statistics} \label{sec: binarystatistics}

The binarity properties of a stellar population are usually quantified in
`binary fractions' \citep[e.g.,][]{reipurth1993}. 
Two common definitions in use are the multiple system
fraction $F_{\rm M}$ and non-single star fraction $F_{\rm NS}$ (since $1-F_{\rm NS}$ is the fraction
of stars that is single). Another frequently used quantity is the companion
star fraction $F_{\rm C}$\footnote{Note that in \cite{kouwenhoven2004a,kouwenhoven2005ao} 
we used an incorrect definition of $F_{\rm C}$.}, which measures the average number 
of companion stars per
primary star. These quantities are defined as
\begin{eqnarray}
  F_{\rm M}       &=& (B+T+\dots) \ /\ (S+B+T+\dots);\\
  F_{\rm NS}      &=& (2B+3T+\dots) \ /\ (S+2B+3T+\dots);\\
  F_{\rm C}       &=& (B+2T+\dots) \ /\ (S+B+T+\dots) \ ,
\end{eqnarray}
where $S$, $B$, and $T$ denote the number of single systems, binary systems
and triple systems in the association. 
Table~\ref{table: statistics_adonis} shows several properties of the three subgroups,
including the binary statistics that are updated with our new findings.

\begin{table}[btp]
  \setlength{\tabcolsep}{0.7\tabcolsep}
  \begin{tabular}{l cc cccc cc c}
    \hline
          & $D$  & Age  & $S$ & $B$ & $T$ & $>3$ & $F_{\rm M}$ & $F_{\rm NS}$ & $F_{\rm C}$ \\
          & (pc) &(Myr) & \\
    \hline
    US    & 145  & 5--6     & 63    & 46    & 7     & 3     & 0.47  & 0.67  & 0.61\\
    UCL   & 140  & 15--22   & 131   & 68    & 17    & 4     & 0.40  & 0.61  & 0.52\\
    LCC   & 118  & 17--23   & 112   & 54    & 13    & 0     & 0.37  & 0.57  & 0.45\\
    \hline
    all   &      &          & 303   & 171   & 37    & 7     & 0.41  & 0.61  & 0.52\\
    \hline
  \end{tabular}
  \caption{Multiplicity among {\it Hipparcos} members of Sco~OB2. The columns
  show the subgroup names (Upper Scorpius; Upper Centaurus Lupus; Lower
  Centaurus Crux), their distances \citep[see][]{dezeeuw1999}, the ages
  (\cite{degeus1989,preibisch2002} for US; \cite{mamajek2002} for UCL and LCC), the number
  of known single stars, binary stars, triple systems and $N>3$ systems, and
  the binary statistics (see \S~\ref{sec: binarystatistics}) . \label{table:
  statistics_adonis} }
\end{table}

\begin{figure}[btp]
  \centering
  \includegraphics[width=0.7\textwidth,height=!]{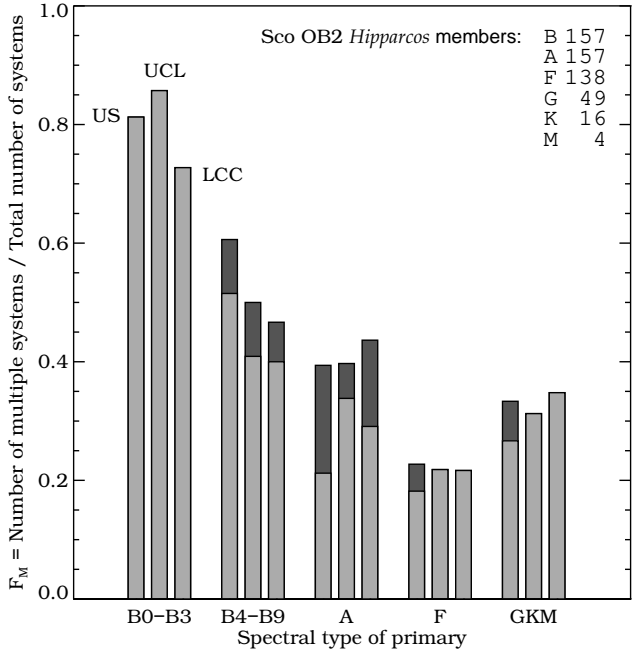}
  \caption{The fraction of stellar systems which is multiple versus the
  spectral type of the primary, for the three subgroups of Sco~OB2. Only
  confirmed Hipparcos member primaries are considered here. 
  The binarity dataset consists of literature data and
  our near-infrared AO survey.
  The light and dark
  gray parts of the bars correspond to literature data and the new data
  presented in this article, respectively. The spectral types of the companion
  stars (not included in this plot) are always later than those of the primary
  stars. Apparently, the multiplicity is a function of spectral type, but this
  conclusion may well be premature when observational biases are not properly
  taken into account. That this is at least partly true, is supported by our
  detection of 41 new close companion stars at later spectral
  types. \label{figure: mulspt}}
\end{figure}

Figure~\ref{figure: mulspt} shows the multiple system fraction $F_{\rm M}$ 
as a
function of spectral type of the primary for the three subgroups. For example,
86\% of the B0--B3 type {\it Hipparcos} member stars of UCL have one or more
known companion stars. Information about the number of companion stars and the
spectral type of the companion stars around the primary is not shown in the
figure. The new companion stars found in our AO survey are indicated with the
darker shaded parts of the bars. 
A trend between multiplicity and the spectral type
of the primary seems to be present, but this conclusion may well be premature
when observational biases are not properly taken into account. Our detection
of 41 new close companion stars shows that this is at least partially true: the
number of A-type member stars which is in a binary/multiple system in US has
doubled as a result of our survey. This strongly supports the statement made
in \cite{brown2001} that the low multiplicity for A- and F-type stars can at
least partially be explained by observational biases.

% ====================================================================
% ====================================================================
% ====================================================================
% ==CONCLUSIONS=======================================================
% ====================================================================
% ====================================================================
% ====================================================================

\section{Conclusions and discussion} \label{sec:conclusions}

% The conclusions start here

We carried out a near-infrared adaptive optics search for companions around
199 (mainly) A and late-B stars in the nearby OB association Sco~OB2. 
Our sample is a selection of the {\it Hipparcos} membership list of Sco~OB2
in \cite{dezeeuw1999}. We find
a total of 151 stellar components around the target stars. We use a simple
brightness criterion to separate candidate companion stars ($K_S < 12~{\rm mag}$) and
probable background stars ($K_S > 12~{\rm mag}$). The validity of this criterion is
verified in several ways (\S~\ref{sec: backgroundstars}). 
Of the detected components, 77 are likely
background stars, and 74 are candidate companion stars.

The 74 candidate companions occupy the full range in $K_S$ magnitude, down to
$K_S=12~{\rm mag}$. At the distance of Sco~OB2, an M5 main sequence star has a $K_S$ magnitude
of approximately $12~{\rm mag}$. The angular separation
between primary and companion ranges from $0.22''$ to $12.4''$.
These values correspond to orbital periods of several decades to
several thousands of years.

The $J$, $H$, and $K_S$ magnitudes for all observed objects are corrected for
distance and interstellar extinction to find the absolute magnitudes $M_J$,
$M_H$, and $M_{K_S}$. 
The subset of the components with multi-color observations
are plotted in a color-magnitude diagram. 
All (except one) candidate companion stars
are positioned close to the 5~Myr (for US) and 20~Myr (for UCL and LCC)
isochrones. The background stars are positioned far away from the isochrones.

For all observed primaries and companion stars the mass is derived from the
absolute magnitude $M_{K_S}$. The mass of the A and late-B primaries ranges
between $1.4~{\rm M}_\odot$ and $7.7~{\rm M}_\odot$ and 
the mass of their secondaries between
$0.1~{\rm M}_\odot$ and $3.0~{\rm M}_\odot$. Most observed 
companion stars are less massive than
their primaries.
Most of the systems with previously undocumented companion 
stars have a mass ratio smaller than 0.25.
The mass ratio distribution for the observed objects peaks around $q=0.1$ and
decreases for higher mass ratios. The minimum and maximum value for the mass
ratio of the companion stars that we observed are $q=0.022 \pm 0.008$ 
and $1.0 \pm 0.34$. 
For systems with mass ratio $q \approx 1$ we cannot 
say with absolute certainty which is the most 
massive and therefore the primary star.
The mass ratio distribution for binaries with late-B and A type primaries follows 
the distribution $f_q(q)=q^{-\Gamma}$, with
$\Gamma=0.33$. This is similar to the mass ratio distribution observed 
by \cite{shatsky2002} for systems with
B type primaries. Relatively few systems with
low mass ratio are found, excluding random pairing between primaries and
companion stars.
The uncertainty in the age of the UCL and LCC subgroups does not
affect the mass ratio distribution significantly.

A cross-check with visual, astrometric and spectroscopic binaries in literature
shows that 41 of the candidate companion stars are new: 14 in US; 13 in UCL 
and 14 in LCC. The other 33 candidate companions are already documented in
literature.
We analyze the presently known data on binarity and multiplicity in Sco~OB2,
including the 41 new companions that are found in our survey. We conclude that
at least 41\% of all {\it Hipparcos} member stars of Sco~OB2 are 
either double or multiple, and we
find a companion star fraction $F_{\rm C} = 0.52$. These values are lower limits
since the dataset is affected by observational and selection biases.

% The discussion stars here.

Our AO observations are close to completeness in the angular
separation range $1'' \lesssim \rho \lesssim 9''$. 
Next to the 199 target stars we find 50 companion stars
in this angular separation range. This corresponds to an $F_{\rm C}$ of 0.25
in this range of angular separation 
for A and late-B type stars in Sco~OB2.
This value (not corrected for incompleteness) is slightly higher than the values for 
B type stars in Sco~OB2 
\citep[$F_{\rm C}=0.20$ per decade of $\rho$;][]{shatsky2002}
and pre-main sequence stars in 
Sco~OB2 \citep[$F_{\rm C}=0.21$ per decade of $\rho$;][]{koehler2000}.

At small angular separation ($\rho \leq 3.75''$) no components other
than the target stars are detected for magnitudes $12~{\rm mag}\lesssim K_S\lesssim
14~{\rm mag}$. This `gap' cannot be explained by observational biases or low-number
statistics. A fully populated mass spectrum for the companions (assuming
random pairing of binary components from the same underlying IMF) is also
incompatible with this gap. This implies a lower limit on the
companion masses of $\sim 0.1$~M$_\odot$. This is consistent with our finding
that the mass ratio distribution points to a deficit of low-mass companions
compared to the random pairing case.
On the other hand, if we assume that the
sources fainter than $K_S\approx14~{\rm mag}$ are actually physical companions, 
a gap may be present in the companion mass distribution.  
We will carry out follow-up multi-color AO observations to further
investigate this issue.

The gap described above might indicate a {\it brown dwarf desert},
such as observed for solar-type stars
in the solar neighbourhood \citep{duquennoy1991}.
The presence of this gap could support the embryo-ejection 
formation scenario for brown dwarfs
\citep[e.g.][]{reipurth2001,bate2003,kroupabouvier2003}. 
This is further supported by the
detection of 28 candidate free-floating brown dwarf members of Sco~OB2 by
\cite{martin2004}.

The observed $F_{\rm M}$ decreases with decreasing primary
mass. Part of this trend was ascribed by \cite{brown2001} to observational and
selection biases. We lend support to this conclusion 
with our new AO observations. In particular, the
multiplicity fraction for A-type {\it Hipparcos} members of US is doubled.
Nevertheless our knowledge of the present day binary population in Sco~OB2 is
still rather fragmentary. None of the multiplicity surveys of Sco~OB2 so far
has been complete due to practical and time constraints, but also because of a
lack of full knowledge of the membership of the association. In addition, each
observational technique used in these surveys has its own biases. For example,
visual binaries are only detected if the angular separation between the components is
large enough with respect to the luminosity contrast, which means that with
this technique one cannot find the very short-period (spectroscopic)
binaries. Moreover the observational biases depend on the way in which a
particular survey was carried out (including how background stars were weeded
out).

We therefore intend to follow up this observational study by a careful
investigation of the effect of selection biases on the interpretation of the
results. This will be done through a detailed modeling of evolving OB
associations using state-of-the-art N-body techniques coupled with a stellar
and binary evolution code. The synthetic OB association will subsequently be
`observed' by simulating in detail the various binary surveys that have been
carried out. This modeling includes simulated photometry, adaptive optics
imaging and {\it Hipparcos} data, as well as synthetic radial velocity
surveys. Examples of this type of approach can be found in \citet{ecology4}
(for photometric data) and \citet{quist2000} (for {\it Hipparcos} data), where
the latter study very clearly demonstrates that an understanding of the
observational selection biases much enhances the interpretation of binarity
data. Finally, the time dependence of the modeling will also allow us to
investigate to what extent stellar/binary evolution and stellar dynamical effects
have altered the binary population over the lifetime of the association.

This combination of the data on the binary population in Sco OB2 and a
comprehensive modeling of both the association and the observations should
result in the most detailed description of the characteristics of the
primordial binary population to date.

% ====================================================================
% ====================================================================
% ====================================================================
% ==FINAL=STUFF=======================================================
% ====================================================================
% ====================================================================
% ====================================================================

\section*{Acknowledgements}

We thank Francesco Palla for providing the near-infrared pre-main sequence isochrones and evolutionary tracks. We thank the anonymous referee and Andrei Tokovinin for their constructive criticism, which helped to improve the paper. This research was supported by NWO under project number 614.041.006 and the Leids Kerkhoven Bosscha Fonds.

% ====================================================================
% ====================================================================
% ====================================================================
% ==FINAL=STUFF=======================================================
% ====================================================================
% ====================================================================
% ====================================================================

%\begin{landscape}

\addcontentsline{toc}{section}{Appendix A: Results of the ADONIS binarity survey}
\section*{Appendix A: Results of the ADONIS binarity survey}
\markright{Appendix A: Results of the ADONIS binarity survey}

\setlength{\LTcapwidth}{1\textwidth}

\begin{longtable}{|l|rll|lll|ll|}
  \caption{Results from the ADONIS adaptive optics survey among 199 Sco~OB2
    member stars. The columns list: {\it Hipparcos} number, observed $J$, $H$ and $K_S$
    magnitude, spectral type of the primary, angular separation (arcsec),
    position angle (degrees; measured from North to East), and status (p = primary; c =
    companion star; nc = new companion star; b = background star). The last 
    column lists the subgroup of which the system is a member (US = Upper
    Scorpius; UCL = Upper Centaurus Lupus; LCC = Lower Centaurus Crux).
    Non-photometric observations are marked with an asterisk.
    The {\it Hipparcos} member stars HIP77315 and HIP77317
    are a common proper motion pair (see the Double Star Catalog).
    In this paper the two stars are treated as separate primaries and not 
    as one system.  \label{table: results}
  }\\
  \hline
  HIP & $J$ mag & $H$ mag & $K_S$ mag & SpT & $\rho$ ($''$)  & $PA$ ($^\circ$) & status & group \\ 
  \endfirsthead
  \hline
  \multicolumn{9}{|l|}{\tablename\ \thetable{} -- continued from previous page} \\
  \hline
  HIP & $J$ mag & $H$ mag & $K_S$ mag & SpT & $\rho$ ($''$)  & $PA$ ($^\circ$) & status & group \\ 
  \hline
  \endhead
  \hline
  \multicolumn{9}{|l|}{{Continued on next page}} \\ 
  \hline
  \endfoot
  \hline 
  \endlastfoot
  \hline
  \hline 
  50520 &  &  & 6.23 & A1V &  &  & p & LCC \\
 &  &  & 6.39 &  & 2.51 & 313.3 & c &  \\
\hline
52357 &  &  & 7.64 & A3IV &  &  & p & LCC \\
 &  &  & 7.65 &  & 0.53 & 73.0 & c &  \\
 &  &  & 11.45 &  & 10.04 & 72.7 & nc &  \\
\hline
53524 &  &  & 6.76 & A8III &  &  & p & LCC \\
 &  &  & 12.67 &  & 4.87 & 316.9 & b &  \\
\hline
53701 & 6.30 & 6.37 & 6.48 & B8IV &  &  & p & LCC \\
 & 9.05 & 8.76 & 8.86 &  & 3.88 & 75.8 & c &  \\
 & 13.06 & 12.93 & 13.04 &  & 6.57 & 120.1 & b &  \\
\hline
54231 &  &  & 6.75 & A0V &  &  & p & LCC \\
\hline
55188 &  &  & 7.43 & A2V &  &  & p & LCC \\
\hline
55899 &  &  & 7.07 & A0V &  &  & p & LCC \\
\hline
56354 &  &  & 5.78 & A9V &  &  & p & LCC \\
\hline
56379 &  &  & 5.27 & B9Vne &  &  & p & LCC \\
\hline
56963 &  &  & 7.46 & A3V &  &  & p & LCC \\
\hline
56993 &  &  & 7.38 & A0V &  &  & p & LCC \\
 &  &  & 11.88 &  & 1.68 & 23.1 & nc &  \\
\hline
57809 &  &  & 6.61 & A0V &  &  & p & LCC \\
\hline
58416 &  &  & 7.03 & A7V &  &  & p & LCC \\
 &  &  & 8.66 &  & 0.58 & 166.1 & nc &  \\
\hline
58452 &  &  & 6.51 & B8/B9V &  &  & p & LCC \\
\hline
58465 &  &  & 6.32 & A2V &  &  & p & LCC \\
\hline
58720 &  &  & 6.05 & B9V &  &  & p & LCC \\
\hline
58859 &  &  & 6.52 & B9V &  &  & p & LCC \\
\hline
59282 &  &  & 7.00 & A3V &  &  & p & LCC \\
\hline
59397 &  &  & 7.01 & A2V &  &  & p & LCC \\
\hline
59413 &  &  & 7.46 & A6V &  &  & p & LCC \\
 &  &  & 8.18 &  & 3.18 & 99.8 & c &  \\
 &  &  & 15.15 &  & 7.22 & 250.3 & b &  \\
\hline
59502 &  &  & 6.80 & A2V &  &  & p & LCC \\
 &  &  & 11.28 &  & 2.94 & 26.4 & nc &  \\
 &  &  & 12.79 &  & 9.05 & 308.6 & b &  \\
\hline
59898 &  &  & 5.99 & A0V &  &  & p & LCC \\
\hline
60084 &  &  & 7.65 & A1V &  &  & p & LCC \\
 &  &  & 10.10 &  & 0.46 & 329.6 & nc &  \\
\hline
60183 &  &  & 6.33 & B9V &  &  & p & LCC \\
\hline
60561 &  &  & 6.59 & A0V &  &  & p & LCC \\
\hline
60851 &  &  & 5.98 & A0Vn &  &  & p & LCC \\
 &  &  & 11.04 &  & 2.01 & 44.1 & nc &  \\
 &  &  & 11.66 &  & 6.92 & 181.1 & nc &  \\
\hline
61257 &  &  & 6.60 & B9V &  &  & p & LCC \\
 &  &  & 12.43 &  & 5.54 & 324.3 & b &  \\
\hline
61265 &  &  & 7.44 & A2V &  &  & p & LCC \\
 &  &  & 11.29 &  & 2.51 & 67.4 & nc &  \\
 &  &  & 14.70 &  & 3.43 & 168.6 & b &  \\
\hline
61639 &  &  & 6.94 & A1/A2V &  &  & p & LCC \\
 &  &  & 7.06 &  & 1.87 & 182.4 & c &  \\
 &  &  & 14.36 &  & 4.21 & 220.6 & b &  \\
\hline
61782 &  &  & 7.56 & A0V &  &  & p & LCC \\
\hline
61796 &  &  & 6.37 & B8V &  &  & p & LCC \\
 &  &  & 11.79 &  & 9.89 & 109.0 & nc &  \\
 &  &  & 11.86 &  & 12.38 & 136.8 & nc &  \\
 &  &  & 13.35 &  & 4.17 & 305.8 & b &  \\
\hline
62002 &  &  & 7.09 & A1V &  &  & p & LCC \\
 &  &  & 7.65 &  & 0.38 & 69.2 & c &  \\
 &  &  & 13.95 &  & 5.33 & 214.6 & b &  \\
\hline
62026 &  &  & 6.49 & B9V &  &  & p & LCC \\
 &  &  & 7.46 &  & 0.22 & 12.5 & nc &  \\
\hline
62058 &  &  & 6.17 & B9Vn &  &  & p & LCC \\
\hline
62179 &  &  & 7.20 & A0IV/V &  &  & p & LCC \\
 &  &  & 7.57 &  & 0.23 & 282.7 & c &  \\
 &  &  & 14.03 &  & 12.21 & 131.3 & b &  \\
 &  &  & 14.60 &  & 4.47 & 255.8 & b &  \\
\hline
63204 &  &  & 6.47 & A0p &  &  & p & LCC \\
 &  &  & 7.36 &  & 1.80 & 46.9 & nc &  \\
\hline
63236 &  &  & 6.66 & A2IV/V &  &  & p & LCC \\
 &  &  & 12.91 &  & 7.56 & 118.4 & b &  \\
 &  &  & 12.85 &  & 11.97 & 316.5 & b &  \\
\hline
63839 &  &  & 6.66 & A0V &  &  & p & LCC \\
 &  &  & 13.78 &  & 5.99 & 358.1 & b &  \\
 &  &  & 14.21 &  & 4.30 & 13.7 & b &  \\
 &  &  & 13.16 &  & 6.31 & 300.9 & b &  \\
\hline
64320 &  &  & 6.22 & Ap &  &  & p & LCC \\
\hline
64515 &  &  & 6.78 & B9V &  &  & p & LCC \\
 &  &  & 6.94 &  & 0.31 & 165.7 & c &  \\
\hline
64892 &  &  & 6.82 & B9V &  &  & p & LCC \\
\hline
64925 &  &  & 6.88 & A0V &  &  & p & LCC \\
\hline
64933 &  &  & 6.29 & A0V &  &  & p & LCC \\
\hline
65021 &  &  & 7.26 & B9V &  &  & p & LCC \\
\hline
65089 &  &  & 7.37 & A7/A8V &  &  & p & LCC \\
\hline
65178 &  &  & 6.71 & B9V &  &  & p & LCC \\
\hline
65219 &  &  & 6.52 & A3/A4III/IV &  &  & p & LCC \\
\hline
65394 &  &  & 7.25 & A1Vn... &  &  & p & LCC \\
\hline
65426 &  &  & 6.78 & A2V &  &  & p & LCC \\
\hline
65822 &  &  & 6.68 & A1V &  &  & p & LCC \\
 &  &  & 11.08 &  & 1.82 & 303.9 & nc &  \\
\hline
65965 &  &  & 7.51 & B9V &  &  & p & LCC \\
 &  &  & 15.21 &  & 10.28 & 41.1 & b &  \\
\hline
66068 &  &  & 7.04 & A1/A2V &  &  & p & LCC \\
\hline
66447 &  &  & 7.16 & A3IV/V &  &  & p & UCL \\
\hline
66454 &  &  & 6.33 & B8V &  &  & p & LCC \\
\hline
66566 &  &  & 7.36 & A1V &  &  & p & LCC \\
\hline
66651 &  &  & 7.35 & B9.5V &  &  & p & LCC \\
 &  &  & 15.36 &  & 7.58 & 173.3 & b &  \\
\hline
66722 &  &  & 6.32 & A0V &  &  & p & UCL \\
\hline
66908 &  &  & 6.86 & A4V &  &  & p & UCL \\
\hline
67036 &  &  & 6.69 & A0p &  &  & p & LCC \\
\hline
67260 &  &  & 7.03 & A0V &  &  & p & LCC \\
 &  &  & 8.37 &  & 0.44 & 228.9 & c &  \\
 &  &  & 14.70 &  & 2.22 & 77.1 & b &  \\
\hline
67919 &  &  & 6.58 & A9V &  &  & p & LCC \\
 &  &  & 9.05 &  & 0.70 & 299.1 & nc &  \\
\hline
68080 &  &  & 6.28 & A1V &  &  & p & UCL \\
 &  &  & 7.19 &  & 1.92 & 10.2 & c &  \\
\hline
68532 &  &  & 7.03 & A3IV/V &  &  & p & UCL \\
 &  &  & 9.50 &  & 3.03 & 288.6 & nc &  \\
 &  &  & 10.53 &  & 3.15 & 291.7 & nc &  \\
\hline
68781 &  &  & 7.38 & A2V &  &  & p & UCL \\
\hline
68867 &  &  & 7.17 & A0V &  &  & p & UCL \\
 &  &  & 11.61 &  & 2.16 & 284.8 & nc &  \\
\hline
68958 &  &  & 6.72 & Ap... &  &  & p & UCL \\
\hline
69113 & 6.25 & 6.32 & 6.43 & B9V &  &  & p & UCL \\
 & 11.14 & 10.51 & 10.25 &  & 5.33 & 64.8 & nc &  \\
 & 11.19 & 10.46 & 10.23 &  & 5.52 & 66.9 & nc &  \\
\hline
69749 &  &  & 6.62 & B9IV &  &  & p & UCL \\
 &  &  & 11.60 &  & 1.50 & 0.8 & nc &  \\
 &  &  & 12.69 &  & 8.10 & 50.0 & b &  \\
 &  &  & 13.40 &  & 5.57 & 352.4 & b &  \\
 &  &  & 14.16 &  & 8.79 & 62.5 & b &  \\
 &  &  & 14.48 &  & 9.13 & 66.8 & b &  \\
 &  &  & 14.51 &  & 4.43 & 278.6 & b &  \\
\hline
69845 &  &  & 7.78 & B9V &  &  & p & UCL \\
\hline
70441 &  &  & 7.31 & A1V &  &  & p & UCL \\
\hline
70455 &  &  & 7.07 & B8V &  &  & p & UCL \\
\hline
70626 &  &  & 6.56 & B9V &  &  & p & UCL \\
\hline
70690 &  &  & 7.71 & B9V &  &  & p & UCL \\
\hline
70697 &  &  & 7.17 & A0V &  &  & p & UCL \\
\hline
70809 &  &  & 6.54 & Ap... &  &  & p & UCL \\
 &  &  & 14.64 &  & 4.97 & 214.4 & b &  \\
 &  &  & 14.60 &  & 8.72 & 297.3 & b &  \\
\hline
70904 &  &  & 6.39 & A6V &  &  & p & UCL \\
 &  &  & 12.08 &  & 6.08 & 120.3 & b &  \\
 &  &  & 14.17 &  & 10.03 & 309.2 & b &  \\
\hline
70918 &  &  & 6.35 & A0/A1V &  &  & p & UCL \\
\hline
70998 &  &  & 7.06 & A1V &  &  & p & UCL \\
 &  &  & 10.83 &  & 1.17 & 354.6 & nc &  \\
\hline
71140 &  &  & 7.13 & A7/A8IV &  &  & p & UCL \\
\hline
71271 &  &  & 7.57 & A0V &  &  & p & UCL \\
\hline
71321 &  &  & 7.17 & A9V &  &  & p & UCL \\
\hline
71724 &  &  & 6.79 & B8/B9V &  &  & p & UCL \\
 &  &  & 9.70 &  & 8.66 & 23.0 & c &  \\
\hline
71727 &  &  & 6.89 & A0p &  &  & p & UCL \\
 &  &  & 7.80 &  & 9.14 & 245.0 & c &  \\
\hline
72140 &  &  & 7.09 & A1IV/V &  &  & p & UCL \\
 &  &  & 12.21 &  & 4.51 & 229.1 & b &  \\
\hline
72192 & 6.71 & 6.71 & 6.71 & A0V &  &  & p & UCL \\
\hline
72627 & 6.54 & 6.54 & 6.53 & A2V &  &  & p & UCL \\
\hline
72940 &  &  & 6.85 & A1V &  &  & p & UCL \\
 &  &  & 8.57 &  & 3.16 & 221.6 & c &  \\
\hline
72984 &  &  & 7.05 & A0/A1V &  &  & p & UCL \\
 &  &  & 14.37 &  & 5.83 & 118.1 & b &  \\
 &  &  & 8.50 &  & 4.71 & 260.3 & c &  \\
\hline
73145 &  &  & 7.54 & A2IV &  &  & p & UCL \\
\hline
73266 &  &  & 7.30 & B9V &  &  & p & UCL \\
\hline
73341 &  &  & 6.67 & B8V &  &  & p & UCL \\
\hline
73393 &  &  & 7.21 & A0V &  &  & p & UCL \\
\hline
73937 &  &  & 6.05 & Ap &  &  & p & UCL \\
 &  &  & 14.05 &  & 3.48 & 30.5 & b &  \\
\hline
74066 &  &  & 6.08 & B8IV &  &  & p & UCL \\
 &  &  & 8.43 &  & 1.22 & 109.6 & nc &  \\
\hline
74100 &  &  & 6.12 & B7V &  &  & p & UCL \\
\hline
74479 &  &  & 6.31 & B8V &  &  & p & UCL \\
 &  &  & 10.83 &  & 4.65 & 154.1 & nc &  \\
\hline
74657 &  &  & 6.97 & B9IV &  &  & p & UCL \\
\hline
74752 &  &  & 6.84 & B9V &  &  & p & UCL \\
 &  &  & 13.13 &  & 9.66 & 21.0 & b &  \\
\hline
74797 &  &  & 7.55 & A2IV &  &  & p & UCL \\
\hline
74985 &  &  & 7.53 & A0V &  &  & p & UCL \\
 &  &  & 13.13 &  & 6.37 & 145.2 & b &  \\
\hline
75056 &  &  & 7.31 & A2V &  &  & p & UCL \\
 &  &  & 11.17 &  & 5.19 & 34.5 & nc &  \\
\hline
75077 &  &  & 6.97 & A1V &  &  & p & UCL \\
\hline
75151 &  &  & 6.65 & A+... &  &  & p & UCL \\
 &  &  & 8.09 &  & 5.70 & 120.9 & c &  \\
\hline
75210 &  &  & 6.82 & B8/B9V &  &  & p & UCL \\
\hline
75476 &  &  & 6.88 & A1/A2V &  &  & p & UCL \\
\hline
75509 &  &  & 7.40 & A2V &  &  & p & UCL \\
\hline
75647 &  &  & 5.86 & B5V &  &  & p & UCL \\
\hline
75915 &  &  & 6.44 & B9V &  &  & p & UCL \\
 &  &  & 8.15 &  & 5.60 & 229.4 & c &  \\
\hline
75957 &  &  & 7.24 & A0V &  &  & p & UCL \\
 &  &  & 13.41 &  & 5.56 & 105.7 & b &  \\
 &  &  & 13.21 &  & 9.21 & 227.1 & b &  \\
\hline
76001 &  &  & 7.60 & A2/A3V &  &  & p & UCL \\
 &  &  & 7.80 &  & 0.25 & 3.2 & c &  \\
 &  &  & 8.20 &  & 1.48 & 124.8 & c &  \\
 &  &  & 12.85 &  & 6.58 & 127.5 & b &  \\
\hline
76048 &  &  & 6.26 & B6/B7V &  &  & p & UCL \\
\hline
76071 & 7.05 & 7.10 & 7.06 & B9V &  &  & p & US \\
 &  & 11.28 & 10.87 &  & 0.69 & 40.8 & nc &  \\
\hline
76310 &  &  & 7.35 & A0V &  &  & p & US \\
\hline
76503 &  &  & 6.25 & B9IV &  &  & p & US \\
\hline
76633 &  &  & 7.51 & B9V &  &  & p & US \\
\hline
77150 &  &  & 7.28 & A2V &  &  & p & UCL \\
\hline
77295 &  &  & 7.64 & A2IV/V &  &  & p & UCL \\
 &  &  & 15.13 &  & 4.63 & 309.9 & b &  \\
\hline
77315 &  &  & 7.24 & A0V   &       &       & p & UCL \\
 &  &  & 7.92 &  & 0.68 & 67.0 & c &  \\
77317 &  &  & 7.37 & B9.5V & 37.37 & 137.3 & p & UCL \\
 &  &  & 14.24 &  & 32.78 & 146.8 & b &  \\
%77317 &  &  & 7.37 & B9.5V &  &  & p & UCL \\
% &  &  & 14.24 &  & 7.37 & 270.3 & b &  \\
%
%
\hline
77457 &  &  & 7.33 & A7IV &  &  & p & US \\
\hline
77523 &  &  & 7.41 & B9V &  &  & p & UCL \\
\hline
77858 &  &  & 5.40 & B5V &  &  & p & US \\
\hline
77859 &  &  & 5.57 & B2V &  &  & p & US \\
\hline
77900 &  &  & 6.44 & B7V &  &  & p & US \\
\hline
77909 &  &  & 6.19 & B8III/IV &  &  & p & US \\
\hline
77911 & 6.67 & 6.71 & 6.68 & B9V &  &  & p & US \\
 & 12.68 & 12.20 & 11.84 &  & 7.96 & 279.3 & nc &  \\
\hline
77939 &  &  & 6.56 & B2/B3V &  &  & p & US \\
 &  &  & 8.09 &  & 0.52 & 119.1 & c &  \\
\hline
77968 &  &  & 7.00 & B8V &  &  & p & UCL \\
 &  &  & 12.64 &  & 6.54 & 344.4 & b &  \\
 &  &  & 14.56 &  & 6.43 & 349.8 & b &  \\
\hline
78099 &  &  & 7.35 & A0V &  &  & p & US \\
\hline
78168 &  &  & 5.91 & B3V &  &  & p & US \\
\hline
78196 &  &  & 7.08 & A0V &  &  & p & US \\
\hline
78246 &  &  & 5.84 & B5V &  &  & p & US \\
\hline
78494 &  &  & 7.11 & A2m... &  &  & p & US \\
\hline
78530 & 6.87 & 6.92 & 6.87 & B9V &  &  & p & US \\
 &  & 14.56 & 14.22 &  & 4.54 & 139.7 & b &  \\
\hline
78533 &  &  & 6.99 & Ap &  &  & p & UCL \\
 &  &  & 12.28 &  & 6.09 & 186.3 & b &  \\
\hline
78541 &  &  & 6.99 & A0V &  &  & p & UCL \\
\hline
78549 &  &  & 7.13 & B9.5V &  &  & p & US \\
 &  &  & 14.62 &  & 11.78 & 47.3 & b &  \\
\hline
78663 &  &  & 7.76 & F5V &  &  & p & US \\
 &  &  & 13.42 &  & 8.88 & 103.0 & b &  \\
 &  &  & 15.41 &  & 6.11 & 184.7 & b &  \\
\hline
78702 &  &  & 7.41 & B9V &  &  & p & US \\
\hline
78754 &  &  & 6.95 & B8/B9V &  &  & p & UCL \\
\hline
78756 &  &  & 7.16 & Ap &  &  & p & UCL \\
 &  &  & 9.52 &  & 8.63 & 216.4 & c &  \\
\hline
78809 & 7.41 & 7.50 & 7.51 & B9V &  &  & p & US \\
 & 11.08 & 10.45 & 10.26 &  & 1.18 & 25.7 & nc &  \\
\hline
78847 &  &  & 7.32 & A0V &  &  & p & US \\
 &  &  & 11.30 &  & 8.95 & 164.0 & nc &  \\
\hline
78853 &  &  & 7.50 & A5V &  &  & p & UCL \\
 &  &  & 8.45 &  & 1.99 & 270.4 & c &  \\
 &  &  & 15.02 &  & 7.12 & 84.2 & b &  \\
\hline
78877 &  &  & 6.08 & B8V &  &  & p & US \\
\hline
78956 & 7.52 & 7.54 & 7.57 & B9.5V &  &  & p & US \\
 & 9.76 & 9.12 & 9.04 &  & 1.02 & 48.7 & nc &  \\
\hline
78968 & 7.40 & 7.41 & 7.47 & B9V &  &  & p & US \\
 &  & 14.59 & 14.47 &  & 2.75 & 321.1 & b &  \\
\hline
78996 &  &  & 7.46 & A9V &  &  & p & US \\
\hline
79031 &  &  & 7.00 & B8IV/V &  &  & p & US \\
\hline
79044 &  &  & 6.91 & B9V &  &  & p & UCL \\
 &  &  & 15.18 &  & 5.02 & 91.6 & b &  \\
\hline
79098 &  &  & 5.90 & B9V &  &  & p & US \\
 &  &  & 14.10 &  & 2.37 & 116.8 & b &  \\
\hline
79124 & 7.16 & 7.14 & 7.13 & A0V &  &  & p & US \\
 & 11.38 & 10.55 & 10.38 &  & 1.02 & 96.2 & nc &  \\
\hline
79156 & 7.56 & 7.56 & 7.61 & A0V &  &  & p & US \\
 & 11.62 & 10.89 & 10.77 &  & 0.89 & 58.9 & nc &  \\
\hline
79250 &  &  & 7.49 & A3III/IV &  &  & p & US \\
 &  &  & 10.71 &  & 0.62 & 180.9 & nc &  \\
\hline
79366 &  &  & 7.47 & A3V &  &  & p & US \\
\hline
79410 &  &  & 7.05 & B9V &  &  & p & US \\
 &  &  & 14.93 &  & 3.17 & 339.8 & b &  \\
\hline
79439 &  &  & 6.97 & B9V &  &  & p & US \\
\hline
79530 &  &  & 6.60 & B6IV &  &  & p & US \\
 &  &  & 8.34 &  & 1.69 & 219.7 & c &  \\
\hline
79599 &  &  & 6.30 & B9V &  &  & p & US \\
\hline
79622 &  &  & 6.34 & B8V &  &  & p & US \\
\hline
79631 &  &  & 7.17 & B9.5V &  &  & p & UCL \\
 &  &  & 7.61 &  & 2.94 & 127.9 & nc &  \\
 &  &  & 14.08 &  & 8.86 & 151.8 & b &  \\
\hline
79739 &  &  & 7.24 & B8V &  &  & p & US \\
 &  &  & 11.60 &  & 0.94 & 118.6 & nc &  \\
\hline
79771 &  &  & 7.09 & B9V &  &  & p & US \\
 &  &  & 10.94 &  & 3.66 & 313.6 & nc &  \\
\hline
79785 &  &  & 6.41 & B9V &  &  & p & US \\
\hline
79860 &  &  & 7.88 & A0V &  &  & p & US \\
\hline
79878 &  &  & 7.06 & A0V &  &  & p & US \\
\hline
79897 &  &  & 6.99 & B9V &  &  & p & US \\
\hline
80019 &  &  & 7.08 & A0V &  &  & p & US \\
\hline
80024 &  &  & 6.73 & B9II/III &  &  & p & US \\
\hline
80059 &  &  & 7.44 & A7III/IV &  &  & p & US \\
\hline
80126 &  &  & 6.44 & B6/B7Vn &  &  & p & US \\
\hline
80142 &  &  & 6.60 & B7V &  &  & p & UCL \\
 &  &  & 12.16 &  & 8.54 & 44.0 & b &  \\
 &  &  & 9.53 &  & 9.32 & 216.2 & c &  \\
\hline
80238 & 7.45 & 7.45 & 7.34 & A1III/IV &  &  & p & US \\
 & 7.96 & 7.66 & 7.49 &  & 1.03 & 318.5 & c &  \\
\hline
80324 &  &  & 7.33 & A0V+... &  &  & p & US \\
 &  &  & 7.52 &  & 6.23 & 152.5 & c &  \\
\hline
80371 &  &  & 6.40 & B5III &  &  & p & US \\
 &  &  & 8.92 &  & 2.73 & 140.6 & c &  \\
 &  &  & 13.36 &  & 9.22 & 32.0 & b &  \\
\hline
80425 &  &  & 7.40 & A1V &  &  & p & US \\
 &  &  & 8.63 &  & 0.60 & 155.8 & c &  \\
\hline
80461 &  &  & 5.92 & B3/B4V &  &  & p & US \\
 &  &  & 7.09 &  & 0.27 & 285.6 & c &  \\
\hline
80474 & 6.13$^\ast$  & 6.08$^\ast$  & 5.76  & B5V &  &  & p & US  \\
         & 13.00$^\ast$ & 12.30$^\ast$ & 11.71 &  & 4.83 & 206.2 & nc & \\
\hline
80493 &  &  & 7.05 & B9V &  &  & p & US \\
\hline
80591 &  &  & 7.82 & A5V &  &  & p & UCL \\
\hline
80799 & 7.68$^\ast$ & 7.75$^\ast$ & 7.46 & A2V &  &  & p & US  \\
         & 10.94$^\ast$ & 10.35$^\ast$ & 9.88 &  & 2.94 & 205.2 & nc &   \\
\hline
80896 &  &  & 7.53 & F3V &  &  & p & US \\
 &  &  & 10.51 &  & 2.28 & 177.0 & nc &  \\
\hline
80897 &  &  & 7.78 & A0V &  &  & p & UCL \\
\hline
81136 &  &  & 5.21 & A7/A8+... &  &  & p & UCL \\
\hline
81316 &  &  & 6.72 & B9V &  &  & p & UCL \\
\hline
81472 &  &  & 5.96 & B2.5IV &  &  & p & UCL \\
 &  &  & 13.23 &  & 5.21 & 357.5 & b &  \\
 &  &  & 13.20 &  & 4.52 & 274.4 & b &  \\
\hline
81474 &  &  & 5.74 & B9.5IV &  &  & p & US \\
\hline
81624 &  &  & 5.80 & A1V &  &  & p & US \\
 &  &  & 7.95 &  & 1.13 & 224.3 & c &  \\
\hline
81751 &  &  & 8.29 & A9V &  &  & p & UCL \\
 &  &  & 12.18 &  & 8.91 & 68.0 & b &  \\
 &  &  & 14.44 &  & 6.12 & 219.6 & b &  \\
\hline
81914 &  &  & 6.34 & B6/B7V &  &  & p & UCL \\
 &  &  & 12.29 &  & 6.18 & 49.5 & b &  \\
 &  &  & 14.28 &  & 9.11 & 39.8 & b &  \\
 &  &  & 14.61 &  & 5.15 & 286.2 & b &  \\
\hline
81949 &  &  & 7.32 & A3V &  &  & p & UCL \\
 &  &  & 13.23 &  & 3.90 & 89.5 & b &  \\
 &  &  & 14.05 &  & 3.42 & 28.7 & b &  \\
 &  &  & 14.70 &  & 9.60 & 76.6 & b &  \\
 &  &  & 14.63 &  & 6.31 & 239.0 & b &  \\
 &  &  & 14.62 &  & 5.69 & 292.2 & b &  \\
 &  &  & 15.15 &  & 5.22 & 340.5 & b &  \\
 &  &  & 15.33 &  & 9.70 & 345.5 & b &  \\
\hline
81972 &  &  & 5.92 & B3V &  &  & p & UCL \\
 &  &  & 10.57 &  & 2.01 & 312.0 & nc &  \\
 &  &  & 10.75 &  & 7.07 & 258.4 & c &  \\
 &  &  & 11.54 &  & 5.03 & 213.7 & nc &  \\
\hline
82154 & 6.89$^\ast$ & 7.22$^\ast$ & 7.05 & B9IV/V &  &  & p & UCL  \\
 & 14.86$^\ast$ & 14.93$^\ast$ & 14.47 &  & 8.39 & 359.0 & b &   \\
\hline
82397 &  &  & 7.28 & A3V &  &  & p & US \\
 &  &  & 15.36 &  & 7.88 & 227.7 & b &  \\
\hline
82430 &  &  & 7.25 & B9V &  &  & p & UCL \\
 &  &  & 12.41 &  & 4.59 & 96.1 & b &  \\
 &  &  & 14.03 &  & 5.98 & 65.3 & b &  \\
 &  &  & 14.29 &  & 6.08 & 329.3 & b &  \\
\hline
82560 & 6.83$^\ast$ & 6.97$^\ast$ & 6.58 & A0V &  &  & p & UCL  \\
 &  & 13.86$^\ast$ & 12.76 &  & 4.73 & 4.6 & b &   \\
 &  & 14.40$^\ast$ & 13.13 &  & 3.94 & 222.0 & b &   \\
 &  & 14.38$^\ast$ & 13.68 &  & 6.20 & 283.3 & b &   \\
\hline
83457 &  &  & 6.49 & A9V &  &  & p & UCL \\
\hline
83542 &  &  & 5.38 & G8/K0III &  &  & p & US \\
 &  &  & 10.01 &  & 8.96 & 196.1 & nc &  \\
\hline
83693 &  &  & 5.69 & A2IV &  &  & p & UCL \\
 &  &  & 9.26 &  & 5.82 & 78.4 & c &  \\
 &  &  & 13.64 &  & 12.69 & 134.9 & b &  \\

\end{longtable}

\setlength{\LTcapwidth}{1.3\textwidth}

\chapter{A brown dwarf desert for intermediate mass stars in Sco~OB2?} \label{chapter: naco}

 \author{M.B.N. Kouwenhoven\inst{1}
          \and
          A.G.A. Brown\inst{2}
          \and
          L. Kaper\inst{1}
          }

\begin{center}
M.B.N. Kouwenhoven, A.G.A. Brown, \& L. Kaper

\vspace{0.2cm}
{\it Astronomy \& Astrophysics}, submitted
\end{center}

% ====================================================================
% ====================================================================
% ====================================================================
% ==ABSTRACT==========================================================
% ====================================================================
% ====================================================================
% ====================================================================

\section*{Abstract}

We present $JHK_S$ observations of 22 intermediate-mass stars in the
Scorpius-Centaurus OB~association, obtained with the NAOS/CONICA system at the
ESO Very Large Telescope. This survey was performed to determine the status of (sub)stellar candidate companions of Sco~OB2 member stars of spectral type A and
late-B. The distinction between companions and background stars 
is made on the basis of a comparison to isochrones and additional statistical arguments. We are sensitive to companions with 
an angular separation of $0.1''-11''$ ($13-1430$~AU) and the detection limit is $K_S=17$~mag.
 We detect 62 stellar components of which 18 turn out to be physical companions, 11~candidate companions, and 33~background stars. Three of the 18 confirmed companions were previously undocumented
as such. The companion masses are in the range $0.03~{\rm M}_\odot
\leq M \leq 1.19~{\rm M}_\odot$, corresponding to mass ratios $0.06 \leq q \leq
0.55$.  
We include in our sample a subset of 9~targets with multi-color ADONIS observations from \cite{kouwenhoven2005}. In the ADONIS survey secondaries with $K_S < 12$~mag were classified as companions; those with $K_S > 12$~mag as background stars. The multi-color analysis in this paper demonstrates that the simple $K_S=12$~mag criterion correctly classifies the secondaries in $\sim 80\%$ of the cases. We reanalyse the total sample (i.e. NAOS/CONICA and ADONIS) and conclude that of the 176~secondaries, 25~are physical companions, 55~are candidate companions, and 96~are background stars. Although we are sensitive (and complete) to brown dwarf companions as faint as $K_S=14$~mag in the semi-major axis range $130-520$~AU, we detect only one, corresponding to a brown dwarf companion fraction of $0.5 \pm 0.5\%$ ($M \ga 30~{\rm M_J}$). 
However, the number of brown dwarfs is consistent with an extrapolation of the (stellar) companion mass distribution into the brown dwarf regime. This suggests that the brown dwarf desert (discovered in planet searches around solar-type stars) has to be explained by an excess of planetary companions, rather than a lack of brown dwarf companions. Such a conclusion indicates that the physical mechanism for the formation of brown dwarf companions is similar to that of stellar companions, rather than that of planetary companions.

% ====================================================================
% ====================================================================
% ====================================================================
% ==INTRODUCTION======================================================
% ====================================================================
% ====================================================================
% ====================================================================

\section{Introduction}

The predominance of star formation in binary or multiple systems inside
stellar clusters makes the binarity and multiplicity of newly born
stars one of the most sensitive probes of the process of star and star
cluster formation \citep[see][ and references
therein]{blaauw1991}. Ideally one would like to have detailed
knowledge of the binary population at the time that the stars are being
formed. However, this is difficult to achieve in practice and therefore we have
embarked on a project to characterize the observationally better accessible
``primordial binary population'', which is defined as {\em the
population of binaries as established just after the gas has been removed from
the forming system, i.e., when the stars can no longer accrete gas from their
surroundings} \citep{kouwenhoven2005}. We chose to focus our efforts on the
accurate characterization of the binary population in nearby OB
associations. The youth and low stellar density of OB~associations ensure that their
binary population is very similar to the primordial binary population. We
refer to \citet{kouwenhoven2005} for a more extensive discussion and
motivation of this project.

Our initial efforts are concentrated on the Sco~OB2 association. Sco~OB2 consists of the three
subgroups Upper Scorpius (US), Upper Centaurus Lupus (UCL), and Lower Centaurus Crux (LCC). The properties of the subgroups are listed in Table~\ref{table: statistics}. 
Its stellar
population is accurately known down to late A-stars thanks to the
\textit{Hipparcos} catalogue \citep{dezeeuw1999}, and extensive
literature data is available on its binary population \citep{brown2001}. In addition
Sco~OB2 has recently been the target of an adaptive optics survey of its
\textit{Hipparcos} B-star members \citep{shatsky2002}. 
We have conducted our own adaptive optics survey of 199 A-type
and late-B type stars in this association \citep{kouwenhoven2005} using 
the ADONIS instrument, which was mounted on the ESO 3.6~meter telescope at La~Silla, Chile. We performed these observations in the $K_S$-band (and for a subset of the targets additionally in the $J$ and $H$ band). We detected 151 stellar
components other than the target stars and used a simple brightness criterion
to separate background stars\footnote{When mentioning ``background star'',
we refer to any stellar object that does not belong to the system, including
foreground stars.} from physical companions. All
components fainter than $K_S=12$~mag were considered background stars; all brighter components were identified as candidate companion stars \citep[see also][]{shatsky2002}. Of the 74
candidate physical companions 33 were known already and 41 were new candidate companions.

In examining the binary properties of our sample of A and late B-stars we
noticed that at small angular separations ($\leq 4$~arcsec) no companions
fainter than $K_S \approx 12$~mag are present, assuming that the sources
fainter than $K_S \approx 14$~mag are background stars (as we had no information on their colors). The absence of companions with $K_S > 12$~mag and $\rho < 4''$ is clearly visible in Figure~3 of \cite{kouwenhoven2005}. 
This result implies that A and B stars do not
have close companions with masses less than about 0.08~M$_{\odot}$, unless the
assumed background stars {\em are} physical companions. In the latter case the
close faint sources would be brown dwarfs \cite[which are known to be present
in Sco~OB2; see][]{martin2004} and a gap would exist in the companion mass
distribution. In either case a peculiar feature would be present in the mass
distribution of companions which has to be explained by the binary
formation history.

We decided to carry out follow-up multi-color observations in order to (1) determine the reliability of our $K_S=12$~mag criterion to separate companions and background stars, (2) investigate the potential gap or lower limit of the companion mass distribution, and (3) search for additional close and/or faint companions.

These follow-up near-infrared observations were conducted with NAOS/CONICA (NACO) on the ESO Very Large Telescope at Paranal, Chile. We obtained $JHK_S$ photometric observations of 22~A and late-B members in Sco~OB2 and their secondaries\footnote{We use the term ``secondary'' for any stellar component in the field near the target star. A secondary can be a companion star or a background star.}. In Section~\ref{section: observationsanddatareduction} we describe our NACO sample, the observations, the data reduction procedures, and photometric accuracy of the observations. 
In Section~\ref{section: detectionlimits} we describe the detection limit and completeness limit of the ADONIS and NACO observations. 
In Section~\ref{section: status} we determine the status (companion or background star) of the secondaries with multi-color observations. We perform the analysis for the secondaries around the 22~NACO targets, and for those around the 9~targets in the ADONIS sample for which we have multi-color observations. In Section~\ref{section: status} we also analyze the background star statistics, and we evaluate the accuracy of the $K_S=12$~mag separation criterion. In Section~\ref{section: massfunction} we derive for each companion its mass and mass ratio.
In Section~\ref{section: gap} we discuss the lack of brown dwarf companions with separations between $1''$ and $4''$ (130$-$520~AU) in our sample, and discuss the existence of the brown dwarf desert for A~and late-B type members of Sco~OB2. 
Finally, we present updated binary statistics of the Sco~OB2 association in Section~\ref{section: binarystatistics} and summarize our results in Section~\ref{section: conclusion}.

\begin{figure}[btp]
  \begin{tabular}{ccc}
    \includegraphics[width=0.3\textwidth,height=!]{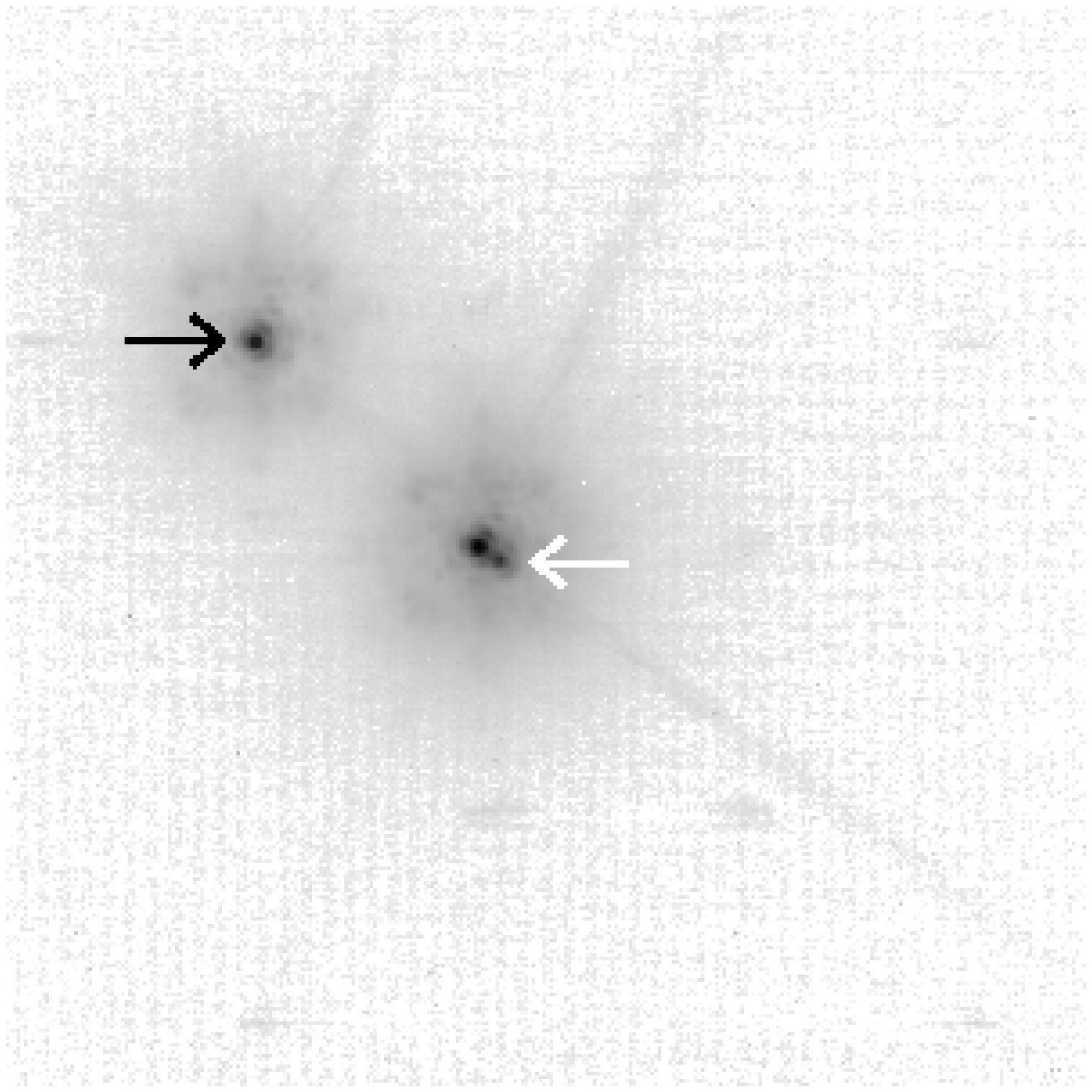} &
    \includegraphics[width=0.3\textwidth,height=!]{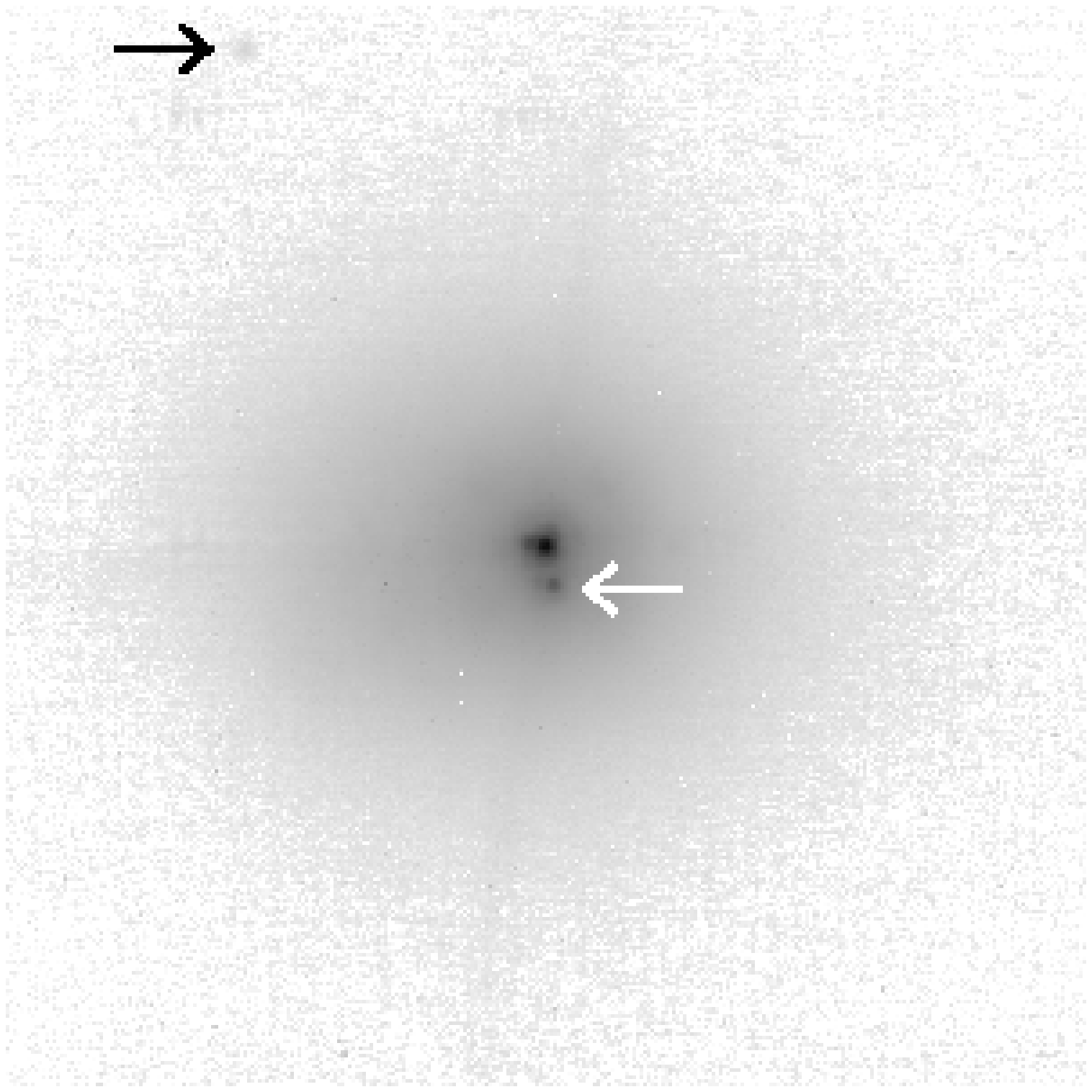} &
    \includegraphics[width=0.3\textwidth,height=!]{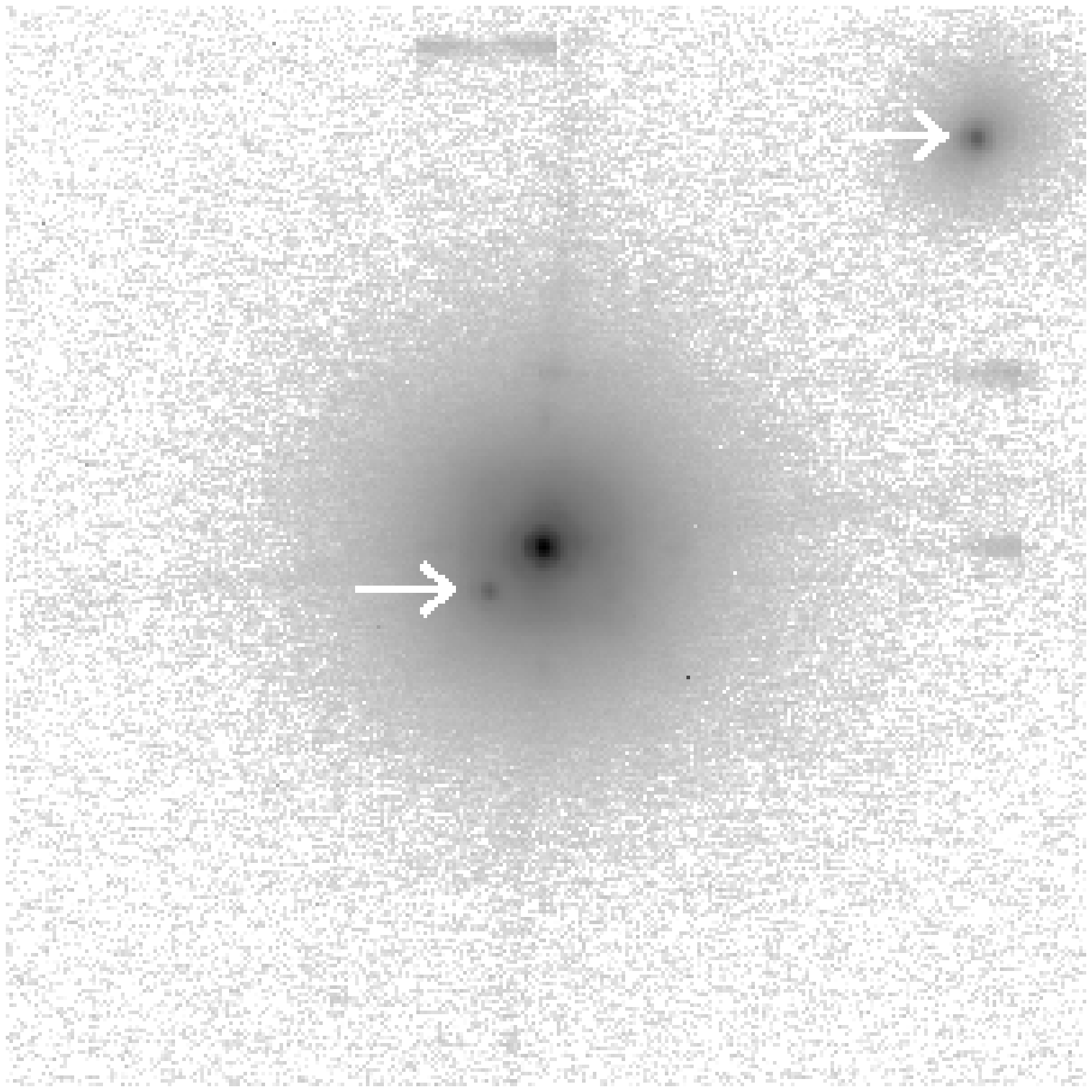} \\
  \end{tabular}
  \caption{With our NACO survey we find three close companions (shown in the
    figure),
    which were not detected in our ADONIS survey. The panels
    ($6.6'' \times 6.6'')$ are centered on the primary
    stars. {\em Left:} The binary HIP63204 in $K_S$, with a companion
    at angular separation $0.15''$ and a background star 
    at angular separation $1.87''$. {\em
    Middle:} The binary HIP73937 in $K_S$, with a close companion at $\rho=0.24''$ and a 
    background star at
    $\rho=3.56''$. {\em Right:} HIP79771 in $K_S$, with two companion stars at
    $\rho=0.44''$ and $\rho=3.67''$. Companions are indicated with white
    arrows and background stars with black arrows.
    Several artifacts are visible in the fields of HIP63204 and HIP79771, which can easily be recognized as such. The panels show a subset of the total field of
    view for each observation, which is $14'' \times 14''$. For these three targets we observe no
    stellar components other than those shown in the panels.  
    \label{figure: triples}}     
\end{figure}

% ====================================================================
% ====================================================================
% ====================================================================
% ==OBSERVATIONS======================================================
% ====================================================================
% ====================================================================
% ====================================================================

\section{Observations and data reduction} \label{section: observationsanddatareduction}

\subsection{Definition of the NACO sample}\label{section: sample}

\begin{table}[tbp]
  \begin{tabular}{|cc|cc|cll|}
    \hline
    HIP \#         &    HD \#         & $\pi$ & $\sigma_\pi$ & $K_S$ & Type & Group \\
       &          & (mas) & (mas) & (mag) &  & \\
\hline
\multicolumn{7}{|l|}{NAOS/CONICA targets} \\ 
\hline  
59502       &   106036   &   10.26   &   0.49   &   6.87   &   A2V   &   LCC   \\
60851       &   108501   &   9.63   &   0.50   &   6.06   &   A0Vn   &   LCC   \\
61265       &   109197   &   7.50   &   0.48   &   7.46   &   A2V   &   LCC   \\
62026       &   110461   &   9.20   &   0.45   &   6.31   &   B9V   &   LCC   \\
63204       &   112381   &   9.07   &   0.49   &   6.78   &   A0p   &   LCC   \\
67260       &   119884   &   8.15   &   0.49   &   6.98   &   A0V   &   LCC   \\
67919       &   121040   &   9.64   &   0.49   &   6.59   &   A9V   &   LCC   \\
68532       &   122259   &   8.09   &   0.43   &   7.02   &   A3\,IV/V   &   UCL   \\
69113       &   123445   &   5.92   &   0.41   &   6.37   &   B9\,V   &   UCL   \\
73937       &   133652   &   8.17   &   0.46   &   6.23   &   Ap\,Si   &   UCL   \\
78968       &   144586   &   5.87   &   0.53   &   7.42   &   B9\,V   &   US    \\
79098       &   144844   &   7.28   &   0.45   &   5.69   &   B9\,V   &   US    \\
79410       &   145554   &   7.00   &   0.52   &   7.09   &   B9\,V   &   US    \\
79739       &   146285   &   6.79   &   0.52   &   7.08   &   B8\,V   &   US    \\
79771       &   146331   &   6.86   &   0.51   &   7.10   &   B9\,V   &   US    \\
80142       &   147001   &   5.82   &   0.44   &   6.66   &   B7\,V   &   UCL   \\
80474       &   147932   &   7.20   &   0.51   &   5.80   &   B5\,V   &   US    \\
80799       &   148562   &   7.91   &   0.52   &   7.45   &   A2\,V   &   US    \\
80896       &   148716   &   7.77   &   0.57   &   7.44   &   F3\,V   &   US    \\
81949       &   150645   &   6.21   &   0.52   &   7.33   &   A3\,V   &   UCL   \\
81972       &   150742   &   5.36   &   0.40   &   5.87   &   B3\,V   &   UCL   \\
83542       &   154117   &   5.00   &   0.53   &   5.38   &   G8/K0\,III   &   US    \\
\hline  \multicolumn{7}{|l|}{ADONIS multi-color subset} \\ 
\hline  
53701       &   95324   &   7.93   &   0.58   &   6.48   &   B8\,IV   &   LCC   \\
76071       &   138343   &   5.96   &   0.56   &   7.06   &   B9\,V   &   US    \\
77911       &   142315   &   6.87   &   0.49   &   6.68   &   B9\,V   &   US    \\
78530       &   143567   &   7.11   &   0.48   &   6.87   &   B9\,V   &   US    \\
78809       &   144175   &   7.20   &   0.51   &   7.51   &   B9\,V   &   US    \\
78956       &   144569   &   5.55   &   0.50   &   7.57   &   B9.5\,V   &   US    \\
79124       &   144925   &   6.41   &   0.53   &   7.13   &   A0\,V   &   US    \\
79156       &   144981   &   6.21   &   0.53   &   7.61   &   A0\,V   &   US    \\
80238       &   147432   &   7.64   &   0.68   &   7.34   &   A1\,III/IV   &   US    \\

    \hline
  \end{tabular}
  \caption{We have obtained follow-up multi-color observations with NACO for 22~targets in the Sco~OB2 association. We include in our analysis 9~targets with multi-color observations in the ADONIS sample. All targets listed above are known to have secondaries in the ADONIS survey.  The table lists for each star the parallax and error \cite[taken from][]{debruijne1999}, the $K_S$ magnitude, and the spectral type of the primary star. The $K_S$ magnitudes are those derived in this paper for the 22~stars observed with NACO, and are taken from \cite{kouwenhoven2005} for the other nine stars. The last column shows the subgroup membership of each star (US~= Upper Scorpius; UCL~= Upper Centaurus Lupus; LCC~= Lower Centaurus Crux), taken from
    \cite{dezeeuw1999}. \label{table: sample}}
\end{table}

% nogmaals het doel

The major goals of our NACO follow-up observations are to determine the validity of the $K_S=12$~mag criterion that we used to separate companions and background stars in our ADONIS sample \citep{kouwenhoven2005}, to study the companion mass distribution near the stellar-substellar boundary, and to search for additional faint and/or close companions.

% het naco sample

Our NACO sample consists of 22~member stars (listed in Table~\ref{table: sample}) in the Sco~OB2 association: 10~of spectral type B, 10~of spectral type~A, and one each of spectral type~F and~G. The targets are more or less equally distributed over the three subgroups of Sco~OB2: 9~in US, 6~in UCL, and 7~in LCC. All 22~targets are known to have secondaries in the ADONIS survey.

% welke sterren gekozen?

We included in our sample all seven target stars with faint ($K_S > 14$~mag) and close ($\leq 4$~arcsec) candidate background stars: HIP61265, HIP67260, HIP73937, HIP78968, HIP79098, HIP79410, and HIP81949. The other 15~targets all have candidate companion stars, for which we will use the multi-color data to further study their nature. Priority was given to target stars with multiple secondaries (candidate companions and candidate background stars) and targets close to the Galactic plane (HIP59502, HIP60851, HIP80142, and HIP81972) because of the larger probability of finding background stars. 

% adonis subsample van 9 sterren

There are 9~targets in the ADONIS dataset with (photometric) multi-color observations. These targets are also listed in Table~\ref{table: sample} and all have secondaries. \cite{kouwenhoven2005} use only the $K_S$ magnitude to determine the status of a secondary, including the secondaries around the 9~targets with $JHK_S$ observations. Later, in Section~\ref{section: status}, we will combine the data of the 22~NACO targets and the 9~ADONIS targets with multi-color observations, and determine the status of the secondaries of these 31~targets using their $JHK_S$ magnitudes. In the remaining part of Section~\ref{section: observationsanddatareduction} we describe the NACO observations, data reduction procedures, and photometric accuracy.

\subsection{NACO observations} \label{section: observations}

The observations were performed using the NAOS/CONICA system, consisting of
the near-infrared camera CONICA \citep{lenzen1998} and the adaptive optics
system NAOS \citep{rousset2000}. NAOS/CONICA is installed at the Nasmyth~B
focus of UT4 at the ESO Very Large Telescope on Paranal, Chile. The
observations were carried out in Service~Mode on the nights of April~6,
April~28-30, May~4-5, June~8, June~19, June~25, June~27-28, July~3, July~24,
and September~10, 2004. Some representative images are shown in Figure~\ref{figure: triples}.

The targets were imaged using the S13 camera, which has a pixel scale of
13.27~mas/pixel, and a field of view of $14''\times 14''$. The CONICA detector
was an Alladin~2 array in the period April~6 to May~5, 2004. The detector was
replaced by an Alladin~3 array in May~2004, which was used for the remaining
observations. We used the readout~mode Double\_RdRstRd and the detector mode
HighDynamic. For both detectors, the rms readout noise was $46.2\ e^-$ and the
gain was $\approx 11 e^-$/ADU. The full-well capacity of the Alladin~2 array
is 4300~ADU, with a linearity limit at about 50\% of this value. For the
Alladin~3 array the full-well capacity is 15000~ADU, with the linearity limit
at about two-thirds of this value.

Each observation block corresponding to a science target includes six
observations. The object is observed with the three broad band filters $J$
($1.253~\mu$m), $H$ ($1.643~\mu$m), and $K_S$ ($2.154~\mu$m). Since our
targets are bright, several of them will saturate the detector, even with the
shortest detector integration time. For this reason we also obtain
measurements in $J$, $H$, and $K_S$ with the short-wavelength neutral density
filter (hereafter NDF). The NDF transmissivity is about 1.4\% in the
near-infrared. The observations {\em with} NDF allow us to study the primary
star and to obtain an accurate point spread function (PSF), while the observations {\em without}
NDF allow us to analyze the faint companions in detail. In order to
characterize the attenuation of the NDF we observed the standard stars GSPC~S273-E and
GSPC~S708-D. These are LCO/Palomar NICMOS Photometric Standards
\citep{persson1998}. The near-infrared magnitudes of these stars are
consistent with spectral types G8V and G1V, respectively. By comparing the
detected fluxes in $J$, $H$, $K_S$, with and without the NDF, we determined the
attenuation of the NDF in the three filters.

Each observation consisted of six sequential exposures of the form $OSSOOS$,
where $O$ is the object, and $S$ is a sky observation. Each exposure was
jittered using a jitter box of 4~arcsec. We used a sky offset of 15~arcsec,
and selected a position angle such that no object was in the sky field.

Each exposure consists of 5 to 35 short observations of integration times in
the range $0.35-5.3$~seconds, depending on the brightness of the source. For
the target observations without NDF we chose the minimum integration time of
0.35~seconds. For all standard star observations and target observations with
NDF we chose the integration time such that the image does not saturate for
reasonable Strehl ratios. We optimized the integration time to obtain the
desired signal-to-noise ratio. The short integrations are combined by taking the
median value.

Visual wavefront sensing was performed directly on the target stars, which
minimized the effects of anisoplanatism. For a subset of the target stars, the
observation block was carried out multiple times. The target stars were
usually positioned in the center of the field. Occasionally we observed the
target off-center to be able to image a companion at large angular separation.

The observations on the nights of 29~April, 28~June, and 3~July, 2004, were
obtained under bad weather or instrumental conditions and were removed from
the dataset. These observations were repeated under better conditions later on
in the observing run. 
The observations in the nights of 6~April and 27~June, 2004,
were partially obtained under non-photometric conditions, and were calibrated using the targets themselves (see \S~\ref{section: photometry}).
All other observations were performed under photometric
conditions. Most observations (85\%) were obtained with a seeing between
0.5~and 1.5~arcsec. For a large fraction of the remaining observations the
seeing was between 1.5~and 2.0~arcsec. The majority (65\%) of the observations were
obtained at an airmass of less than 1.2, and for 98\% of the observations the
airmass was less than 1.6.

\subsection{Data reduction procedures}

The primary data reduction was performed with the ECLIPSE package
\citep{devillard1997}. Calibration observations, including dark images, flat
field images, and standard star images, were provided by ESO Paranal. Twilight
flat fields were used to create a pixel sensitivity map. For several
observations, no twilight flats were available. In these cases we used the
lamp flats. The dark-subtracted observations were flatfielded and
sky-subtracted. Finally, the three jittered object observations were combined.

\subsection{Component detection} \label{section: componentdetection}

The component detection is performed with the STARFINDER package
\citep{diolaiti2000}. The PSF of the target star is
extracted from the background subtracted image. The flux of the primary star
is the total flux of the extracted PSF. A scaled-down version of the PSF is
compared to the other signals in the field with a peak flux larger than
$2.5-3$ times the background noise\footnote{Note that a peak flux of
$2.5-3$~times the background noise corresponds to a total flux with a much
larger significance, since the flux of a faint companion is spread out over
many pixels. All detected components in our survey have a signal-to-noise ratio larger than $12.8$.}. 
The profile of these signals is then cross-correlated with the
PSF. Only those signals with a profile very similar to that of the PSF star
(i.e., a correlation coefficient larger than $\approx 0.7$) are considered
as real detections. Finally, the angular separation, the position angle, and the
flux of the detected component are derived.

\begin{SCfigure}[][btp]
  \centering
  \includegraphics[width=0.5\textwidth,height=!]{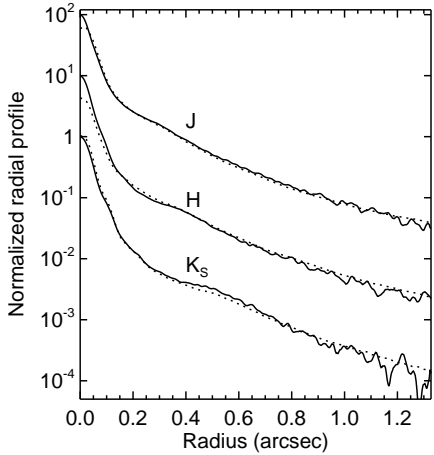}
  \caption{The radial profile of the PSF for the target star HIP78968. The
    observations are obtained using the NACO system in the night of May~5,
    2004 in $J$, $H$, and $K_S$. The corresponding Strehl ratios for these
    observations are 6.5\% in $J$, 15.6\% in $H$, and 23.6\% in $K_S$.
    Observations are obtained with neutral density filter (NDF; solid curves) 
    and without NDF 
    (dotted curves). The profiles are normalized such that the
    peak flux for the images obtained with NDF is 100 in $J$, 10 in $H$, and 1
    in $K_S$. The observations with NDF allow us to extract the PSF and
    measure the flux of the primary star. The images obtained without NDF are
    much deeper, but the primary is usually saturated. Assuming the PSF is
    similar, we use the non-saturated PSF to analyze the secondaries in the
    saturated image. \label{figure: psfprofile}}
\end{SCfigure}

As discussed in \S~\ref{section: observations}, we observe each target in
$JHK_S$ {\em with} NDF to obtain an accurate PSF template, and {\em without}
NDF to do accurate photometry on the faint companions. The observations
without NDF are often saturated, which makes PSF extraction impossible. None
of the observations with NDF are saturated. Since these observations are
carried out close in time and close in airmass, we assume that the PSFs of the
observations with and without NDF are not significantly different. This is
illustrated in Figure~\ref{figure: psfprofile}, where we plot the radial
profiles of the extracted PSF of HIP78968. For the saturated images we use the
PSF that was extracted from the corresponding non-saturated image for analysis
of the secondaries.

Artifacts are present in the image when observing a bright
object. The location of the artifacts is well-defined (CONICA manual), and
since the artifacts look significantly non-circular, they can be easily
recognized as non-stellar components (see Figure~\ref{figure: triples}). Additionally, observations obtained with
the NDF show a faint artifact at $\sim 2$~arcsec to the North-East of the
target star. Care was taken that the extracted PSF and the analysis of
companions in that area were not affected by the presence of this artifact.

\subsection{Photometry} \label{section: photometry}

Observations of standard stars from the \cite{persson1998} catalog were
provided by ESO. The standard stars are used to determine the magnitude
zeropoints for each night and each filter individually. No standard stars are
available in the nights of 28~April and 10~September, 2004. For 28~April we
determine the zero point magnitude using 2MASS as a reference system, and the
target stars HIP67260, HIP67919, HIP68532 as substitute standard stars. For
10~September, we use the 2MASS data for HIP83542 to calibrate $J$ and $H$. For
the $K_S$ filter we use the HIP83542 measurement of \cite{kouwenhoven2005}
since its 2MASS $K_S$ magnitude is inaccurate due to confusion with 
the diffraction spike of a nearby star. 
The $H$ magnitude of HIP59502 was
obtained under non-photometric conditions on 6~April, 2004, so instead we used
the corresponding 2MASS measurement of this star. All observations in the
night of 27~June, 2004 were obtained under non-photometric conditions; the
fluxes of HIP80474 and HIP81972 are therefore calibrated using 2MASS.
We derive the calibrated magnitudes using the mean
extinction coefficients for Paranal: $k_J=0.11$, $k_H=0.06$, and
$k_{K_S}=0.07$.

The attenuation $m_{\rm NDF}$ (in magnitudes) of the NDF is determined for
$J$, $H$, and $K_S$ using the standard stars GSPC~S273-E and GSPC~S708-D and
the non-saturated target stars. All observations are performed in pairs (with
and without NDF), which allows us to determine the NDF attenuation. For each
filter $m_{\rm NDF}$ is calculated as the median
difference in magnitude:
$m_{\rm NDF} = \left< m_{\star,{\rm without\ NDF}} - m_{\star,{\rm with\ NDF}} \right>
$.
We find $m_{\rm NDF,J} = 4.66\pm 0.03$, $m_{\rm
NDF,H} = 4.75\pm 0.02$, and $m_{\rm NDF,K_S} = 4.85\pm 0.02$, respectively. We
do not measure a significant difference between the values found for the
early-type program stars and the late-type standard stars. All observations
done with the NDF are corrected with the values mentioned above.

% ====================================================================
% ====================================================================
% ====================================================================
% ==OBSERVATIONS======================================================
% ====================================================================
% ====================================================================
% ====================================================================

\subsection{Photometric precision and accuracy of the NACO observations} \label{section: photometricaccuracy}

We estimate the photometric uncertainty of the NACO observations using simulations 
(see \S~\ref{section: starfinderprecision}) and the comparison with other datasets (\S~\ref{section:
2masscomparison}-\ref{section: tokovinin}). For primaries, the external $1\sigma$ error
in $J$, $H$, and $K_S$ is $\sim 0.04$~mag, corresponding to an error of $\sim
0.06$~mag in the colors. Typical $1\sigma$ external errors in magnitude and color for
the bright companions ($8 \la K_S/{\rm mag} \la 13$) are 0.08~mag and
0.11~mag, respectively. For the faintest sources ($K_S \ga 13$~mag) the
errors are 0.12~mag in magnitude and 0.17~mag in color. In the following subsections we will 
discuss the analysis of our photometric errors.

\subsubsection{Algorithm precision} \label{section: starfinderprecision}

The instrumental magnitudes of all objects are obtained using STARFINDER. We
investigate the precision of the STARFINDER algorithm using simulated
observations. We create simulations of single and binary systems with varying
primary flux ($10^4 - 10^6$~counts), flux ratio ($\Delta K_S = 0-10$~mag),
angular separation ($0''-13''$), Strehl ratio ($1\%-50\%$), and position
angle. We estimate the flux error by comparing the
input flux with the flux measured by STARFINDER for several realizations.

Using the simulations we find that the precision of the STARFINDER fluxes is
$\sim 1\%$ ($\sim 0.01$~mag) for most primary stars in our sample, for all
relevant Strehl ratios and as long as the PSF of the primary star is not
significantly influenced by the presence of a companion. The 1\% error is due to the
tendency of STARFINDER to over-estimate the background underneath bright
objects \citep{diolaiti2000}. For the fainter primaries in our sample (flux
between $5 \times 10^4$ and $5 \times 10^5$ counts), the flux error is $1-3\%$
($\sim 0.01-0.03$~mag).

For secondaries outside the PSF-halo of the primary, the error is typically
$\sim 0.01$~mag if the flux difference with the primary is less than
5~mag. Fainter secondaries have a larger flux error, ranging from $0.01$~mag
to $0.1$~mag, depending on the brightness of primary and companion.

If the secondary is in the halo of the primary, its flux error is somewhat
larger. For example, for a companion at $\rho=2''$ which is less than
4~magnitudes fainter than the primary, the flux error is 4\% ($\sim
0.04$~mag) or smaller. Deblending the PSF of a primary and close secondary does not
introduce a much larger error, as long as the magnitude difference is less than $\sim 5$~magnitudes. No close companions with a magnitude difference larger than 5~mag are detected in our NACO observations.

For several fields the observations without NDF are saturated. In order to
analyze the faint companions in the field we use the PSF of the corresponding
non-saturated observation obtained with the NDF (see \S~\ref{section:
observations}). These observations are performed close in airmass and time, so that
their PSFs are similar. We estimate the flux error by
comparing PSFs corresponding to non-saturated images obtained with NDF and
without NDF. These comparisons show that the resulting error ranges from 0.02 to
0.5~magnitudes, depending on the brightness of the secondary. We therefore
minimize flux calculations using this method, and only use measurements
obtained with the PSF of the non-saturated image when no other measurements
are available. In the latter case, we place a remark in Table~\ref{table:
longtable}.

\subsubsection{Comparison with 2MASS} \label{section: 2masscomparison}

We compare the near-infrared measurements of the 22~targets in our NACO survey with the measurements in 2MASS
\citep{2mass} to get an estimate of the external errors. We only select those
measurements in 2MASS that are not flagged. Since the resolution in our
observations is higher than the $\approx 4''$ resolution of 2MASS, we combine
the observed fluxes of the primaries and close companions before the comparison with
2MASS. For the observations {\em not} calibrated with the 2MASS measurements,
the rms difference between our measurements and those of 2MASS are $0.055$~mag in $J$, $0.040$~mag in $H$, and $0.049$~mag in $K_S$.

\subsubsection{Comparison with the ADONIS survey of \cite{kouwenhoven2005}} \label{section: adoniscomparison}

We detect all but two of the stellar components found by
\cite{kouwenhoven2005} around the 22 target stars in our NACO survey. We
do not observe the faint companions of HIP80142 at $\rho=8.54''$ and HIP81949
at $\rho=9.70''$ because they are not within our NACO field of view. 
We find three bright companions at small angular separation of HIP63204 ($\rho=0.15''$), HIP73937 ($\rho=0.34''$), and
HIP79771 ($\rho=0.44''$). Since these objects are not found in the ADONIS observations of 
\cite{kouwenhoven2005}, their fluxes and those of the corresponding primaries
are summed for comparison with \cite{kouwenhoven2005}.

The rms difference between the 22 primaries observed with NACO and those described 
in \cite{kouwenhoven2005} is 0.055~mag in $K_S$. HIP69113 and HIP78968
additionally have multi-color observations in \cite{kouwenhoven2005}, which
are in good agreement with the measurements presented in this paper. The $J$
and $H$ measurements of HIP80474 and HIP80799 are flagged ``non-photometric''
in \cite{kouwenhoven2005}, and are not discussed here.

Our dataset and that of \cite{kouwenhoven2005} have 35 stellar components
other than the target stars in common. The rms difference between the $K_S$
magnitude of these objects in the two papers is 0.26~magnitudes. The
differences are similar for the objects that have common $J$ and $H$
measurements in both papers.

\subsubsection{Comparison with  \cite{shatsky2002}} \label{section: tokovinin}

Three targets in our NACO survey are also included in the binarity survey amongst B-stars in
Sco~OB2 by \cite{shatsky2002}: HIP79098, HIP80142, and HIP81972. Seven
secondaries are detected both their survey and in our NACO survey (1 for HIP79098; 2 for HIP80142; 4 for
HIP81972). \cite{shatsky2002} classify these seven secondaries all as `definitely optical'
or `likely optical'. They performed their observations
in both coronographic and non-coronographic mode, and were therefore able to find five~faint secondaries which do not appear in our NACO sample.

The $J$ and $K_S$ magnitudes of HIP79098 and HIP80142 and their companions are
in good agreement with our measurements. Our measurements of HIP81972 are in
good agreement with those in 2MASS as well as the measurements in
\cite{kouwenhoven2005}, but there is a discrepancy between our measurements of
HIP81972 and those in \cite{shatsky2002}. The magnitude difference between
HIP81972 and its companions in our observations and in \cite{shatsky2002} are
similar. The observations of HIP81972 are flagged `likely photometric' in
\cite{shatsky2002}, but since they disagree with those in this paper and those
in 2MASS, we assume they are non-photometric, and ignore them for the
magnitude comparison.

\subsection{General properties of the NACO observations}\label{section: generalproperties}

In the fields around the 22~targets we observed with NACO, we find 62 components other than the target stars. The properties of these targets and their secondaries are listed in Table~\ref{table: longtable}.  The 22~primaries have $5.5~{\rm mag} < J < 7.8~{\rm mag}$, $5.0~{\rm mag} < H < 7.7~{\rm mag}$, and $4.9~{\rm mag} < K_S < 7.7~{\rm mag}$. The brightest companions observed are $\sim 7.5$~mag in the three filters, while the faintest secondaries found have $J=16.6$~mag, $H=17.3$~mag, and $K_S=17.3$~mag. 

With NACO we detect components in the angular separation range $0.15'' < \rho < 11.8''$. The lower limit on $\rho$ depends on the Strehl ratio and the magnitude difference between primary and companion (see also \S~\ref{section: detectionlimits}). The upper limit is determined by the size of the field-of-view. The median formal error in angular separation is 4~mas for bright components ($8~{\rm mag} \la K_S \la 13~{\rm mag}$) and 10~mas for faint components ($K_S \ga 13~{\rm mag}$). Position angles are measured from North to East. The median formal error in the position angle is $0^\circ.007$. 

We find 27~stellar components that are not detected by
\cite{kouwenhoven2005}. Three close secondaries are found at small angular
separation from HIP63204-2 ($\rho=0.15''$; $K_S=8.40$~mag), HIP73937-1 ($\rho=0.34''$; $K_S=8.37$~mag), and HIP79771-2 ($\rho=0.44''$; $K_S=11.42$~mag). The former two have been reported as candidate companions \citep{wds1997}; the latter was previously undocumented. The other 25 new secondaries are all faint ($K_S \ga 12$~mag). Two of these 25~secondaries were also reported by \cite{shatsky2002}.

% ====================================================================
% ====================================================================
% ====================================================================
% ==COMPLETENESS======================================================
% ====================================================================
% ====================================================================
% ====================================================================

\section{The completeness and detection limit of the ADONIS and NACO surveys} \label{section: detectionlimits}

\begin{figure}[tbp]
  \centering
  \includegraphics[width=0.7\textwidth,height=!]{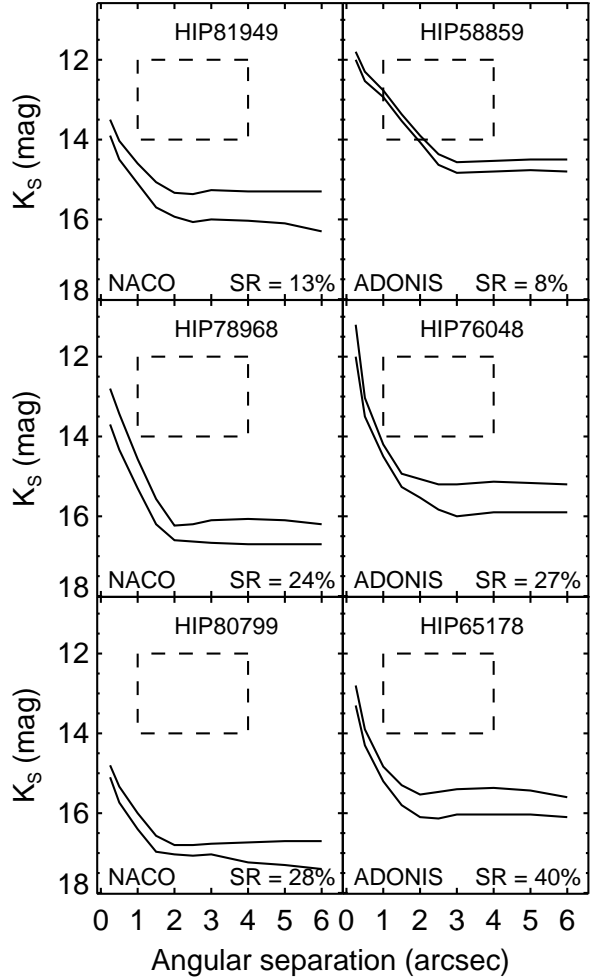}
  \caption{The detection limit and completeness limit for several targets in our ADONIS survey \citep{kouwenhoven2005} and NACO survey (this paper). The target star, Strehl ratio, and instrument are indicated in each panel. The lower and upper curve show the detection limit and the completeness limit, respectively. The detection and completeness limits shown above are representative for the ADONIS and NACO observations.
    The target stars have $K_S$ magnitudes of 7.33~mag (HIP81949), 7.42~mag (HIP78968), 7.45~mag (HIP80799), 6.52~mag (HIP58859), 6.26~mag (HIP76048), and 6.71~mag (HIP65178). The completeness limit is generally $\sim 0.3$~mag brighter than the detection limit. 
    At close angular separation a higher Strehl ratio results in a fainter detection limit. The dashed rectangle encloses the region with $12~\mbox{mag} \leq K_S \leq 14~\mbox{mag}$ and $1'' \leq \rho \leq 4''$, which is relevant for our analysis of the substellar population in Sco~OB2 (see \S~\ref{section: gap}). In this region the NACO observations are complete and ADONIS observations are more than 95\% complete.
    \label{figure: detectionlimits}
  }
\end{figure}

% introductie

We cannot detect sources fainter than a certain magnitude because of the background noise in the images. The faintest detectable magnitude additionally depends on the angular distance to the primary star, the primary star magnitude, and the Strehl ratio. For a correct interpretation of the results of the survey it is therefore important to characterize the limiting magnitude of the observations (the detection limit) and the magnitude at which a star is likely detected (the completeness limit). 

% sample

We study the completeness limit and detection limit as a function of angular separation from the primary for six stars. These are HIP58859, HIP65178, and HIP76048 from our ADONIS survey and HIP80799, HIP78968, and HIP81949 from our NACO survey. These stars are selected to cover the range in Strehl ratio of the observations, so that the completeness and detection limits are representative for the other targets in the ADONIS and NACO surveys.

% methode

For each observation STARFINDER extracts the PSF from the image (see \S~\ref{section: componentdetection}). We simulate observations by artificially adding a scaled and shifted copy of the PSF to the observed image. We reduce the simulated image as if it were a real observation. We repeat this procedure twenty times for simulated secondaries with different angular separation and magnitude. We define the detection limit and completeness limit as the magnitude (as a function of angular separation) at which respectively 50\% and 90\% of the simulated secondaries are detected.
%
% resultaten
%
The curves in Figure~\ref{figure: detectionlimits} show the completeness and detection limit for the six stars mentioned above. Due to our sampling the magnitude error of the completeness and detection limit is $~\sim 0.15$~mag. 
The figure clearly shows that a high Strehl ratio facilitates the detection of closer and fainter objects as compared to observations with lower Strehl ratio. For all stars the completeness limit is $\sim 0.3$~mag above the detection limit.

% 12-14 en 1-4

In Section~\ref{section: gap} we analyze the substellar companion population in Sco~OB2 in the angular separation range $1'' \leq \rho \leq 4''$ and magnitude range $12~\mbox{mag} \leq K_S \leq 14~\mbox{mag}$.  The NACO observations of 22~targets are complete in this region. Only a few targets in the ADONIS sample are incomplete in this region. Assuming a flat semi-major axis distribution, we estimate that  about 5\% of the faint companions at small angular separation are undetected for the 177~targets that are {\em only} observed with ADONIS (see  Figure~\ref{figure: detectionlimits}). For the combined NACO and ADONIS sample this means that we are more than 95\% complete in the region $12~\mbox{mag} \leq K_S \leq 14~\mbox{mag}$ and $1'' \leq \rho \leq 4''$.

% examples of what we can see

With NACO we are sensitive down to brown dwarfs and massive planets. To estimate the mass corresponding to the faintest detectable magnitude as a function of angular separation, we use the models of \cite{chabrier2000}. We assume a distance of 130~pc, the mean distance of Sco~OB2, and an age of 5~Myr for the US subgroup and 20~Myr for the UCL and LCC subgroups (cf. \S~\ref{section: cmd}). At a distance of 130~pc, the brightest brown dwarfs have an apparent magnitude of $K_S \approx 12$~mag.
With NACO we are able to detect brown dwarfs at an angular separation larger than $\approx 0.3''$. The magnitude of the faintest detectable brown dwarf increases with increasing angular separation between $\rho = 0.3''-2''$. 
The minimum detectable mass is a function of age due to the cooling of the brown dwarfs. For angular separations larger than $\approx 2''$ the halo of the primary PSF plays a minor role. For these angular separations we are sensitive (but not complete) down to $K_S \approx 16.5$~mag with NACO, corresponding to planetary masses possibly as low as $\sim 5~{\rm M_J}$ for US and $\sim 10~{\rm M_J}$ for UCL and LCC. However, because of the large number of background stars and the uncertain location of the isochrone for young brown dwarfs and planets, we will not attempt to identify planetary companions in this paper (see \S~\ref{section: separation} for a further discussion).

% lower limits for magnitudes

Several secondaries are not detected in one or two filters. For the filter(s) in which the secondary is not observed we determine a lower limit on the magnitude of the missing star using simulations. We perform the simulations as described in \cite{kouwenhoven2005}. The position of the secondary is known from observations in other filters. This position is assigned to the simulated companion, so that the detection limit as a function of angular separation is taken into account. Images are created with simulated secondaries of different magnitude. We reduce the images as if they were real observations. The lower limit on the magnitude is then determined by the faintest detectable simulated secondary. Two secondaries (HIP73937-1 and HIP76071-1) have a lower limit in $J$ because they are unresolved in the wings of the primary star PSF. The other secondaries with a lower magnitude limit for a filter have a flux below the background noise.

% ====================================================================
% ====================================================================
% ====================================================================
% ==COMPANION=STAR=PROPERTIES=========================================
% ====================================================================
% ====================================================================
% ====================================================================

\section{Status of the stellar components} \label{section: status}

In this section we determine the status (companion star or background star) of the secondaries. We analyze the secondaries detected around the 22~targets observed with NACO, as well as the secondaries around the 9~targets with multi-color observations in the ADONIS sample. For the 31~targets analyzed in this section we detect 72~secondaries in total. Sco~OB2 members and their companions should be located near the isochrone in the color-magnitude diagram, while the background stars should show a much larger spread. We use this property in \S~\ref{section: separation} to separate companions and background stars. As a consistency check we study in \S~\ref{section: backgroundstarpopulation} how our results compare to the expected number of background stars in our observations.

\subsection{Color-magnitude diagram and isochrones} \label{section: cmd}

\begin{figure}[btp]
  \begin{tabular}{cc} 
    \includegraphics[width=0.48\textwidth,height=!]{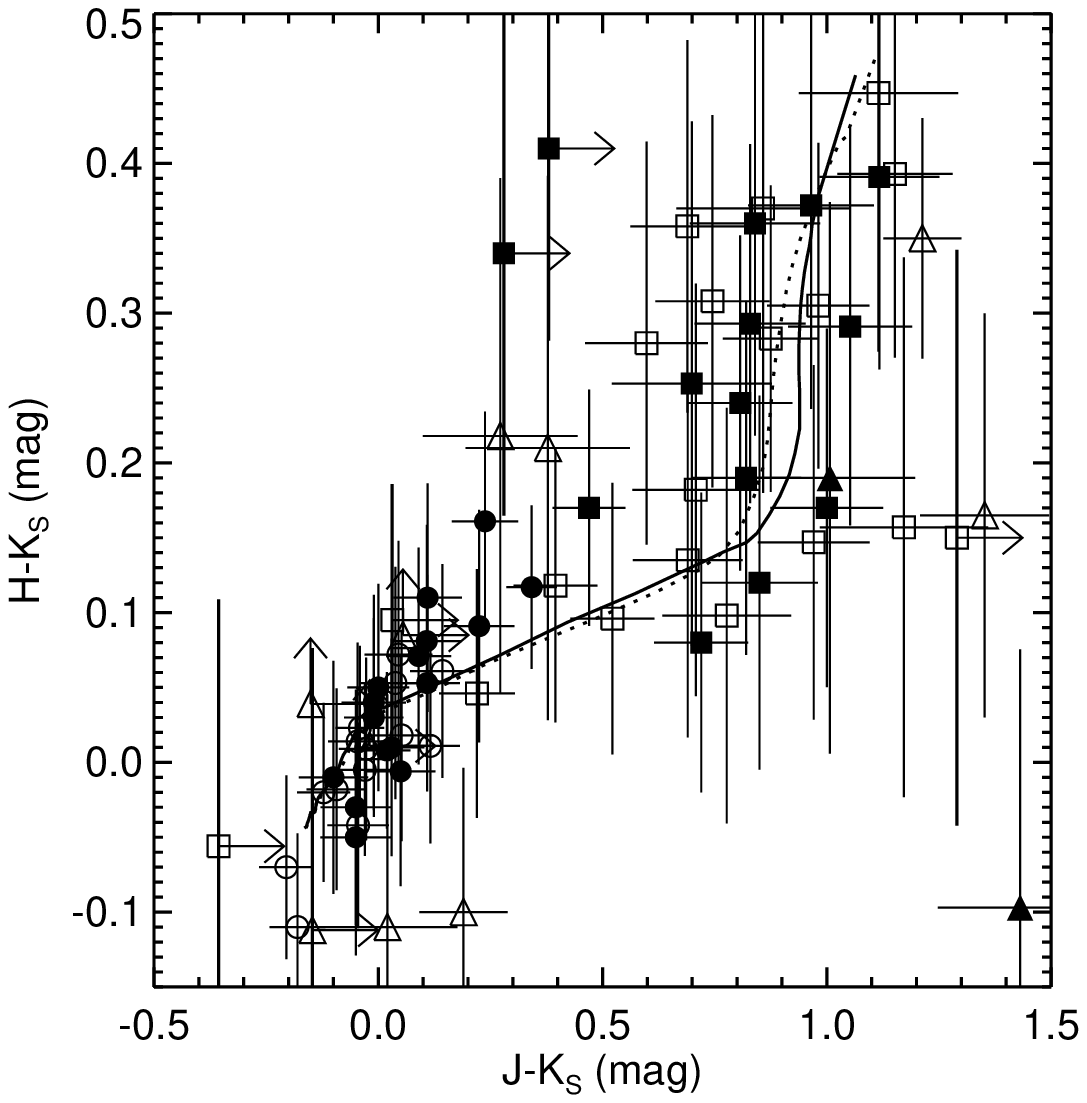} &
    \includegraphics[width=0.48\textwidth,height=!]{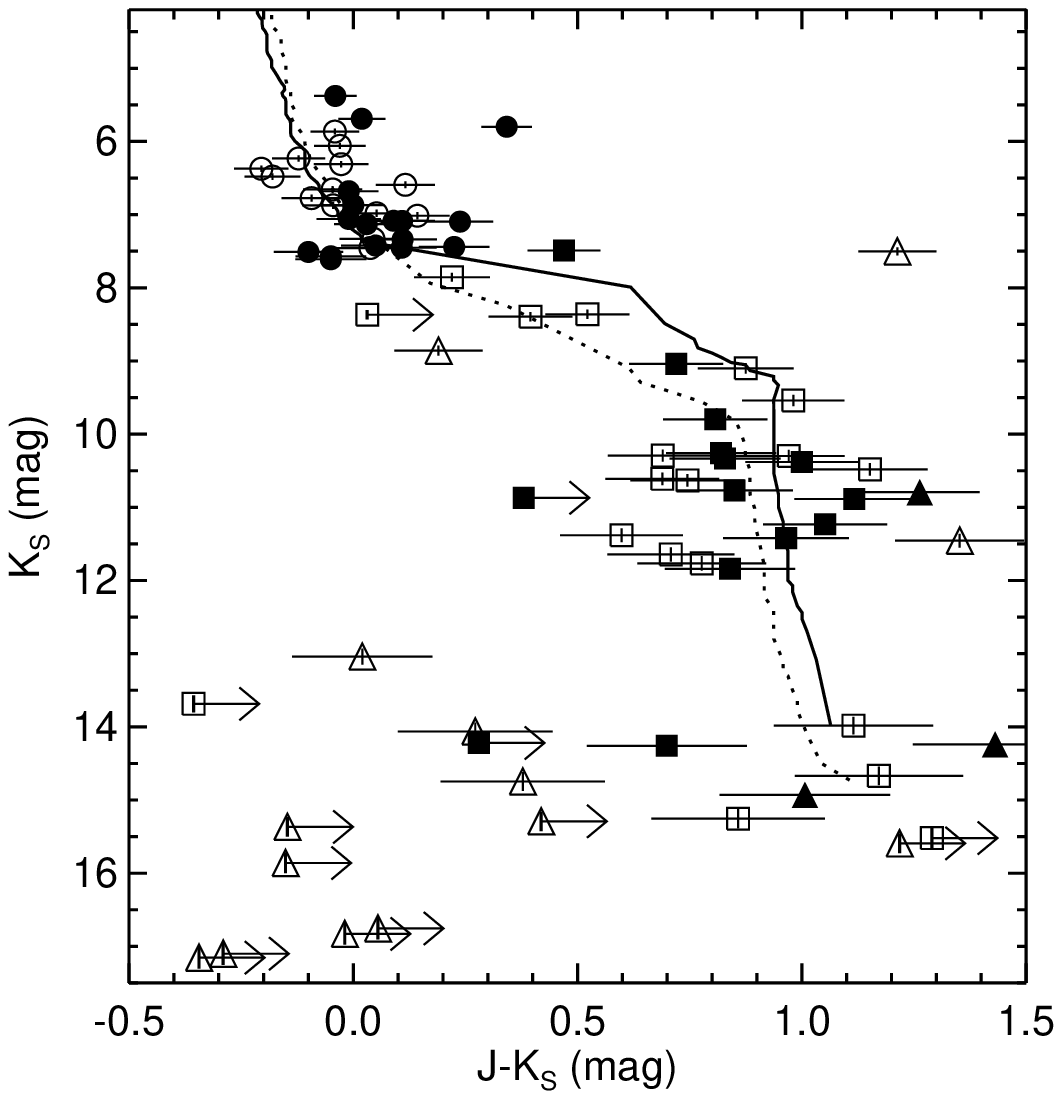} \\
  \end{tabular}
  \caption{
    The color-color diagram ({\em left}) and color-magnitude diagram ({\em right}) of the objects in our sample. Measurements are shown for the 22~targets observed with NACO and for the 9~targets with multi-color observations in the ADONIS sample. Both panels show target stars (circles), confirmed and candidate companions (squares), and background stars (triangles). The target stars and secondaries in the US subgroup are indicated with filled symbols; those from UCL and LCC are indicated with open symbols. 
The $1\sigma$ error bars are indicated for all data points. Lower limits are given for objects that are not detected in all three filters. Several detected objects are outside the ranges of the figures; these are all background stars. The status (companion or background star) of the secondaries is discussed in \S~\ref{section: separation}. The 5~Myr isochrone for US and the 20~Myr isochrone for UCL and LCC are indicated with the solid and dotted curves, respectively. 
    \label{figure: ccdiagrams}}     
\end{figure}

For the 22~targets in the NACO sample and the 9~targets in the ADONIS multi-color subset we have magnitudes in three filters, as well as for most of their secondaries. Several of the faintest secondaries are undetected in one or two filters, as they are below the detection limit. Color-color and color-magnitude diagrams with the 31~targets and 72~secondaries are shown in Figure~\ref{figure: ccdiagrams}. Lower limits are indicated for objects that are not detected in all three filters. Several secondaries in our sample are either very red or very blue. These secondaries fall outside the plots in Figure~\ref{figure: ccdiagrams}, and are all background stars.

For our analysis we adopt the isochrones described in \cite{kouwenhoven2005}, which consist of models from \cite{chabrier2000} for $0.02~\mbox{M}_\odot \leq M < 1~\mbox{M}_\odot$, \cite{palla1999} for $1~\mbox{M}_\odot \leq M < 2~\mbox{M}_\odot$, and  \cite{girardi2002} for $M > 2~\mbox{M}_\odot$.  For members of the US subgroup we use the 5~Myr isochrone, and for UCL and LCC members we use the 20~Myr isochrone. 

Absolute magnitudes $M_J$, $M_H$, and $M_{K_S}$ are derived from the apparent magnitudes using for each star individually the parallax and interstellar extinction $A_V$ from \cite{debruijne1999}; see \cite{kouwenhoven2005} for details. The error on the parallax is $5-10\%$ for all targets, and can therefore be used to derive reliable distances to individual stars \citep[e.g.,][]{brown1997}.
The median fractional error on the distance is 6.5\%, which introduces an additional error of 0.15~mag on the absolute magnitudes. Combining this error with the error in the apparent magnitude (\S~\ref{section: photometricaccuracy}) we obtain $1\sigma$ error estimates for the absolute magnitudes of $0.16$~mag for the primaries, $0.17$~mag for the bright companions, and $0.19$~mag for the faint companions. The colors are directly calculated from the apparent magnitudes, and are not affected by parallax errors.

\begin{figure}[btp]
  \centering
  \includegraphics[width=\textwidth,height=!]{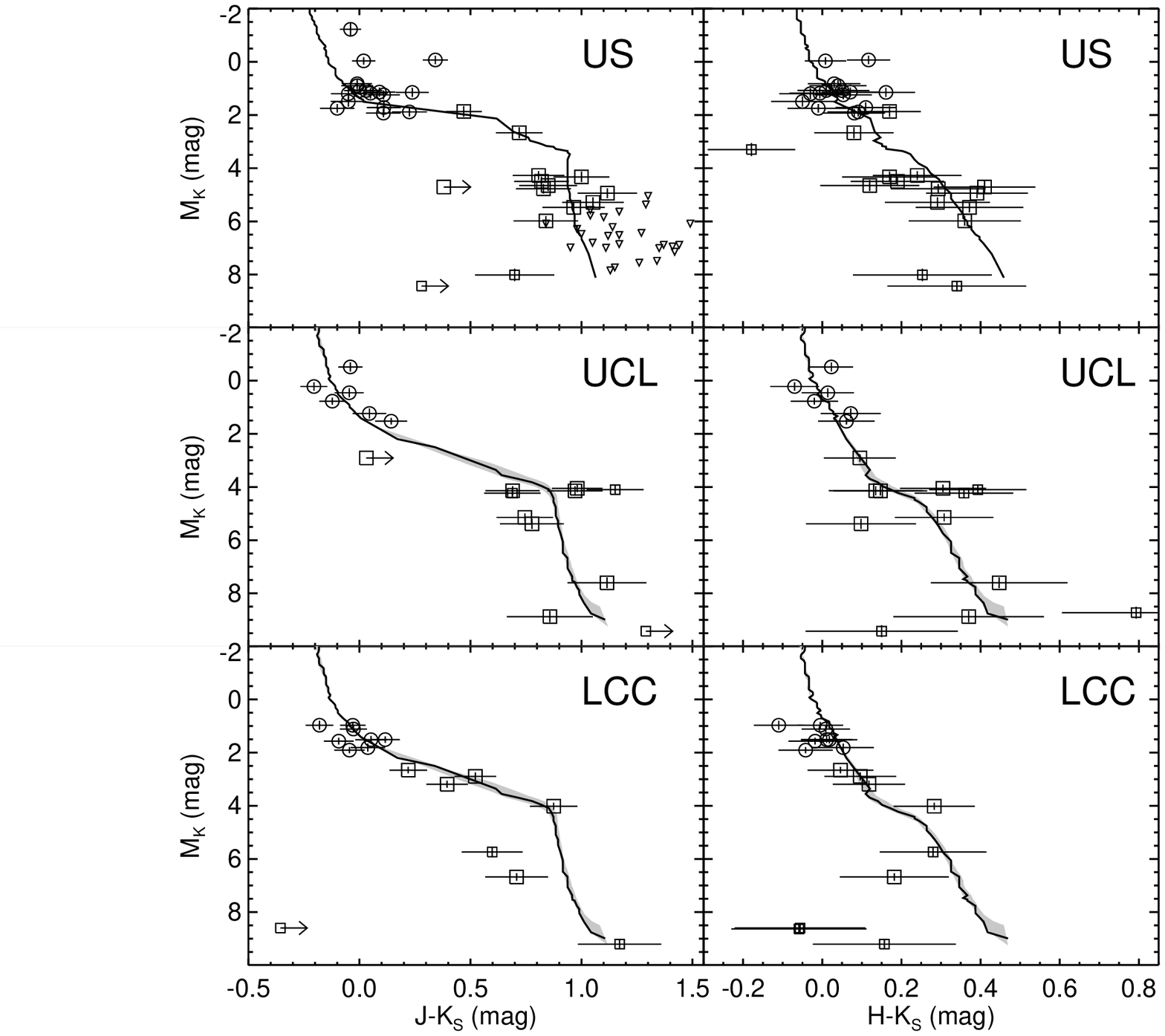}
  \caption{The (absolute) color-magnitude diagrams for the 22~targets in our NACO sample and the 9~targets with multi-color observations in the ADONIS sample. 
The results are split into the three subgroups US ({\em top}), UCL ({\em middle}), and LCC ({\em bottom}). The primary stars are indicated with circles; the confirmed companions with large squares, and the candidate companions with small squares. The $M_{K_S}$ magnitude is derived from the $K_S$ magnitude by correcting for distance and extinction for each target individually. The solid curves represent isochrones of 5~Myr (for US) and 20~Myr (for UCL and LCC).  The 15~Myr and 23~Myr isochrones enclose the gray-shaded area   and represent the uncertainty in the age of the    UCL and LCC subgroups. For each data point we indicate the $1\sigma$ error. 
The photometry of the observed objects cannot be used to distinguish between the subgroups of Sco~OB2, due to the errors and the small sample.   
The free-floating brown dwarfs in US    identified by \cite{martin2004} are indicated with triangles, adopting a distance of 145~pc. 
    \label{figure: hrdiagram2}}
\end{figure}

The color-magnitude diagrams (with {\em absolute} magnitudes) for the subgroups are shown in Figure~\ref{figure: hrdiagram2}. The measurements for US are in the top panels, and those of UCL and LCC are in the middle and bottom panels, respectively. The curves represent the 5~Myr (for US) and 20~Myr (for UCL and LCC) isochrones. The gray-shaded area indicates the inaccuracy in isochrone placement due to the age uncertainty in the subgroups (see Table~\ref{table: statistics}). Due to our small sample and the errors in the photometry we cannot see a difference between the magnitude and color distributions of the three subgroups.
For comparison we have included the free-floating brown dwarf candidates in US reported by \cite{martin2004}.

\subsection{Distinction between companions and background stars} \label{section: separation}

Individual distances to Sco~OB2 member stars are available \citep{debruijne1999}, as well as the ages of the three subgroups of Sco~OB2. A companion of a Sco~OB2 member star has (by definition) practically the same distance as its primary. It is very likely that a primary and companion in Sco~OB2 are coeval. The probability that a companion is captured dynamically is very small, since this involves either a multiple-star interaction or significant tidal dissipation.
It is even less likely that the companion is a captured field star, i.e., that the primary and companion have a different age. Background stars generally have a different age, distance, or luminosity class than Sco~OB2 members. In principle it is possible that an ADONIS or NACO field contains two members of the Sco~OB2 association at different distances due to projection effects, but \cite{kouwenhoven2005} showed that this effect can be neglected. 
Physical companion stars thus have the same age and distance as their primary, and therefore should fall on the isochrone for the subgroup to which the primary belongs, contrary to background stars. We use this property to separate physical companions and background stars.

For each stellar component we determine in the color-magnitude diagram (Figure~\ref{figure: hrdiagram2}) the point on the isochrone that corresponds best to the measured position. The differences in color and magnitude of the star and the nearest point on the isochrone are then compared to the observational errors. We use the $\chi^2$ test to determine how compatible the observed color and magnitude of a secondary are with the isochrone.
For example, if the best-fitting value in the $(J-K_S,M_{K_S})$-diagram deviates $\Delta(J-K_S)$ in color and $\Delta M_{K_S}$ in magnitude from the isochrone, the $\chi^2$ value is given by
\begin{equation} 
\chi^2 = \frac{ [\, \Delta(J-K_S) \, ]^2 }{  \sigma^2_{J-K_S} + \sigma^2_{J-K_S,{\rm iso}}}  
         + \frac{ \Delta M^2_{K_S}}{ \sigma^2_{M_{K_S}} + \sigma^2_{M_{K_S},{\rm iso}} } 
\end{equation}
where $\sigma_{J-K_S}$ and $\sigma_{M_{K_S}}$ are the observational errors in color and absolute magnitude, respectively. The errors on the location of the isochrone (due to age and metallicity uncertainty) are denoted with $\sigma_{J-K_S,{\rm iso}}$ and $\sigma_{M_{K_S},{\rm iso}}$. 
% errors on isochrone itself
The age uncertainty is $\sim 1$~Myr for members of US and $\sim 4$~Myr for members of UCL and LCC (see Table~\ref{table: statistics}).  The error in the placement of the corresponding isochrones due to age uncertainty is shown in Figure~\ref{figure: hrdiagram2}, and is small compared to the photometric errors. 
We assume a solar metallicity for all observed stellar components. The metallicity $[M/H]$ of Sco~OB2 has not been studied in detail. In their metallicity study of Ori~OB2, \cite{cunha1994} found a metallicity slightly ($\sim 0.2$~dex) lower than solar for this association. Using the models of \cite{siess2000} we estimate that the metallicity uncertainty $\Delta [M/H]=0.2$ results in an additional isochronal error of $0.05$~mag in $JHK_S$ and $0.06$~mag in the colors of low-mass ($M \la 0.5~{\rm M}_\odot$) companions. 

The $\chi^2$ values for the secondaries in each of the three color-magnitude diagrams are listed in Table~\ref{table: criteria} (as long as they are available). For the classification into companions and background stars we consider the {\em largest} of the three $\chi^2$ values available for each secondary. We choose this strategy (instead of, e.g., taking the average $\chi^2$ value), because background stars may be consistent with the isochrone for e.g. $J-K_S$, but not for $H-K_S$. The physical companions, however, should be consistent with the isochrone in all color-magnitude diagrams (within the error bars).

\begin{table}
  \begin{tabular}{lcl}
    \hline
    Secondary status    & Symbol   & Criterion\\
    \hline
    Confirmed companion &c & $0.00 <\chi^2\leq 2.30$ \\
    Candidate companion &? & $2.30 <\chi^2\leq 11.8$ \\
    Background star     &b & $11.8 <\chi^2 < \infty$ \\
    \hline
  \end{tabular}
  \caption{Criteria adopted to separate the secondaries with multi-color observations into confirmed companions, candidate companions, and background stars. The $\chi^2$ values of 2.30 and 11.8 correspond to the $1\sigma$ and $3\sigma$ levels. This means that (statistically) 68.3\% of the physical companions have $\chi^2 < 2.30$ and 99.73\% of the physical companions have $\chi^2 < 11.8$. A secondary with $\chi^2 < 2.30$ is very likely a companion star. A secondary with $\chi^2 > 11.8$ is almost certainly a background star. The secondaries with $2.30 < \chi^2 < 11.8$ may be companions or background stars.  \label{table: classification}}
\end{table}

% classification

Table~\ref{table: classification} lists the criteria we adopt to classify the secondaries into three groups: confirmed companions, candidate companions, and background stars. The $\chi^2$ value of 2.30 corresponds to the $1\sigma$ confidence level, which means that statistically 68.3\% of the companion stars have $\chi^2 < 2.30$. Similarly, 99.73\% of the companions have $\chi^2 < 11.8$ ($3\sigma$ confidence level). The above confidence levels are for a dataset with two degrees of freedom, under the assumption that the errors are Gaussian \citep{numericalrecipies}. 

% lower limits

For several objects we have lower limits on the magnitudes in one or two of the filters. Using the lower limits on $J$, $H$, or $K_S$ we calculate upper or lower limits on $\chi^2$ and are able to classify several additional objects as background stars. 

% planets all as bg stars

Even though we are sensitive to massive planets (although not complete) around our NACO targets, we will not classify the very faint secondaries in this paper, for two reasons. First, many faint background stars are expected in the ADONIS and NACO field of view (see \S~\ref{section: backgroundstarpopulation}). Due to the larger error bars for faint secondaries, many background stars may be consistent with the isochrone, and the vast majority of the ``candidate planets'' most likely are background stars. Second, the presently available evolutionary models for massive planets are not very reliable for young ages \citep[see e.g.,][for a review]{chabrier2005}. Throughout our analysis we do not consider objects with a mass below $0.02~{\rm M}_\odot$. Consequently, all secondaries inconsistent with a companion mass of $\geq 0.02~{\rm M}_\odot$ are classified as background stars. 

% summary van classificatie

The status given to each secondary is listed in Table~\ref{table: criteria}. The secondaries with $\chi^2 < 2.30$ are very likely companions because of their proximity to the isochrone. As statistically only 1~out of 370 companions have $\chi^2 > 11.8$, we claim with high confidence that the secondaries with $\chi^2>11.8$ are background stars. We cannot confirm the status of the secondaries with $2.30 < \chi^2 < 11.8$. These secondaries are consistent with the isochrone within the $3\sigma$ error bars. However, due to the large number of faint background stars, it is likely that several background stars also satisfy this criterion. As we cannot confirm either their companion status or their background star status, we will refer to the secondaries with  $2.30 < \chi^2 < 11.8$ as candidate companions.

Table~\ref{table: 13sigma} lists the distribution of confirmed and candidate companions over angular separation $\rho$ and $K_S$ magnitude. The largest fraction of candidate companions (relative to the number of confirmed companions) is seen for faint secondaries at large angular separation. In Section~\ref{section: gap} we will study the virtual absence of companions with   $1''\leq\rho\leq 4''$ and $12 \leq K_S \leq 14$~mag. Table~\ref{table: 13sigma} shows that only one confirmed companion is detected in this region; no candidate companions are found. 

\begin{table}
  \begin{tabular}{lrrrr}
    \hline
    & $\rho < 1''$ & $1''\leq\rho\leq 4''$ & $\rho > 4''$ & Total \\
    \hline
    $K_S < 12$~mag & 9 \ ($-$)   & 10 \ (2)  & 4 \ (2)  & 23 \ (4) \\ 
    $12 \leq K_S \leq 14$~mag &  $-$ \ ($-$) &  1 \ ($-$)  & $-$ \ (1)  & 1 \ (1) \\ 
    $K_S > 14$~mag &  $-$ \ ($-$) &  $-$ \ (3) &  1 \ (3) & 1 \ (6) \\
    no $K_S$  & $-$ \ ($-$) & $-$ \ ($-$) & $-$ \ (1) & $-$ \ (1) \\
    \hline
    Total & 9 \ ($-$) & 11 \ (5) & 5 \ (7) & 25 \ (12) \\
    \hline
  \end{tabular}
  \caption{The distribution of companion stars over angular separation $\rho$ and $K_S$ magnitude for our sample of 31~targets with multi-color observations. Each entry lists the number of confirmed companion stars, i.e., the secondaries with $\chi^2 < 2.30$. Between brackets we list the number of candidate companion stars, which have $2.30 < \chi^2 < 11.8$. We have also included the candidate companion HIP80142-2 ($\rho=5.88''$), for which no $K_S$ measurement is available. Several candidate companions are likely to be background stars, especially faint candidates at large separation. In the region $1''\leq \rho \leq 4''$ and $12~{\rm mag} \leq K_S \leq 14$~mag (which we will study in \S~\ref{section: gap}) we find one confirmed companion and no candidate companions.  \label{table: 13sigma}}
\end{table}

\subsection{The background star population} \label{section: backgroundstarpopulation}

% introductie - wat gaan we doen?

Our method to separate companions and background stars is based on a comparison between the location of the secondaries in the color-magnitude diagrams and the isochrone. The number of background stars identified with this method should be comparable to the {\em expected} number of background stars in the fields around the targets. In this section we make a comparison between these numbers, where the expected number of background stars is based on (1) the Besan\c{c}on model of the Galaxy, and (2) the background star study in Sco~OB2 performed by \cite{shatsky2002}.

% besanson - hoe kom je aan de data

We use the Besan\c{c}on model of the Galaxy \citep{besancon} to characterize the statistical properties of the background star population. We obtain star counts in the direction of the centers of the three subgroups, as well as for $(l,b)=(300^\circ,0^\circ)$, where LCC intersects the Galactic plane. We include objects of any spectral type, luminosity class, and population, up to a distance of 50~kpc, and convert the model $K$ magnitude to $K_S$ magnitude.
As expected, the Besan\c{c}on model shows a strong variation in the number of background stars with Galactic latitude. Most background stars are found in the direction of the Galactic plane.
For a given numerical value of the magnitude limit, more background stars are expected to be found in the $K_S$ band than in the $J$ and $H$ bands. 
Although the Besan\c{c}on model is in good agreement with the observed properties of the Galaxy, it cannot be used to make accurate predictions for individual lines of sight. For example, the high variability of the interstellar extinction with the line-of-sight, which is known to be important for the background star statistics \citep{shatsky2002}, is not taken into account in the Besan\c{c}on model. A high interstellar extinction reduces the observed number of background stars significantly, which is especially important for the US subgroup, which is located near the $\rho$~Oph star forming region.

\begin{figure}[btp]
  \centering
  \includegraphics[width=0.6\textwidth,height=!]{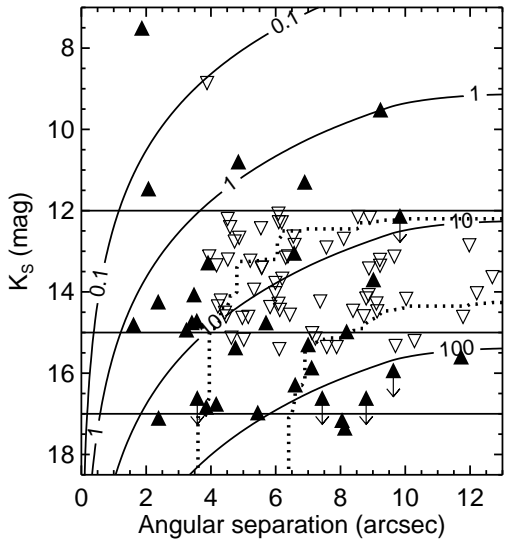}
  \caption{The {\em expected} number of background stars brighter than $K_S$ at angular separation smaller than $\rho$ for the combined sample of 199~targets (solid contours). Overplotted are the background stars detected around the 22~NACO targets (filled triangles), and those found around the 177~targets observed with ADONIS {\em only} (open triangles). The {\em observed} cumulative number of background stars is indicated with the dotted contours, for values of 10 and 50~background stars, respectively. In \cite{kouwenhoven2005} we classify secondaries brighter than $K_S=12$~mag (upper horizontal line) as candidate companions, and those fainter than $K_S=12$~mag as probable background stars. The faintest objects we detect in our ADONIS and NACO surveys have $K_S$ magnitude of approximately 15~and 17~magnitudes, respectively (lower horizontal lines). 
    \label{figure: besancon}}
\end{figure}

% besancon - de fit 

Let $F(K_S)$ be the number of background stars brighter than $K_S$, per unit of surface. Above we mentioned that the number of background stars in the Besan\c{c}on model, i.e., the normalization of $F(K_S)$, varies strongly with Galactic coordinates. The profile of $F(K_S)$, however, is very similar for different lines of sight, and can be approximated with a function of the form $F(K_S) = C_i  \times 10^{\gamma \cdot K_S}$, with $\gamma = 0.32 \pm 0.01$ for $5 \leq K_S \leq 20$~mag. The constant $C_i$ defines the normalization of $F(K_S)$, which depends on the Galactic coordinates.

% cumulative area

The number of background stars within a certain angular separation $\rho$ is proportional to the enclosed area $A(\rho)$ in the field of view within that angular separation. For our NACO observations we have a square detector of size $L_{\rm NACO}=14$~arcsec; for the observing strategy we used for the ADONIS observations we effectively have $L_{\rm ADO}=\frac{3}{2} \cdot 12.76$~arcsec \citep[see][]{kouwenhoven2005}. For a given $\rho$ (in arcsec) the enclosed area (in arcsec$^2$) is then given by
\begin{equation} \label{equation: area}
  A_{i}(\rho) =  \left\{ 
  \begin{array}{lll}
    \rho^2 & \mbox{for} & \rho \leq L_i/2 \\
    \multicolumn{3}{l}{ 
      \pi \rho^2 - 4\rho^2 \arccos \left( L_i/2\rho \right) 
      + L_i \sqrt{4\rho^2 - L_i^2} 
      } \\
    \quad\quad\quad\quad\quad\quad\quad\quad 
    & \mbox{for} & L_i/2 < \rho \leq L_i/\sqrt{2} \\
    L_i^2 & \mbox{for} & \rho > L_i/\sqrt{2} \\
  \end{array}
  \right. \,,
\end{equation}
where the subscript $i$ refers to either the ADONIS or the NACO observations.
Here we make the assumption that the target star is always in the center of the field of view. In our NACO survey we occasionally observe the target star off-axis in order to study a secondary at angular separation larger than $L_{\rm NACO}/\sqrt{2} = 9.9''$, but we ignore this effect here.

% background star distribution

We now have expressions for the quantity $N(K_S,\rho)$, the expected number of background stars with magnitude brighter than $K_S$ and angular separation smaller than $\rho$, as a function of $K_S$ and $\rho$:
\begin{equation} \label{equation: backgroundstars}
  N_{i}(K_S,\rho) = F(K_S) \cdot A_{i}(\rho)  = C_{i} \cdot 10^{\gamma \cdot K_S} \cdot  A_{i}(\rho) \,,
\end{equation}
where $\gamma= 0.32 \pm 0.01$, $A(\rho)$ is defined in Equation~\ref{equation: area}, and $C_{i}$ is a normalization constant.

\begin{table}[btp]
  \begin{tabular}{lcccc}
    \hline
     & \multicolumn{2}{l}{\ \quad ADONIS} & \multicolumn{2}{l}{\ \quad NACO} \\
    \hline
    Region          & stars per field          & $N_{\rm fields}$ & stars per field &  $N_{\rm fields}$ \\
    \hline
    US             & $0.06\quad^{+0.06}_{-0.02}$ & 51 &$0.10\quad^{+0.11}_{-0.04}$ & 9 \\
    UCL            & $0.27\quad^{+0.06}_{-0.02}$ & 64 &$0.47\quad^{+0.11}_{-0.10}$ & 3 \\
    LCC            & $0.43 \pm 0.08$        & 40 &$0.73 \pm 0.14$        & 5 \\
    GP             & $1.51 \pm 1.03$        & 22 &$2.59 \pm 1.76$        & 5 \\
    \hline
    Total          &                        & 177&                       & 22 \\
    \hline
  \end{tabular}
  \caption{The number of background stars {\em expected} in our ADONIS and NACO fields of view, based on the background star study of \cite{shatsky2002}. Column~1 lists the four regions for which we study the background statistics. The targets with Galactic latitude $|b|\leq 5^\circ$ are included in the region GP. The other targets are grouped according to their membership of US, UCL, and LCC. For the 177~targets {\em only} observed with ADONIS and the 22~targets observed with NACO we list the expected number of background stars per field of view, and the number of targets observed in the four regions. In total we expect to find $70.75 \pm 4.90$ background stars in the ADONIS sample, and $18.96 \pm 3.96$ in the NACO sample.  \label{table: backgroundstars}}
\end{table}

% shatsky introductie

For the normalization of Equation~\ref{equation: backgroundstars} we compare our observations with the background star study of \cite{shatsky2002}. Apart from their target observations, the authors additionally obtained sky images centered at 21~arcsec from each target in order to characterize the background star population. 
%
% shatsky normalization
%
From the background star study of \cite{shatsky2002} we derive the expected number of background stars in the ADONIS and NACO field of view. We assume a detection limit of $K_S = 15$~mag for ADONIS and $K_S=17$~mag for NACO \citep[which roughly corresponds to the completeness limit of the background star study of][]{shatsky2002}. Using the data from the background star study of \cite{shatsky2002} we calculate the expected number of background stars for each ADONIS and NACO field, for the three subgroups and for targets close to the Galactic plane ($|b|<5^\circ$). Table~\ref{table: backgroundstars} lists the expected number of background stars with corresponding Poisson errors. Additionally, we list the number of targets observed in each of the four regions. For the 177~targets observed with ADONIS {\em only} we expect $\approx 71$ background stars, and for the 22~targets observed with NACO we expect $\approx 19$ background stars. 
In our combined ADONIS and NACO dataset, the expected number of background stars is $90 \pm 6$. This gives normalization factors  $C_{\rm ADONIS}=1.74 \times 10^{-8}$ arcsec$^{-2}$ and  $C_{\rm NACO}=1.60 \times 10^{-8}$ arcsec$^{-2}$ for Equation~\ref{equation: backgroundstars}.  
% figuur beschrijving

Figure~\ref{figure: besancon} shows the expected number of background stars brighter than $K_S$ and closer than $\rho$ as a function of $K_S$ and $\rho$, for the combined ADONIS and NACO dataset. Since most of our targets are observed with ADONIS only, the shape of the expected background star density distribution is dominated by that of ADONIS. 
As we have a square field of size $L_{\rm ADO}=19.1$~arcsec, the cumulative number of background stars rises rapidly between $\rho=0''$ and $\rho=L_{\rm ADO}/\sqrt{2} =13.5''$ and becomes flat for larger separations. The background stars from the NACO survey are represented with the filled triangles, and those detected {\em only} in the ADONIS survey are represented with open triangles. Figure~\ref{figure: besancon} shows that the $(\rho,K_S)$ distribution of the observed 97~background stars is in good agreement with that of the expected 90~background stars. In the extreme case that all 12~candidate companions are actually background stars, the observed number of background stars is~110. In this extreme case there are 22\% more background stars than expected, which suggests that a significant part of the candidate companions may indeed be physical companions.

\begin{figure}[btp]
  \centering
  \includegraphics[width=0.6\textwidth,height=!]{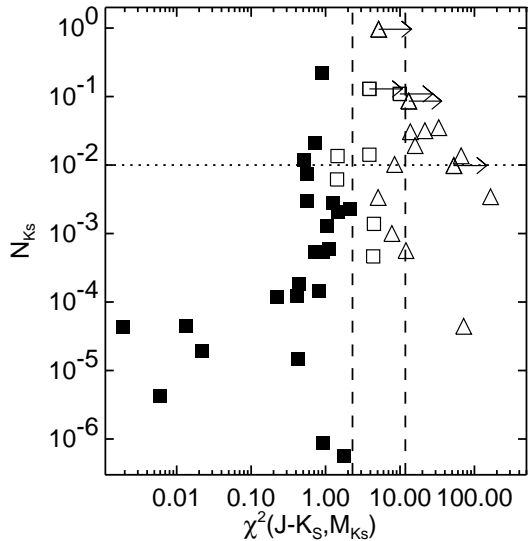}
  \caption{The $\chi^2$ distance to the isochrone in the $(J-K_S,M_{K_S})$ diagram versus $N(K_S,\rho)$. The latter quantity is the probability to detect a background star with magnitude $K_S$ (or brighter) at separation $\rho$ (or closer).
The vertical dashed lines correspond to $\chi^2=2.30$ ($1\sigma$; {\em left}) and $\chi^2=11.8$ ($3\sigma$; {\em right}). The symbols represent the confirmed companions (filled squares), the candidate companions (open squares), and background stars (triangles). Note that the classification of a secondary is based on the $\chi^2$ values for the different color-magnitude diagrams; not only for the $\chi^2$ of the $(J-K_S,M_{K_S})$-diagram, which is shown above. The expected number of background stars in a field (vertical axis) is used as a consistency check. The horizontal dotted line represents the 1\%~filter used by \cite{poveda1982} to separate companion stars (below the line) and background stars (above the line). The 1\% filter is a reasonable method when multi-color observations are not available, but is not used in our study. The triangle in the lower-right quadrant represents HIP63204-1 (see~\S~\ref{section: individualnotes}). \label{figure: bgchanceandisochrone}}
\end{figure}

% correlatie met X2 - klopt het een beetje?

Background stars are by definition not associated with the target star, and therefore generally have (1) a random position in the field of view of the observation, and (2) a position in the color-magnitude diagram that is likely to be inconsistent with the isochrone.
Figure~\ref{figure: bgchanceandisochrone} shows the relation between $N(K_S,\rho)$ and the $\chi^2$ value derived from the location of the secondary in the ($J-K_S,M_{K_S}$) diagram with the isochrone . In other words, Figure~\ref{figure: bgchanceandisochrone} shows the probability of detecting a background star at separation $\rho$ (or smaller) and magnitude $K_S$ (or brighter) versus how far away the secondary is from the isochrone.
The vertical dashed lines in Figure~\ref{figure: bgchanceandisochrone} are at $\chi^2=2.30$ and $\chi^2=11.8$, the values used to determine the status of the companions (see Table~\ref{table: classification}).
This correlation provides additional support to the method we use to separate companions and background stars. 

% correlatie met X2 - 1% filter

\cite{poveda1982} performed a statistical study of binary stars in the Index Catalogue of Visual Double Stars. They showed that it is statistically plausible to assume that components with $N(\rho,m_2) > 0.01$ are background stars, where $m_2$ is the magnitude of the secondary. This technique is referred to as the ``1\%~filter''. The horizontal line in Figure~\ref{figure: bgchanceandisochrone} represents the 1\%~filter used by \cite{poveda1982}. Secondaries below this line would be classified as companions using the 1\%~filter, and those above would be classified as background stars. Figure~\ref{figure: bgchanceandisochrone} shows that the 1\%~filter is a reasonable technique, but not as accurate as the multi-color technique used in this paper.

\subsection{Notes on some individual secondaries} \label{section: individualnotes}

% beschrijving van een ster

We detect two hierarchical triple systems: HIP68532 and HIP69113. The two companions of HIP68532 \citep[previously reported in][]{kouwenhoven2005} have a mass of 0.73~M$_\odot$ and 0.39~M$_\odot$, respectively. HIP68532 has a companions-to-primary mass ratio of $(0.73+0.39)/1.95 = 0.57$. The companions are separated $0.23''$ ($\sim28$~AU) from each other and $3.11''$ ($\sim385$~AU) from the primary, giving an estimate of 0.073 for the ratio between the semi-major axes of the inner and outer orbits. The two companions of HIP69113 \citep[previously reported in][]{huelamo2001} have a mass of 0.77~M$_\odot$ and 0.72~M$_\odot$, respectively, corresponding to a companions-to-primary mass ratio of 0.39. The companions are separated $0.26''$ ($\sim44$~AU) from each other and $5.43''$ ($\sim917$~AU) from the primary, giving an estimate of 0.048 for the ratio between the semi-major axes of the inner and outer orbits. 

% beschrijving van een ster

For HIP62026-1 we find a significant difference in position angle between the ADONIS and NACO observations. With the ADONIS observations, obtained on 8~June 2001, we find $(\rho,\varphi)=(0.22'',12.5^\circ)$. With NACO we measure $(\rho,\varphi)=(0.23'',6.34^\circ)$, 1033~days later. As the angular separation between HIP62026-1 and its primary is small, the observed position angle difference may well be the result of orbital motion. Assuming a circular, face-on orbit, we estimate an orbital period of 165~year for the system HIP62026. The total mass of the system (taken from Table~\ref{table: longtable}) is $3.64\pm 0.25~{\rm M}_\odot$, which gives via Kepler's law a semi-major axis of 46~AU. This value is in agreement with the projected semi-major axis of $\sim 24$~AU derived from the angular separation of $0.22''$ between the components (adopting a distance of 109~pc).

% beschrijving van een ster

HIP63204-1 is a bright and red object separated only $1.87$~arcsec from the LCC member HIP63204. The isolated location of HIP63204-1 in the bottom-right quadrant of Figure~\ref{figure: bgchanceandisochrone} shows that the probability of finding a background star of this magnitude (or brighter) at this angular separation (or closer) is small. According to its location in the color-magnitude diagrams, HIP63204-1 is a background star and hence we classify it as such. HIP63204 and its companion HIP63204-2 at $\rho=0.15''$ have masses of $2.05~{\rm M}_\odot$ and $1.06~{\rm M}_\odot$, respectively. If HIP63204-1 (at $\rho=1.87''$) would be a companion, its mass would be approximately $1~{\rm M}_\odot$, in which case HIP63204 would be an unstable triple system.  The colors of HIP63204-1 are consistent with a $0.075~{\rm M}_\odot$ brown dwarf with an age of 10~Gyr at a distance of 60~pc, and are also consistent with those of an M5~III giant at a distance of $\sim 5.6$~kpc \citep[using the models of][]{allen2000,chabrier2000}. 

% beschrijving van een ster

HIP81972-3, HIP81972-4, and HIP81972-5 fall on the 20~Myr isochrone in all three color-magnitude diagrams. These objects are likely low-mass companions of HIP81972 (see \S~\ref{section: massfunction}).  HIP81972-5 is the topmost companion (black square) in Figure~\ref{figure: bgchanceandisochrone}. HIP81972-5 is the faintest companion in our sample, and the expected number of background stars with a similar or brighter magnitude and a similar or smaller separation is large ($\sim 16$ for the ADONIS sample). The secondary HIP81972-2 (at $\rho=7.02''$) was reported before as a possible companion of HIP81972 in \cite{wds1997}, but the secondary is classified as a background star by \cite{shatsky2002}. With our NACO multi-color observations we cannot determine the nature of this secondary with high confidence. The LCC member HIP81972 is located very close to the Galactic equator ($b=3^\circ 11'$), so care should be taken; background stars with a magnitude similar to that of the secondaries of HIP81972 are expected in the field around this star.

% close background stars

\cite{kouwenhoven2005} identified seven ``close background stars'' ($K_S > 14~\mbox{mag}; \rho < 4''$), for which the background star status was derived using the $K_S$ magnitude only. These are objects next to the targets HIP61265, HIP67260, HIP73937, HIP78958, HIP79098, HIP79410, and HIP81949. The ADONIS observations are incomplete in this region (see Figure~\ref{figure: detectionlimits}). More low-mass companions with $K_S > 14$~mag and $1'' < \rho < 4''$ may be present for the 177~A and late-B stars {\em only} observed with ADONIS.

With our NACO multi-color observations we confirm that five of the ``close background stars'' (HIP61265-2, HIP73937-2, HIP79098-1, 79410-1, and HIP81949-2) are background stars. 
For the other two secondaries, HIP67260-3 and HIP78969-1, we cannot determine whether they are background stars or brown dwarf companions. As many background stars with similar magnitudes are expected in the field, these are likely background stars. However, follow-up spectroscopic observations are necessary to determine the true nature of these close secondaries.

\subsection{Accuracy of the $K_S=12$ separation criterion} \label{section: k12}

\begin{table}[btp]
  \begin{tabular}{c cc cc c}
    \hline
    Status & \multicolumn{2}{l}{$K_S < 12$~mag} & \multicolumn{2}{l}{$K_S > 12$~mag} & Total \\
    \hline
    c & 23 &(70\%) & 2& (6\%)& 25 \\
    ? & 4 &(12\%)& 7 &(19\%)& 11 \\
    b & 6 &(18\%)& 27& (75\%) & 33 \\
    \hline
    Total  & 33 &(100\%)& 36& (100\%)& 69 \\  
    \hline
  \end{tabular}
  \caption{Accuracy of the $K_S=12$~mag criterion to separate companions and background stars. This table contains 69~out of the 72~secondaries with multi-color observations in the ADONIS or NACO dataset. Three secondaries (1~candidate companion; 2~background stars) for which no $K_S$ magnitudes are available, are not included. The first column shows the status of the secondary (c = confirmed companion, ? = candidate companion, b = background star). Columns~2 to~5 list the distribution over status for secondaries with $K_S<12$~mag and $K_S>12$~mag. Depending on the true nature of the candidate companion stars, the $K_S=12$~mag criterion correctly classifies the secondaries in $\sim 80\%$ of the cases. \label{table: k12comparison}}
\end{table}

One of the goals of our study is to evaluate the accuracy of the $K_S=12$~mag criterion that we used to separate companions and background stars in the ADONIS survey. This is possible, now that we have performed a multi-color analysis of 72~secondaries around 31~members of Sco~OB2.
Table~\ref{table: k12comparison} shows the distribution of secondary status for the secondaries with $K_S <12$~mag and those with $K_S>12$~mag (three secondaries without $K_S$ measurements are not included). According to the $K_S=12$~criterion, all secondaries brighter than $K_S=12$~mag are companions, while all fainter secondaries are background stars.

If we consider only the confirmed companions and background stars, we see that the $K_S=12$~mag criterion correctly classifies the secondaries in $f = (23+27)/(25+33) = 86\%$ of the cases. If all candidate companions are indeed companions, we have $f=78\%$, while if all candidate companions are background stars, we have $f=82\%$. This indicates that  $\sim 80\%$ of the candidate companions identified by \cite{kouwenhoven2005} are indeed companion stars.

The $K_S=12$~mag criterion is accurate for the measured set of secondaries {\em as a whole}. It is obvious that for the lowest-mass companions the criterion is not applicable, as virtually all brown dwarf and planetary companions have $K_S > 12$~mag at the distance of Sco~OB2. Out of the 25~confirmed companions we find with our multi-color analysis, 23 indeed have $K_S < 12$~mag, but two have $K_S > 12$~mag. These are the brown dwarf companions of HIP81972 (see \S~\ref{section: massfunction} for a discussion).  

Now that we have confirmed the validity of the $K_S=12$~mag criterion, many of the candidate companions found in the ADONIS survey very likely are physical companion stars. Table~\ref{table: adonisnaco} gives an overview of all candidate and confirmed companion stars identified in the ADONIS and NACO surveys.

% ====================================================================
% ====================================================================
% ====================================================================
% ==MASSES============================================================
% ====================================================================
% ====================================================================
% ====================================================================

\section{Masses and mass ratios} \label{section: massfunction}

For each primary and companion star we derive the mass using its color and magnitude. We find the best-fitting mass by minimizing the $\chi^2$ difference between the isochrone and the measurements, while taking into account the errors in the measurements:
\begin{equation}
\chi^2 = \left( \frac{\Delta (J-K_S)}{\sigma_{J-K_S}} \right)^2 +  \left( \frac{\Delta (H-K_S)}{\sigma_{H-K_S}} \right)^2 +  \left( \frac{\Delta M_{K_S}}{\sigma_{M_{K_S}}} \right)^2 .
\end{equation}
The masses of all primaries and confirmed companions are listed in Table~\ref{table: longtable}. We additionally list the masses of the candidate companions, assuming that they are indeed companions, but we do not include these masses in our analysis. We find primary masses between $1.1~{\rm M}_\odot$ and $4.9~{\rm M}_\odot$. The confirmed companion star masses range between $0.03~{\rm M}_\odot$  and $1.19~{\rm M}_\odot$, with mass ratios $0.006 < q < 0.55$.  The average error in the mass as a result of the error in the color and magnitude is $8.5\%$ for the primaries and $12\%$ for the companion stars. The average error in the mass ratio is $12.5\%$.
Although accurate $B$ and $B-V$ measurements are available for the primaries, we do not use these. The $B$ and $V$ measurements often include the flux of unresolved close companions, and therefore lead to overestimating the primary masses.

\cite{kouwenhoven2005} derived masses from $K_S$ magnitudes only. For the primary stars these are close to those obtained from multi-color observations in our current analysis. The rms difference between the masses derived using the two methods is 6.6\%. No systematic difference is present for the primaries. \cite{kouwenhoven2005} overestimated the companion star masses with $\sim 2.2\%$ ($\sim 0.01~{\rm M}_\odot$) on average.

The companions with the lowest mass are those of HIP81972, which have masses of $0.35~{\rm M}_\odot$ ($370~{\rm M_J}$), $0.06~{\rm M}_\odot$ ($63~{\rm M_J}$), and $0.03~{\rm M}_\odot$ ($32~{\rm M_J}$). The latter two are likely brown dwarfs. With an angular separation of $\rho=2.79''$, the $63~{\rm M_J}$ component is the only observed brown dwarf in the $1''-4''$ angular separation interval (see~\S~\ref{section: gap}). The other two companions of HIP81972 have a larger angular separation.

% ====================================================================
% ====================================================================
% ====================================================================
% ==GAP=======================================================
% ====================================================================
% ====================================================================
% ====================================================================

\section{The lower end of the companion mass distribution} \label{section: gap}

\begin{SCfigure}[][btp]
  \centering
  \includegraphics[width=0.5\textwidth,height=!]{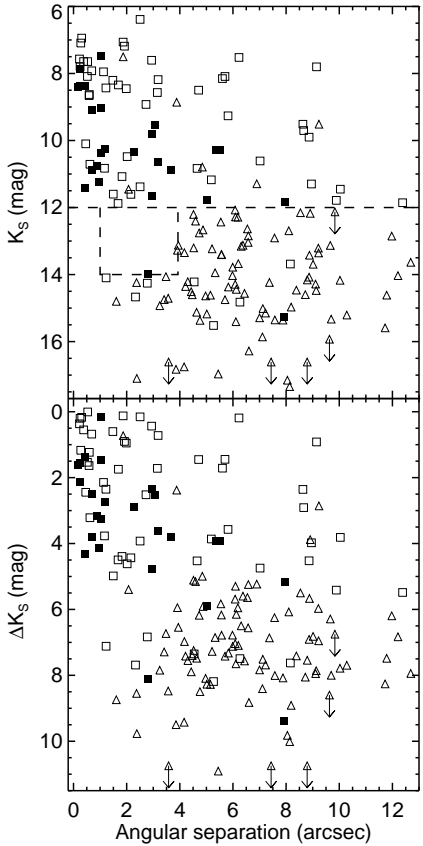}
  \caption{Companion star magnitude $K_S$ ({\em top}) and magnitude difference $\Delta K_S$ ({\em bottom}) versus angular separation for the combined ADONIS and NACO datasets.
The symbols represent confirmed companions (filled squares), candidate companions (open squares), and background stars (triangles); see Section~\ref{section: separation} for further information on this classification.
The horizontal line corresponds to $K_S=12$~mag, the criterion used by \cite{kouwenhoven2005} to separate companion stars and background stars. Typical detection and completeness limits corresponding to the observations are shown in Figure~\ref{figure: detectionlimits}. For a given $K_S$ magnitude, the number of background stars closer than angular separation $\rho$ is given by Equation~\ref{equation: backgroundstars}. This figure clearly shows the dearth of brown dwarf companions with $1'' < \rho < 4''$ around A and late-B stars in Sco~OB2 (region indicated with the dashed rectangle).  \label{figure: detlimk} } 
\end{SCfigure}

\cite{kouwenhoven2005} discussed the potential lack of substellar companions to A and late-B members of Sco~OB2. With our NACO follow-up observations we confirm the very low number of brown dwarf companions with respect to the number of stellar companions found around these stars (\S~\ref{section: gapexistence}). In Section~\ref{section: realdesert} we will discuss whether a brown dwarf desert exists among A and late-B members of Sco~OB2. In Section~\ref{section: origin} we will briefly discuss the potential origin of such a brown dwarf desert. We will show that the small brown dwarf companion fraction among A- and B stars in Sco~OB2 can be explained by an extrapolation of the {\em stellar} companion mass distribution, i.e., there is no need to eject brown dwarf companions from binary systems at an early stage of the formation process \cite[the embryo ejection scenario;][]{reipurth2001}.

\subsection{The brown dwarf ``gap'' for $1'' \leq \rho \leq 4''$} \label{section: gapexistence}

\cite{kouwenhoven2005} observed a gap in the $(\rho,K_S)$ distribution of the stellar companions in the Sco~OB2 binary population: no secondaries with a magnitude $12~\mbox{mag} \leq K_S \leq 14~\mbox{mag}$ and an angular separations $\rho < 4''$ were detected. These secondaries should have been detected, had they existed, since the ADONIS survey is almost complete in this region (see Figure~\ref{figure: detectionlimits}). Figure~\ref{figure: detlimk} shows the distribution of $K_S$ and $\Delta K_S$ as a function of $\rho$ for the ADONIS and NACO observations combined. The ``gap'' in the $(\rho,K_S)$ distribution described above is clearly visible. With our NACO survey we detect one companion at the bottom of this region: the brown dwarf companion HIP81972-4 ($K_S = 13.98 \pm 0.12$~mag; see also \S~\ref{section: separation} and~\ref{section: massfunction}). No other secondaries are present in this region.

The stellar companion fraction is the fraction of stars with stellar companions. Among A and late-B stars in Sco~OB2 in the semi-major axis range $1''-4''$ ($130-520$~AU) we find a stellar companion fraction of $14\pm 3\%$. Similarly, the brown dwarf companion fraction is the fraction of stars with brown dwarf companions. The brown dwarf companion fraction for the stars in this separation range is $0.5 \pm 0.5\%$\footnote{Note that we only find one brown dwarf companion with $12~{\rm mag} \leq K_S \leq 14~{\rm mag}$. We find no background stars in this region, so accidental misclassification of companions as background stars is not an issue here.} (for brown dwarfs with $K_S < 14$~mag). The substellar-to-stellar companion ratio $R$ is defined as
\begin{equation} \label{equation: substel-to-stel}
R = \frac{ \mbox{number of brown dwarf companions} }{ \mbox{number of stellar companions} } \,.
\end{equation}
In our study we cannot calculate $R$ because we do not know how many faint ($K_S > 14$~mag) brown dwarf companions are missing. We therefore calculate the {\em restricted} substellar-to-stellar companion ratio $R_\star$, using only the brown dwarf companions brighter than $K_S=14$~mag\footnote{In this section we denote the {\em observed} quantities with a star as a subscript. For example, $R$ denotes the substellar-to-stellar companion ratio (including all brown dwarfs), while $R_\star$ indicates the {\em observed} substellar-to-stellar companion ratio, including only the brown dwarfs brighter than $K_S=14$~mag.}.  In the angular separation range $1''-4''$ we observe $R_{\rm \star,IM} = 0.036 \pm 0.036$ for intermediate mass stars in Sco~OB2.
We cannot make similar statements for companions with properties other than those described above. For angular separations smaller than $1''$ our survey is significantly incomplete for $12~\mbox{mag} \leq K_S \leq 14~\mbox{mag}$. For $\rho > 4''$ many objects with $12~\mbox{mag} \leq K_S~\leq 14~\mbox{mag}$ are likely background stars, of which the status still needs to be confirmed. Finally, the ADONIS survey is incomplete for $K_S > 14$~mag.

\subsection{A real brown dwarf desert?} \label{section: realdesert}

% BDD definition

The brown dwarf desert is defined as a deficit (not necessarily a total absence) of brown dwarf companions, either relative to the frequency of companion stars or relative to the frequency of planetary companions \citep{mccarthy2004}. 
%
% measurables
%
In this paper the brown dwarf desert for A~and late-B members of Sco~OB2 is characterized by a small number of observed companions $N_{\rm \star,BD}$ with $12~\mbox{mag} \leq K_S \leq 14$~mag and $1'' \leq \rho \leq 4''$ and a small substellar-to-stellar companion ratio $R_\star$. In general, the quantities $N_{\rm BD}$ and $R$ depend on  (1) the mass distribution, (2) the pairing properties of the binary systems, and (3) the spectral type of the stars in the sample. Unlike $R$, the value of $N_{\rm BD}$ also depends on (4) the multiplicity fraction $F_{\rm M}$ (Equation~\ref{equation: binarystatistics}), and (5) the semi-major axis (or period) distribution. We use simulated observations and compare the observed values with those predicted for various models, in order to roughly estimate the mass distribution and pairing properties. For comparison between observations and simulations we only consider the companions brighter than $K_S=14$~mag in the angular separation range $1'' \leq \rho \leq 4''$.

\begin{landscape}
\begin{table}
  \begin{tabular}{llcccc}
    \hline
    \# & Model & $N_{\rm \star,BD,IM}$ & $R_{\rm \star,IM}$  & $N_{\rm \star,BD,LM}$ & $R_{\rm \star,LM}$ \\
    \hline
    0 & ADONIS/NACO observations                                                                    & $1   \pm 1  $    & $0.036 \pm 0.036$ & unknown             & unknown \\
    \hline
    1 &  extended Preibisch  MF, $\alpha=-0.9$, random pairing             & $      5.50\pm       0.48 $&$     0.34\pm     0.03 $&$       7.19\pm      0.17 $&$      0.50 \pm    0.01  $ \\ 
    2 &  extended Preibisch  MF, $\alpha=-0.3$, random pairing             & $      4.50\pm       0.41 $&$     0.24\pm     0.03 $&$       5.08\pm      0.13 $&$      0.30 \pm    0.01  $ \\ 
    3 &  extended Preibisch  MF, $\alpha=+2.5$, random pairing            & $      1.07\pm       0.18 $&$     0.05\pm     0.01 $&$       1.42\pm      0.07 $&$      0.07 \pm    0.01  $ \\ 
    4 &  Salpeter MF, random pairing                                      & $     15.31\pm       2.79 $&$     6.00\pm     2.90 $&$      17.18\pm      0.88 $&$      3.95 \pm    0.45  $ \\ 
    5 &  extended Preibisch MF, $\alpha=-0.9$, $f_q(q) \propto q^{-0.33}$    & $      0.72\pm       0.24 $&$     0.04\pm     0.01 $&$       3.42\pm      0.14 $&$      0.18 \pm    0.01  $ \\
    6 &  extended Preibisch MF, $\alpha=-0.3$, $f_q(q) \propto q^{-0.33}$    & $      0.71\pm       0.22 $&$     0.04\pm     0.01 $&$       3.35\pm      0.14 $&$      0.18 \pm    0.01  $ \\ 
    7 &  extended Preibisch MF, $\alpha=+2.5$, $f_q(q) \propto q^{-0.33}$   & $      1.19\pm       0.27 $&$     0.06\pm     0.01 $&$       3.30\pm      0.13 $&$      0.18 \pm    0.01  $ \\ 
    8 &  Salpeter MF, $f_q(q) \propto q^{-0.33}$                            & $      1.00\pm       0.57 $&$     0.05\pm     0.02 $&$       3.70\pm      0.57 $&$      0.20 \pm    0.03  $ \\ 
    \hline
   \end{tabular}
  \caption{The observed and expected number of brown dwarfs with $1'' \leq \rho \leq 4''$ and $12 \leq K_S \leq 14$~mag for the sample of 199~target stars. The left columns shows the various models for which we simulated observations. Each model has a semi-major axis distribution $f_a(a) \propto a^{-1}$ with $15~$R$_\odot \leq a \leq 5\times 10^6$~R$_\odot$, and a multiplicity fraction of $F_{\rm M} =100\%$. 
Columns 3 and 4 show for a survey of intermediate mass stars (late-B and A stars; $1.4~{\rm M}_\odot < M < 7.7~{\rm M}_\odot$) the expected number of brown dwarfs $N_{\rm \star,BD,IM}$ and the substellar-to-stellar companion ratio $R_{\rm \star,IM}$, both with $1\sigma$ errors. 
By comparing the predicted values of $N_{\rm \star,BD,IM}$ and $R_{\rm \star,IM}$ with the observations, we can exclude models 1, 2, and 4. In \cite{kouwenhoven2005} we exclude random pairing from the Preibisch mass distribution (models 1$-$3) since these models are inconsistent with the observed mass ratio distribution of stellar companions. We additionally list the values $N_{\rm \star,BD,LM}$ and $R_{\rm \star,LM}$ that are expected for a survey amongst 199~low-mass stars ($0.3~{\rm M}_\odot < M < 1.4~{\rm M}_\odot$) in columns 5 and 6. For models with $F_{\rm M} < 100\%$ the expected number of brown dwarfs reduces to $F_{\rm M} \times N_{\rm \star,BD}$, while $R$ remains unchanged. Models with a smaller semi-major axis range and models with the log-normal period distribution found by \cite{duquennoy1991} have a larger expected value of $N_{\rm \star,BD,IM}$, $N_{\rm \star,BD,LM}$. Under the assumption that companion mass and semi-major axis are uncorrelated, the values of  $R_{\rm \star,IM}$ and  $R_{\rm \star,LM}$ are equal to those listed above, if the log-normal period distribution is chosen.
    \label{table: modelbackgroundstars}}
\end{table}
\end{landscape}

\begin{table}
  \begin{tabular}{lcccc}
    \hline
    \# & $N_{\rm BD,IM}$ & $R_{\rm IM}$  & $N_{\rm BD,LM}$ & $R_{\rm LM}$ \\
    \hline
    0  &  unknown      & unknown & unknown            & unknown \\
    \hline
    1  & $      6.68 \pm      0.53  $&$      0.44 \pm     0.04  $&$       8.50 \pm      0.18  $&$      0.65 \pm     0.02  $\\ 
    2  & $      5.46 \pm      0.45  $&$      0.31 \pm     0.03  $&$       6.42 \pm      0.15  $&$      0.42 \pm     0.01  $\\ 
    3  & $      2.01 \pm      0.25  $&$      0.10 \pm     0.01  $&$       2.65 \pm      0.09  $&$      0.14 \pm     0.01  $\\
    4  & $     15.82 \pm      2.84  $&$      7.75 \pm     4.12  $&$      18.26 \pm      0.91  $&$      5.60 \pm     0.72   $\\ 
    5  & $      1.20 \pm      0.31  $&$      0.06 \pm     0.02  $&$       4.26 \pm      0.16  $&$      0.24 \pm     0.01   $\\ 
    6  & $      1.13 \pm      0.28  $&$      0.06 \pm     0.02  $&$       4.30 \pm      0.16  $&$      0.25 \pm     0.01   $\\ 
    7  & $      1.42 \pm      0.29  $&$      0.07 \pm     0.01  $&$       4.18 \pm      0.14  $&$      0.24 \pm     0.01   $\\ 
    8  & $      1.19 \pm      0.63  $&$      0.06 \pm     0.03  $&$       5.02 \pm      0.66  $&$      0.30 \pm     0.04   $\\
    \hline
   \end{tabular}
  \caption{The observed and expected number of brown dwarfs with $1'' \leq \rho \leq 4''$ for the sample of 199~target stars. In this table we show the results for the full brown dwarf mass range ($0.02$~M$_\odot \leq M \leq 0.08$~M$_\odot$), unlike in Table~\ref{table: modelbackgroundstars}, where we show the results for the brown dwarfs restricted to $12 \leq K_S \leq 14$~mag.
    \label{table: modelbackgroundstars2}}
\end{table}

% cluster opbouwen / distributions

We simulate models using the STARLAB package \citep[e.g.,][]{ecology4}. The primary mass is drawn in the mass range $0.02~{\rm M}_\odot \leq M \leq 20~{\rm M}_\odot$, either the Salpeter mass distribution 
\begin{equation} \label{equation: imfsalpeter}
f_M(M) = \frac{dM}{dN} \propto M^{-2.35} \,,
\end{equation}
or from the the extended Preibisch mass distribution
\begin{equation} \label{equation: imfpreibisch}
  f_M(M) = \frac{dM}{dN} \propto \left\{
  \begin{array}{llll}
    M^{\alpha}   & {\rm for \quad } 0.02 & \leq M/{\rm M}_\odot & < 0.08 \\
    M^{-0.9}     & {\rm for \quad } 0.08 & \leq M/{\rm M}_\odot & < 0.6 \\
    M^{-2.8}     & {\rm for \quad } 0.6  & \leq M/{\rm M}_\odot & < 2   \\
    M^{-2.6}     & {\rm for \quad } 2    & \leq M/{\rm M}_\odot & < 20  \\
  \end{array}
  \right.\,.
\end{equation}
The extended Preibisch mass distribution  \citep[see][]{kouwenhoven2005} is based on the study by \cite{preibisch2002}, who derived $f_M(M)$ with $M > 0.1$~M$_\odot$ for the US subgroup of Sco~OB2.  Since our current knowledge about the brown dwarf
population in OB~associations (particularly Sco~OB2) is incomplete \citep[e.g., Table 2 in][]{preibisch2003}
we simulate associations with three different slopes for the mass distribution
in the brown dwarf regime. 
We extend the Preibisch mass distribution down to $0.02~{\rm M}_\odot$ with $\alpha=-0.9$, $\alpha=-0.3$, or $\alpha=+2.5$. The mass distribution with $\alpha=-0.9$ has the same slope in the brown dwarf regime as for the low-mass stars. The simulations with $\alpha=-0.3$ and $\alpha=+2.5$ bracket the values for $\alpha$ that are observed in various clusters and the field star population \citep[see][for a summary]{preibisch2003}.
The companion mass is obtained via randomly pairing the binary components from the mass distribution or via a mass ratio distribution of the form $f_q(q) \propto q^{-0.33}$ with $0 < q < 1$ and the requirement that any companion has a mass larger than $0.02~{\rm M}_\odot$. The latter mass ratio distribution was derived  from the observed mass ratio distribution in our ADONIS survey \citep{kouwenhoven2005}. For the models with random pairing, the primary star and companion mass are drawn independently from the mass distribution, and switched, if necessary, so that the primary is the most massive star.

% cluster opbouwen / overige parameters
Each simulated association consists of 100\,000 binaries, has a distance of 130~pc and a homogeneous density distribution with a radius of 20~pc, properties similar to those of the subgroups in Sco~OB2. We assume a thermal eccentricity distribution and a  semi-major axis distribution of the form $f_a(a) \propto a^{-1}$, which is equivalent to $f_{\log a}(\log a) = \mbox{constant}$ (\"{O}piks law). The lower limit of $a$ is set to $15~$R$_\odot$. The upper limit is set to $5\times 10^6~$R$_\odot \approx 0.1~\mbox{pc}$, the separation of the widest observed binaries in the Galactic disk \citep[e.g.,][]{close1990,chaname2004}. 

% tabel en comparison 

Table~\ref{table: modelbackgroundstars} lists for eight models the predicted value of $N_{\rm \star,BD}$, the expected number of brown dwarfs with $1'' \leq \rho \leq 4''$ and $12 \leq K_S \leq 14$~mag, normalized to a sample of 199~stars. The table also lists $R_{\star}$, the ratio between the number of brown dwarf companions with $K_S \leq 14$~mag and the number of stellar companions in the separation range $1''-4''$.
Table~\ref{table: modelbackgroundstars2} lists $N_{\rm BD}$ (the intrinsic number of brown dwarfs with $1''\leq \rho \leq 4''$), and corresponding ratio $R$. In this table {\em all} companions are taken into account, including those with $K_S > 14$~mag. The values in Table~\ref{table: modelbackgroundstars} can be compared directly with the observations, while those in Table~\ref{table: modelbackgroundstars2} represent the intrinsic properties of each association model.
A brown dwarf with $K_S=14$~mag in the US subgroup has a mass of less than $0.02~$M$_\odot$  (21~${\rm M_J}$), and a brown dwarf with a similar brightness in the UCL and LCC subgroup has a mass of $\sim 0.038~$M$_\odot$ (40~${\rm M_J}$). 

% B en G samples

The definition of the brown dwarf desert given in the beginning of this section is generally used for binarity studies of late-type stars. In our study the primaries are intermediate mass stars, allowing companion stars over a larger mass range than for low mass primaries. This naturally leads to lower values of $N_{\rm BD}$ and $R$. 
We therefore also list the results for a simulated survey of 199~low mass stars. Tables~\ref{table: modelbackgroundstars} and~\ref{table: modelbackgroundstars2} show that indeed the expected values $N_{\rm BD}$ and $R$ for low-mass stars are higher than those for intermediate-mass stars by $\sim 30\%$ for the random pairing models, and by $\sim 250\%$ for the models with $f_q(q) \propto q^{-0.33}$.

% multiplicity fraction

A multiplicity fraction of $F_{\rm M}=100\%$  is assumed in each model. For A and late~B members of Sco~OB2 the {\em observed} multiplicity fraction $F_{\rm M}$ is $\approx 50\%$ for late~B and A stars \citep{kouwenhoven2005}. This is a lower limit of the {\em true} multiplicity fraction due to the presence of unresolved companions, and hence we have $50\% \la F_{\rm M} \leq 100\%$. For a multiplicity fraction smaller than $100\%$, the expected number of brown dwarfs is given by $F_{\rm M} \times N_{\rm \star,BD}$, while the values of $R$ remain unchanged. In each model we adopted \"{O}piks law, with $15~$R$_\odot \leq a \leq 5\times 10^6~$R$_\odot$. In reality, the upper limit for $a$ may be smaller, as Sco~OB2 is an expanding association \citep{blaauw1964A,brown1999}. If this is true, the values for $N_{\rm \star,BD}$ are underpredicted, as less companions are expected to have very large separations. 
Furthermore, instead of \"{O}piks law, it may also be possible that the log-normal period distribution found by \cite{duquennoy1991} holds. For a model with \"{O}piks law at a distance of 130~pc, 11\% of the companions have separations between $1''-4''$, while for a model with the log-normal period distribution, 13\% of the companions have separations between $1''-4''$. If the log-normal period distribution holds, the values for $N_{\rm \star,BD}$ in Tables~\ref{table: modelbackgroundstars} and~\ref{table: modelbackgroundstars2} are underpredicted. The value of $R$ (and $R_\star$) does not change for the possible adaptations described here, under the assumption that the stellar and substellar companions have the same semi-major axis (or period) distribution.

% tabel en comparison / resultaten : mass ratios

\begin{figure}[btp]
  \centering
  \includegraphics[width=1\textwidth,height=!]{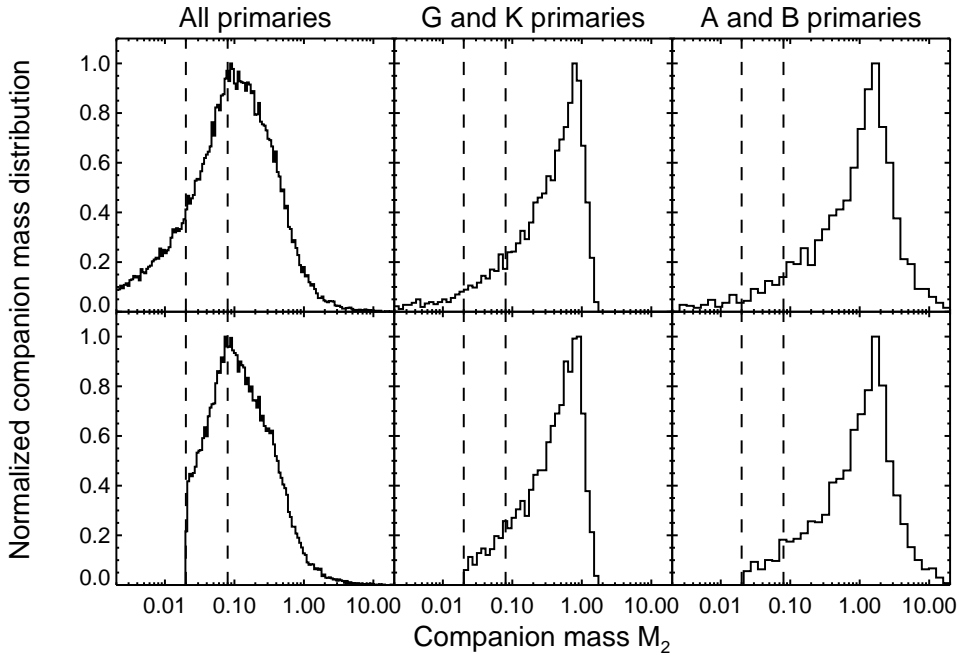}
  \caption{The companion mass distribution $f_{M_2}(M_2)$ for a simulated association. In each panel we show the distribution $f_{M_2}(M_2)$ of an association consisting of 50\,000 binaries for which the primary mass distribution is given by Equation~\ref{equation: imfpreibisch}. From left to right, the panels show the distribution of companion mass in all binaries with stellar primaries, for those of G and K primaries, and for those of A and B primaries, respectively. In the top panels we adopted a mass ratio distribution $f_q(q) \propto q^{-0.33}$ with $0 < q \leq 1$. In the bottom panels we adopt the same distribution, but with the additional constraint that $M_2 \geq 0.02$~M$_\odot$. The brown dwarf regime is indicated with the dashed lines. This figure shows that, even though the mass ratio distribution is strongly peaked to low values of $q$, the substellar-to-stellar companion ratio among intermediate- and high-mass stars is very low. 
  \label{figure: fm2_distribution} } 
\end{figure}

Table~\ref{table: modelbackgroundstars} shows that the models with $f_q(q) \propto q^{-0.33}$ are in good agreement with our observations for any value of $\alpha$. The reason for this is that $N_{\rm \star,BD,IM}$ and $R_{\rm \star,IM}$ are independent of $\alpha$ for these models, as only the primary is chosen from the mass distribution.
For the models with random pairing, the two components of each binary system are {\em independently} chosen from the mass distribution. Only those models with a turnover in the mass distribution in the brown dwarf regime are consistent with the observations (for a multiplicity fraction of $0.5 \la F_{\rm M} \leq 1$).
However, in \cite{kouwenhoven2005} we excluded random pairing by studying the observed mass ratio distribution for stellar companions. The remaining models that are consistent with our observations have an extended Preibisch mass distribution and a mass ratio distribution of the form $f_q(q) \propto q^{-0.33}$. Although this distribution is peaked to low values of $q$, the number of brown dwarf companions is significantly smaller than the number of stellar companions. For example, for a sample of binaries with a primary mass of $3$~M$_\odot$, the substellar-to-stellar companion mass ratio $R$ (see Equation~\ref{equation: substel-to-stel}) resulting from $f_q(q) \propto q^{-0.33}$ is given by
\begin{equation}
R = \frac{\displaystyle\int_{0.02/3}^{0.08/3} f_q(q)\,dq}{\displaystyle\int_{0.08/3}^{1} f_q(q)\,dq} 
  = \frac{\displaystyle\left[\ q^{0.67}\ \right]_{\ 0.02/3}^{\ 0.08/3}}{\displaystyle\left[\ q^{0.67}\ \right]_{\ 0.08/3}^{\ 1}} 
  = 0.059 \,,
\end{equation}
Figure~\ref{figure: fm2_distribution} further illustrates that a small value for $R$ is expected among binaries with an intermediate-mass or solar-type primary, even if $f_q(q)$ is strongly peaked to low values of $q$. For an OB~association with 100\% binarity and a mass ratio distribution of the form $f_q(q) \propto q^{-0.33}$, only a fraction $(1+R^{-1})^{-1}$ of the binaries have a brown dwarf companion. Among primaries with a mass of 1~M$_\odot$ and 3~M$_\odot$, this fraction is $\sim 14\%$ and $\sim 6\%$, respectively. For this mass ratio distribution, the number of brown dwarf companions is significantly smaller than the number of stellar companions, even if observational biases are not taken into account. If binary formation truly results in a mass ratio distribution similar to $f_q(q)\propto q^{-0.33}$, the brown dwarf desert (in terms of the ``deficit'' of brown dwarf companions relative to stellar companions) is a natural outcome of the star forming process.

% vergelijking met totale rho range

The observed number of brown dwarfs (with 12~mag $\leq K_S \leq 14$~mag) is $N_{\rm \star,BD,IM} = 1\pm 1$. After correction for unseen low-mass brown dwarfs (with $K_S>14$~mag) this translates to $N_{\rm BD,IM} = 1.6 \pm 1.6$ brown dwarfs (cf. Tables~\ref{table: modelbackgroundstars} and~\ref{table: modelbackgroundstars2}). If we assume a semi-major axis distribution of the form $f_a(a) \propto a^{-1}$ with $15~$R$_\odot < a < 5\times 10^6~$R$_\odot$ and a distance of 130~pc, we expect $\sim 11\%$ of the companions to be in the angular separation range $1''-4''$. Assuming that companion mass and semi-major axis are uncorrelated, this also means that $11\%$ of the brown dwarfs are in this range. Extrapolation gives an estimate of $(1.6 \pm 1.6)/0.11 = 14.5 \pm 14.5$~brown dwarf companions around the 199~target stars, or a brown dwarf companion fraction of $7.3 \pm 7.3\%$ for intermediate mass stars in Sco~OB2. On the other hand, if we assume the log-normal period distribution found by \cite{duquennoy1991}, we find a corresponding brown dwarf companion fraction of $6.2 \pm 6.2\%$ (for a model binary fraction of 100\%).
Note that if a mass ratio distribution $f_q(q)$ is adopted, these values are independent of the slope $\alpha$ of the mass distribution in the brown dwarf regime.

% vergelijking met McCarthy

\cite{mccarthy2004} find a companion frequency of $0.7 \pm 0.7 \%$ for brown dwarf companions ($M > 30~{\rm M_J}$) to F, G, K, and M stars in the separation range $120-1200$~AU. Assuming a semi-major axis distribution of the form $f_a(a) \propto a^{-1}$, our brown dwarf companion frequency of $0.5 \pm 0.5\%$ ($M \ga 30~{\rm M_J}$) for the range $130-520$~AU translates to a value of $0.83 \pm 0.83\%$ for the range $120-1200$~AU, which is in good agreement with the frequency found by \cite{mccarthy2004}. This ``extrapolated'' brown dwarf companion frequency may underestimate the true value, if the brown dwarf desert does not exist at larger separations \citep[which may be the case for low-mass stars in the solar neighbourhood; e.g.,][]{gizis2001}.

\subsection{The origin of the brown dwarf desert} \label{section: origin}

Most stars are formed and reside in binary or multiple stellar systems. Knowledge about binary and multiple systems in young stellar groupings is of fundamental importance for our understanding of the star formation process. The formation of brown dwarfs and the dearth of brown dwarf companions has attained much interest over the last decade.
Theories have been developed, explaining the existence of the brown dwarf desert using migration \citep[][most effective at $a \la 5$~AU]{armitage2002} or ejection \citep{reipurth2001} of brown dwarfs.
The most popular theory that explains the brown dwarf desert is the {\em embryo ejection scenario}. 
This scenario predicts ejection of brown dwarfs soon after their formation by dynamical interactions in unstable multiple systems \citep{reipurth2001}. Hydrodynamical calculations \citep[e.g.,][]{bate2003} suggest that star formation is a highly dynamic and chaotic process. Brown dwarfs are ejected during or soon after their formation. In this theory brown dwarfs can be seen as failed stellar companions.

\begin{table}
  \begin{tabular}{ccccc}
    \hline
    \# & $F_{1''-4''}~(\%)$ & $F_{\rm IM}~(\%)$ & $N_{\rm BD,IM,i}$ & $N_{\rm BD,IM,i,total}$ \\
    \hline
    1  & 10.9 & 4.93  & $ 6.26  \pm 1.19  $  & $ 7.93 \pm 2.05 $\\
    2  & 10.9 & 5.34  & $ 6.78  \pm 1.29  $  & $ 8.45 \pm 2.11 $\\
    3  & 10.9 & 6.33  & $ 8.04  \pm 1.52  $  & $ 9.71 \pm 2.26 $\\
    4  & 10.9 & 0.40  & $ 0.51  \pm 0.10  $  & $ 2.18 \pm 1.67 $\\
    5  & 10.9 & 2.45  & $ 3.11  \pm 0.59  $  & $ 4.78 \pm 1.77 $\\
    6  & 10.9 & 2.70  & $ 3.43  \pm 0.65  $  & $ 5.10 \pm 1.79 $\\
    7  & 10.9 & 3.19  & $ 4.05  \pm 0.77  $  & $ 5.72 \pm 1.84 $\\
    8  & 10.9 & 0.20  & $ 0.25  \pm 0.05  $  & $ 1.92 \pm 1.67 $\\
    \hline
  \end{tabular}
  \caption{An estimate of the number of {\em primordial} binaries in Sco~OB2 with A and late-B primaries and brown dwarf companions  (with $0.02$~M$_\odot \leq M \leq 0.08$~M$_\odot$) in the angular separation range $1'' \leq \rho \leq 4''$. Columns~1 and~2 list the model number (cf. Table~\ref{table: modelbackgroundstars}), and the fraction of binaries with angular separation $1'' \leq \rho \leq 4''$ (assuming \"{O}pik's law). Column~3 lists the fraction $F_{\rm IM}$ of primaries in the simulated association that is of type A or late-B. Column 4 shows the contribution of brown dwarf companions in Sco~OB2 with A and late-B primaries in the angular separation range $1'' \leq \rho \leq 4''$, inferred from the 28~free-floating brown dwarfs in Sco~OB2 found by \cite{martin2004}, assuming that {\em all} brown dwarfs were formed as companions. Column 5 shows the total number of primordial brown dwarf companions with $1'' \leq \rho \leq 4''$ and A or late-B primaries, with the observed brown dwarf companions (corrected for unseen brown dwarfs with $K>14$~mag) included. The values of $N_{\rm BD,IM,i,total}$ are upper limits, as it is likely that not {\em all} free-floating brown dwarfs were formed as companions. \label{table: martin}}
\end{table}

In the section above we have shown that the small number of brown dwarfs among A and late-B members of Sco~OB2 can be explained with an extrapolation of the mass ratio distribution for stellar companions into the brown dwarf regime. There is thus no need for a mechanism to remove brown dwarfs.
On the other hand, the embryo ejection scenario predicts that (at least a fraction of) the free-floating brown dwarfs in Sco~OB2 have been formed as companions to association members. Below, we study the consequences in the case that embryo ejection has affected the binary population, making use of the detection of 28~free-floating brown dwarfs in Upper Scorpius by \cite{martin2004}.
Under the assumption that this is what happened, we roughly estimate the number of primordial binaries with an A or late-B primary and a brown dwarf companion. For comparison between model predictions and observations we consider only those companions with an angular separation $1'' \leq \rho \leq 4''$ and $K_S < 14$~mag, for which our ADONIS and NACO observations are complete.

\cite{martin2004} present a sample of 104~candidate very low mass members, based on DENIS $IJK$ photometry, in a region of 60~square degrees in US. The authors report spectroscopic observations of 40~of these candidates and show that 28~are indeed strong candidate members of the US subgroup. 
Under the assumption that \cite{martin2004} randomly selected their 40~observed targets out of the sample of 104~candidates, we estimate the total number of brown dwarfs in US to be $73 \pm 14$ in the 60~square degrees region in US.
The projected area of the three subgroups of Sco~OB2 is approximately 960~square degrees, which gives us an estimate of $1165 \pm 221$ free-floating brown dwarfs in Sco~OB2. 

For our estimate of the number of primordial binaries we assume that all free-floating brown dwarfs were once companions. The number of systems with an A or late-B primary constitutes a small fraction $F_{\rm IM}$ of the total number of binaries, depending on the mass distribution (see Table~\ref{table: martin}). Assuming a primordial semi-major axis distribution of the form $f_a(a) \propto a^{-1}$ with $15$~R$_\odot < a < 5\times 10^6$~R$_\odot$, about $11\%$ of the brown dwarf companions are in the angular separation range $1'' \leq \rho \leq 4''$. 

For each model in Table~\ref{table: martin} we calculate how many of the $1165 \pm 221$ free-floating have the properties $1'' \leq \rho \leq 4''$ and $K_S < 14$~mag, and obtain the contribution $N_{\rm BD,IM,i} = F_{1''-4''} \times F_{\rm IM} \times (1165 \pm 221)$ of the free-floating brown dwarfs found by \cite{martin2004} to the number of primordial binaries in Sco~OB2 with A and late-B primaries and brown dwarf companions in the angular separation range $1'' \leq \rho \leq 4''$. We estimate the total number of primordial brown dwarf companions $N_{\rm BD,IM,i,total}$ with $1'' \leq \rho \leq 4''$ of A and late-B primaries by adding the observed number of brown dwarf companions, corrected for unseen companions with $K_S> 14$~mag (i.e., $1.67 \pm 1.67$).

We have assumed that all free-floating brown dwarfs were once companion stars, and therefore obtained upper limits for $N_{\rm BD,IM,i,total}$. By comparing $N_{\rm BD,IM}$ in Table~\ref{table: modelbackgroundstars2} with $N_{\rm BD,IM,i,total}$ in Table~\ref{table: martin} we can derive which primordial mass and mass ratio distributions are consistent with the predictions. As not necessarily all free-floating brown dwarfs have their origin in a binary system, all models with $N_{\rm BD,IM} \la N_{\rm BD,IM,i,total}$ are consistent with the predictions (i.e., the {\em current} number of brown dwarf companions should be less or equal to the {\em primordial} number of brown dwarf companions). A comparison shows that all models are consistent, except model 4 (random pairing from the Salpeter mass distribution). Under the hypothesis that embryo ejection has affected Sco~OB2, the current mass ratio distribution is slightly shallower than the primordial mass ratio distribution, due to the ejection of brown dwarf companions.
The above derivation gave an estimate of the number of primordial binaries with brown dwarf companions, under the assumption that the origin of the free-floating brown dwarfs in Sco~OB2 can be explained with the embryo ejection scenario. With our observations we cannot exclude firmly that this happened for several of the free-floating brown dwarfs. 

However, if binary formation would result in a mass ratio distribution similar to $f_q(q) \propto q^{-0.33}$, the ``brown dwarf desert'', if defined as a deficit of brown dwarf companions relative to stellar companions, would be a {\em natural outcome of star formation}. The embryo ejection scenario is not necessary to explain the small observed brown dwarf companion fraction in this case. 
Our analysis indicates that brown dwarf companions are formed like stellar companions. This suggests that the brown dwarf desert is defined by an excess of planetary companions, rather than by a lack of brown dwarf companions.

% ====================================================================
% ====================================================================
% ====================================================================
% ==BINARITY SCOOB2============================================================
% ====================================================================
% ====================================================================
% ====================================================================

\section{Binarity and multiplicity in Sco~OB2} \label{section: binarystatistics}

In \cite{kouwenhoven2005} we provided a census on binarity in Sco~OB2, consisting of all available data on visual, spectroscopic, eclipsing, and astrometric binaries and multiples. In Table~\ref{table: statistics} we present an update on the binary statistics in Sco~OB2. The statistics have been updated with the new results presented in this paper, as well as with the binaries recently discovered by \cite{nitschelm2004}, \cite{jilinski2006}, and \cite{chen2006}.

The multiple system fraction $F_{\rm M}$, the non-single star fraction  $F_{\rm NS}$, and companion star fraction $F_{\rm C}$ are defined as:
\begin{equation}  \label{equation: binarystatistics}
  F_{\rm M}       = \frac{ B+T+\dots   }{  S+B+T+\dots } \quad\ 
  F_{\rm NS}      = \frac{ 2B+3T+\dots }{  S+2B+3T+\dots} \quad\ 
  F_{\rm C}       = \frac{ B+2T+\dots  }{  S+B+T+\dots} \,,
\end{equation}
where $S$, $B$, and $T$ denote the number of single systems, binary systems
and triple systems in the association. 
In the Sco~OB2 association at least $40\%$ of the systems are multiple. Of the individual stars at least $60\%$ is part of a multiple system. Each system contains on average $F_{\rm C} \approx 0.5$ known companion stars. The updated values of $F_{\rm M}$, $F_{\rm NS}$, and  $F_{\rm C}$ are slightly larger than the values mentioned in \cite{kouwenhoven2005}, respectively. Note that these frequencies are lower limits due to the presence of undiscovered companion stars.

\begin{table}[btp]
  \setlength{\tabcolsep}{0.7\tabcolsep}
  \begin{tabular}{l cc cccc cc c}
    \hline
          & $D$  & Age  & $S$ & $B$ & $T$ & $>3$ & $F_{\rm M}$ & $F_{\rm NS}$ & $F_{\rm C}$ \\
          & (pc) &(Myr) & \\
    \hline
    US    & 145  & 5--6     & 64    & 44    & 8     & 3     & 0.46  & 0.67  & 0.61\\
    UCL   & 140  & 15--22   & 132   & 65    & 19    & 4     & 0.40  & 0.61  & 0.52\\
    LCC   & 118  & 17--23   & 112   & 57    & 9     & 1     & 0.37  & 0.56  & 0.44\\
    \hline
    all   &      &          & 308   & 166   & 36    & 8     & 0.41  & 0.61  & 0.51\\
    \hline
  \end{tabular}
  \caption{Multiplicity among \textit{Hipparcos} members of Sco~OB2. The columns
  show the subgroup names (Upper Scorpius; Upper Centaurus Lupus; Lower
  Centaurus Crux), their distances \citep[see][]{dezeeuw1999}, the ages
  (\cite{degeus1989,preibisch2002} for US; \cite{mamajek2002} for UCL and LCC), the number
  of known single stars, binary stars, triple systems and $N>3$ systems, and
  the binary statistics (see \S~\ref{section: binarystatistics}).  \label{table: statistics} }
\end{table}

% ====================================================================
% ====================================================================
% ====================================================================
% ==CONCLUSIONS============================================================
% ====================================================================
% ====================================================================
% ====================================================================

\section{Conclusions} \label{section: conclusion}

We have carried out near-infrared $JHK_S$ observations of 22 A and late-B stars in the Sco~OB2 association. The observations were performed with the NAOS/CONICA system at the ESO Very Large Telescope at Paranal, Chile. The observations resulted from a follow-up program of our previous work \citep{kouwenhoven2005}, in which we surveyed 199~A and late-B Sco~OB2 members for binarity with ADONIS. The data were obtained with the goal of (1) determining the validity of the $K_S=12$~mag criterion we used in our ADONIS survey to separate companions and background stars, (2) studying the behaviour of the companion mass distribution in the low-mass regime, and (3) search for additional companion stars.  We have included in our analysis the multi-color observations of 9~targets observed with ADONIS. In our ADONIS survey, these targets were analyzed using their $K_S$ magnitude only. The main results of our study are:

\begin{itemize}\addtolength{\itemsep}{-0.5\baselineskip}
\item[--] We detect 72~secondaries around the 31~target stars in our analysis. By comparing the near-infrared colors with the isochrones in the color-magnitude diagram, we find 25~confirmed companion stars, 12~candidate companion stars, and 35~background stars.   
\item[--] For most objects in our ADONIS survey \citep{kouwenhoven2005} only the $K_S$ magnitude was available. We used a magnitude criterion to separate companion stars ($K_S < 12$~mag) and background stars ($K_S > 12$~mag). With our analysis of the 22~NACO targets and 9~ADONIS targets with multi-color observations, we estimate the accuracy of the $K_S=12$~mag criterion. We find that the $K=12$~mag criterion is a very useful tool, correctly classifying the secondaries in $\sim 80\%$ of the cases. 
\item[--] We report two candidate brown dwarf companions of HIP81972. From their near-infrared magnitudes we infer masses of $32~{\rm M_J}$ and $63~{\rm M_J}$. The objects are located at an angular separation of $7.92''$ (1500~AU) and $2.79''$ (520~AU) from HIP81972, respectively. Follow-up spectroscopy is necessary to determine the true nature of these secondaries. Although we are sensitive (but incomplete) to massive planets, we classify the faintest secondaries as background stars (irrespective of their location in the color-magnitude diagram), because of isochronal uncertainty and the large number of faint background stars. 
\item[--] In our combined survey of 199~A and late-B members of Sco~OB2 we detect one confirmed companion star with $12~{\rm mag} \leq K_S \leq 14$~mag in the angular separation range $1''-4''$. In this region we detect no other secondary, while both the ADONIS and NACO observations are complete. This indicates a very low frequency of brown dwarf companions in the separation range $130-520$~AU for late-B and A type stars in Sco~OB2.   
\item[--] Our results are in good agreement with a mass ratio distribution of the form $f_q(q) \propto q^{-0.33}$. We find a brown dwarf companion fraction (for $M \ga 30~{\rm M_J}$) of $0.5 \pm 0.5\%$ for A and late-B stars in Sco~OB2.  After correction for unseen faint companions ($M \la 30~{\rm M_J}$), we estimate a substellar-to-stellar companion ratio of $R=0.06\pm 0.02$. 
\end{itemize}
The number of brown dwarfs among A and late-B members of Sco~OB2 is consistent with an extrapolation of the (stellar) companion mass distribution into the brown dwarf regime, suggesting that the formation mechanism for stars and brown dwarfs is the same. 
The embryo ejection mechanism does not need to be invoked to explain the small number of brown dwarf companions among intermediate mass stars in Sco~OB2.
The brown dwarf desert (which has been discovered in planet searches among solar-type stars) should therefore be explained by an excess of planetary companions, rather than a lack of brown dwarf companions.

% ====================================================================
% ====================================================================
% ====================================================================
% ==FINAL=STUFF=======================================================
% ====================================================================
% ====================================================================
% ====================================================================

\section*{Acknowledgments}

We thank ESO and the Paranal Observatory staff for efficiently conducting the Service-Mode observations and for their support. We thank Simon Portegies Zwart and the anonymous referee for their constructive criticism, which helped to improve the paper. This publication makes use of data products from the Two Micron All Sky Survey, which is a joint project of the University of Massachusetts and the Infrared Processing and Analysis Center/California Institute of Technology, funded by the National Aeronautics and Space Administration and the National Science Foundation. This research is supported by NWO under project number 614.041.006.

\begin{landscape}

\markright{Appendix A: Results of the NACO binarity survey}
\addcontentsline{toc}{section}{Appendix A: Results of the NACO binarity survey}
\section*{Appendix A: Results of the NACO binarity survey}

\setlength{\LTcapwidth}{1.3\textwidth}
{\small
\begin{longtable}{| l | rrr rr | rrr r | cl |}
  \caption{Properties of the 25~confirmed companion stars found around the 22~members in our NACO survey and the 9~members with multi-color observations in the ADONIS survey.
The columns show the secondary designation, the $J$, $H$, and $K_S$ magnitudes, the angular separation, and the position angle (measured from North to East). Magnitude lower limits are given if a secondary is not detected in a filter. We list the absolute magnitude and corresponding mass in columns $7-10$. 
The 11th column lists the status of the object (c = confirmed companion star, nc = new confirmed companion star). The last column shows additional remarks. A ``J'', ``H'', or ``K'' means that the secondary flux in this filter was obtained from the image obtained {\em without} the NDF, using the PSF from the corresponding image that was obtained {\em with} NDF (see \S~\ref{section: componentdetection}). Properties of the observed primaries, candidate companions, and background stars are not shown here; these are listed in Table~\ref{table: longtable}.  \label{table: companionstable}}\\
  \hline
  Star & \multicolumn{1}{c}{$J$} & \multicolumn{1}{c}{$H$} & \multicolumn{1}{c}{$K_S$} & \multicolumn{1}{c}{$\rho$} & \multicolumn{1}{c}{PA}  & \multicolumn{1}{c}{$M_J$}  & \multicolumn{1}{c}{$M_H$}  & \multicolumn{1}{c}{$M_{K_S}$}  & \multicolumn{1}{c}{Mass}      & \multicolumn{1}{c}{Status} &  \multicolumn{1}{c|}{Remarks} \\
  \hline
       & \multicolumn{1}{c}{mag} & \multicolumn{1}{c}{mag} & \multicolumn{1}{c}{mag}   & \multicolumn{1}{c}{arcsec} & \multicolumn{1}{c}{deg}   & \multicolumn{1}{c}{mag}    & \multicolumn{1}{c}{mag}    &  \multicolumn{1}{c}{mag}       & \multicolumn{1}{c}{M$_\odot$} &        &          \\
  \hline
  \multicolumn{12}{|l|}{NACO targets} \\ 
  \endfirsthead
  \hline
  \multicolumn{12}{|l|}{\tablename\ \thetable{} -- continued from previous page} \\
  \hline
  Star & \multicolumn{1}{c}{$J$} & \multicolumn{1}{c}{$H$} & \multicolumn{1}{c}{$K_S$} & \multicolumn{1}{c}{$\rho$} & \multicolumn{1}{c}{PA}  & \multicolumn{1}{c}{$M_J$}  & \multicolumn{1}{c}{$M_H$}  & \multicolumn{1}{c}{$M_{K_S}$}  & \multicolumn{1}{c}{Mass}      & \multicolumn{1}{c}{Status} &  \multicolumn{1}{c|}{Remarks} \\
  \hline
       & \multicolumn{1}{c}{mag} & \multicolumn{1}{c}{mag} & \multicolumn{1}{c}{mag}   & \multicolumn{1}{c}{arcsec} & \multicolumn{1}{c}{deg}   & \multicolumn{1}{c}{mag}    & \multicolumn{1}{c}{mag}    &  \multicolumn{1}{c}{mag}       & \multicolumn{1}{c}{M$_\odot$} &        &          \\
  \hline
  \endhead
  \hline
  \multicolumn{12}{|l|}{{Continued on next page}} \\ 
  \hline
  \endfoot
  \hline 
  \hline
  \endlastfoot
  \hline
  HIP59502    -1   &   12.35   &   11.83   &   11.64   &   2.94   &   26.39   &   7.39   &   6.86   &   6.68   &   0.14   &   c   &          \\
HIP62026    -1   &   8.08   &   7.90   &   7.86   &   0.23   &   6.34   &   2.88   &   2.71   &   2.66   &   1.19   &   c   &          \\
HIP63204    -2   &   8.79   &   8.51   &   8.40   &   0.15   &   236.56   &   3.59   &   3.31   &   3.19   &   1.06   &   c   &          \\
HIP67260    -1   &   8.88   &   8.46   &   8.36   &   0.42   &   229.46   &   3.42   &   2.99   &   2.90   &   1.10   &   c   &          \\
HIP67919    -1   &   9.98   &   9.38   &   9.10   &   0.69   &   296.56   &   4.89   &   4.30   &   4.02   &   0.75   &   c   &          \\
HIP68532    -1   &   10.52   &   9.85   &   9.54   &   3.05   &   288.50   &   5.03   &   4.36   &   4.05   &   0.73   &   c   &          \\
HIP68532    -2   &   11.38   &   10.94   &   10.63   &   3.18   &   291.92   &   5.89   &   5.45   &   5.14   &   0.39   &   c   &          \\
HIP69113    -1   &   10.98   &   10.43   &   10.29   &   5.34   &   65.15   &   4.83   &   4.28   &   4.14   &   0.77   &   c   &          \\
HIP69113    -2   &   11.27   &   10.45   &   10.30   &   5.52   &   67.17   &   5.12   &   4.29   &   4.15   &   0.72   &   c   &          \\
HIP73937    -1   &   $>8.40$   &   8.46   &   8.37   &   0.24   &   190.58   &   $>2.94$   &   3.00   &   2.91   &   1.11   &   c   &          \\
HIP79739    -1   &   12.28   &   11.52   &   11.23   &   0.96   &   118.33   &   6.34   &   5.58   &   5.29   &   0.16   &   c   &          \\
HIP79771    -1   &   12.00   &   11.28   &   10.89   &   3.67   &   313.38   &   6.06   &   5.33   &   4.94   &   0.19   &   c   &          \\
HIP79771    -2   &   12.39   &   11.79   &   11.42   &   0.44   &   128.59   &   6.44   &   5.85   &   5.47   &   0.13   &   nc   &          \\
HIP80799    -1   &   10.60   &   10.04   &   9.80   &   2.94   &   205.02   &   5.08   &   4.51   &   4.27   &   0.34   &   c   &          \\
HIP80896    -1   &   11.16   &   10.63   &   10.33   &   2.28   &   177.23   &   5.60   &   5.07   &   4.77   &   0.24   &   c   &          \\
HIP81972    -3   &   12.54   &   11.86   &   11.77   &   5.04   &   213.45   &   6.16   &   5.48   &   5.39   &   0.35   &   c   &   J       \\
HIP81972    -4   &   15.10   &   14.43   &   13.98   &   2.79   &   106.94   &   8.72   &   8.05   &   7.60   &   0.06   &   nc   &   JHK       \\
HIP81972    -5   &   16.11   &   15.63   &   15.26   &   7.92   &   229.27   &   9.73   &   9.25   &   8.88   &   $\approx$0.03   &   nc   &   JHK       \\
\hline  \multicolumn{12}{|l|}{ADONIS multi-color subset} \\ \hline  HIP76071    -1   &   $>11.25$   &   11.28   &   10.87   &   0.69   &   40.85   &   $>5.09$   &   5.12   &   4.71   &   0.23   &   c   &          \\
HIP77911    -1   &   12.68   &   12.20   &   11.84   &   7.96   &   279.25   &   6.82   &   6.34   &   5.98   &   0.09   &   c   &          \\
HIP78809    -1   &   11.08   &   10.45   &   10.26   &   1.18   &   25.67   &   5.32   &   4.69   &   4.50   &   0.30   &   c   &          \\
HIP78956    -1   &   9.76   &   9.12   &   9.04   &   1.02   &   48.67   &   3.39   &   2.75   &   2.67   &   1.16   &   c   &          \\
HIP79124    -1   &   11.38   &   10.55   &   10.38   &   1.02   &   96.18   &   5.33   &   4.50   &   4.33   &   0.33   &   c   &          \\
HIP79156    -1   &   11.62   &   10.89   &   10.77   &   0.89   &   58.88   &   5.50   &   4.77   &   4.65   &   0.27   &   c   &          \\
HIP80238    -1   &   7.96   &   7.66   &   7.49   &   1.03   &   318.46   &   2.34   &   2.04   &   1.87   &   1.67   &   c   &          \\

  \hline
\end{longtable}
}

{\small
\begin{longtable}{| l | rrr rr | rrr r | cl |}
  \caption{Results from our multi-color binarity study among 22~Sco~OB2 member stars observed with NACO ({\em top part of the table}) and the subset of 9~members with multi-color observations in in the ADONIS survey ({\em bottom part of the table}). The columns show the \textit{Hipparcos} number (for the targets) and the secondary designation, the $J$, $H$, and $K_S$ magnitudes, the angular separation, and the position angle (measured from North to East). Lower limits to the magnitudes are given if an object is not detected in the NACO survey, unless the ADONIS measurement was available (marked with a $\star$). Entries marked with $\star\star$ have no available measurement, e.g., because the object is not in the field of view for that filter. For each primary and companion star we list the absolute $JHK_S$ magnitudes and the mass in columns $7-10$. We additionally provide absolute magnitudes and a mass estimate for the candidate companions, {\em under the assumption} that these are indeed companions. We stress that a significant number of the candidate companions may actually be background stars. The 11th column lists the status of the object (p = primary, c = confirmed companion star, nc = new confirmed companion star, ? = candidate companion star; b = background star). The last column provides additional remarks. A remark ``J'', ``H'', or ``K'' means that the secondary flux in this filter was obtained from the image obtained {\em without} the NDF, using the PSF from the corresponding image that was obtained {\em with} NDF (see \S~\ref{section: componentdetection}). If the secondary status was obtained without color information, an exclamation mark is placed in the last column. The results for the 9~targets with multi-color information in the ADONIS survey are marked with ``ADO''. \label{table: longtable}}\\
  \hline
  Star & \multicolumn{1}{c}{$J$} & \multicolumn{1}{c}{$H$} & \multicolumn{1}{c}{$K_S$} & \multicolumn{1}{c}{$\rho$} & \multicolumn{1}{c}{PA}  & \multicolumn{1}{c}{$M_J$}  & \multicolumn{1}{c}{$M_H$}  & \multicolumn{1}{c}{$M_{K_S}$}  & \multicolumn{1}{c}{Mass}      & \multicolumn{1}{c}{Status} &  \multicolumn{1}{c|}{Remarks} \\
  \hline
       & \multicolumn{1}{c}{mag} & \multicolumn{1}{c}{mag} & \multicolumn{1}{c}{mag}   & \multicolumn{1}{c}{arcsec} & \multicolumn{1}{c}{deg}   & \multicolumn{1}{c}{mag}    & \multicolumn{1}{c}{mag}    &  \multicolumn{1}{c}{mag}       & \multicolumn{1}{c}{M$_\odot$} &        &          \\
  \endfirsthead
  \hline
  \multicolumn{12}{|l|}{\tablename\ \thetable{} -- continued from previous page} \\
  \hline
  Star & \multicolumn{1}{c}{$J$} & \multicolumn{1}{c}{$H$} & \multicolumn{1}{c}{$K_S$} & \multicolumn{1}{c}{$\rho$} & \multicolumn{1}{c}{PA}  & \multicolumn{1}{c}{$M_J$}  & \multicolumn{1}{c}{$M_H$}  & \multicolumn{1}{c}{$M_{K_S}$}  & \multicolumn{1}{c}{Mass}      & \multicolumn{1}{c}{Status} &  \multicolumn{1}{c|}{Remarks} \\
  \hline
       & \multicolumn{1}{c}{mag} & \multicolumn{1}{c}{mag} & \multicolumn{1}{c}{mag}   & \multicolumn{1}{c}{arcsec} & \multicolumn{1}{c}{deg}   & \multicolumn{1}{c}{mag}    & \multicolumn{1}{c}{mag}    &  \multicolumn{1}{c}{mag}       & \multicolumn{1}{c}{M$_\odot$} &        &          \\
  \hline
  \endhead
  \hline
  \multicolumn{12}{|l|}{{Continued on next page}} \\ 
  \hline
  \endfoot
  \hline 
  \endlastfoot
  \hline
  \hline 
  HIP59502       &   6.83   &   6.83   &   6.87   &          &          &   1.86   &   1.87   &   1.91   &   1.80   &   p   &          \\
HIP59502    -1   &   12.35   &   11.83   &   11.64   &   2.94   &   26.39   &   7.39   &   6.86   &   6.68   &   0.14   &   c   &          \\
HIP59502    -2   &   $>15.22$   &   15.26   &   15.37   &   4.76   &   101.87   &      &      &      &          &   b   &   HK       \\
HIP59502    -3   &   $^{\star\star}$   &   $^{\star\star}$   &   13.69   &   9.02   &   309.01   &      &      &      &          &   b   &   K!       \\
\hline
HIP60851       &   6.03   &   6.06   &   6.06   &          &          &   0.94   &   0.97   &   0.97   &   2.63   &   p   &          \\
HIP60851    -1   &   12.81   &   11.62   &   11.46   &   2.07   &   45.30   &      &      &      &          &   b   &   J       \\
HIP60851    -2   &   $>13.33$   &   11.68   &   11.29   &   6.89   &   180.38   &      &      &      &          &   b   &          \\
HIP60851    -3   &   $>13.33$   &   13.63   &   13.69   &   8.16   &   231.46   &   ($>8.24$)   &   (8.54)   &   (8.60)   &   (0.04)   &   ?   &   HK       \\
HIP60851    -4   &   $>13.33$   &   14.82   &   14.80   &   1.61   &   280.38   &      &      &      &          &   b   &   HK       \\
HIP60851    -5   &   $>13.33$   &   15.53   &   14.97   &   8.19   &   69.19   &      &      &      &          &   b   &   HK       \\
HIP60851    -6   &   $>13.33$   &   15.83   &   $^{\star\star}$   &   7.65   &   153.67   &      &      &      &          &   b   &   H!       \\
HIP60851    -7   &   $>13.33$   &   16.67   &   $^{\star\star}$   &   7.47   &   287.03   &      &      &      &          &   b   &   H!       \\
HIP60851    -8   &   $>13.33$   &   16.87   &   16.97   &   5.45   &   76.38   &      &      &      &          &   b   &   HK       \\
%\hline
HIP61265       &   7.49   &   7.51   &   7.46   &          &          &   1.85   &   1.87   &   1.81   &   1.82   &   p   &          \\
HIP61265    -1   &   11.98   &   11.66   &   11.38   &   2.51   &   67.15   &   (6.34)   &   (6.02)   &   (5.74)   &   (0.27)   &   ?   &   J       \\
HIP61265    -2   &   15.13   &   14.96   &   14.75   &   3.41   &   167.27   &      &      &      &          &   b   &   J       \\
HIP61265    -3   &   $>15.71$   &   16.30   &   15.29   &   7.00   &   24.46   &      &      &      &          &   b   &          \\
HIP61265    -4   &   $>15.71$   &   16.80   &   16.28   &   6.60   &   31.84   &      &      &      &          &   b   &          \\
HIP61265    -5   &   $>15.71$   &   $>15.90$   &   15.86   &   7.11   &   344.55   &      &      &      &          &   b   &   !       \\
\hline
HIP62026       &   6.28   &   6.32   &   6.31   &          &          &   1.09   &   1.12   &   1.11   &   2.45   &   p   &          \\
HIP62026    -1   &   8.08   &   7.90   &   7.86   &   0.23   &   6.34   &   2.88   &   2.71   &   2.66   &   1.19   &   c   &          \\
\hline
HIP63204       &   6.68   &   6.76   &   6.78   &          &          &   1.48   &   1.55   &   1.57   &   2.05   &   p   &          \\
HIP63204    -1   &   8.72   &   7.85   &   7.50   &   1.87   &   47.44   &      &      &      &          &   b   &          \\
HIP63204    -2   &   8.79   &   8.51   &   8.40   &   0.15   &   236.56   &   3.59   &   3.31   &   3.19   &   1.06   &   c   &          \\
\hline
HIP67260       &   7.03   &   7.00   &   6.98   &          &          &   1.57   &   1.53   &   1.52   &   2.00   &   p   &          \\
HIP67260    -1   &   8.88   &   8.46   &   8.36   &   0.42   &   229.46   &   3.42   &   2.99   &   2.90   &   1.10   &   c   &          \\
HIP67260    -2   &   $^{\star\star}$   &   14.04   &   14.10   &   1.23   &   355.65   &   ($^{\star\star}$)   &   (8.57)   &   (8.63)   &   (0.04)   &   ?   &          \\
HIP67260    -3   &   15.84   &   14.83   &   14.67   &   2.33   &   77.25   &   (10.38)   &   (9.36)   &   (9.20)   &   ($\approx$0.02)   &   ?   &   JHK       \\
\hline
HIP67919       &   6.71   &   6.60   &   6.59   &          &          &   1.63   &   1.52   &   1.51   &   1.97   &   p   &          \\
HIP67919    -1   &   9.98   &   9.38   &   9.10   &   0.69   &   296.56   &   4.89   &   4.30   &   4.02   &   0.75   &   c   &          \\
\hline
HIP68532       &   7.16   &   7.08   &   7.02   &          &          &   1.67   &   1.59   &   1.53   &   1.95   &   p   &          \\
HIP68532    -1   &   10.52   &   9.85   &   9.54   &   3.05   &   288.50   &   5.03   &   4.36   &   4.05   &   0.73   &   c   &          \\
HIP68532    -2   &   11.38   &   10.94   &   10.63   &   3.18   &   291.92   &   5.89   &   5.45   &   5.14   &   0.39   &   c   &          \\
\hline
HIP69113       &   6.17   &   6.30   &   6.37   &          &          &   0.02   &   0.15   &   0.22   &   3.87   &   p   &          \\
HIP69113    -1   &   10.98   &   10.43   &   10.29   &   5.34   &   65.15   &   4.83   &   4.28   &   4.14   &   0.77   &   c   &          \\
HIP69113    -2   &   11.27   &   10.45   &   10.30   &   5.52   &   67.17   &   5.12   &   4.29   &   4.15   &   0.72   &   c   &          \\
\hline
HIP73937       &   6.11   &   6.21   &   6.23   &          &          &   0.65   &   0.75   &   0.77   &   2.94   &   p   &          \\
HIP73937    -1   &   $>8.40$   &   8.46   &   8.37   &   0.24   &   190.58   &   $>2.94$   &   3.00   &   2.91   &   1.11   &   c   &          \\
HIP73937    -2   &   $>11.41$   &   14.32   &   14.71   &   3.56   &   31.24   &      &      &      &          &   b   &   HK       \\
\hline
HIP78968       &   7.47   &   7.42   &   7.42   &          &          &   1.23   &   1.17   &   1.18   &   2.33   &   p   &          \\
HIP78968    -1   &   14.96   &   14.51   &   14.26   &   2.78   &   322.13   &   (8.71)   &   (8.27)   &   (8.01)   &   ($\approx$0.02)   &   ?   &   JHK       \\
\hline
HIP79098       &   5.71   &   5.70   &   5.69   &          &          &   -0.02   &   -0.03   &   -0.04   &   4.30   &   p   &          \\
HIP79098    -1   &   15.67   &   14.14   &   14.24   &   2.37   &   116.63   &      &      &      &          &   b   &   JK       \\
\hline
HIP79410       &   7.20   &   7.14   &   7.09   &          &          &   1.35   &   1.29   &   1.24   &   2.24   &   p   &          \\
HIP79410    -1   &   15.94   &   15.12   &   14.93   &   3.24   &   340.93   &      &      &      &          &   b   &   J       \\
\hline
HIP79739       &   7.17   &   7.16   &   7.08   &          &          &   1.23   &   1.21   &   1.14   &   2.32   &   p   &          \\
HIP79739    -1   &   12.28   &   11.52   &   11.23   &   0.96   &   118.33   &   6.34   &   5.58   &   5.29   &   0.16   &   c   &          \\
\hline
HIP79771       &   7.33   &   7.26   &   7.10   &          &          &   1.39   &   1.31   &   1.15   &   2.14   &   p   &          \\
HIP79771    -1   &   12.00   &   11.28   &   10.89   &   3.67   &   313.38   &   6.06   &   5.33   &   4.94   &   0.19   &   c   &          \\
HIP79771    -2   &   12.39   &   11.79   &   11.42   &   0.44   &   128.59   &   6.44   &   5.85   &   5.47   &   0.13   &   nc   &          \\
\hline
HIP80142       &   6.61   &   6.67   &   6.66   &          &          &   0.41   &   0.47   &   0.46   &   3.33   &   p   &          \\
HIP80142    -1   &   12.01   &   10.59   &   9.51   &   9.23   &   216.16   &      &      &      &          &   b   &   J       \\
HIP80142    -2   &   16.64   &   15.88   &   $^{\star\star}$   &   5.88   &   119.94   &   (10.44)   &   (9.68)   &   ($^{\star\star}$)   &   ($\approx$0.02)   &   ?   &   HJ       \\
\hline
HIP80474       &   6.14   &   5.92   &   5.80   &          &          &   0.27   &   0.05   &   -0.07   &   3.78   &   p   &          \\
HIP80474    -1   &   12.06   &   12.34   &   10.79   &   4.85   &   206.36   &      &      &      &          &   b   &   JHK       \\
\hline
HIP80799       &   7.56   &   7.53   &   7.45   &          &          &   2.04   &   2.01   &   1.93   &   1.86   &   p   &          \\
HIP80799    -1   &   10.60   &   10.04   &   9.80   &   2.94   &   205.02   &   5.08   &   4.51   &   4.27   &   0.34   &   c   &          \\
\hline
HIP80896       &   7.67   &   7.53   &   7.44   &          &          &   2.11   &   1.97   &   1.88   &   1.81   &   p   &          \\
HIP80896    -1   &   11.16   &   10.63   &   10.33   &   2.28   &   177.23   &   5.60   &   5.07   &   4.77   &   0.24   &   c   &          \\
\hline
HIP81949       &   7.38   &   7.40   &   7.33   &          &          &   1.28   &   1.31   &   1.23   &   2.26   &   p   &          \\
HIP81949    -1   &   15.73   &   14.11   &   13.28   &   3.91   &   88.47   &      &      &      &          &   b   &          \\
HIP81949    -2   &   14.34   &   14.28   &   14.06   &   3.48   &   28.46   &      &      &      &          &   b   &          \\
HIP81949    -3   &   $>16.81$   &   15.26   &   14.75   &   5.70   &   292.80   &      &      &      &          &   b   &          \\
HIP81949    -4   &   $>16.81$   &   15.67   &   15.52   &   5.27   &   340.72   &   ($>10.71$)   &   (9.57)   &   (9.42)   &   ($\approx$0.02)   &   ?   &          \\
HIP81949    -5   &   $>16.81$   &   16.52   &   $>15.93$   &   9.63   &   76.17   &      &      &      &          &   b   &   !       \\
HIP81949    -6   &   $>16.81$   &   15.62   &   14.82   &   6.26   &   239.37   &   ($>10.71$)   &   (9.52)   &   (8.73)   &   ($\approx$0.02)   &   ?   &          \\
HIP81949    -7   &   $>16.81$   &   $>16.84$   &   15.59   &   11.72   &   40.80   &      &      &      &          &   b   &   !       \\
HIP81949    -8   &   $>16.81$   &   $>16.84$   &   16.75   &   4.16   &   236.05   &      &      &      &          &   b   &   !       \\
HIP81949    -9   &   $>16.81$   &   $>16.84$   &   16.83   &   3.86   &   105.30   &      &      &      &          &   b   &   !       \\
HIP81949    -10   &   $>16.81$   &   $>16.84$   &   17.10   &   2.38   &   48.01   &      &      &      &          &   b   &   !       \\
HIP81949    -11   &   $>16.81$   &   $>16.84$   &   17.15   &   8.05   &   96.11   &      &      &      &          &   b   &   !       \\
HIP81949    -12   &   $>16.81$   &   $>16.84$   &   17.34   &   8.13   &   36.64   &      &      &      &          &   b   &   !       \\
\hline
HIP81972       &   5.82   &   5.89   &   5.87   &          &          &   -0.56   &   -0.49   &   -0.51   &   4.92   &   p   &          \\
HIP81972    -1   &   11.63   &   10.87   &   10.48   &   2.02   &   313.69   &   (5.25)   &   (4.49)   &   (4.10)   &   (0.67)   &   ?   &          \\
HIP81972    -2   &   11.30   &   10.97   &   10.61   &   7.02   &   258.81   &   (4.92)   &   (4.59)   &   (4.23)   &   (0.68)   &   ?   &          \\
HIP81972    -3   &   12.54   &   11.86   &   11.77   &   5.04   &   213.45   &   6.16   &   5.48   &   5.39   &   0.35   &   c   &   J       \\
HIP81972    -4   &   15.10   &   14.43   &   13.98   &   2.79   &   106.94   &   8.72   &   8.05   &   7.60   &   0.06   &   nc   &   JHK       \\
HIP81972    -5   &   16.11   &   15.63   &   15.26   &   7.92   &   229.27   &   9.73   &   9.25   &   8.88   &   $\approx$0.03   &   nc   &   JHK       \\
HIP81972    -6   &   $>16.58$   &   16.25   &   $>16.61$   &   8.79   &   167.71   &      &      &      &          &   b   &   H!       \\
HIP81972    -7   &   $>16.58$   &   17.12   &   $>16.61$   &   3.58   &   33.65   &      &      &      &          &   b   &   H!       \\
HIP81972    -8   &   $>16.58$   &   17.28   &   $>16.61$   &   7.44   &   265.65   &      &      &      &          &   b   &   H!       \\
\hline
HIP83542       &   5.34   &   4.91   &   5.38   &          &          &   -1.26   &   -1.69   &   -1.22   &   1.10   &   p   &          \\
HIP83542    -1   &   $^{\star\star}$   &   9.72   &   9.90   &   8.86   &   196.21   &   ($^{\star\star}$)   &   (3.12)   &   (3.30)   &   (0.91)   &   ?   &          \\
HIP83542    -2   &   $>15.54$   &   15.65   &   $>12.13$   &   9.84   &   156.45   &      &      &      &          &   b   &   H!       \\
\hline
  \multicolumn{12}{|l|}{{ADONIS targets with multi-color observations}} \\ 
\hline
HIP53701       &   6.30   &   6.37   &   6.48   &          &          &   0.79   &   0.86   &   0.97   &   2.84   &   p   &   ADO       \\
HIP53701    -1   &   9.05   &   8.76   &   8.86   &   3.88   &   75.81   &      &      &      &          &   b   &   ADO       \\
HIP53701    -2   &   13.06   &   12.93   &   13.04   &   6.57   &   120.05   &      &      &      &          &   b   &   ADO       \\
\hline
HIP76071       &   7.05   &   7.10   &   7.06   &          &          &   0.89   &   0.94   &   0.90   &   2.70   &   p   &   ADO       \\
HIP76071    -1   &   $>11.25$   &   11.28   &   10.87   &   0.69   &   40.85   &   $>5.09$   &   5.12   &   4.71   &   0.23   &   c   &   ADO       \\
\hline
HIP77911       &   6.67   &   6.71   &   6.68   &          &          &   0.81   &   0.85   &   0.82   &   2.80   &   p   &   ADO       \\
HIP77911    -1   &   12.68   &   12.20   &   11.84   &   7.96   &   279.25   &   6.82   &   6.34   &   5.98   &   0.09   &   c   &   ADO       \\
\hline
HIP78530       &   6.87   &   6.92   &   6.87   &          &          &   1.08   &   1.13   &   1.08   &   2.48   &   p   &   ADO       \\
HIP78530    -1   &   $>14.50$   &   14.56   &   14.22   &   4.54   &   139.69   &   ($>8.71$)   &   (8.77)   &   (8.43)   &   ($\approx$0.02)   &   ?   &   ADO       \\
\hline
HIP78809       &   7.41   &   7.50   &   7.51   &          &          &   1.65   &   1.74   &   1.75   &   2.03   &   p   &   ADO       \\
HIP78809    -1   &   11.08   &   10.45   &   10.26   &   1.18   &   25.67   &   5.32   &   4.69   &   4.50   &   0.30   &   c   &   ADO       \\
\hline
HIP78956       &   7.52   &   7.54   &   7.57   &          &          &   1.15   &   1.17   &   1.20   &   2.40   &   p   &   ADO       \\
HIP78956    -1   &   9.76   &   9.12   &   9.04   &   1.02   &   48.67   &   3.39   &   2.75   &   2.67   &   1.16   &   c   &   ADO       \\
\hline
HIP79124       &   7.16   &   7.14   &   7.13   &          &          &   1.11   &   1.09   &   1.08   &   2.48   &   p   &   ADO       \\
HIP79124    -1   &   11.38   &   10.55   &   10.38   &   1.02   &   96.18   &   5.33   &   4.50   &   4.33   &   0.33   &   c   &   ADO       \\
\hline
HIP79156       &   7.56   &   7.56   &   7.61   &          &          &   1.44   &   1.44   &   1.49   &   2.09   &   p   &   ADO       \\
HIP79156    -1   &   11.62   &   10.89   &   10.77   &   0.89   &   58.88   &   5.50   &   4.77   &   4.65   &   0.27   &   c   &   ADO       \\
\hline
HIP80238       &   7.45   &   7.45   &   7.34   &          &          &   1.83   &   1.83   &   1.72   &   1.94   &   p   &   ADO       \\
HIP80238    -1   &   7.96   &   7.66   &   7.49   &   1.03   &   318.46   &   2.34   &   2.04   &   1.87   &   1.67   &   c   &   ADO       \\

  \hline
\end{longtable}
}

\end{landscape}

\setlength{\LTcapwidth}{1.0\textwidth}

\begin{longtable}{| l| c | rrr | }
  \caption{Criteria used to determine whether a secondary is a companion star or a background star. Results are listed for secondaries found around the 22~targets observed with NACO ({\em top part of the table}) and the 9~targets with multi-color observations in the ADONIS dataset ({\em bottom part of the table}). Columns 1 and 2 show the secondary designation and the status of the component as determined in this paper (c = companion star; ? = candidate companion star; b = background star). Columns $3-5$ show the compatibility of the location of the object in the color-magnitude diagrams with the isochrones in terms of $\chi^2$. Confirmed companions have $\chi^2 < 2.30$ and (confirmed) background stars have $\chi^2 > 11.8$. The other secondaries have $2.30 < \chi^2 < 11.8$ and are labeled ``candidate companion''. A substantial fraction of these candidate companions may in fact be background stars. Several faint ($K_S > 14$~mag) secondaries are only detected in one filter (thus have no $\chi^2$), and are all assumed to be background stars. \label{table: criteria}}\\  
  \hline
  Star & Status & $\chi^2_{J-K_S,M_{K_S}}$ & $\chi^2_{H-K_S,M_{K_S}}$ & $\chi^2_{J-H,M_J}$  \\
  \hline
  \endfirsthead
  \hline
  \multicolumn{5}{|l|}{\tablename\ \thetable{} -- continued from previous page} \\
  \hline
  Star & Status & $\chi^2_{J-K_S,M_{K_S}}$ & $\chi^2_{H-K_S,M_{K_S}}$ & $\chi^2_{J-H,M_J}$  \\
  \hline
  \endhead
  \hline
  \multicolumn{5}{|l|}{{Continued on next page}} \\ 
  \hline
  \endfoot
  \hline 
  \endlastfoot
  \hline
  \hline 
  HIP59502    -1   &   c  &   $2.11$    &    $1.09$    &    $0.19$        \\
HIP59502    -2   &   b  &   ---    &    $75.23$    &    ---        \\
HIP59502    -3   &   b  &   ---    &    ---    &    ---        \\
HIP60851    -1   &   b  &   $7.77$    &    $1.29$    &    $15.51$        \\
HIP60851    -2   &   b  &   $>52.81$    &    $0.19$    &    $>45.54$        \\
HIP60851    -3   &   ?  &   ---    &    $7.54$    &    ---        \\
HIP60851    -4   &   b  &   ---    &    $22.34$    &    ---        \\
HIP60851    -5   &   b  &   ---    &    $26.37$    &    ---        \\
HIP60851    -6   &   b  &   ---    &    ---    &    ---        \\
HIP60851    -7   &   b  &   ---    &    ---    &    ---        \\
HIP60851    -8   &   b  &   ---    &    $237.30$    &    ---        \\
HIP61265    -1   &   ?  &   $4.46$    &    $0.02$    &    $3.83$        \\
HIP61265    -2   &   b  &   $13.77$    &    $1.81$    &    $5.29$        \\
HIP61265    -3   &   b  &   ---    &    $13.70$    &    ---        \\
HIP61265    -4   &   b  &   ---    &    $66.72$    &    ---        \\
HIP61265    -5   &   b  &   ---    &    ---    &    ---        \\
HIP62026    -1   &   c  &   $0.91$    &    $0.09$    &    $0.24$        \\
HIP63204    -1   &   b  &   $71.62$    &    $9.64$    &    $17.02$        \\
HIP63204    -2   &   c  &   $1.76$    &    $0.01$    &    $1.38$        \\
HIP67260    -1   &   c  &   $0.01$    &    $0.00$    &    $0.01$        \\
HIP67260    -2   &   ?  &   ---    &    $7.20$    &    ---        \\
HIP67260    -3   &   ?  &   $1.45$    &    $3.59$    &    $5.27$        \\
HIP67919    -1   &   c  &   $0.02$    &    $0.86$    &    $0.39$        \\
HIP68532    -1   &   c  &   $0.93$    &    $1.10$    &    $0.00$        \\
HIP68532    -2   &   c  &   $1.06$    &    $0.02$    &    $1.70$        \\
HIP69113    -1   &   c  &   $1.26$    &    $0.09$    &    $0.63$        \\
HIP69113    -2   &   c  &   $0.56$    &    $0.04$    &    $1.38$        \\
HIP73937    -1   &   c  &   ---    &    $0.01$    &    ---        \\
HIP73937    -2   &   b  &   ---    &    $22.37$    &    ---        \\
HIP78968    -1   &   ?  &   $3.88$    &    $1.26$    &    $0.73$        \\
HIP79098    -1   &   b  &   $8.45$    &    $14.45$    &    $39.65$        \\
HIP79410    -1   &   b  &   $21.58$    &    $23.51$    &    $19.09$        \\
HIP79739    -1   &   c  &   $0.44$    &    $0.09$    &    $0.91$        \\
HIP79771    -1   &   c  &   $1.46$    &    $0.24$    &    $0.52$        \\
HIP79771    -2   &   nc  &   $0.00$    &    $0.04$    &    $0.02$        \\
HIP80142    -1   &   b  &   $164.54$    &    $59.81$    &    $33.27$        \\
HIP80142    -2   &   ?  &   ---    &    ---    &    $2.61$        \\
HIP80474    -1   &   b  &   $5.08$    &    $73.31$    &    $36.05$        \\
HIP80799    -1   &   c  &   $1.11$    &    $0.12$    &    $0.62$        \\
HIP80896    -1   &   c  &   $0.73$    &    $0.02$    &    $0.53$        \\
HIP81949    -1   &   b  &   $66.30$    &    $7.46$    &    $27.51$        \\
HIP81949    -2   &   b  &   $15.93$    &    $0.91$    &    $8.99$        \\
HIP81949    -3   &   b  &   $>13.09$    &    $0.02$    &    $>19.08$        \\
HIP81949    -4   &   ?  &   $>3.88$    &    $5.69$    &    $>5.59$        \\
HIP81949    -5   &   b  &   ---    &    ---    &    ---        \\
HIP81949    -6   &   ?  &   $>9.95$    &    $0.98$    &    $>6.88$        \\
HIP81949    -7   &   b  &   $>5.16$    &    $>11.84$    &    ---        \\
HIP81949    -8   &   b  &   ---    &    ---    &    ---        \\
HIP81949    -9   &   b  &   ---    &    ---    &    ---        \\
HIP81949    -10   &   b  &   ---    &    ---    &    ---        \\
HIP81949    -11   &   b  &   ---    &    ---    &    ---        \\
HIP81949    -12   &   b  &   ---    &    ---    &    ---        \\
HIP81972    -1   &   ?  &   $4.36$    &    $2.27$    &    $0.57$        \\
HIP81972    -2   &   ?  &   $1.42$    &    $1.08$    &    $6.03$        \\
HIP81972    -3   &   c  &   $0.56$    &    $1.65$    &    $0.20$        \\
HIP81972    -4   &   nc  &   $0.51$    &    $0.11$    &    $0.13$        \\
HIP81972    -5   &   nc  &   $0.90$    &    $0.07$    &    $0.40$        \\
HIP81972    -6   &   b  &   ---    &    $>16.02$    &    ---        \\
HIP81972    -7   &   b  &   ---    &    ---    &    ---        \\
HIP81972    -8   &   b  &   ---    &    ---    &    ---        \\
HIP83542    -1   &   ?  &   ---    &    $8.25$    &    ---        \\
HIP83542    -2   &   b  &   ---    &    ---    &    ---        \\
\hline HIP53701    -1   &   b  &   $12.01$    &    $3.72$    &    $0.80$        \\
HIP53701    -2   &   b  &   $33.11$    &    $8.54$    &    $8.16$        \\
HIP76071    -1   &   c  &   ---    &    $0.57$    &    ---        \\
HIP77911    -1   &   c  &   $0.73$    &    $0.00$    &    $0.74$        \\
HIP78530    -1   &   ?  &   ---    &    $3.10$    &    ---        \\
HIP78809    -1   &   c  &   $0.82$    &    $0.74$    &    $0.02$        \\
HIP78956    -1   &   c  &   $0.01$    &    $0.25$    &    $0.33$        \\
HIP79124    -1   &   c  &   $0.22$    &    $0.83$    &    $1.88$        \\
HIP79156    -1   &   c  &   $0.41$    &    $1.94$    &    $0.45$        \\
HIP80238    -1   &   c  &   $0.43$    &    $0.89$    &    $0.06$        \\

  \hline
\end{longtable}
\setlength{\LTcapwidth}{1.3\textwidth}

\begin{landscape}
\begin{longtable}{| l | rrr | rr | l|l |}
  \caption{All companion stars identified in our ADONIS and NACO binarity surveys among A and late-B stars in Sco~OB2 \citep[][and this paper]{kouwenhoven2005}. The columns show the \textit{Hipparcos} number of the primary star, the $JHK_S$ magnitudes, the angular separation, the position angle, the current status of the companion, and the date of observation (dd/mm/yy). If measurements are performed in both the ADONIS and NACO surveys, the NACO data are provided.
    The wide companion of HIP77315 at $\rho=37.37''$ is HIP77317, another member of Sco~OB2. These stars are found to be a common proper motion pair \citep{wds1997}, and were both observed in our ADONIS survey. The confirmed and candidate companions for which the status is determined using their $JHK_S$ photometry, are indicated with ``confirmed'' and ``inconclusive'', respectively. The candidate companions identified by \cite{kouwenhoven2005}, for which the status is determined using the $K_S=12$~mag criterion, and indicated with ``candidate'' here. Background stars are not listed here. \label{table: adonisnaco}}\\
  \hline
  Host primary & $J$ (mag) & $H$ (mag) & $K_S$ (mag) & $\rho$ (``) & PA ($^\circ$) & Companion status & Date \\
  \hline
  \endfirsthead
  \hline
  \multicolumn{8}{|l|}{\tablename\ \thetable{} -- continued from previous page} \\
  \hline
  Host primary & $J$ (mag) & $H$ (mag) & $K_S$ (mag) & $\rho$ (``) & PA ($^\circ$) & Companion status & Date \\
  \hline
  \endhead
  \hline
  \multicolumn{8}{|l|}{{Continued on next page}} \\ 
  \hline
  \endfoot
  \hline 
  \endlastfoot
  \hline
  HIP50520 &          &          &   6.39   &   2.51   &   313.32   &   candidate &            06/06/01        \\
\hline
HIP52357 &          &          &   11.45   &   10.04   &   72.69   &   candidate &           06/06/01         \\
HIP52357 &          &          &   7.65   &   0.53   &   73.01   &   candidate &             06/06/01       \\
\hline
HIP56993 &          &          &   11.88   &   1.68   &   23.07   &   candidate &            06/06/01        \\
\hline
HIP58416 &          &          &   8.66   &   0.58   &   166.12   &   candidate &            06/06/01        \\
\hline
HIP59413 &          &          &   8.18   &   3.18   &   99.83   &   candidate &             06/06/01       \\
\hline
HIP59502 &   12.35   &   11.83   &   11.64   &   2.94   &   26.39   &   confirmed      &    06/04/04   \\
\hline
HIP60084 &          &          &   10.10   &   0.46   &   329.64   &   candidate &           06/06/01         \\
\hline
HIP60851 &          &   13.63   &   13.69   &   8.16   &   231.46   &   inconclusive      &    06/04/04   \\
\hline
HIP61265 &   11.98   &   11.66   &   11.38   &   2.51   &   67.15   &   inconclusive      &    06/04/04   \\
\hline
HIP61639 &          &          &   7.06   &   1.87   &   182.40   &   candidate &            07/06/01        \\
\hline
HIP61796 &          &          &   11.79   &   9.89   &   108.98   &   candidate &           07/06/01         \\
HIP61796 &          &          &   11.86   &   12.38   &   136.77   &   candidate &          07/06/01          \\
\hline
HIP62002 &          &          &   7.65   &   0.38   &   69.24   &   candidate &             08/06/01       \\
\hline
HIP62026 &   8.08   &   7.90   &   7.86   &   0.23   &   6.34   &   confirmed      &    06/04/04   \\
\hline
HIP62179 &          &          &   7.57   &   0.23   &   282.75   &   candidate &            08/06/01        \\
\hline
HIP63204 &   8.79   &   8.51   &   8.40   &   0.15   &   236.56   &   confirmed      &    06/04/04   \\
\hline
HIP64515 &          &          &   6.94   &   0.31   &   165.69   &   candidate &            08/06/01        \\
\hline
HIP65822 &          &          &   11.08   &   1.82   &   303.87   &   candidate &           08/06/01         \\
%\hline
HIP67260 &          &   14.04   &   14.10   &   1.23   &   355.65   &   inconclusive      &    28/04/04   \\
HIP67260 &   15.84   &   14.83   &   14.67   &   2.33   &   77.25   &   inconclusive      &    28/04/04   \\
HIP67260 &   8.88   &   8.46   &   8.36   &   0.42   &   229.46   &   confirmed      &    28/04/04   \\
\hline
HIP67919 &   9.98   &   9.38   &   9.10   &   0.69   &   296.56   &   confirmed      &    28/04/04   \\
\hline
HIP68080 &          &          &   7.19   &   1.92   &   10.20   &   candidate &             05/06/01       \\
\hline
HIP68532 &   10.52   &   9.85   &   9.54   &   3.05   &   288.50   &   confirmed      &    28/04/04   \\
HIP68532 &   11.38   &   10.94   &   10.63   &   3.18   &   291.92   &   confirmed      &    28/04/04   \\
\hline
HIP68867 &          &          &   11.61   &   2.16   &   284.76   &   candidate &           08/06/01         \\
\hline
HIP69113 &   10.98   &   10.43   &   10.29   &   5.34   &   65.15   &   confirmed      &    30/04/04   \\
HIP69113 &   11.27   &   10.45   &   10.30   &   5.52   &   67.17   &   confirmed      &    30/04/04   \\
\hline
HIP69749 &          &          &   11.60   &   1.50   &   0.84   &   candidate &             08/06/01       \\
\hline
HIP70998 &          &          &   10.83   &   1.17   &   354.60   &   candidate &           06/06/01         \\
\hline
HIP71724 &          &          &   9.70   &   8.66   &   23.02   &   candidate &             08/06/01       \\
\hline
HIP71727 &          &          &   7.80   &   9.14   &   244.96   &   candidate &            08/06/01        \\
\hline
HIP72940 &          &          &   8.57   &   3.16   &   221.58   &   candidate &            06/06/01        \\
\hline
HIP72984 &          &          &   8.50   &   4.71   &   260.35   &   candidate &            06/06/01        \\
\hline
HIP73937 &          &   8.46   &   8.37   &   0.24   &   190.58   &   confirmed      &    30/04/04   \\
\hline
HIP74066 &          &          &   8.43   &   1.22   &   109.62   &   candidate &            08/06/01        \\
\hline
HIP74479 &          &          &   10.83   &   4.65   &   154.15   &   candidate &           08/06/01         \\
\hline
HIP75056 &          &          &   11.17   &   5.19   &   34.51   &   candidate &            08/06/01        \\
\hline
HIP75151 &          &          &   8.09   &   5.70   &   120.87   &   candidate &            08/06/01        \\
\hline
HIP75915 &          &          &   8.15   &   5.60   &   229.41   &   candidate &            05/06/01        \\
\hline
HIP76001 &          &          &   7.80   &   0.25   &   3.17   &   candidate &              08/06/01      \\
HIP76001 &          &          &   8.20   &   1.48   &   124.82   &   candidate &            08/06/01        \\
\hline
HIP76071 &          &   11.28   &   10.87   &   0.69   &   40.85   &   confirmed      &    02/06/00, 07/06/01    \\
\hline
HIP77315 &          &          &   7.12   &   37.37   &   137.32   &   candidate &           08/06/01         \\
HIP77315 &          &          &   7.92   &   0.68   &   67.01   &   candidate &             05/06/01       \\
\hline
HIP77911 &   12.68   &   12.20   &   11.84   &   7.96   &   279.25   &   confirmed      &    02/06/00, 07/06/01    \\
\hline
HIP77939 &          &          &   8.09   &   0.52   &   119.13   &   candidate &            31/05/00        \\
\hline
HIP78530 &          &   14.56   &   14.22   &   4.54   &   139.69   &   inconclusive      &    02/06/00, 07/06/01    \\
\hline
HIP78756 &          &          &   9.52   &   8.63   &   216.40   &   candidate &            02/06/00        \\
\hline
HIP78809 &   11.08   &   10.45   &   10.26   &   1.18   &   25.67   &   confirmed      &    03/06/00, 07/06/01    \\
\hline
HIP78847 &          &          &   11.30   &   8.95   &   164.02   &   candidate &           03/06/00         \\
\hline
HIP78853 &          &          &   8.45   &   1.99   &   270.39   &   candidate &            08/06/01        \\
\hline
HIP78956 &   9.76   &   9.12   &   9.04   &   1.02   &   48.67   &   confirmed      &    03/06/00, 07/06/01    \\
\hline
HIP78968 &   14.96   &   14.51   &   14.26   &   2.78   &   322.13   &   inconclusive      &    04/05/04   \\
\hline
HIP79124 &   11.38   &   10.55   &   10.38   &   1.02   &   96.18   &   confirmed      &    03/06/00, 07/06/01    \\
\hline
HIP79156 &   11.62   &   10.89   &   10.77   &   0.89   &   58.88   &   confirmed      &    03/06/00, 07/06/01    \\
\hline
HIP79250 &          &          &   10.71   &   0.62   &   180.92   &   candidate &           03/06/00         \\
\hline
HIP79530 &          &          &   8.34   &   1.69   &   219.66   &   candidate &            31/05/00        \\
\hline
HIP79631 &          &          &   7.61   &   2.94   &   127.85   &   candidate &            05/06/01        \\
\hline
HIP79739 &   12.28   &   11.52   &   11.23   &   0.96   &   118.33   &   confirmed      &    19/06/04   \\
\hline
HIP79771 &   12.00   &   11.28   &   10.89   &   3.67   &   313.38   &   confirmed      &    19/06/04   \\
HIP79771 &   12.39   &   11.79   &   11.42   &   0.44   &   128.59   &   confirmed      &    19/06/04   \\
\hline
HIP80142 &   16.64   &   15.88   &          &   5.88   &   119.94   &   inconclusive      &    04/05/04   \\
\hline
HIP80238 &   7.96   &   7.66   &   7.49   &   1.03   &   318.46   &   confirmed      &    02/06/00, 07/06/01    \\
\hline
HIP80324 &          &          &   7.52   &   6.23   &   152.46   &   candidate &            31/05/00, 03/06/00        \\
\hline
HIP80371 &          &          &   8.92   &   2.73   &   140.65   &   candidate &            02/06/00, 03/06/00        \\
\hline
HIP80425 &          &          &   8.63   &   0.60   &   155.77   &   candidate &            08/06/01        \\
\hline
HIP80461 &          &          &   7.09   &   0.27   &   285.64   &   candidate &            31/05/00        \\
\hline
HIP80799 &   10.60   &   10.04   &   9.80   &   2.94   &   205.02   &   confirmed      &    05/05/04   \\
\hline
HIP80896 &   11.16   &   10.63   &   10.33   &   2.28   &   177.23   &   confirmed      &    08/06/04   \\
\hline
HIP81624 &          &          &   7.95   &   1.13   &   224.28   &   candidate &            05/06/01      \\
\hline
HIP81949 &          &   15.62   &   14.82   &   6.26   &   239.37   &   inconclusive      &    04/05/04, 05/05/04, 08/06/04, 25/06/04   \\
HIP81949 &          &   15.67   &   15.52   &   5.27   &   340.72   &   inconclusive      &    04/05/04, 05/05/04, 08/06/04, 25/06/04   \\
\hline
HIP81972 &   11.30   &   10.97   &   10.61   &   7.02   &   258.81   &   inconclusive      &    27/06/04   \\
HIP81972 &   11.63   &   10.87   &   10.48   &   2.02   &   313.69   &   inconclusive      &    27/06/04   \\
HIP81972 &   12.54   &   11.86   &   11.77   &   5.04   &   213.45   &   confirmed      &    27/06/04   \\
HIP81972 &   15.10   &   14.43   &   13.98   &   2.79   &   106.94   &   confirmed      &    27/06/04   \\
HIP81972 &   16.11   &   15.63   &   15.26   &   7.92   &   229.27   &   confirmed      &    27/06/04   \\
\hline
HIP83542 &          &   9.72   &   9.90   &   8.86   &   196.21   &   inconclusive      &    10/09/04   \\
\hline
HIP83693 &          &          &   9.26   &   5.82   &   78.35   &   candidate &             06/06/01       \\

\end{longtable}
\end{landscape}

\newcommand\unitvec[1]{\ensuremath{\mathbf{\hat#1}}}
\newcommand\inprod[2]{\ensuremath{\langle\vect{#1},\vect{#2}\rangle}}
\newcommand\inprodu[2]{\ensuremath{\langle\unitvec{#1},\unitvec{#2}\rangle}}
\newcommand{\vect}[1]{\ensuremath{\mbox{\boldmath $#1$}}}
 
% ====================================================================
% ====================================================================
% ====================================================================
% ==START=HERE========================================================
% ====================================================================
% ====================================================================
% ====================================================================

\chapter{Recovering the true binary population from observations} \label{chapter: sos}

 \author{M.B.N. Kouwenhoven\inst{1}
          \and
          A.G.A. Brown\inst{2}
          \and
          S.F. Portegies Zwart\inst{1,3}
          \and
          L. Kaper\inst{1}
          }

\begin{center}
M.B.N. Kouwenhoven, A.G.A. Brown, S.F. Portegies Zwart, \& L. Kaper

\vspace{0.2cm}
{\it Astronomy \& Astrophysics}, to be submitted
\end{center}

% ====================================================================
% ====================================================================
% ====================================================================
% ==ABSTRACT==========================================================
% ====================================================================
% ====================================================================
% ====================================================================

\section*{Abstract}

Knowledge on the binary population in OB~associations and star clusters provides important information about the outcome of the star forming process in different environments. Binarity is also a key ingredient in stellar population studies, and is a prerequisite to calibrate the binary evolution channels. The current binary population can be used to constrain the primordial binary population, i.e. the number of stars that have formed in binary and multiple systems. Characterizing the current binary population poses significant challenges, as our knowledge is limited due to selection effects.
In this paper we address the issue of deriving the true binary population from the observed binary population, by comparing the observations with simulated observations of association models. We investigate and discuss the different selection effects present in surveys of visual, spectroscopic, and astrometric binaries.
We present an overview of several commonly used parameter distributions. We identify five possibilities for pairing the components of a binary system, and discuss the consequences of these on the observables. In particular, we study the dependence of the specific mass ratio distribution for samples of binaries with different spectral types. We also discuss the trend between primary spectral type and binary fraction, as a result of pairing binary components.
By varying the association distance and the properties of the binary population, we derive which fraction of the binary systems can be detected as visual, spectroscopic, or astrometric binaries. We show that practically all binarity surveys produce an observed binary fraction that increases with increasing primary mass, even though this trend is not present in the model. Observations of binary systems are additionally biased to high mass ratios, in several cases even when the true mass ratio distribution is decreasing towards high mass ratios. 
In this paper we illustrate how selection effects can mislead the observer in his/her interpretation of the observed binary population. In order to find the true binary population, it is generally necessary to compare the observations with simulated observations, rather than applying incompleteness corrections.
%

% ====================================================================
% ====================================================================
% ====================================================================
% ==INTRODUCTION======================================================
% ====================================================================
% ====================================================================
% ====================================================================

\section{Introduction}

%%% primordial binary population

Observations have shown that most --- possibly even all --- stars form in binary
or multiple systems. Hence detailed knowledge of a young binary population can
be employed to study the outcome of star formation, and consequently the star
formation process itself. This paper is one in a series addressing the issues
involved when deriving the primordial binary population (PBP), which is defined
as {\em``the population of binaries as established just after the gas has been
removed from the forming system, i.e., when the stars can no longer accrete gas
from their surroundings''} \citep{kouwenhoven2005}.  The term refers to the
point in time beyond which the freshly formed binary population is affected by
stellar/binary evolution and stellar dynamical effects only. Interactions with a
surrounding gaseous medium are no longer of importance.  Once the PBP is known,
it should be used as the end point for hydrodynamical simulations of star
cluster formation \citep[e.g.,][]{bate2003}. The PBP should also be used as the
initial condition for N-body simulations of evolving stellar groupings
\citep[e.g.,][]{ecology4}, for studies of interacting close binaries
\cite[e.g.,][]{spzverbunt1996}, stellar mergers \citep[e.g.,][]{sills2002},
the formation of high-mass X-ray binaries \citep[e.g.,][]{terman1998}, and for
identifying the progenitors of type~Ia supernovae \citep[e.g.,][]{yungelson1998,hillebrandt2000}.

Ideally, one would like to observe the PBP directly. In practice this is not
feasible due to high interstellar extinction in star forming regions. However, it is
possible to study a binary population which is very close to primordial in young
(i.e. freshly exposed), low-density stellar groupings. The prime targets
satisfying these conditions are OB~associations \citep[see][for a more detailed
description]{blaauw1991,brown2001,kouwenhoven2005}. OB~associations are young ($\la 50$~Myr) stellar
groupings of low density --- such that they are likely to be unbound ---
containing a significant population of B~stars, with projected dimensions
ranging from $\sim 10-100$~pc \citep{brown1999}. The youth and low stellar
density of OB~associations imply that stellar and dynamical evolution have only
modestly affected the binary population. The PBP can be derived from the {\em
current} binary population using inverse dynamical population synthesis
\citep[see, e.g.,][]{kroupa1995a,kroupa1995c}, taking into account the effects
of stellar and dynamical evolution.

%%% explain selection effects

The first step towards finding the PBP of an OB~association is to establish a
census of its actual binary population. The association membership issue
is a complication that arises already during the definition of the observing
program: which stars are bona-fide members of the association, and which stars
are foreground or background objects?  \cite{dezeeuw1999} used
{\em Hipparcos} positions, proper motions and parallaxes to separate the
members of nearby ($\la 1$~kpc) OB~associations from the field star population.
This is the most reliable method to identify association members. Other methods
are based on the position of the stars relative to the cluster center, the age
of individual stars, or photometric distances to individual stars. Methods that
are used to determine membership are often more accurate for the more luminous
stars in the cluster or association. 

A large number of binary systems is detected in imaging surveys, radial
velocity surveys, and astrometric surveys of the member stars. A smaller number
of binaries is discovered with photometric techniques (e.g. eclipsing binaries
and lunar occultation surveys). Each technique has its own limitations, which
are dependent on e.g. the instrument used, the observing strategy, and the
atmospheric conditions at the time of observation. Each technique is thus only
sensitive to a sample of binaries with specific properties. For
example, imaging surveys favour the detection of binaries with a large
orbital period and a mass ratio near unity. Spectroscopic and astrometric
surveys are more sensitive to short-period ($P \la 100$~yr) systems for which the
orbital motion can be seen in the observations. 
Due to observational limitations it is practically impossible to directly
observe the {\em true} binary population. The matter is complicated even further
by contamination with background stars that project close to association
members, and so constitute optical (fake) binaries. Consequently, the {\em
observed} binary population consists of a (often heavily) biased subset of the
{\em true} binary population.

%%% sos history

It is of course well known that observational selection effects play an
important role in the derivation of the binarity properties of a stellar
population. Over the last century most authors have focused on deriving the
binary population properties in the solar neighborhood,
often by using extrapolations and analytical corrections to remove the selection
effects. In the 1930s Kuiper worked on the interpretation of large samples of
available data on binarity among stars in the solar neighborhood
\citep{kuiper1935a,kuiper1935b}. He derives an estimate for the binary fraction,
after having corrected for unseen binaries. A similar, but more extensive study
was performed by \cite{heintz1969}. After about 1970 several authors have
employed computer simulations to study the properties of a generated ensemble of
binaries, and used simulated observations to correct for observational biases
using Monte Carlo techniques \citep[e.g.,][]{halbwachs1981, halbwachs1983,
halbwachs1986, vereshchagin1987, hogeveen1990, trimble1990, hogeveen1992a,
hogeveen1992b, kobulnicky2006}. \cite{kroupa1991} analyze the effect of
unresolved binary stars on the stellar luminosity function and consequently on
the mass distribution. 

%%% verschil van ons paper met de rest.

There are two major differences between this study and most previous studies
in literature.
First, we model realistic, state-of-the art stellar clusters, for which the
properties are projected onto the observable space for imaging, spectroscopic,
and astrometric techniques. With an extension of existing simulation software we
generate star clusters quickly, and the observables are derived. No
simplifications are necessary, such as circular orbits, random orbit
orientation, equal masses, point-like clusters etc. The second major difference
with previous studies is that our analysis focuses on associations and star
clusters, rather than the stellar population in the solar neighborhood.

%%% application

Throughout this paper we will describe the simulated observations of modeled
associations with various properties. First we will consider observations
without including selection effects. Due to the nature of the observing
techniques, even these observations cannot provide us with all data on the
binary population, unless the system is both a spectroscopic and an astrometric
binary. For example, a single-epoch imaging survey for binarity (henceforth
referred to as ``visual binary survey'') barely provides information on the
eccentricity distribution of the binary population. Following this analysis,
we will focus on the importance of the
different selection effects. In many surveys for binarity, only a small fraction
of the binaries is actually observed. What does this ``tip of the iceberg'' tell
us about the underlying distribution?

In this paper we focus mainly on OB~associations since these are our prime
targets of interest. Many of the results presented in this article are
applicable to other stellar groupings as well, such as the stellar population in
the solar neighborhood, T~associations, open clusters, and globular clusters.
This paper is organized as follows: 
\begin{itemize}\addtolength{\itemsep}{-0.5\baselineskip}
\item[--] In \S~\ref{section: method} we discuss the method of simulating
  observations, and how to compare the results to observations.
\item[--] In \S~\ref{section: conversiontoobservables} we describe the projection of
  binary orbits onto the space of observables for visual, spectroscopic, and
  astrometric binaries. 
\item[--] In \S~\ref{section: associationandbinarypopulationproperties} we introduce
  the properties of the ``default association'' model we use as a basis in our
  analysis. In this section we also discuss the problems that arise when
  choosing a pairing relation between primary stars and companions.
\item[--] In \S~\ref{section: observationalselectioneffects} we describe the
  selection effects and observational errors related to the observations of
  visual, spectroscopic, and astrometric binaries.
\item[--] In \S~\ref{section: simulatedobservations} we study how the association
  and binary population properties are reflected in the synthetic observations,
  depending on the observational technique.
\item[--] In \S~\ref{section: inverseproblem} we discuss the inverse problem, i.e.,
  deriving association properties and binary population properties from
  observations.
\item[--] In \S~\ref{section: example1} we give a detailed example of the usefulness
  and necessity of the method of simulating observations.
\item[--] Finally, in \S~\ref{section: summaryandoutlook}, we describe the
  implications of our results on the interpretation of binary star observations,
  and summarize our findings.
\end{itemize}

% ====================================================================
% ====================================================================
% ====================================================================
% ==METHOD=TERMINOLOGY======================================================
% ====================================================================
% ====================================================================
% ====================================================================

\section{Method and terminology} \label{section: method}

With the increase of computer power over the last decade, computer simulations
have become an integral part of astronomical research. Physical processes are
modeled with increasing detail. On the other hand, large surveys and
more powerful telescopes have provided us with a wealth of observational
data. A comparison between observational data and the results of
simulations is important for validating the simulations (are the assumptions
valid?), determining initial conditions, and finding discrepancies between
theory and observations, which could lead to new ideas. The theoretical
community has acknowledged the necessity for comparisons between observations
and simulations \citep[e.g.,][]{modest1,modest2}, but detailed comparisons have
been scarce so far. 

\begin{figure}[!tbp]
  \centering
  \includegraphics[width=1\textwidth,height=!]{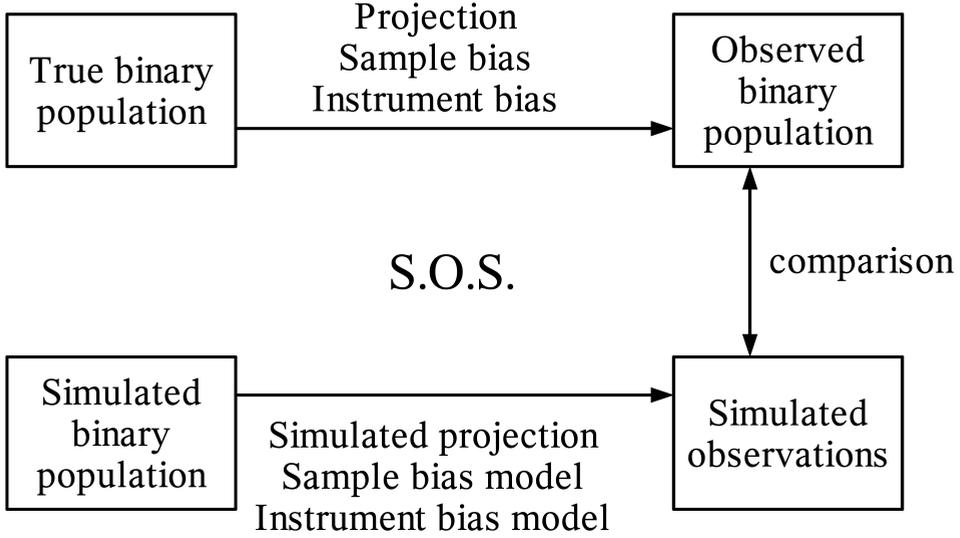}
  \caption{A schematic illustration of the S.O.S. (Simulating Observations of
  Simulations) method to recover the {\em true} binary population from the {\em
  observed} binary population. With the S.O.S. approach we simulate observations
  of simulated stellar associations by imposing selection effects. By applying
  this method to simulated associations with different properties, and comparing the
  simulated observations with the real observations, we can derive the intrinsic
  properties (or the set of possibilities) that are consistent with the
  observations.
  \label{figure: sosmethod} }
\end{figure}

The main difference between observational and simulated datasets is that the
former are often heavily biased due to selection effects. In order to properly
compare the two datasets it is of crucial importance to understand these biases.
These biases can be modeled and applied to the simulated dataset. The
simulated observations are then reduced as if they were real data. By varying
the properties of the simulated cluster, and comparing the simulated
observations with the observational data, the properties of the real cluster can
be retrieved. This procedure is known as ``Simulating Observations of
Simulations'' (SOS), and is illustrated in Figure~\ref{figure: sosmethod}.

Let ${\mathcal T}$ be the set containing all information about the binary
population in a star cluster (such as the binary fraction, mass ratio distribution,
\dots). In a survey for binarity only a subset $\tilde{\mathcal T}$ is observed
due to sampling and observational limitations. We define the function ${\mathcal
B}$ that maps ${\mathcal T}$ onto $\tilde{\mathcal T}$:
\begin{equation} \label{equation: method_projection1}
{\mathcal T} \overset{\mathcal B}{\longrightarrow} \tilde{\mathcal T}
\end{equation}
For a given sample and well-defined observational limitations, ${\mathcal B}$ is
unique. The projection ${\mathcal B}$ can be divided into three more or less
independent projections. These are the projection onto observable space
${\mathcal B}_{\rm obs}$, the selection of the sample ${\mathcal B}_{\rm
sample}$, and the effect of observational limitations ${\mathcal B}_{\rm
instrument}$:
\begin{equation} \label{equation: method_projection2}
{\mathcal T} \ 
\overset{{\mathcal B}_{\rm obs       }}{\longrightarrow}  \ \ 
\overset{{\mathcal B}_{\rm sample    }}{\longrightarrow}  \ \ 
\overset{{\mathcal B}_{\rm instrument}}{\longrightarrow}  \ 
\tilde{\mathcal T}.
\end{equation}
Not all intrinsic properties of a binary system (such as the orbital energy $E$
and angular momentum $\vec{L}$) are directly observable. The properties of a
binary system are first {\em projected onto the space of observables} via the
projection $\mathcal{B}_{\rm obs}$. Properties such as $E$ and $\vec{L}$ are
then reflected in the observables, such as in the angular separation $\rho$ at
an instant of time, and the shape of the radial velocity curve. No information
is lost during the projection onto observable space, as all properties of a
binary system can be derived with a combination of astrometric and spectroscopic
observations. In practice, however, only a subset of the information is
retrieved, as many binaries are only observed either in a visual, a
spectroscopic, or an astrometric survey. For a single-epoch imaging survey in a
single filterband, four independent properties are left: the angular
separation, the position angle, and the luminosity of both components in the
system. This is less than the seven independent physical system parameters and hence
information is lost. By combination of different observing techniques, all seven
orbital parameters may be recovered.

With the {\em sample bias} ${\mathcal B}_{\rm sample}$ we refer to the process
of selecting the targets of interest. The selected sample consists often of a
group of stars with specific properties, such as the solar-type stars in the
solar neighborhood, or the B~stars in an association. In the case of a binary
survey among the members of an OB~association, the observer may erroneously
include a bright background star, assuming that it is an association member. We
consider this part of the sample bias, although we do not discuss this aspect in this paper.

When stars in a sample are surveyed for binarity, observational constraints are
responsible for the {\em instrument bias} ${\mathcal B}_{\rm instrument}$. For
example, the minimum and maximum detectable angular separation of binary stars
is determined by the properties of the telescope and the detector. We include in
the instrument bias the selection effects imposed by the telescope-instrument
combination and atmospheric conditions. We additionally include the bias that
results from the difficulties of identifying companions. For example, faint
companions at a large separation of their primary may not always be identified
as such, due to the confusion with background stars.

Due to each of the projections ${\mathcal B}_{\rm sample}$ and ${\mathcal
B}_{\rm instrument}$, a piece of information contained in the original dataset
${\mathcal T}$ is lost. We refer to the combined effect ${\mathcal B}_{\rm
sample} {\mathcal B}_{\rm instrument}$ of sample selection and selection effects
as {\em observational biases}. In practice the sample bias ${\mathcal B}_{\rm
sample}$ and instrument bias ${\mathcal B}_{\rm instrument}$ cannot always be
completely separated, but throughout this paper we will consider them
independently.

The inverse ${\overline{\mathcal B}}$ of ${\mathcal B}$ is not unique: different
binary distributions ${\mathcal T}$ can produce the same set of observations
$\tilde{\mathcal T}$. For this reason it is dangerous to apply a simple
``incompleteness correction'' to the observed dataset $\tilde{\mathcal T}$ in
order to recover the true $\mathcal T$ \citep[see, e.g.,][]{mazeh1992}.
In order to find $\overline{{\mathcal B}}$ and to derive the various binary
populations ${\mathcal T}$ that are able to produce observations
$\tilde{\mathcal T}$, we simulate star clusters and observations thereof. Let
${\mathcal S}$ be a {\em simulated} star cluster. For a given set of selection
effects ${\mathcal B_S}$ we obtain {\em simulated observations} $\tilde{\mathcal
S}$:
\begin{equation} \label{equation: method_projection3}
{\mathcal S} \ 
\overset{{\mathcal B}_{\rm S,obs       }}{\longrightarrow}  \ \ 
\overset{{\mathcal B}_{\rm S,sample    }}{\longrightarrow}  \ \ 
\overset{{\mathcal B}_{\rm S,instrument}}{\longrightarrow}  \ 
\tilde{\mathcal S}.
\end{equation}
The procedure described above allows us to get a grip on how observational
biases mislead us in our interpretation of observations $\tilde{\mathcal T}$. A
good understanding of the sample bias and instrument bias is necessary to model
them with ${\mathcal B}_{\rm S,sample}$ and ${\mathcal B}_{\rm S,instrument}$,
so that the observations $\tilde{\mathcal{T}}$ can be compared with the
simulated observations $\tilde{\mathcal{S}}$.

In this paper we apply the S.O.S. method in order to analyze (1) the effect of
changes in the binary population of the simulated OB~association ${\mathcal S}$
on the simulated observations $\tilde{\mathcal S}$ for a given set of selection
effects ${\mathcal B_S}$; (2) the effect of using different selection effects
${\mathcal B_S}$ (i.e., different observing strategies or instruments) on the
properties of the simulated observations $\tilde{\mathcal S}$; and (3) how the
observed binary population $\tilde{\mathcal T}$ can be used to derive the true
binary population ${\mathcal T}$. To this end we simulate a large set of
associations with properties ${\mathcal S}$ and compare the simulated
observations $\tilde{\mathcal S}$ with the observations $\tilde{\mathcal T}$. In
a following paper we will apply this method to the Sco~OB2 association.

\subsection{Simulating observations of an OB~association} \label{section: sos_sosmethod}

A binary orbit can be completely described by seven orbital parameters: the
primary star mass $M_1$, the companion star mass $M_2$, the semi-major axis $a$
and eccentricity $e$ of the orbit, the inclination $i$ and argument of
periastron $\omega$ of the orbit, and the position angle of the line of nodes
$\Omega$. The seven orbital parameters, together with the orbital phase
$\mathcal{V}$ define the binary system at an instant of time. Usually the
properties of a binary population are described with a parameter distributions,
such as $f_a(a)$ for the semi-major axis and $f_e(e)$ for the eccentricity. These
distributions are either derived from observations, or from theoretical
arguments. 
Additionally, the properties of the association itself also affect the
observables. The most relevant parameters include the distance and the binary
fraction. The size, the age, and the metallicity also play a role, although not
as important as the distance.

When simulating an OB~association, we perform the following sequence of operations.
\begin{enumerate}\addtolength{\itemsep}{-0.5\baselineskip}
\item An association of $N = S+B$ ``systems'' is created, where $S$ and $B$ are
  the number of single stars and binaries, respectively. For simplicity, we do
  not include higher-order multiple systems. The systems are given a position
  according to a stellar density distribution function. Each system is given a
  velocity, in such a way that the association is in virial equilibrium.
\item Each single star and primary is assigned a mass. The companion is assigned
  a mass according to the pairing prescriptions. In the case that the primary
  and companions are randomly paired from the mass distribution\footnote{We
  refer to the distribution of stellar masses as the ``mass distribution''. In
  literature the term ``mass function'' is often used to describe this
  distribution. In this paper, we use the former term, in order to avoid
  confusion with the mass function $\mathcal{F}(M)$, an observable for
  spectroscopic binaries.}, primary and companion are swapped if the companion
  is more massive.
\item Orbits are generated for each binary system: each orbit is assigned a
  semi-major axis (from which the period is derived), an eccentricity, and an
  orientation. Each orbit is given a random orbital phase.
\item The association is assigned a distance, coordinates on the sky, an age and
  a metallicity.
\item The other intrinsic properties of stars and binary systems are derived,
  for example the absolute magnitude for a star in different filterbands, and
  the period of a binary system through Kepler's third law.
\item The simulated OB~association is projected onto the space of observables
  using $\mathcal{B}_{S,obs}$; the observables for visual, spectroscopic, and
  astrometric surveys are derived. These include for example the angular
  separation $\rho$ between the binary components, the colour and apparent
  magnitude of each component in various filters (using the isochrone for the
  association), the radial velocity curve, and the projection of the orbit on
  the plane of the sky.
\item Those single stars and binary systems satisfying the sample bias
  $\mathcal{B}_{S,sample}$ are selected. For example, if a survey among B~stars
  is simulated, only the B~stars are selected.
\item The instrumental bias $\mathcal{B}_{\rm S,instrument}$ is imposed on the
  observables. This means for example that several companions will be removed
  from the dataset, or that the magnitudes of the components of an unresolved
  binary are combined.
\item Finally, the observed properties of the model OB~association are obtained
  from the simulated observations. These include the observed distribution
  functions, and the observed binary fraction.
\end{enumerate}

We create association models using the STARLAB simulation package \citep[see,
e.g.,][]{ecology4}. The properties of the stellar and binary population are
projected onto the space of observables, using an extension of the STARLAB
package. Details on the projection are given in Section~\ref{section: conversiontoobservables} and Appendix~A. The
sample and instrument bias are imposed using a ``synthetic observation'' tool,
developed in IDL.

In our models we assume that most binary parameters are independent. In terms of
distribution functions, this means
\begin{equation} \label{equation: independent_parameters}
  f_{\rm BP}(M_1,M_2,a,e,i,\omega,\Omega,\mathcal{M})  =
  f_{M_1,M_2}(M_1,M_2)\,f_a(a)\,f_e(e)\,f_i(i)\,f_\omega(\omega)\,f_\Omega(\Omega)\,f_\mathcal{M}(\mathcal{M})\,,
\end{equation}
where $\mathcal{M}$ is the mean anomaly at a certain instant of time. The
primary and companion mass distribution are never independent, $f_{M_1,M_2}(M_1,M_2) \neq
f_{M_1}(M_1)f_{M_2}(M_2)$, as by definition $M_1 \geq M_2$.

In our models the overall binary fraction for the association can be described
with a single number. In most cases, this corresponds to a binary fraction that
is independent of primary spectral type. However, in Section~\ref{section:
pairingfunction} we describe two cases (random pairing and PCP-II) where as a
result of the pairing properties the binary fraction does depend on primary
spectral type even though this dependency is not included explicitly.

The independence of $a$ with respect to most other parameters may be a good
assumption, as observations suggest that these parameters are only mildly
correlated for solar-type stars in the solar neighbourhood
\citep[e.g.,][]{duquennoy1991,heacox1997}. However, one should keep in mind that
selection effects are present in the datasets explored in these two papers. We
assume that the inclination $i$, the argument of periastron $\omega$, the
position angle of the ascending node $\Omega$, and the true anomaly
$\mathcal{V}$ at any instant of time are independent of each other and of all
other parameters. 

Throughout this paper we will refer with ``star'', ``primary star'', and
``companion star'' to both stellar and substellar objects (brown dwarfs and
planets), unless stated otherwise. A ``companion star'' refers to a physically
bound object, and a ``background star'' to a stellar object at close separation
from the target star due to projection effects. A ``secondary star'' can be
either a companion or a background star. 

Our models of OB associations do not include the background population of field
stars and we assume that membership is perfectly known for the association.
However, the confusion with background stars is included in our model for the
instrument bias in imaging surveys.

\subsection{Distribution functions}

We describe each binary parameter $x$ using its distribution function $f$, where
$f_x(x)\, dx$ is the probability that the binary parameter $x$ is in the interval
$[x,x+dx]$. The distribution function is normalized according to:
\begin{equation}
\int^{x_{\rm max}}_{x_{\rm min}} f_{x}(x')\, {\rm d}x' = 1\,,
\end{equation}
where $x_{\rm min}$ and $x_{\rm max}$ are the minimum and maximum possible
values of $x$. The corresponding cumulative distribution function $F_x(x)$ is
defined as:
\begin{equation}
F_x(x) = \int^{x}_{x_{\rm min}} f_x(x')\, {\rm d}x'
\end{equation}
and takes values between $0$ and $1$. $F_x(x)$ represents the probability that the
parameter is smaller than $x$.

\subsection{Statistical tests and error calculation}

In our method we compare the observations with the simulated
observations. This comparison is done with statistical tests. Parameter
distributions are compared using the Kolmogorov-Smirnov test, while individual
measurements (such as the binary fraction) are compared with the $\chi^2$ test.

\subsubsection{Comparison between two distributions}

When comparing distributions resulting from two different models, we use the
Kolmogorov-Smirnov (KS) test for testing the null hypothesis that the two
distributions are drawn from a common underlying distribution. The test is
parameterized by a probability $0 \leq p \leq 1$. 

Large values of $p$ mean that no significant difference between the two
distributions is observed. For small values of $p$ we can reject the null
hypothesis (thus, the distributions differ significantly). We can reject the
hypothesis that the two distributions are drawn from the same underlying
distribution with a confidence of $1\sigma$, $2\sigma$, or $3\sigma$, if $p=0.317$, $p=0.046$, or $p=0.0027$, respectively.

\subsubsection{Comparison between two binary fractions}

To test whether the binary fraction in the real population and the model
observations are statistically different, 
we use a Pearson $\chi^2$-test \citep[e.g.,][]{miller1990}. Let
$N_1$ and $N_2$ denote the number of targets in each of the two models, and
$d_1$ and $d_2$ the corresponding number of binaries found. With the Pearson
$\chi^2$-test we estimate the probability that the resulting binary fractions
$d_1/N_1$ and $d_2/N_2$ are significantly different. The Pearson $\chi^2$ is
defined as
\begin{equation}
\chi^2 = \left( 
\frac{ d_1/N_1 - d_2/N_2 }
     { \sqrt{\tilde{f}\ \left(1-\tilde{f} \right) \ \left( N_1^{-1} +N_2^{-1} \right) } } 
\right)^2 \,,
\end{equation}
where $\tilde{f} \equiv (d_1+d_2)/(N_1+N_2)$ is the common binary fraction. The
Pearson $\chi^2$ is a $\chi^2$-distribution with one degree of freedom and can
be associated with a probability $p$ that the two binary fractions are drawn
from a common underlying binary fraction:
\begin{equation} \label{equation: pearsonprobability}
p = 1 - \frac{ \gamma \left( \tfrac{1}{2},\tfrac{1}{2}\chi^2 \right) }{ \sqrt{\pi} },
\end{equation}
where $\gamma$ is the lower incomplete gamma function and $\sqrt{\pi}$ the
normalization. The Pearson $\chi^2$ is a good approximation
if $d_1$, $(N_1-d_1)$, $d_2$, and $(N_2-d_2)$ are all larger than~$\sim 5$.
If this condition is not satisfied, a more sophisticated
statistical test, such as Fisher's exact test \citep{fisher1925} or
Barnard's exact test \citep{barnard1945} should be employed.

\subsection{The error on the binary fraction} \label{section: errorbinaryfraction}

Suppose that a sample of $N$ targets is surveyed for binarity, and $d$ binaries are
found. The observed binary fraction is then $\tilde{F}_{\rm M} = d/N$. We
calculate the statistical error on $\tilde{F}_{\rm M} = d/N$ using the formalism
described in \cite{burgasser2003}. The confidence interval
$[\tilde{F}_{\rm M-},\tilde{F}_{\rm M+}]$ of $\tilde{F}_{\rm M}$ corresponding
to a $k\sigma$ limit for a Gaussian error distribution can be found by
numerically solving:
\begin{equation}
  \sum_{i=0}^d \frac{(N+1)!}{i!(N+1-i)!} \, x^i (1-x)^{N+1-i} = 
  \left\{
  \begin{tabular}{ll}
    $\mbox{CDF}\,(k)$,  & $x=\tilde{F}_{\rm M-}$ \\
    $\mbox{CDF}\,(-k)$, & $x=\tilde{F}_{\rm M+}$ \\
  \end{tabular}
  \right. ,
\end{equation}
where 
\begin{equation}
 \mbox{CDF}(k) \equiv \frac{1}{2}+\frac{1}{2}\,\mbox{erf}\, \left(\frac{k}{\sqrt{2}}\right)\,.
\end{equation}
The $1\sigma$ confidence interval corresponds to values of CDF$(1) \approx 0.84$
and CDF$(-1) \approx 0.16$. For large samples ($N \ga 100$), the $1\sigma$ error
$\sigma_{\rm F}$ on $\tilde{F}_{\rm M}$ may be approximated with
\begin{equation}
  \frac{ \sigma_{\rm F} }{ \tilde{F}_{\rm M} } \approx \sqrt{\frac{1}{d} +\frac{1}{N}}\,.
\end{equation}
The $k\sigma$ confidence interval as calculated above takes into account the
{\em statistical} errors. It does not include the {\em systematic} errors that
may result from the choice of the sample or instrumental constraints.

% ====================================================================
% ====================================================================
% ====================================================================
% =CONVERSION TO OBSERVABLES======================================================
% ====================================================================
% ====================================================================
% ====================================================================

\section{Conversion to observables} \label{section: conversiontoobservables}

In this section we discuss the projection of the intrinsic properties of a
binary system onto the space of observables.
In this paper, each binary system is described as a visual binary, a
spectroscopic binary, or an astrometric binary. These designations do not refer
to any intrinsic properties of the binary systems, but to the observational
properties alone. Visual binaries are those binaries that are detected in a
single-epoch imaging survey, spectroscopic binaries are those detected in a
(multi-epoch) radial velocity survey, and astrometric binaries are those detected in a
(multi-epoch) astrometric survey. A binary system may be classified in more than one of these
groups, if it is detectable with multiple techniques. Note that, as we allow
overlap between the groups, the definition is slightly different from that
adopted in \cite{kouwenhoven2005, kouwenhoven2006a}. Binaries that are not
detected with any of these techniques are referred to as unresolved binaries.

\subsection{Visual binary observables}

Visual binaries are systems for which the presence of the companion is derived from (single-epoch) imaging.
For a visual binary observed using a single filter band, there are four
independent observables: the angular separation $\rho$ between primary and
companion, the position angle $\varphi$ of the vector connecting primary and
companion, and the magnitudes of both components. 

If the primary and companion have coordinates $(\alpha_1,\delta_1)$ and
$(\alpha_2,\delta_2)$, respectively, the angular separation $\rho$ and position
angle $\varphi$ are defined by
\begin{eqnarray}
\quad \cos \rho              &=& 
\sin \delta_1 \sin \delta_2 + \cos(\alpha_2-\alpha_1) \cos \delta_1 \cos \delta_2 \\
\quad \sin \varphi \sin \rho &=& 
\sin(\alpha_2-\alpha_1) \cos \delta_2,
\end{eqnarray}
where $\varphi$ is measured from North to East.
For an ensemble of binary systems with a semi-major axis $a$, eccentricity $e$,
and a random orbit orientation, the mean angular separation is given by:
\begin{equation} \label{equation: projecteda}
\langle \rho \rangle = \frac{\pi}{4} \, \hat{a} \, \left( 1 + \frac{e^2}{2}\right)\,,
\end{equation}
where $\hat{a}$ is the projection of the semi-major axis $a$ at the distance of
the binary system \citep{leinert1993}. For an association with purely circular
orbits ($e = 0$) and for an association with a thermal eccentricity distribution
$f_{2e}(e) = 2e$ equation~\ref{equation: projecteda} reduces to:
\begin{equation} \label{equation: projecteda2}
\langle \rho_{e=0} \rangle = \frac{\pi \hat{a}}{4} \approx 0.79\, \hat{a}
\quad {\rm and} \quad
\langle \rho_{2e} \rangle \approx \frac{5\pi \hat{a}}{16} \approx 0.98\,
\hat{a}\,,
\end{equation}
respectively \citep[e.g.,][]{bahcall1985}. If the distance $D$ to the binary
system is known, an estimate for the semi-major axis can be obtained from the
measured value of $\rho$ and the equations above using $a=D\,\tan \hat{a}$.
Although equations~(\ref{equation: projecteda}) and~(\ref{equation:
projecteda2}) are only valid for a sample of targets, they are often used to
obtain an estimate of the semi-major axis $a$ of an individual binary with
angular separation $\rho$.

For a visual binary, the primary mass $M_1$ and companion mass $M_2$ may be
derived if the brightness of the components (usually at multiple wavelengths),
the distance to the binary system, the age and metallicity of each component,
and the interstellar extinction along the line-of-sight to each component are
known (see Section~\ref{section: isochrones}). Finally, the mass ratio $q \equiv
M_2/M_1$ of the binary system can be computed.

\subsection{Spectroscopic binary observables}

The projected orbital motion of a binary system onto the line-of-sight can be
measured through the Doppler shift of the spectral lines of the components of the
binary. Repeated measurements reveal the radial velocity curve of the primary
star. In several cases the spectral lines of the companion star are also
resolved, and the radial velocity curve of the companion may thus be observed as well.
The radial velocity of the primary $V_1(t)$ and companion $V_2(t)$ as a function
of time $t$ are usually expressed as
\begin{equation} \label{equation: radialvelocitycurve}
V_1(t) = K_1 \left( e \cos \omega + \cos (\mathcal{V}(t) - \omega)\right) 
\quad {\rm and } \quad 
V_2(t) = - \frac{K_2}{K_1} V_1(t),
\end{equation}
where $\mathcal{V}$ is the true anomaly of the binary system at time $t$. The
semi-amplitude of the radial velocity curve of the primary ($K_1$) and of
the companion ($K_2$) are given by
\begin{equation} \label{equation: radialvelocitycurve_k}
K_1 = \frac{M_2}{M_1+M_2}\frac{2\pi a \sin i}{P \sqrt{1-e^2}}
\quad {\rm and } \quad 
K_2 = \frac{M_1}{M_2} K_1,
\end{equation}
In Section~\ref{section: sb-w} we study the detectability of a binary orbit as a function of the eccentricity $e$ and the argument of periastron $\omega$. We will then use the quantities $L_1$ and $L_2$, which we define as
\begin{equation} \label{equation: L}
L_1 = K_1 \sqrt{1-e^2}
\quad {\rm and } \quad 
L_2 = K_2 \sqrt{1-e^2},
\end{equation}
which are independent of both $e$ and $\omega$.

For single-lined spectroscopic binaries (SB1) the following observables can potentially be derived from the shape of the radial velocity curve: $K_1$, $P$, $e$, $a_1\sin i$, and the mass function ${\mathcal F}(M)$, which is defined as
\begin{equation} \label{equation: massfunction}
\mathcal{F}(M) = \frac{\left( M_2 \sin i\right)^3}{\left( M_1+M_2\right)^2} = 1.035\times 10^{-7} K_1^3 P \left( 1-e^2 \right)^{3/2} \ \mbox{M}_\odot .
\end{equation}
The orbital elements can in practice be derived only if the radial velocity curve is sampled with at least several measurements per orbital period. If the radial velocity curve is under-sampled, binarity may be inferred from the radial velocity variations, but the orbital elements cannot be derived.

The mass ratio $q$ can be calculated from these observables if an assumption
about --- or a measurement of --- the inclination $i$ is made. There are two
cases for which all orbital parameters of a binary system can be observed
directly: (1) an eclipsing binary, and (2) a binary system which is both
spectroscopically and astrometrically resolved. 

In several cases, spectral lines of both primary and companion are detected in
the spectrum of a binary system. In this case, binaries are classified as
double-lined spectroscopic binaries (SB2). More information can be derived for
the orbit of an SB2 system than that of an SB1 system, as both radial velocity
curves are observed. For SB2 systems the observables are
$K_1$, $K_2$, $P$, $e$, $a_1 \sin i = a M_2 \sin i / (M_1+M_2)$, $a_2 \sin i$,
$M_1 \sin^3 i$, and $M_2 \sin^3 i$. From these parameters the mass ratio can be
calculated as $q = K_1 / K_2$.

\subsection{Astrometric binary observables}

For an astrometric binary the presence of a companion is derived from the
projected orbital motion on the plane of the sky. In several cases a ``wobble''
of the primary star due to the orbital motion reveals binarity, in other cases
the companion is actually seen orbiting the primary star. For an astrometric
binary, the position and projected velocity of the primary and/or companion can
be measured as a function of time.

Astrometric observations are very useful in the sense that potentially six
orbital parameters can be resolved: the projected semi-major axis $\hat{a}$, the
eccentricity $e$, the absolute value of the inclination $|i|$, the reduced
position angle of the ascending node $\Omega^*$, the argument of periastron
$\omega$, and the time of periastron passage.

If the distance $D$ to the binary system is known, the true semi-major axis
$a=D\tan \hat{a}$ can be calculated. Due to projection effects, it is not
possible to obtain the sign of $i$ and the true value of $\Omega$ from
astrometric observations alone, unless the observations are combined with radial
velocity measurements. 
Orbits with $0^\circ < i < 90^\circ$ are called prograde if the position
angle of the companion increases with time, and retrograde if the position angle
decreases with time.

\subsection{Photometry} \label{section: isochrones}

To simulate photometric data we obtain the magnitude of each
star in the $UBVRIJHK_S$ bands using the isochrones described in
\cite{kouwenhoven2005}. These isochrones consist of models from
\cite{chabrier2000} for $0.02~\mbox{M}_\odot \leq M < 1~\mbox{M}_\odot$,
\cite{palla1999} for $1~\mbox{M}_\odot \leq M < 2~\mbox{M}_\odot$, and
\cite{girardi2002} for $M > 2~\mbox{M}_\odot$.  We derive the {\em Hipparcos}
magnitude $H_p$ for each object from its $V$ magnitude and $V-I$ color, using
the tabulated values listed in the {\em Hipparcos} Catalog (ESA 1997, Vol.~1,
\S~14.2). 

We convert the absolute magnitude in each filterband to apparent magnitude using
the distance to each individual binary system. In our analysis we neglect the
interstellar extinction. OB~associations are (by definition) practically cleared
of gas. For many (but not all) nearby associations the interstellar extinction
is therefore very small (e.g., for Sco~OB2).

% ====================================================================
% ====================================================================
% ====================================================================
% ==STANDARD MODEL======================================================
% ====================================================================
% ====================================================================
% ====================================================================

\section{Association and binary population properties} \label{section: associationandbinarypopulationproperties}

In order to study how the properties of an association and its binary population
are reflected in the observations and the interpretation thereof we define a
``default model''. The latter is a simulated association with a binary
population that approximates the binary population in real associations.
Throughout this paper we will change one or more parameters in the default
model, while keeping the others fixed to those listed in Tables~\ref{table:
modelcluster} and~\ref{table: modelpopulation}. We study the effect of these
changes on the observed binary parameters, and identify the parameters that can
and cannot be constrained using the various observational techniques. In the
remaining part of this section we motivate the choice of the default model
parameters.

\begin{table}[!tbp]
  \begin{tabular}{lcl}
    \hline
    Quantity                 & Symbol & Value \\
    \hline
    Number of single stars   & $S$ & 0\\
    Number of binary systems & $B$ & 10\,000\\
    Density distribution     & DF & Homogeneous, $R=20$~pc \\
    Distance                 & $D$ & 145~pc\\
    Age                      & $\tau$ & 5~Myr\\
    \hline
  \end{tabular}   
  \caption{Association properties of the ``default model''. Together with the
  binary parameter distributions listed in Table~\ref{table: modelpopulation},
  these constitute the reference model for the OB~association simulations in
  this paper. The density distribution, characteristic radius, distance, and age
  are typical for the nearby Sco~OB2 association. The number of
  (stellar and substellar) systems $S+B$ is similar to that of Sco~OB2. The
  binary fraction $F_{\rm M} = B/(S+B)$ is set to 100\% and is lowered when
  appropriate. \label{table: modelcluster}}
\end{table}

\begin{table}[!tbp]
  \begin{tabular}{p{3.5cm}p{0.1cm}lcc}
    \hline
    Quantity  &  & Distribution & Min & Max \\
    \hline
    Single/Primary mass       & $M$  & Preibisch, $\alpha=-0.9$ & 0.02 M$_\odot$ & 20 M$_\odot$ \\
    Mass ratio                & $q$    & $f_q(q)=q^{-0.33}$, PCP-III & $M_2 \geq 0.02~\mbox{M}_\odot$  & 1    \\
    Semi-major axis           & $a$    & $f_a(a)=1/a$     & $15$ R$_\odot$ & $10^6$ R$_\odot$ \\
    Eccentricity              & $e$    & $f_e(e)=2e$ (thermal)    & 0 & 1 \\
    Inclination               & $i$    & $f_{\rm \cos i}(\cos i) ={\rm flat}$  & $-90^\circ$ & $\ 90^\circ$ \\
    Argument~of~periastron    & $\omega$& $f_\omega(\omega)={\rm flat}$  & $\ \ \ 0^\circ$ & $360^\circ$ \\
    Angle of ascending node   & $\Omega$& $f_\Omega(\Omega)={\rm flat}$  & $\ \ \ 0^\circ$ & $360^\circ$ \\
    Mean anomaly              & ${\mathcal M}$ & $f_{\mathcal M}({\mathcal M})={\rm flat}$       & $\!-180^\circ$ & $180^\circ$ \\
    \hline
  \end{tabular}   
  \caption{Binarity properties for the model which we will refer to as the
  ``default model''. Throughout this paper we will study the effect of changes
  in the above quantities on the observed binary population. The mass
  distribution, defined by equation~\ref{equation: sos_preibischimf}, is the most
  accurate determination for OB associations to date. The mass ratio
  distribution is a power-law, where the exponent $\gamma_q$ is taken from
  \cite{kouwenhoven2005}. We use \"{O}piks law for the semi-major axis
  distribution (a flat distribution in $\log a$), and a thermal eccentricity
  distribution. Orbits are assumed to have a random orientation and orbital
  phase. \label{table: modelpopulation}}
\end{table}

\subsection{Choice of binary fraction}

Throughout this article we let $N=S+B$ be the number of systems in an association, where $S$ is the number of single stars and $B$ is the number of binary systems. We assume that there are no triple systems and higher-order multiples present in the association. In the default model we let $S=0$ and $B=10\,000$. The value of $N$ is chosen here to be of the same order as the number of systems in the Sco~OB2 association\footnote{\cite{preibisch2002} estimate the number of stars in the Upper Scorpius subgroup of Sco~OB2 to be 2525. Including systems with a brown dwarf primary, this number increases with $\sim 30\%$ (see Table~\ref{table: massfractions}). Assuming the two other subgroups of Sco~OB2 contain a similar number of members, we estimate the number of systems in Sco~OB2 to be of order 10\,000.}.

The multiplicity fraction $F_{\rm M} = B/(S+B)$ is set to 100\% in the default association. As no triples and higher-order multiples are present in the association, we will often refer to $F_{\rm M}$ as the binary fraction. The {\em observed} binary fraction in OB~associations is of the order 50\%. This is a lower limit to the {\em true} binary fraction, as many binary systems remain unresolved. Several authors have suggested that the binary fraction in OB~associations may be close to 100\%, hence our choice for $F_{\rm M}$ in the default model.

In the context of star formation, the number of binary systems relative to the number of single stars is the most important parameter. To zeroth order, this ratio describes the binary population. Other properties of the binary population, such as the distributions over mass ratio, period, and eccentricity, provide further information. For these distributions the number of binary systems are merely a normalization of the distribution functions. However, statistical errors become increasingly important for stellar populations with smaller $B$.

\begin{figure}[!tbp]
  \centering
  \includegraphics[width=1\textwidth,height=!]{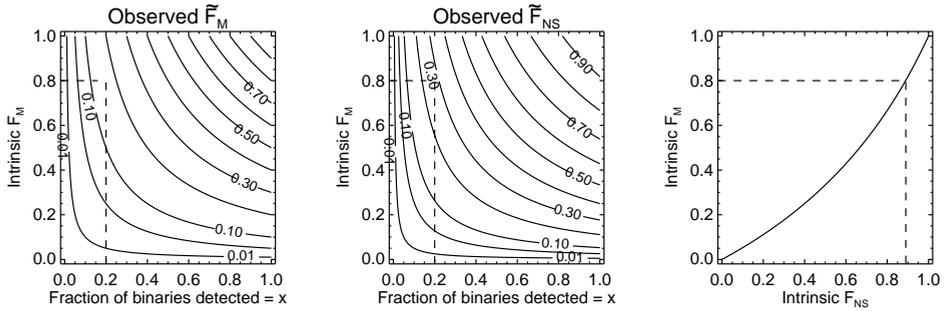}
  \caption{The relation between the intrinsic multiplicity fraction $F_{\rm M}$  and non-single star fraction $F_{\rm NS}$ and the observed fractions $\tilde{F}_{\rm M}$ and $\tilde{F}_{\rm NS}$ (see equations~\ref{equation: multiplicityfractions} and~\ref{equation: multiplicitylimits}). The ratio between the number of observed binaries and the true number of binaries is denoted with $x$. No triples or higher-order multiples are assumed to be present. The contours in the left and middle panels show  $\tilde{F}_{\rm M}$ and $\tilde{F}_{\rm NS}$ as a function of  $F_{\rm M}$ and $x$. Each contour is labeled with a value of $\tilde{F}_{\rm M}$ and $\tilde{F}_{\rm NS}$, respectively.
The curve in the right panel shows the relation between $F_{\rm M}$ and $F_{\rm NS}$. The dashed lines indicate an example, where $F_{\rm M}=0.8$ and $x=0.2$. In this case, $F_{\rm NS}=0.89$,  $\tilde{F}_{\rm M}=0.16$, and $\tilde{F}_{\rm NS}=0.28$. 
 \label{figure: observed_intrinsic_fm} }
\end{figure}

Three common expressions to quantify the multiplicity of a stellar population are the multiplicity fraction $F_{\rm M}$, the non-single star fraction $F_{\rm NS}$ and the companion star fraction $F_{\rm C}$ \citep{kouwenhoven2005}, which are defined as
\begin{equation} \label{equation: multiplicityfractions}
  F_{\rm M}       = \frac{ B+T+\dots   }{  S+B+T+\dots } \quad\ 
  F_{\rm NS}      = \frac{ 2B+3T+\dots }{  S+2B+3T+\dots} \quad\ 
  F_{\rm C}       = \frac{ B+2T+\dots  }{  S+B+T+\dots} \,,
\end{equation}
where $S$ is the number of single stars, $B$ the number of binaries, and $T$ the number of triple systems.
In this paper we do not consider triple and higher-order systems. In this case we have
$F_{\rm M}=F_{\rm C}$. The total number of
(individual) stars is $S+2B = N (F_{\rm M} + 1)$.
Let $x$ be the fraction of binary systems that are actually detected. In this
case, the detected number of binaries is $xB$ and the detected number of single
stars (including the unresolved binaries) is $S+(1-x)B$. The {\em observed}
multiplicity fraction $\tilde{F}_{\rm M}$ and  {\em observed} non-single star
fraction $\tilde{F}_{\rm NS}$ are then
\begin{equation} \label{equation: multiplicitylimits}
\tilde{F}_{\rm M} = x  F_{\rm M} \quad \quad   \tilde{F}_{\rm NS} = \frac{2xB}{S+(1+x)B} = \frac{2}{\tilde{F}_{\rm M}^{-1} + 1}.
\end{equation}
Figure~\ref{figure: observed_intrinsic_fm} shows the dependence of $\tilde{F}_{\rm M}$ and $\tilde{F}_{\rm NS}$ on $x$ and $F_{\rm M}$.

\subsection{Choice of association distance, size, and age}

The association properties that are of importance are its stellar content (number of singles, binaries, and multiples), the distance to the association $D$, the stellar density distribution $\mbox{DF}$ (in particular the radius $R$), and the age $\tau$ and metallicity of the association, and the Galactic coordinates of the association.

The distance $D$ to the association is a parameter that very strongly determines
whether a binary system is detected visually, spectroscopically, or
astrometrically. The separation between two binary components on the plane of the sky, 
and their corresponding projected velocities scale with $D^{-1}$. 
The brightness of each star scales with $D^{-2}$. Since the stellar density
distribution and characteristic radius of an association determine the locations of the individual
stars with respect to the center of the association, these parameters are of
(mild) importance to the distributions of angular size, angular velocity, and
stellar brightness. Crowding becomes more important for associations at larger distance,
and depends on the density distribution and the observational limitations. For
the default model we adopt a distance of 145~pc and a homogeneous stellar
density distribution with a radius of 20~pc, values typical for nearby
OB~associations \citep[e.g.,][]{dezeeuw1999}. The homogeneous-sphere approximation
is appropriate as OB~associations do not have a significant central
concentration. 

We adopt a solar metallicity and an age of 5~Myr, parameters that are typical
for young OB~associations \citep[see, e.g.,][]{brown1999}. The color and
magnitude of a star are related to its mass through the isochrone of the
OB~association. The isochrone adopted in this paper is described in
Section~\ref{section: isochrones}.  Since the luminosity of a star depends only
mildly on the age for young OB~associations we will not model associations with
a different age. For a discussion of the photometric evolution of star clusters
we refer to \cite{ecology4}.

In reality, observable properties of an OB~association additionally depend on
the position of the association on the sky, for several reasons. For the
analysis of an association, it is important to know which stars belong to the
association, and which stars are background stars. Membership is easier to
determine for an association at high Galactic latitude than for targets near the
Galactic plane. Also, the confusion between companion stars and
background stars in an imaging survey for binarity increases when the
association is closer to the Galactic plane. The coordinates of the association
therefore impose a maximum separation at which companion stars can be reliably
distinguished from background stars. The interstellar extinction may vary
strongly with position \citep[e.g.,][]{shatsky2002}, which introduces an
additional uncertainty. In this paper we do not study the effects of incomplete
knowledge of membership and interstellar extinction. We (indirectly) take into
account the confusion between physical companion stars and  background stars,
although we do not model the dependence on the coordinates of the association.
The actual coordinates of the association are therefore not of relevance for the
purpose of this paper.

\subsection{Choice of mass distribution} \label{section: massdistribution}

\begin{figure}[!tbp]
  \centering
  \includegraphics[width=0.8\textwidth,height=!]{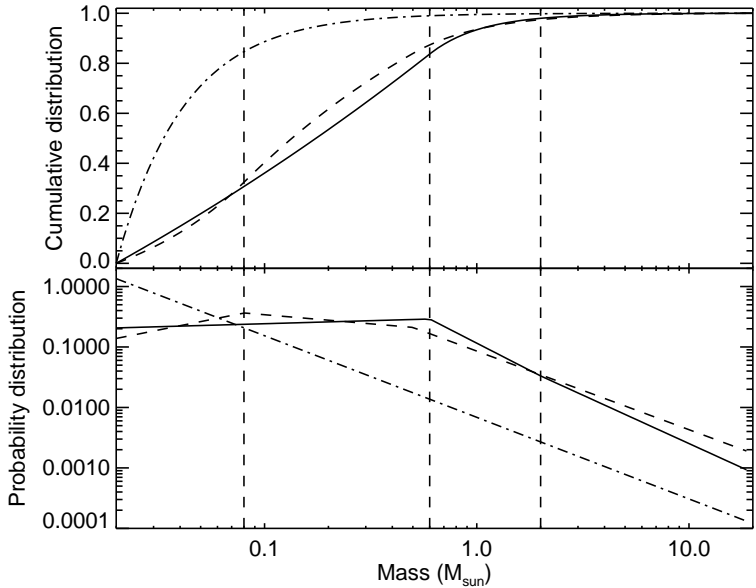}
  \caption{The cumulative mass distributions ({\em top}) and mass distributions
  ({\em bottom}) used in our analysis. We use an extended version of the
  Preibisch mass distribution as defined in equation~\ref{equation:
  sos_preibischimf} with $\alpha=-0.9$ as a default value (solid curves). We study the
  effect of the mass distribution on observations using the Salpeter mass
  distribution (equation~\ref{equation: salpeterimf}; dash-dotted curve) and the
  Kroupa mass distribution (equation~\ref{equation: kroupaimf}; dashed curve;
  truncated above $20~$M$_\odot$). The three distributions have the same
  normalization. The points at which the slope of the extended Preibisch
  mass distributions changes are indicated with the vertical dashed lines.
 \label{figure: sos_massdistributions} }
\end{figure}

\begin{table}
  \small
  \begin{tabular}{l r rr rr rr}
    \hline
    \hline
    Spectral type &   B (\%) & A (\%) & F (\%)& G (\%)& K (\%)& M (\%)& BD (\%)\\
    Mass range (M$_\odot$) &  3-20 & 1.5-3 & 1-1.5 & 0.8-1 & 0.5-0.8 & 0.08-0.5 & 0.02-0.08\\
    \hline
    Kroupa                     &  1.44 &  2.31  & 2.70  & 2.22  & 7.41  & 51.50   & 32.41 \\
    Preibisch, $\alpha=-0.9$   &  1.03 &  2.24  & 3.33  & 3.18  &11.76  & 47.73   & 30.72 \\
    Preibisch, $\alpha=2.5$    &  1.36 &  2.95  & 4.38  & 4.18  &15.48  & 62.80   &  8.85\\
    Preibisch, no BDs          &  1.49 &  3.24  & 4.81  & 4.59  &16.98  & 69.89   & $-$ \\
    Salpeter                   &  0.11 &  0.18  & 0.21  & 0.18  & 0.61  & 14.01   & 84.62  \\
    \hline
    \hline
  \end{tabular}
\caption{The fraction of stars (in per\,cent) for each spectral type for several mass distributions given in column~1 (see text). Columns~$2-8$ list the fraction of stars for each spectral type. The mass ranges (in $\mbox{M}_\odot$) for each spectral type are listed in the second row, and are approximate.  \label{table: massfractions}}
\end{table}

In this section we discuss our choice of the mass distribution, followed by a
discussion in \S~\ref{section: pairingfunction} on how we choose the primary and
companion masses from the mass distribution. Briefly summarized, in the default
model we draw the primary mass from an extended version of the Preibisch mass
distribution (equation~\ref{equation: sos_preibischimf}). We draw the mass ratio
from the mass ratio distribution $f_q(q) \propto q^{-0.33}$, which represents the mass
ratio distribution for intermediate mass stars in the nearby Sco~OB2 association
\citep{kouwenhoven2005}. We impose a minimum companion mass of
$0.02~\mbox{M}_\odot$ (pairing function PCP-III; see \S~\ref{section:
pairingfunction}), which roughly corresponds to the deuterium burning limit and
separates planetary companions from brown dwarf companions. We adopt a maximum
mass of $20~\mbox{M}_\odot$, which is typically the upper mass limit of the
present day mass function for $5-20$~Myr old OB~associations.

The mass distribution $f_M(M)$ defines the spectrum of masses in a stellar population, and is usually expressed as a multi-component power law. Throughout this paper we will consider several mass distributions and analyze the main differences in an observational context. 

\cite{preibisch2002} provide the most extensive description to date of the
(single star) mass distribution in OB~associations. They studied the stellar
population of the US subgroup of Sco~OB2. They combine their observations of
pre-main sequence stars with those of \cite{preibisch1999} and
\cite{dezeeuw1999} and derive an empirical mass distribution in the mass range
$0.1~\mbox{M}_\odot \leq M \leq 20~\mbox{M}_\odot$. As we study binarity in
OB~associations in this paper, we use the Preibisch mass distribution as
default. We also study the effect of brown dwarfs on the interpretation of the
observations, and therefore we extrapolate the Preibisch MF
\citep[see][]{kouwenhoven2005} down to the brown dwarf regime:
\begin{equation} \label{equation: sos_preibischimf}
  f_{\rm Preibisch}(M) \propto \left\{
  \begin{array}{llll}
    M^\alpha  & {\rm for \quad } 0.02 & \leq M/{\rm M}_\odot & < 0.08 \\
    M^{-0.9}  & {\rm for \quad } 0.08 & \leq M/{\rm M}_\odot & < 0.6 \\
    M^{-2.8}  & {\rm for \quad } 0.6  & \leq M/{\rm M}_\odot & < 2   \\
    M^{-2.6}  & {\rm for \quad } 2    & \leq M/{\rm M}_\odot & < 20 \\
  \end{array}
  \right. \,.
\end{equation}
The slope $\alpha$ of the power law for substellar masses is poorly constrained by observations. \cite{preibisch2003} give an overview of the slope $\alpha$ in different stellar populations, where $\alpha$ ranges from $-0.3$ for the Galactic field \citep{kroupa2002} to $+2.5$ for the young open cluster IC~348 \citep{preibisch2003}. A steeper slope of $\alpha=-0.6$ is found by \cite{moraux2003} for the Pleiades cluster. The mass distribution in the brown dwarf regime has not been reliably determined to date for OB~associations. Free-floating brown dwarfs are known to be present in OB~associations \citep[e.g.,][]{martin2004,slesnick2006}. We therefore choose to extrapolate the Preibisch mass distribution into the brown dwarf regime down to $0.02~\mbox{M}_\odot$, i.e., $\alpha=-0.9$. 
It must be noted that, although $\alpha$ has not been constrained observationally for OB~associations, its value can have a pronounced effect on the observed mass ratio distribution, depending on the pairing algorithm used to pair primaries and companions (see \S~\ref{section: pairingfunction}).

Later in our analysis we additionally consider the Salpeter mass distribution and the Kroupa mass distribution.
The classical Salpeter mass distribution \citep{salpeter1955} is defined as
\begin{equation} \label{equation: salpeterimf}
f_{\rm Salpeter}(M) \propto M^{-2.35}
\end{equation}
This mass distribution is derived for massive stars in the Galactic field. Although it is known that the Salpeter mass distribution is incorrect for masses below $\sim 1~\mbox{M}_\odot$, we use this classical mass distribution for comparison with other mass distributions. 
The universal mass distribution derived by \cite{kroupa2001} is given by
\begin{equation} \label{equation: kroupaimf}
  f_{\rm Kroupa}(M) \propto \left\{
  \begin{array}{llll}
    M^{-0.3}  & {\rm for \quad } 0.01 & \leq M/{\rm M}_\odot & < 0.08 \\
    M^{-1.3}  & {\rm for \quad } 0.08 & \leq M/{\rm M}_\odot & < 0.5   \\
    M^{-2.3}  & {\rm for \quad } 0.5  & \leq M/{\rm M}_\odot & < \infty   \\
  \end{array}
  \right. .
\end{equation}
For most stellar populations the mass distribution can be described accurately with the Kroupa universal mass distribution. 
The three mass distributions described above are plotted in Figure~\ref{figure: sos_massdistributions}. The figure shows the similarity between the extended Preibisch and Kroupa mass distributions. Table~\ref{table: massfractions} lists the fraction of stars of a given spectral type for the three mass distributions described above.

\subsection{Choice of pairing function and mass ratio distribution} \label{section: pairingfunction}

The primary mass $M_1$ and companion mass $M_2$ in a binary system are related
through the mass ratio $q \equiv M_2/M_1$. We define the primary star always as
the most massive (which is usually the most luminous) in a binary system, so
that $0<q\leq 1$.  We describe the relationship between the primary mass and
companion mass for a group of binaries as the ``pairing function'' of that
group. In Section~\ref{section: simulations_massdistribution_pairingfunction} we
identify five commonly used pairing functions, which we refer to as RP (random
pairing), PCRP (primary-constrained random pairing), PCP-I (primary-constrained pairing type~I), PCP-II (primary-constrained pairing type~II), and PCP-III (primary-constrained pairing type~III). The latter three pairing functions
are associated with a mass ratio distribution $f_q(q)$. Each of these pairing
functions generally give significantly (and measurably) different results. In
any population synthesis paper where the authors use a mass ratio distribution,
it should therefore be clear which pairing relation is used.

For our default model we adopt pairing function PCP-III 
(see \S~\ref{section: pairingfunction3}), with a minimum companion 
mass of $M_{\rm 2,min} = 0.02$~M$_\odot$, and a mass ratio distribution of the form $f_q(q) \propto
q^{-0.33}$. The choice for these properties is made based on the observations of
\cite{shatsky2002} and \cite{kouwenhoven2005}. In the latter paper it is shown that
the observed mass ratio distribution for A~and late~B stars in Sco~OB2 is
consistent with $f_q(q) \propto q^{-0.33}$. \cite{shatsky2002} find that the mass
ratio distribution for B~stars in Sco~OB2 has a roughly similar behaviour: $f_q(q) \propto
q^{-0.5}$. In section~\ref{section: simulations_massdistribution_pairingfunction} 
we will show that these observations are inconsistent with pairing functions RP and PCRP. 

The mass ratio distribution $f_q(q)$ is usually obtained by fitting a function
with one or more free parameters to the observed mass ratio distribution for a
sample of stars. Often, the observed distribution is assumed to be a power-law,
\begin{equation} \label{equation: q_powerlaw}
f_{\gamma_q}(q) \propto q^{\gamma_q} \quad \mbox{for} \quad 0 \leq q \leq 1,
\end{equation}
and the exponent $\gamma_q$ is fitted. Distributions with $\gamma_q =0$ are
flat, while those with $\gamma_q < 0$ and $\gamma_q > 0$ are falling and rising
with increasing $q$, respectively. Usually the adopted minimum value for the fit
$q_0$ is the value of $q$ below which the observations become 
incomplete or biased. For distributions with $\gamma_q \leq -1$, a minimum value
$q_0 > 0$ is necessary to avoid a diverging mass ratio distribution at $q=0$.
Sometimes the necessity of $q_0$ is avoided by fitting a mass ratio distribution
of the form
\begin{equation} \label{equation: 1+q_powerlaw} f_{\Gamma_q}(q) \propto (1+q)^{\Gamma_q}
  \quad \mbox{for} \quad 0 \leq q \leq 1 \end{equation}
to the data. This distribution is more commonly used to describe the mass
ratio distribution of high-mass spectroscopic binaries, while the distribution
in equation~\ref{equation: q_powerlaw} is often used for visual binaries. As equations~\ref{equation:
q_powerlaw} and~\ref{equation: 1+q_powerlaw} show a similar behaviour (both are
either falling, flat, or rising), we will only consider the distribution of
equation~\ref{equation: q_powerlaw} in this paper. 

To allow for a peak in the mass ratio distribution in the range $0 < q < 1$, we 
consider the Gaussian mass ratio distribution:
\begin{equation} \label{equation: q_gaussian}
f_{\rm Gauss}(q) \propto \exp \left\{ - \frac{(q-\mu_q)^2}{2\sigma_q} \right\} \quad \mbox{for} \quad 0 \leq q \leq 1.
\end{equation}
Models with $\mu_q < 0$ and $\mu_q > 1$ show a falling, respectively rising distribution $f_{\rm Gauss}(q)$. When $\mu_q \ll 0$ and $\mu_q \gg 1$, the distribution may be approximated by a power-law (equation~\ref{equation: q_powerlaw}). Models with $\sigma_q \gg 0.5$ may be approximated with a flat mass ratio distribution. Note that the values of
$\mu_q$ and $\sigma_q$ do not necessarily reflect physical properties.
A value $\mu_q \ll 0$, for example, merely means that the mass ratio distribution in the
interval $0 < q \leq 1$ can be described by equation~\ref{equation: q_gaussian} in this
interval.

Note that the choices made for the default model do not imply that
OB~associations indeed have these properties. It is not known how binary stars are formed, so that no reliable predictions of their properties can be made. Different
binary formation mechanisms may produce different mass ratio distributions,
possibly varying with primary mass, period, or eccentricity.  In addition
dynamical evolution (during the formation process) may alter the mass and mass
ratio distribution \citep[e.g., embryo ejection;][]{reipurth2001}.

% ---------------------------------------------------------------------------------------
% ---------------------------------------------------------------------------------------
% ---------------------------------------------------------------------------------------

\subsection{The semi-major axis and period distribution} \label{section: smaandperiod}

\begin{figure}[!tbp]
  \centering
  \includegraphics[width=0.8\textwidth,height=!]{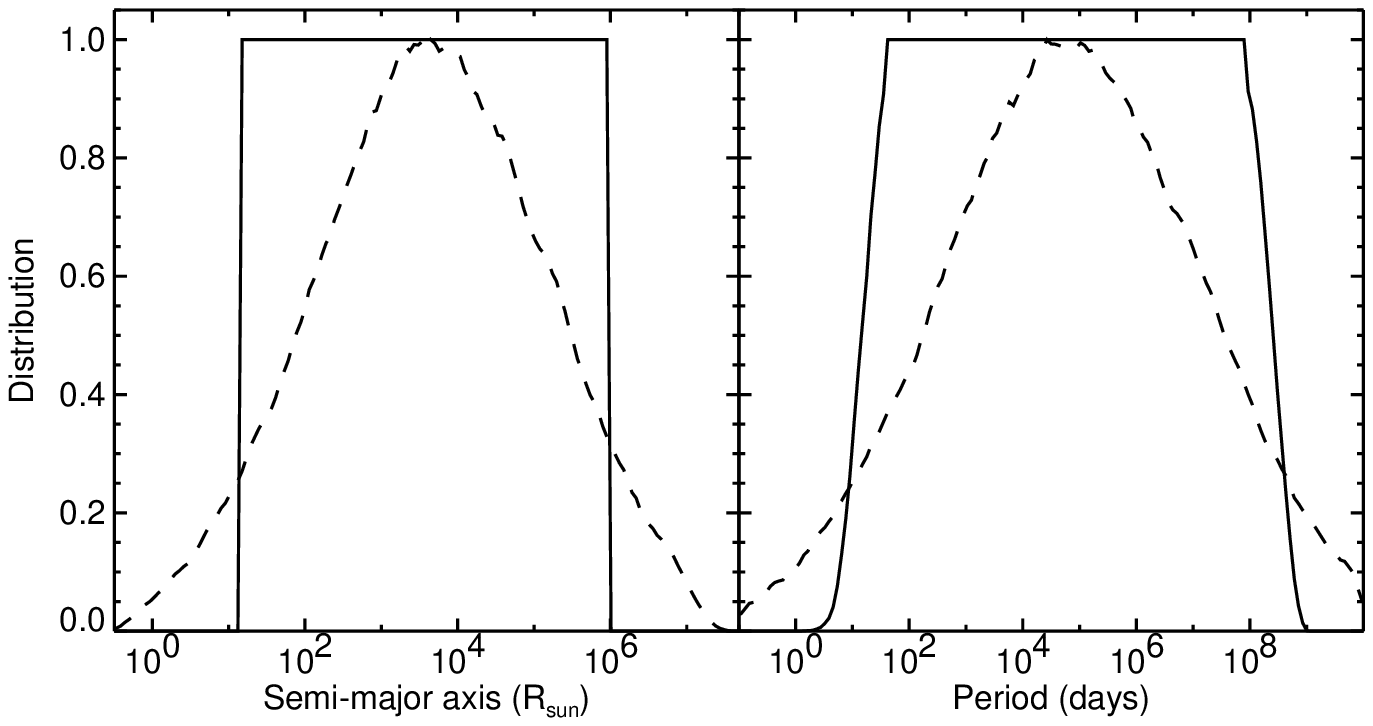}\\
  \includegraphics[width=0.8\textwidth,height=!]{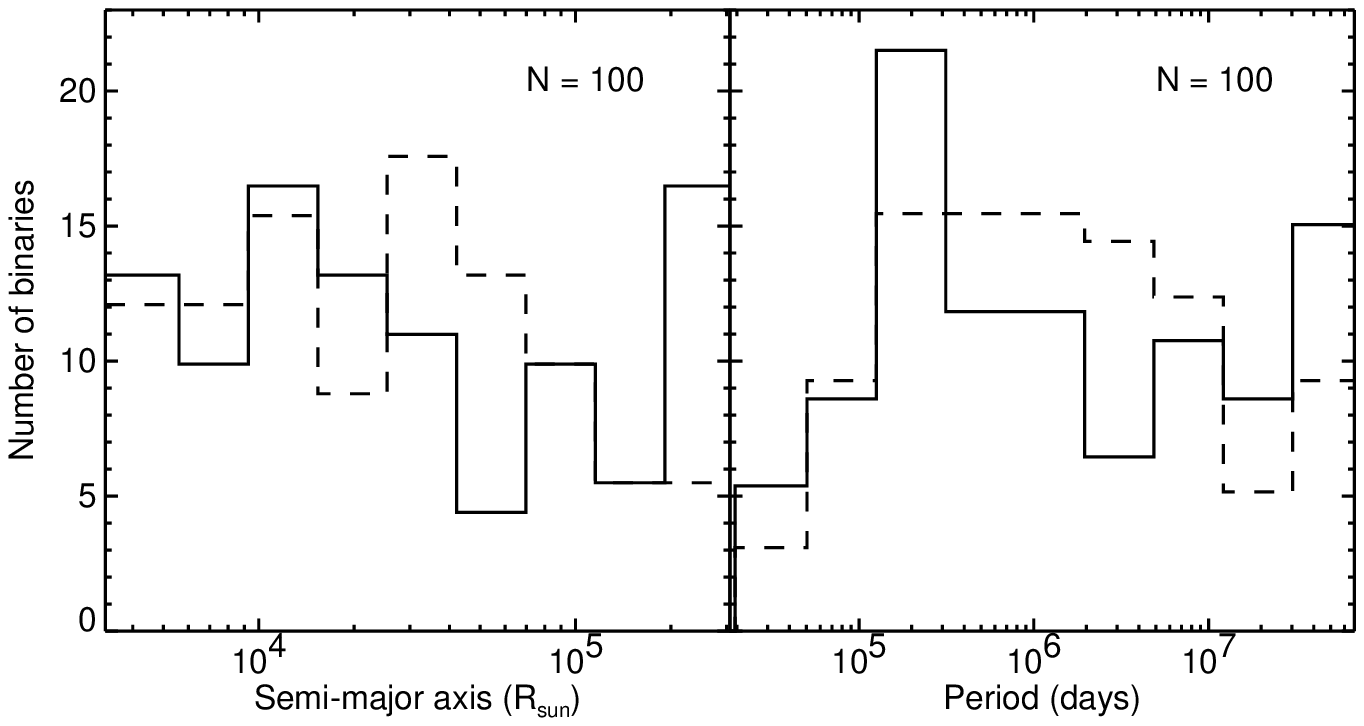}
  \caption{The semi-major axis distribution $f_a(a)$ and the period distribution $f_P(P)$, resulting from \"{O}pik's law (solid curves) and the log-normal period distribution (dashed curves) found by \cite{duquennoy1991}. {\em Top:} both models, each consisting of 100\,000 binaries, with a Preibisch mass distribution with $\alpha=-0.9$, and pairing function PCP-III. In general, a flat distribution in $\log a$ results in an approximately flat distribution in $\log P$, and vice versa. The same holds approximately for a Gaussian distribution in $\log a$ or $\log P$. No selection effects have been applied. The above statements are approximately valid, as the results depend mildly on the total mass distribution $f_{M_T}(M_T)$ of the binary systems. This effect is for example visible in the distribution $f_P(P)$ resulting from \"{O}pik's law (top-right panel; solid curve), near the edges of the distribution. The distributions are normalized so that their maximum equals unity. {\em Bottom:} the results for a simulated set of 100~binaries with $3\,200$~R$_\odot < a < 320\,000~$R$_\odot$. These are the typical limits that can be obtained with an imaging survey for binarity for the nearest OB~associations ($0.1'' < \rho < 10''$). Even though the top panels show a clear difference between  \"{O}pik's law and the log-normal period distribution, the differences in the bottom panel are not significant.
  \label{figure: fa_fp_difference} }
\end{figure}

\begin{figure}[!tbp]
  \centering
  \includegraphics[width=1\textwidth,height=!]{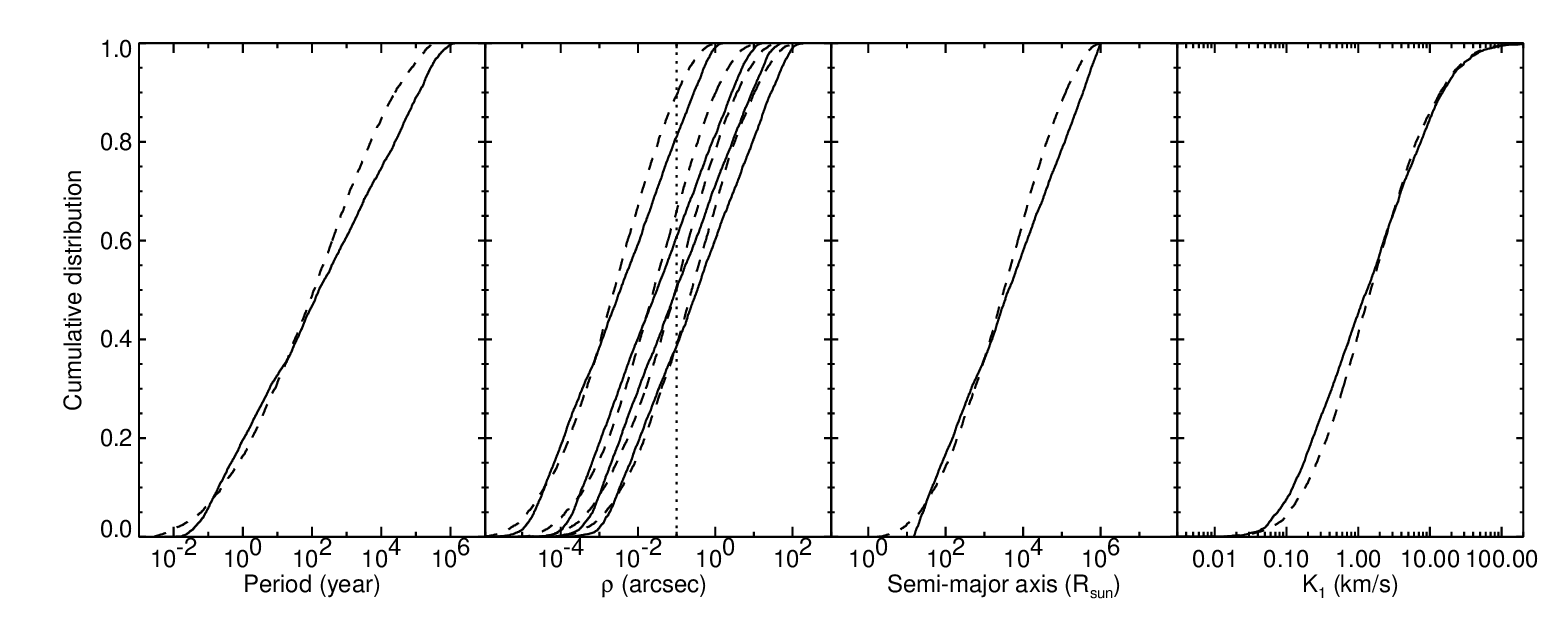}
  \caption{The differences and similarities between \"{O}piks law, with $f_a(a)
  \propto a^{-1}$ (solid curves) and the log-normal period distribution of
  \cite{duquennoy1991} (dashed curves). Model associations are created at
  distances of 50, 150, 500, and 5000~pc (from left to right in the second
  panel). Primaries are drawn from the extended Preibisch mass distribution with
  $\alpha=-0.9$; the mass ratio of each binary system is drawn from $f_q(q)
  \propto q^{-0.33}$ (PCP-III). The typical minimum angular separation 
  for visual binary surveys (0.1~arcsec) 
  is indicated with the vertical dotted line.
    \label{figure: opik_dm_difference} }
\end{figure}

The motion of the stars in the orbital plane can be described by the total mass of the binary system $M_T= M_1+M_2$, the semi-major axis $a$ and eccentricity $e$ of the orbit. In this paper we construct the orbit by choosing the stellar mass (see~\S~\ref{section: massdistribution}), the eccentricity, and the semi-major axis independently from distribution functions: 
\begin{equation}
f_{M_T,a,e}(M_T,a,e) = f_{M_1+M_2}(M_1+M_2)f_a(a)f_e(e) \,.
\end{equation}
As the semi-major axis $a$ and the period $P$ are related through Kepler's third law,
\begin{equation} \label{equation: keplerslaw}
  P^2 = \left( \frac{4\pi^2}{GM_T} \right) a^3 \,,
\end{equation}
it is also possible to generate orbits by drawing values of $f_P(P)$ instead of
$f_a(a)$. However, due to a correlation with the total mass of the system, this
may give a different result. Alternatively,
it is possible to choose the orbital energy $E$ and angular momentum $L$ from
$f_E(E)$ and $f_L(L)$ instead of (indirectly) from $f_a(a)$ and $f_e(e)$.

In our default model we use a semi-major axis distribution of the form:
\begin{equation} \label{equation: opikslaw}
f_a(a) \propto a^{-1} \quad \quad a_{\rm min} \leq a \leq a_{\rm max} \,,
\end{equation}
which is equivalent to $f_{\log a}(\log a) =~\mbox{constant}$. This distribution, commonly known as \"{O}pik's law, has been observed for a wide range of stellar populations \citep[e.g.,][]{vanalbada1968,vereshchagin1988,poveda2004}. After the semi-major axis of a binary system is drawn, its period is derived using Kepler's third law.

A value of $15~$R$_\odot$ for the minimum semi-major axis $a_{\rm min}$ is adopted, so that close spectroscopic binaries with a period of the order of a day are also included in the model. A physical minimum semi-major axis $a_{\rm min}$ is defined by the stellar radius, or more precisely, the semi-major axis at which one of the binary components fills its Roche lobe. In this paper we therefore set the value of $a_{\rm min}$ to either $15~$R$_\odot$ or the semi-major axis at which Roche lobe overflow starts, depending on which of the two is larger. As a result of this choice for $a_{\rm min}$, the shortest possible period ranges between a day for the most massive binaries and a month for the lowest-mass brown-dwarf binaries.

The maximum semi-major axis $a_{\rm max}$ is set by the argument of tidal disruption of
binary systems. In the Galactic field the maximum observed semi-major axis is of
the order of $\sim 0.1$~pc \citep[$5\times
10^6$~R$_\odot$][]{close1990,chaname2004}. The analysis of \cite{close1990}
shows that $\sim 3\%$ of the Galactic disk binaries has a separation larger than 0.01~pc
($0.5\times 10^6$~R$_\odot$). OB~associations are expanding groups and are
likely unbound \citep{blaauw1964A,brown1999}, so that they will dissolve in the
field star population within a few tens of Myr. As a significant fraction of the
field stars is expected to be born in OB~associations, we set the upper limit
$a_{\rm max} = 10^6$~R$_\odot$.

We additionally simulate associations with the log-normal period distribution
found by \cite{duquennoy1991}, who performed a multiplicity
study among solar-type stars in the solar neighborhood. They find a log-normal
period distribution of the form:
\begin{equation} \label{equation: duquennoyperiods}
f_{\log P}(\log P) \propto \exp \left\{ - \frac{(\log P - \overline{\log P})^2 }{ 2 \sigma^2_{\log P} }  \right\},
\end{equation}
where $\overline{\log P} = 4.8$, $\sigma_{\log P} = 2.3$, and $P$ is in days.
In our association models, we use a lower limit of $P_{\rm min} = 1$~day, as
main sequence stars with periods shorter than $P_{\rm min}$ generally have a
semi-major axis smaller than 2.5~R$_\odot$. We impose a maximum orbital period
of $P_{\rm max} = 1.1\times 10^8$~days (0.3~Myr). Orbits with a period $P_{\rm
max}$ have a semi-major axis of $\sim 10^6$~R$_\odot$, the maximum value adopted
for \"{O}pik's law, for reasons mentioned above.  Figure~\ref{figure:
fa_fp_difference} shows the semi-major axis distribution and period distribution
corresponding to \"{O}pik's law and the log-normal period distribution. In
theory the models can easily be discriminated. In practice, this is more
difficult due to the limited range in $a$ or $P$ that can be studied, and the
finite number of observations. The bottom panels in Figure~\ref{figure:
fa_fp_difference} show the results for a simulated imaging survey among the nearby
OB~associations ($D \approx 145$~pc). The resulting observed semi-major axis
distributions (measured indirectly through $\rho$), and the observed period
distributions (not measured for visual binaries) are very similar for both
models, and cannot be discriminated. In \S~\ref{section: opikdm_difference} we
will discuss in detail under which conditions \"{O}pik's law and the log-normal
period distribution can be discriminated. 

Figure~\ref{figure: opik_dm_difference} shows the {\em cumulative} distributions $F_P(P)$, $F_\rho(\rho)$, $F_a(a)$, and $F_{K_1}(K_1)$ corresponding to \"{O}pik's law and the log-normal period distribution. The cumulative angular separation distribution is shown for association distances of 50, 150, 500, and 5000~pc. The latter figure clearly shows that \"{O}pik's law, corresponding to a flat distribution in $\log a$, also corresponds approximately to a flat distribution in $\log P$, in $\log \rho$, and in $\log K_1$ (see Appendix~D for further details).

We adopt \"{O}pik's law for our default model, with the limits $a_{\rm min}$ and $a_{\rm max}$. In Section~\ref{section: sma_and_period} we discuss the consequences of these choices, and of possible other choices for $f_a(a)$ and $f_P(P)$. In Section~\ref{section: opikdm_difference} we show which observations are necessary to determine whether the binary population in an OB~association follows \"{O}pik's law or the log-normal period distribution.

% ---------------------------------------------------------------------------------------
% ---------------------------------------------------------------------------------------
% ---------------------------------------------------------------------------------------

\subsection{The eccentricity distribution} \label{section: eccentricitychoice}

Until now there is no knowledge about the exact shape of the eccentricity distribution
$f_e(e)$ in OB~associations. Knowledge about $f_e(e)$ of binaries in the solar
neighborhood is also scarce. \cite{duquennoy1991} have shown that for solar-type
stars in the solar neighborhood the eccentricity distribution tends to be
thermal for binaries with a period larger than $\sim 3$~year. For binaries with
smaller orbital periods they find that the eccentricity distribution peaks at $e
\approx 0.3$. 

From theory there are several predictions for $f_e(e)$. The most common distribution used in simulations is the thermal eccentricity distribution, which has the form $f_{2e}(e) \propto 2e$ \citep{heggie1975}. The thermal eccentricity distribution results from energy equipartition, which is expected in relaxed stellar systems. 
In our default model we will adopt the thermal eccentricity distribution. In Section~\ref{section: eccentricityobserved} we study the consequences of this assumption.

% ---------------------------------------------------------------------------------------
% ---------------------------------------------------------------------------------------
% ---------------------------------------------------------------------------------------

\subsection{Choice of orientation and phase parameters}

The orientation of an orbit is defined by the inclination $i$, the argument of
periastron $\omega$, and the angle of the ascending node $\Omega$. The
definition of $i$, $\omega$, and $\Omega$ for a given orbit is given in
Section~\ref{section: orbitorientationvectors}. The orbital phase as a function
of time is given by the mean anomaly $\mathcal{M}$, or alternatively by the true
anomaly $\mathcal{V}$.

If there is no preferred orientation of binary systems, the inclination distribution is given by $f_{\cos i}(\cos i)=\frac{1}{2}$ for $-1 \leq \cos i \leq 1$. An inclination $i=0^\circ$ corresponds to a face-on view, and $i=\pm 90^\circ$ corresponds to an edge-on view. Orbits with inclination $+i$ and $-i$ have the same projection, but the velocity vectors of the components are opposite. The distributions of the argument of periastron and the angle of the ascending node for random orientation are given by $f_\omega(\omega) = $~constant for $0^\circ \leq \omega < 360^\circ$, and $f_\Omega(\Omega) = $~constant for $0^\circ \leq \Omega < 360^\circ$, respectively. 

We sample the orbital phase of each binary system at an instant of time using the mean anomaly ${\mathcal M}$, which is the fraction of orbital period that has elapsed since the last passage at periastron, expressed as an angle. We assign a mean anomaly to each orbit by drawing ${\mathcal M}$ from a flat distribution $f_{\mathcal M}$ for $-180^\circ \leq {\mathcal M} \leq 180^\circ$. The true anomaly ${\mathcal V}$ is derived from the mean anomaly ${\mathcal M}$ and the eccentricity $e$. 
Throughout this paper we adopt the distributions for $i$, $\omega$, $\mathcal{M}$ as listed above. We will always use random orientation of the orbits, except in Section~\ref{section: orientationoftheorbits}, where we will briefly discuss the effect of a preferred orientation on the observables.

% ====================================================================
% ====================================================================
% ====================================================================
% ==OBSERVATIONAL BIASES======================================================
% ====================================================================
% ====================================================================
% ====================================================================

\section{Observational selection effects} \label{section: observationalselectioneffects}

Although several other techniques exist, most binaries are found with imaging,
radial velocity, and astrometric surveys: the visual, spectroscopic, and
astrometric binaries. Each of these techniques has its own limitations,
introducing different selection effects. These selection effects depend
on the observing strategy, the instrumental properties,
and the atmospheric conditions. Furthermore, confusion may be introduced, so
that it is difficult to decide whether an object is truly a binary or just a
single star. Well-known examples of this confusion are the presence of
background stars in imaging surveys and stellar pulsations in radial velocity
surveys.

In this section we discuss models that describe the most important selection
effects for these three techniques. We refer to ``visual biases'', ``spectroscopic
biases'', and ``astrometric biases'' when addressing the selection effects (i.e., sample
bias and instrument bias) of a visual, spectroscopic, and astrometric binarity
survey, respectively. 
Below we will describe below simplified  models
of the selection effects, so that these can be used for the analysis of a wide
range of datasets.

\begin{table}
  \small
  \begin{tabular}{p{0.2cm}lp{0.8cm}ll}
    \hline
    \hline
              & Constraint & Type & Section & Examples and remarks  \\
    \hline
    All       & Observer's choice          & Sample        & \ref{section: observerschoice}             & The B~stars; the M~stars \\
    \hline
    VB        & Brightness constr.      & Sample        & \ref{section: visual_brightnessconstraint} & $V < V_{\rm max}$ \\
    VB        & Separation constr.      & Instr.    & \ref{section: visual_separationconstraint} & $\rho_{\rm min} \leq \rho \leq \rho_{\rm max}$ \\
    VB        & Contrast constr.        & Instr.    & \ref{section: visual_contrastconstraint}   & equation~\ref{equation: classical_detectionlimit1} \\
    VB        & Confusion constr.       & Instr.    & \ref{section: visual_confusionconstraint}  & equation~\ref{equation: classical_detectionlimit2} \\
    \hline
    SB        & Brightness constr.      & Sample        & \ref{section: spectroscopic_brightnessconstraint} & $V_{\rm comb} < V_{\rm max}$ \\
    SB        & Amplitude constr.       & Instr.    & \ref{section: sb-s}--4 & SB-S, SB-C, SB-W ($\sigma_{\rm RV} = 2$~km\,s$^{-1}$) \\
    SB        & Temporal constr.        & Instr.    & \ref{section: sb-s}--4 & $P<30$~yr (SB-S and SB-C), implicit (SB-W) \\
    SB        & Aliasing constr.        & Instr.    & \ref{section: sb-w}                                           & SB-W  \\
    SB        & Sampling constr.        & Instr.    & \ref{section: spectroscopic_samplingconstraint}               & Not applied  \\
    \hline
    AB        & Brightness constr.      & Sample        & \ref{section: astrometric_brightnessconstraint} & {\em Hipparcos} detection limit \\
    AB        & Amplitude constr.       & Instr.    & \ref{section: astrometric_instrumentconstraint} & \quad Classification into the  categories  \\
    AB        & Temporal constr.        & Instr.    & \ref{section: astrometric_instrumentconstraint} & \quad (X), (C), (O), or (G) depending on the \\
    AB        & Aliasing constr.        & Instr.    & \ref{section: astrometric_instrumentconstraint} & \quad observables of each binary system \\
    AB        & Sampling constr.        & Instr.    & \ref{section: astrometric_instrumentconstraint} & \quad (see Table~\ref{table: hipparcosbiases}) \\
    \hline
    \hline
  \end{tabular}
  \caption{The selection effects for visual binaries (VB), spectroscopic binaries (SB), and astrometric binaries (AB), as modeled in this paper (see \S~\ref{section: observationalselectioneffects}). The first column gives the observing technique (VB, SB, or AB). The second and third columns list the constraints for the various types of binaries, and the type of constraint (sample or instrument bias). The fourth column lists the section in which each constraint is described. Finally, the last column lists a remark or example for each constraint. The instrumental bias for spectroscopic binaries is modeled in three different ways (SB-S, SB-C, and SB-W), which are explained in Section~\ref{section: spectroscopic_biases}. \label{table: biasesoverview}  }
\end{table}

% ==================================================================
% ==================================================================

\subsection{The observer's choice}\label{section: observerschoice}

The observer may be interested in a specific group of targets for his or her binarity survey. For example, the observer may wish to quantify the mass ratio distribution for B~type stars \citep{shatsky2002}, the binary fraction for late~B and A~type stars \citep{kouwenhoven2005}, or multiplicity among very low-mass and brown dwarf members of an OB~association \citep{bouy2006}. We refer to this a-priori choice of the sample as the ``observer's choice''. The observing technique is often chosen in such a way that most of the targets can actually be observed. Due to the brightness constraint and time limitations, not all targets can be observed. We define the sample bias as the combination of the observer's choice and the brightness constraint. The decision of an observer to survey a specific group of targets for binarity may lead to observed distributions that are not representative for the association as a whole.

% ==================================================================
% ==================================================================

\subsection{Visual biases} \label{section: visual_biases}

Due to observational limitations not all binaries are detected as visual binaries. The most important properties of a binary system that are required for detection and proper analysis are:
\begin{itemize}\addtolength{\itemsep}{-0.5\baselineskip}
\item[--] {\em Brightness constraint} (sample bias). The primary and companion
  magnitude should be within the range of detection. Stars that are too bright
  saturate the detector; stars that are too faint are undetected due to noise.
\item[--] {\em Separation constraint} (instrument bias). The angular separation
  between primary and companion should be larger than the spatial resolution,
  and should fit within the field of view of the detector. The spatial
  resolution is determined by the detector, the telescope, and atmospheric
  conditions. The maximum detectable separation additionally depends on the
  position angle of a binary system for a non-circular field of view. 
\item[--] {\em Contrast constraint} (instrument bias). The contrast between primary
  and companion should be small enough for detection of the companion. The maximum
  magnitude difference for which a companion can be detected is often a function
  of the angular separation between the two components. Faint stars at large
  separation from the primary may be detected, while they remain undetected if
  near the primary.
\item[--] {\em Confusion constraint} (instrument bias). When two stellar components
  are detected, it is not always possible to determine whether the two components
  are physically bound. The binary may be optical, i.e., it may result from
  projection effects. The probability of detecting background stars in the field
  of view increases with the separation between the components and decreases
  with increasing brightness of the companion. The number of background stars in a
  field of view additionally depends on the position of the binary in the sky,
  most importantly on the Galactic latitude.
\end{itemize}
Each of these constraints is discussed in the subsections below. An overview of the
constraints that are applied the simulated observations of OB~associations is given
in Table~\ref{table: biasesoverview}.

\subsubsection{The brightness constraint for visual binaries} \label{section: visual_brightnessconstraint}

In this paper we study the general properties of visual binary stars in
OB~associations. With the currently available techniques and instruments, it is
possible to perform imaging surveys of targets over a large brightness range. We
model the brightness constraint by setting the magnitude of the faintest
observable target to $V_{\rm max} = 17$~mag. Stars fainter than $V= 17$~mag are
rarely surveyed for binarity. In the Washington Double Star Catalog, for example,
only 1\% of the binaries has $V>17$~mag. For young ($5-20$~Myr) populations at
a distance of 145~pc, the limit $V_{\rm max} = 17$~mag corresponds to low-mass
brown dwarfs. 

Multiplicity surveys among brown dwarfs are rare, certainly when compared to
surveys among more massive objects. We wish to stress that this is not a result
of the brightness constraint (as it is possible to image faint objects), but
primarily a result of the observer's choice of the stars in the sample (see
\S~\ref{section: observerschoice}).

\subsubsection{The separation constraint for visual binaries} \label{section: visual_separationconstraint}

Once a target sample for the visual binary survey is composed, and the targets
are observed, the dominant constraint for the detection of a companion is often
their separation. A lower limit $\rho_{\rm min}$ is imposed by the
spatial resolution obtained in the survey. The spatial resolution depends
on the size of the telescope mirror, the properties of the instrument (e.g., whether adaptive
optics is used), and the atmospheric conditions at the time of the observations.
The maximum detectable angular separation $\rho_{\rm min}$ is determined by the
size of the field of view. For a rectangular field of view, the maximum detectable
separation is also a function of the position angle. 

In our model for the separation constraint of visual binaries we adopt
$\rho_{\rm min} = 0.1$~arcsec. We choose this value as this is the limiting
spatial resolution for imaging for the majority of the observations over the
past decades. Although a typical seeing of 1~arcsec would prevent the
detection of close binaries, with adaptive optics binaries as close as 0.1~arcsec
can be resolved. It is possible to obtain observations with a spatial resolution
even better than 0.1~arcsec with interferometry, space-based telescopes, and
large ground-based telescopes. However, we do not consider these here, as only a
very small fraction of the visual binaries have been detected this way.

We do not adopt a maximum value to the detectable angular separation due to the
separation constraint. Usually the actual detection of the companion is not an
issue. Due to the presence of background stars, the confusion constraint
(Section~\ref{section: visual_confusionconstraint}) sets an 
effective upper limit to the observed angular separation, as it is
often difficult to separate faint, wide companions from background stars.

\subsubsection{The contrast constraint for visual binaries} \label{section: visual_contrastconstraint}

The maximum brightness difference at which a binary is detectable is a function
of the angular separation between the components (and the observing wavelength).
The faintest detectable magnitude as a function of projected distance to the
primary star for a certain filter band is called the {\em detection limit} for
that filter.  Several attempts to quantify the detection limit have been carried out
in the last century.  These formulae were derived before adaptive optics and
space missions became available. \cite{hogeveen1990}
obtained a formula which describes the relation between angular separation
$\rho_{\rm min}$ and the maximum visual brightness difference $\Delta V$:
\begin{equation} \label{equation: classical_detectionlimit1}
  \log \left( \frac{\rho_{\rm min}}{\rm \ arcsec \ } \right) = \frac{\Delta V}{6}  -1.
\end{equation}
Adopting this model for the detection limit, the minimum detectable separation occurs at $\rho_{\rm min} = 0.1''$, where $\Delta V = 0$, in good agreement with the separation constraint adopted in \S~\ref{section: visual_separationconstraint}. The maximum contrast $\Delta V$ at separations of $1''$ and $10''$ is 6~magnitudes and 12~magnitudes, respectively. 
Equation~\ref{equation: classical_detectionlimit1} is less restrictive than that of \cite{heintz1969}, for which $0.22\Delta V - \log \rho_{\rm min} = 0.5 \wedge \rho_{\rm min} > 0.3''$, and that of \cite{halbwachs1983}, for which $\Delta V < 2.45 \rho_{\rm min}^{0.6} - 0.5$. As the sample studied by \cite{hogeveen1990} is larger and more recent, we choose to use the expression for $\rho_{\rm min}$ given in his paper. 

The contrast constraint imposes limits on the range of mass ratios $q$ that is observed. The minimum and maximum observable magnitude of a star, respectively, impose a maximum magnitude difference $\Delta V_{\rm max}$ of a binary system. An estimate for the minimum observable mass ratio $\tilde{q}_{\rm min}$ can be obtained using an approximation for the main sequence mass-luminosity relation $\log q_{\rm min} = -\Delta V/10$ \citep[e.g.,][]{halbwachs1983}. 

\subsubsection{The confusion constraint for visual binaries} \label{section: visual_confusionconstraint}

Due to the presence of background stars, single-epoch, single-filter
imaging surveys for binarity can only reliably identify companion stars up to a
maximum angular separation $\rho_{\rm max}$. The value of  $\rho_{\rm max}$ can
be obtained from a statistical analysis of the background population, and is a
function of magnitude and the coordinates of the target (in particular the
Galactic latitude).

Several models have been proposed for $\rho_{\rm max}$ as a function of the observable properties of the binary system (brightness and separation) and those of the background star population \citep[e.g., the 1\% filter,][]{poveda1982}. As we try to keep our models simple, we use Aitken's separation criterion \citep{aitken1932}. The formula relating the maximum angular separation $\rho_{\rm max}$ and the combined magnitude of the components $V_{\rm comb}$ of the components is given by
\begin{equation} \label{equation: classical_detectionlimit2}
\log \rho_{\rm max} = 2.8 - 0.2 \, V_{\rm comb},
\end{equation}
where $\rho_{\rm max}$ is in arcseconds. For an OB~association such as Sco~OB2,
at a distance of 145~pc, the upper limit for the angular separation due to
confusion is approximately $\rho_{\rm max}=60''$ for a B-type star, $22''$ for
an A-type star, and $3.5''$ for a G-type star.

\subsubsection{The errors on the visual binary observables} \label{section: visual_errormodeling}

In our simulations we ignore the errors on the visual binary observables.
However, in Section~\ref{section: powerlawaccuracy} we discuss the consequences
of neglecting these observational errors, notably the errors on the
component masses.

% ==================================================================
% ==================================================================

\subsection{Spectroscopic biases} \label{section: spectroscopic_biases}

For a spectroscopic binary the presence of a companion is derived from the
spectrum. As a large fraction of the spectroscopic binaries are single-lined spectroscopic binaries (SB1s), and as the double-lined spectroscopic binaries (SB2s)
can be considered as a subset of the SB1 systems, we only discuss SB1s
in the remaining part of this section. The properties of
a binary system that determine whether it can be spectroscopically resolved as
an SB1 are the following:
\begin{itemize}\addtolength{\itemsep}{-0.5\baselineskip}
\item[--] {\em Brightness constraint} (sample bias). The brightness of the primary
  should be such that the signal-to-noise ratio of the spectrum is large enough
  for a proper spectral analysis.
\item[--] {\em Contrast constraint} (instrument bias). This constraint is only
  relevant for the detection of SB2 systems. For an SB2, the spectra of both
  targets have to be analyzed in detail to obtain the radial velocity curve of
  both the primary and the companion, hence the brightness contrast between the two
  components cannot be too large.
\item[--] {\em Amplitude constraint} (instrument bias). The radial velocity
  variations should have {\em large} enough amplitude, so that the variations are
  significantly larger than the radial velocity precision of the spectrograph
  and analysis technique.
\item[--] {\em Temporal constraint} (instrument bias). The radial velocity
  variations should be {\em fast} enough, so that the radial velocity change is
  measurable within the length of the observing campaign.   
\item[--] {\em Aliasing constraint} (instrument bias). If the radial velocity
  measurements are obtained at regular time intervals, or if very few radial
  velocity measurements are available, binaries with orbital periods similar to
  the mean difference in time between the measurements remain undetected.
\item[--] {\em Sampling constraint} (instrument bias). Even though a binary system
  satisfies the constraints above, and the presence of the companion is
  detected, the radial velocity curve is not necessarily sampled well enough that the
  orbital elements (such as the period) can be measured accurately. The orbital elements
  can be derived only if (1) the orbital period is not much shorter
  than the smallest measurement interval, (2) the orbital period is not
  much larger than the observing run, and (3) the radial velocity curve
  is well-sampled.
\end{itemize}

\subsubsection{The brightness constraint for spectroscopic binaries}
\label{section: spectroscopic_brightnessconstraint}

The brightness constraint imposes a maximum magnitude $V_{\rm max}$ on the
targets in the sample. We adopt a value of $V_{\rm max} = 10$~mag for the
primary star. This value is rather arbitrary, as the maximum
magnitude strongly depends on the observing strategy: the telescope, the
integration time, and the observed wavelength range. The adopted value for
$V_{\rm max}$ is a compromise between the classical spectroscopic surveys
\citep[see, e.g.,][]{levato1987,hogeveen1992a} which have $V_{\rm max} \approx
5$~mag, and the brightness limit of $V_{\rm max} \ga 15$~mag that can be
achieved with modern instruments such as UVES and FLAMES at the ESO Very Large
Telescope in Paranal, Chile. 

In the following subsections we discuss the models
for the observations of spectroscopic binaries. In these models we include the
magnitude limits motivated above.

\subsubsection{The instrument bias for spectroscopic binaries: method~1 (SB-S)} \label{section: sb-s}

In the simplest model SB-S for the instrument bias of spectroscopic binarity
surveys, we only consider the effect of the amplitude constraint and the
temporal constraint for spectroscopic binaries as an instrument bias. The
amplitude constraint is the strongest component in the instrument bias. Binary
systems with a large primary radial velocity amplitude are easily detected as such. For
binary systems with a small radial velocity amplitude, the variations in the
radial velocity cannot be discriminated from the statistical noise.

For the amplitude constraint we require that the variation in the radial
velocity $V_1(t)$ of the primary is sufficiently large, so that the orbital
motion can be detected in the spectrum of the primary star. In our model SB-S
systems with $K_1 < \sigma_{\rm RV}$ remain unresolved. We adopt $\sigma_{\rm
RV}=2$~km\,s$^{-1}$. In \S~\ref{section: error_rv} we further motivate our
choice of $\sigma_{\rm RV}$.

As a temporal constraint we impose a maximum period of 30~year. For the default
model, about 6\% of the binary systems have $K_1 > 2$~km\,s$^{-1}$ and a period
larger than 30~year. The majority of these systems have a very high
eccentricity (50\% have $e>0.9$), significantly increasing $K_1$. As stars in highly
eccentric systems spend most of their time near apastron, the probability of
actually detecting the maximum radial velocity variation (which occurs near
periastron) is small. 

Modeling selection effects for radial velocity surveys using method SB-S is
rather simplistic. Method SB-S neglects the dependence of the detection
probability on the length of the observing run, the measurement interval, and
the fact that binaries with a high eccentricity or $\cos \omega \approx 1$ are
more difficult to detect. Despite these shortcomings, this method is sometimes used
in literature to estimate the number of unseen spectroscopic binaries
\citep[e.g.,][]{abtlevy1976}. In the following Sections we discuss two more
sophisticated methods: SB-C and SB-W.

\subsubsection{The instrument bias for spectroscopic binaries: method~2 (SB-C)} \label{section: sb-c}

\begin{SCfigure}[][!tbp]
  \centering
  \includegraphics[width=0.5\textwidth,height=!]{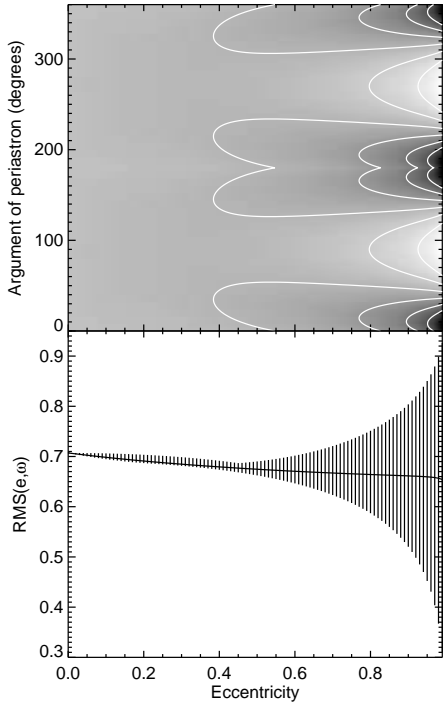}
  \caption{{\em Top:} the quantity $\mathcal{R}(e,\omega)$ as a function of
  eccentricity $e$ and argument of periastron $\omega$ (see
  equation~\ref{equation: rmsdefinition}). The lowest values of
  $\mathcal{R}(e,\omega)$ are black, and the highest values are white. 
  Binary systems with a large value of $\mathcal{R}(e,\omega)$ are more easily detected than binaries with a small value. The contours represent $\mathcal{R}(e,\omega)$ values of 0.45, 0.525, 0.6. 0.675,
  0.75, and 0.825, respectively. Orbits with $e=0$ have $\mathcal{R}(0,\omega) = \tfrac{1}{2}\sqrt{2}$.
  For orbits with $e > 0$, $\mathcal{R}(e,\omega)$ is largest for $\omega =
  90^\circ$ and $\omega=270^\circ$, so that orbits with these values for
  $\omega$ are easiest to detect. For orbits with $e > 0$,
  $\mathcal{R}(e,\omega)$ is smallest for $\omega=0^\circ$ and $\omega =
  180^\circ$; these orbits are thus most difficult to detect for a given $e>0$.
  {\em Bottom:} The possible values of $\mathcal{R}(e,\omega)$ as a function of
  $e$. The solid curve represents the mean $\mathcal{R}(e,\omega)$, averaged over
  all values of $\omega$. The vertical lines indicate the spread in
  $\mathcal{R}(e,\omega)$ for given $e$, as a result of the variation in $\omega$.
 \label{figure: rmsdefinition} }
\end{SCfigure}

\begin{figure}[!tbp]
  \centering
  \includegraphics[width=1\textwidth,height=!]{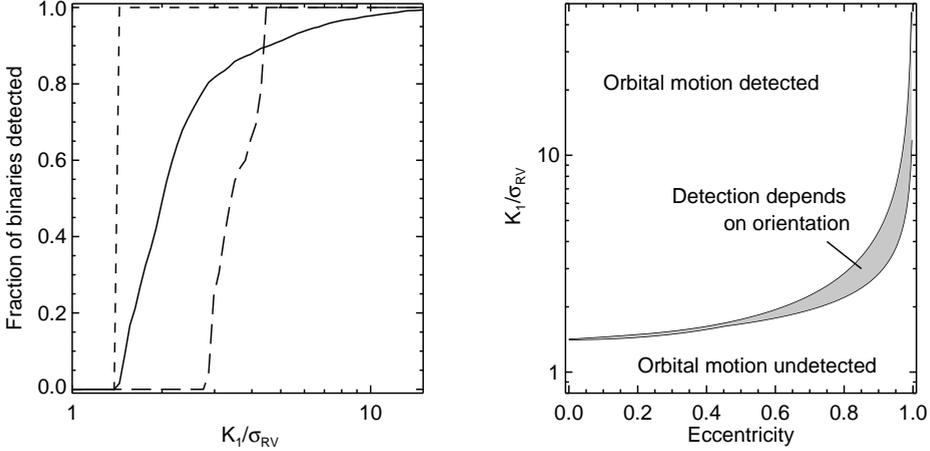}
  \caption{
    {\em Left:} The detection probability for binary systems as a function of
    $K_1/\sigma_{\rm RV}$, where $K_1$ is the semi-amplitude of the primary
    radial velocity curve, and $\sigma_{\rm RV}$ is the accuracy of the radial
    velocity determination. The binary systems are assumed to be randomly
    oriented (i.e., they have a flat distribution in $\omega$). Results are
    shown for binaries with $e=0$ (short-dashed curve), binaries with a thermal
    eccentricity distribution (solid curve), and binaries with $e=0.9$
    (long-dashed curve). Binary systems are detected spectroscopically if their
    properties satisfy equation~\ref{equation: rms}. The required value of
    $K/\sigma_{\rm RV}$ for a binary system generally increases with increasing
    eccentricity, but additionally depends on the argument of periastron. {\em
    Right:} Whether a binary is detected depends on $K_1/\sigma_{\rm RV}$, the
    eccentricity $e$, and the argument of periastron $\omega$. This Figure
    indicates the regions in the $(e,K_1/\sigma_{\rm RV})$ diagram for which
    binarity is not, likely, or always detected (assuming that all binaries are
    bright enough for a proper analysis). \label{figure: rms_requirement_k} 
  }
\end{figure}

A more accurate model for the instrumental bias of radial velocity surveys takes
into account the effects related to the period, eccentricity, and argument of periastron as
well. In this section we outline model SB-C, which describes the amplitude
constraint in a more accurate way the previous method. As
in Section~\ref{section: sb-s}, we adopt as the temporal
constraint a maximum orbital period of 30~years.

The detectability of variations in the primary radial velocity not only
depends on the amplitude $K_1$ (as assumed in SB-S),
but also on the eccentricity $e$ and the argument of periastron
$\omega$. As $K_1$ is proportional to $(1-e^2)^{-1/2}$, this implies that
orbits with high eccentricity are easier to detect than those with a low
eccentricity. However, highly eccentricity binaries are usually difficult to detect
in a spectroscopic campaign. The components of the binary system spend most of their time
near apastron, where their radial velocity change over time is very small. Only
for a short instant of time, near periastron, the primary star reaches its
extreme velocities $V_{1,\mathrm{min}}$ and $V_{1,\mathrm{max}}$. The
probability that observations are performed near periastron is small. A model
for the amplitude constraint using $\Delta V_1$ is therefore inaccurate.

The root-mean-square (rms) of the radial velocity $V_{\rm 1,rms}$ can be used to describe the amplitude constraint more accurately than the amplitude itself does. The quantity $V_{\rm 1,rms}$ depends on the amplitude $K_1$, the eccentricity $e$, and the argument of periastron $\omega$. We describe the rms variations due to $e$ and $\omega$ with the dimensionless quantity $\mathcal{R}(e,\omega)$, which we define as follows
\begin{equation} \label{equation: rmsdefinition}
V_{\rm 1,rms} = K_1 \sqrt{1-e^2} \, \mathcal{R}(e,\omega) = L_1 \, \mathcal{R}(e,\omega). 
\end{equation}
Note that $L_1$ is independent of both $e$ and $\omega$ (see equation~\ref{equation: L}). The expression for $\mathcal{R}(e,\omega)$ is derived in Section~\ref{section: vradstat}. 

The behaviour of $\mathcal{R}(e,\omega)$ as a function of eccentricity and argument of periastron is shown in Figure~\ref{figure: rmsdefinition}. For circular orbits the expression reduces to $\mathcal{R}(0,\omega) = \tfrac{1}{2}\sqrt{2}$. Eccentric orbits with $\cos\omega \approx 0^\circ$ have $\mathcal{R}(e,\omega) > \tfrac{1}{2}\sqrt{2}$, meaning that they are easier to detect than circular orbits. Eccentric orbits with $\sin\omega \approx 0^\circ$ have $\mathcal{R}(e,\omega) < \tfrac{1}{2}\sqrt{2}$, and are therefore more difficult to detect than circular orbits.

Let $\sigma_{\rm RV}$ be the accuracy of a radial velocity measurement. The second requirement for detection (the amplitude constraint) in the SB-C is given by
\begin{equation} \label{equation: rms} 
  V_{\rm 1,rms}  >  \sigma_{\rm RV}\,.
\end{equation}
This criterion assumes random time sampling of the radial velocity curves. The quantities $K_1$ and $\sigma_{\rm RV}$ are usually known from observations. We show in Figure~\ref{figure: rms_requirement_k} the fraction of binaries that satisfy equation~\ref{equation: rms} as a function of $K_1/\sigma_{\rm RV}$ $=  \left( \sqrt{1-e^2} \, \mathcal{R}(e,\omega) \right)^{-1}$. The orbits are assumed to be randomly oriented, i.e. they have a flat distribution in $\omega$. Results are shown for circular orbits ($e=0$), for highly-eccentric orbits ($e=0.9$), and for a thermal eccentricity distribution ($f_{2e}(e) = 2e$).  As the thermal eccentricity distribution has a large fraction of orbits with high eccentricity, about 3\% of the orbits with $K_1/\sigma_{\rm RV} = 10$ remain undetected.

\subsubsection{The instrument bias for spectroscopic binaries: method~3 (SB-W)} \label{section: sb-w}

\begin{figure}[!tbp]
  \includegraphics[width=0.9\textwidth,height=!]{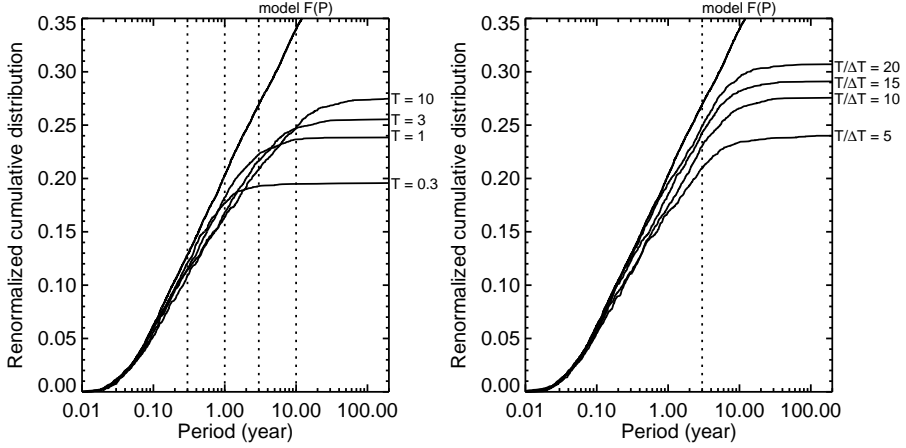}
  \caption{The observed cumulative distribution of the period $\tilde{F}_P(P)$ for spectroscopic binaries, for method SB-W, as a function of the time-span $T$ of the observations, and the interval $\Delta T$ at which radial velocity measurements are obtained. Each model contains 5\,000 binary systems. The radial velocity error is $\sigma_{\rm RV}=2$~km\,s$^{-1}$ in each model. {\em Left:} the cumulative period distribution for observing runs of $T=0.3$, 1, 3, and 10~year (drawn lines). During each run, we obtain $T/\Delta T=8$ radial velocity measurements, and use the measurements to determine whether the target is a spectroscopic binary. Note that the cumulative distributions in this figure are normalized with respect to the {\em total} number of binaries in the association. {\em Right:} the cumulative period distributions for an observing run of $T=3$~year, with $T/\Delta T = 5$, 10, 15, and 20 radial velocity measurements. The figure indicates the importance of the choice of $T$ and $\Delta T$. For a large value of $T$, orbits with larger $P$ are detected, while the detected fraction of orbits with period $P < T$ depends on the number of observing intervals $T/\Delta T$.\label{figure: tdt_period} }
\end{figure}

Window-sampled spectroscopic biasing is closest to reality. In SB-W we
obtain radial velocity measurements of all binary systems in the simulated
association, at regular intervals $\Delta T$ for a time-span $T$. We assume a
value for the error $\sigma_{\rm RV}$, which is constant over the time of observations.
If the radial velocity measurements show a spread significantly larger than the
error, the binarity is detected.

For each time-step we obtain the radial velocity of the primary star from the
simulated association, and Gaussian observational errors are added to each radial
velocity measurement.

For each single star and binary system we test the hypothesis that the observed velocity measurements $\{v_i\}$ result from a constant velocity. We calculate the $\chi^2$ of the set of $N_v$ radial velocity measurements:
\begin{equation}
\chi^2 = \sum_i \frac{(v_i-\overline{v})^2}{\sigma_{\rm RV}^2}\,,
\end{equation}
where $\overline{v}$ is the mean of the measurements $v_i$. We then calculate the probability $p$ that $\chi^2$ is drawn from the $\chi^2$-distribution:
\begin{equation}
  p = 1 - \Gamma \left( \tfrac{1}{2}\nu,\tfrac{1}{2}\chi^2 \right)\,,
\end{equation}
where $\nu = N-1$ is the number of degrees of freedom. High values ($p\approx 1$) indicate that our hypothesis (that the radial velocity is constant) is true, and that the measurements are likely the result of statistical noise. Values of $p$ close to zero show that the observed variations in the radial velocity are real. We classify objects with radial velocity sets with $p \leq 0.0027$ (corresponding to the $3\sigma$ confidence level) as binary systems. The other systems are marked as single stars.

The method described above is often used to detect the presence of companion stars \citep[see, e.g.,][]{trumpler1953,demarco2004}. With this method, binaries have a very low detection probability if the radial velocity amplitude is small ($K_1 \ll \sigma_{\rm RV}$) or if the orbital period is very large ($P \ga 2T$). Also, orbits with a period $P$ very close to the observing interval $\Delta T$ are not detected due to aliasing. In order to avoid aliasing, observers often obtain their measurements with varying observing intervals, the optimal scheme of which is a geometrical spacing of the observations in time \citep[see][for details]{saunders2006}.

The properties of the detected binaries depend on the choice of $T$ and $\Delta
T$. Figure~\ref{figure: tdt_period} illustrates the observed cumulative period
distribution, for fixed $T/\Delta T$ and varying $T$ (left), and for fixed $T$
and variable $T/\Delta T$ (right). For a radial velocity survey with a fixed
number of measurements $T/\Delta T$, the observed period range, and the observed
number of binaries increase with $T$. However, several orbits with $P \la T$ are
missing for large $T$, as they are under-sampled, and thus remain undetected. The
right-hand panel in Figure~\ref{figure: tdt_period} shows how the detected fraction
of binaries depends on the number of measurements $T/\Delta T$, for a fixed
value of $T=3$~year. An increasing number of measurements, as expected,
increases the fraction of binaries detected. For $T/\Delta T \rightarrow
\infty$, all orbits with $P \la T$ are detected.

\subsubsection{A comparison between SB-S, SB-C, and SB-W} \label{section: choice_sb_model}

Each of the three models SB-S, SB-C, and SB-W for the instrument bias of spectroscopic binaries has its own approximations and limitations. The three models approximate the instrument bias by taking into account the amplitude constraint and the temporal constraint. Of these three models, SB-W approximates most accurately the true selection effects, and also takes into account the aliasing constraint.

\begin{figure}[!tbp]
  \centering
  \includegraphics[width=1\textwidth,height=!]{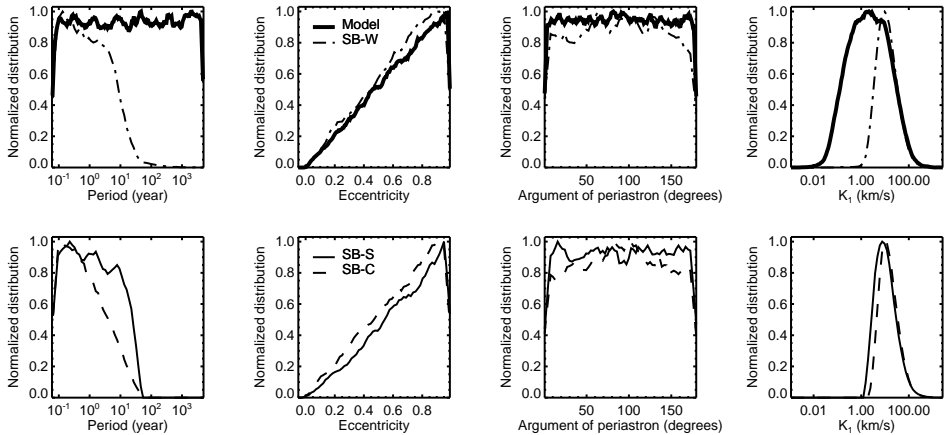}
  \caption{The ``observed'' period distribution $\tilde{f}_P(P)$, eccentricity distribution $\tilde{f}_e(e)$, argument of periastron distribution $\tilde{f}_\omega(\omega)$, and distribution over semi-amplitude of the primary radial velocity curve $\tilde{f}_{K_1}(K_1)$ for simulated observations of spectroscopic binaries. The three bias models are listed in Table~\ref{table: biasesoverview} and are indicated in the Figure. For the models SB-S and SB-C we adopted a radial velocity accuracy $\sigma_{\rm RV} = 2$~km\,s$^{-1}$ and a maximum orbital period of 30~years. For SB-W we obtained simulated observations at an interval of $\Delta T=6$~months for $T=5$~year. The true distributions (thick solid curves in the top panels) are shown for comparison. The simulations are made for the default model, consisting of $N=25\,000$ binaries. Each distribution is normalized so that its maximum is unity. See \S~\ref{section: choice_sb_model} for a discussion of the results.  \label{figure: sb_model} }
\end{figure}

In order to characterize the accuracy of the SB-S and SB-C models relative to
that of SB-W, we simulate a spectroscopic survey using each of these bias
models. For SB-S, SB-C, and SB-W we adopt a a radial velocity error $\sigma_{\rm
RV} = 2$~km\,s$^{-1}$. In our model, $\sigma_{\rm RV}$ is independent of
spectral type (see \S~\ref{section: error_rv} for a discussion). For both SB-S and
SB-C we adopt a maximum orbital period of 30~year. For SB-W we adopt an
observing interval of $\Delta T = 0.5$~year, and observing run of $T = 5$~year. As
a model for the association we use the one outlined in Tables~\ref{table: modelcluster} 
and~\ref{table: modelpopulation}, with $N=25\,000$ binary systems.
Figure~\ref{figure: sb_model} shows the results of the simulated spectroscopic
binary surveys. As the period distribution, the eccentricity distribution, the
distribution over the argument of periastron, and the distribution over $K_1$
are most sensitive to the spectroscopic selection effects, we show the
respective observed distributions for each of the spectroscopic bias models. For
comparison, we also show the corresponding true distributions in the
model association.

Figure~\ref{figure: sb_model} shows that the overall shape of the observed
distributions $\tilde{f}_P(P)$,  $\tilde{f}_e(e)$, $\tilde{f}_\omega(\omega)$, and
$\tilde{f}_{K_1}(K_1)$  are similar for the three spectroscopic bias models. With the
three bias models all short-period orbits ($P \la 1$~year) are detected. Due to
the definition of SB-S and SB-C, binaries with orbits with $P > 30$~year are
unresolved. Even though the observing time-span of SB-W is $T=5$~year, several
orbits with $P > 30$~year are still detected. The SB-S model is biased to the
detection of highly eccentric orbits, as these have a larger radial velocity
amplitude than circular orbits with similar total mass and semi-major axis. SB-C
and SB-W are slightly biased towards low-eccentricity orbits. The distribution
over the argument of periastron is flat for the model and for the observations
with SB-S. Both SB-C and SB-W are biased to detect orbits with $\sin \omega
\approx 1$, as a result of the correlation between the shape of the radial
velocity profile and the argument of periastron. With the three spectroscopic
bias models we detect all orbits with $K_1 \ga 10$~km\,s$^{-1}$. For the orbits
with $K_1 \la 10$~km\,s$^{-1}$, the three bias models give similar results.

Method SB-S, often used in literature, is suitable for obtaining a rough estimate
of the number of unseen binary systems. Methods SB-C and SB-W represent the
observations of spectroscopic binary stars more accurately. Method SB-C is a
more convenient method from a computational point of view, and can be used to
rapidly simulate the observed binary fraction in a radial velocity survey. In
this paper we use method SB-W, which most accurately approaches the radial
velocity technique to find spectroscopic binaries.

\subsubsection{The sampling constraint for spectroscopic binaries} \label{section: spectroscopic_samplingconstraint} 

The presence of a companion star may be inferred from the radial velocity variations, if the constraints mentioned above are satisfied. This does not necessarily mean that the orbital elements can be derived. Whether or not the spectroscopic elements of a binary can be derived, depends on the properties of the binary and on the sampling of the observations.

 The radial velocity curve of a spectroscopic binary is defined by six independent parameters: the period $P$, the eccentricity $e$, the argument of periastron $\omega$, the system velocity $V_{\rm sys}$, the semi-amplitude of the radial velocity curve $K_1$, and the time of periastron passage. Therefore, at least six radial velocity measurements are necessary in order to determine these parameters. Furthermore, a constraint is imposed by the Nyquist theorem, which requires that at least two measurements should be obtained per orbital period. Additional constraints are imposed by the properties of the binary system: the spectral type of the star (in particular the number of spectral lines), the brightness of the system, and the values of $K_1$ (relative to $\sigma_{\rm RV}$), $e$, and $\omega$.

The sampling constraint is sophisticated, and highly dependent on the properties of the targeted binary system and on the observing strategy. In this paper we therefore do not attempt to classify the spectroscopic binaries in the three subcategories, keeping in mind that the spectroscopic elements are in reality only available for the SB1 and SB2 systems.

\subsubsection{The errors on the spectroscopic binary observables} \label{section: error_rv}

In models SB-S, SB-C, and SB-W we adopt a constant radial velocity error
$\sigma_{\rm RV} = 2$~km\,s$^{-1}$. This error is partially caused by the
observing strategy: the telescope, the spectrograph, the wavelength calibration,
and the properties of the atmosphere at the time of observation. The properties
of the targeted star also contribute to the radial velocity error. This
intrinsic error (which we do not model explicitly) is caused by line broadening
due to (projected) rotation, turbulence in the stellar atmosphere, and the number of
observable spectral lines in the observed wavelength range. Early type stars
often have high rotational velocities, so that  $\sigma_{\rm RV}$ is higher
than for low-mass stars. For low-mass stars it is often possible to determine
the radial velocity with an accuracy $\sigma_{\rm RV} < 1$~km\,s$^{-1}$. In
this paper we adopt $\sigma_{\rm RV} = 2$~km\,s$^{-1}$, as due to the brightness
constraint, practically all spectroscopic targets are of spectral type~A or
earlier, at the distance of the most nearby OB~associations.

Under certain conditions it is possible to derive the orbital elements of a
spectroscopic binary from the radial velocity curve. The orbital period can
usually be derived with high accuracy, while the errors on $K_1$, $e$, $\omega$,
and $a_1\sin i$ may still be significant. The calculation of the errors on the
spectroscopic observables is a non-trivial issue. These errors are specific to
each set of binary properties, each observing strategy, and on the reduction
techniques that are applied by the observer. In this paper we ignore the errors
on the spectroscopic observables.

% ==================================================================
% ==================================================================

\subsection{Astrometric biases} \label{section: astrometric_biases}

\begin{table}
  \small
  \begin{tabular}{llll}
    \hline
    Constraints on $\rho$ and $\Delta H_p$ & Period constraints & Solution  & Elements \\
    \hline
    $2 \leq \langle \rho \rangle \leq 100$~mas  or $\Delta H_p > 4$   & $P \leq 0.1$~year      & Stochastic (X)   & no \\
    $2 \leq \langle \rho \rangle \leq 100$~mas  or $\Delta H_p > 4$   & $0.1 < P \leq 10$~year & Orbital (O)      & yes \\
    $2 \leq \rho \leq 100$~mas  or $\Delta H_p > 4$                   & $5 < P \leq 30$~year   & Acceleration (G) & no \\
    $0.1 \leq \rho \leq 100$~arcsec and $\Delta H_p \leq 4$           & $P > 30$~year          & Resolved  (C)    & no \\
    \hline
  \end{tabular}
  \caption{A model for the instrumental bias of the {\em Hipparcos} catalog, based on the analysis of \cite{lindgren1997}. The binary systems satisfying the above constraints are resolved with {\em Hipparcos}. No orbital motion is detected for the binaries classified as resolved visual systems; binaries in this group are similar to ``visual binaries''. Binary systems that do not satisfy the constraints remain undetected with {\em Hipparcos} in our model. For the comparison between the true and simulated observations, we use the total astrometric binary fraction, and do not discriminate between the different subcategories.  \label{table: hipparcosbiases} }
\end{table}

Astrometric binaries are binaries for which the presence of the binary can be inferred from the orbital motion observed in the plane of the sky. The general requirements for a binary to be an astrometric binary are:
\begin{itemize}\addtolength{\itemsep}{-0.5\baselineskip}
\item[--] {\em Brightness constraint} (sample bias). The primary should be within the range of detection. Stars that are too bright saturate the detector; stars that are too faint remain undetected.
\item[--] {\em Amplitude constraint} (instrument bias). The projected orbital motion variations should be large enough, so that the movement of the photocenter (for an unresolved binary) or companion (for a resolved binary) is detectable.
\item[--] {\em Temporal constraint} (instrument bias). The projected orbital motion variations should be fast enough, so that a significant change in the position of the primary is measurable within the observing campaign. 
\item[--] {\em Aliasing constraint} (instrument bias). If the astrometric measurements are obtained at regular time intervals, or if very few astrometric measurements are available, binaries with orbital periods similar to the mean difference in time between the measurements remain undetected.
\item[--] {\em Sampling constraint} (instrument bias). Even though a binary system satisfies the constraints above, and the presence of the companion is detected, the projected orbit is not necessarily sampled well enough that the orbital elements (such as the period) can be detected. The orbital elements can be derived only if (1) the orbital period is not significantly shorter than the smallest measurement interval, (2) the orbital period is not significantly larger than the observing run, and (3) the projected orbit is sampled with enough measurements.
\end{itemize}
Most of the astrometric binaries currently known in Sco OB2 were found
with the {\em Hipparcos} satellite \citep{brown2001}. In this paper we therefore study
the astrometric binaries using a simplified model of the {\em Hipparcos} mission
biases. Between 1989 and 1993 the {\em Hipparcos} satellite measured positions,
proper motions, and trigonometric parallaxes of over 100\,000 stars. More than
18\,000 binary and multiple systems were detected and listed in the {\em
Hipparcos} Catalog of Double and Multiple Systems Annex. The majority ($\sim
13\,000$) of these systems is visual (i.e., no orbital motion was detected),
while astrometric variability is detected for $\sim 5\,000$ systems, indicating
binarity.

\subsubsection{The brightness constraint for astrometric binaries} \label{section: astrometric_brightnessconstraint}

The {\em Hipparcos} photometric system measures the brightness in the $H_{\rm p}$ band (see \S~\ref{section: isochrones}). We model the detection limit of {\em Hipparcos} using the prescription given by \cite{soderhjelm2000}. The 50\% detection limit of {\em Hipparcos} is given by
\begin{equation}
  H_{\rm p,50\%} = \\ 
  \left\{
  \begin{tabular}{lr}
    $8.20 + 2.5c\,(c-0.3)         +1.2 \sin |b|$     &  $c < 0.30 $\\
    $8.20 + 0.45\,(c-0.3)\,(0.9-c)+1.2 \sin |b|$     &  $0.30 < c < 1.45$ \\
    $7.53 + 0.35\,(c-2.5)^2       +1.2 \sin |b|$     &  $c > 1.45 $\\
  \end{tabular}
  \right.,
\end{equation}
where $c \equiv V-I$ and $b$ is the Galactic latitude.
A description of these dependencies is given in
\cite{soderhjelm2000} which are incorporated in our model of the {\em Hipparcos}
biases. We assume a Galactic latitude of $b=15^\circ$.

For each binary system with a given $H_{\rm p}$ and $V-I$ we calculate the detection probability for {\em Hipparcos}. We use this probability to determine whether the target is in our simulated {\em Hipparcos} sample.

\subsubsection{The instrument bias for astrometric binaries} \label{section: astrometric_instrumentconstraint}

Our model for the {\em Hipparcos} instrumental bias is based on the analysis of \cite{lindgren1997}. They classify binaries in several groups, where the classification is based on the detectability with {\em Hipparcos}, and classify the observed astrometric binaries into several groups, depending on the properties of these binaries. In our simplified model, we consider binaries in their groups 1, 2d, 2e, 4, and 5 as single stars. Binaries in their groups 2a, 2b, 2c are assumed to have stochastic solutions (X), orbital solutions (O), and acceleration solutions (G), respectively. The binaries in group 3 are resolved with {\em Hipparcos}, and are classified as component solutions (C). {\em Hipparcos} binaries in the categories (X) and (G) have astrometric variations that indicate the presence of a companion, but orbital elements cannot be derived. For binaries in category (O) the orbital elements can be derived. In category (C) the individual stellar components are resolved, but the orbital solution cannot be found. This category includes the common proper motion pairs found with {\em Hipparcos}. The properties of each of the binaries in each of these categories according to \cite{lindgren1997} are listed in Table~\ref{table: hipparcosbiases}. 

The classification of \cite{lindgren1997} describes the properties of the {\em Hipparcos} binaries in each category. However, this does not necessarily mean that each binary system satisfying the constraints listed in  Table~\ref{table: hipparcosbiases} is actually detected and classified in this category. This problem is already obvious in the overlap between two of the four categories. For this reason, we only consider whether a binary system is detected with {\em Hipparcos} in our simulated observations. We do not discriminate between the four categories (O), (X), (G), and (C).

\subsubsection{The errors on the astrometric binary observables}

Given that we make no distinction between the different classes of astrometric
binaries we do not consider the errors on the astrometric observables. For a discussion on the photometric and astrometric errors of the {\em Hipparcos} satellite, we refer to \cite{hipparcos}.

% ====================================================================
% ====================================================================
% ====================================================================
% ==INTRODUCTION======================================================
% ====================================================================
% ====================================================================
% ====================================================================

\section{Simulated observations} \label{section: simulatedobservations}

In this section we study the dependence of the observables on the association properties and the binary parameter distributions.

% ====================================================================
% ====================================================================
% ====================================================================
% ==INTRODUCTION======================================================
% ====================================================================
% ====================================================================
% ====================================================================

\subsection{The distance to the association}

\begin{figure}[!tbp]
  \centering
  \includegraphics[width=0.7\textwidth,height=!]{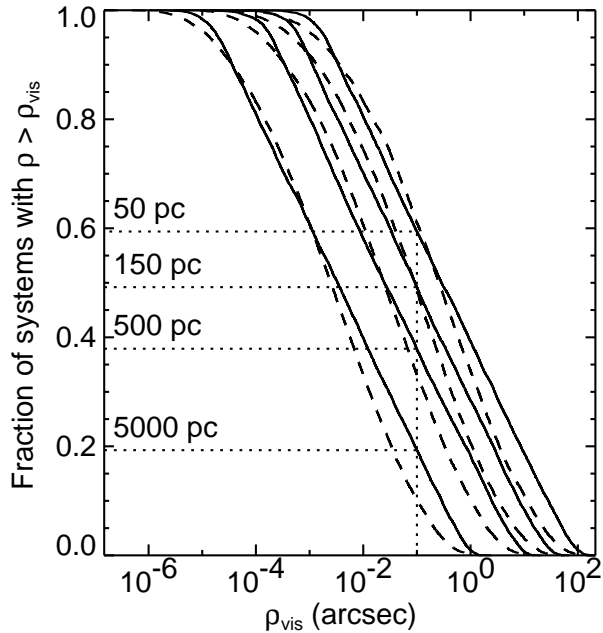}
  \caption{
%
% NB - het groter-dan teken is hier correct.
The fraction of binary systems with angular separation $\rho > \rho_{\rm vis}$
as a function of $\rho_{\rm vis}$. Each model association consists of $10\,000$ binaries,
has a thermal eccentricity distribution, and has a semi-major axis distribution
of the form $f_a(a) \propto \log a$ between 15~R$_\odot$ and  $10^6~$R$_\odot$
(solid curves) or a log-normal period distribution (dashed curves).  
From left to right, the solid curves represent the ``observed'' cumulative
distributions $F_\rho(\rho)$ for associations at a distance of 50, 150, 500, and
5000~pc, respectively. At these distances, the fraction of visual binaries
($\rho > 0.1''$) is 60\%, 49\%, 35\%, and 15\%, respectively, depending on the
semi-major axis distribution.  This simulation shows that for stellar groupings
at moderate distances ($D < 5$~kpc), a significant fraction of the companions
could potentially be resolved with direct imaging observations.
 \label{figure: visual_distance} }
\end{figure}

\begin{figure}[!tbp]
  \centering
  \includegraphics[width=1\textwidth,height=!]{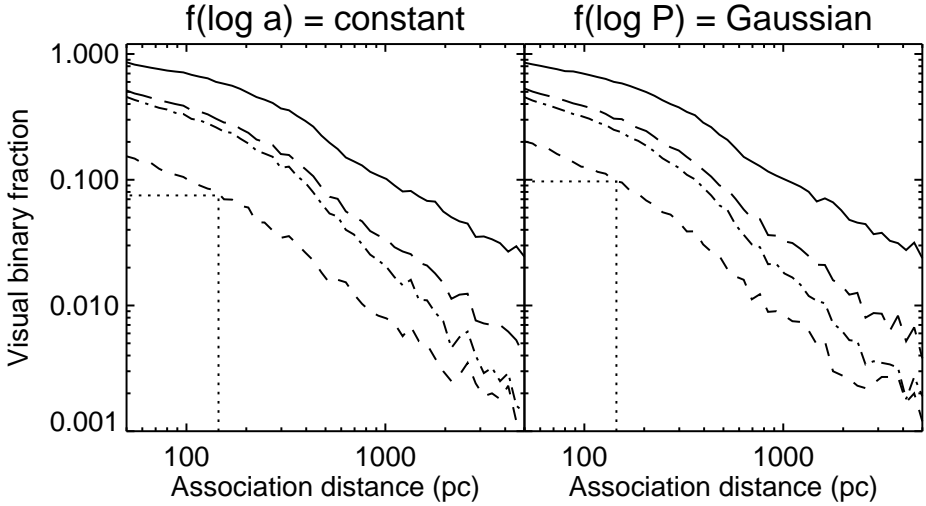}
  \caption{The ``observed'' fraction of visual binaries as a function of association
  distance, for \"{O}pik's law ({\em left}) and the log-normal period
  distribution of \cite{duquennoy1991} ({\em right}). The curves in each panel
  represent the fraction of binary systems detected with imaging surveys. Moving from
  top to bottom the curves represent a more realistic model for the
  observational biases. The top curve represents the brightness constraint ($V <
  17$~mag) for the target star. The long-dashed line also includes the
  separation constraint ($\rho \geq 0.1''$). The dash-dotted curve represents
  the fraction of binaries with the above constraints, and additionally the
  detection limit as a function of angular separation (equation~\ref{equation:
  classical_detectionlimit1}). Finally, we model the confusion with background
  stars (equation~\ref{equation: classical_detectionlimit2}), which is shown
  with the dashed curve. With this model for the visual biases, the fraction of
  binaries observed as visual binaries for an association at 145~pc is $\sim
  8\%$ for a model with \"{O}pik's law, and $\sim 10\%$ for a model with the
  \cite{duquennoy1991} period distribution.
 \label{figure: visual_distance_splitup} }
\end{figure}

\begin{figure}[!tbp]
  \centering
  \includegraphics[width=1\textwidth,height=!]{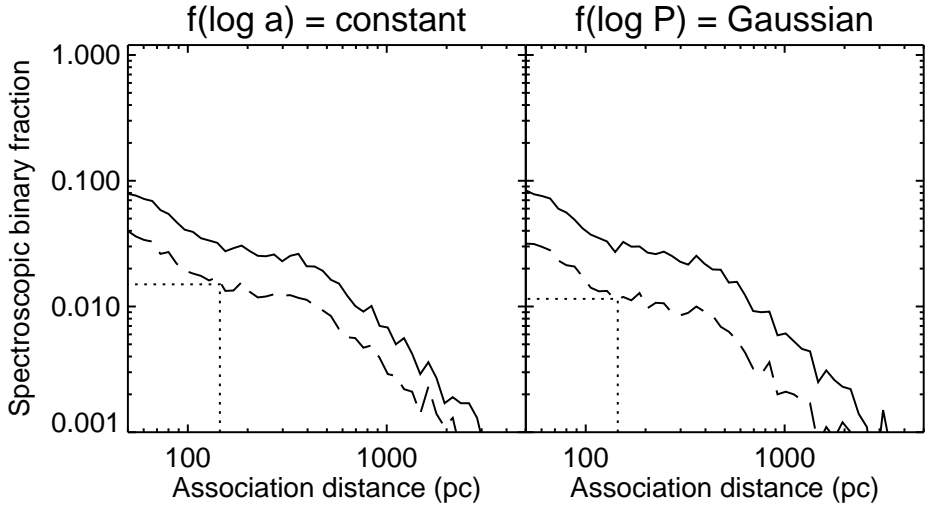}
  \caption{The fraction of spectroscopic binaries (relative to the total number
  of systems) as a function of association distance, for \"{O}pik's law ({\em
  left}) and the log-normal period distribution of \cite{duquennoy1991} ({\em
  right}). The curves in each panel represent the fraction of binary systems
  detected with radial velocity surveys. The binary fraction of each model is
  100\%. The solid curve respresents the fraction of binary systems that are
  bright enough for a spectroscopic survey ($V < 10$~mag). The dashed curve
  represents the fraction of binaries that also satisfies the instrument
  bias, i.e., the radial velocity amplitude is large enough, and the period is
  small enough, so that the presence of a companion can be inferred from the
  measured radial velocity variations. For a typical association at 145~pc
  about $1-2\%$ of all binary systems can be resolved spectroscopically in our
  model; these are mostly close and massive binaries. Among the systems with a
  {\em stellar} primary, the observed spectroscopic binary fraction is $2-5\%$. 
 \label{figure: spectroscopic_distance_splitup} }
\end{figure}

\begin{table}[!tbp]
  \begin{tabular}{l ccc}
    \hline
    \hline
    $D$     & Sample & $\tilde{F}_{\rm M,sample}$ & $\tilde{F}_{\rm M}$\\
    \hline
    50~pc   & 7.56\% & 40\% & 2.00\% \\
    150~pc  & 2.42\% & 45\% & 1.17\% \\
    500~pc  & 0.82\% & 48\% & 0.38\% \\
    5000~pc & 0.01\% & 80\% & $<0.01$\% \\
    \hline
    \hline
  \end{tabular}
  \caption{The effect of distance (column~1) on the results of an astrometric binarity survey with {\em Hipparcos}. Column~2 lists the fraction of systems that satisfy the brightness constraint (i.e., the {\em Hipparcos} detection limit). Column~3 lists the fraction of targets {\em in the sample} that is in one of the categories (X), (O), (G), or (C). The last column shows the observed fraction of binaries {\em in the complete association} that is detected as a binary star by {\em Hipparcos}. For each simulation we use default OB~association model, with $N=50000$ systems and binary fraction of 100\%. This table shows that, although {\em Hipparcos} observations are very useful for studying individual systems, only a very small fraction of the binaries in nearby OB~associations is discovered with this space mission.  \label{table: hipparcos_distance} 
  }
\end{table}

The distance $D$ to an OB association is not of physical importance.
Observationally, however, it is one of the most important parameters. A small
distance to an association facilitates a detailed observational study. The
projected angular separation of a binary system, and the projected velocity of a
star on the plane of the sky, scale with $D^{-1}$. The brightness of a star is proportional to $D^{-2}$. 

In the default model we assume an association distance of 145~pc. At this
distance, the lower limit $a_{\rm min}=15~$R$_\odot$ and upper limit $a_{\rm
max}=10^6~$R$_\odot$ correspond to a projected separation of 0.48~mas and
32~arcsec, respectively.  The projected angular separation can be arbitrarily
smaller than 0.48~mas due to projection effects. Observed angular separations
can also be a factor $(1+e)$ larger than $a/D$ when the binary system is in
apastron. For $a_{\rm min}/D \la \rho \la a_{\rm max}/D$, a good
approximation for $f_a(a)$ can be derived from the angular separation distribution, using
equation~\ref{equation: projecteda}. Outside these limits, this approximation is
no longer valid (see Section~\ref{section: sos_recovering_sma_and_period} 
and Appendix~D for a further discussion).

As a first order approximation for the number of binaries as a function of
distance, we define all binaries with $\rho \geq \rho_{\rm min} = 0.1''$ as
visual binaries (see~\ref{section: visual_separationconstraint}). This value for
$\rho_{\rm min}$ corresponds to the diffraction limit of a
10~meter telescope at infrared wavelengths. Figure~\ref{figure: visual_distance}
shows the cumulative fraction of simulated binary stars with $\rho < \rho_{\rm vis}$ as a
function of $\rho_{\rm vis}$. We show the results for the default model with
$f_{\log a}(\log a)=$~constant (solid curves), and for the log-normal period distribution
found by \cite{duquennoy1991} (dashed curves). For associations at a distance of
50, 150, 500, and 5000~pc, the fraction of visual binaries ($\rho > 0.1''$) in
the association is 59\%, 49\%, 38\%, and 19\%, respectively, for the default
model. The results for the model with the log-normal period distribution found
by \cite{duquennoy1991} are slightly lower for close ($D\la 200$~pc)
associations, and slightly higher for far ($D\ga 200$~pc) associations. A
semi-major axis distribution $f_{\log a}(\log a) =$~constant corresponds well to an
angular separation distribution $f_{\log \rho}(\log\rho)=$~constant. This is the reason for
the relatively shallow fall-off of the number of visual binaries with distance.

A more accurate calculation of the fraction of detected visual binaries is
obtained by also taking into account the brightness constraint, the contrast
constraint, and the confusion constraint. Figure~\ref{figure:
visual_distance_splitup} shows how each of these constraints reduce the
fraction of binaries observed as a visual binary. The left-hand panel shows the
results for the model with $f_{\log a}(\log a)=$~constant and the right-hand panel shows the
model with the log-normal period distribution. From top to bottom, the curves
indicate the observed binary fraction after one of the constraints has been added. The
model binary fraction for each association is 100\%. The solid line represents
the visual binaries satisfying the brightness constraint ($V \leq 17$~mag). For
distant associations, this is the dominant constraint. The long-dashed curve
indicates the fraction of detected visual binaries after the separation
constraint ($\rho \geq 0.1''$) has been taken into account. The lower two curves
represent the visual binary fraction when the contrast constraint and the
confusion constraint are also taken into account. For nearby
associations, many candidate companions are removed from the analysis as they
are confused with background stars. For associations at a large distance, the
confusion constraint is relatively unimportant as most companions are not
detected anyway.

All observables of a spectroscopic binary system are distance-independent, except the brightness of the binary system. As the brightness of a binary system is the strongest constraint for the detection of a binary system, the distance is an important parameter for radial velocity surveys. Figure~\ref{figure: spectroscopic_distance_splitup} shows the observed binary fraction as a function of association distance. We used the default model, and applied spectroscopic biases. The two panels show that the observed spectroscopic binary fraction for a model with \"{O}pik's law is very similar to that for model with a log-normal period distribution. Due to the brightness constraint, most binary systems cannot be observed (solid curves). When the other constraints are added (dashed curves), the spectroscopic binary fraction drops with another factor $2-3$. The resulting spectroscopic binary fraction is approximately inversely proportional to the association distance, dropping from $\sim 10\%$ for associations at 50~pc, to $0\%$ for assciations at 2000~pc.

The distance to the association is also an important parameter for astrometric surveys. Not only the brightness of the target, but also the angular size of the orbit and the angular velocities decrease with increasing distance. Table~\ref{table: hipparcos_distance} illustrates the distance dependence for our model of the {\em Hipparcos} mission. The majority of the systems in an association is not in the sample, as their brightness is below the {\em Hipparcos} detection limit. Depending on the distance to the association and the properties of its binary population, $\sim 25 \%$ of the binary systems that are in the {\em Hipparcos} sample, are indeed detected as such. The total fraction of binaries detected astrometrically, ranges from $\sim 2\%$ at a distance of 50~pc, down to 0\% for associations at $\sim 1\,500$~pc.

% ---------------------------------------------------------------------------------------
% ---------------------------------------------------------------------------------------
% ---------------------------------------------------------------------------------------

\subsection{The size of the association} \label{section: size}

In the default association model we assume that the stellar systems are homogeneously distributed in a sphere of radius $R=20$~pc at a distance of $D=145$~pc. Due to the extended nature of OB~associations, a natural spread in distance-dependent observables is expected; the distance to a binary is in the range between $D-R$ and $D+R$, the near and far side of the association, respectively. The angular separation difference $\Delta \rho$ relative to an angular separation $\rho$ in the association center, is given by
\begin{equation}
\frac{\Delta \rho}{\rho} = \frac{2DR}{D^2-R^2}\,,
\end{equation}
and the magnitude difference between stars at distances $D-R$ and $D+R$ is 
\begin{equation}
\Delta m = 5 \log \left( \frac{D+R}{D-R} \right)\,.
\end{equation}
The values of $\Delta\rho/\rho$ and $\Delta m$ are important for the interpretation of the observations if $R$ and $D$ have the same order of magnitude. Close visual binaries and faint companions are more easily found at the near-side of the association. If the spread in the observables due to the size of the association is large, this becomes an important selection effect. 

OB~associations have typical dimensions of $10-100$~pc \citep{brown2001}. For the nearby OB~associations listed in Table~1 of \cite{dezeeuw1999} the effects of association depth are $\Delta\rho/\rho = 1\%-100\%$ and $\Delta m = 0.02~{\rm mag}-2$~mag. The most nearby associations have a distance $D\approx 140$~pc and a size of $\sim 20$~pc, so that $\Delta\rho/\rho = 30\%$ and $\Delta m = 0.6$~mag. For a detailed analysis of these nearby associations, the size of the association should therefore be taken into account. The more distant associations have $R/D \ll 1$, for which $\Delta \rho/\rho$ and $\Delta m$ can be neglected.

\begin{figure}[!tbp]
  \centering
  \includegraphics[width=0.8\textwidth,height=!]{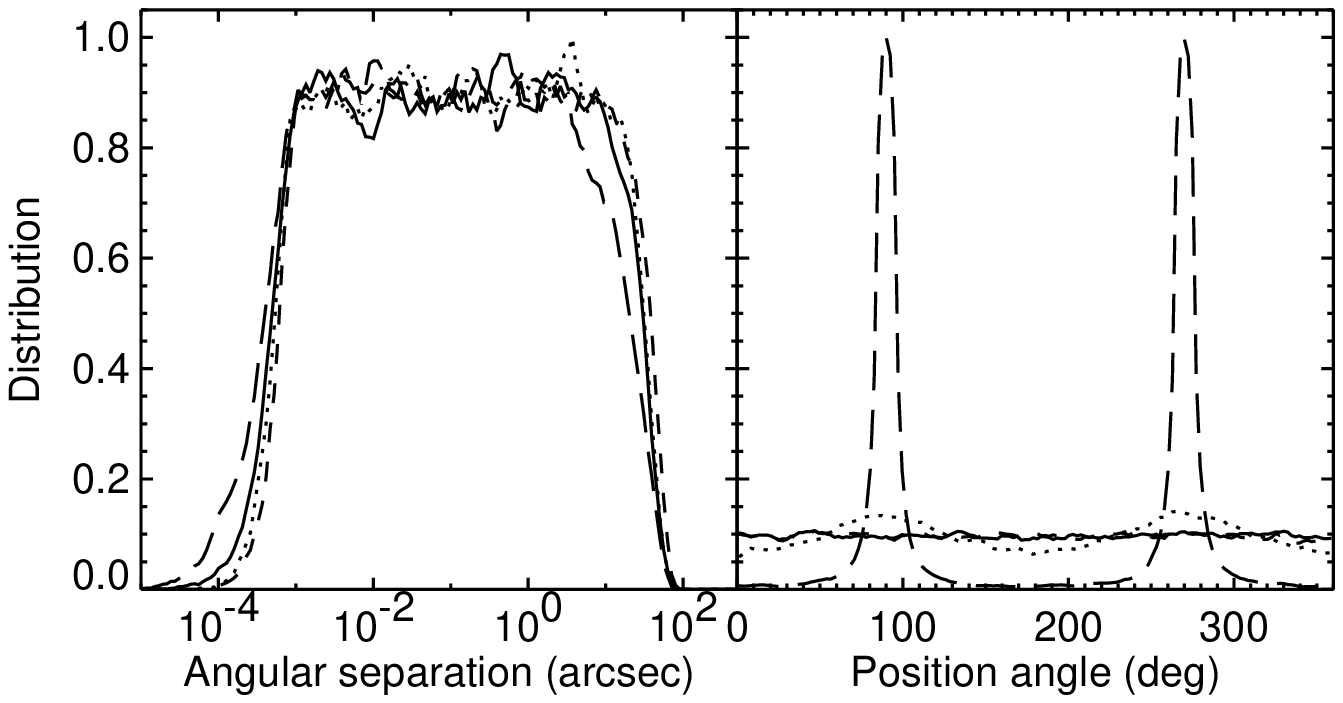}\\
  \includegraphics[width=1\textwidth,height=!]{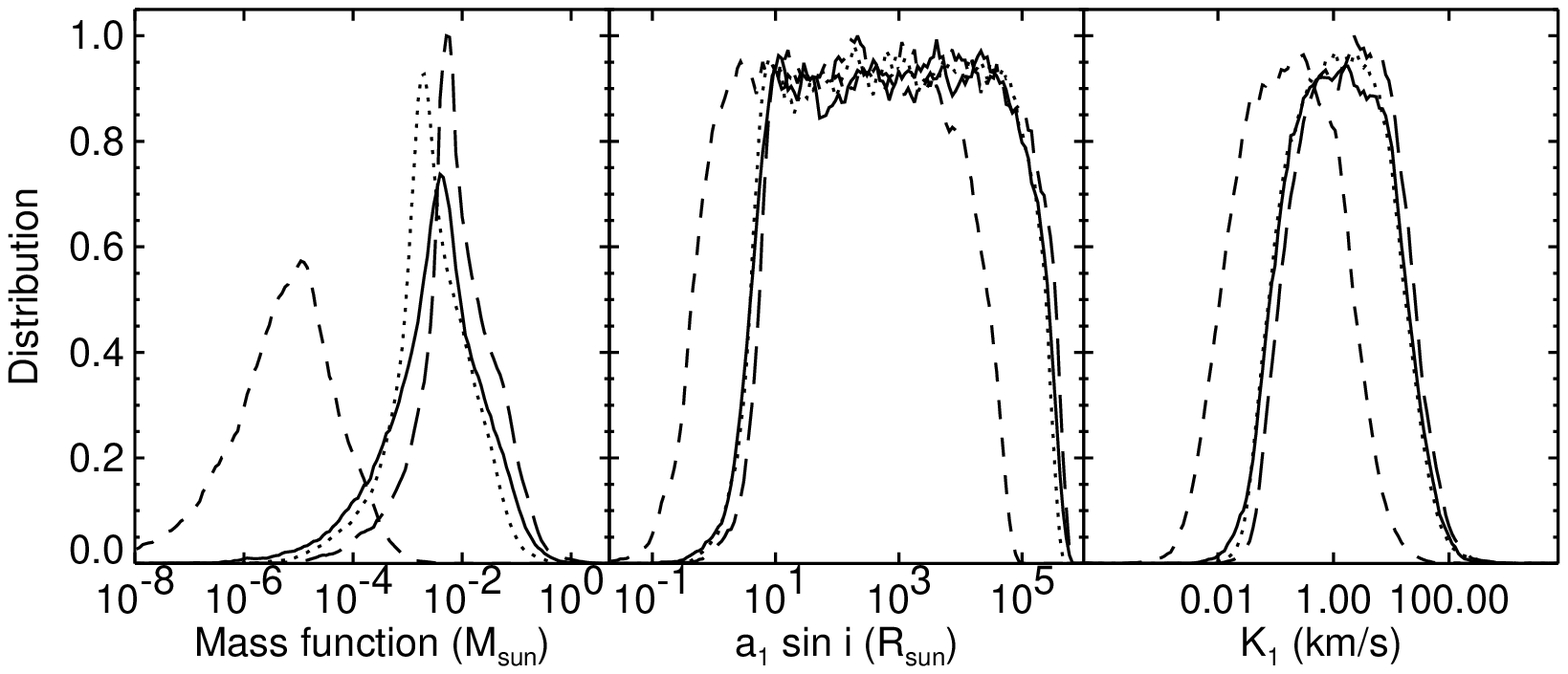}
  \caption{The effect of the orientation on the observables of visual ({\em
  top}) and spectroscopic ({\em bottom}) binaries, for the default association
  model ($D=145$~pc, $R=20$~pc) with $20\,000$ binary systems. The figure shows
  the curves for random orientation (solid curves), and for alignment angles of
  $\theta_a=0^\circ$ (dashed curves), $\theta_a=45^\circ$ (dotted curves), and
  $\theta_a=90^\circ$ (long-dashed curves).  No selection effects have been
  applied. This figure shows that generally, even with a very large number of
  measurements, the observables of visual and spectroscopic binaries cannot be
  used to discriminate between random orientation or preferred orientation
  distribution with any $\theta_a$. Only in those special cases for which an
  association has $\sin (2\theta_a) \approx 0$ (long-dashed and short-dashed
  curves) the effects of non-randomness of the orientation can be measured.
 \label{figure: effect_of_orientation} }
\end{figure}

% ====================================================================
% ====================================================================
% ====================================================================
% ==INTRODUCTION======================================================
% ====================================================================
% ====================================================================
% ====================================================================

\subsection{The orientation of the orbits} \label{section: orientationoftheorbits}

Throughout this paper we assume random orientation of binary systems. Only in this section we briefly discuss the hypothetical case that the binary systems in an association have a preferred orientation. This is important for theories of star and binary formation and dynamical evolution \citep[e.g.,][]{glebocki2000}. For the formation of binaries by capture, stellar motion and star streams might determine the orientation of binary orbits. In the case of binary formation by fragmentation, a more random orientation of orbits is expected. However, the conservation of angular momentum and the presence of a global magnetic field in the natal cloud may play a role in generating a preferred orientation of binary systems. Several investigations have been carried out, searching for a preferred orientation of binary systems in the solar neighborhood \citep[e.g.,][]{batten1967,gillett1988,glebocki2000}. These studies are hampered by (1) the difficulty of determining the orientation of a binary system, (2) observational selection effects, and (3) low-number statistics, and have not been able to discriminate the orientation distribution from random orientation.

Now suppose that all binary orbits in an association are not randomly oriented, but aligned along an ``alignment axis'', so that for all orbits the normal vector of the orbital plane is parallel to this alignment axis. We define the alignment angle $\theta_a$ as the angle between the line-of-sight to the association center and the alignment axis. All binaries now have an inclination $i \approx \theta_a$. Due to the finite size $R$ of an association, each binary system has a slightly different inclination.

A preferred orientation leaves a signature in the observables. However, in the
case that the orbits are aligned along an angle $\theta_a$, the signature is
often barely detectable, as its effect is smeared out due to projection effects, 
so that that the inclination distribution is
broadened. The distributions are further smeared out due to the spread in
semi-major axis, eccentricity, and orbital phase for the binaries. Only in the
special cases that $i$ and $\Omega$ can be measured directly for each system, or
$\sin (2\theta_a) \approx 0$, several possibilities can be excluded. Unless 
$\sin (2\theta_a) \approx 0$ visual and spectroscopic surveys cannot
distinguish between a specific orientation or a random orientation of the
orbits.

The observable distributions of visual and spectroscopic binaries are weakly
dependent on the preferred orientation of visual binaries. In practice these
distributions contain insufficient information to exclude random orientation,
which is illustrated in Figure~\ref{figure: effect_of_orientation}. Given the
observable distributions, only in the special cases that $\sin (2\theta_a) \approx 0$ 
a statement can be made regarding the orientation of the orbits. 
For example, in the case that all orbits in an
association are seen edge-on ($\cos\theta_a=0$), the position angle distribution
$f_\varphi(\varphi)$ peaks at $\varphi=\Omega$ and $\varphi=\Omega+180^\circ$. The
observables of spectroscopic binaries that contain information on the
orientation of binary systems are $K_1$, $a_1 \sin i$, and the mass function
$\mathcal{F}(M)$. The figure shows that (only) a preferred orientation with
$\sin\theta_a \approx 0$ can be identified, as all three observables
depend on $\sin i$, which is close to zero if $\cos \theta_a \approx 1$.  For
astrometric binaries, the absolute value of the inclination $|i|$ is directly
observed. Astrometric binaries can thus be used to constrain the orientation
distribution, provided that enough measurements are available, and that the
observational selection effects are well-understood. For random orientations the
inclination distribution has the form $f_{\cos i}(\cos i) = $~constant. If a preferred
orientation exists, $f_i(i)$ peaks at $i=\theta_a^*$, where $\theta_a^*$ is
the value of $\theta_a$ reduced to the interval $[0^\circ,90^\circ]$.

% ====================================================================
% ====================================================================
% ====================================================================
% ==INTRODUCTION======================================================
% ====================================================================
% ====================================================================
% ====================================================================

\subsection{The eccentricity distribution} \label{section: eccentricityobserved}

\begin{figure}[!tbp]
  \centering
  \includegraphics[width=0.7\textwidth,height=!]{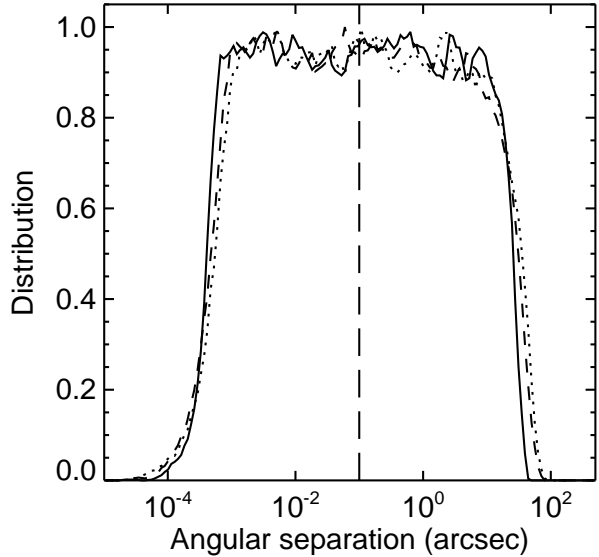}
  \caption{
The effect of the eccentricity distribution $f_e(e)$ on the angular separation distribution $f_\rho(\rho)$. The curves represent three eccentricity distributions: $e=0$ for all systems ({\it solid curve}), a thermal distribution $f_{2e}(e)=2e$ for $0\leq e \leq 1$ ({\it dashed curve}), and $e=0.9$ for all systems ({\it dotted curve}). We use the default association model with $20\,000$ binaries; no selection effects are applied. The vertical line indicates the typical limit  $\rho_{\rm lim} = 0.1''$ for visual binary observations. In this figure the different eccentricity distributions result in small, though measurable differences in $f_\rho(\rho)$. In practice, however, these differences cannot be seen, due to a significantly smaller number number of observations.  \label{figure: effect_of_eccentricity} }
\end{figure}

In Section~\ref{section: eccentricitychoice} we adopted the thermal eccentricity distribution in our default model. As the eccentricity distribution in OB~associations has not been accurately determined thus far, we study in this section the difference between the observations of associations with different eccentricity distributions.

The most common eccentricity distributions are the thermal eccentricity distribution 
\begin{equation}
f_{2e}(e) = 2e \quad \quad (0 \leq e < 1) 
\end{equation}
and the circular eccentricity distribution  $f_0(e) = \delta(e)$ ($e \equiv 0$). We describe the observational differences between models with these distributions. In order to study the effects of highly eccentric orbits, we also consider the highly-eccentric distribution $f_{0.9}(e) = \delta(e-0.9)$ ($e \equiv 0.9$). The eccentricity distribution is generally derived from observations of spectroscopic binaries. Below we describe the derivation, and also describe the possibilities of using visual and astrometric binaries to determine the eccentricity distribution.

% visual binaries - theory

As shown in Figure~\ref{figure: effect_of_eccentricity} it is in principle possible
to constrain the eccentricity distribution using only visual binaries. However,
this would require unrealistically large numbers of observations and an accurate
knowledge of $f_a(a)$. Thus, observational data of visual binary systems cannot be
used to constrain the eccentricity distribution.

% spectroscopic and astrometric binaries - theory

The eccentricity of an orbit can be measured directly with spectroscopic and astrometric observations. The observed distribution $\tilde{f}_e(e)$ can thus be used to derive the true eccentricity distribution $f_e(e)$. We will discuss the accuracy of this derivation in Section~\ref{section: recovering_eccentricity}.

% ====================================================================
% ====================================================================
% ====================================================================

\subsection{The mass distribution and pairing function} \label{section: simulations_massdistribution_pairingfunction}

\begin{figure}[!tbp]
  \centering
  \includegraphics[width=1\textwidth,height=!]{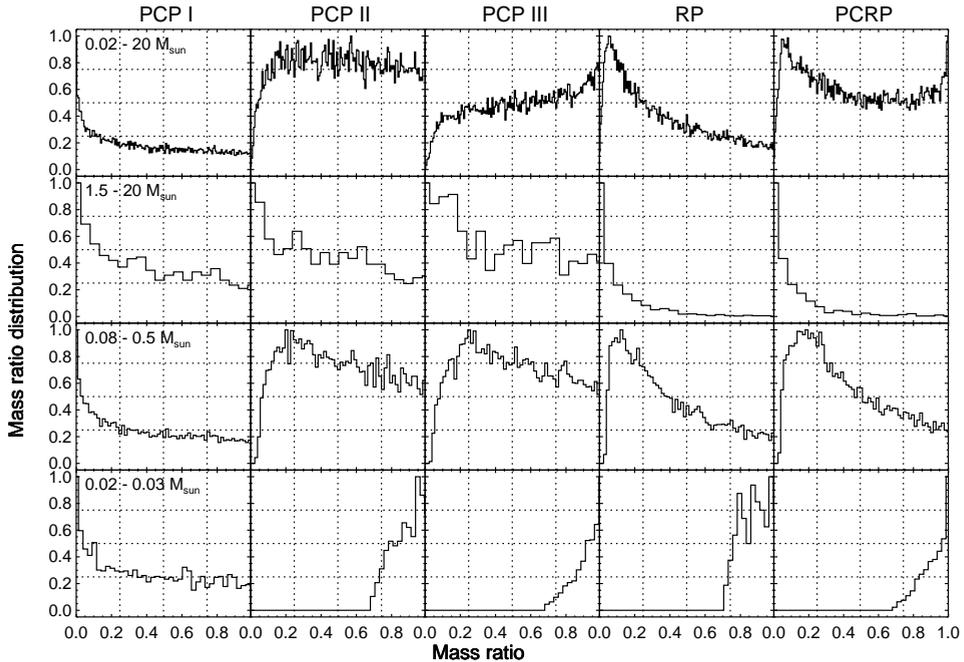}
  \caption{The mass ratio distribution for different pairing functions and subsamples. The panels show from left to right the mass ratio distribution resulting from the five pairing functions described in Section~\ref{section: pairingfunction}. The mass ratio distribution is shown for the complete sample ({\em top row}), the subsample with A and B-type primaries ({\em second row}), for the subsample with K and M-type primaries ({\em third row}), and for those with very low mass brown dwarf primaries ({\em fourth row}). The top panels show the {\em overall} mass ratio distribution; the middle and bottom panels show the {\em specific} mass ratio distributions for the corresponding primary mass range. Each mass ratio distribution is normalized so that the maximum is unity. Each model consists of $10^4$ binaries, and has a Preibisch mass distribution (equation~\ref{equation: sos_preibischimf}) with $\alpha=-0.9$. For PCP-I, PCP-II, and PCP-III the generating mass ratio distribution is $f_q(q) \propto q^{-0.33}$. This figure illustrates the importance of the choice of the pairing function. For PCP-I all subsamples show a mass ratio distribution similar to the generating one. The mass ratio distribution for PCP-II and PCP-III are significantly different from the generating distribution. For the subsample of high-mass stars, the three PCP pairing functions give similar results. For all pairing functions except PCP-I, binaries with low primary mass have on average a higher mass ratio than binaries with high primary mass.
 \label{figure: concept_pairing} }
\end{figure}

\begin{figure}[!tbp]
  \centering
  \includegraphics[width=1\textwidth,height=!]{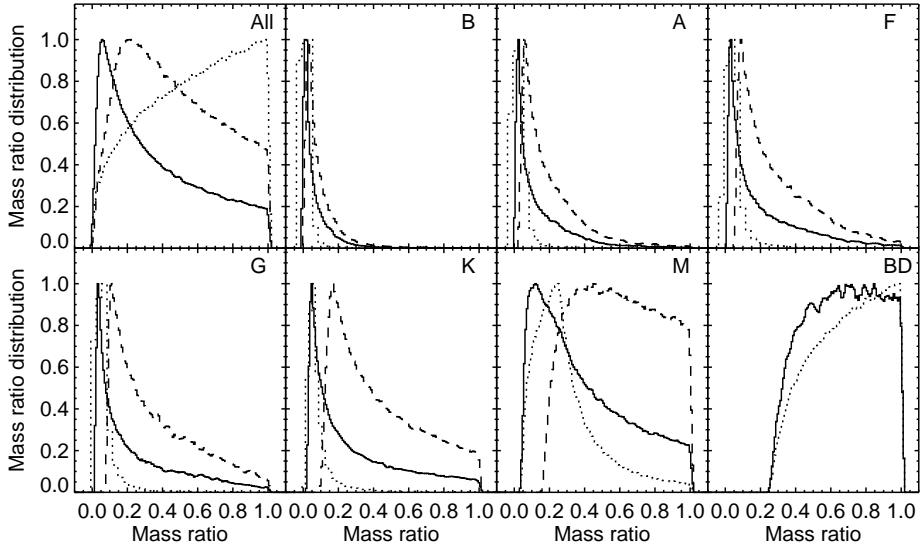}
  \caption{The mass ratio distribution for different subsamples as a result of random pairing. For each binary the two components are randomly drawn from the mass distribution, and swapped, if necessary, to make the primary the most massive star. In this figure we show the resulting mass ratio distribution for the extended Preibisch mass distribution with $\alpha=-0.9$ (solid curves), the extended Preibisch mass distribution without brown dwarfs (dashed curves) and the Salpeter mass distribution (dotted curves). The top-left panel shows the mass ratio distribution for the complete association. The other panels show the mass ratio distribution for the binaries with a primary of the spectral type shown in the top-right corner of each panel (All = all binaries included; BD = systems with brown dwarf primaries). All distributions are normalized so that their maximum is unity. Each model contains $5\times 10^5$ binaries. This figure illustrates that the shape of the mass ratio distribution strongly depends on the considered range in primary mass, as well as on the lower limit of the mass distribution and the shape of the mass distribution in the brown dwarf regime.
 \label{figure: rp_different_samples} }
\end{figure}

\begin{figure}[!tbp]
  \centering
    \includegraphics[width=\textwidth,height=!]{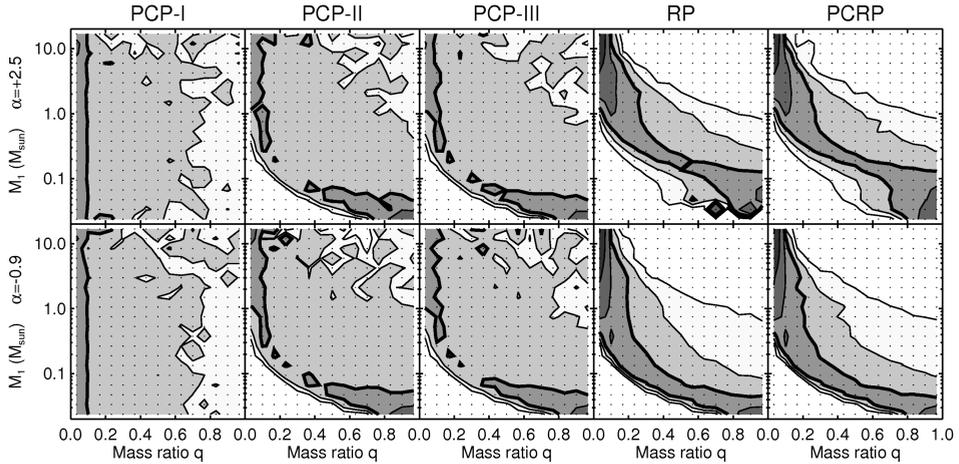} 
    \caption{Each panel in this figure shows the same as Figure~\ref{figure: concept_pairing}, the
    {\em specific} mass ratio distribution $f_{M_1}(q)$, but now for all
    primary masses between $0.02$ and $20$ solar masses. The grey scale and
    contours indicate the values of $f_{M_1}(q)$ for each combination of $M_1$
    (vertical axis) and $q$ (horizontal axis). That is, each horizontal cut
    across a panel corresponds to a mass ratio distribution as shown in
    Figure~\ref{figure: concept_pairing}.
  From left to right, the panels show the results for
  pairing functions PCP-I, PCP-II, PCP-III, RP, and PCRP. The top and bottom
  panels show the results for mass distributions with $\alpha=+2.5$ and $\alpha=-0.9$, respectively. In
  each panel the largest
  values of $f_{M_1}(q)$ correspond to the darkest colors. Contours are shown
  for $f_{M_1}(q) = 0.02, 0.05, 0.1$ (thick curve) $0.2, 0.5, 0.9$. Each
  association model contains $10^5$ binaries. Each dot represents a bin in mass
  ratio and logarithmic primary mass. No selection effects have been applied.
  For large values of $M_1$, the specific mass
  ratio distributions resulting from the three PCP pairing functions are
  approximately equal. The same holds for the specific mass ratio distributions
  resulting from RP and PCRP. Differences between $\alpha=+2.5$ and
  $\alpha=-0.9$ are only seen for RP and PCRP, as $f_{M_1}(q)$ is
  independent of $\alpha$ for pairing functions PCP-I, PCP-II, and PCP-III. 
  \label{figure: massratio_2d_fq} }
\end{figure}

\begin{figure}[!tbp]
  \centering
    \includegraphics[width=\textwidth,height=!]{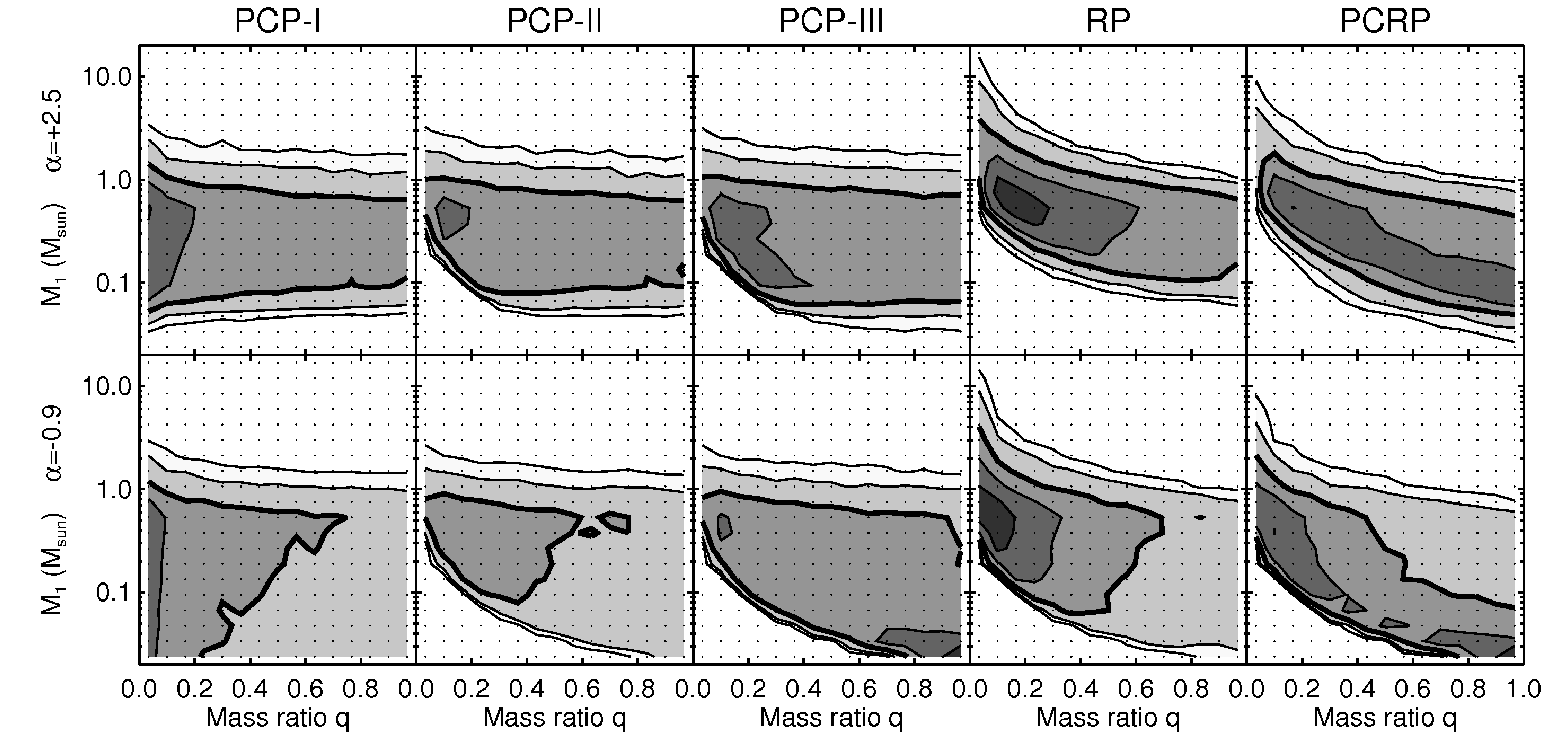}
  \caption{The distribution of binary systems over primary mass $M_1$ and mass
  ratio $q$, for different pairing functions and values of $\alpha$. Note that
  in contrast to Figure~\ref{figure: massratio_2d_fq} this shows the
  two-dimensional distribution in the $(q,\log M_1)$ plane, and that the vertical
  scale is logarithmic.  From left to right, the panels show the results for
  pairing functions PCP-I, PCP-II, PCP-III, RP, and PCRP. The top and bottom
  panels show the results for
  $\alpha=+2.5$ and $\alpha=-0.9$, respectively. The colors represent the
  fraction of binary systems in each $(q,\log M_1)$ bin. Darker colors
  correspond to larger fractions. The contours represent values of 0.01\%,
  0.02\%, 0.05\% (thick curve), 0.1\%, and 0.2\% of the binaries. Each
  association model contains $10^5$ binaries. Each dot represents a bin in mass
  ratio and logarithmic primary mass. No selection effects have been applied.
  For pairing functions PCP-I, PCP-II, and PCP-II, the distribution of binary
  systems over $(q,M_1)$ is identical for massive primaries. The figure also
  shows the importance/unimportance (depending on the pairing function) 
  of $\alpha$ in the analysis of mass ratio distributions. 
  \label{figure: massratio_2d_fm} }
\end{figure}

In previous simulations the mass of a star is usually drawn from a mass distribution $f_M(M)$ in the mass range $M_{\rm min} \leq M \leq M_{\rm max}$. This mass distribution is often assumed to be the {\em single star} mass distribution (e.g., equations~\ref{equation: sos_preibischimf}--\ref{equation: kroupaimf}). The slope $\alpha$ of the mass distribution in the brown dwarf regime is poorly constrained by observations. In this paper we will therefore simulate associations with different values of $\alpha$. 
For the extended Preibisch mass distribution with $\alpha=-1.97$, the number of stellar objects is equal to the number of substellar objects. For $\alpha > +2.5$, the fraction of brown dwarfs among the stars is less than $8.5\%$. 
These fractions do not depend on the pairing function, unless random pairing is selected, although in the latter case the whole set of stars (singles/primaries plus companions) has the corresponding substellar-to-stellar ratio.

After the members of an association have been drawn from the mass distribution, a fraction of the stars is assigned a companion star, the mass of which is defined according to some recipe. There are many ways to obtain $M_1$ and $M_2$ using mass distribution $f_M(M)$, of which we list some below.
\begin{itemize}\addtolength{\itemsep}{-0.5\baselineskip}
\item[--] {\em Random pairing } (RP). The masses of the two binary components are randomly drawn from the mass distribution $f_M(M)$. For each system, the most massive component is labeled ``primary'', the other component ``companion'' (see \S~\ref{section: pairingfunction_rp}).
\item[--] {\em Primary-constrained random pairing} (PCRP). The primary mass $M_1$ is drawn from the mass distribution $f_M(M)$. Subsequently, the companion mass $M_2$ is chosen from the same mass distribution, with the constraint $M_2 \leq M_1$ (see \S~\ref{section: pairingfunction_pcrp}).
\item[--] {\em Primary-constrained pairing} (PCP). The primary mass is drawn from the mass distribution $f_M(M)$. The companion mass is then determined by a mass ratio that is drawn from a mass ratio distribution $f_q(q)$, where $0 \leq q_{\rm min} \leq q \leq q_{\rm max} \leq 1$ (see below). 
\end{itemize}
In the case of PCP there is another complication, which occurs if $M_{\rm 2,min}/M_{\rm 1,max} < q_{\rm min}$ (where $q_{\rm min}$ may be constant or dependent on $M_1$). If this is the case, the mass ratio distribution generates companion masses smaller than the permitted value $M_{\rm 2,min}$. There are three obvious choices to handle such companions:
\begin{itemize}\addtolength{\itemsep}{-0.5\baselineskip}
\item[--] {\em Accept all companions} (PCP-I). If $M_{\rm 2,min}/M_{\rm 1,max} \geq q_{\rm min}$, all companions have mass $M_2 \geq M_{\rm 2,min}$. All companions are accepted. The resulting mass ratio distribution obtained with this method is equivalent to the generating mass ratio distribution (see \S~\ref{section: pairingfunction1}).
\item[--] {\em Reject low-mass companions} (PCP-II). If $M_{\rm 2,min}/M_{\rm 1,max} < q_{\rm min}$, a fraction of the companions has $M_2 < M_{\rm 2,min}$. All companions with  $M_2 < M_{\rm 2,min}$ are rejected in the case of PCP-II, and the corresponding ``primaries'' remain single (see \S~\ref{section: pairingfunction2}). 
\item[--] {\em Redraw low-mass companions} (PCP-III). If $M_{\rm 2,min}/M_{\rm 1,max} < q_{\rm min}$, a fraction of the companions has $M_2 < M_{\rm 2,min}$. For all companions with $M_2 < M_{\rm 2,min}$, the mass ratio is redrawn from $f_q(q)$. This procedure is repeated until $M_2 \geq M_{\rm 2,min}$. This method is equivalent to drawing a mass ratio from the distribution $f_q(q)$ with limits $M_{\rm 2,min}/M_1 \leq q \leq 1$ (see \S~\ref{section: pairingfunction3}). 
\end{itemize} 
Above we have made the assumption that the distributions $f_M(M)$ are equal to the single star mass distribution. As currently no generally accepted (theoretical or observational) prescription for the mass distribution of binary stars is available, we restrict ourselves to using the single star mass distribution and a recipe to obtain the primary and companion mass of a binary system from $f_M(M)$.  The above-mentioned five pairing functions (RP, PCRP, and the three variations of PCP) are described in detail in Appendix~B. The main differences between the five pairing functions are shown with the examples in Figures~\ref{figure: concept_pairing} and~\ref{figure: rp_different_samples}, and in a more general way in Figures~\ref{figure: massratio_2d_fq} and~\ref{figure: massratio_2d_fm}. Below, we summarize several important conclusions.  

Each of the five pairing functions results in different properties of the binary population. The top panels in Figure~\ref{figure: concept_pairing} show the significant difference between the overall mass ratio distribution for the different pairing functions. The other panels in Figure~\ref{figure: concept_pairing} show for each pairing function the {\em specific} mass ratio distribution for target samples of different spectral types.

For pairing functions RP, PCRP, PCP-II, and PCP-III, the specific mass ratio distribution is a function of primary spectral type (see Figure~\ref{figure: concept_pairing}). For each of these pairing functions, high-mass binaries have on average a lower mass ratio than low-mass binaries. The lowest-mass binaries have a mass ratio distribution peaked to $q=1$. Only for pairing function PCP-I, the mass ratio distribution is independent of spectral type.

For pairing functions RP, PCRP, PCP-II, and PCP-III, the {\em overall} mass ratio distribution depends on the mass distribution. The {\em specific} mass ratio distribution is a function of primary spectral type for these pairing functions. As the number of binaries with a certain spectral type depends on the mass distribution, the {\em overall} mass ratio distribution, which is often dominated by the low-mass binaries, depends on the mass distribution.

For pairing functions RP and PCRP, the {\em specific} mass ratio distribution depends on the mass distribution (see, e.g., Figure~\ref{figure: rp_different_samples}). For RP and PCRP we draw both components from the mass distribution (albeit with the additional constraint $M_2 \leq M_1$ for PCRP), which results in a mass distribution dependence. As the specific mass ratio distribution for RP and PCRP depends on $f_M(M)$, it can in principle be used to constrain the properties of $f_M(M)$ in the substellar regime, under the assumption that RP or PCRP are truly applicable. 

For pairing functions RP and PCP-II, the binary fraction is a function of spectral type, which is illustrated in Figures~\ref{figure: binaryfraction_randompairing} (for RP) and~\ref{figure: binaryfraction_pcp2} (for PCP-II). For both pairing functions, the binary fraction increases with increasing primary mass.  Note that the trend between binary fraction and primary mass for RP and PCP-II is merely a result of the choice of the pairing method. In principle, the binary fraction can therefore be used to identify pairing functions RP and PCP-II.

Figure~\ref{figure: massratio_2d_fq} shows a generalized version of Figure~\ref{figure: concept_pairing}. In each panel, this figure shows the specific mass ratio distribution for different primary masses. The grayscale represents the value of $f_{q;M_1}(q)$. For a given value of $M_1$ (horizontal line) one gets the mass ratio distribution $f_{q;M_1}(q)$. The figure illustrates that for the three pairing functions PCP-I, PCP-II, and PCP-III the specific mass ratio distribution for binaries with high-mass primaries is approximately equal to the overall mass ratio distribution:
\begin{equation}\label{equation: pcp_fq_highm1}
f_q(q) \approx f_{{\rm I;high\,}M_1}(q) \approx f_{{\rm II;high\,}M_1}(q) \approx f_{{\rm III;high\,}M_1}(q) \,.
\end{equation}
For lower mass primaries ($M \la 0.5$~M$_\odot$) the specific mass ratio distributions differ. The results show only a minor dependence on $\alpha$.
Figure~\ref{figure: massratio_2d_fm} shows how the binaries in an association
are distributed over primary mass $M_1$ and mass ratio $q$. For pairing functions PCP-I, PCP-II, and PCP-III, the specific binary fraction for binaries with high-mass stars is approximately equal to the overall binary fraction:
\begin{equation}\label{equation: pcp_bf_highm1}
F_{\rm M} \approx F_{{\rm M,I;high}\,M_1} \approx F_{{\rm M,II;high\,}M_1} \approx F_{{\rm M,III;high\,}M_1} \,.
\end{equation}

Figures~\ref{figure: concept_pairing} and~\ref{figure: rp_different_samples} illustrate the strong dependence of the specific mass ratio distribution on the targeted sample in the survey. Caution should therefore be taken when extrapolating the results to the association as a whole. The interpretation of the observations is further complicated by the instrument bias and observational errors, an effect we will discuss in Section~\ref{section: recovering_fm_fq}.

The total mass of an OB~association or star cluster is usually derived from the mass distribution for single stars, often extrapolated into the brown dwarf regime. If a large fraction of the members are binary or multiple systems, the total mass calculated using the single star mass distribution underestimates the true total mass. In Appendix~C we discuss for each of the five pairing functions the implications of not properly taking into account the binary systems in the total mass calculation.

Note that it is possible that pairing mechanisms other than those described above operate during the star forming process. It is for example possible to draw the {\em total mass} of the binary system from the single-star mass distribution, and divide this mass over the masses of the companions. A detailed analysis of observations may reveal the ``natural'' pairing function associated with the outcome of star formation.

% ====================================================================
% ====================================================================
% ====================================================================
% ==INTRODUCTION======================================================
% ====================================================================
% ====================================================================
% ====================================================================

\subsection{The semi-major axis and period distributions} \label{section: sma_and_period}

\begin{figure}[!tbp]
  \centering
  \includegraphics[width=1\textwidth,height=!]{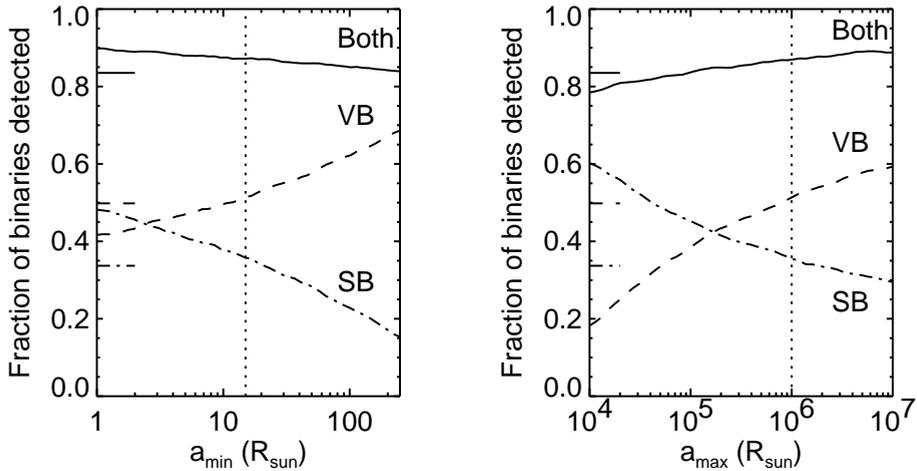}
  \caption{The fraction of visual and spectroscopic binary systems, as a function of the minimum semi-major axis $a_{\rm min}$ and maximum semi-major axis $a_{\rm max}$, for an OB~association at a distance of 145~pc. The semi-major axis distribution of each model is given by \"{O}pik's law, with $a_{\rm min} \leq a \leq a_{\rm max}$. In the left-hand panel, we vary $a_{\rm min}$, while we keep $a_{\rm max}$ fixed at $10^6$~R$_\odot$. In the right-hand panel, we vary $a_{\rm max}$, while we keep $a_{\rm min}$ fixed at $15$~R$_\odot$. The dotted lines in each panel represent the values of $a_{\rm min}$ and $a_{\rm max}$ in our default model. In this figure a visual binary is defined as a binary with angular separation $\rho \geq 0.1''$. A spectroscopic binary has $K_1 \geq 2$~km\,s$^{-1}$ and $P \leq 30$~year. The curves in each panel represent the fraction of visual binaries (dashed curve), the number of spectroscopic binaries (dash-dotted curve), and the total number of binaries with the properties mentioned above (solid curve). The short lines on the left represent the results for an association with the log-normal period distribution of \cite{duquennoy1991}, with $a_{\rm min} = 15$~R$_\odot$ and $a_{\rm max} = 10^6$~R$_\odot$.
  \label{figure: amin_amax} }
\end{figure}

\begin{figure}[!tbp]
  \centering
  \includegraphics[width=0.9\textwidth,height=!]{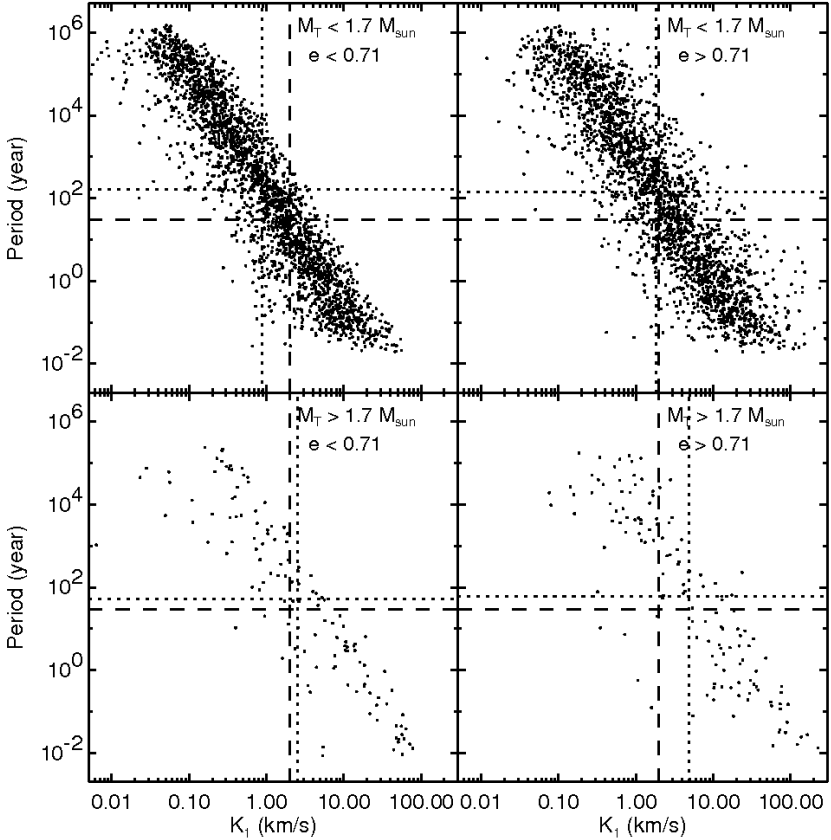}
  \caption{Only a small fraction of the binary systems in an association are resolved as spectroscopic binaries. In this figure we illustrate the importance of the observational constraints for an association consisting of 5\,000 binary systems. Each panel shows for a subgroup of systems the distribution of orbital period $P$ and semi-amplitude of the radial velocity $K_1$. The top and bottom panels show the binary systems with total mass smaller and larger than $1.7$~M$_\odot$, respectively. At a distance of 145~pc, a star of this mass has $V\approx 10$~mag, so that for such an association only the targets in the bottom panels are bright enough to be included in a spectroscopic binarity survey. Typically, radial velocity surveys can only resolve binary systems with $K_1 \ge 2$~km\,s$^{-1}$ (vertical dashed line) and $P \le 30$~year (horizontal dashed line), i.e., those in the bottom-right quadrant of the bottom panels. Several binaries these quadrants are still unresolved, due to an unfavourable combination of ($e$,$\omega$). In each panel we also plot the average $\log K_1$ (horizontal dotted line) and the average $\log P$ (vertical dotted line), illustrating that more massive binaries tend to have a shorter orbital period, and that the value of $K_1$ tends to increase with increasing eccentricity.
  \label{figure: sb-c_k_p_diagram} }
\end{figure}

\begin{figure}[!tbp]
  \centering
  \includegraphics[width=1\textwidth,height=!]{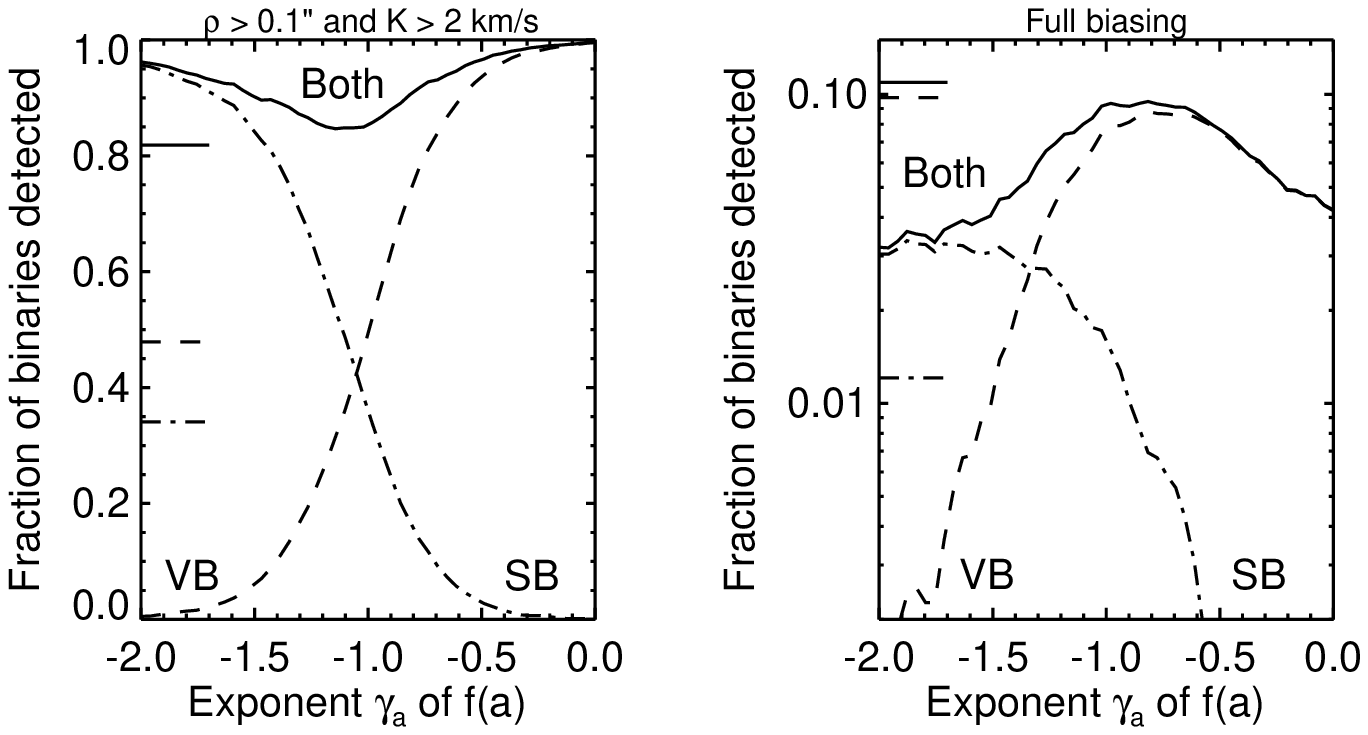}
  \caption{The number of detected visual and spectroscopic binaries as a function of semi-major axis distribution $f_a(a)$. We create models with a semi-major axis distribution of the form $f_a(a) \propto a^{\gamma_a}$, with different values of $\gamma_a$, and simulate observations. The left-hand panel shows the model with the ``simple'' simulated observations ($\rho > 0.1''$ for visual binaries, $K_1 > 2$~km\,s$^{-1}$). The right-hand panel shows the correctly modeled selection effects for the visual and spectroscopic binaries (note that the scale is logarithmic). The dashed and dash-dotted curves correspond to the visual and spectroscopic binary fractions, respectively; the solid curve corresponds to the combined dataset. In each panel we show on the left the results for the model with a log-normal period distribution. This figure shows that the observed spectroscopic and visual binary fraction strongly depend on the shape of $f_a(a)$, and can thus be used to constrain this distribution.
  \label{figure: visual_gamma_a_numvisbin} }
\end{figure}

In this section we discuss how the shape of the semi-major axis distribution $f_a(a)$, or equivalently the period distribution $f_P(P)$ propagates into the observations. We demonstrate how the minimum and maximum orbital dimensions of a binary population are revealed in the observed visual and spectroscopic binary fractions. We will focus on two distributions that determine the orbital dimensions of a binary system: a power-law semi-major axis distribution $f_{\gamma_a}(a) \propto a^{\gamma_a}$, and the log-normal period distribution found by \cite{duquennoy1991}, which is given in equation~\ref{equation: duquennoyperiods}.

In Section~\ref{section: smaandperiod} we adopted for the default model a mass ratio distribution $f_a(a) \propto a^{-1}$, with $a_{\rm min} = 15~$R$_\odot$ and $a_{\rm max} =  10^6~$R$_\odot$. We argued that the true values of $a_{\rm min}$ and  $a_{\rm max}$ are likely near the adopted values, but yet unknown. The values of $a_{\rm min}$ and  $a_{\rm max}$ determine how many binary systems are observed as close spectroscopic binaries and as wide visual binaries. Figure~\ref{figure: amin_amax} shows the visual and spectroscopic binary fraction as a function of $a_{\rm min}$ and  $a_{\rm max}$, for an association at a distance of 145~pc. For this figure we consider all binaries with $\rho \geq 0.1''$ as visual binaries, and all systems with $K_1 \geq 2$~km\,s${-1}$ and $P \leq 30$~year as spectroscopic binaries. The figure clearly shows that the number of spectroscopic binaries decreases with increasing $a_{\rm min}$ and  $a_{\rm max}$. The number of visual binaries increases strongly with increasing $a_{\rm min}$ and  $a_{\rm max}$. The fraction of binaries observable with either visual or spectroscopic techniques, is more or less constant. The figure also shows that the fraction of visual and spectroscopic binaries in our default model (dotted lines), is very similar to that for a model with the log-normal period distribution proposed by \cite{duquennoy1991}.

For a visual binary system, an estimate for the semi-major axis can be obtained from the angular separation $\rho$ between the binary components, and equation~\ref{equation: projecteda}. The orbital period can then be derived through Kepler's third law, using an estimate of the total mass of the binary system. For a spectroscopic binary, the orbital period can be measured directly, and the semi-major axis can be derived using mass estimates. For astrometric binaries the orbital period and semi-major axis can be derived independently (if the distance to the binary is known).

We first focus on how the distributions $f_a(a)$ and $f_P(P)$ affect the angular separation distribution of visual binaries. The other observables of visual binaries (the luminosity of both components and the position angle) are not affected by the choice of $f_a(a)$. 

The fraction of visual binaries $F_{\rm M, vis}$ as a function of $\gamma_a$, where we define a visual binary as any binary with an angular separation $\rho \geq \rho_{\rm lim} \approx 0.1''$ at the time of observation, depends strongly on the value of $\gamma_a$, and can be estimated with
\begin{equation}
F_{\rm M, vis} = 
\frac{ F_a(a_{\rm max})-F_a(D\,\rho_{\rm lim})   }{ F_a(a_{\rm max})  }
= \frac{ \left( 1-  D\,\rho_{\rm lim}/ a_{\rm max} \right)^{\gamma_a+1}  }
       { \left( 1-  a_{\rm min} / a_{\rm max}      \right)^{\gamma_a+1}  },
\end{equation}
where $F_a(a)$ is the cumulative distribution of $a$, $D$ is the distance to the association, and $\gamma_a \neq -1$. This approximation is only valid when $a_{\rm min} \la D \,\rho_{\rm lim} \la a_{\rm max}$. 

Spectroscopic binary systems provide information on the period distribution $f_P(P)$ for very small values of $P$. For a nearby OB~association, many candidate binaries cannot be surveyed spectroscopically, as their luminosity is too low to perform proper spectral analysis. The high-mass (i.e., bright) targets can be surveyed for binarity. For these targets, the most important parameter for the detection of orbital motion is the semi-amplitude of the radial velocity $K_1$ (equation~\ref{equation: radialvelocitycurve_k}). Using Kepler's third law (equation~\ref{equation: keplerslaw}), $K_1$ can be written as
\begin{equation} \label{equation: keplerslaw_alternative}
K_1 = M_2 \sin i \sqrt{\frac{G}{aM_T(1-e^2)}}\,.
\end{equation}
For an association the variation in $a$ is much larger than that of $M_T$, $M_2$, and $e$, making $K_1$ an important parameter for spectroscopic binary detections.

Figure~\ref{figure: sb-c_k_p_diagram} shows the distribution over $K_1$ and orbital period $P$ for an association consisting of 5\,000 binary systems. The figure clearly shows that $K_1$ increases with decreasing values of $P$ and $e$, making these orbits more easily detectable. The figure also shows the very small fraction of binary systems that can be detected spectroscopically, under the detection constraints adopted in this paper.

Above we have seen that the observable distributions of visual and spectroscopic
binaries depend strongly on $f_a(a)$ and $f_P(P)$. The fraction of binary systems
that is observed as a visual or spectroscopic binary also sensitively depends on
$f_a(a)$ and $f_P(P)$, which is illustrated in Figure~\ref{figure:
visual_gamma_a_numvisbin} for a power-law semi-major axis distribution and the
log-normal period distribution. The figure shows the fraction of visual and
spectroscopic binaries for different distributions $f_a(a)$ and $f_P(P)$, for an
association at a distance of 145~pc. The left-hand panel shows the case where all
companions with $\rho > 0.1''$ in the association are detected with imaging
surveys, and all binaries with $K_1 > 2$~km\,s$^{-1}$ with a spectroscopic
survey. Under these (optimistic) conditions, over 80\% of the binaries are
detected.  The right-hand panel in Figure~\ref{figure: visual_gamma_a_numvisbin}
shows the result for more realistic biasing, where the brightness, contrast,
and confusion constraints are taken into account. For high values of
$\gamma_a$ many companions are wide, and cannot be identified anymore as such.
The visual binary fraction therefore drops for high $\gamma_a$. The observed
overall binary fraction is $2.5\%-10\%$, depending on the shape of $f_a(a)$. The
specific binary fraction of {\em stellar} targets ($M_1 \ga 0.08$~M$_\odot$) is
$7\%-15\%$, while for the intermediate and high mass targets ($M_1 \ga
1$~M$_\odot$) the observed specific binary fraction is $40\%-55\%$.

% ====================================================================
% ====================================================================
% ====================================================================
% ==INTRODUCTION======================================================
% ====================================================================
% ====================================================================
% ====================================================================

\subsection{The binary fraction}

\begin{figure}[!tbp]
  \centering
  \includegraphics[width=0.8\textwidth,height=!]{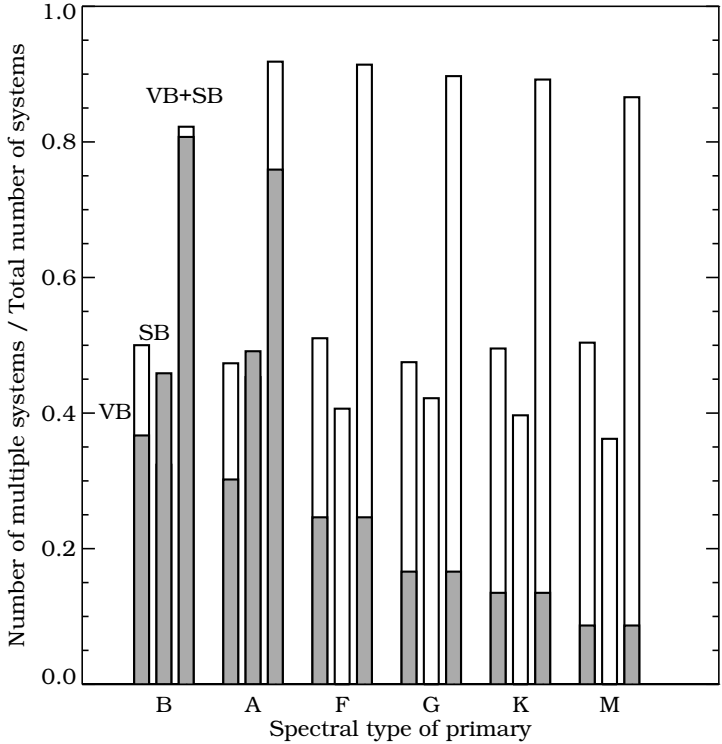}
  \caption{The observed binary fraction as a function of primary spectral type, for an association at a distance of 145~pc, with a binary fraction of 100\%. For each primary spectral type, the bars show the fraction of visual binaries (left), the fraction of spectroscopic binaries (middle), and the combined binary fraction (right). The grey part of each bar represents the detected binary fraction in our simulated observations. The white part of each bar represents the visual binaries with $\rho \geq 0.1''$ and spectroscopic binaries with $K_1 \geq 2$~km\,s$^{-1}$ that remain undetected due to their low brightness. The {\em observed} dataset (gray bars) suggests a trend in binary fraction with the primary spectral type, while in reality, no such trend is present.
  \label{figure: mulspt_sbs_vbc} }
\end{figure}

The binary fraction $F_{\rm M}$ of an association is the most fundamental parameter that describes the properties of the binary population. Even though the binary fraction is of such importance, it can (generally) only be derived when all other properties of the binary population are known. It is practically impossible to detect {\em all} binary systems in a stellar grouping. For this reason, the observed binary fraction $\tilde{F}_{\rm M}$ has to be corrected for the ``missing'' binaries, in order to obtain the true binary fraction $F_{\rm M}$. 
For example, if we do not know $a_{\rm max}$ or $f_q(q)$ for a population, we do not know how many binaries are undetected in our survey, being too wide, or too faint.

If no detailed information on the binary population is available, a lower limit on the binary fraction can be obtained from observations. The number of {\em detected} binary systems, relative to the total number of systems (singles and binaries), provides an absolute lower limit to the binary fraction. In practice, the absolute lower limit may be somewhat lower, due to false binaries, e.g., background stars falsely assumed to be companion stars. The absolute upper limit of the binary fraction is 100\% binarity. Note that, when discussing ``the binary fraction'', it is very important to mention how it is defined. The observed binary fraction among the {\em stellar} members of an association is generally higher than the {\em overall} binary fraction (including the brown dwarfs), as binarity among the low-mass members is generally difficult to observe.

Figure~\ref{figure: mulspt_sbs_vbc} shows the observed binary fraction as a function of spectral type for the default model. The gray bars show the visual and spectroscopic binary fraction with full biasing. For reference, we also show the fraction of visual binaries with $\rho \geq 0.1''$ and spectroscopic binaries with $K_1 \geq 2$~km\,s$^{-1}$ (white bars). Many of the binaries with these properties are undetected due to the low brightness. The figure shows an increasing {\em observed} binary fraction with increasing primary mass. The observed trend between binary fraction and spectral type, often mentioned in literature, is possibly merely a result of observational biases.

% ====================================================================
% ====================================================================
% ====================================================================
% ==INTRODUCTION======================================================
% ====================================================================
% ====================================================================
% ====================================================================

\section[The inverse problem: deriving properties from observations]{The inverse problem: deriving binary population properties from observations} \label{section: inverseproblem}

In the previous section we discussed the impact of projection effects, sampling, and observational selection effects on the observed binary fraction and  binary parameter distributions. In this section we discuss the inverse problem, i.e., how to derive the {\em true} binary population from the {\em observed} binary population. We will discuss the limitations imposed by observations, and the uncertainty that will be present in the estimate for the true binary population stemming from observational biases. The binary parameter distributions and the binary fraction itself are discussed in the sections below. Throughout our analysis we assume random orientation.

% ====================================================================
% ====================================================================

\subsection{Recovering the mass distribution and pairing function} \label{section: recovering_fm_fq}

\begin{figure}[!tbp]
  \centering
  \includegraphics[width=0.9\textwidth,height=!]{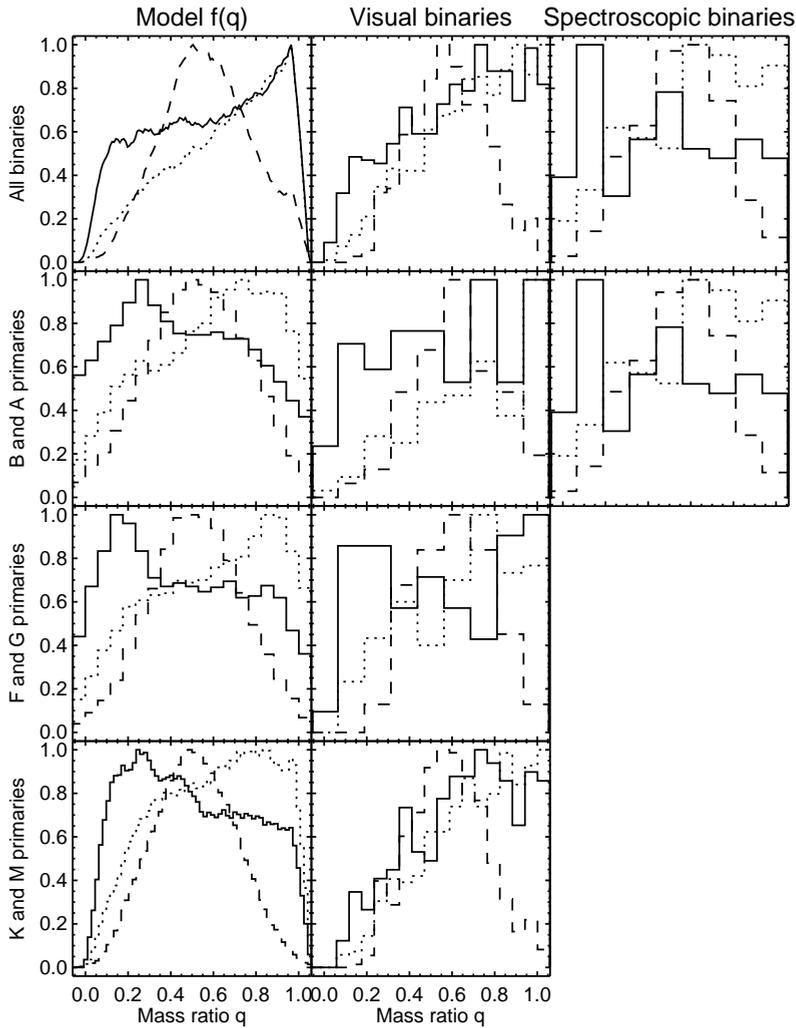}
  \caption{The ``observed'' mass ratio distribution $\tilde{f}_q(q)$ for different model mass ratio distributions $f_q(q)$ and samples. Each model contains 10\,000 binaries. 
The left-hand panels show the intrinsic mass ratio distribution, the middle panels show $\tilde{f}_q(q)$ for the visual binaries, and the right-hand panels show $\tilde{f}_q(q)$ for the spectroscopic binaries. The top panels show the results for all binaries. The other panels show the results for different primary mass ranges, which are indicated on the left. No spectroscopic binaries with primary spectral type later than A are ``detected''. The solid and dotted histograms represent the results for the generating mass ratio distributions $f_q(q)\propto q^{-0.33}$ and $f_q(q)\propto q^{+0.33}$, respectively. The dashed histograms represent the results for a Gaussian mass ratio distribution (equation~\ref{equation: q_gaussian}) with $\mu_q=0.5$ and $\sigma_q=0.2$. This figure shows that, even though the intrinsic mass ratio distributions are significantly different, the observed mass ratio distributions appear quite similar. In this example, simulated observations should be compared with the observations in order to derive the true mass ratio distribution. A simple fit to the observed mass ratio distribution will, in this case, not work.
  \label{figure: different_fq_in_observations} }
\end{figure}

The mass distribution of an association is often derived after the members of an association have been securely identified \citep[e.g.,][]{preibisch2002}. The upper limit of the mass distribution is defined by the member with the largest mass. The behaviour of the mass distribution for low-mass stars and substellar objects is often not strongly constrained by observations. A minimum mass $M_{\rm min}$ in the stellar or substellar region is assumed or measured.  This minimum mass is very important for the study of the binary population, as most stars are of low mass. The low-mass binaries therefore dominate practically all binary parameter distributions. 

Moreover, for the pairing functions RP and PCRP, the lower limit on $f_M(M)$ is of great importance. If the shape of the mass distribution is known, the lower mass limit can in principle be derived from observations of high-mass stars for the pairing functions RP and PCRP. If plenty of measurements are available, and if the selection effects are well understood, the observed mass ratio distribution could in principle be used to characterize the behaviour of the mass distribution in the brown dwarf regime, provided that, of course, the underlying pairing function is RP or PCRP.
For the pairing functions PCP-II, PCP-III, and RP, the lower mass limit may be inferred from the trend between binary fraction and primary spectral type. The methods described above apply to the ideal case, when the shape of the mass distribution is known a-priori.

The pairing function and the overall mass ratio distribution can be derived from the observed mass ratio for visual binary systems. We refer to the specific mass ratio distribution for high-mass primaries, intermediate mass primaries, and low-mass primaries as $f_{\rm high}(q)$, $f_{\rm med}(q)$ and $f_{\rm low}(q)$, respectively. We refer to the specific binary fraction for high-mass primaries, intermediate mass primaries, and low-mass primaries as $F_{\rm M,high}$, $F_{\rm M,med}$ and $F_{\rm M,low}$, respectively. Below we list several criteria that could be used to rule out the different pairing functions, by comparing simulated observations with observations.
\begin{itemize}\addtolength{\itemsep}{-0.5\baselineskip}
\item[--] If $f_{\rm high}(q)$, $f_{\rm med}(q)$ and $f_{\rm low}(q)$ are not equal, PCP-I can be ruled out.
\item[--] If the companion mass distributions $f_{\rm high}(M_2)$, $f_{\rm med}(M_2)$ and $f_{\rm low}(M_2)$ are different from the primary mass distribution, PCRP can be ruled out. 
\item[--] If the distributions $f_{\rm high}(q)$, $f_{\rm med}(q)$ and $f_{\rm low}(q)$ are inconsistent with the predictions for random pairing, RP can be ruled out. 
\item[--] If the average mass ratio does not decrease with increasing primary mass, PCP-II, PCP-III, RP, and PCRP can be ruled out.
\item[--] If the minimum mass ratio does not decrease with increasing primary mass, PCP-II and PCP-III can be ruled out.
\item[--] If $F_{\rm M,high}$, $F_{\rm M,med}$ and $F_{\rm M,low}$ are not identical, PCRP and PCP-I can be ruled out. 
\item[--] If $F_{\rm M,low} \ll F_{\rm M,med} < F_{\rm M,high}$ does not hold, PCP-II can be ruled out. 
\item[--] If $F_{\rm M,high} = F_{\rm M,med} = F_{\rm M,low}$ does not hold, PCP-II and PCP-III can be ruled out. RP can be ruled out, unless the binary fraction is 0\% or 100\%.
\end{itemize}
The spectroscopic and astrometric binary systems may provide further constraints on the pairing function. If the resulting pairing function is PCP-I, PCP-II, or PCP-III, the overall mass ratio distribution is approximately given by $f_{\rm high}(q)$. The mass ratio distribution of systems with high-mass primaries can therefore be used to recover the overall mass ratio distribution.

A significant complication is introduced by the selection effects. The above criteria are valid for the {\em true} binary population, and generally not for the {\em observed} binary population. An example is shown in Figure~\ref{figure: different_fq_in_observations}. In this figure we show simulated observations of three associations, all with pairing function PCP-III, and each with a different mass ratio distribution $f_q(q)$. The figure clearly shows that, even if the true specific mass ratio distributions are very different, the observed specific mass ratio distributions can be very similar. The mean mass ratio of the resolved binary systems is significantly larger than that of the unresolved systems, as a result of the selection effects. Moreover, the observed mass ratio distributions are significantly different as compared to the true mass ratio distributions. A simple fit to the observations will, in this example, give the wrong mass ratio distribution. This figure illustrates well the necessity of modeling the selection effects and simulating observations.

% ====================================================================
% ====================================================================

\subsubsection{The accuracy of a power-law fit to a mass ratio distribution} \label{section: powerlawaccuracy}

\begin{figure}[!tbp]
  \centering
  \includegraphics[width=1\textwidth,height=!]{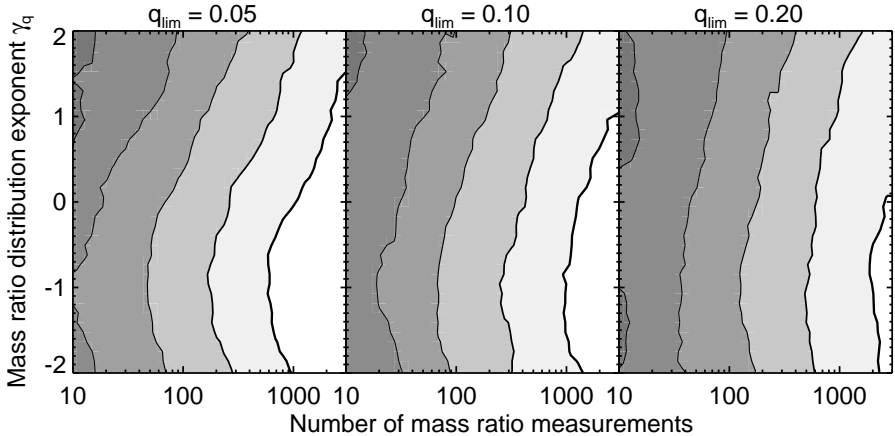}
  \caption{The $1\sigma$ error $\Delta\tilde{\gamma}_q$ on the fitted exponent $\tilde{\gamma}_q$ to the mass ratio distribution (in levels of gray). A set of mass ratio measurements is generated using a mass ratio distribution $f_q(q) \propto q^{\gamma_q}$. The resulting error $\Delta\tilde{\gamma}_q$ is shown as a function of the true exponent $\gamma_q$, the number of measurements, and the minimum mass ratio $q_{\rm lim}$ considered in the fit. See Section~\ref{section: powerlawaccuracy} for a discussion of the results. From left to right in each panel, each plot shows contours at values for $\Delta\tilde{\gamma}_q$ of 0.8, 0.4, 0.2, 0.1, and 0.05. For a study on the mass ratio distribution, with typically $50-100$ measurements, the error on the derived exponent is of order~0.2. 
  \label{figure: massratio_gamma_accuracy} }
\end{figure}

For visual binary systems, the component masses are often derived using the magnitude of each component in one or more filters. An error in the mass is introduced via the uncertainty of the magnitude, which is of order 0.01~mag in the optical and of order 0.1~mag in the near-infrared. The value of the magnitude error depends on the observing strategy (filter, integration time, atmospheric conditions, etc.) and the properties of the binary system (apparent magnitude, magnitude difference, and angular separation). The error in the mass also has a contribution from systematic errors due to the uncertainties in age, distance, metallicity, and interstellar extinction. The error can be reduced by observing a binary system with multiple filters, so that multi-color information can further constrain the mass.

Below we briefly describe how the magnitude uncertainty propagates into the mass distribution and mass ratio distribution. We adopt a constant magnitude error of each binary component of $0.1$~mag in the $K_S$ band, and ignore the systematic error described above.

For a 5~Myr isochrone, an error of 0.1~mag in the $K_S$ band corresponds to a fractional error $\Delta M / M$ in the mass of each component of 4\% for early-type stars ($M \ga 1~\mbox{M}_\odot$) and 6\% for late-type stars ($M \la 1~\mbox{M}_\odot$). For simplicity, we therefore assume a mass error estimate of $\Delta M / M$ of 5\%. The resulting fractional error $\Delta q / q$ of the mass ratio $q = M_2/M_1$ of the binary system is then approximately 7\%.

Due to the error on the magnitude of each component, the measured mass of each component differs from the true mass. For binary systems with $q \approx 1$ this leads to a complication, as the {\em measured} mass of the primary may be smaller than the {\em measured} mass of the companion. In this case, the companion is interpreted by the observer as the primary, and vice-versa. The {\em measured} mass ratio is therefore derived directly from the mass measurements, assuming that the component with the largest {\em measured} mass is the primary.

What is the accuracy that can be obtained with fitting exponent $\gamma_q$ to $\tilde{f}_q(q)$? The mass ratio measurements used in the fit are in the range $[q_{\rm lim},1]$, under the assumption that the instrument bias has not affected this range. We show how the results depend on the minimum mass ratio $q_{\rm lim}$ in the observations, and on the number of observations $N_{\rm obs}$.

For each $\gamma_q$ we draw $N_{\rm obs}$ mass ratios $\{q_i\}$ from a mass ratio distribution $f_q(q) \propto q^{\gamma_q}$, with $q$ in the range $[q_{\rm lim},1]$. We then compare the obtained distribution with analytical distributions $f_q(q) \propto q^{\gamma_q'}$. For each exponent $\gamma_q'$ we test the hypothesis that  $\{q_i\}$ is drawn from a mass ratio distribution with exponent $\gamma_q'$, using the KS test. Finally, we derive the $1\sigma$ error bars $\Delta\tilde{\gamma}_q$ corresponding to the best-fitting exponent $\tilde{\gamma}_q$.

Figure~\ref{figure: massratio_gamma_accuracy} shows the $1\sigma$ error $\Delta\tilde{\gamma}_q$ of the fitted value $\tilde{\gamma}_q$ as a function of the true exponent $\gamma_q$, of the number of measurements $N_{\rm obs}$, and of the limiting mass ratio $q_{\rm lim}$ of the observations. The figure illustrates that (1) for given values of $\gamma_q$ and $q_{\rm lim}$, a better accuracy in $\tilde{\gamma}_q$ is obtained with increasing $N_{\rm obs}$, (2) for given values of $N_{\rm obs}$ and $q_{\rm lim}$, the most accurate fits are obtained for $-2 \la \gamma_q \la 0$, and (3) for given values of $\gamma_q$ and $N_{\rm obs}$, a better fit is obtained for low values of $q_{\rm lim}$, as the covered range in $q$ is larger. In literature, the mass ratio distribution is often derived using a set of $50-100$ measurements. The typical accuracy of the exponent of a power-law mass ratio distribution is then of order 0.2 (under the assumptions that the true mass ratio distribution is indeed a power-law, that selection effects (other than $q_{\rm lim}$) are not present, and that if $\gamma_q \leq -1$, a minimum value $q_{\rm min}$ for the mass ratio distribution exists, with $0 < q_{\rm min} \leq q_{\rm lim}$).

In the above example we have fitted a power-law to the mass ratio distribution. We have used the true values $\{q_i\}$, and we ignored the observational error. In reality, the observed values $\{\tilde{q}_i\}$ have an associated error $\Delta q$. In the case that the masses of the components of a binary system are derived from their brightness, the error $\Delta q$ results from the uncertainty in the brightness of the components, and from the shape of the isochrone. 

In order to study the effect of the error $\Delta q$ on the fitted value $\gamma_q'$ and $\Delta\tilde{\gamma}_q$ , we repeat the experiment mentioned above, but with a relative error of 7\% for each measurement. We convert these value to measurements $\{\tilde{q}_i\}$ with the description given in \S~\ref{section: visual_errormodeling}, adopting a fractional error of 5\% in the binary component masses. Again, we fit a power-law to the measured data and obtain the best-fitting exponent $\gamma_q'$. Our results indicate that the $1\sigma$ error $\Delta\tilde{\gamma}_q$ of the fitted value $\tilde{\gamma}_q$ is very similar to the error obtained in the experiment with fixed values of $q$. The systematic bias $\gamma_q-\gamma_q'$ can be neglected, compared to the statistical error $\Delta\tilde{\gamma}_q$, for all values of $\gamma_q$, $N_{\rm obs}$, and $q_{\rm lim}$ discussed above.

% ====================================================================
% ====================================================================

\subsection{Recovering the semi-major axis and period distributions} \label{section: sos_recovering_sma_and_period}

\begin{figure}[!tbp]
  \centering
  \includegraphics[width=0.8\textwidth,height=!]{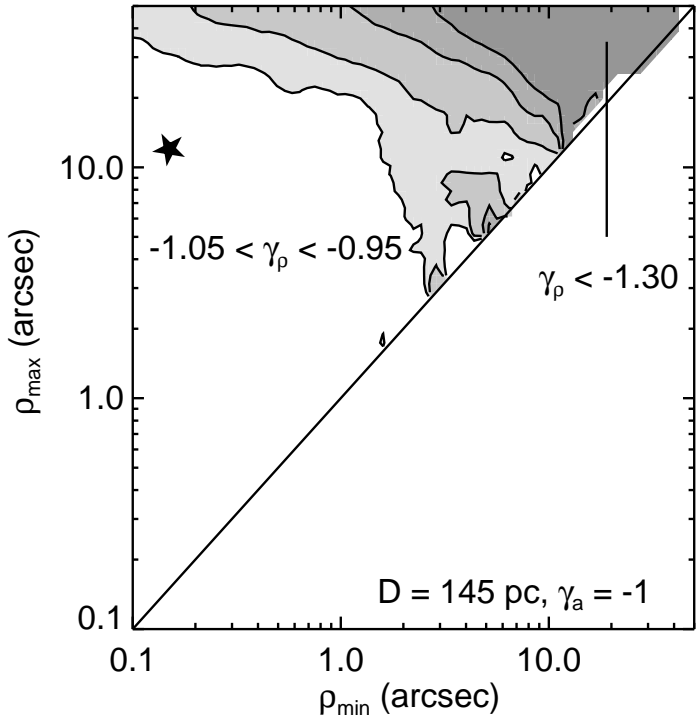}
  \caption{The systematic error on $\gamma_a$ if the assumption is made that the fitted $\gamma_\rho$ of the angular separation distribution $f_\rho(\rho)$ is equal to $\gamma_a$ of the semi-major axis distribution $f_a(a)$. The contours in this figure represent from left to right fitted values for $\gamma_\rho$ of $-1.05$, $-1.1$, $-1.2$, and $-1.3$. The model association has a distance of 145~pc and a semi-major axis distribution $f_a(a) \propto a^{-1}$ for $10^2$ R$_\odot \leq a \leq 10^6$ R$_\odot$. An imaging survey has a typical lower limit $\rho_{\rm min} = 0.15''$ and an upper limit of $\rho_{\rm max} = 12''$ due to confusion with background stars. For such a survey, indicated with the filled star, the figure shows that the systematic error on $\gamma_a$ is less than 0.05 for an association at a distance of 145~pc. For associations at a smaller or larger distance, the star is located more to the top-right, or bottom-left, respectively. 
  \label{figure: gamma_rho_vs_gamma_a} }
\end{figure}

Given the total mass of a binary, the dimensions of the orbit are defined by the semi-major axis, or alternatively, the period. A visual binary provides information on the orbital dimensions through the angular separation. The orbital period can be measured in a straightforward manner for a spectroscopic binary. As the information provided by the close spectroscopic binaries and wide visual binaries is complementary, the distribution over orbital dimensions, thus, $f_a(a)$ or $f_P(P)$, can be recovered. The distributions $f_a(a)$ and $f_P(P)$ can be constrained further by measurements of astrometric binaries, for which $a$ (provided that the distance to the binary is known) and $P$ can be measured independently. 

For nearby associations, the number of (known) visual binaries is generally much larger than the number of (known) spectroscopic binaries. The distribution of the orbital dimensions can therefore most easily be recovered from the observed angular separation distribution $\tilde{f}_\rho(\rho)$.  Below, we illustrate how this can be done, and under which conditions the results are incorrect.

Equation~\ref{equation: projecteda} provides the relation between the mean angular separation $\langle \rho \rangle$ of an ensemble of binary systems with a certain semi-major axis $a$. As a result of this, a semi-major axis distribution $f_a(a) \propto a^{\gamma_a}$ results in an angular separation distribution $f_\rho(\rho) \propto \rho^{\gamma_\rho}$, with $\gamma_\rho \approx \gamma_a$. For a visual binary system $a$ cannot be measured directly, while $\rho$ can. For this reason, the distribution $f_\rho(\rho)$ is often used to study the properties of $f_a(a)$. However, the above approximation is only valid in the semi-major axis range $a_{\rm min} \ll D\,\rho \ll a_{\rm max}$. For angular separations close to $a_{\rm min}/D$ and $a_{\rm max}/D$ projection effects become important, so that $\gamma_\rho \neq \gamma_a$. This projection effect can easily be seen in Figure~\ref{figure: opik_dm_difference}. In this figure we adopted $\gamma_a=-1$, so that the cumulative distribution $F_a(a)$ (solid line) is straight. The cumulative angular separation distribution $F_\rho(\rho)$ is straight over a large range in $\rho$, but shows deviations at its extremes, indicating that $f_\rho(\rho)$ cannot be used directly to characterize $f_a(a)$ in this regime. The method of simulating observations should therefore be used for the smallest and largest separations. The limit $\rho \rightarrow a_{\rm min}/D$ is not of importance for visual binaries, but the limit $\rho \rightarrow a_{\rm max}/D$ is, especially when studying the behaviour of the $f_a(a)$ for large $a$.

Below we show under which conditions we can use the approximation $\gamma_\rho \approx \gamma_a$ for a power-law semi-major axis distribution, and in which cases the not. We create several large associations, all at a distance of 145~pc, using the properties of the default model. Each model has an angular separation distribution $f_a(a) \propto a^{-1}$, with $15$~R$_\odot \leq a \leq 10^6$~R$_\odot$. We project the association properties onto observable space to obtain the angular separation distribution $f_\rho(\rho)$. We select the angular separations with $\rho_{\rm min} \leq \rho \leq \rho_{\rm max}$, and fit a distribution of the form $f_\rho(\rho)\propto \rho^{\gamma_\rho}$ to these values. We repeat this procedure for different values of $\rho_{\rm min}$ and $\rho_{\rm max}$, and find under which conditions the approximation  $\gamma_\rho \approx \gamma_a$ holds. For each fit, we use a simulated dataset of $2\,000$ angular separation measurements.

Figure~\ref{figure: gamma_rho_vs_gamma_a} shows the fitted value $\gamma_\rho$ as a function of $\rho_{\rm min}$ and $\rho_{\rm max}$, for an association at 145~pc. At this distance, the projected maximum semi-major axis corresponds to approximately $a_{\rm max}/D\approx 32$~arcsec. The systematic error $|\gamma_a-\gamma_\rho|$ is largest for large values of both $\rho_{\rm min}$ and $\rho_{\rm max}$, as in this regime the projection effects become important (see, e.g., Figure~\ref{figure: opik_dm_difference}). The systematic error for $\rho_{\rm min} = 0.15\, a_{\rm max}/D = 5''$ and $\rho_{\rm max} = a_{\rm max}/D=32''$, for example, is $|\gamma_a-\gamma_\rho| = 0.3$. For visual binary surveys among the nearest associations, the limits are of order $\rho_{\rm min} = 0.15''$ and $\rho_{\rm max} = 12''$ (due to confusion with background stars), so that the systematic error on $\gamma_a$ is small compared to the random error.
The figure shows that the systematic error due to the assumption $\gamma_\rho \approx \gamma_a$ has important consequences for the study of wide binary systems, where $\rho_{\rm min}$ and $\rho_{\rm max}$ are large. Several studies on the behaviour of $f_a(a)$ for large $a$ have been carried out, and often the assumption $\gamma_\rho \approx \gamma_a$ is made.
Two examples where the systematic error on $\gamma_a$ may have biased the results are the studies of the wide binary population by \cite{close1990} and \cite{chaname2004}.  They find $\gamma_a=-1.3$ and $\gamma_a=-1.67 \pm 0.10$ for the Galactic disk, respectively. In these studies, the authors derive the value of $\gamma_a$ by fitting a power-law to the observed angular separation distribution of the widest binaries, and using $\gamma_\rho \approx \gamma_a$. Their inferred values for  $\gamma_a$ may therefore be underestimated.

% ====================================================================
% ====================================================================

\subsubsection{Discriminating between \"{O}pik's law and a log-normal period distribution} \label{section: opikdm_difference}

\begin{figure}[!tbp]
  \centering
  \includegraphics[width=1\textwidth,height=!]{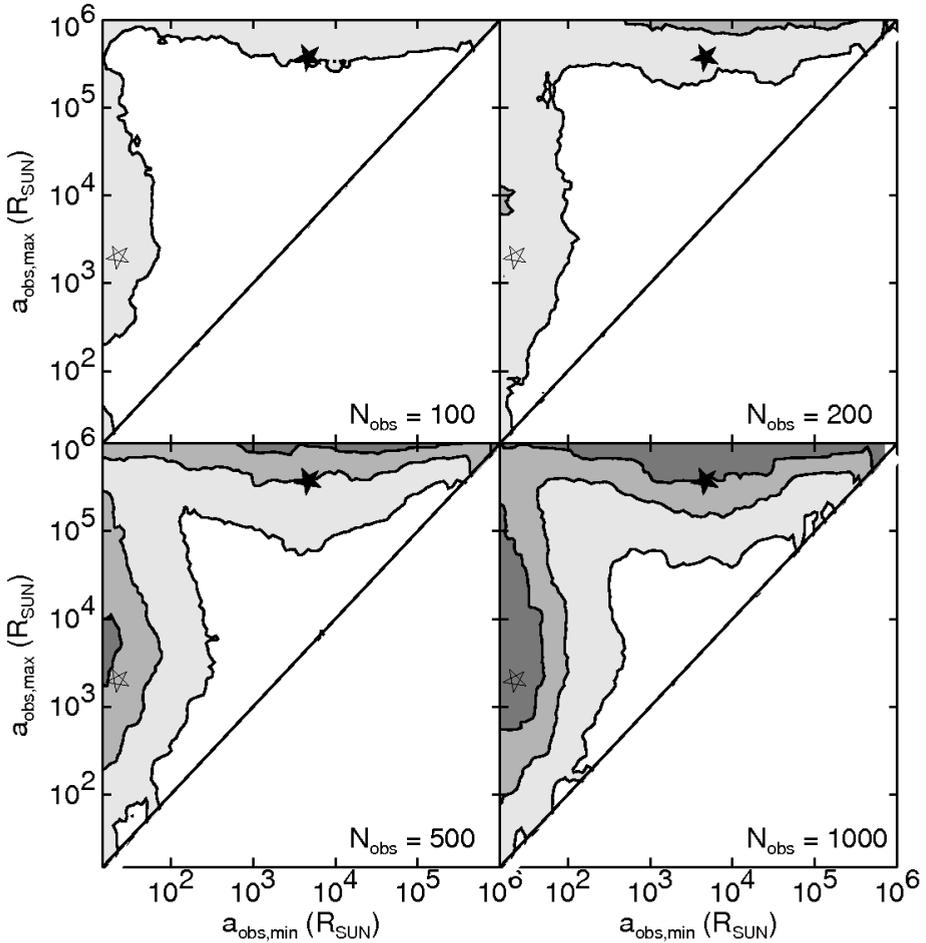}
  \caption{How well can observations be used to discriminate a semi-major axis distribution of the form $f_a(a)\propto a^{-1}$ from a distribution $f_a(a)$ resulting from the log-normal period distribution found by \cite{duquennoy1991}? We have simulated these distributions and compared the results in the semi-major axis range $[a_{\rm obs,min},a_{\rm obs,max}]$, and for different numbers of observations $N_{\rm obs}$ (shown in the bottom-right corner of each panel). The dark areas in the panels indicate the conditions under which the distributions can reliably be discriminated; the three contours represent the $1\sigma$, $2\sigma$, and $3\sigma$ confidence limits. The solid line indicates the line $a_{\rm obs,min}=a_{\rm obs,max}$. The filled star indicates the typical observable semi-major axis range for an imaging survey of an association at 145~pc. The open star indicates the typical properties of a radial velocity survey. 
  \label{figure: opik_dm_kstest_difference} }
\end{figure}

Finding out what is the distribution of orbital dimensions in an association is best done using the angular separation distribution of visual binaries. Additional information may be provided by the period distribution of spectroscopic binaries, and the ratio between the number of visual and spectroscopic binaries found. In this section we determine how accurately one can discriminate between different orbital dimension distributions based on the observations. We consider in this section two well-known distributions: \"{O}pik's law or the log-normal period distribution of \cite{duquennoy1991}. In this section we generate artificial datasets and compare these. We ignore the selection effects, other than the separation constraint for visual binaries. We generate one set of models in which the period distribution is described by equation~\ref{equation: duquennoyperiods}, the log-normal distribution found by \cite{duquennoy1991}. The second set of models has a semi-major axis distribution of the form $f_a(a) \propto a^{-1}$ (\"{O}pik's law). We use the mass distribution and pairing function of the default model (Table~\ref{table: modelpopulation}), which relate the period and semi-major axis for each binary system. We use the KS test to compare the semi-major axis distributions resulting from the two models. We compare the distributions for all binaries with $a_{\rm obs,min} \leq a \leq a_{\rm obs,max}$, with varying $a_{\rm obs,min}$ and $a_{\rm obs,max}$. 

Figure~\ref{figure: opik_dm_kstest_difference} shows the results of the comparison. The best way to find the difference between the two models is by comparing distributions of the very close, or very wide binaries, as is suggested in Figure~\ref{figure: fa_fp_difference}. 
In order to discriminate between the two models at the $3\sigma$ level using the angular separation or period distribution, a sample of $500-1000$ binary systems is necessary. However, the difference between the two models also affects the observed fraction of visual and spectroscopic binary systems. These quantities can be used to discriminate between \"{O}pik's law or the log-normal period distribution based on a smaller number of systems.

Often, in order to describe the observations, the hypothesis that ``the data can be described by a log-normal period distribution'' is tested, and in many cases the data is indeed found to be consistent with this distribution. 
This leads to the common understanding that the period distribution indeed has a log-normal form. However, as shown above (see Figure~\ref{figure: opik_dm_kstest_difference}), the statistical test only tells the observer that it is not possible to reject the hypothesis (a log-normal period distribution) with confidence. It does not necessarily mean that the period distribution is indeed log-normal. The data are possibly also consistent with \"{O}pik's law, and only if more than $\sim 1\,000$ systems are observed will one be able to distinguish between the two using the observed angular separation or period distribution.

% ====================================================================
% ====================================================================

\subsection{Recovering the eccentricity distribution} \label{section: recovering_eccentricity}

The eccentricity distribution cannot be inferred from observations of visual binaries (\S~\ref{section: eccentricityobserved}). With observations of spectroscopic and astrometric binaries, however, it is possible to derive $f_e(e)$, as long as the measurements are accurate enough, as long as the eccentricity is determined for a large number of systems, and as long as the selection effects are well understood.

The three most often applied eccentricity distributions are the flat eccentricity distribution $f_{\rm flat}(e) = 1$, the thermal eccentricity distribution $f_{2e}(e) = 2e$, and the single-value eccentricity distribution $f_{e_0}(e) = \delta(e-e_0)$ (i.e., all orbits have $e=e_0$).

At first glance the single-value eccentricity distribution is the easiest to recognize in the observed eccentricity distribution $\tilde{f}_e(e)$, as it produces a peak near $e_0$. The width of the observed distribution is defined by the typical error in the derived eccentricities. However, if such a distribution is observed, this does not necessarily mean that $f_e(e)$ is indeed single-valued. Due to selection effects it is possible that another distribution $f_e(e) \neq \delta(e-e_0)$ produces the {\em same} $\tilde{f}_e(e)$. On the other hand, it is possible to exclude the single-value eccentricity distribution if $\tilde{f}_e(e)$ is significantly different from a peaked distribution around $e_0$ (taking into account the error on $e$). 

It is more difficult to determine whether the eccentricity distribution is flat or thermal, based on the observed eccentricity distribution $\tilde{f}_e(e)$. Due to the difficulty of detecting highly-eccentric orbits, and due to low-number statistics, both intrinsic eccentricity distributions often result in a very similar observed distribution. 

To find out with how much confidence we can discriminate between a flat eccentricity distribution and a thermal eccentricity distribution, we first consider the ideal case, where no selection effects are present. We draw $N$ eccentricities from both eccentricity distributions. We use the KS test to test the hypothesis that both datasets are drawn from the same underlying distribution. Our analysis shows that in order to claim with $1\sigma$ confidence that the two distributions are different, one needs at least 12~measurements in each dataset. For $2\sigma$ and $3\sigma$ confidence, one needs at least 50 and 120~measurements in each dataset, respectively. The above values apply for an unbiased dataset with a zero error, so that the values above are lower limits. In reality, {\em at least} 120~accurate and unbiased eccentricity measurements are necessary to discriminate between a flat and a thermal eccentricity distribution at the $3\sigma$ confidence level. 

\cite{duquennoy1991} analyze the eccentricity distribution for a sample of 29~binaries with a period larger than $\sim 3$~year, and find an eccentricity distribution that ``tends to be thermal''. According to our simulations, the confidence level of this statement is of order $1.5\sigma$. It is clear that further measurements and a detailed analysis is necessary to determine $f_e(e)$. Once $f_e(e)$ is accurately known, it can be used to place further constraints on the star forming process.

% ====================================================================
% ====================================================================

\subsection{Recovering the binary fraction}

One of the most important parameters that describes the binarity properties of an OB~association is the binary fraction $F_{\rm M}$. The binary fraction cannot be measured directly, but a value $\tilde{F}_{\rm M} < F_{\rm M}$ is obtained due to selection effects.

Equation~\ref{equation: multiplicitylimits} and Figure~\ref{figure: observed_intrinsic_fm} show the relationship between the {\em observed} multiplicity fraction $\tilde{F}_{\rm M}$ and the {\em true} multiplicity fraction $F_{\rm M}$, given the ratio $x$ between the observed number of binaries and the true number of binaries in the sample. Combining the measurement of $\tilde{F}_{\rm M}$ with the predicted value of $x$ (from the simulated observations) gives $F_{\rm M}$. The statement above is only valid for the surveyed sample (e.g., intermediate mass association members), as $F_{\rm M}$, $\tilde{F}_{\rm M}$, and $x$ may vary depending on the sample choice.

Suppose that two large binarity surveys are done: a visual and a spectroscopic survey. Suppose also that companion stars at a separation larger than $\rho_{\rm max}$ (say 10~arcsec) cannot be reliably identified due to confusion with background stars. The available angular separation measurements for visual binaries give the shape of the semi-major axis distribution $f_a(a)$ for $D\,\rho_{\rm min} \la a \la D\,\rho_{\rm max}$. The minimum value $a_{\rm min}$ for the semi-major axis distribution can be derived from the observed period distribution of the spectroscopic binaries. The maximum value $a_{\rm max}$, however, cannot be derived due to the confusion of companions with background stars at large separations. The overall binary fraction of the association depends on the value of $a_{\rm max}$, as $a_{\rm max}$ determines what fraction of the companion stars is ``hiding'' between the background stars. For a distribution $f_a(a) \propto a^{-1}$, and for a given set of observations with $D\,\rho_{\rm min} \gg a_{\rm min}$ and $D\,\rho_{\rm max} \ll a_{\rm max}$, the true binary fraction $F_{\rm M}$ is related to the observed binary fraction $\tilde{F}_{\rm M}$ and $a_{\rm max}$ as follows:
\begin{equation} \label{equation: binaryfraction_vs_amax}
  F_{\rm M} = \frac{\tilde{F}_{\rm M} }{F_a(D\,\rho_{\rm max})-F_a(D\,\rho_{\rm min})} = \tilde{F}_{\rm M} \times \frac{ \ln \left(  a_{\rm max}/a_{\rm min} \right) }{ \ln \left(  \rho_{\rm max}/\rho_{\rm min} \right)}\,,
\end{equation}
where $F_a(a)$ is the cumulative distribution function of the semi-major axis, evaluated at semi-major axis $a$. The above equation is only valid if $\tilde{F}_{\rm M}$ refers to {\em all} companions with $D\,\rho_{\rm min} \la a \la D\,\rho_{\rm max}$.

A typical upper limit for $a_{\rm max}$ may be obtained using dynamical arguments: binaries with a semi-major axis larger than a certain value $a_{\rm tidal}$ are unlikely to have survived over the association lifetime, due to interactions with other objects in the association or in the Galactic population, or due to disruption by the Galactic tidal field. A lower limit for $a_{\rm max}$ is given by the semi-major axis $a_{\rm widest} \approx \rho_{\rm max}D/(1+e)$ of the widest binary that can reliably be identified as a true, physical binary. The maximum semi-major axis can therefore be constrained with 
\begin{equation}
a_{\rm widest} \leq a_{\rm max} \leq a_{\rm tidal} \,.
\end{equation}
A strict constraint for the binary fraction can be obtained by counting all detected binary systems in the association, and finding the maximum possible number of systems (singles and binaries) in the association, which gives the observed binary fraction $F_{\rm M,obs}$. The true binary fraction is then constrained by 
\begin{equation}
F_{\rm M,obs} \leq F_{\rm M} \leq 1 \,.
\end{equation}
By simulating observations with different values for $a_{\rm max}$ and $F_{\rm M}$, and comparing the observed properties with the number of visual, spectroscopic, and astrometric binaries detected, the true values of $a_{\rm max}$ and $F_{\rm M}$ are obtained.

The above discussion on the relation between $F_{\rm M}$ and $a_{\rm max}$ was made under the assumption that the other properties of the binary population are derived with high accuracy. If this is not the case, the model associations have to be varied within the degrees of freedom available. For example, if the mass ratio distribution is poorly constrained by the observations, simulated observations of models with different mass ratio distributions, different values of $a_{\rm max}$, and different values of $F_{\rm M}$ have to be made, and compared to the observations.

% ====================================================================
% ====================================================================

\subsection{A strategy to derive the true binary population from observations} \label{section: strategy}

In this section we propose a strategy to derive the true binary population from the observations of visual, spectroscopic, and astrometric binary systems. The strategy below is developed under several assumptions, and should be adapted if one of these assumptions does not hold:
\begin{itemize}\addtolength{\itemsep}{-0.5\baselineskip}
\item[--] The association consists of single stars and binary systems; no higher order multiples are present.
\item[--] All association members have been identified.
\item[--] The distance, size, age, and metallicity of the association are known. 
\item[--] The mass distribution $f_M(M)$ for single stars is known in the stellar mass regime.
\item[--] The pairing function is PCP-I, PCP-II, PCP-III, RP, or PCRP. In the case of PCP-I, PCP-II, PCP-III, and PCRP, the primary mass $M_1$ is drawn from $f_M(M)$. In the case of PCRP, the companion mass is also drawn from $f_M(M)$, with $M<M_1$. In the case of RP, both components of a binary system are independently drawn from $f_M(M)$. 
\item[--] The binary parameters in the association are independent (equation~\ref{equation: independent_parameters}). The strategy is also valid if, instead of $f_a(a)$, the distribution $f_P(P)$ is chosen independently of the other parameters.
\item[--] The binary systems are assumed to have a random orientation in space.
\end{itemize}
Deriving the true binary population $\mathcal{T}$ from observations is an iterative process. Below we describe one of many strategies. Depending on the properties of the available dataset, several adaptions to this strategy could be made.
\begin{enumerate}\addtolength{\itemsep}{-0.5\baselineskip}
\item Collect all relevant information on binarity in the association.
\item Identify for each dataset the observer's choice, the sample bias, and the instrument bias.
\item Find a rough estimate $\mathcal{T}_1$ for the true binary population. This estimate can be obtained by adopting parameter distributions that result from observations of other stellar groupings, or from theoretical predictions. Construct the ``observed'' binary population for $\mathcal{T}_1$.
\item For different pairing functions and mass ratio distributions, compare the simulated observations with the real observations. This could include a comparison between the observed mass ratio distribution, the magnitude difference distribution, the mass distribution, and the binary fraction, each as a function of primary spectral type. Several important differences that can be used to identify the pairing function are listed in \S~\ref{section: recovering_fm_fq}. Update $\mathcal{T}$ with the best-fitting pairing function and mass ratio distribution, this yields $\mathcal{T}_2$. \label{item: pairingfunction}
\item For different semi-major axis distributions (or period distributions), compare the simulated observations $\tilde{\mathcal{S}}$ with the real observations $\tilde{\mathcal{T}}$. The most useful distributions for this purpose are $f_\rho(\rho)$ for visual binaries, $f_P(P)$ for spectroscopic binaries, and $f_P(P)$ and $f_{a_1\sin i}(a_1\sin i)$ for astrometric binaries. Update $\mathcal{T}$ with the best-fitting $f_a(a)$ or $f_P(P)$, this yields $\mathcal{T}_3$.
\item Given the current estimate for $f_a(a)$ and $f_P(P)$, find the minimum value $a_{\rm min}$ or $P_{\rm min}$ that is most consistent with the observations of spectroscopic binaries. Update $\mathcal{T}$ with the best-fitting $P_{\rm min}$ or $P_{\rm min}$, this yields $\mathcal{T}_4$.
\item For different eccentricity distributions $f_e(e)$, find which is most consistent with the measured eccentricities of spectroscopic and astrometric binaries. Update $\mathcal{T}$ with the best-fitting $f_e(e)$, this yields $\mathcal{T}_5$.
\item Constrain the maximum semi-major axis $a_{\rm max}$ using the widest observed binaries and the Galactic tidal limit.  
\item For model $\mathcal{T}_5$, vary $a_{\rm max}$ between the limits found above, and vary the binary fraction  $F_{\rm M}$ between 0\% and 100\%. Find which combination of these parameters results in the correct number of visual, spectroscopic, and astrometric binaries. Update $\mathcal{T}$ with the best-fitting combination of $a_{\rm max}$ and $F_{\rm M}$, yielding $\mathcal{T}_6$.  \label{item: binaryfraction}
\item Repeat steps~\ref{item: pairingfunction} to~\ref{item: binaryfraction} with the updated model population $\mathcal{T}_6$, until the solution converges. If none of the models is compatible with the observations within the $1\sigma$ level, this might indicate that the model assumptions are incorrect.
\item Vary all properties of the final model $\mathcal{T}_f$ in order to find the $1\sigma$ confidence limit of each parameter.
\item The best-fitting $\mathcal{T}_f$ is now a good estimate for the true binary population $\mathcal{T}$, within the errors derived above.
\end{enumerate}
If the assumptions listed above hold, the true binary population $\mathcal{T}$ is equal to that of the final model $\mathcal{T}_f$, within the freedom defined by the errors. In several cases this freedom, which we describe above in terms of the $1\sigma$ confidence interval, needs a more general description. It is for example possible that the model is consistent with \"{O}pik's law and also with a log-normal period distribution.

In practice, parameter distributions are often represented with a functional form. For example, for the default model we adopted a mass ratio distribution of the form $f_q(q)\propto q^{\gamma_q}$. Suppose we use this functional form to derive the true binary population, and the final model $\mathcal{T}$ has $\gamma_q = -0.5 \pm 0.1$. Then this does not necessarily mean that the true binary population has indeed the form $f_q(q)\propto q^{\gamma_q}$. In this example we have {\em added another assumption}, i.e., that the mass ratio distribution has the form $f_q(q)\propto q^{\gamma_q}$. However, if the observed binary population is consistent with this functional form for $f_q(q)$, we have, given the available dataset, a good approximation $\mathcal{T}_f$ for the true binary population.

% ====================================================================
% ====================================================================
% ====================================================================
% ==INTRODUCTION======================================================
% ====================================================================
% ====================================================================
% ====================================================================

\section{An example: recovering the true binary population from simulated data} \label{section: example1}

\begin{table}
  \begin{tabular}{lrrrr}
    \hline
    \hline
    Target group & \# stars & $\tilde{F}_{\rm VB,M_1}$ &  $\tilde{F}_{\rm SB,M_1}$ &  $\tilde{F}_{\rm VB \vee SB ,M_1}$ \\ 
    \hline
    Visual targets               & 5\,992   & 12.02   & $-$     & $-$    \\
    Spectroscopic targets        &  358   & $-$     & 27.37   & $-$    \\
    All targets                  & 5\,992   & $-$     & $-$     & 13.65  \\  
    \hline
    All primaries                &10\,000   & 7.20   &  0.98  &  8.18 \\
    $M_1 > 1$~M$_\odot$          &  664   & 26.96   & 14.76   &  41.72 \\
    Stellar primaries            & 6\,900   & 10.43   & 1.42   &  11.86 \\
    Substellar primaries         & 3\,100   & none   & none   & none  \\
    \hline
    \hline
  \end{tabular}
  \caption{The {\em observed} binary fraction (in per\,cent) for various subsets of the population, for the example model discussed in Section~\ref{section: example1}. The first column gives the properties of the subset; the second column the number of stars in each subset. The following three columns give the observed {\em specific} binary fraction for the visual binaries, the spectroscopic binaries, and the combined observations, respectively. \label{table: sosexample_fractions} } 
\end{table}

In this section we show an example how the true binary population can be recovered from the observations. The model we adopt for our example is the default model (Tables~\ref{table: modelpopulation} and~\ref{table: modelcluster}) for an OB~association at a distance of 145~pc. The association has a binary fraction $F_{\rm M}=100\%$ and consists of $N=B=10\,000$ binary systems.

\subsection{Simulated observations}

We obtain simulated observations of a large imaging survey and a large radial velocity survey of the association. The visual binary observations are hampered by the brightness constraint ($V \leq 17$~mag), the separation constraint ($\rho \geq 0.1''$), the contrast constraint (equation~\ref{equation: classical_detectionlimit1}), and the confusion constraint (equation~\ref{equation: classical_detectionlimit2}). Only binary systems brighter than $V=10$~mag are targeted with our radial velocity survey. We use model SB-W to model the instrument bias for the spectroscopic binaries, with $T=10$~year, $\Delta T=1$~year, and adopt a radial velocity error $\sigma_{RV} = 2$~km\,s$^{-1}$. 

In total we observe 5992~targets, i.e., about 60\% of the total number of association members. All of these targets are observed in the imaging survey, and a subset of 358~targets additionally in the radial velocity survey. In total we detect 720~visual binaries and 98~spectroscopic binaries. There is no overlap between the detected visual and spectroscopic binaries. Table~\ref{table: sosexample_fractions} provides an overview of the observed binary fraction for several subsets of the population.

\begin{figure}[!tbp]
  \centering
  \includegraphics[width=1\textwidth,height=!]{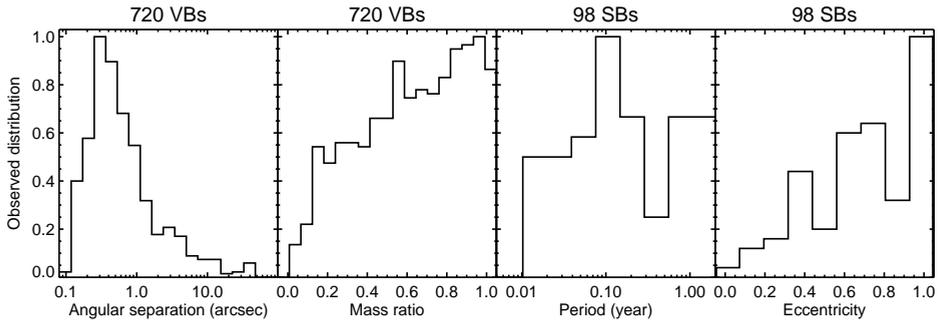}
  \caption{The observed binary population for the example model discussed in Section~\ref{section: example1}. The panels show the observed angular separation and mass ratio distribution for the 720~visual binaries, and the observed period and eccentricity distributions for the 98~spectroscopic binaries. Due to selection effects the observed distributions are significantly different from the intrinsic distributions. In order to recover the true binary population, a comparison between observations and simulated observations is necessary. 
  \label{figure: sosexample_observed} }
\end{figure}

\begin{SCfigure}[][!tbp]
  \centering
  \includegraphics[width=0.7\textwidth,height=!]{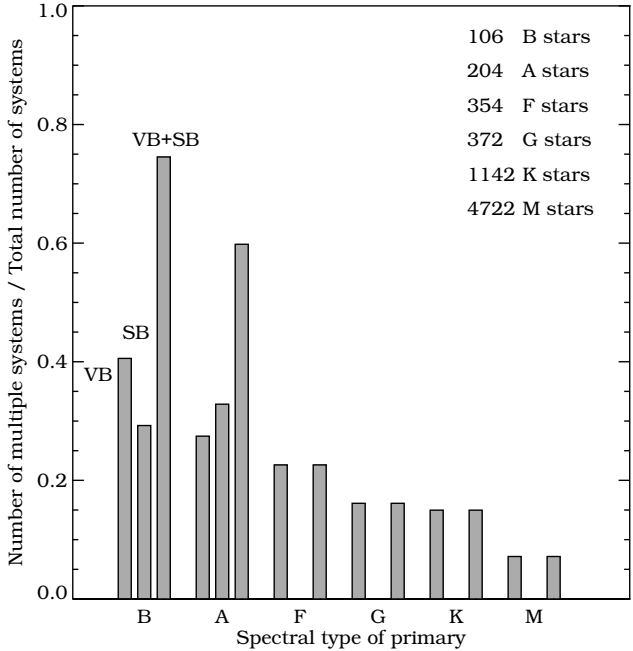}
  \caption{The observed binary fraction as a function of primary spectral type, for the example model discussed in Section~\ref{section: example1}. The {\em observed} specific binary fraction as a function of spectral type is shown here for the visual binaries ({\em left bar}), the spectroscopic binaries ({\em middle bar}), and the combined dataset ({\em right bar}). The default model with $N=10\,000$ binaries is used. The {\em true} specific binary fraction is 100\% for each spectral type.
  \label{figure: sosexample_mulspt} }
\end{SCfigure}

The most important observed binary parameter distributions are shown in Figure~\ref{figure: sosexample_observed}. The observed binary fraction as a function of primary spectral type is shown in Figure~\ref{figure: sosexample_mulspt}. Although the {\em true} binary fraction is 100\% for all spectral types, the latter figure clearly illustrates that the {\em observed} binary fraction decreases with decreasing primary mass.

Below we recover the true binary population from the observed binary population, using the available observations. We adopt the assumptions listed in Section~\ref{section: strategy}. We do not describe the full iterative process of the strategy described in Section~\ref{section: strategy}, but summarize the most important derivations and tests.

\subsection{Finding the pairing function and mass ratio distribution}

We derive the pairing function and the mass ratio distribution, using the mass ratio measurements for the detected visual binary systems. 
First we look at the pairing functions RP and PCRP. For these pairing functions, practically all systems with an early-type primary have a low-mass companion. We exploit this property to prove that our observations are inconsistent with both RP and PCRP. We compare observations and simulations using the quantity $\tilde{Q}$, which we define as
\begin{equation}
  \tilde{Q} \equiv \frac{\mbox{number of observed binaries with $M_1 \geq 1.5$~M$_\odot$ and $q \geq 0.8$}}
                        {\mbox{number of targets with $M_1 \geq 1.5$~M$_\odot$}} \ .
\end{equation}
In our visual binary survey we have $\tilde{Q}_{\rm obs} = 22/310 = 7.1 \pm 1.5\%$. The corresponding values for the models with RP and PCRP depend on the binary fraction and on the slope $\alpha$ of the mass distribution in the brown dwarf regime. For these pairing functions, $\tilde{Q}$ is largest for a binary fraction of 100\% and $\alpha=\infty$ (i.e., no brown dwarfs). Under the latter conditions, $\tilde{Q}_{\rm RP} = 6/980 = 0.6 \pm 0.2\%$ and $\tilde{Q}_{\rm PCRP} = 6/930 = 0.6 \pm 0.2\%$ for the simulated visual binary survey. For any other choice of $F_{\rm M}$ or $\alpha$, both $\tilde{Q}_{\rm RP}$ and $\tilde{Q}_{\rm PCRP}$ will have lower values. As the derived upper limits for  $\tilde{Q}_{\rm RP}$ and $\tilde{Q}_{\rm PCRP}$ are significantly lower (see \S~\ref{section: errorbinaryfraction}) than $\tilde{Q}_{\rm obs}$, we can exclude the pairing functions RP and PCRP with high confidence.

As RP and PCRP are excluded, the pairing function is one of PCP-I, PCP-II, or PCP-III. According to equation~\ref{equation: pcp_fq_highm1} the specific mass ratio distribution for systems with high-mass primaries is equal to the generating mass ratio distribution $f_q(q)$. We therefore use the visual binaries with high-mass primaries to derive $f_q(q)$. Due to the observational biases, $f_q(q)$ cannot be derived directly from $\tilde{f}_q(q)$ (see, e.g., Figure~\ref{figure: different_fq_in_observations}). We use simulated observations to find the $f_q(q)$ that corresponds best to $\tilde{f}_q(q)$. We adopt a mass ratio distribution of the form $f_q(q) \propto q^{\gamma_q}$, and compare the simulated observations with the real observations for targets with $M_1 \geq 1.5$~M$_\odot$. The comparison shows that $\gamma_q = -0.39\pm0.25$ for high-mass targets. 

Now that $f_q(q)$ is known, we can determine the pairing function (PCP-I, PCP-II, and PCP-III) by comparing $\tilde{f}_{\rm M_1}(q)$ and $\tilde{F}_{{\rm M},M_1}$ for systems with high-mass and low-mass primaries. Unfortunately, few low-mass stars are available, making it impossible to discriminate between PCP-I, PCP-II, and PCP-III. Each of these pairing functions is consistent with the observations. Adopting pairing function PCP-III for the model, while keeping in mind that pairing functions PCP-I and PCP-II are also consistent with the observations, we pursue the derivation of the true binary population.

\subsection{Finding $f_a(a)$} \label{section: example_sma}

The next step is to find the shape of the semi-major axis distribution $f_a(a)$, or, alternatively, the period distribution $f_P(P)$. We restrict ourselves to two possibilities: a semi-major axis distribution of the form $f_a(a) \propto a^{\gamma_a}$, and the log-normal period distribution of \cite{duquennoy1991}. We use the angular separation distribution $f_\rho(\rho)$ for the visual binaries. Alternatively, if plenty of spectroscopic binaries would have been available, we could have chosen $f_P(P)$ for this calibration.

The observed angular separation $\tilde{f}_\rho(\rho)$ is very different from the true distribution $f_\rho(\rho)$. At small $\rho$, the distribution has a cutoff due to the separation and contrast constraints, while at large $\rho$ a cutoff is present as a result of the confusion constraint. The location of each cutoff depends on the luminosity of the companion relative to that of the primary, and hence on the mass ratio $q$ (for which the distribution was identified in the previous section).

We first test whether $\tilde{f}_\rho(\rho)$ is consistent with the log-normal period distribution. We simulate an association with the log-normal period distribution, pairing function PCP-III, and a power-law mass ratio distribution with $\gamma_q= -0.39$. The comparison between $\tilde{f}_\rho(\rho)$ for observations and simulated observations shows that the log-normal period distribution cannot be excluded with more than $1\sigma$ confidence. 

We additionally test which distribution $f_a(a) \propto a^{\gamma_a}$ fits the observations best. With the comparison we find $\gamma_a = -1.02 \pm 0.08$ ($1\sigma$ error), indicating that $\tilde{f}_\rho(\rho)$ is consistent with \"{O}pik's law. It is not surprising that both \"{O}pik's law and the log-normal period distribution are consistent with the observed $\tilde{f}_\rho(\rho)$. More than 90\% of the measured angular separations are in the range $0.1''-3''$ corresponding to $3\,200-93\,500$~R$_\odot$. Figure~\ref{figure: opik_dm_kstest_difference} shows that, given these limits for the 720~angular separation measurements, the log-normal period distribution can barely be discriminated from \"{O}pik's law at the $1\sigma$-level. 

We will now determine the minimum values $a_{\rm min}$ and $P_{\rm min}$. Binaries with the smallest orbital periods are easiest to detect in a spectroscopic survey. We therefore use $\tilde{f}_P(P)$ for the spectroscopic binaries (see Figure~\ref{figure: sosexample_observed}) to find $P_{\rm min}$ directly. In our observations we have spectroscopic elements for 98~targets. A cutoff in the observed period distribution is seen near $P=3-4$~days; we therefore adopt a value of $P_{\rm min} \approx 3.5 \pm 0.5$~days.
For  spectroscopic binaries we cannot  measure the total system mass directly. We therefore estimate the total mass using the primary spectral type. Most of the observed spectroscopic binaries are of early type, and have a mass of $2-4$~M$_\odot$. If we estimate the average mass ratio for these binaries to be $q\approx 0.5$, we obtain a typical mass of $4.5$~M$_\odot$. Under these conditions the minimum period converts to a minimum semi-major axis $a_{\rm min} = 10-20$~R$_\odot$.

The next step is to find the maximum semi-major axis $a_{\rm max}$ or period $P_{\rm max}$. It is unlikely that $a_{\rm max}$ is above the Galactic tidal limit of 0.1~pc ($5\times 10^6$~R$_\odot$). The widest observed binaries have an angular separation of $\sim 30''$, which corresponds to a projected separation of  $0.93\times 10^6$~R$_\odot$ at the distance of the association. Using equation~\ref{equation: projecteda2}, we estimate that the semi-major axis of the widest observed binaries corresponds to $0.73\times 10^6$~R$_\odot$ if all orbits are circular, and $0.91\times 10^6$~R$_\odot$ in the case of a thermal eccentricity distribution. The derived value of $a_{\rm max}$ depends on the assumption for the binary fraction (or vice-versa), and will be discussed in \S~\ref{section: example_bf}.

\subsection{Finding $f_e(e)$}

The observed distribution $\tilde{f}_e(e)$ in Figure~\ref{figure: sosexample_observed} shows that an intrinsic eccentricity distribution of the form $f_e(e) \approx \delta(e-e_0)$ can be ruled out for any $e_0$ (here we assume that the observational errors on $e$ are negligible). 
A comparison between $\tilde{f}_e(e)$ and the simulated observations for the above distribution, indicates that, as suggested by the observations, the thermal eccentricity distribution is fully consistent with the observations, while the flat eccentricity distribution can be excluded with $2\sigma$ confidence.

In the hypothetical case that the form of the eccentricity distribution is $f_e(e) \propto e^{\gamma_e}$, where the flat and thermal eccentricity distributions correspond to the special cases with $\gamma_e=0$ and $\gamma_e=1$, respectively, we find that the observations are consistent with $\gamma_e = 0.96\pm 0.08$ ($1\sigma$ error). This analysis shows that the true $f_e(e)$ is likely thermal, even though we cannot rule out the flat eccentricity distribution with high confidence.

\begin{table}
  \begin{tabular} {ll}
    \hline
    \hline
    True binary population & Derived ``true binary population''  \\
    \hline
    Pairing function PCP-III                          & PCP-I, PCP-II, or PCP-III \\
                                                      & RP and PCRP excluded  \\
    $f_q(q) \propto q^{\gamma_q}$, $\gamma_q = -0.33$ & $\gamma_q = -0.39 \pm 0.25$  \\
    \hline
    $f_e(e) = 2e$                                     &  best fit: $f_e(e) \approx 2e$ \\
                                                      & $f_e(e) \approx \delta(e-e_0)$ excluded  \\
                                                      & $f_e(e) = 1$ excluded with $2\sigma$ confidence \\
    \hline
    $f_a(a) \propto a^{\gamma_a}$, $\gamma_a = -1$    & $\gamma_a = -1.02 \pm 0.08$ \\
                                                      & or the log-normal period distribution $f_{\log P}(\log P)$ \\
    $a_{\rm min} = 15$~R$_\odot$                      & $a_{\rm min} =  10-20$~R$_\odot$     \\
    $a_{\rm max} = 10^6$~R$_\odot$                    & $a_{\rm max} =  0.5\times 10^6$~R$_\odot - 3 \times 10^6$~R$_\odot$  for $\gamma_a \approx -1.02$ \\
                                                      & ($a_{\rm max} =  0.8\times 10^6$~R$_\odot - 5 \times 10^6$~R$_\odot$ $f_{\log P}(\log P)$) \\
    \hline
    $F_{\rm M} = 100\%$                               & $F_{\rm M} = 95\%-100\%$ for $\gamma_a \approx -1.02$ \\
                                                      & ($F_{\rm M} = 85\%-100\%$ for $f_{\log P}(\log P)$)   \\
    \hline
    \hline
  \end{tabular}
  \caption{The results for our example model, where we derive the true binary population from the observed binary population of a simulated association. The left-hand column shows the properties of the model association, which contains 10\,000 binary systems. The right-hand column shows the inferred binary population, derived from the properties of the 720~visual binaries and 98~spectroscopic binaries in the simulated observations. In reality, the derivation of the properties of an association may not be as accurate as above, since (1) the available dataset is often smaller, and (2) the selection effects for true observations are often less-well understood. \label{table: sosexample_results}}
\end{table}

\subsection{Finding the binary fraction} \label{section: example_bf}

With our imaging and radial velocity survey we have detected 818~binaries among the 10\,000 association members, which gives us a strict lower limit of 8.18\% for the binary fraction. The total number of binaries in the association can be derived from the observed number of binaries if the number of missing binaries is known. The estimate for the number of missing binaries depends on the assumption for $a_{\rm max}$, which is thus far unknown, although we have constrained it in \S~\ref{section: example_sma}. The number of visual binaries and spectroscopic binaries detected in a simulated survey depends on the values of both $a_{\rm max}$ and $F_{\rm M}$ in the model. A smaller value for $a_{\rm max}$, or a larger value for $F_{\rm M}$ will result in a larger number of detected binaries (see equation~\ref{equation: binaryfraction_vs_amax}). The predicted number of detected binaries for a model also depends on whether either $f_a(a) \propto a^{-1.02}$ holds, or the log-normal period distribution. We therefore simulate observations, varying $a_{\rm max}$ in the range $0.5-10\times 10^6$~R$_\odot$ and $F_{\rm M}$ in the range $0-100\%$, and compare the number of detected binaries with the predictions. We perform this comparison for models with $f_a(a) \propto a^{-1.02}$, and for models with the log-normal period distribution. 

For the models with $f_a(a) \propto a^{-1.02}$, the inferred binary fraction based on the detected number of visual binaries ranges from $\sim 95$ (adopting $a_{\rm max} = 0.5\times 10^6$~R$_\odot$) to $\sim 100\%$ (adopting $a_{\rm max} = 2\times 10^6$~R$_\odot$). Models with $a_{\rm max}$ larger than $\sim 3\times 10^6$~R$_\odot$ contain too few visual binaries, even if the binary fraction is 100\%. The observed number of spectroscopic binaries is consistent with the values of $a_{\rm max}$ and $F_{\rm M}$ derived above.

For models with a log-normal period distribution, the value of $a_{\rm max}$ is of less importance. Since $a_{\rm max}$ lies in the upper tail of the log-normal period distribution, adopting an incorrect value for  $a_{\rm max}$  will only mildly affect the inferred properties of the binary population. If a log-normal period distribution is assumed, the inferred binary fraction as derived from the visual binaries is lower than the corresponding value for the power-law $f_a(a)$. The reason for this is that in the log-normal period distribution, most binaries are expected near the ``visual regime'', while relatively few binaries have a very large or small period. 

A model with a log-normal period distribution produces the observed number of {\em visual} binaries, if the binary fraction is between $\sim 75\%$ (adopting $a_{\rm max}=0.5\times 10^6$~R$_\odot$) and $\sim 100\%$ if (adopting $a_{\rm max}=5\times 10^6$~R$_\odot$). However, if we adopt a binary fraction in the lower part of this range, we obtain a rather low number of {\em spectroscopic} binaries. In order to produce enough spectroscopic binaries, the binary fraction should be larger than $\sim 85\%$, and $a_{\rm max} \ga 0.8\times 10^6$~R$_\odot$.

\subsection{The recovered binary population}

The properties of the true binary population as derived from the observations are summarized in Table~\ref{table: sosexample_results}. 
In this example we have been able to recover most of the properties of the true binary population, using observations of visual and spectroscopic binaries. As all targets in our survey are more massive than $\sim 1$~M$_\odot$, we have to rely on the assumption that the derived properties are also applicable to the low-mass population. Obviously, binarity studies of the low-mass population are necessary to confirm that this assumption is correct. Additionally, with a study of the mass ratio distribution of the low-mass population we will be able to discriminate between the pairing functions PCP-I, PCP-II, and PCP-III. A deep imaging survey for binarity among the low-mass members of the association should be sufficient to constrain the most important properties of the low-mass population. Spectroscopic and astrometric surveys can provide even more detailed information, but are observationally more complicated for very faint targets.

In this example we have derived the eccentricity distribution and minimum semi-major axis using the orbital elements for the 98~spectroscopic binaries. Here we have neglected the fact that in reality only a subset of these would be SB1s and SB2s, so that the number of available periods and eccentricities would be smaller. In reality, the derived $f_e(e)$ and $a_{\rm min}$ would therefore have a larger statistical error.

In our example we fully understood the selection effects, and were therefore able to accurately generate simulated observations. In reality, however, the selection effects are usually not fully understood. The observer's choice and sample bias are often well-known, as the observer compiled the list of targets. Modeling the instrument bias is more complicated. For example, the contrast bias depends on the atmospheric observations, which vary during the observing run. In spectroscopic surveys, the error in the radial velocity depends on atmospheric conditions and the spectral type of the primary. Due to these complications, it is possible that a systematic error in the modeled selection effects is present. Unlike in our example, the inferred true binary population could therefore be slightly biased.

% ====================================================================
% ====================================================================
% ====================================================================
% ==INTRODUCTION======================================================
% ====================================================================
% ====================================================================
% ====================================================================

\section{Summary and outlook} \label{section: summaryandoutlook}

We have conducted a study on the interpretation of binary star observations in OB~associations. We describe the method of simulating observations, which is a reliable method to derive the properties of the binary population using the available observations. The method of simulating observations fully takes into account the selection effects (as long as these are well-understood). 
We demonstrate with several examples how a naive interpretation of the observations (by directly correcting for the selection effects) may lead to a biased conclusion. We describe in \S~\ref{section: strategy} a strategy on how to recover the true binary population from observations of visual, spectroscopic, and astrometric stars in OB~associations, and demonstrate the method in \S~\ref{section: example1}. 
The main results of our study relate to how the observed binary fraction depends on the properties of the association and the selection effects. Briefly summarized, these are the following:
\begin{itemize}\addtolength{\itemsep}{-0.5\baselineskip}
\item[--] Due selection effects, surveys of visual, spectroscopic, and astrometric binaries generally result in an {\em observed} binary fraction that increases with increasing primary mass, even if in reality this trend is not present.
\item[--] For the nearest OB~associations ($\sim 145$~pc), 10\% of the binaries are detected as visual binaries, and 3\% as spectroscopic binaries. These fractions depend strongly on the number of binaries with a low-mass primary, as these systems are often not resolved (or not observed at all). Most of the detected binaries have an intermediate- or high-mass primary. For the binaries with a primary star more massive than a solar mass, 25\% and 23\% of the binaries are detected as visual and spectroscopic binaries, respectively.
\item[--] For many stellar groupings, the number of known visual and spectroscopic binaries is significantly larger than the number of astrometric binaries. Although astrometric observations can provide more information on individual systems, the properties of the binary population are in practice best studied using observations of visual and spectroscopic binaries.
\item[--] The binary fraction $F_{\rm M}$ is the most important parameter describing the properties of a binary population. It can only be accurately derived if {\em all} other properties of the binary population are known, as the number of ``missing binaries'' depends on these properties. In particular, the derived binary fraction depends strongly on the adopted maximum semi-major axis $a_{\rm max}$ or period $P_{\rm max}$. The combination ($F_{\rm M},a_{\rm max}$) or ($F_{\rm M},P_{\rm max}$) for the association can be derived using the detected number of visual, spectroscopic, and astrometric binaries.
\end{itemize}

In this paper we describe in detail five methods of pairing the binary components: RP, PCRP, PCP-I, PCP-II, and PCP-III. For random pairing (RP) both components are randomly drawn from the mass distribution $f_M(M)$. For primary-constrained random pairing (PCRP), both components are drawn from $f_M(M)$, with the constraint that the companion is less massive than the primary. For primary-constrained pairing (PCP-I, PCP-II, and PCP-III), the primary is drawn from $f_M(M)$, and the companion mass is determined using a mass ratio distribution $f_q(q)$. The difference between the three PCP pairing functions lies in the treatment of low-mass companions.
 Although these five pairing functions are common in literature, this does not necessarily imply that one of these pairing functions is the natural outcome of the star forming process.
The five pairing functions give significantly different results. The binary fraction and mass ratio distribution strongly depend on the number of substellar objects in the association, and on the properties of the surveyed sample. The binary fraction and mass ratio distribution can thus be used to discriminate between the different pairing functions.
This also means that the choice of the observational sample may mislead the observer in deriving the overall properties of a stellar population. If the binary fraction or mass ratio distribution of two samples (e.g., systems with B-type primaries and those with M-type primaries) are different, this does {\em not} necessarily mean that the underlying pairing function is different. 
The pairing functions RP and PCRP can be excluded if more than a $1-2\%$ of the A- and B-type stars are 'twins' ($q \geq 0.8$). 
The mass ratio distribution and binary fraction for the pairing functions PCP-I, PCP-II, or PCP-III are very similar for high-mass primaries, but differ significantly for low-mass primaries.
Only in a limited number of cases the mass ratio distribution for a sample of binaries can be described accurately with a functional form such as $f_q(q) \propto q^\gamma$ or $f_q(q)\propto (1+q)^\gamma$. The form of the mass ratio distribution depends strongly on (1) the overall mass distribution, (2) the lower mass limit and the behaviour of the mass distribution for the lowest-mass objects, (3) the pairing function, and (4) the mass range of the primary stars in the surveyed sample.
For each of the five pairing functions the binarity is described with a single parameter, the overall binary fraction. Despite this choice, the resulting population can have a binary fraction which depends on primary spectral type. This trend (which is present for RP and PCP-II) is a result of pairing the binary components.
The {\em observed} mass ratio distribution is biased to high values of $q$ for practically any pairing function and mass ratio distribution.

The distance to the nearest OB~associations is of order 145~pc. At this distance, about $70-80\%$ of the binary systems with an early type primary can be resolved by combining a radial velocity survey with (adaptive optics) imaging observations.
Visual binaries cannot be used to constrain the eccentricity distribution $f_e(e)$. The eccentricity distribution can be recovered from the measured eccentricities of spectroscopic and astrometric binaries by comparing the observed distribution with simulated observations. In general, at least $\sim 100$ eccentricity measurements are necessary to discriminate between a flat and a thermal eccentricity distribution with $3\sigma$ confidence.
It is difficult to discriminate between random or preferred orbit orientation. Only in the hypothetical case that all binary systems in an association have an inclination $i$ with $\sin 2i \approx 0$, the preferred orientation can be derived using observations of visual and spectroscopic binaries. Astrometric binaries can be used to identify a preferred orientation of the binary systems, but only if many measurements are available.

In literature, the most popular distributions for the orbital size are \"{O}pik's law (a flat distribution in $\log a$) and the log-normal period distribution found by \cite{duquennoy1991}. For nearby OB~associations, the period distribution for spectroscopic binaries is most useful to discriminate between these. The difference can also be found by comparing the number of visual and spectroscopic binaries detected with the predictions. The observed angular separation distribution $\tilde{f}_\rho(\rho)$ for visual binaries is not very useful, as \"{O}pik's law and the log-normal period distribution give very similar results in the visual binary regime for nearby associations. 
A flat distribution in $\log a$ results in flat distributions in $\log P$, $\log \rho$, $\log K_1$, $\log E$, and $\log L$. A Gaussian distribution in $\log P$ results in a Gaussian distribution in $\log a$.
Over a large range of separations, exponent $\gamma$ of a power-law $f_a(a) \propto a^\gamma$ is equal to that of the fitted exponent of $f_\rho(\rho)$. However, for very wide separations, the exponent fitted to $\tilde{f}_\rho(\rho)$ is significantly different from the exponent of $f_a(a)$. The distribution $\tilde{f}_\rho(\rho)$ for very wide binaries cannot be used directly to characterize $f_a(a)$. Instead, the observations should be compared to simulated observations.

In our analysis we have focused on OB~associations, but many of the results should be applicable to dense stellar clusters and the field star population as well. In the next chapter we will apply the method of simulating observations to the nearby OB~association Sco~OB2. 

% ====================================================================
% ====================================================================
% ====================================================================
% ==INTRODUCTION======================================================
% ====================================================================
% ====================================================================
% ====================================================================

\section*{Acknowledgements}

This research was supported by NWO under project number 614.041.006, the Royal Netherlands Academy of Sciences (KNAW) and  the Netherlands Research School for Astronomy (NOVA).

% ====================================================================
% ====================================================================
% ====================================================================
% ==APPENDIX==========================================================
% ====================================================================
% ====================================================================
% ====================================================================

\markright{Appendix A: The projection of a binary orbit}
\addcontentsline{toc}{section}{Appendix A: The projection of a binary orbit}
\section*{Appendix A: The projection of a binary orbit} \label{section: projection_binaryorbit}

In this Appendix we discuss the projection of a binary orbit onto the space of observables. In particular, we discuss the properties of the radial velocity curve (for spectroscopic binaries) and the projection of the orbital motion on the plane of the sky (for astrometric binaries). We focus on the Equatorial coordinate system $(\alpha,\delta)$, but the results are applicable to other systems, such as the ecliptic and Galactic coordinate systems.

\subsection{The Equatorial coordinate system} \label{section: equatorialsystem}

\begin{figure}[tb]
  \begin{center}
    \includegraphics[width=0.9\textwidth]{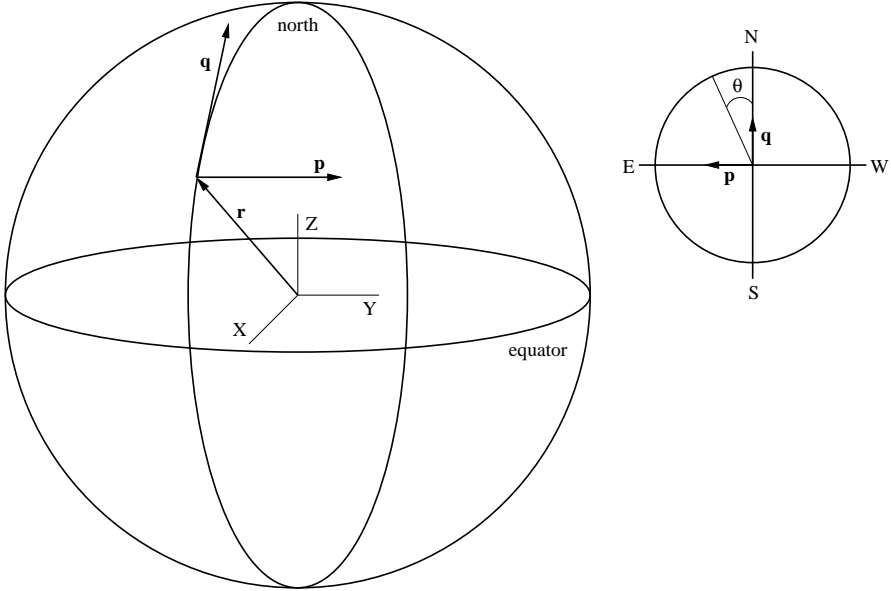}
  \end{center}
  \caption{The left figure shows the Cartesian Equatorial coordinate system defined by
    the axes $X$, $Y$, and $Z$. The vector \vect{r} points towards a position on
    the sky and the vectors \vect{p} and \vect{q} span a plane tangent to the
    celestial sphere, i.e., the plane of the sky. The vectors
    $[\vect{p},\vect{q},\vect{r}]$ are not normalized in this example. The
    direction towards ``North'' is defined by \vect{q} and the direction to ``East''
    is defined by \vect{p}. This is illustrated from the point of view of the
    observer in the right figure, where the traditional definition of the position
    angle $\theta$ with respect to \vect{q} is also shown.}
  \label{fig:eqsys}
\end{figure}

The Equatorial coordinate system is defined by the spherical coordinates ($\alpha,\delta,r$) which indicate directions on the sky, where $0\le\alpha\le2\pi$, $-\pi/2\le\delta\le+\pi/2$, and $r>0$. The associated Cartesian Equatorial system is given by the unit vectors
$[\unitvec{x},\unitvec{y},\unitvec{z}]$, where \unitvec{x} points towards
$(\alpha,\delta)=(0,0)$, \unitvec{y} points to $(\alpha,\delta)=(\pi/2,0)$,
and \unitvec{z} points to $(\alpha,\delta)=(0,\pi/2)$, so that the unit
vectors define a right-handed coordinate system.
The coordinates of a star are usually measured with $(\alpha,\delta)$, and the parallax
$\varpi = A_p/r$, where $A_p=1000$~mas\,pc, and $r$ the distance in parsec. The relation
between the Cartesian position vector \vect{b} (with $|\vect{b}|=r$) and
its distance and position on the sky is given by
\begin{equation}
  \vect{b}=\left(\begin{array}{c}b_x \\ b_y \\ b_z\end{array}\right) =
    \left(\begin{array}{c}r\cos\alpha\cos\delta \\ r\sin\alpha\cos\delta \\
      r\sin\delta\end{array}\right) =
      \frac{A_p}{\varpi} \left(\begin{array}{c}\cos\alpha\cos\delta \\ \sin\alpha\cos\delta \\
	\sin\delta\end{array}\right)\,.
\end{equation}
The velocity vector \vect{v} of a star is measured with the proper motions $\mu_\alpha$ and $\mu_\delta$, and the radial velocity $V_\mathrm{rad}$. The proper motion component $\mu_\alpha$ is often expressed as $\mu_{\alpha*}=\mu_\alpha\cos\delta$. The observed velocity components are thus measured with respect
to the normal triad $[\unitvec{p},\unitvec{q},\unitvec{r}]$, where
\begin{equation}
[\unitvec{p},\unitvec{q},\unitvec{r}] =
\left(
\begin{array}{ccc}
-\sin\alpha & -\sin\delta\cos\alpha & \cos\delta\cos\alpha \\
\phantom{-}\cos\alpha & -\sin\delta\sin\alpha & \cos\delta\sin\alpha \\
 0 & \cos\delta & \sin\delta
\end{array}
\right) \equiv {\bf R} \,.
\end{equation}
For a given position $(\alpha,\delta)$ on the sky the vector \unitvec{p}
points in the direction of increasing right ascension, \unitvec{q} in the
direction of increasing declination, and \unitvec{r} along the line of
sight (see Figure~\ref{fig:eqsys}). The vectors \unitvec{p} and \unitvec{q} define a plane tangential to
the celestial sphere and thus locally define the plane of the sky.  From the view of the observer, the vectors 
\unitvec{p} and \unitvec{q} point in the directions ``East'' and ``North'', respectively (see Figure~\ref{fig:eqsys}).
The transformation between  the astrometric parameters and radial velocity to the Cartesian Equatorial system is given by
\begin{equation}
  \label{eq:astromtob}
  \left(
  \begin{array}{c}
    b_x \\ b_y \\ b_z
  \end{array} \right) = {\bf R}
  \left(
  \begin{array}{c}
    0 \\ 0 \\ A_p/\varpi
  \end{array} \right)
\end{equation}
and
\begin{equation}
  \label{eq:astromtov}
  \left(
  \begin{array}{c}
    V_x \\ V_y \\ V_z
  \end{array} \right) = {\bf R}
  \left(
  \begin{array}{c}
    \mu_{\alpha*}A_V/\varpi \\ \mu_\delta A_V/\varpi \\ V_\mathrm{rad}
  \end{array} \right) \,,
\end{equation}
where $A_V=4.74047$~yr\,km\,s$^{-1}$. The calculation of the sky positions from \vect{b} proceeds as follows:
\begin{gather}
  \label{eq:sphe2cart1}
  \alpha = \mathrm{arctan}(b_x,b_y)\,, \\
  \label{eq:sphe2cart2}
  |\vect{b}|=\sqrt{b_x^2+b_y^2+b_z^2}=r\,, \\
  \label{eq:sphe2cart3}
  \delta=\arcsin(b_z/r)\,,
\end{gather}
where the function arctan is assumed to return the angle defined by
the point $(b_x,b_y)$ in the $XY$-plane, assigned to the proper quadrant. The proper motions and radial velocity can be obtained with
\begin{equation}
  \varpi=\frac{A_p}{|\vect{b}|}
\quad \mbox{and} \quad
\left(
\begin{array}{l}
V_{\alpha*} \\ V_\delta \\ V_\mathrm{rad}
\end{array} \right) = {\bf R}^T
\left(
\begin{array}{c}
V_x \\ V_y \\ V_z
\end{array} \right) \,,
\end{equation}
where $V_{\alpha*}=\mu_{\alpha*}A_V/\varpi$ and $V_\delta=\mu_\delta A_V/\varpi$ 
are the velocity components perpendicular to the line of sight to the
star, and ${\bf R}^T$ is the transpose of {\bf R}.

\subsection{The orbit orientation vectors} \label{section: orbitorientationvectors}

\begin{figure}[t]
  \begin{center}
    \includegraphics[width=0.9\textwidth]{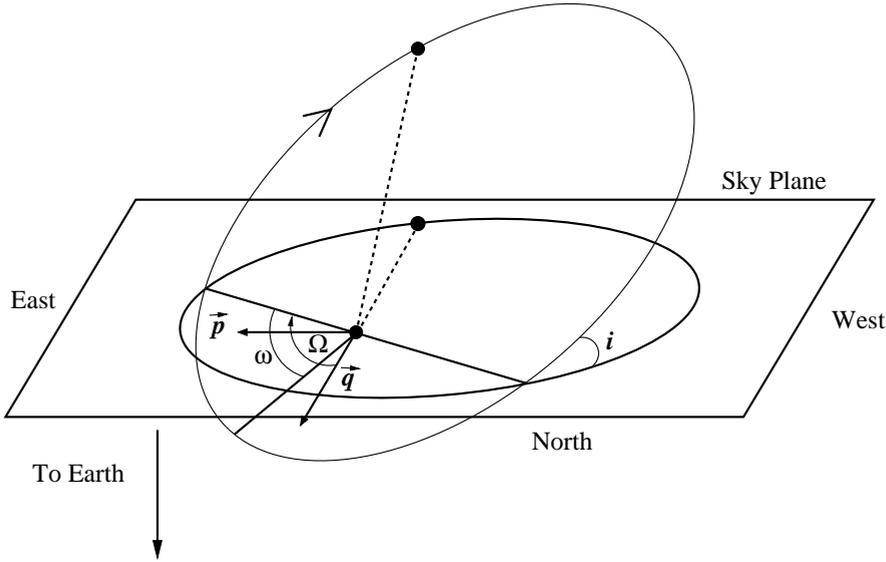}
  \end{center}
  \caption{The orbit of a binary with respect to the sky plane. The large ellipse
    indicates the actual orbit, and the small ellipse indicates the orbit
    projected in the plane of the sky. The local tangent plane $\cal{S}$ (the ``sky plane'') is defined by the vectors \unitvec{p} and \unitvec{q}. The orbital plane
    $\cal{O}$ is inclined with respect to the tangent plane with an angle $i$.
    The line of nodes is the intersection between $\cal{S}$ and $\cal{O}$. The thick line connects the focal point of the elliptical orbit with the true periastron. The angle $\Omega$ is the
    position angle of the line of nodes with respect to \unitvec{q}. The angle $\omega$
    is the angle between the line of nodes and the periastron as measured along
    the true orbit. The orientation of the orbit on the sky is completely specified by the angles $i$, $\Omega$ and $\omega$.
}
  \label{fig:orbit}
\end{figure}

The orientation of the orbit in space is defined by
$[\unitvec{x},\unitvec{y},\unitvec{z}]$. The
unit orientation vectors of the orbit are (see Fig.~\ref{fig:orbitorient}):
\begin{itemize}\addtolength{\itemsep}{-0.5\baselineskip}
\item[--] the longitudinal vector \unitvec{l}; pointing along the major axis from
  the focal point (where the primary is located) to the pericenter of the
  companion's orbit.
\item[--] the transverse vector \unitvec{t}; pointing parallel to the minor axis
  from the focal point to the true anomaly of $90^\circ$ ($\mathcal{V}=\pi/2$).
\item[--] the normal vector \unitvec{n}; defined as
  $\unitvec{n}=\unitvec{l}\times{\unitvec{t}}$. This is the vector normal to
  the plane of the binary orbit.
\end{itemize}
These vectors are defined in the Cartesian Equatorial system
$[\unitvec{x},\unitvec{y},\unitvec{z}]$, and, combined with the semi-major axis $a$ and the eccentricity $e$, define the projection of the orbit on the sky. The relation of the binary orbit
to the sky plane is shown in Figure~\ref{fig:orbit}. In practice, the orientation of an orbit is usually expressed using the orbital elements $i$, $\Omega$, and $\omega$, which we describe below.

The inclination is the angle between the orbit and sky planes (see Figure~\ref{fig:orbit}) and is defined as:
\begin{equation}
  \cos i = \inprodu{n}{r}\,.
\end{equation}
When \unitvec{n} is parallel or anti-parallel to \unitvec{r} we have
$\inprodu{n}{r}=\pm1$, and $i=0^\circ$ or $i=180^\circ$, respectively. The orbit
is then seen face-on. When \unitvec{n} is perpendicular to \unitvec{r} we
have $\inprodu{n}{r}=0$, and $i=\pm 90^\circ$, and the orbit is seen edge-on.

The inclination angle is not sufficient to specify the orientation of the orbit
because we have only defined the angle between the sky-plane $\cal{S} =
\{\unitvec{p},\unitvec{q}\}$ and the orbital plane
$\cal{O}=\{\unitvec{l},\unitvec{t}\}$. The line of nodes \unitvec{L} defines the intersection of $\cal{O}$ and $\cal{S}$ with respect to $\cal{S}$, and is given by:
\begin{equation}
\unitvec{L}=\frac{\unitvec{n}\times\unitvec{r}}{||\unitvec{n}\times\unitvec{r}||}\,
= \frac{ \unitvec{n}\times(\unitvec{p}\times\unitvec{q}) }{\sin i}
= \frac{ \inprodu{n}{q}\unitvec{p} - \inprodu{n}{p}\unitvec{q} }{\sin i}\,.
\end{equation}
The right-hand side of the equation brings out the fact that the
line of nodes lies in the plane of the sky and is oriented at some position
angle with respect to \unitvec{q}. This angle $\Omega$ is the longitude or
position angle of the line of nodes, for which
\begin{equation}
  \cos\Omega=-\frac{\inprodu{n}{p}}{\sin i}
  \quad\text{and}\quad
  \sin\Omega=\frac{\inprodu{n}{q}}{\sin i}\,.
\end{equation}
In terms of \unitvec{l} and \unitvec{t}, the expression for \unitvec{L} is
\begin{equation}
  \unitvec{L}=\frac{\unitvec{n}\times\unitvec{r}}{\sin i}=
  \frac{1}{\sin i}(\unitvec{l}\times\unitvec{t})\times\unitvec{r} =
  \frac{1}{\sin i}\left(-\inprodu{r}{t}\unitvec{l} + \inprodu{r}{l}\unitvec{t}\right)\,.
\end{equation}
This expression shows that \unitvec{L} has a position angle $\omega$ in
$\cal{O}$ defined with respect to \unitvec{l} (measured over \unitvec{t}) and
given by:
\begin{equation}
  \cos\omega=-\frac{\inprodu{r}{t}}{\sin i} \quad\text{and}\quad
  \sin\omega=\frac{\inprodu{r}{l}}{\sin i}\,.
  \label{eq:defomega}
\end{equation}
The angle $\omega$ is also the angle measured along the orbit from the periastron to the point at the intersection between \unitvec{L} and the orbit, and is known as argument of periastron. The orientation of a binary orbit is now defined by the three orientation parameters $i$, $\Omega$ and $\omega$.

\subsection{The radial velocity curve}

\begin{figure}[t]
\begin{center}
\includegraphics[width=0.7\textwidth]{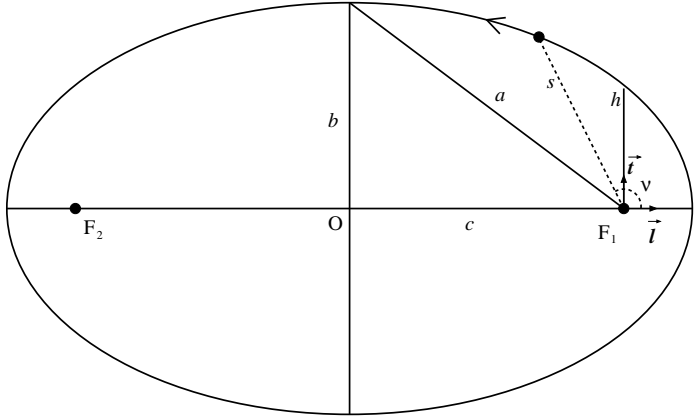}
\end{center}
\caption{The orbit of a binary seen face-on. In a coordinate system centered on
$\vect{b}_1$ the primary is located at the focal point $F_1$ of the ellipse and
the motion of the secondary is along the ellipse in the direction indicated by
the arrow. The semi-major and semi-minor axes have lengths $a$ and $b$,
respectively, and $c$ is the distance from $F_1$ to $O$, where $a^2=b^2+c^2$.
The distance $h$ is given by $h=a(1-e^2)$ and is used in the description of the
orbit in polar coordinates centered on $F_1$ (see text). The unit vectors
\unitvec{l} and \unitvec{t} defining the orientation of the orbit in space are
indicated. The third orientation vector \unitvec{n} is not shown
($\unitvec{n}=\unitvec{l}\times{\unitvec{t}}$). The dashed line indicates the
position of the secondary expressed in polar coordinates $(s,\mathcal{V})$ centered on
$F_1$. The angle $\mathcal{V}$ is known as the true anomaly.}
\label{fig:orbitorient}
\end{figure}

The orbit of a binary system is well-described in literature, and will not be discussed here in detail. The orbital motion of a star in a binary system can be described in polar coordinates $(s,\mathcal{V})$, where $s$ is the distance to the focal point, and $\mathcal{V}$ is the true anomaly (see Figure~\ref{fig:orbitorient}). For the orbit of a star in a binary system, these are related as
\begin{equation}
\label{eq:snu}
s=\frac{a(1-e^2)}{1+e\cos\mathcal{V}} \, ,
\end{equation}
where $a$ is the semi-major axis of the orbit, and $e$ the eccentricity. The dependence of $\mathcal{V}$ on time is given by
\begin{equation}
  \frac{d\mathcal{V}}{dt}=\frac{2\pi}{P(1-e^2)^{3/2}}(1+e\cos\mathcal{V})^2,
  \label{eq:dnudt}
\end{equation}
where $P$ is the orbital period. The dependence of the radial velocity on time can now be worked out from the expression above:
\begin{equation}
  t=\int_0^\mathcal{V} \left(\frac{d\mathcal{V}'}{dt}\right)^{-1}d\mathcal{V}' =
  \frac{P(1-e^2)^{3/2}}{2\pi}\int_0^\mathcal{V}\frac{1}{(1+\cos\mathcal{V}')^2}d\mathcal{V}'\,.
\end{equation}
The solution for this integral can be looked up in a table of integrals, and is given by
\begin{equation}
  t=\frac{P\sqrt{1-e^2}}{2\pi}\left(
  \frac{2}{\sqrt{1-e^2}}\arctan\left(\frac{\sqrt{1-e^2}\tan(\mathcal{V}/2)}{1+e}\right) - 
  \frac{e\sin\mathcal{V}}{1+e\cos\mathcal{V}}\right) \,.
\end{equation}
With this expression $t$ varies between $-P/2$ and $+P/2$ when $\mathcal{V}$ varies from
$0$ to $2\pi$. This expression cannot be inverted and thus no explicit
expression for the radial velocity curve $V(t)$ can be given. However, a graphical representation of $V(t)$ can be obtained using the expressions derived above, by plotting $V(\mathcal{V})$ versus $t(\mathcal{V})$.

% ======================================
% ======================================
% ======================================

\subsection{The full projection on the sky} \label{section: astrometric_projection}

The orientation vectors defined and the expression for $s$ in Equation~\ref{eq:snu} can be used to find the complete projection of a binary orbit on the sky. In the $[\unitvec{l},\unitvec{t},\unitvec{n}]$
system the position of the companion is given by:
\begin{equation}
\vect{s} = (s\cos\mathcal{V})\unitvec{l} + (s\sin\mathcal{V})\unitvec{t}\,.
\label{eq:slt}
\end{equation}
To find the position in the plane of the sky we use the fact that \vect{s} is
defined with respect to the position of the primary. For a primary with sky
coordinates $(\alpha,\delta)$ and parallax $\varpi$ we then obtain the position
of the star expressed in terms of \unitvec{p} and \unitvec{q} as
\begin{equation}
\label{eq:project}
\vect{s}=\left(\begin{array}{c} s_p \\ s_q\end{array}\right) = \varpi \times s \times
\left(\begin{array}{cc}\inprodu{l}{p} & \inprodu{t}{p}
\\
\inprodu{l}{q} & \inprodu{t}{q}\end{array}\right)
\left(\begin{array}{c}\cos\mathcal{V} \\ \sin\mathcal{V}\end{array}\right) \,,
\end{equation}
where $s=\sqrt{s_s^2+s_q^2}$. If the units of $\varpi$ are arcsec and the units of $s$
are AU, then the units of $s_p$ and $s_q$ are in arcsec. Note that we assume that
the separation $s$ is small enough so that a pure
projection from $\cal O$ onto $\cal S$ is a good approximation of the angle
between the primary and the secondary on the sky.

% ======================================
% ======================================
% ======================================

\subsection{Radial velocity statistics} \label{section: vradstat}

The radial velocity curve for a star in a binary system of mass $M_T = M_1+M_2$ is given by
\begin{equation}
  V=K(e\cos\omega+\cos(\mathcal{V}-\omega)) \quad \mbox{with} \quad K=\frac{M}{M_T}\frac{2\pi a\sin i}{P\sqrt{1-e^2}} \,,
\end{equation}
where $M=M_2$ or $M=-M_1$. For Monte Carlo modeling of the detection of spectroscopic binary stars
through the variations in $V$ it is of interest to derive the root-mean-square (rms)
variations of $V$ under the condition of random time sampling. 
It is sufficient to consider the distribution
$f_v(v)\,dv$ of $v=V/K$. We know $v(\mathcal{V})=e\cos\omega+\cos(\mathcal{V}-\omega)$ so we can
derive the distribution of $v$ from that of $\mathcal{V}$:
\begin{equation}
  f_v(v)=f_{\mathcal V}(\mathcal{V})\left|\frac{d\mathcal{V}}{dv}\right| =f_{\mathcal V}(\mathcal{V}) \left|\frac{d\mathcal{V}}{dv}\right|=\frac{1}{\sqrt{1-(v-e\cos\omega)^2}}\,.
  \label{eq:dnudv}
\end{equation}
An expression for $f_{\mathcal V}(\mathcal{V})$ can be obtained from equation~\ref{eq:dnudt} 
by assuming that the times at
which we observe the binary (modulo $P$) are randomly distributed between $t=0$
and $t=P$ with a uniform distribution, i.e., $f_t(t)=P^{-1}$ ($0 \leq t \leq P$). It then follows that:
\begin{equation}
  \label{eq:nuprob}
  f_{\mathcal V}(\mathcal{V})d\mathcal{V} 
  = f_t(t) \left| \frac{dt}{d\mathcal{V}} \right| d\mathcal{V} 
  = \frac{(1-e^2)^{3/2}}{2\pi(1+e\cos\mathcal{V})^2}d\mathcal{V}\,.
\end{equation}
The distribution is peaked toward $\mathcal{V}=\pi$ which
reflects the fact that the stars spend (much) more time near apastron than
near periastron. So the value of $\mathcal{V}$ is more likely to be close to $\pi$ for
non-zero eccentricity. Note that $f_{\mathcal V}(\mathcal{V})d\mathcal{V}$ as given above is normalized to unity.
The distribution of $v$ (not yet normalized to 1) can now be found by combining
equations~\ref{eq:dnudv} and~\ref{eq:nuprob}.
From the geometry of the
problem it follows that the only difference between viewing an orbit with
argument of periastron $\omega$ or $\omega+\pi$ is in the sign of the observed
velocities. The distribution of their absolute values remains the same. That is:
\begin{equation}
  f_v(v)|_{\omega=x+\pi}=f_v(-v)|_{\omega=x}\,,
  \label{eq:fvsymm}
\end{equation}
To make the expression for $f_v(v)$ work the value of $\omega$ should be
restricted between $0$ and $2\pi$ and two cases should be distinguished,
$\omega\le\pi$ and $\omega>\pi$. This leads to the final expression for the
distribution of $v$:
\begin{equation}
  f_v(v)\,dv=
  \left\{
  \begin{tabular}{l}
    $\displaystyle\frac{(1-e^2)^{3/2}}{2\pi\,\big(1+e\cos(\arccos(v-e\cos\omega)+\omega)\big)^2
      \sqrt{1-(v-e\cos\omega)^2}}\ dv$ \\
      \multicolumn{1}{r}{$\mbox{if}~0\le\omega\le\pi$} \\
    $\displaystyle\frac{(1-e^2)^{3/2}}{2\pi\,\big(1-e\cos(\arccos(e\cos\omega-v)+\omega)\big)^2
      \sqrt{1-(e\cos\omega-v)^2}}\ dv$ \\
      \multicolumn{1}{r}{$\mbox{if}~\pi<\omega\le2\pi$}
  \end{tabular}
  \right.\,.
\end{equation}
This distribution is non-zero on the interval $[v_\mathrm{min},v_\mathrm{max}]$
and has a $(1-x^2)^{-1/2}$ singularity at both ends of the interval,
which makes it easy to calculate the moments of the distribution through numerical
integration.

For $0\le\omega\le\pi$ the integrals for the moments of $f_v(v)$ can be
transformed to a simpler form by using the substitution
\begin{equation}
  v=e\cos\omega+\cos(z-\omega),
\end{equation}
for which $dv=-\sin(z-\omega)dz$ (a similar
transformation can be used for $\pi<\omega\le2\pi$). The zeroth moment
(normalization) becomes:
\begin{equation}
  h_0=\int_{v_\mathrm{min}}^{v_\mathrm{max}}f_v(v)dv =
  \int_\omega^{\omega+\pi}\frac{(1-e^2)^{3/2}}{2\pi(1+e\cos z)^2}dz\,.
\end{equation}
which, after integration, is given by
\begin{equation}
  h_0=\left[\frac{\sqrt{1-e^2}}{2\pi}\left(
  \frac{2}{\sqrt{1-e^2}}\arctan\left(\frac{\sqrt{1-e^2}\tan(z/2)}{1+e}\right) - 
  \frac{e\sin z}{1+e\cos z}\right)\right]_\omega^{\omega+\pi}\,.
\end{equation}
Substituting the integral limits in this case involves making the right choice
for the value of $\arctan(0)$ (i.e., either $0$ or $\pm\pi$). From the physics of the problem we already know that the first moment or mean $h_1$ of $v$ is zero. The integral for the second moment is:
\begin{equation}
  h_2=\frac{1}{h_0}\int_{v_\mathrm{min}}^{v_\mathrm{max}}v^2f_v(v)dv = 
  \frac{1}{h_0}\int_\omega^{\omega+\pi}\frac{(1-e^2)^{3/2}}{2\pi(1+e\cos z)^2}
  (e\cos\omega+cos(z-\omega))^2dz\,.
\end{equation}
The rms variation of $v$ is $\sqrt{h_2}$. Finally, the rms variation of the radial velocity $V$ is given by:
\begin{equation}
  V_{\rm rms} = K\sqrt{h_2} \equiv K\sqrt{1-e^2}\,\mathcal{R}(e,\omega) = L\,\mathcal{R}(e,\omega)\,,
\end{equation}
where $L$ is defined in equation~\ref{equation: L}. In Figure~\ref{figure: rmsdefinition} the value of the function
$\mathcal{R}(e,\omega)$ is shown for the entire $(e,\omega)$ plane. This
figure shows how the rms value of the radial velocity of one of the binary
components varies due to variations in $e$ and $\omega$ only. Note the expected
symmetry about $\omega=\pi$.

% ======================================
% ======================================
% ======================================

\markright{Appendix B: The pairing function}
\addcontentsline{toc}{section}{Appendix B: The pairing function}
\section*{Appendix B: The pairing function} \label{section: appendix_pairingfunctions}

\begin{table}
  \small
  \begin{tabular}{l p{0.9cm} p{0.9cm}p{0.9cm} p{0.9cm}p{0.9cm} p{0.9cm}p{0.9cm}}
    \hline
    \hline
    Primary spectral type          & \multicolumn{1}{c}{B} & \multicolumn{1}{c}{A} & \multicolumn{1}{c}{F} & \multicolumn{1}{c}{G} & \multicolumn{1}{c}{K} & \multicolumn{1}{c}{M} & \multicolumn{1}{c}{BD} \\
    and mass range (M$_\odot$) & \multicolumn{1}{c}{3-20} & \multicolumn{1}{c}{1.5-3} & \multicolumn{1}{c}{1-1.5} & \multicolumn{1}{c}{0.8-1} & \multicolumn{1}{c}{\mbox{0.5-0.8}} & \multicolumn{1}{c}{\mbox{0.08-0.5}} & \multicolumn{1}{c}{\mbox{0.02-0.08}} \\
    \hline
    \hline
    \multicolumn{8}{l}{Random pairing}\\
    \hline
    Preibisch, $\alpha=-0.9$   & $-2.40$ & $-1.64$ & $-1.28$ & $-1.08$ & $-0.87$ & $-0.72$ & $+1.01$ \\
    Preibisch, $\alpha=2.5$    & $-2.49$ & $-1.59$ & $-1.28$ & $-1.07$ & $-0.81$ & $+0.02$ & $+2.60$ \\
    Preibisch, no BD           & $-2.32$ & $-1.68$ & $-1.26$ & $-1.08$ & $-0.60$ & $+0.39$ & \multicolumn{1}{c}{$-    $} \\
    Kroupa                     & $-2.09$ & $-1.85$ & $-1.54$ & $-1.38$ & $-1.24$ & $-0.57$ & $+1.08$ \\
    Salpeter                   & \multicolumn{1}{c}{$-    $} & $-2.87$ & $-2.50$ & $-2.39$ & $-2.17$ & $-1.26$ & $+1.25$ \\
    \hline
    \multicolumn{8}{l}{Primary-constrained random pairing}\\
    \hline
    Preibisch, $\alpha=-0.9$   & $-2.13$ & $-1.61$ & $-1.32$ & $-1.00$ & $-0.94$ & $-0.63$ & $+1.78$ \\
    Preibisch, $\alpha=2.5$    & $-2.33$ & $-1.68$ & $-1.30$ & $-1.01$ & $-0.79$ & $+0.31$ & $+2.80$ \\
    Preibisch, no BD           & $-2.43$ & $-1.67$ & $-1.22$ & $-1.03$ & $-0.56$ & $+1.07$ & \multicolumn{1}{c}{$-    $} \\
    Kroupa                     & $-2.21$ & $-1.59$ & $-1.55$ & $-1.38$ & $-1.23$ & $-0.47$ & $+1.70$ \\
    Salpeter                   & \multicolumn{1}{c}{$-    $} & \multicolumn{1}{c}{$-    $} & $-2.37$ & $-2.20$ & $-2.00$ & $-1.23$ & $+2.76$ \\
    \hline
    \multicolumn{8}{l}{Primary-constrained pairing, PCP-I, $M_{\rm 2,min}=0~\mbox{M}_\odot$, Preibisch mass distribution, $\alpha=-0.9$}\\
    \hline
    $f_q(q) \propto q^{0}$ = flat & $+0.17$ & $+0.08$ & $-0.06$ & $-0.03$ & $-0.02$ & $-0.01$ & $+0.02$ \\
    $f_q(q) \propto q^{-0.33}$    & $-0.39$ & $-0.34$ & $-0.31$ & $-0.29$ & $-0.33$ & $-0.32$ & $-0.35$ \\
    $f_q(q) \propto q^{-0.5}$     & $-0.48$ & $-0.51$ & $-0.54$ & $-0.50$ & $-0.51$ & $-0.51$ & $-0.50$  \\
    \hline
    \multicolumn{8}{l}{Primary-constrained pairing, PCP-II, $M_{\rm 2,min}=0.02~\mbox{M}_\odot$, Preibisch mass distribution, $\alpha=-0.9$}\\
    \hline
    $f_q(q) \propto q^{0}$ = flat & $+0.30$ & $-0.03$ & $-0.10$ & $+0.04$ & $-0.01$ & $+0.11$ & $+1.56$ \\
    $f_q(q) \propto q^{-0.33}$    & $-0.39$ & $-0.33$ & $-0.37$ & $-0.35$ & $-0.31$ & $-0.18$ & $+1.39$ \\
    $f_q(q) \propto q^{-0.5}$     & $-0.30$ & $-0.57$ & $-0.47$ & $-0.47$ & $-0.53$ & $-0.36$ & $+1.21$ \\
    \hline
    \multicolumn{8}{l}{Primary-constrained pairing, PCP-II, $M_{\rm 2,min}=0.02~\mbox{M}_\odot$, Salpeter mass distribution, $\alpha=-2.35$}\\
    \hline
    $f_q(q) \propto q^{0}$ = flat & $-0.16$ & $+0.45$ & $+0.16$ & $-0.36$ & $+0.04$ & $+0.23$ & $+2.15$ \\
    $f_q(q) \propto q^{-0.33}$    & $-0.32$ & $-0.23$ & $+0.12$ & $-0.11$ & $-0.33$ & $-0.04$ & $+1.89$ \\
    $f_q(q) \propto q^{-0.5}$     & $-0.59$ & $-0.74$ & $-0.17$ & $-0.63$ & $-0.44$ & $-0.16$ & $+1.81$ \\
    \hline
    \multicolumn{8}{l}{Primary-constrained pairing, PCP-III, $M_{\rm 2,min}=0.02~\mbox{M}_\odot$, Preibisch mass distribution, $\alpha=-0.9$}\\
    \hline
    $f_q(q) \propto q^{0}$ = flat & $-0.05$ & $-0.02$ & $-0.04$ & $+0.10$ & $-0.05$ & $+0.15$ & $+2.27$ \\
    $f_q(q) \propto q^{-0.33}$    & $-0.26$ & $-0.44$ & $-0.38$ & $-0.36$ & $-0.33$ & $-0.17$ & $+2.12$ \\
    $f_q(q) \propto q^{-0.5}$     & $-0.56$ & $-0.54$ & $-0.53$ & $-0.44$ & $-0.49$ & $-0.32$ & $+1.98$ \\
    \hline
    \multicolumn{8}{l}{Primary-constrained pairing, PCP-III, $M_{\rm 2,min}=0.02~\mbox{M}_\odot$, Salpeter mass distribution, $\alpha=-2.35$}\\
    \hline
    $f_q(q) \propto q^{0}$ = flat & $+0.16$ & $-0.09$ & $-0.04$ & $+0.33$ & $+0.05$ & $+0.26$ & $+4.08$ \\
    $f_q(q) \propto q^{-0.33}$    & $-0.49$ & $+0.04$ & $-0.27$ & $-0.38$ & $-0.25$ & $+0.02$ & $+3.91$ \\
    $f_q(q) \propto q^{-0.5}$     & $-0.47$ & $-0.41$ & $-0.56$ & $-0.50$ & $-0.49$ & $-0.10$ & $+3.79$ \\
    \hline
    \hline
  \end{tabular}
\caption{The value of the {\em fitted} exponent $\gamma_q$ of the specific mass ratio distribution of the different primary spectral types, for the different mass distributions and pairing properties. Column~1 lists the mass ratio distribution $f_q(q)$, or, in the case of random pairing, the mass distribution. Each model consists of 50\,000 binary systems. The other columns list the {\em fitted} values of $\gamma_q$ for each spectral type. We only include the systems with $q \geq 0.1$ in the fit. Note that the best-fitting value of $\gamma_q$ is shown, even though a power-law is generally not the natural form of the specific mass ratio distribution (see, e.g., Figures~\ref{figure: concept_pairing} and~\ref{figure: rp_different_samples}). 
In several cases no acceptable fit could be obtained, due to the absence of binaries with the properties required.
This table shows that the shape of the specific mass ratio distribution may depend strongly on (1) the pairing function, (2) the mass distribution, (3) the generating mass ratio distribution, and (4) the spectral type of the primaries in the sample.  \label{table: massratioexponent}}
\end{table}

In this appendix we discuss in detail the properties of each of the five pairing functions used in our analysis: PCP-I, PCP-II, PCP-III, RP, and PCRP.  Each of these pairing functions result in a significantly different mass ratio distribution and binary fraction, which is illustrated in Figures~\ref{figure: rp_different_samples}--\ref{figure: massratio_2d_fm} and Table~\ref{table: massratioexponent}. We denote the {\em specific} mass ratio distribution for binary systems of which the primary is of mass $M_1$ with $f_{M_1}(q)$.  Similarly, $F_{{\rm M};M_1}$ represents the {\em specific} binary fraction for systems with primary mass $M_1$.

\subsection{Pairing function RP} \label{section: pairingfunction_rp}

\begin{SCfigure}[][!tbp]
  \centering
  \includegraphics[width=0.55\textwidth,height=!]{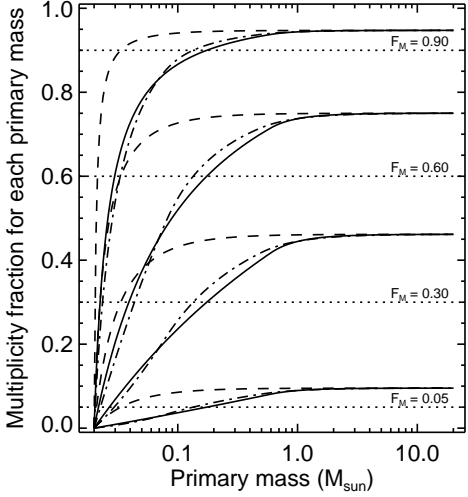}
  \caption{The binary fraction as a function of primary mass for random pairing (RP). Models with a binary fraction $F_{\rm M}$ of (from bottom to top) 5\%, 30\%, 60\%, and 90\% are shown, for the Preibisch mass distribution (solid curves), the Kroupa mass distribution (dash-dotted curves), and the Salpeter mass distribution (dashed curves). The horizontal dotted lines represent the overall binary fraction. No selection effects have been applied.  The figure shows that, even though we have applied the pairing function regardless of the primary spectral type, the binary fraction for binaries with intermediate mass primaries differs significantly from that of binaries with low mass primaries.
 \label{figure: binaryfraction_randompairing} }
\end{SCfigure}

In the case of random pairing (RP), the primary and companion mass are both independently drawn from $f_M(M)$, and swapped, if necessary, so that the most massive star is the primary. As a result of this swapping, neither the resulting primary mass distribution $f_{M_1}(M_1)$, nor the companion mass distribution $f_{M_2}(M_2)$, nor the system mass distribution $f_{M_1+M_2}(M_1+M2)$ is equal to the generating mass distribution $f_M(M)$ \citep[see, e.g.,][]{malkovzinnecker2001}. On the other hand, the mass distribution of {\em all} stars (i.e., all individual primaries and companions) is equal to $f_M(M)$.  The mass ratio distribution for an association of randomly paired binaries is bounded by the limits
\begin{equation} \label{equation: rp_massratioconstraints}
  0 < q_{\rm min} = \frac{M_{\rm 2,min}}{M_{\rm 1,max}} \leq q \leq 1.
\end{equation}
The overall mass ratio distribution (i.e., the mass ratio distribution of the complete association) for the stellar grouping resulting from random pairing is given by:
\begin{equation} \label{equation: randompairingmassratiodistribution}
f_{\rm RP,overall}(q) = 2\,\frac{d}{dq}\, \iint_{(S)} f_M(M') f_M(M'') \,dM' \,dM'' ,
\end{equation}
where $(S)$ is the region defined by
\begin{equation}
(S) = \left\{ \ 
(M',M''): M' \leq M_{\rm max} \wedge M'' \geq M_{\rm min} \wedge M'' \leq qM' 
\ \right\}.
\end{equation}
In the special case where the mass distribution is of the form $f_M(M) \propto M^{-\alpha}$ with $\alpha > 1$, the overall mass ratio distribution is given by
\begin{equation}
f_{\rm RP,overall}(q) \propto q^{\alpha-2} - \left( M_{\rm min}/M_{\rm max} \right)^{2(\alpha-1)} q^{-\alpha} 
\quad
\overset{\underrightarrow{M_{\rm min} \ll M_{\rm max}}}{}
\quad
q^{\alpha-2}
\end{equation}
\citep{piskunov1991}. In this special case, with $M_\mathrm{min} \ll
M_\mathrm{max}$, the overall mass ratio distribution resulting from random
pairing can thus be described by a power-law. 

The {\em specific} mass ratio distribution $f_{\rm RP;M_1}(q)$ for systems with primaries in the mass range
$[M_{\rm 1,low},M_{\rm 1,high}]$ can be calculated using equation~\ref{equation:
randompairingmassratiodistribution}, but now the integration is performed over
region $(S')$, which is given by
\begin{equation} (S') = \left\{ \ (M',M''): (M',M'') \in (S) \wedge M_{\rm
  1,low} \leq M' \leq M_{\rm 1,high} \ \right\}.
\end{equation}
For pairing function RP the specific mass ratio distribution depends on $M_{\rm 1,low}$ and $M_{\rm 1,high}$, which is illustrated in Figure~\ref{figure: concept_pairing}. Binaries with a high-mass primary tend to have very low mass ratios, while the lowest-mass binaries tend to have a mass ratio close to unity.
The specific mass ratio distribution for a sample of stars is additionally
dependent on the behaviour of the mass distribution in the brown dwarf regime (see Figure~\ref{figure: rp_different_samples}). As a result of this dependence, it is in theory possible to derive the properties of the mass distribution in the substellar regime by analyzing the mass ratio distribution for binaries with stellar targets.

For random pairing, the {\em specific} binary fraction $F_{{\rm M,RP;}M_1}$
(i.e., the binary fraction for a sample of stars with mass $M_1$) is generally also 
dependent on the choice of the sample, which can be explained as follows.
For random pairing, the primary star is drawn from the mass distribution. A
fraction $F_{\rm M}$ of the primaries is assigned a candidate, which is drawn
from the same mass distribution, and primary and companion are swapped, if
necessary, so that the primary is the most massive star. As a result, systems
with massive primaries have $F_{{\rm M,RP;}M_1} \geq F_{\rm M}$, while systems
with low-mass primaries have $F_{{\rm M,RP;}M_1} \leq F_{\rm M}$, even if no
selection effects are present. 

A target sample for a binarity survey is usually constructed and analyzed for a group of stars with a certain
spectral type. For pairing function RP, this choice introduces a trend between binary fraction and spectral type. The relation between specific binary fraction and primary mass for random pairing is given by
\begin{equation}
F_{{\rm M,RP;}M_1} = \frac{ 2 B\, F_M(M_1) }{ 2 B\, F_M(M_1) + S  } = \left( \frac{F_{\rm M}^{-1} -1 }{2 F_M(M_1)} + 1 \right)^{-1},
\end{equation}
where $F_M(M)$ is the cumulative distribution function evaluated for mass $M$.
Clearly we have $F_{{\rm M,RP;}M_1} \equiv 1$ when $S=0$ and $F_{{\rm M,RP;}M_1}
\equiv 0$ when $B=0$. Also, $F_{{\rm M,RP;}M_{\rm max}} = 2B/(2B+S) \geq F_{\rm
M}$ and $F_{{\rm M,RP;}M_{\rm min}} = 0\leq F_{\rm M}$. Figure~\ref{figure:
binaryfraction_randompairing} shows the binary fraction as a function of primary
mass, for several binary fractions and mass distributions.
Note that the variation of binary fraction with primary mass is purely a result
of the choice of paring function. No explicit variation of binary fraction with
primary mass is included in the simulations.

\subsection{Pairing function PCRP} \label{section: pairingfunction_pcrp}

For primary-constrained random pairing (PCRP), each primary mass $M_1$ is drawn from the mass distribution. The companion mass $M_2$ is also drawn from the mass distribution, but with the additional constraint that $M_2 \leq M_1$. The limits on the mass ratio are equivalent to those of random pairing (equation~\ref{equation: rp_massratioconstraints}). 

The overall mass ratio distribution for an association with pairing function PCRP and mass distribution $f_M(M)$ is given by 
\begin{equation} \label{equation: pcrpmassratiodistribution}
f_{\rm PCRP,overall}(q) = \frac{d}{dq}\, \iint_{(S)} f_M(M') 
\left( \frac{f_M(M'')}{\int_{M_{\rm min}}^{M'}f_M(M''')\,dM'''} \right) \, dM' \,dM'' \,,
\end{equation}
where the term in the denominator is a re-normalization, and $(S)$ is the region defined by
\begin{equation}
(S) = \left\{ \ 
(M',M''): M' \leq M_{\rm max} \wedge M'' \geq M_{\rm min} \wedge M'' \leq qM' 
\ \right\} \,.
\end{equation}
The {\em specific} mass ratio distribution $f_{\rm PCRP;M_1}(q)$ for systems with primaries in the mass range $[M_{\rm 1,low},M_{\rm 1,high}]$ is given by equation~\ref{equation: pcrpmassratiodistribution}, but now the integration is performed over region $(S')$, which is defined as
\begin{equation}
(S') = \left\{ \ 
(M',M''): (M',M'') \in (S) \wedge M_{\rm 1,low} \leq M' \leq M_{\rm 1,high}
\ \right\},
\end{equation}
resulting in specific mass ratio distributions that are generally different for
different choices of $M_{\rm 1,low}$ and $M_{\rm 1,high}$. For a sample containing
high-mass primaries (e.g., A~or B~stars), the {\em specific} mass ratio
distributions  $f_{\rm PCRP;M_1}(q)$ and $f_{\rm RP;M_1}(q)$ are approximately
equal, as the denominator in equation~\ref{equation: pcrpmassratiodistribution}
approaches unity.

For pairing function PCRP, the companion mass distribution
$f_{M_2}(M_2)$ is equal to the single-star mass distribution $f_M(M)$ for masses
smaller than $M_1$. The companion mass distribution can thus in principle be used to derive the
properties of the mass distribution. 
For example, in a set of primary stars of mass 1~M$_\odot$, the
companions with mass ratio $q<0.08$ are brown dwarfs. If the observations are of
good enough quality to study the mass ratio distribution below 0.08, and it is
known a-priori that the pairing function is PCRP, the mass ratio distribution can be used
to constrain the slope $\alpha$ of the mass distribution in the brown dwarf
regime. 

The pairing algorithms RP and PCRP appear similar, but their difference is most pronounced in the primary mass distribution, which can be understood as follows. Suppose a primary mass $M_1$ of $5~\mbox{M}_\odot$ is drawn from the mass distribution. For PCRP, the companion mass $M_2$ is always smaller than the primary mass, while for RP the companion mass can take any value permitted by the mass distribution (i.e. also $M_2 > M_1$, after which the components are switched). For RP there is consequently a larger number of binary systems with high-mass primaries. Another difference is that for PCRP the binary fraction is independent of primary spectral type, unlike for RP.

\subsection{Pairing function PCP-I} \label{section: pairingfunction1}

For pairing function PCP-I, each binary is generated by drawing a primary mass from $f_M(M)$, and a mass ratio from $f_q(q)$. The companion mass is then calculated from $M_2 = qM_1$. As a result, the {\em specific} mass ratio distribution $f_{{\rm I};M_1}(q)$ and overall mass ratio distribution $f_{\rm I}(q)$ are equal to the generating mass ratio distribution $f_q(q)$:
\begin{equation}
f_{\rm I}(q;M_1) = f_{\rm I}(q) = f_q(q) \quad 0 < q \leq 1 \,,
\end{equation}
and a similar statement can be made for the binary fraction:
\begin{equation}
F_{{\rm M,I};M_1} = F_{{\rm M,I}} = F_{\rm M}  \,.
\end{equation}
Pairing function PCP-I is conceptually the simplest. The minimum companion mass for PCP-I is given by 
$M_{\rm 2,min}=q_{\rm min}M_{\rm 1,min}$, which could be significantly lower than the minimum mass of brown dwarfs. Clearly, even planetary companions are to be included if we wish to use pairing
function PCP-I. However, if we do include planets, we make the implicit
assumption that the star formation process is scalable down to planetary masses.
This assumption is in contradiction with the theories that suggest that stars
(and probably brown dwarfs) form by fragmentation, while massive planets form by
core-accretion. Although the theories of star and planet formation are
incomplete, this is a point to keep in mind when using PCP-I.

If we would {\em require} that none of the companions have a planetary mass, the condition  $M_{\rm 2,min}/M_{\rm 1,max} \geq q_{\rm min}$ makes PCP-I impractical to use. If we wish to study for
example the mass distribution between $M_{\rm 1,min} = 0.08~\mbox{M}_\odot$ and
$ M_{\rm 1,max} = 20~\mbox{M}_\odot$, and further require that all companions
are stellar ($M_2 \geq M_{\rm 2,min} = 0.08~\mbox{M}_\odot)$, the PCP-I method
requires $q_{\rm min} = 1$. If we allow brown dwarf companions ($M_{\rm 2,min} =
0.02~\mbox{M}_\odot)$, a minimum value $q_{\rm min} = 0.25$ is required. This issue is addressed by pairing functions PCP-II and PCP-III, which we describe below.

\subsection{Pairing function PCP-II} \label{section: pairingfunction2}

\begin{SCfigure}[][!tbp]
  \centering
  \includegraphics[width=0.55\textwidth,height=!]{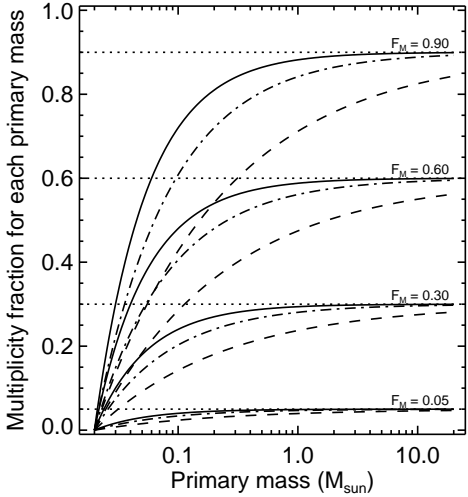}
  \caption{The binary fraction as a function of primary mass for pairing function PCP-II. Models with a binary fraction $F_{\rm M}$ of (from bottom to top) 5\%, 30\%, 60\%, and 90\% are shown, for mass ratio distributions $f_q(q)\propto q^\gamma$, with $\gamma=0$ (solid curves), $\gamma=-0.3$ (dash-dotted curves), and $\gamma=-0.6$ (dashed curves). The minimum companion mass is set to $M_{\rm 2,min}=0.02~$M$_\odot$. The horizontal dotted lines represent the overall binary fraction. No selection effects have been applied. The figure illustrates the fact that, as a result of pairing function PCP-II, the binary fraction for binaries with intermediate mass primaries differs significantly from that of binaries with low mass primaries.
 \label{figure: binaryfraction_pcp2} }
\end{SCfigure}

When generating a binary with pairing function PCP-II, the primary mass $M_1$ is drawn from $f_M(M)$, and the mass ratio from $f_q(q)$, with the additional constraint that the resulting companion mass is larger than a certain minimum mass $M_{\rm 2,min}$. One may wish to use this prescription if a minimum mass $M_{\rm 2,min}$ is expected from theory
(e.g., the Jeans mass). Another, more observational approach may be to ``ignore'' the planets, by not considering objects below a certain
mass $M_2 < M_{\rm 2,min}$ as companions. Although this seems somewhat
artificial, this method is often used in practice. Planets are usually not
considered as companions (the Sun is a ``single star''), which implies a limit
$M_2 \approx 0.02~$M$_\odot$. 

The specific mass ratio distribution for systems with primary mass $M_1$ is given by
\begin{equation}
f_{\rm II}(q;M_1) = f_q(q) \quad {\rm with} \quad q_{\rm min}(M_1)= \frac{M_{\rm 2,min}}{M_1}\leq q \leq q_{\rm max}.
\end{equation}
Note that $f_{\rm II}(q)$ is not normalized, i.e., not all primaries are
assigned a companion.  The overall mass ratio distribution $f_{\rm II}(q)$ is
given by integrating over the primary mass range, weighing with the mass distribution:
\begin{equation}
f_{\rm II}(q) = \int_{M_{\rm 1,min}}^{M_{\rm 1,max}} f_M(M_1) \, f_{\rm II}(q;M_1) \,dM_1 \,.
\end{equation}
The distribution $f_{\rm II}(q)$ has a higher average mass ratio than the generating mass ratio distribution $f_q(q)$.

The pairing function PCP-II results in an overall binary fraction $F_{\rm M,II}$ which is smaller than the generating binary fraction $F_{\rm M}$. The specific binary fraction for systems with primary mass $M_1$ is given by
\begin{equation} \label{equation: binaryfraction_pcp2}
F_{{\rm M,II;}M_1} = F_{\rm M} \int^{q_{\rm max}}_{q_{\rm min}(M_1)} f_q(q)\,dq \,.
\end{equation}
As a result, the overall binary fraction $B/(S+B)$ {\em after} rejection of the low-mass companion is unequal to the generating binary fraction $F_{\rm M,II}$. It may therefore be more appropriate to refer to $F_{\rm M,II}$ as the ``binarity parameter''. In the special case of a mass ratio distribution of the form $f_q(q) \propto q^\gamma$ 
with $0\leq q\leq 1$, equation~\ref{equation: binaryfraction_pcp2} can be written
\begin{equation}
F_{{\rm M,II;}M_1} = F_{\rm M} \left( 1-\left(\frac{M_{\rm 2,min}}{M_1}\right)^{\gamma+1} \right) \quad \quad \gamma \neq -1 \,.
\end{equation}
Figure~\ref{figure: binaryfraction_pcp2} shows the specific binary fraction $F_{{\rm M,II;}M_1}$ as a function of the overall binary fraction, the primary mass, and the mass ratio distribution. The specific binary fraction depends on the shape of the mass ratio distribution, and is independent of the mass distribution. 

The high-mass primaries are hardly affected by the rejection, as $M_{\rm 2,min}/M_{\rm 1,max} \ll 1$; their binary fraction is therefore similar to the generating binary fraction. For binaries with high-mass primaries, the mass ratio distribution obtained with this method approximates the generating mass ratio distribution $f_q(q)$. For the very low-mass primaries, however, a large fraction of the companions is rejected, and therefore their binary fraction is low. The remaining companions have a mass comparable to that of their primary. The resulting mass ratio distribution for the lowest-mass binaries is peaked to unity (see Figure~\ref{figure: concept_pairing}). For example, if we let $F_{\rm M}=100\%$, $\gamma=-0.3$, and $M_{\rm 2,min}=0.02~$M$_\odot$,
then systems with B-type primaries and systems with M-type primaries have a
resulting binary fraction of 98\% and 80\%, respectively. If we also do not
consider brown dwarfs as companions (so $M_{\rm 2,min}=0.08~$M$_\odot$), then
the resulting binary fractions are 96\% and 47\%, respectively.  Note that the
trend between binary fraction and spectral type for PCP-II is merely a result of
the pairing method.

% ==================================
% ==================================
% ==================================

\subsection{Pairing function PCP-III} \label{section: pairingfunction3}

For pairing function PCP-III, the primary mass is chosen from $f_M(M)$, and the mass ratio is chosen from a generating mass ratio distribution $f_q(q)$. If the resulting companion star mass is below a certain limit $M_{\rm 2,min}$, the companion mass is redrawn from $f_q(q)$. This procedure is repeated until a companion more massive than $M_{\rm 2,min}$ is obtained. This procedure can be considered as a re-normalization of $f_q(q)$ in the range $q_{\rm min}(M_1) \leq q \leq q_{\rm max}$. The resulting specific mass ratio distribution $f_{\rm III}(q)$ as a function of primary mass $M_1$ is given by
\begin{equation} \label{equation: massratiodistribution_pcp3}
f_{\rm III}(q;M_1) = \frac{f_q(q)}{ \int^{q_{\rm max}}_{q_{\rm min}(M_1)} f_q(q)\,dq} \quad {\rm with} \quad q_{\rm min}(M_1)\leq q \leq q_{\rm max}
\end{equation}
As a result, the mass ratio distribution for low-mass stars is peaked to high values of $q$. In the special case of a power-law mass ratio distribution $f_q(q) \propto q^\gamma$ with $0\leq q\leq 1$, equation~\ref{equation: massratiodistribution_pcp3} can be written
\begin{equation}
f_{\rm III}(q;M_1) = \frac{(\gamma+1)\,q^\gamma}{1- q_{\rm min(M_1)}^{\gamma+1}}\quad {\rm with} \quad q_{\rm min}(M_1)\leq q \leq 1, \quad \gamma \neq -1 \,.
\end{equation}

The high-mass primaries are hardly affected by the redrawing of companions, as for these stars $M_{\rm 2,min}/M_{\rm 1,max} \ll 1$, and therefore the mass ratio distribution obtained with this method approximates the generating mass ratio distribution $f_q(q)$. For the very low-mass primaries, however, a large fraction of the companions is redrawn until $M_2 \geq M_{\rm 2,min}$. Consequently, all binaries with a low-mass primary have a mass ratio close to unity. The resulting mass ratio distribution for the lowest-mass binaries is therefore peaked to unity (see Figure~\ref{figure: concept_pairing}). For pairing function PCP-III the resulting binary fraction $F_{\rm M,III}$ is
equal to the generating binary fraction $F_{\rm M}$.

%%%%%%%%%%%%%%%%%%%%%%%%%%%%%%%%%%%%%%%%%%%%%%%%%%%%%%%%%%%%%%%%%%%%%%%%%%%
%%%%%%%%%%%%%%%%%%%%%%%%%%%%%%%%%%%%%%%%%%%%%%%%%%%%%%%%%%%%%%%%%%%%%%%%%%%
%%%%%%%%%%%%%%%%%%%%%%%%%%%%%%%%%%%%%%%%%%%%%%%%%%%%%%%%%%%%%%%%%%%%%%%%%%%

\markright{Appendix C: The total mass of a stellar grouping}
\addcontentsline{toc}{section}{Appendix C: The total mass of a stellar grouping}
\section*{Appendix C: The total mass of a stellar grouping} \label{section: totalmass}

\begin{table}[!tbp]
  \small
  \begin{tabular}{l ccc rccc}
    \hline
    \hline
    Pairing & $\alpha$ & $F_{\rm M}$ & $\gamma_q$ & $M_{\rm tot}$ & $\langle M \rangle$ & $\langle M_1 \rangle$ & $\langle M_2 \rangle$  \\
    \hline
    none   & $+2.5$  &   0\% & $-$    &  9\,460  & 0.38  & 0.38  & $-$ \\
    none   & $-0.9$  &   0\% & $-$    & 11\,905  & 0.48  & 0.48  & $-$ \\
    \hline
    \hline
    RP     & $+2.5$  &  50\% & $-$    & 13\,946  & 0.37  & 0.49  & 0.14 \\
    RP     & $-0.9$  &  50\% & $-$    & 17\,628  & 0.47  & 0.60  & 0.22 \\
    \hline
    PCRP   & $+2.5$  &  50\% & $-$    & 10\,484  & 0.28  & 0.37  & 0.10 \\
    PCRP   & $-0.9$  &  50\% & $-$    & 13\,785  & 0.37  & 0.47  & 0.17 \\
    \hline
    PCP-I  & $+2.5$  &  50\% & $-0.5$ & 11\,018  & 0.29  & 0.38  & 0.13 \\
    PCP-I  & $-0.9$  &  50\% & $-0.5$ & 13\,941  & 0.37  & 0.48  & 0.16 \\
    PCP-I  & $+2.5$  &  50\% & $+0.5$ & 12\,061  & 0.32  & 0.37  & 0.22 \\
    PCP-I  & $-0.9$  &  50\% & $+0.5$ & 15\,497  & 0.41  & 0.48  & 0.28 \\
    \hline
    PCP-II & $+2.5$  &  50\%$^\ast$ & $-0.5$ & 11\,018  & 0.34  & 0.38  & 0.21 \\
    PCP-II & $-0.9$  &  50\%$^\ast$ & $-0.5$ & 13\,937  & 0.41  & 0.48  & 0.23 \\
    PCP-II & $+2.5$  &  50\%$^\ast$ & $+0.5$ & 12\,218  & 0.34  & 0.38  & 0.26 \\
    PCP-II & $-0.9$  &  50\%$^\ast$ & $+0.5$ & 15\,427  & 0.42  & 0.48  & 0.30 \\
    \hline
    PCP-III& $+2.5$  &  50\% & $-0.5$ & 11\,167  & 0.30  & 0.37  & 0.15 \\
    PCP-III& $-0.9$  &  50\% & $-0.5$ & 14\,293  & 0.38  & 0.47  & 0.19 \\
    PCP-III& $+2.5$  &  50\% & $+0.5$ & 12\,060  & 0.32  & 0.37  & 0.23 \\
    PCP-III& $-0.9$  &  50\% & $+0.5$ & 15\,472  & 0.41  & 0.47  & 0.29 \\
    \hline
    \hline
    RP     & $+2.5$  & 100\% & $-$    & 18\,849  & 0.38  & 0.61  & 0.14 \\
    RP     & $-0.9$  & 100\% & $-$    & 24\,176  & 0.48  & 0.75  & 0.22 \\
    \hline
    PCRP   & $+2.5$  & 100\% & $-$    & 11\,687  & 0.23  & 0.37  & 0.10 \\
    PCRP   & $-0.9$  & 100\% & $-$    & 16\,113  & 0.32  & 0.48  & 0.17 \\
    \hline
    PCP-I  & $+2.5$  & 100\% & $-0.5$ & 12\,657  & 0.25  & 0.38  & 0.13 \\
    PCP-I  & $-0.9$  & 100\% & $-0.5$ & 15\,902  & 0.32  & 0.48  & 0.16 \\
    PCP-I  & $+2.5$  & 100\% & $+0.5$ & 14\,901  & 0.30  & 0.37  & 0.22 \\
    PCP-I  & $-0.9$  & 100\% & $+0.5$ & 19\,012  & 0.38  & 0.48  & 0.29 \\
    \hline
    PCP-II & $+2.5$  & 100\%$^\ast$ & $-0.5$ & 12\,405  & 0.31  & 0.37  & 0.21 \\
    PCP-II & $-0.9$  & 100\%$^\ast$ & $-0.5$ & 15\,878  & 0.37  & 0.48  & 0.23 \\
    PCP-II & $+2.5$  & 100\%$^\ast$ & $+0.5$ & 14\,701  & 0.32  & 0.37  & 0.26 \\
    PCP-II & $-0.9$  & 100\%$^\ast$ & $+0.5$ & 18\,472  & 0.38  & 0.46  & 0.29 \\
    \hline
    PCP-III& $+2.5$  & 100\% & $-0.5$ & 13\,315  & 0.27  & 0.37  & 0.16 \\
    PCP-III& $-0.9$  & 100\% & $-0.5$ & 16\,765  & 0.34  & 0.48  & 0.19 \\
    PCP-III& $+2.5$  & 100\% & $+0.5$ & 14\,920  & 0.30  & 0.37  & 0.23 \\
    PCP-III& $-0.9$  & 100\% & $+0.5$ & 19\,114  & 0.38  & 0.48  & 0.29 \\
    \hline
    \hline
  \end{tabular}
  \caption{The total mass, the average stellar mass, the average single/primary mass, and the average companion mass, as a function of mass distribution and pairing function (in units of M$_\odot$). Each model consists of 25\,000 systems. The pairing function is listed in the first column. The mass distribution in each model is the extended Preibisch mass distribution with a slope $\alpha$ (second column) in the brown dwarf regime. The binary fraction $F_{\rm M}$ of each model is listed in column~3. The mass ratio distribution for pairing functions PCP-I, PCP-II, and PCP-III has the form $f_q(q) \propto q^{\gamma_q}$, with $\gamma_q$ listed in the fourth column. For the pairing function PCP-II, the binary fraction (marked with an asterisk) is the {\em generating} binary fraction. The true binary fraction for PCP-II is lower, as the very low-mass companions are rejected. \label{table: totalmass_differences} }
\end{table}

An estimate for the total mass of an OB~association or star cluster is often derived by integrating the single star mass distribution over the mass range for which the stellar grouping is known to have members:
\begin{equation}\label{equation: totalmass_singles}
M_{\rm tot,singles} = N \int^{M_{\rm max}}_{M_{\rm min}} M \, f_M(M) \, dM,
\end{equation}
where $M_{\rm min}$ and $M_{\rm max}$ are the minimum and maximum mass of the stellar population, and $N$ is the total number of systems ($N=S+B+\dots$). The total mass obtained in this way underestimates the true total mass of the stellar grouping, as the mass of the companion stars is not included. Below, we list how the true total (stellar) mass of the stellar grouping can be obtained, by taking into account the companions, for the five pairing functions used in this paper. In these calculations we assume that the stellar grouping consists of single stars and binary systems only.

The total mass of a stellar grouping consisting of $N$ stars, with a binary fraction $F_{\rm M}$ for different pairing functions is:
\begin{eqnarray}
M_{\rm tot,RP}      &=& N \int^{M_{\rm max}}_{M_{\rm min}} \left( 1+F_{\rm M} \right) M \, f_M(M) \, dM \label{equation: totalmass_rp} \\
M_{\rm tot,PCRP}    &=& N \int^{M_{\rm max}}_{M_{\rm min}} \left( 1+F_{\rm M} \left( \frac{\int^{M_1}_{M_{\rm min}} M_2 \,  f_M(M_2) \, dM_2}{M_1} \right) \right) M_1 \, f_M(M_1) \, dM_1 \label{equation: totalmass_pcrp} \\
M_{\rm tot,PCP-I}   &=& N  \left( 1+ F_{\rm M} \int^{q^{\rm max}}_{q_{\rm min}} q\, f_q(q)\, dq \right)\int^{M_{\rm max}}_{M_{\rm min}} M\,f_M(M) \, dM \label{equation: totalmass_pcp1} \\
M_{\rm tot,PCP-II}  &=& N  \int^{M_{\rm max}}_{M_{\rm min}} \left( 1+ F_{\rm M} \int^{q_{\rm max}}_{q_{\rm min}(M)} q\, f_q(q)\, dq \right) \, M\,f_M(M) \, dM \label{equation: totalmass_pcp2}  \\
M_{\rm tot,PCP-III} &=& N  \int^{M_{\rm max}}_{M_{\rm min}} \left( 1+ F_{\rm M} \int^{q_{\rm max}}_{q_{\rm min}(M)} q\, f_q(q;M)\, dq \right) \, M\,f_M(M) \, dM \label{equation: totalmass_pcp3} 
\end{eqnarray}
For the pairing function PCP-II, $f_q(q)$ is not normalized to unity, but for PCP-I and PCP-III it is.  For PCP-II and PCP-III, $q_{\rm min}(M)=M_{\rm min}/M$. 

The total mass depends on the behaviour of the mass distribution, the pairing function, the binary fraction, and (in the case of PCP-I, PCP-II, and PCP-III) on the mass ratio distribution. The differences in total mass also have consequences for the average stellar mass and the average system mass. In Table~\ref{table: totalmass_differences} we list these differences, for several stellar groupings consisting of $N=25\,000$ systems. As a companion star by definition has a lower mass than its primary, the total mass as calculated from the single-star mass distribution cannot underestimate the true total mass by more than a factor two.

The total mass of an association or star cluster defines its gravitational potential, and is therefore important for its evolution in a tidal field, and for the interaction with other stellar groupings. Even though the stellar ($M > 0.08$~M$_\odot$) mass distribution is known, uncertainty in the total mass can be significant, due to unknown binarity and the unknown slope of the mass distribution in the brown dwarf regime. Table~\ref{table: totalmass_differences} shows for example that the total mass of a grouping of 25\,000 single stars with $\alpha=+2.5$ is $0.95\times 10^4$~M$_\odot$. For a grouping of  25\,000 randomly paired binary systems, with $\alpha=-0.9$, the total mass is $2.42\times 10^4$~M$_\odot$, i.e., a factor 2.6~lower higher.

% ---------------------------------------------------------------------------------------
% ---------------------------------------------------------------------------------------
% ---------------------------------------------------------------------------------------

\markright{Appendix D: Orbital size and shape distributions}
\addcontentsline{toc}{section}{Appendix D: Orbital size and shape distributions}
\section*{Appendix D: Orbital size and shape distributions} \label{section: otherchoices}

In this section we discuss how the choice of an orbital parameter distribution (e.g., the semi-major axis distribution or period distribution) affects the other parameter distributions. Below we illustrate that for several parameters ``log-normal leads to log-normal'' and ``log-flat leads to log-flat''.

\"{O}pik's law says that the semi-major axis distribution for binary systems has the form $f_a(a)\propto a^{-1}$, which is equivalent to $f_{\log a}(\log a)=$~constant. In the more general case, the semi-major axis distribution can be expressed as $f_a(a) \propto a^\gamma$. The period of each binary system can be derived from the semi-major axis if the total mass $M_T$ of the system is known. For a sample of binaries with system mass $M_T$, the corresponding {\em specific} period distribution is given by
\begin{equation}
   f_{P;M_T}(P;M_T) \propto M_T^{(\gamma+1)/3} \, P^{(2\gamma-1)/3}.
\end{equation}
A power-law $f_a(a)$ thus leads to a power-law {\em specific} period distribution $f_{P;M_T}(P;M_T)$ for each $M_T$. The overall period distribution is obtained by  integrating over $M_T$. 
\begin{equation}
   f_P(P) = \int f_{M_T}(M_T) \, f_{P;M_T}(P;M_T) \, dM_T \quad \quad P_{\rm min} \leq P \leq P_{\rm max},
\end{equation}
where $f_{M_T}(M_T)$ is the system mass distribution. The resulting $f_P(P)$ is not a power-law anymore, as each $f_{P;M_T}(P;M_T)$ is given a different weight $f_P(M_T)$. The reason for this is that the semi-major axis ranges approximately between $15~{\rm R}_\odot$ and $10^6~{\rm R}_\odot$, i.e., 4.8~orders of magnitude. The mass of substellar and stellar objects ranges approximately between $0.02$~M$_\odot$ and $20$~M$_\odot$, which results in a variation in the total mass of 3~orders of magnitude. However, as most binaries are of low mass, the average mass is $\sim 0.6$~M$_\odot$, so that effectively the total mass varies $\sim 1.5$~orders of magnitude. For any reasonable mass distribution, where the variation in $M_T$ (1.5~dex) can be neglected with respect to that in $P$ (4.8~dex), the distribution $f_P(P)$ can be approximated by $f_P(P) \propto P^{(2\gamma-1)/3}$, in the period range $P_{\rm min} \ll P \ll P_{\rm max}$. We illustrate these results with a simulation of the default model, with $20\,000$ binaries. For \"{O}pik's law, the best-fitting exponent $\gamma_P$ of for the resulting period distribution is $\gamma_P = -1.00 \pm 0.002$ ($1\sigma$ errors), where the error is partially caused by the finite number of binaries used in the fit. Simulations with other mass distributions, pairing functions, and mass ratio distributions, show that the result is relatively insensitive to changes in these properties.

In a similar way, the approximate shape of the angular separation distribution $f_\rho(\rho)$ and the distribution over the semi-amplitude of the radial velocity can be derived. Using the approximation for $\rho$ in equation~\ref{equation: projecteda} and the definition of $K_1$ in equation~\ref{equation: keplerslaw_alternative}, it can be shown that for a distribution $f_a(a) \propto a^\gamma$, the following expressions hold approximately: 
\begin{equation}
f_\rho(\rho) \propto \rho^\gamma, \quad \quad f_{K_1}(K_1) = K_1^{-2\gamma-3}
\end{equation}
In the case of \"{O}pik's law ($\gamma=-1$), this therefore results in $f_{\log \rho}(\log \rho) \approx$ constant and $f_{\log K_1}(\log K_1) \approx$~constant. The latter approximations are clearly seen as the straight lines in the cumulative distributions functions in Figure~\ref{figure: opik_dm_difference}.

The energy $E$ and angular momentum $L$ of a binary system are given by
\begin{equation}
E = \frac{qGM_1^2}{2a}, \quad \quad L = q\sqrt{\frac{Ga(1-e^2)M_1^3}{1+q}}.
\end{equation}
The energy and angular momentum depend strongly on $a$, as the variation in $a$ is much larger than in the other parameters ($M_1$, $e$, and $q$). In the case that the semi-major axis distribution is given by  $f_a(a) \propto a^\gamma$, for the same reasons as above, the distributions $f_E(E)$ and $f_L(L)$ may be approximated by
\begin{equation}
f_L(L) \propto L^{2\gamma+1}, \quad \quad f_E(E) \propto E^{-\gamma-2}.
\end{equation}
In the case of \"{O}pik's law ($\gamma=-1$), the distributions $f_{\log L}(\log L)$ and $f_{\log E}(\log E)$ are constant. The result for $\gamma=-1$ was previously reported by \cite{heacox1997}. In numerical simulations the energy is often expressed in units of $kT$, for which the equation above also holds approximately. Note, however, that for extremely evolved or extremely dense stellar systems, the hypothesis $f_a(a) \propto a^{-1}$ may not hold, due to the effect of dynamical interactions.

The top panels in Figure~~\ref{figure: fa_fp_difference} show $f_a(a)$ and $f_P(P)$ for a model with \"{O}pik's law and for a model with the \cite{duquennoy1991} log-normal period distribution. The figure clearly shows that a flat distribution in $f_a(a)$ leads to a flat distribution in $f_P(P)$, as shown above. It also shows that a log-normal distribution in $f_P(P)$ leads approximately to a log-normal distribution in $f_a(a)$. Below we show that the  \cite{duquennoy1991} log-normal period distribution indeed leads to an approximately log-normal $f_a(a)$.

Suppose that the binary population in a stellar grouping satisfies the \cite{duquennoy1991} log-normal period distribution. First consider the subset of binary systems with a total mass $M_T$. For this subset of binaries,  the {\em specific} semi-major axis distribution resulting from equation~\ref{equation: duquennoyperiods} is also log-normal:
\begin{equation} \label{equation: duquennoysma}
f_{\log a; M_T}(\log a; M_T) \propto \exp \left\{ - \frac{(\log a - \overline{\log a_{M_T}})^2 }{ 2 \sigma^2_{\log a_{M_T}} }  \right\},
\end{equation}
where $a$ is in astronomical units and
\begin{equation}
  \begin{tabular}{lll}
    $\overline{\log a_{M_T}}$ &$= \tfrac{2}{3} \overline{\log P} - \tfrac{1}{3} \log\left(\frac{4\pi^2}{GM_T}\right) $&$= 1.49 + \tfrac{1}{3} \log \left( \frac{M_T}{{\rm M}_\odot} \right)$ \\
    $\sigma_{\log a_{M_T}}   $&$= \tfrac{2}{3}  \sigma_{\log P} $&$= 1.53$ \\
  \end{tabular} \, .
\end{equation}
The value of $\overline{\log a_{M_T}}$ increases as the total system mass $M_T$ increases. For the lowest-mass systems ($M_T = 2\times 0.02$~M$_\odot$), $\overline{\log a_{M_T}} \approx 1.02$ (10.5~AU), while for the systems consisting of two G-stars ($M_T = 2\times 1$~M$_\odot$), $\overline{\log a_{M_T}}$ equals $1.59$ (39~AU). 

The complete stellar grouping contains binaries with a wide range in $M_T$. The overall semi-major axis distribution resulting from equation~\ref{equation: duquennoyperiods} is obtained by integrating over, and weighing by, the total system mass distribution
\begin{equation} \label{equation: duquennoysma2}
f_{\log a}(\log a) = \int f_{M_T}(M_T) \,f_{\log a; M_T}(\log a; M_T) \, dM_T \,,
\end{equation}
where $f_{M_T}(M_T)$ is the system mass distribution. This distribution approximates a Gaussian distribution. The resulting $\sigma_{\log a}$ is generally larger than $\sigma_{\log a_{M_T}}$. The fitted values for the default model (Figure~\ref{figure: fa_fp_difference}) are $\overline{\log a} = 1.36 \pm 0.06$  and  $\sigma_{\log a} = 1.48 \pm 0.05$. These results show that that a log-normal period distribution leads to an approximately log-normal semi-major axis distribution.

\chapter[Recovering the binary population in Sco~OB2]{Recovering the young binary population in the nearby OB association Scorpius~OB2} \label{chapter: true}

\begin{center}
M.B.N. Kouwenhoven, A.G.A. Brown, S.F. Portegies Zwart, \& L. Kaper

\vspace{0.2cm}
{\it Astronomy \& Astrophysics}, to be submitted
\end{center}

% ====================================================================
% ====================================================================
% ====================================================================
% ==ABSTRACT==========================================================
% ====================================================================
% ====================================================================
% ====================================================================

\section*{Abstract}

We characterize the binary population in the Scorpius~OB2 (Sco~OB2) association using all available observations of visual, spectroscopic, and astrometric binaries. We take into account observational biases by comparison with simulated observations of model populations. The available data indicate a large binary fraction ($> 70\%$ with $3\sigma$ confidence), with a large probability that all stars in Sco~OB2 are in binaries. In our analysis we do not consider systems of higher multiplicity.
The binary systems have a mass ratio distribution of the from $f_q(q) \propto q^{\gamma_q}$, with $\gamma_q = -0.4\pm 0.2$. 
Sco~OB2 has a semi-major axis distribution of the from $f_a(a) \propto a^{\gamma_a}$ with $\gamma_a = -1.0 \pm 0.15$ (\"{O}pik's law), in the range $5$~R$_\odot \la a \la 5\times 10^6$~R$_\odot$. The log-normal period distribution of \cite{duquennoy1991} results in too few spectroscopic binaries, even if the model binary fraction is 100\%, and can thus be excluded.
Sco~OB2 is a young association with a low stellar density; its current population is expected to be very similar to the primordial population. The fact that practically all stars in Sco~OB2 are part of a binary (or multiple) system demonstrates that multiplicity is a fundamental factor in the star formation process.

% ====================================================================
% ====================================================================
% ====================================================================
% ==INTRODUCTION======================================================
% ====================================================================
% ====================================================================
% ====================================================================

\section{Introduction}

Over the past decades observations have indicated that a large fraction of the stars are part of a binary or multiple system. Apparently, multiplicity is an important aspect of the star formation process. Binaries also play a vital role in explaining many spectacular phenomena in astrophysics (supernovae type Ia, (short) gamma-ray bursts, X-ray binaries, millisecond pulsars, etc.) and strongly affect the dynamical evolution of dense stellar clusters.
Therefore it is of crucial importance to characterize the outcome of the star forming process in terms of multiplicity and binary parameters.

In this chapter we aim to recover the population of binaries that results from the formation process: the primordial binary population, which is defined as {\em the population of binaries as established just after the gas has been removed from the forming system, i.e., when the stars can no longer accrete gas from their surroundings} \citep{kouwenhoven2005}. 
The dynamical evolution of stars of a newly born stellar population is influenced by the presence of gas. After the gas has been removed, the binary population is only affected by stellar evolution and pure N-body dynamics. From a numerical point of view, the primordial binary population can be considered as a boundary between hydrodynamical simulations and N-body simulations. Hydrodynamical simulations of a contracting gas cloud \citep[e.g.,][]{bate2003} produce stars. After the gas is removed by accretion and the stellar winds of the most massive O~stars, pure N-body simulations \citep[e.g.,][]{ecology4} can be used to study the subsequent evolution of star clusters and the binary population.

OB~associations are well suited for studying the primordial binary population. They are young ($5-50$~Myr) and low-density ($< 0.1~{\rm M}_\odot\,{\rm pc}^{-3}$) aggregates of stars. Their youth implies that only a handful of the most massive systems have changed due to stellar evolution. The effects of dynamical evolution are expected to be limited due to the young age and low stellar density of the association. Moreover, in contrast to, e.g., the T~associations, OB associations cover the full range of stellar masses.

The association under study in this chapter is Scorpius~Centaurus (Sco~OB2), the nearest young OB~association, and thus the prime candidate for studying the binary population. The proximity of Sco~OB2 ($118-145$~pc) facilitates observations, and the young age ($5-20$~Myr) ensures that dynamical evolution has not significantly altered the binary population since the moment of gas removal. 
The membership and stellar content of the association was established by \cite{dezeeuw1999} using {\em Hipparcos} parallaxes and proper motions, and its binary population is relatively well-studied.

Due to selection effects, it is not possible to observe the binary population in Sco~OB2 directly. The observed binary population is biased, making it difficult to draw conclusions about the binary population. However, by using the method of simulating observations of modeled stellar populations (see Chapter~4), it is possible to put constraints on the binary population. We accurately model the selection effects of the six major binarity surveys in Sco~OB2, and compare simulated observations with the true observations, to determine the properties of the current binary population in Sco~OB2. The organization of this chapter is as follows.
In \S~\ref{section: true_method} we briefly describe the method and terminology that we use to recover the true binary population.
In \S~\ref{section: scoob2properties} we discuss the Sco~OB2 association and membership issues.
In \S~\ref{section: true_observations} we describe the available datasets with visual, spectroscopic, and astrometric binaries in Sco~OB2, and outline our models for the respective selection effects.
\S~\ref{section: recovering} is the main part of this chapter, where we recover the binary population in Sco~OB2 from observations.
In \S~\ref{section: true_discussion} we compare our results with those of others, discuss the brown dwarf desert in Sco~OB2, and evaluate the possible differences between the current binary population and primordial binary population in Sco~OB2.
Finally, we summarize our main results in \S~\ref{section: summary}.

% ====================================================================
% ====================================================================
% ====================================================================
% ==INTRODUCTION======================================================
% ====================================================================
% ====================================================================
% ====================================================================

\section{Method and terminology} \label{section: true_method}

We recover the binary population in Sco~OB2 from observations using the method of simulating observations of modeled stellar populations. This method is extensively described in Chapter~4, and can be briefly summarized as follows. 

With the increasing computer power, it has become possible to create sophisticated models of star clusters and OB~associations. One would like to compare these simulated associations with real associations in order to constrain the properties of the latter. However, this is not possible, as the observational dataset is hampered by selection effects. Only a small (and biased) subset of the binary population is known. In the method of simulating observations of simulations \citep[SOS][]{modest1,modest2} one characterizes the selection effects, and applies these to the simulated association. The simulated observations that are then obtained can be compared directly with the real observations.

In order to recover the binary population in OB~associations, we simulate OB~association models with different properties. We compare each model with the observational data, by simulating observations for each major binarity survey in Sco~OB2. With this comparison we identify which association model is consistent with the observations, and thus constrain the binary population in Sco~OB2.

In Chapter~4 we have shown that this is a very safe method to derive the binary population. As long as the parameter space (of the binary population) is fully searched, and as long as the selection effects are well-modeled, this method provides the intrinsic binary population, as well as the uncertainties on each derived property. We also show in Chapter~4 that the traditional method of correcting for selection effects (using a ``correction factor'') may lead to erroneous or unphysical results.

In this paper we make several assumptions when recovering the binary population in Sco~OB2. In our model we consider only single stars and binary systems; no higher order multiples are assumed to be present. We assume the distributions of the different observed parameters to be independent of each other:
\begin{equation} \label{equation: true_independence}
  \begin{tabular}{rr}
    $f_{\rm BP}(M_1,M_2,a,e,i,\omega,\Omega,\mathcal{M})  = $ &
    $f_{M_1,M_2}(M_1,M_2)\,f_a(a)\,f_e(e)\,f_i(i)\,f_\omega(\omega)\,f_\Omega(\Omega)\,f_\mathcal{M}(\mathcal{M})$ \, \\
    or \quad &
    $f_{M_1,M_2}(M_1,M_2)\,f_P(P)\,f_e(e)\,f_i(i)\,f_\omega(\omega)\,f_\Omega(\Omega)\,f_\mathcal{M}(\mathcal{M})$ \,, \\
    \end{tabular}
\end{equation}
where $M_1$ and $M_2$ are the primary and companion mass, $a$ the semi-major axis, $P$ the orbital period, $e$ the eccentricity, $i$ the inclination, $\omega$ the argument of periastron, $\Omega$ the position angle of the line of nodes, and $\mathcal{M}$ the mean anomaly at an instant of time. For a motivation of the assumption of independent distributions in Equation~\ref{equation: true_independence} we refer to Section~4.2.1. Finally, we assume that the binary systems have a random orientation in space. Even in the unlikely case that binary systems do not have a random orientation, the results do not change measurably (see Section~4.6.3).

%
% ============================================================================
% ============================================================================
% ============================================================================
% ============================================================================

\section{The Sco~OB2 association } \label{section: scoob2properties}

% ============================================================================
% ============================================================================
% ============================================================================
% ============================================================================

\begin{table}
  \small
  \begin{tabular}{p{0.55cm}c cc cc cp{0.3cm}cc p{0.3cm}p{0.3cm}p{0.3cm}}
    \hline \hline
    Group & $D$ & $R$ & Age & $A_V$ & $N_\star$ & $S_\star$ & $B_\star$ & $T_\star$ & $>3$ & $F_{\rm M,\star}$ & $F_{\rm NS,\star}$ & $F_{\rm C,\star}$ \\
          & pc  & pc  & Myr & mag \\
    \hline 
    US  &  145$^1$ & $\sim 20^5$  & 5$-$6$^{2,3}$ & 0.47$^5$  & 120$^1$ & 64  & 44 & 8  & 3 & 0.46 & 0.67 & 0.61 \\
    UCL &  140$^1$ & $\sim 35^5$  & 15$-$22$^4$   & 0.06$^5$  & 221$^1$ & 132 & 65 & 19 & 4 & 0.40 & 0.61 & 0.52\\
    LCC &  118$^1$ & $\sim 35^5$  & 17$-$23$^4$   & 0.05$^5$  & 180$^1$ & 112 & 57 & 9  & 1 & 0.37 & 0.56 & 0.44 \\
    \hline 
    Model& 130     & 20           & 5  & 0.00 & 12\,000 & \multicolumn{2}{c}{varying}       & 0  & 0 & \multicolumn{3}{c}{varying} \\
    \hline 
    \hline 
  \end{tabular}
  \caption{Properties of the subgroups Upper Scorpius (US), Upper Centaurus Lupus (UCL), and Lower Centaurus Crux (LCC) of Sco OB2, and of our model for Sco~OB2. Columns 2--4 list for each subgroup its distance, effective radius, and age. Column~5 lists the median interstellar extinction towards each subgroup. Column~6 lists the number of confirmed {\em Hipparcos} members of each subgroup, and is followed by the {\em observed} number of singles, binaries, triples, and higher-order systems among the confirmed members in columns 7--10. Finally, columns 11--13 list the observed multiplicity fraction, non-single star fraction, and companion star fraction among the confirmed members \citep[see][for a definition of these fractions]{kouwenhoven2005}. Note that the latter quantities are lower limits due to the presence of unresolved binary and multiple systems. 
In the bottom row we list the properties of our Sco~OB2 model. The number of systems (i.e., singles and binaries) used in our model includes substellar objects with masses down to the 0.02~M$_\odot$. References: $^1$ \cite{dezeeuw1999}, $^2$ \cite{degeus1989}, $^3$ \cite{preibisch2002}, $^4$ \cite{mamajek2002}, $^5$ \cite{debruijne1999}. 
 \label{table: subgroups}}
\end{table}

Sco~OB2 is a well-studied OB~association. It consists of three subgroups \citep[e.g.,][]{blaauw1964A,dezeeuw1999}. These three subgroups are likely the result of triggered star formation \citep[e.g.,][]{blaauw1991,preibisch1999}, and in turn may have triggered star formation in the $\rho$~Ophiuchus region.  Table~\ref{table: subgroups} lists several properties of the three subgroups of Sco~OB2. 

\cite{preibisch2002} studied the single star population of the US subgroup of Sco~OB2. They combine their observations of PMS-stars with those of \cite{preibisch1999} and \cite{dezeeuw1999}. They derive an empirical mass distribution in the mass range 0.1~M$_\odot \leq M \leq 20~$M$_\odot$. In this paper we adopt an extended version of the Preibisch MF \citep[see, e.g.,][]{kouwenhoven2005}, which goes down to masses of 0.02~M$_\odot$:
\begin{equation} \label{equation: preibischimf}
  f_M(M) \propto \left\{
  \begin{array}{llll}
    M^\alpha  & {\rm for \quad } 0.02 & \leq M_P/{\rm M}_\odot & < 0.08 \\
    M^{-0.9}  & {\rm for \quad } 0.08 & \leq M_P/{\rm M}_\odot & < 0.6 \\
    M^{-2.8}  & {\rm for \quad } 0.6  & \leq M_P/{\rm M}_\odot & < 2   \\
    M^{-2.6}  & {\rm for \quad } 2    & \leq M_P/{\rm M}_\odot & < 20 \\
  \end{array}
  \right. \,.
\end{equation}
The slope $\alpha$ of the power law for substellar masses is poorly constrained. \cite{preibisch2003} give an overview of the slope $\alpha$ in different populations, where $\alpha$ ranges from $-0.3$ for the Galactic field \citep{kroupa2002} to $+2.5$ for the young stellar cluster IC~348 \citep{preibisch2003}.

\subsection{The model for Sco~OB2}

In our model for Sco~OB2 we adopt a distance of 130~pc (the median distance of the confirmed members of Sco~OB2) and an age of 5~Myr. Although the subgroups UCL and LCC are older than 5~Myr, the systematic error introduced by our choice of the age is small. In our models we slightly overestimate the luminosity of stars in the UCL and LCC subgroups, but this affects only the stars close to the detection limit (see below), and does not affect the properties of our simulated observations significantly. The error in the age neither affects the interpretation of the observed mass ratio distribution, as each observed mass and mass ratio is derived from the brightness of the star, assuming the correct age for the subgroup, and the {\em Hipparcos} parallax to each star individually. 

We adopt the extended Preibisch mass distribution in our model for Sco~OB2. We make the assumptions that (1) the mass distribution for the subgroups UCL and LCC is identical to that of US, (2) the mass distribution in the brown dwarf regime can be approximated with a power-law with slope $\alpha=-0.9$ (i.e., we extend the stellar mass distribution into the brown dwarf regime), and (3) the lower limit to the mass distribution is 0.02~M$_\odot$. 

\cite{preibisch2002} estimate that the US subgroup contains approximately 2525 single/primary stars in the mass range $0.1-20$~M$_\odot$.
 With the extension to lower mass in Equation~\ref{equation: preibischimf} the number of singles/primaries is higher, as we also include the very low mass stars and brown dwarfs. For a slope $\alpha=-0.9$, and assuming that the UCL and LCC subgroups have an equal number of singles/primaries, the total number of singles/primaries in Sco~OB2 is approximately $12\,000$. We will therefore adopt $N=S+B=12\,000$ in our simulations, where $S$ is the number of single stars, and $B$ the number of binary systems.  In this chapter we compare simulated observations with real observations of the three subgroups of Sco~OB2 combined. We vary the binary fraction $F_{\rm M} = B/(S+B)$ and the other binary parameter distributions of the model, in order to find the combination of binary parameters that fits the observations best.

\subsection{Photometry}

We obtain the magnitude of each simulated star in the optical and near-infrared bands using the isochrones described in \cite{kouwenhoven2005}. These isochrones consist of models from \cite{chabrier2000} for $0.02~\mbox{M}_\odot \leq M < 1~\mbox{M}_\odot$, \cite{palla1999} for $1~\mbox{M}_\odot \leq M < 2~\mbox{M}_\odot$, and  \cite{girardi2002} for $M > 2~\mbox{M}_\odot$.  We adopt the isochrone of 5~Myr and solar metallicity. The {\em Hipparcos} magnitude $H_p$ for each star is derived from its $V$ magnitude and $V-I$ color, using the tabulated values listed in the {\em Hipparcos} Catalog (ESA 1997, Vol.~1, \S~14.2). For each star we convert the absolute magnitude into the apparent magnitude using the distance to each star individually. We do not include interstellar extinction in our models. Sco~OB2 is practically cleared of gas, so that the interstellar extinction is negligible for our purposes.

% ====================================================================
% ====================================================================
% ====================================================================

\subsection{Sco~OB2 membership} \label{section: membership}

\begin{figure}[!btp]
  \centering
  \includegraphics[width=0.5\textwidth,height=!]{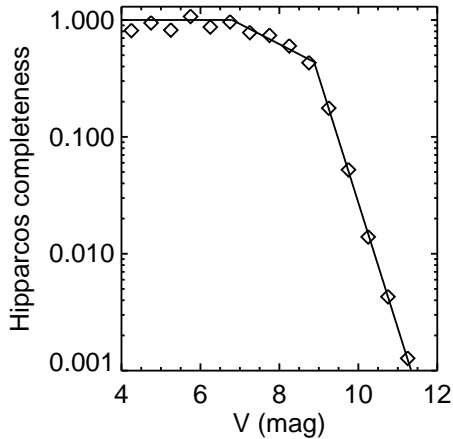}
  \caption{The completeness of the {\em Hipparcos} database in the Sco~OB2 region, as a function of $V$ magnitude. The diamonds represent the ratio between the number of stars in the {\em Hipparcos} catalogue and the number of stars in the {\em TYCHO-2} catalogue, in each $V$ magnitude bin. The comparison above is made for the Sco~OB2 region, and is similar for each of the three subgroups of Sco~OB2. The solid line represents the model for the completeness adopted in this paper (Equation~\ref{equation: hipparcos_fraction}).  
\label{figure: hipparcos_fraction} }
\end{figure}

De~Zeeuw et al. (1999) have published a census of the stellar content and membership of nearby ($\la 1$~kpc) OB~associations. They present a list of 521~members of the Sco~OB2 association, based on the {\em Hipparcos} position, proper motion, and parallax of each star. Of these members, 120~are in the US subgroup, 221~in UCL, and 180~in LCC. Due to the {\em Hipparcos} completeness limit, most of the confirmed members are bright  ($V \la 8$~mag) and mostly of spectral type B, A, and~F. 

In the analysis of the observational data (\S~\ref{section: true_observations}) we consider only the confirmed members of Sco~OB2 \citep[i.e., those identified by][]{dezeeuw1999}, all of which are in the {\em Hipparcos} catalog. Among the stars observed by {\em Hipparcos} it is unlikely that a Sco~OB2 member star is not identified as such. On the other hand, it is possible that non-members are falsely classified as members of Sco~OB2; the so-called interlopers. The fraction of interlopers among the ``confirmed'' Sco~OB2 members stars is estimated to be $\sim 6\%$ for B~stars, $\sim 13\%$ for A~stars, and $\sim 22\%$ for F~and G~stars \citep[see Tables~A2 and~C1 in][]{dezeeuw1999}. The interlopers among B and A~stars are likely Gould Belt stars, which have a distance and age comparable to that of the nearby OB~associations. In our analysis we assume that all confirmed members in the list of \cite{dezeeuw1999} are truly member stars, and do not attempt to correct for the presence of interlopers.

The {\em Hipparcos} completeness limit is studied in detail by \cite{soderhjelm2000}. His prescription for the completeness is based on all entries in the {\em Hipparcos} catalog. However, many OB~associations were studied in more detail by {\em Hipparcos}, based on candidate membership lists. Due to the {\em Hipparcos} crowding limit of 3~stars per square degree, only a selected subset of the candidate members of Sco~OB2 was observed \citep[see][for details]{dezeeuw1999}, which significantly complicates the modeling of the {\em Hipparcos} completeness.
We therefore calibrate the completeness of {\em Hipparcos} in the Sco~OB2 region by comparing the number of {\em Hipparcos} entries with the number of stars of a given magnitude in the same region. We use the {\em TYCHO-2} catalog for this comparison. The {\em TYCHO-2} catalog is complete to much fainter stars than {\em Hipparcos}. In Figure~\ref{figure: hipparcos_fraction} we show the proportion $P$ of stars that is in the {\em Hipparcos} catalog, relative to the number of stars in the {\em TYCHO-2} catalog, as a function of $V$ magnitude. We model the proportion $P$ as a function of $V$ with three line segments:
\begin{equation} \label{equation: hipparcos_fraction}
  \log P(V) = \left\{
  \begin{array}{cll}
    0                & \mbox{for}\quad                   & V \leq 6.80~\mbox{mag} \\
    1.18 - 0.17\, V & \mbox{for}\quad 6.80~\mbox{mag} < & V \leq 8.88~\mbox{mag} \\
    9.18 - 1.07\, V & \mbox{for}\quad 8.88~\mbox{mag} < & V                      \\
  \end{array}
  \right. \,.
\end{equation}
The {\em TYCHO-2} catalog is 99\% complete down to $V=11$~mag and 90\% complete down to $V=11.5$. The completeness of the {\em Hipparcos} catalog for $V\ga 11$~mag is therefore not accurately described by Equation~\ref{equation: hipparcos_fraction}. However, this does not affect our results, as the surveys under study only include the brightest members of Sco~OB2.  In this model we ignore the fact that the {\em Hipparcos} completeness is also depending on color. We additionally assume that the membership completeness does not depend on spectral type.

% ====================================================================
% ====================================================================
% ====================================================================
% ==INTRODUCTION======================================================
% ====================================================================
% ====================================================================
% ====================================================================

\section{Observations of binary systems in Sco~OB2} \label{section: true_observations}

\begin{SCtable}[][!btp] 
  \begin{tabular}{ll}
    \hline 
    Reference & Detection method \\
    \hline
    \cite{alencar2003}     & Spectroscopic \\
    \cite{andersen1993}    & Combination \\
    \cite{balega1994}      & Visual \\
    \cite{barbier1994}     & Spectroscopic\\
    \cite{batten1997}      & Spectroscopic \\
    \cite{buscombe1962}    & Spectroscopic \\
    \cite{chen2006}        & Visual \\
    \cite{couteau1995}     & Combination\\
    The Double Star Library & Combination\\
    \cite{duflot1995}      &Spectroscopic\\
    \cite{hartkopf2001}    &Visual\\
    \cite{jilinski2006}    & Spectroscopic \\
    \cite{jordi1997}       &Eclipsing\\
    The {\em Hipparcos} and {\em TYCHO} Catalogues &Astrometric\\ 
    \cite{kraicheva1989}   &Spectroscopic\\ 
    \cite{kouwenhoven2005} & Visual \\
    \cite{kouwenhoven2006a}& Visual \\
    \cite{lindroos1985}    & Visual \\
    \cite{malkov1993}      &Combination \\
    \cite{mason1995}       &Visual \\
    \cite{mcalister1993}   &Visual\\
    Miscellaneous, e.g. SIMBAD &Combination\\
    \cite{miura1992}       &Visual\\
    \cite{nitschelm2004}   & Spectroscopic \\
    \cite{pedoussaut1996}  &Spectroscopic\\ 
    \cite{shatsky2002}     &Visual\\
    \cite{sowell1993}      &Visual\\
    \cite{svechnikov1984}  &Combination\\
    \cite{tokovininmsc}    &Combination\\
    \cite{wds1997}         &Combination\\
    \hline 
  \end{tabular}
  \caption{References to literature data with spectroscopic, astrometric, eclipsing, and visual binaries in Sco~OB2. The data for a number of binary systems in Sco~OB2 is taken from several catalogues. This table is similar to that presented in \cite{kouwenhoven2005}, but is updated with recent discoveries.
    \label{table: referencelist}}
\end{SCtable}

\begin{figure}[!btp]
  \centering
  \includegraphics[width=0.8\textwidth,height=!]{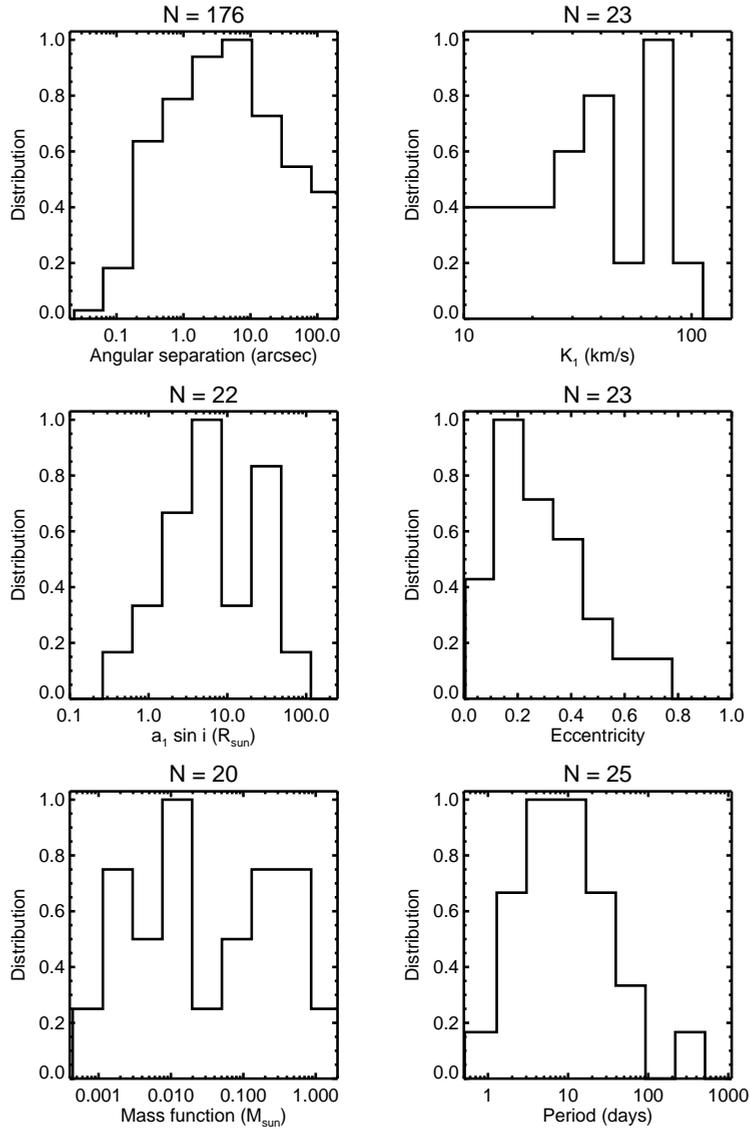}
  \caption{Properties of the {\em observed} binary population in Sco~OB2. Only the 521~confirmed members of Sco~OB2 are considered. The top-left panel shows the angular separation distribution for visual binaries. The other panels show the distributions over radial velocity amplitude $K_1$, the projected semi-major axis $a_1 \sin i$, the eccentricity $e$, the mass function $\mathcal{F}(M)$, and the period $P$, for the spectroscopic binaries and for HIP78918, the only astrometric binary in Sco~OB2 with an orbital solution. The measurements shown in this figure include those of multiple systems. Above each panel we indicate the number of companions for which the corresponding orbital element is available. Spectroscopic and astrometric binaries with a stochastic solution (i.e., stars for which the radial or apparent velocity change indicates binarity, even though the orbital parameters cannot be derived) are not shown here.\label{figure: observed_distributions} }
\end{figure}

A large fraction of the Sco~OB2 member stars is known to be part of a binary or multiple system. In Table~\ref{table: subgroups} we have provided an overview of the {\em observed} binary fraction in the association, for which we included all known binary and multiple systems in Sco~OB2. In total there are 266~known companions among the 521~confirmed members of Sco~OB2, so that the multiplicity fraction in Sco~OB2 is {\em at least} 40\% (assuming that all proposed companions are physical companions). The references for these binary and multiple systems are listed in Table~\ref{table: referencelist}. 
Figure~\ref{figure: observed_distributions} shows several observed parameter distributions. These {\em observed} distributions are not representative of the {\em intrinsic} distributions, as selection effects prohibit the detection of a significant fraction of the companion stars. Furthermore, it is possible that several of the reported companions are spurious; these may be background stars that are projected close to a Sco~OB2 member star.

The binary systems in Sco~OB2 were discovered by different observers, using various techniques and instruments. As each of these observing runs is characterized by specific selection effects, it is difficult to study each of these in detail. We therefore focus primarily on a subset of the observed binaries, i.e., those observed in \cite{kouwenhoven2005,kouwenhoven2006a,shatsky2002,levato1987,brownverschueren}, and those detected by {\em Hipparcos} \citep{esa1997}. We refer to these papers and the corresponding datasets hereafter as KO5, KO6, SHT, LEV, BRV, and HIP, respectively. An overview of the number of observed targets and detected binary systems in each of these models is listed in Table~\ref{table: observations_overview}. 
Combined, these datasets contain a large fraction of the known binary and multiple systems in Sco~OB2. The selection effects for each of these datasets can be modeled, making it possible to use the method of simulated observations. In the following sections we describe these five datasets, and discuss our model for the selection effects affecting their observations. 

\begin{figure}[!btp]
  \centering
  \includegraphics[width=0.5\textwidth,height=!]{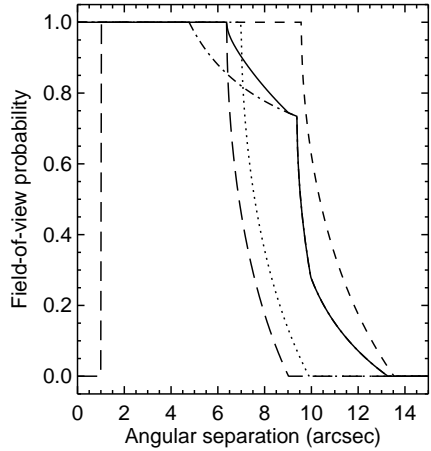}
  \caption{For the three imaging surveys for binarity discussed in this paper, the field of view is non-circular. Whether a secondary is in the field of view, depends therefore not only on its separation $\rho$, but also on its position angle $\varphi$. This figure shows the probability that a secondary is in the field-of-view, as a function of $\rho$, assuming random orientation of the binary systems, for the KO5 observations (short-dashed curve), the KO6 observations (dotted curve), the non-coronographic SHT observations (dash-dotted curve), the coronographic SHT observations (long-dashed curve), and the combined SHT observations (solid curve). Whether a secondary is detected or not, depends additionally on its brightness and on the brightness difference with the primary (see Figure~\ref{figure: probability_detlim_ka}). \label{figure: probability_rho_vs_pa} }
\end{figure}

\begin{figure}[!btp]
  \includegraphics[width=0.95\textwidth,height=!]{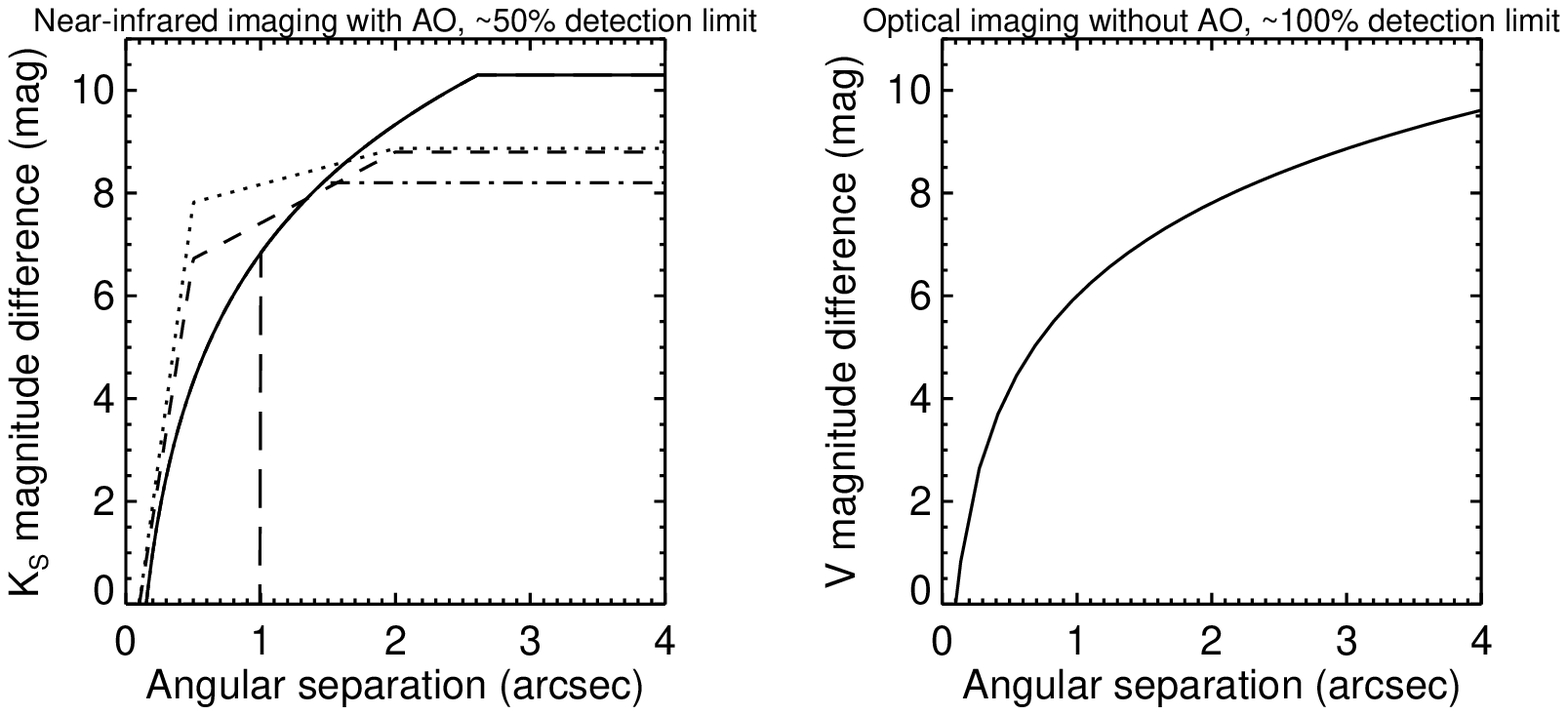}
  \caption{{\em Left}: the 50\% detection limit $\Delta K_S$ as a function of angular separation, for the KO5 observations (short-dashed curve), the KO6 observations (dotted curve), the non-coronographic SHT observations (dash-dotted curve), the coronographic SHT observations (long-dashed curve), and the combined SHT observations (solid curve). For the KO5 and KO6 observations the curves represent those for average Strehl ratios of 30\% and 24\%, respectively. {\em Right}: for comparison we additionally show the maximumum $\Delta V$ as a function of $\rho$ that \cite{hogeveen1990} derived for visual binaries in the 1990 version of the WDS catalog. \label{figure: probability_detlim_ka} }
\end{figure}

A summary of the modeled selection effects is given in Table~\ref{table: sos_biases}. For the three imaging surveys discussed below, we show in Figure~\ref{figure: probability_rho_vs_pa} the probability that a secondary is in the field of view, as a function of its separation from the primary. In Figure~\ref{figure: probability_detlim_ka} we show the detection limit as a function of separation for the three visual binary surveys.

\begin{table}
  \begin{tabular}{p{0.7cm}p{4.1cm}l rr rr}
    \hline
    \hline
    Abbrev.& Reference & Dataset & $N_{\rm orig}$ & $B_{\rm orig}$ & $N_{\rm used}$ & $B_{\rm used}$ \\
    \hline
    KO5    & \cite{kouwenhoven2005}   & Visual         & 199 & 74 & 199 & 60 \\
    KO6    & \cite{kouwenhoven2006a}  & Visual         & 22  & 29 & 22  & 18 \\
    SHT    & \cite{shatsky2002}       & Visual         & 115 & 19 & 80  & 23 \\
    LEV    & \cite{levato1987}        & Spectroscopic  & 81  & 61  & 52  & 38 \\
    BRV    & \cite{brownverschueren}  & Spectroscopic  & 156 & 91  & 71  & 47 \\
    HIP    & \cite{hipparcos}         & Astrometric    & 521 & 125 & 521 & 125 \\
    \hline
    \hline
  \end{tabular}
  \caption{An overview of the datasets used to derive the properties of the binary population in Sco~OB2. Columns~$1-3$ list the dataset acronym, the reference, and the type of binary studied in the dataset. Columns~4 and~5 list the number of targets in the original dataset, and the number of companions found for these targets. Columns~6 and~7 list the number of targets and companions used in our analysis. This dataset is smaller than the original dataset, as we do not include the non-members of Sco~OB2 in our analysis and as we include at most one companion per targeted star (i.e., we do not include multiple systems). We list in this table the total number of spectroscopic binaries, including the radial velocity variables (RVVs; irrespective of their true nature), SB1s, and SB2s. For the {\em Hipparcos} observations we list the number of entries in the categories (X), (O), (G), (C), and (S), among the confirmed members of Sco~OB2. \label{table: observations_overview}}
\end{table}

\begin{table}
  \small
  \begin{tabular}{ll}
    \hline
    \hline
    \multicolumn{2}{l}{KO5 --- \citep{kouwenhoven2005} --- Visual binaries} \\
    \hline
    Observer's choice         & A and late-B members of Sco~OB2 (incl. {\em Hipparcos} completeness) \\
    Brightness constraint     & $5.3~\mbox{mag} \leq V_1 \leq 9.5~\mbox{mag}$, $M_1 \geq 1.4$~M$_\odot$ \\
    Separation constraint     & Equation~\ref{equation: adonis_separationconstraint} \\
    Contrast constraint       & Equation~\ref{equation: adonis_constrastconstraint} \\
    Confusion constraint      & $K_{S,2} \leq 12$~mag \\    
    \hline
    \multicolumn{2}{l}{KO6 --- \citep{kouwenhoven2006a} --- Visual binaries} \\
    \hline
    Observer's choice         & A selection (11\%) of the KO5 sample  \\
    Brightness constraint     & $5.3~\mbox{mag} \leq V_1 \leq 9.5~\mbox{mag}$, $M_1 \geq 1.4$~M$_\odot$ \\
    Separation constraint     & Equation~\ref{equation: naco_separationconstraint} \\
    Contrast constraint       & Equation~\ref{equation: naco_constrastconstraint}  \\
    Confusion constraint      & Not applicable \\
    \hline
    \multicolumn{2}{l}{SHT --- \cite{shatsky2002}  --- Visual binaries}\\
    \hline
    Observer's choice         & B members of Sco~OB2  (incl. {\em Hipparcos} completeness) \\
    Brightness constraint     & $V_1 \leq 7.0~\mbox{mag}$, $M_1 \geq 3.5$~M$_\odot$  \\
    Separation constraint     & Non-coronographic mode: Equation~\ref{equation: tokovinin_separationconstraint_noncoro} \\
    \quad \quad (idem)        & Coronographic mode:  Equation~\ref{equation: tokovinin_separationconstraint_coro} \\
    Contrast constraint       & Non-coronographic mode: Equation~\ref{equation: tokovinin_contrastconstraint_noncoro}  \\
    \quad \quad (idem)        & Coronographic mode:  Equation~\ref{equation: tokovinin_contrastconstraint_coro}  \\
    Confusion constraint      & $K_{S,2} \leq 12$~mag, $J_2 \leq 13$~mag, and $J_2-K_{S,2} < 1.7$~mag   \\
    \hline
    \multicolumn{2}{l}{LEV --- \cite{levato1987}  --- Spectroscopic binaries}\\
    \hline
    Observer's choice         & B members of Sco~OB2  (incl. {\em Hipparcos} completeness) \\
    Brightness constraint     & $V_{\rm comb} \leq 8.1$~mag, $M_1 \geq 3$~M$_\odot$  \\
    Contrast constraint       & Not applicable \\
    Amplitude constraint      & \quad Spectroscopic bias model SB-W \\
    Temporal constraint       & \quad with $T=2.74$~year, $\Delta T = 0.38$~year, \\
    Aliasing constraint       & \quad and $\sigma_{\rm RV} = 3.1$~km\,s$^{-1}$. \\
    Sampling constraint       & Not applied  \\
    \hline
    \multicolumn{2}{l}{BRV --- \cite{brownverschueren} --- Spectroscopic binaries}\\
    \hline
    Observer's choice         & B members of Sco~OB2  (incl. {\em Hipparcos} completeness) \\
    Brightness constraint     & $V_1 \leq 7.0~\mbox{mag}$, $M_1 \geq 3.5$~M$_\odot$  \\
    Contrast constraint       & Not applicable \\
    Amplitude constraint      & \quad Spectroscopic bias model SB-W, \\
    Temporal constraint       & \quad with $T=2.25$~year, $\Delta T = 0.75$~year, \\
    Aliasing constraint       & \quad and $\sigma_{\rm RV} = 1.4$~km\,s$^{-1}$. \\
    Sampling constraint       & Not applied  \\
    \hline
    \multicolumn{2}{l}{HIP --- {\em Hipparcos} mission --- Astrometric binaries}\\
    \hline
    Brightness constraint     & {\em Hipparcos} completeness \\
    Amplitude constraint      & \quad Classification into the \\
    Temporal constraint       & \quad categories (X), (C), (O), or (G) \\
    Aliasing constraint       & \quad depending on the observables \\
    Sampling constraint       & \quad of each binary system (see Table~\ref{table: true_hipparcosbiases}) \\
    \hline
    \hline
  \end{tabular}
  \caption{An overview of the models for the selection effects used to generate simulated observations of simulated OB~associations, for the six major datasets discussed in Sections~\ref{section: adonisobservations} to~\ref{section: hipparcosobservations}.  \label{table: sos_biases}}
\end{table}

%

% ====================================================================
% ====================================================================
% ====================================================================

\subsection{KO5 --- \cite{kouwenhoven2005} observations} \label{section: adonisobservations}

\cite[][Chapter~2]{kouwenhoven2005} performed a near-infrared adaptive optics binarity survey among A and late~B members of Sco OB2. Their observations were obtained with the ADONIS/SHARPII+ system on the ESO~3.6~meter telescope at La~Silla, Chile. The adaptive optics technique was used to obtain high spatial resolution, in order to bridge the gap between the known close spectroscopic and wide visual binaries. 
The survey was performed in the near-infrared, as in this wavelength regime the contrast between the components of a binary system with a high mass ratio is less than in the visual regime.
All targets were observed in the $K_S$ band, and several additionally in the $J$ and $H$ bands. 
KO5 selected their sample of A and late~B targets from the list of confirmed {\em Hipparcos} members that were identified by \cite{dezeeuw1999}. All targets have $6~\mbox{mag} \la V \la 9~\mbox{mag}$, which corresponds to similar limits in the $K_S$ band.

With their observations KO5 are sensitive to companions as faint as $K_S \approx 15.5$~mag. Due to the large probability of finding faint background stars in the field of view, KO5 classify all secondaries with $K_S > 12$~mag as background stars, and those with $K_S \leq 12$~mag as candidate companion stars. The $K_S = 12$~criterion separates companion stars and background stars in a statistical way, and is based on the background star study of SHT. A member of Sco~OB2 with $K_S=12$~mag has a mass close to the hydrogen-burning limit. The follow-up study of KO6 with VLT/NACO (see \S~\ref{section: nacoobservations}) has shown that the $K_S=12$~criterion correctly classifies secondaries as companions in $80-85\%$ of the cases. 
With their survey, KO5 find 151~secondaries around the 199~target stars. Out of these 151~secondaries, 74 are candidate companions ($K_S \leq 12$~mag), and 77~are background stars ($K_S > 12$~mag).  
KO5 find that the mass ratio distribution $f_q(q)$ for late-B and A~type stars in Sco~OB2 is consistent with $f_q(q) \propto q^{-0.33}$, and exclude random pairing.

% ====================================================================
% ====================================================================
% ====================================================================

\subsubsection{Treatment of the KO5 dataset} \label{section: treatment_adonis}

All 199~targets in the KO5 dataset are confirmed members of the Sco~OB2 association, and are therefore included in our analysis. We use in our analysis a subset of the companions identified in KO5.
Several targets in the ADONIS survey have more than one candidate or confirmed companion. In this paper we do not study triples and higher-order multiples; we therefore consider at most one companion per target star. For each of these candidate multiple systems we include the (candidate) companion that is most likely a physical companion. 
For HIP52357 we include the companion with $(\rho,K_S) = (0.53'',7.65~\mbox{mag})$, as it is brighter and closer to the target star than the candidate companion with $(\rho,K_S) = (10.04'',11.45~\mbox{mag})$.
For the same reason, we do not include the wide and faint candidate companion of HIP61796 with $(\rho,K_S) = (12.38'',11.86~\mbox{mag})$ in our analysis. KO5 find two bright and close companions of HIP76001, with $(\rho,K_S) = (0.25'',7.80~\mbox{mag})$ and $(\rho,K_S) = (1.48'',8.20~\mbox{mag})$, respectively. Although HIP76001 is likely a physical triple, we choose to retain only the innermost candidate companion. 
KO5 find a bright secondary separated 1.8~arcsec from HIP63204. With their follow-up study, KO6 find a close companion separated 0.15~arcsec from HIP63204. KO6 show that this close companion is physical, while the secondary at 1.8~arcsec is optical. We therefore do not consider the secondary at 1.8~arcsec in our analysis.

HIP68532 and HIP69113 are both confirmed triple systems, each with a primary and a ``double companion''. For both HIP68532 and HIP69113, the two stars in the ``double companion'' have a similar separation and position angle with respect to the primary, and a similar magnitude. In physical terms, the double companions of HIP68532 and HIP69113 could have originated from a more massive companion that fragmented into a binary. We therefore model the double companions of these stars as single companions, taking the average $\rho$ and $\varphi$, the combined $K_S$ magnitude, and the total mass of each double companion.

For the comparison with the simulated observations the targets HIP77315 and HIP77317 are both treated as individual stars. The star HIP77317 is known to be a companion of HIP77315 at $\rho=37.37''$, and is for that reason listed as such in KO5. This binary system is far too wide to be detected with the observing strategy of KO5; both stars are therefore treated as individual stars.
With the ADONIS survey we find three candidate companions of HIP81972. Of these three, only the secondary at separation $5.04''$ is a confirmed companion in the follow-up study of \cite{kouwenhoven2006a}. As HIP81972 is near the Galactic plane, the other two secondaries are likely background stars, and are thus not included in the dataset. 

KO5 separated the secondaries into candidate companions and background stars using the $K_S$~magnitude of each secondary. The follow-up study of KO6, using multi-color analysis, has shown that several of these candidate companions are background stars. We therefore do not consider in our analysis these secondaries, indicated with HIP53701-1 ($K_S=8.9$), HIP60851-1 ($K_S=11.5$), HIP60851-2 ($K_S=11.3$), HIP80142-1 ($K_S=9.51$), and HIP80474-1 ($K_S=10.8$) in KO6.

The resulting KO5 dataset that we use for our analysis contains data on 60 companion stars. For each of these targets and their companions we use the measurements given in KO5, unless more recent (and more accurate) measurements for these stars are presented in the follow-up study of KO6.
For each of the targets HIP63204, HIP73937, and HIP79771 a new close companion is resolved by KO6, which was unresolved in the observations of KO6. For these three targets, we use the $K_S$ magnitude of the primary, as provided by KO6.

The mass of each primary and companion is derived from the near-infrared magnitude. If available, the mass of each star is taken from KO6, who use the $JHK_S$ magnitude to derive the mass. In the other cases, the mass is taken from KO5, who derive the mass from the $K_S$ magnitude only. Finally, the mass ratio $q = M_2/M_1$ is calculated for each binary system. In Table~\ref{table: data_adonis} we list the properties of the binaries used for comparison with simulated observations. 

% ====================================================================
% ====================================================================
% ====================================================================

\subsubsection{Modeling the observational bias of KO5}

We model the sample bias in KO5 as follows. The authors selected the A~and late~B members of Sco~OB2. We therefore first impose the {\em Hipparcos} completeness (see \S~\ref{section: membership}) on the simulated association. Based on the properties of the target list of KO5, we model the observer's choice and brightness constraint by removing all targets with $V < 5.3$~mag, all targets with $V > 9.5$~mag, and all targets with $M < 1.4$~M$_\odot$ from the sample. 

We model the detection limit of the KO5 observations using the analysis presented in KO6. We study the 50\% detection limit, and find its dependence on angular separation and Strehl ratio (SR). We parametrize the dependence of the detection limit $\Delta K_{\rm S,det}(\rho)$ on Strehl ratio SR as 
\begin{equation} \label{equation: adonis_constrastconstraint}
  \Delta K_{\rm S,det}(\rho) = \left\{
  \begin{array}{lrl}
    0                                         &  \quad                        \rho &< \rho_{\rm lim,A}\\
    (22.0 - 3.75\, s({\rm SR}))\ (\rho-0.1'') &  \quad \rho_{\rm lim,A}  \leq \rho &< 0.5''\\
    8.8 + s({\rm SR})\, (\rho-2'')            &  \quad 0.5''             \leq \rho &< 2''\\
    8.8                                       &  \quad                        \rho &\geq 2'' \\
  \end{array}
  \right. \,,
\end{equation}
where $\rho_{\rm lim,A}$ is the angular resolution of the KO5 observations.
We assume the parameter $s({\rm SR})$ of the detection limit to vary linearly with SR. Following the properties of the KO5 observations, we model $ s({\rm SR})$ with
\begin{equation} \label{equation: adonis_constrastconstraint_slope}
  s({\rm SR})= 2.54 - 3.85 \times {\rm SR} \,. 
\end{equation}
As an example we plot the detection limit $\Delta K_{\rm S,det}(\rho)$ in Figure~\ref{figure: probability_detlim_ka} for observations with SR~=~30\%.
We simulate the distribution over SR by drawing for each target the SR randomly from the observed distribution $\tilde{f}_{\rm SR}({\rm SR})$, which is approximated with
\begin{equation} \label{equation: adonis_constrastconstraint_strehl}
  \tilde{f}_{\rm SR}({\rm SR}) \propto \exp \left( - \frac{({\rm SR}-\mu_{\rm SR})^2}{2\sigma_{\rm SR}^2} \right) \quad \quad 5\% < {\rm SR} < 50\% \,,
\end{equation}
where $\mu_{\rm SR}=30\%$ and $\sigma_{\rm SR}=5\%$.

KO5 considered only the secondaries with $K_S \leq 12$~mag as physical companions. The follow-up study of KO6 has shown that this $K_S=12$ criterion indeed correctly classifies most of the companions and background stars. Following this model for the confusion with background stars, we identify in our simulated observations only the companions with $K_S \leq 12$~mag as true companions.

Each measurement is assigned a detection probability $D_A(\rho)$ as a function of angular separation $\rho$. This detection probability refers solely to whether or not a companion is projected in the field of view. As the field of view is not circular, the detection probability of a companion is a function of the angular separation. For ADONIS we have a square field of view sized $12.76 \times 12.76$ arcsec. KO5 observed each target four times, each time with the target in another quadrant of the field of view, so that the effective field of view is $L_{\rm A}=\frac{3}{2} \cdot 12.76$~arcsec $=19.14$~arcsec. The probability $D_A$ that a secondary with an angular separation $\rho$ is in the field-of-view is then given by:
\begin{equation} \label{equation: adonis_separationconstraint}
  D_A(\rho)  =
  \left\{ \begin{array}{lll}
    1                                       & {\rm for} &                         \rho < L_{\rm A}/2 \\
    1- (4/\pi) \arccos (L_{\rm A}/2\rho)    & {\rm for} & L_{\rm A}/2        \leq \rho < L_{\rm A}/\sqrt{2} \\
    0                                       & {\rm for} & L_{\rm A}/\sqrt{2} \leq \rho
  \end{array} \right. \,.
\end{equation}
%

% ====================================================================
% ====================================================================
% ====================================================================

\subsection{KO6 --- \cite{kouwenhoven2006a} observations} \label{section: nacoobservations}

The results of the ADONIS binarity survey performed by KO5 raised several questions, in particular on the absence of faint secondaries in the $1''-4''$ separation range, and on the validity of the $K_S =12$~mag criterion that KO5 used to separate secondaries into companion stars and background stars. Although SHT and KO5 argue that the latter criterion statistically classifies the background stars correctly, the correct classification of the individual companion stars with $K_S \approx 12$~mag was still uncertain.
To this end, KO6 performed follow-up multi-color $JHK_S$ observations of a subset of the ADONIS targets. With multi-color observations, each secondary can be placed in the color-magnitude diagram, and compared with the isochrone of the Sco~OB2 subgroups. Companion stars are expected to be near the isochrone, while background stars are (generally) expected to be far from the isochrone.

 The observations were carried out with the adaptive optics instrument NAOS-CONICA (NACO), mounted on the ESO Very Large Telescope on Paranal, Chile. A subset of 22 (out of 199) KO5 targets were selected for follow-up observations. The subset was not randomly selected, but preference was given to faint and close background stars, to secondaries with $K_S \approx 12$~mag, and to newly discovered candidate companions.
KO6 analyzed the $JHK_S$ observations of these 22~stars observed with NACO, including the multi-color ADONIS observations of 9~targets. With their observations KO6 additionally found three close companions (of HIP63204, HIP73937, and HIP79771) that were unresolved in the survey of KO5.

% ====================================================================
% ====================================================================
% ====================================================================

\subsubsection{Treatment of the KO6 dataset}

We consider the 22~targets observed with NACO, all of which are confirmed members of Sco~OB2. The 9~ADONIS targets that were also studied in KO6 are not considered here (as they were not observed with NACO). Around these 22~targets, KO6 find 62~secondaries, of which they classify 18 as confirmed companions (c), 11~as possible companions (?), and 33~as background stars (b).

For the analysis we use most 15~(out of 18) confirmed companions, and 5~(out of 8) candidate companions. Both HIP68532 and HIP69113 have a tight ``double companion''. We treat each of these as a single companion, by combining the separation and mass of the individual companions (see \S~\ref{section: treatment_adonis}). The targets HIP67260, HIP79771, and HIP81949 all have three companions, for which we only include the inner companion in our analysis. We do not include the very faint secondary of HIP80142, as this is likely a background star. For HIP81972 we only include the companion HIP81972-3, which is by far the most massive companion, in our analysis.

The final KO6 dataset used in this paper contains 22~targets with 18~companions. Note that when this dataset is compared with simulated observations, a discrepancy may be present, as the sample was composed to study candidate companions and background stars with particular properties. In Table~\ref{table: data_naco} we list the properties of the binaries used for comparison with simulated observations.

% ====================================================================
% ====================================================================
% ====================================================================

\subsubsection{Modeling the observational bias of KO6}

The observer's choice and brightness constraint for the KO6 observations are equivalent to those of the KO5 survey. The only difference is that a subset of $22/199=11\%$ of the KO5 targets are in the KO6 survey. We model the KO6 sample by randomly drawing 11\% of the targets in the simulated KO5 target sample. Note that in reality, the subset was not random (see above); instead, the targets were selected based on the properties of their secondaries. The simulated KO6 observations therefore cannot be compared with the results of the KO6 observations as such. However, they can be used to find the expected number of close and/or faint companions with KO6; companions that could not be found with the KO5 survey.

We use the 50\% detection limit from the analysis presented in KO6, and parameterize it with the Strehl ratio (SR) of the observations. The 50\% detection limit as a function of $\rho$, for targets with a different brightness is derived using simulations (see \S~3.3 for details).  From the observational data we derive a detection limit $ \Delta K_{\rm S,det}(\rho)$, consisting of four line segments:
\begin{equation} \label{equation: naco_constrastconstraint}
  \Delta K_{\rm S,det}(\rho) = \left\{
  \begin{array}{ll}
    0                                         &  \quad \mbox{for\ } \rho < \rho_{\rm lim,N} \\
    (2.5\, B({\rm SR}) - 2.63 )\ (\rho-0.1'') &  \quad \mbox{for\ } \rho_{\rm lim,N} \leq \rho < 0.5''\\
    B({\rm SR}) + 0.70\, (\rho-2'')           &  \quad \mbox{for\ } 0.5'' \leq \rho < 2''\\
    B({\rm SR})                               &  \quad \mbox{for\ } \rho \geq 2''\\
  \end{array}
  \right.\,,
\end{equation}
where $\rho_{\rm lim,N} = 0.1''$ is the angular resolution of the KO6 observations, and
\begin{equation} \label{equation: naco_constrastconstraint_slope}
  B({\rm SR}) = 6.86 + 8.37\times\mbox{SR} 
\end{equation}
is the magnitude difference of the faintest detectable source, for a given Strehl ratio SR. As an example we show the detection limit $\Delta K_{\rm S,det}(\rho)$ in Figure~\ref{figure: probability_detlim_ka} for observations with SR~=~24\%. We simulate the distribution over SR by drawing for each target the SR randomly from the observed distribution $\tilde{f}_{\rm SR}({\rm SR})$, which we approximate with Equation~\ref{equation: adonis_constrastconstraint_strehl},
with $\mu_{\rm SR}=24\%$ and $\sigma_{\rm SR}=7\%$. The field of view for the observations of KO6 is $14\times 14$~arsec. As the field of view is non-circular, the detection limit is a function of both angular separation $\rho$ and position angle $\varphi$. For our simulated observations, each measurement is assigned a detection probability $D(\rho)$ as a function of angular separation $\rho$, given by
\begin{equation} \label{equation: naco_separationconstraint}
  D(\rho)  =
  \left\{ \begin{array}{lll}
    1                               & {\rm for} &  \rho < L_{\rm N}/2 \\
    1- (4/\pi) \arccos (L_{\rm N}/2\rho)    & {\rm for} & L_{\rm N}/2 < \rho < L_{\rm N}/\sqrt{2} \\
    0                               & {\rm for} & L_{\rm N}/\sqrt{2} < \rho
  \end{array} \right. \,,
\end{equation}
where $L_{\rm N}=14$~arcsec.

% ====================================================================
% ====================================================================
% ====================================================================

\subsection{SHT --- \cite{shatsky2002} observations} \label{section: tokovininobservations}

\cite{shatsky2002} performed an imaging binarity survey among 115~B~type stars in the Sco~OB2 region. Their observations were carried out in 2000 with the near-infrared adaptive optics instrument ADONIS at the ESO 3.6~meter telescope on La Silla, Chile. 
Their sample is based on the study of \cite{brownverschueren}; see \S~\ref{section: brownverschuerenobservations}. Among the 115 B-type stars surveyed by SHT,  80~are confirmed members of Sco~OB2 according to \cite{dezeeuw1999}.  Near these 115~stars, SHT find 96~secondaries in the angular separation range $0.3''-6.4''$, of which they identify 10~as new physical companions. 
The authors find that the mass ratio distribution $f_q(q)$ for B-type stars in Sco~OB2 is consistent with $f_q(q) \propto q^{-0.5}$, and exclude random pairing.

% ====================================================================
% ====================================================================
% ====================================================================

\subsubsection{Treatment of the SHT dataset}

Near the 80 confirmed members of Sco~OB2 targeted by SHT, they find 80~secondaries, of which 61~likely optical, and 19~likely physical companions. Of this set of 19~physical companions, we use a subset of 17~for our analysis. The target HD132200 is likely a physical triple system. As we consider in this paper only single and binary system, we do not include the widest and faintest component of HD132200, and retain the component with $\rho=0.128''$ and $K_S=5.46$~mag. The secondary HD133937P is incorrectly reported in SHT. For this secondary $\rho=0.57''$ and $J-K_S=2.06$~mag (N.~Shatsky \& A.~Tokovinin, private communication). Due to its large $J-K_S$ value, and as $J>13$~mag, this secondary is likely a background star. We therefore do not consider HIP133937P in our analysis.

Several targets were not included in the observed sample of SHT. These targets were known to have close companions, and were therefore not suitable for wavefront sensing. Five of these are confirmed members of Sco~OB2. We include these non-observed targets in our analysis, either as single or as a binary system, depending on whether their companions would have been detected with the SHT observing strategy. Technically, the non-inclusion of a set of stars falls under the ``observer's choice'', and should be modeled in this way. In this case however, we manually add these stars to the list of observed targets, as the properties of these stars and their companions are well-understood (making detailed models of the observer's choice redundant). The stars HIP57851, HIP23223, HIP74117, HIP76371, and HIP77840 were reported as visually resolved (C)-binaries in the {\em Hipparcos} catalog. We use the angular separation and magnitude of these components as given in the catalog, and include the stars in the sample. HIP64425 is a known triple system \citep{tokovininmsc}, for which we use the massive inner binary in our analysis. We treat the non-observed star HIP53701 as a single star, as KO6 have shown that its secondary is a background star.

For each star we derive the mass using its visual and near-infrared magnitude. If the mass of a target or companion star is derived in the multi-color study of KO6, we use this mass. For the stars observed by SHT, we derive the mass of target and companion star from the $K_S$ magnitude, using the evolutionary models described in Section~\ref{section: scoob2properties}. For the stars that are in the SHT dataset, but not observed by these authors, we derive the mass using the $V$ band magnitude and {\em Hipparcos} $H_p$ magnitude. For each star we use the distance given by the {\em Hipparcos} parallax, and the age of the subgroup of which the target is a member.

The final dataset from the SHT survey that we use in our analysis of the binary population in Sco~OB2 comprises 80~targets with 23~physical companions. The properties of these 23~companions are listed in Table~\ref{table: data_tokovinin}.

% ====================================================================
% ====================================================================
% ====================================================================

\subsubsection{Modeling the observational bias of SHT}

We require that all targets are confirmed Sco~OB2 members, and therefore first impose the {\em Hipparcos} detection limit on the simulated association. We model the brightness constraint of the SHT observations by adopting a minimum mass of 3.5~M$_\odot$, and a minimum brightness of $V=7$~mag for the targets.

SHT show the typical detection limit of their observations in their Figure~3. The detection limit is obviously different for the observations with and without the coronograph. The observations with coronograph are deeper, and the observations without the coronograph provide a larger range in angular separation. A companion star is detected if it is observed {\em either} in the coronographic mode {\em or} in the non-coronographic mode.

In the non-coronographic observations, each companion is assigned a detection probability $D_{\rm NC}(\rho)$ as a function of its separation $\rho$. In the non-coronographic observations, SHT observed each target twice in the non-coronographic mode, both times with the target in a quadrant of the detector. Due to the square shape of the detector, and due to the observing strategy, the position angle is of importance to whether a companion at separation $\rho$ is in the field-of-view. We model this dependence by assigning a probability $D_{\rm NC}(\rho)$ that a companion is in the field of view, depending on $\rho$. For a square field-of-view of a detector with linear size $L$, and a separation $K$ between the two observations (along the diagonal of the field-of-view), the probability is given by
\begin{equation} \label{equation: tokovinin_separationconstraint_noncoro}
  D_{\rm NC}(\rho)  =
  \left\{ \begin{array}{lll}
    1                                                                      & {\rm for} & \rho < r_1 \\ 
    \tfrac{1}{2} + \frac{2}{\pi} \arcsin \left(\frac{r_1}{\rho}\right)                                & {\rm for} & r_1    < r_2/\sqrt{2}  \\    
    \tfrac{1}{2} + \frac{2}{\pi} \arcsin \left(\frac{r_1}{\rho}\right) - \frac{4}{\pi} \arccos \left(\frac{r_2}{\rho}\right)     & {\rm for} & r_2\sqrt{2}   < \rho < \sqrt{\tfrac{1}{2}(r_1^2+r_2^2)} \\
     \tfrac{1}{2} - \frac{2}{\pi} \arccos \left(\frac{r_2}{\rho}\right)                               & {\rm for} & \sqrt{\tfrac{1}{2}(r_1^2+r_2^2)} < \rho < r_2 \\

    0                                                                      & {\rm for} &  r_2  < \rho
  \end{array} \right. \,,
\end{equation}
where 
\begin{equation}
r_1=\tfrac{1}{2}L\sqrt{2}  - \tfrac{1}{2} K 
\quad \quad \mbox{and} \quad \quad 
r_2=\tfrac{1}{2}L\sqrt{2} + \tfrac{1}{2} K \,. 
\end{equation}
For the non-coronographic observations of SHT, the linear dimension of the detector is $L=12.76$~arcsec, and the translation along the diagonal of the field of view is $K=8.5''$ (see Figure~1 in SHT, for details). 
We model the detection limit of the observations {\em without} the coronograph with
\begin{equation} \label{equation: tokovinin_contrastconstraint_noncoro}
  \Delta K_{\rm S,det} = \left\{
  \begin{array}{ll}
    0                     & \quad \mbox{for\ } \rho < \rho_{\rm lim,S} \\
    8.32 \log \rho + 6.83 & \quad \mbox{for\ } \rho_{\rm lim,S} \leq \rho < 1.46'' \\
    8.2                   & \quad \mbox{for\ } \rho \geq 1.46''\\
  \end{array}
  \right. \,,
\end{equation}
where $\rho_{\rm lim,S} = 0.1''$ is the angular resolution of the SHT observations (based on Figure~3 in SHT).

SHT additionally observe each target using the coronograph. They do not perform their coronographic observations in mosaic-mode; only one pointing is used. Each measurement is therefore assigned a detection probability $D_{\rm C}(\rho)$ as a function of angular separation $\rho$, given by
\begin{equation} \label{equation: tokovinin_separationconstraint_coro}
  D_{\rm C}(\rho)  =
  \left\{ \begin{array}{lll}
    0                               & {\rm for} & \rho < d_{\rm C} \\ 
    1                               & {\rm for} & d_{\rm C} < \rho < L/2 \\
    1- \frac{4}{\pi} \arccos \left(\frac{L}{2\rho}\right)    & {\rm for} & L/2 < \rho < L/\sqrt{2} \\
    0                               & {\rm for} & L/\sqrt{2} < \rho
  \end{array} \right. \,,
\end{equation}
where $L=12.76$~arcsec and $d_{\rm C} = 1''$ is the radius of the coronograph. 
Based on Figure~3 in SHT, we model the detection limit of the observations {\em with} the coronograph with
\begin{equation} \label{equation: tokovinin_contrastconstraint_coro}
  \Delta K_{\rm S,det} = \left\{
  \begin{array}{ll}
    0                     & \quad \mbox{for\ } \rho < d_{\rm C} \\
    8.32 \log \rho + 6.83 & \quad \mbox{for\ } d_{\rm C} \leq \rho < 2.62'' \\
    10.3                  & \quad \mbox{for\ } \rho \geq 2.62 \\
  \end{array}
  \right. \,.
\end{equation}

Finally, we combine the simulated observations in coronographic and non-coronographic mode. We consider a binary system as detected, if it is observed in at least one of the two modes.
SHT additionally studied the background star population in the Sco~OB2 region. Due to the large number of background stars, it is likely that a very faint or red secondary is a background star. SHT classify a secondary as a background star if $J>13$~mag, if $K_S>12$~mag, or if $J-K_S > 1.7$~mag, unless the secondary is a known companion. In our model for the selection effects, we adopt this procedure when obtaining the simulated observations.

% ====================================================================
% ====================================================================
% ====================================================================

\subsection{LEV --- \cite{levato1987} observations} \label{section: levatoobservations}

\cite{levato1987} performed a large radial velocity survey for binarity among early-type stars in the Sco~OB2 region. They performed their observations in May 1974 with the 0.9-m and 1.5-m CTIO telescopes, and in 1976 with the 2.1-m telescope at KPNO. 
Their sample consists of 81~candidate members of Sco~OB2, and is based on that of \cite{slettebak1968} who composed a list of suspected Sco~OB2 members for a study on stellar rotation. All except 4 of the 82~targets of \cite{slettebak1968}, and 3~additional targets were observed by LEV. The spectral type of the observed targets ranges from B0\,V to A0\,V. The targets in the sample have $2.5~\mbox{mag} < V < 8.1~\mbox{mag}$. 

On average, each star is observed over an interval of $\langle T \rangle = 2.74$~year, with a spread of $\sigma_T = 0.68$~year. Each target is observed $5-12$ times, with an average observing interval $\langle \Delta T \rangle = 0.38$~year and a corresponding spread of $\sigma_{\Delta T} = 0.14$~year. 
For each target LEV list the internal error in the radial velocity measurements. Averaged over all targets, this error is $\langle \sigma_{\rm RV}\rangle = 3.1$~km\,s$^{-1}$, with a spread of 1.0~km\,s$^{-1}$; approximately 90\% of the targets have $\sigma_{\rm RV} > 2$~km\,s$^{-1}$.

In their Table~3, LEV list their conclusions on binarity. Of the 52~confirmed members of Sco~OB2 that they observed, 14~have a constant radial velocity (within the measurement errors), 23 have a variable radial velocity (RVV), 8~are SB1, and 7~are SB2. Given these observations, the spectroscopic binary fraction is {\em at least} $(8+7)/52 = 29\%$. If all reported RVV targets are indeed binaries, the observed binary fraction is $73\%$.

% ====================================================================
% ====================================================================
% ====================================================================

\subsubsection{Treatment of the LEV dataset}

For the comparison between the observational data and the simulated observations, we only consider the 52~confirmed members \citep[according to][]{dezeeuw1999} of Sco~OB2 that LEV observed.
In their Tables~3 and~4, LEV include the star HIP76945 (HD140008), a confirmed member of the UCL subgroup. LEV did not observe this target, but take the orbital elements from \cite{thackeray1965}; we do not include this star in our analysis. For a subset of the targets the orbital elements are derived. In their Table~4, LEV list the elements of 22~targets, of which 15~are confirmed members of Sco~OB2. 
In Table~\ref{table: data_levato} we list the properties of these 15~SB1 and SB2 systems from the LEV dataset that are confirmed members of Sco~OB2. We additionally list the 23~radial velocity variables (RVVs), for which the orbital elements are unavailable. The LEV dataset consists of 52~targets, of which $23+15= 38$~are detected as binary systems.

% ====================================================================
% ====================================================================
% ====================================================================

\subsubsection{Modeling the observational bias of LEV}

In this paper we consider only the known members of Sco~OB2. We therefore first impose the {\em Hipparcos} detection limit on the association. We model the choice of the sample of LEV by removing all binary systems with a combined magnitude fainter than $V=8.1$~mag from the simulated observations. 

We model the instrument bias of LEV using windowed sampling \citep[SB-W; see][]{kouwenhoven2006b}, with an observing run of $T=2.74$~year and an observing interval $\Delta T = 0.38$~year. In our model we assume a radial velocity accuracy of $\sigma_{\rm RV} = 3.1$~km\,s$^{-1}$. The latter assumption is a simplification, as the value of $\sigma_{\rm RV}$ is slightly different for each observation in the LEV dataset (with a spread of $\sim 1$~km\,s$^{-1}$). A star is more easily detected if $\sigma_{\rm RV} < 3.1$~km\,s$^{-1}$, and less easy if $\sigma_{\rm RV}$ is larger. Our simulations show, however, that our assumption of a constant $\sigma_{\rm RV}$ introduces an error significantly smaller than the error introduced by low-number statistics, justifying our assumption.

We do not attempt to classify the simulated binaries in the categories RVV, SB1 and SB2. Whether a binary system belongs to as one of these categories, depends strongly on the properties of the binary and on the observing strategy \citep[see][]{kouwenhoven2006b}; it is not trivial to model this.

% ====================================================================
% ====================================================================
% ====================================================================

\subsection{BRV --- \cite{brownverschueren} observations} \label{section: brownverschuerenobservations}

\cite{brownverschueren} studied stellar rotation among members of the Sco~OB2 association. The observations were carried out between 1991 and 1993 using the ECHELEC spectrograph at the ESO 1.52~meter telescope on La~Silla, Chile. The sample of BRV contains the pre-{\em Hipparcos} candidate and established members of Sco~OB2, based on the studies of \cite{blaauw1964A}, \cite{bertiau1958}, and \cite{degeus1989}. 
The observations and data reduction procedure are described in detail in \cite{verschueren1997}, and the results on duplicity are described in \cite{verschueren1996} and \cite{brownverschueren}. 
Their sample consists of 156~targets in the Sco~OB2 region, mostly of spectral type~B.
They find that $\sim 60\%$ of the binary systems exhibit a significant radial velocity variation. After combination of their data with those of LEV and those of the Bright Star Catalogue \citep{hoffleit1982,hoffleit1983}, they obtain a binary fraction of 74\%.

\subsubsection{Treatment of the BRV dataset}

Among the 156~observed targets there are 71~confirmed members of Sco~OB2 (18~in US, 30~in UCL, and 23~in LCC). Among these 71~targets, 7~are SB1, 10~are SB2, 30~are RVV, and 12~have a constant radial velocity (CON). For 12~targets, insufficient measurements are available to make a statement on duplicity. 

Two out of the 30~RVV binaries are known to exhibit radial velocity variation due to line profile variability. HD120324 is a non-radial pulsator, and HD136298 is a $\beta$~Cephei variable. For both stars, this is likely the reason  that they are classified as RVV. These stars are therefore not considered as binary systems in our analysis.  

The final BRV dataset used in our analysis consists of 71~confirmed members. Of these targets, 12~are spectroscopically single, 7~are SB1, 10~are SB2, 28~are RVV, and 12~have insufficient data to decide on duplicity. The binary fraction is therefore {\em at least} $(7+10)/71 \approx 24\%$, if {\em none} of the RVV targets are binary, and $(71-12)/71 = 83\%$ if {\em all} RVV targets are binary. Among the target stars with sufficient data to make a statement on duplicity, the observed binary fraction is $\sim 65\%$ at most.

\subsubsection{Modeling the observational bias of BRV}

We model the choice of the BRV sample in a similar way as we did for the LEV dataset. Each target is observed three times over an interval of $T=2.25$~year, so that $\Delta T = 0.75$~year. Following the reduction of the original data \citep{verschueren1996}, we classify each target with a radial velocity variation larger than $3\sigma_{\rm RV} = 4.2$~km\,s$^{-1}$ as a RVV (see Section~\ref{section: sb-w} for details).

% ====================================================================
% ====================================================================
% ====================================================================

\subsection{HIP --- {\em Hipparcos} observations} \label{section: hipparcosobservations}

In the {\em Hipparcos} catalog, the (possible) presence of a binary system is indicated in field H59 with the flags (X), (O), (G), (C), or (V), and in field H61 with the flag (S). For with a (C) flag, both stars in the binary system are resolved, but no orbital motion is detected. These systems can therefore be considered as visual binaries.
{\em Hipparcos} entries with an (X)-flag have a stochastic solution. These stars exhibit an apparent motion significantly larger than the statistical uncertainties, although no double star solution could be found. For entries with an (O)-flag, at least one of the orbital elements could be derived from the apparent motion. Entries with a (G)-flag show a significant acceleration in the apparent motion, but no solution could be found. These are likely long-period binaries. The (V)-flagged entries are variability-induced movers. For this group of binaries, the photocenter exhibits apparent motion due to variability of one of the components. Finally, (S)-flagged entries are suspected non-single stars.  These targets are effectively single as observed by {\em Hipparcos} \cite[][\S~2.1]{esa1997}, although no convincing single-star solution could be found for these stars.  Several of the (S) binaries are also (X) binaries. 

Among the 521~confirmed members of Sco~OB2, 46~are candidate or confirmed astrometric binaries --- i.e., those in the categories (X), (O), (G), and (S) --- in the {\em Hipparcos} catalog: 12~in US, 17~in UCL, and 17~in LCC. An additional 79~Sco~OB2 members are classified as (C)-binaries; these are visually resolved binaries. Table~\ref{table: hipparcos_observations} lists for the three subgroups of Sco~OB2 the number of entries in each of the {\em Hipparcos} categories. 
As for the binaries in the category (C) no orbital motion is detected, we will consider this group as visual binaries. (V)-binaries are not present among the confirmed Sco~OB2 members. HIP78918 is the only member of Sco~OB2 with an orbital solution (O). The three stochastic (X) binaries in Sco~OB2 are also flagged as suspicious binaries (S).

\begin{table}
  \small
  \begin{tabular}{p{0.8cm}r ccccc p{1.45cm}p{1.45cm}p{1.45cm}}
    \hline
    \hline
    Group& $N_\star$&(X)& (O) &(G) & (C)& (S)             & $\tilde{F}_{\rm M,XOG}$ & $\tilde{F}_{\rm M,XOGS}$ & $\tilde{F}_{\rm M,C}$ \\
    \hline
    US      & 120   & 1 & 0   & 4  & 15 & 8  (1)  & $4.2 \pm 1.5\%$      &  $10.0\pm 2.1\%$      & $ 12.5\pm 3.0\%$    \\
    UCL     & 221   & 0 & 1   & 9  & 36 & 7  (0)  & $4.5 \pm 1.5\%$      &  $\ 7.7\pm 1.9\%$      & $ 16.3\pm 3.0\%$   \\
    LCC     & 180   & 2 & 0   & 6  & 28 & 11  (2)  & $4.4 \pm 1.6\%$      &  $\ 9.4\pm 2.4\%$      & $ 15.6\pm 3.0\%$    \\
    \hline
    Sco~OB2 & 520 & 3 & 1   & 19 & 79 & 26  (3)  & $4.4 \pm 1.0\%$      &  $\ 8.8\pm 1.4\%$      & $ 15.2\pm 1.8\%$   \\
    \hline
    \hline
  \end{tabular}
  \caption{Candidate and confirmed astrometric binaries in the {\em Hipparcos} catalog. For each subgroup we list the number $N_\star$ of known members, the number of stochastic (X), orbital (O), acceleration (G), component (C), and suspected (S) binaries in the {\em Hipparcos} catalog. For each (S) binary we list between brackets how many of these also have an (X) flag. The last three columns list the ``astrometric binary fraction'' --- including the (X), (O), (G) binaries --- without the (S) binaries and with the (S) binaries included, and the {\em Hipparcos} ``visual'' binary fraction, for the (C) binaries only. (V) binaries are not present in Sco~OB2.
 \label{table: hipparcos_observations} }
\end{table}

% ====================================================================
% ====================================================================
% ====================================================================

\subsubsection{Treatment of the {\em Hipparcos} dataset}

In our analysis we consider each target in the categories (X), (O), (G), and (C) in Table~\ref{table: hipparcos_observations} as a binary system. Binarity among the stars in the (S) category (the ``suspected non-single'' targets) is rather uncertain. We therefore compare our results with the {\em Hipparcos} data, with and without the suspected (S) binaries included. 
Note that not all targets in the category (X) are necessarily binary systems. For example, the flag (X) of HIP80763 ($\alpha$~Sco), may be due to the extended nature of the star, which is surrounded by a dust-shell \citep{cruzalebes1998}. This may induce photocentric motions that are not related to binarity.

% ====================================================================
% ====================================================================
% ====================================================================

\subsubsection{Modeling the observational bias of {\em Hipparcos}}

The observer's choice and sample bias for the {\em Hipparcos} member list of Sco~OB2 are discussed in Section~\ref{section: membership}; our adopted model to describe these biases is given in Equation~\ref{equation: hipparcos_fraction}. The instrument bias for {\em Hipparcos} was described in detail in Chapter~4, and is summarized in Table~\ref{table: true_hipparcosbiases}. 

The binaries in category (C) are considered as visual binaries, as these binaries are visually resolved, but astrometrically unresolved. The (C)-binaries are modeled and directly compared with the observations.
The binaries in the categories (X), (O), and (G) are considered as astrometric binaries. As we do not model the {\em Hipparcos} observations in detail, we are unable to accurately predict in which of these categories each {\em Hipparcos} target falls. The simple model that we adopt for the {\em Hipparcos} biases results in an overlap between the properties of the stars in these categories, and furthermore, we over-predict the number of stars observed in these categories. \cite{lindgren1997} have analyzed the properties of the binary systems in each of the categories. But this does not mean that any binary system with these properties is always observed as such. In our model, we make the latter assumption, resulting in an overestimation of the number of binaries detected by {\em Hipparcos}, and an overlap between the categories (O) and (G) in our models.
We do not include stellar variability in our model, and are therefore unable to model the (V)-flagged binaries (field H59). We do not model the (S)-flagged binaries (field H61) either.

\begin{table}
  \small
  \begin{tabular}{llp{1cm}cl}
    \hline
    Constraints on $\rho$ and $\Delta H_p$ & Period constraints & Solution & Symbol & Elements \\
    \hline
    $2 \leq \langle \rho \rangle \leq 100$~mas  or $\Delta H_p > 4$   & $P \leq 0.1$~year      & Stochastic  & (X) & no \\
    $2 \leq \langle \rho \rangle \leq 100$~mas  or $\Delta H_p > 4$   & $0.1 < P \leq 10$~year & Orbital     & (O) & yes \\
    $2 \leq \rho \leq 100$~mas  or $\Delta H_p > 4$                   & $5 < P \leq 30$~year   & Acceleration& (G) & no \\
    $0.1 \leq \rho \leq 100$~arcsec and $\Delta H_p \leq 4$              & $P > 30$~year       & Resolved    & (C) & no \\
    \multicolumn{2}{l}{Not modeled}                                                            & Suspected   & (S) & no \\
    \multicolumn{2}{l}{Not modeled}                                                            & VIM         & (V) & no \\
    \hline
  \end{tabular}
  \caption{A model for the instrument bias of the {\em Hipparcos} catalog, based on the analysis of \cite{lindgren1997}. The binary systems satisfying the above constraints are resolved with {\em Hipparcos} and flagged in fields H59 and H61 of the catalog. For the comparison between the observations and the simulated observations, we consider two sets of {\em Hipparcos} binaries: the visual binaries and the astrometric binaries. No orbital motion is detected for the (C) binaries; these are visually resolved and therefore technically visual binaries. The {\em Hipparcos} astrometric binaries contain the targets with (X), (O), (G), and optionally (S) entries. No difference between the latter categories is made for the comparison with the astrometric binaries. Binary systems that do not satisfy the constraints listed in this table remain undetected in our simulated observations for {\em Hipparcos}. We do not model the (V)-binaries (variability-induced movers; VIMs) and (S)-binaries (suspected non-single stars). \label{table: true_hipparcosbiases} }
\end{table}

% ====================================================================
% ====================================================================
% ====================================================================
% ==INTRODUCTION======================================================
% ====================================================================
% ====================================================================
% ====================================================================

\section{Recovering the true binary population} \label{section: recovering}

In this section we discuss the determination of the binary population in Sco~OB2 from observations. We adopt the strategy described in Chapter~4. This strategy is an iterative process, which finally leads to binary population(s) that is/are consistent with the observations. In the subsections below we only describe the final steps of the process. First we derive the pairing function and mass ratio distribution in \S~\ref{section: true_pairingfunction}. We recover the semi-major axis and period distribution in \S~\ref{section: true_smadistribution}, and the eccentricity distribution in \S~\ref{section: true_eccentricitydistribution}. Finally, we derive the binary fraction in \S~\ref{section: true_binaryfraction}.

% ============================================================================
% ============================================================================
% ============================================================================
% ============================================================================

\subsection{Recovering the pairing function and mass ratio distribution} \label{section: true_pairingfunction}

\begin{figure}[!btp]
  \centering
  \includegraphics[width=1\textwidth,height=!]{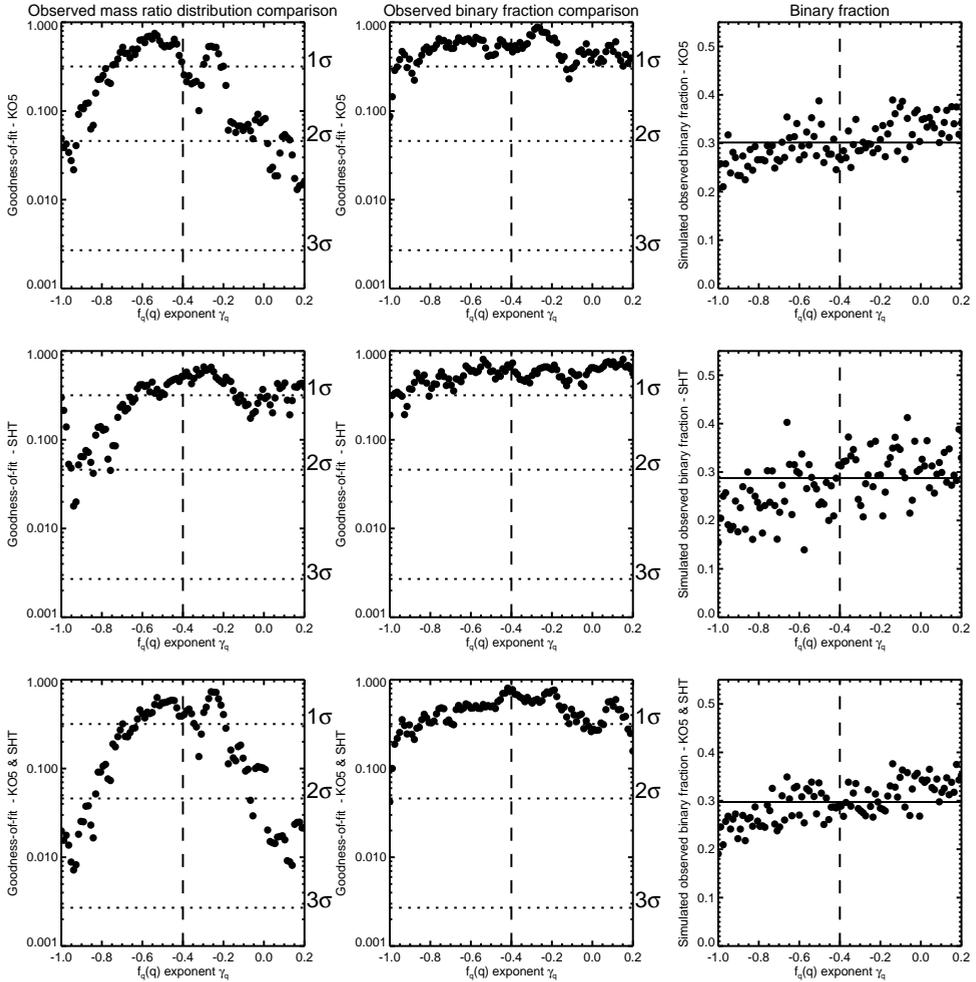}
  \caption{The mass ratio distribution among intermediate mass stars in Sco~OB2 is of the form  $f_q(q) \propto q^{\gamma_q}$ with exponent $\gamma_q = -0.4 \pm 0.2$ (dashed line in each panel). Simulated observations for models with this value of $\gamma_q$ and with a binary fraction 100\% are consistent with the observed mass ratio distribution of the KO5 and SHT datasets (left-hand panels), and with the observed binary fraction in these datasets (middle and right-hand panels). The top, middle, and bottom panels correspond to the analysis of the KO5 dataset, the SHT dataset, and the combined KO5/SHT datasets, respectively. The observed binary fraction in each of the datasets is indicated with the solid line in the right panels. See~\ref{section: true_pairingfunction} for a further discussion of this figure.  \label{figure: bestfit_q} }
\end{figure}

In Chapter~4 we discussed five possible ways of pairing the components of a binary system. We will show below that the observations are inconsistent with both random pairing (RP) or primary-constrained random pairing (PCRP). We show that the other three pairing functions (PCP-I, PCP-II, and PCP-III) are consistent with the observations, and derive the mass ratio distribution from the observations.

For the pairing functions RP and PCRP, the mass ratio distribution $f_q(q)$ depends strongly on the mass distribution $f_M(M)$, in particular on the shape of the mass distribution in the brown dwarf regime. The mass distribution in the stellar regime is known for Sco~OB2 (Equation~\ref{equation: preibischimf}), but is poorly constrained in the brown dwarf regime. However, the mass distribution can still be used to rule out pairing functions RP and PCRP for Sco~OB2. In Chapter~4 we have shown that if more than $1-2\%$ percent of the binaries with a high-mass primary have a mass ratio larger than $q \geq 0.8$, the pairing functions RP and PCRP can be excluded with high confidence. In formula, this fraction is given by:
\begin{equation}
  \tilde{Q} = \frac{\mbox{observed number of binaries with $q>0.8$}}{\mbox{number of targets}} \,.
\end{equation}
Among the 199~targets in the KO5 sample, 10~binary systems with $q \geq 0.8$ are detected. The observed fraction of binaries with $q>0.8$ is therefore $\tilde{Q}_{\rm KO5} = 5.0 \pm 1.6\%$. Due to selection effects KO5 have certainly missed several binaries with $q>0.8$ at separations smaller than the spatial resolution, or outside the field of view, so that the true value is $Q_{\rm KO5} \geq 5.0 \pm 1.6\%$. 

In order to compare these results with those expected from RP and PCRP, we simulate associations with the Preibisch mass distribution with $\alpha=-0.9$, and a binary fraction of 100\%. The simulated KO5 sample for this simulated association has $Q_{\rm RP} \approx 0.1\%$ and  $Q_{\rm PCRP} \approx 0.1\%$, thus excluding pairing functions RP and PCRP with this mass distribution. The values of $Q_{\rm RP}$ and $Q_{\rm PCRP}$ increase when the number of brown dwarfs decreases. Even in the limit that no brown dwarfs are present ($\alpha = \infty$), the ratios are $Q_{\rm RP} \approx 0.2\%$ and  $Q_{\rm PCRP} \approx 0.2\%$.  Even in this extreme case, with 100\% binarity and $\alpha=\infty$, these pairing functions do not reproduce the observed value of $\tilde{Q}$. RP and PCRP are therefore excluded with high confidence.

The pairing function is therefore either PCP-I, PCP-II, or PCP-III. The unknown slope $\alpha$ of the mass distribution (Equation~\ref{equation: preibischimf}) is now irrelevant, as the distribution over companion masses among intermediate mass stars is independent of $\alpha$. These three pairing functions are all characterized by a mass ratio distribution $f_q(q)$. The mass ratio distribution and binary fraction for high-mass and intermediate mass targets is the same for the three pairing functions. As we do not have detailed information on binarity among low-mass stars in Sco~OB2, we cannot discriminate between PCP-I, PCP-II, and PCP-III. A detailed membership study for low-mass stars in Sco~OB2, followed by a detailed binary study, is necessary to find the difference. In Chapter~4 we have also shown that for the three PCP pairing functions, the mass ratio distribution for binaries with a high-mass primary is approximately equal to the generating mass ratio distribution $f_q(q)$ of an association. In section~\ref{section: true_binaryfraction} we will return to this issue, and discuss the inferred properties of the association as a whole, for each of the three pairing functions. Note that the results derived in the sections below do not depend on whether pairing function PCP-I, PCP-II, or PCP-III is adopted in the model association.

The mass ratio distribution among intermediate mass stars in Sco~OB2 was studied in KO5 and SHT. SHT find a mass ratio distribution of the form $f_q(q) \propto q^{\gamma_q}$ with $\gamma_q=-0.5$ in their B~star survey, and KO5 find $\gamma_q=-0.33$ in their A and late-B star survey. 

Adopting a mass ratio distribution of the form $f_q(q) \propto q^{\gamma_q}$, we combine the KO5 and SHT datasets, and study for which value of $\gamma_q$ the simulated observations correspond best to the observations. We compare the observed mass ratio distribution $\tilde{f}_q(q)$ with that of the simulated observations using the Kolmogorov-Smirnov (KS) test, and we compare the observed binary fraction $\tilde{F}_{\rm M}$ with the predictions using the Pearson $\chi^2$ test. In both cases, we test the hypothesis that the observations and simulated observations are realizations of the same underlying association model.

Figure~\ref{figure: bestfit_q} shows the results of this comparison for models with a varying value of $\gamma_q$. The three left-hand panels show the probability associated with the KS comparison between the observed $\tilde{f}_q(q)$ and that of the simulated observations. The middle panels show the probability associated with the Pearson $\chi^2$ comparison between the observed binary fraction $\tilde{F}_{\rm M}$ and that of the simulated observations. The right-hand panels show the observed binary fraction, and that of the simulated observations. The comparison for the KO5 dataset is shown in the top panels, that for the SHT dataset in the middle panels, and that for the combined KO5 and SHT datasets in the bottom panels.

The figure shows that the mass ratio distribution among intermediate mass stars in Sco~OB2 is of the form $f_q(q) \propto q^{\gamma_q}$ with $\gamma_q=-0.4\pm 0.2$ ($1\sigma$ errors). The $1\sigma$ errors correspond to the set of models that are consistent with the combined KO5/SHT observations, both for $\tilde{f}_q(q)$ (bottom-left panel) and $\tilde{F}_{\rm M}$ (bottom-middle panel). This set of consistent models is centered around the value $\gamma_q=-0.4$. 

The observed binary fraction $\tilde{F}_{\rm M}$ is 30\% for the KO5 dataset, 29\% for the SHT dataset, and 30\% for the combined KO5/SHT dataset. These values are in good agreement with the simulated observations of associations with a binary fraction of 100\% (right-hand panels), indicating that the intrinsic binary fraction among binaries with intermediate mass primaries in Sco~OB2 is close to $100\%$.

% ============================================================================
% ============================================================================
% ============================================================================
% ============================================================================

\subsection{Recovering the semi-major axis and period distribution} \label{section: true_smadistribution}

In this section we recover the properties of the semi-major axis and period distribution of binaries in Sco~OB2. We determine the lower limits $a_{\rm min}$ and $P_{\rm min}$ in \S~\ref{section: recover_amin}, the upper limits $a_{\rm max}$ and $P_{\rm max}$ in \S~\ref{section: recover_amax}, and finally the best-fitting distribution in \S~\ref{section: recover_ashape}. We show in \S~\ref{section: fa_vs_fq} that our {\em independent} derivation of the distributions $f_a(a)/f_P(P)$ and $f_q(q)$ results in the best-fitting combination of these.

\subsubsection{The minimum semi-major axis and period} \label{section: recover_amin}

\begin{table}
  \small
  \begin{tabular}{p{1cm}p{0.5cm} p{0.6cm}p{0.6cm} cc p{0.5cm}l}
    \hline
    Member   & SpT   & $P$    & $M_1$           & \multicolumn{2}{c}{$a$} & Group & Reference \\
             &       &        &       & ($q=0$) & ($q=1$) \\ 
    \hline
    HIP74449 & B3\,IV& 0.90d & $6.2$~M$_\odot$  &  7~R$_\odot$ &  9~R$_\odot$ & UCL  & \cite{buscombe1962} \\
    HIP77911 & B9\,V & 1.26d & $2.8$~M$_\odot$ &  7~R$_\odot$ &  9~R$_\odot$ & US   & \cite{levato1987} \\
    HIP78265 & B1\,V & 1.57d & $6.8$~M$_\odot$ & 11~R$_\odot$ & 14~R$_\odot$ & US   & \cite{levato1987} \\
    HIP74950 & B9\,IV& 1.85d & $2.9$~M$_\odot$ &  9~R$_\odot$ & 11~R$_\odot$ & UCL  & \cite{andersen1993} \\
    HIP77858 & B5\,V & 1.92d & $5.3$~M$_\odot$ & 11~R$_\odot$ & 14~R$_\odot$ & US   & \cite{levato1987} \\
    \hline
    \hline
  \end{tabular}
  \caption{The five known binaries in Sco~OB2 with an orbital period less than two days. We list the primary star, the primary spectral type, the period, and an estimate for the primary mass. Columns~5 and~6 list extremes for the semi-major axis of the binary, as derived using Kepler's third law, under the assumption of a mass ratio $q=0$ (5th column), and $q=1$ (6th column). Finally, columns 7 and 8 list the subgroup and the reference. It is possible that several closer, yet undiscovered binaries in Sco~OB2 exist. \label{table: recovering_amin}}
\end{table}

The smallest value $P_{\rm min}$ for the orbital period distribution can be constrained using the observations of spectroscopic binaries. Table~\ref{table: recovering_amin} lists the five known binaries in Sco~OB2 with an orbital period less than two days. These data indicate that $P_{\rm min} \la 1$~day.  Binary systems with an orbital period shorter than 1~day may be present in Sco~OB2, but as this orbital period is close to the physical minimum period (corresponding to Roche Lobe overflow). Only a very small fraction of systems is expected to have such a short orbital period, as stars in many of such binaries would have been in physical contact during their contraction phase.

For each short-period binary in  Table~\ref{table: recovering_amin} we have obtained an estimate for the semi-major axis, using Kepler's third law. The mass estimate of the primary star is obtained from its absolute near-infrared and visual magnitude, based on the {\em Hipparcos} parallax. As the mass ratio of each of these binaries is unknown, we show the inferred semi-major axis for the extremes $q=0$ and $q=1$. As the five binary systems are all spectroscopic binaries, the value of the mass ratio is likely of order 0.5 or larger. From the semi-major axis estimates in  Table~\ref{table: recovering_amin} we thus conclude that $a_{\rm min} \la 10$~R$_\odot$. 

As a result of selection effects, the observations cannot be used to constrain $P_{\rm min}$ and $a_{\rm min}$ any further. However, physical arguments can be used to obtain lower limits for $P_{\rm min}$ and $a_{\rm min}$. 
The value of $a_{\rm min}$ cannot be significantly smaller than the semi-major axis at which Roche lobe overflow occurs for one of the components of a binary system. One of the components such a tight binary may fill its Roche Lobe if the binary separation $a$ is less than about 2--3 times the radius of that star \citep[see, e.g.,][]{hilditch2001}. This suggests that $a_{\rm min} \ga 2$~R$_\odot$, corresponding to a typical period of 12~hours for intermediate mass stars.  Note that binary systems with shorter orbital periods are known, such as cataclysmic variables, double pulsars, WU~majoris binaries, and symbiotic stars. We do not consider compact companions, as Sco~OB2 is a young association, and although several of its members may have evolved into one of these tight systems, our conclusions on the {\em primordial} binary population will not be affected by this assumption.

Using observations and physical limitations, we have constrained the values 2~R$_\odot \la a_{\rm min} \la 10$~R$_\odot$ and $0.5~\mbox{day} \la P_{\rm min} \la 1$~day. Note that these are rather ``rough'' constraints, imposed by physical limitations and the observations. The true values of $a_{\rm min}$ and  $P_{\rm min}$ are somewhere in between the limits given above. In Section~\ref{section: true_binaryfraction} we will return to this issue in the context of the binary fraction of Sco~OB2.

% ============================================================================
% ============================================================================
% ============================================================================
% ============================================================================

\subsubsection{The maximum semi-major axis and period} \label{section: recover_amax}

\begin{table}
  \begin{tabular}{p{1.2cm}c p{0.4cm}c p{0.6cm}p{1.2cm}l}
    \hline
    \hline
    Member & $\rho$ & \multicolumn{1}{c}{$\pi$} & $a_{\rm est}$    & Group  &  SpT & Reference \\
           & arcsec & \multicolumn{1}{c}{mas}   & $10^6$~R$_\odot$ \\
    \hline
HIP64004 & 25.1 & 7.92 & 0.68 & LCC & B1.5V    &   \cite{lindroos1985}   \\
HIP71860 & 27.6 & 5.95 & 0.99 & UCL & B1.5III    & \cite{worley1978}     \\
HIP69113 & 28.6 & 4.57 & 1.34 & UCL & B9V    &     \cite{lindroos1985} \\
HIP75647 & 30.0 & 7.79 & 0.82 & UCL & B5V    &     \cite{wds1997}  \\
HIP69749 & 30.2 & 4.07 & 1.59 & UCL & B9IV    &    \cite{oblak1978}  \\
HIP60320 & 32.4 & 9.71 & 0.71 & LCC & Am    &      \cite{wds1997} \\
HIP69618 & 33.9 & 6.71 & 1.08 & UCL & B4Vne    &   \cite{wds1997}   \\
HIP77315 & 34.7 & 7.64 & 0.97 & UCL & A0V    &     \cite{wds1997}  \\
HIP63003 & 34.8 & 8.64 & 0.86 & LCC & B2IV-V    &  \cite{lindroos1985}    \\
HIP72192 & 35.3 & 7.72 & 0.98 & UCL & A0V    &     \cite{lindroos1985}  \\
HIP78104 & 38.3 & 7.97 & 1.03 & US & B2IV/V    &   \cite{wds1997}    \\
HIP72984 & 39.0 & 5.93 & 1.41 & UCL & A0/A1V    &  \cite{wds1997}     \\
HIP79374 & 41.4 & 7.47 & 1.19 & US & B2IV    &     \cite{tokovininmsc} \\
HIP83693 & 43.3 & 7.73 & 1.20 & UCL & A2IV    &    \cite{wds1997} \\
HIP80024 & 46.7 & 6.98 & 1.43 & US & B9II/III    & \cite{oblak1978}       \\
HIP67472 & 48.0 & 6.19 & 1.66 & UCL & B2IV/Ve    & \cite{wds1997}      \\
HIP78265 & 49.2 & 7.10 & 1.48 & US & B1V+B2V    &  \cite{wds1997}    \\
HIP64661 & 60.0 & 8.04 & 1.60 & LCC & B8V    &     \cite{lindroos1985}    \\
HIP65271 & 60.0 & 9.20 & 1.40 & LCC & B3V    &     \cite{tokovininmsc}   \\
HIP78384 & 115.0 & 6.61 & 3.73 & UCL & B2.5IV    & \cite{tokovininmsc}    \\
    \hline
    \hline
  \end{tabular}
  \caption{The widest known binary systems in Sco~OB2. For each of these binary systems we list the angular separation, the {\em Hipparcos} parallax, an estimate for the semi-major axis $a_{\rm est} = D\tan \rho$, the subgroup, and the spectral type of the primary. The last column lists the reference. Note that this list must be incomplete, as very wide binaries are difficult to detect. Furthermore, several of these binaries may be optical, i.e., not physically bound. \label{table: recovering_amax} }
\end{table}

It is difficult to measure $a_{\rm max}$ and $P_{\rm max}$ directly. Wide binaries are difficult to identify, and due to the presence of a background population, confusion will occur between these companions and background stars. 
However, the widest observations can be used to constrain $a_{\rm max}$. Table~\ref{table: recovering_amax} lists the known binaries in Sco~OB2 with angular separation larger than 25~arcsec. Under the assumption that these wide binary systems are all physically bound, we additionally list a first-order estimate for the semi-major axis $a_{\rm est} = D\tan \rho$, where $D$ is the distance to the binary, and $\rho$ the angular separation. The maximum semi-major axis is likely of order $2\times 10^6$~R$_\odot$ or larger. According to Kepler's third law, the orbital periods of most binary systems listed in Table~\ref{table: recovering_amax} range between $\sim 0.150$~Myr to $\sim 0.25$~Myr (assuming a system mass of 5~M$_\odot$ for each binary). 

An upper limit for $a_{\rm max}$ can be obtained from stability in the Galactic tidal field. Binary systems with a semi-major axis larger than $a_{\rm tidal} \approx 0.2$~pc ($=9\times 10^6$~R$_\odot$) are unstable in the Galactic tidal field and are ionized quickly. If Sco~OB2 is indeed an expanding association, the association must have been denser in the past. The upper limit for the semi-major axis may therefore be smaller than $a_{\rm tidal}$. For binary systems with a total mass of 5~M$_\odot$, this tidal limit corresponds to $\sim 4$~Myr. Note that if such a binary system would exist in Upper Scorpius, it would have completed only one revolution since its birth. For lower-mass binaries, the period corresponding to $a=0.2$~pc would be significantly larger.

We wish to note that the goal of our study is to find the primordial binary population, so that theories on star formation can be constrained with observations. In this context, the exact value of $a_{\rm max}$ is not of great importance. Binary systems with a semi-major axis of order $a=0.2$~pc have an orbital period larger than $\sim 4$~Myr. 
In the context of star formation, the components of these wide binaries may be considered as single, as both stars have formed practically independent from each other (assuming the orbital elements have not significantly changed since the time of formation)

Combining the information above, we have constrained $2 \times 10^6$~R$_\odot \la a_{\rm max} \la 8.9 \times 10^6$~R$_\odot$, and $\sim 0.15~\mbox{Myr} \la P_{\rm max} \la 4~\mbox{Myr}$. In Section~\ref{section: true_binaryfraction} we will return to this issue in the context of the binary fraction of Sco~OB2.

% ============================================================================
% ============================================================================
% ============================================================================
% ============================================================================

\subsubsection{The distributions $f_a(a)$ and $f_P(P)$} \label{section: recover_ashape}

\begin{figure}[!btp]
  \centering
  \includegraphics[width=1\textwidth,height=!]{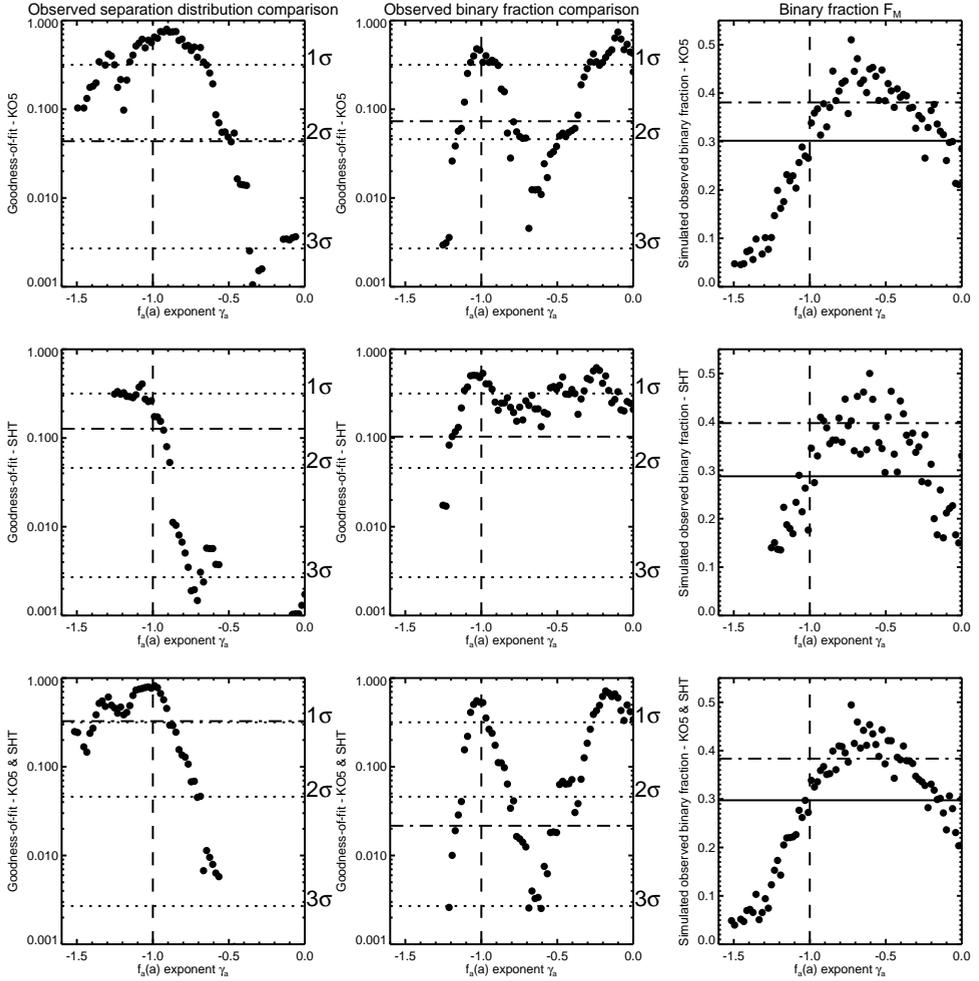}
  \caption{The semi-major axis distribution for intermediate mass stars in Sco~OB2 is of the form  $f_a(a) \propto a^{\gamma_a}$ with exponent $\gamma_a = -1.0 \pm 0.15$ (dashed line in each panel), commonly known as \"{O}pik's law. Simulated observations for models with this value of $\gamma_a$ and with a binary fraction of 100\% are consistent with the observed angular separation distribution of the KO5 and SHT datasets (left-hand panels), and with the observed binary fraction in these datasets (middle and right-hand panels). The top, middle, and bottom panels correspond to the analysis of the KO5 dataset, the SHT dataset, and the combined KO5/SHT datasets, respectively. The observed binary fraction in each of the datasets is indicated with the solid line in the right panels. In each panel we additionally show the results for the log-normal period distribution (Equation~\ref{equation: true_duquennoyperiods}) with the dash-dotted line. The latter distribution is consistent with the observed properties of the visual binaries (with a binary fraction of $\sim 80\%$, but is later in this chapter shown to be inconsistent with the observed number of spectroscopic binaries. See~\ref{section: recover_ashape} for a further discussion of this figure. \label{figure: bestfit_sma} }
\end{figure}

The semi-major axis and period distribution of different binary populations are currently under debate. Observations have brought forward two distributions: \"{O}pik's law, and the log-normal period distribution of \cite{duquennoy1991}. 
\"{O}pik's law is characterized by a flat distribution in $\log a$; the semi-major axis distribution is given by
\begin{equation} \label{equation: true_opikslaw}
f_{\gamma_a}(a) \propto a^{\gamma_a} 
\quad \quad a_{\rm min} \leq a \leq a_{\rm max} \,,
\end{equation}
with $\gamma_a=-1$. The log-normal period distribution of \cite{duquennoy1991} is given by:
\begin{equation} \label{equation: true_duquennoyperiods}
f_{\rm DM}(P) \propto \exp \left\{ - \frac{(\log P - \overline{\log P})^2 }{ 2 \sigma^2_{\log P} }  \right\} 
\quad \quad P_{\rm min} \leq P \leq P_{\rm max} \,,
\end{equation}
where $\overline{\log P} = 4.8$, $\sigma_{\log P} = 2.3$, and $P$ is in days. The difference between binary populations resulting from Equations~\ref{equation: true_opikslaw} and~\ref{equation: true_duquennoyperiods} was discussed in detail in Chapter~4, where we have shown that for nearby OB~associations it is practically impossible to see the difference between the two distributions using the observed angular separation distribution $\tilde{f}_\rho(\rho)$ alone. On the other hand, the distributions $f_{\rm DM}(P)$ and $f_{\gamma_a}(a)$ produce a significantly different visual and spectroscopic binary fraction; these can thus be used to discriminate between \"{O}pik's law and the log-normal period distribution.

Adopting a semi-major distribution of the form $f_a(a) \propto a^{\gamma_a}$, we combine the KO5 and SHT datasets, and determine for which value of $\gamma_a$ the simulated observations correspond best to the observations. We compare the observed angular separation distribution $\tilde{f}_{\rho}(\rho)$ with that of the simulated observations using the KS test, and we compare the observed binary fraction $\tilde{F}_{\rm M}$ with the predictions using the Pearson $\chi^2$ test. In both cases, we test the hypothesis that the observations and simulations are realizations of the same underlying association model.

Figure~\ref{figure: bestfit_sma} shows the results of this comparison for models with a varying value of $\gamma_a$. The three left-hand panels show the probability associated with the KS comparison between the observed $\tilde{f}_\rho(\rho)$ and that of the simulated observations. The middle panels show the probability associated with the Pearson $\chi^2$ comparison between the observed binary fraction $\tilde{F}_{\rm M}$ and that of the simulated observations. The right-hand panels show the observed binary fraction, and that of the simulated observations. The comparison for only the KO5 dataset is shown in the top panels, that for only the SHT dataset in the middle panels, and that for the combined KO5 and SHT datasets in the bottom panels.

The figure shows that the semi-major axis distribution for intermediate mass stars in Sco~OB2 is of the form $f_a(a) \propto a^{\gamma_a}$ with $\gamma_a=-1.0\pm 0.15$ ($1\sigma$ errors). The observations are therefore consistent with \"{O}pik's law ($\gamma_a = -1$). The observed binary fraction $\tilde{F}_{\rm M}$ is 30\% for the KO5 dataset, 29\% for the SHT dataset, and 30\% for the combined KO5/SHT dataset. These values are only reproduced by models with an intrinsic binary fraction of $\sim 100\%$. 

Note that a second peak near $\gamma_a = -0.1$ is present in the middle panels of Figure~\ref{figure: bestfit_sma}. This means that the observed binary fraction for these values of $\gamma_a$ is consistent with the predictions. However, this value of $\gamma_a$ can be excluded, as the observed angular separation distribution is inconsistent with the observations (left panels). 

In Figure~\ref{figure: bestfit_sma} we additionally show in each panel the results for models with the log-normal period distribution (Equation~\ref{equation: true_duquennoyperiods}). The observed angular separation distribution is consistent with the simulated observations of a model with the log-normal period distribution, which is not surprising, as we have shown in Chapter~4 that these models produce a similar distribution $\tilde{f}_\rho(\rho)$ in the visual regime for nearby OB~associations. Models with a binary fraction of 100\% and the log-normal period distribution over-predict the observed binary fraction. For these models the binary fraction should be of order $75\%-85\%$. However, in Section~\ref{section: true_binaryfraction} we will show that the log-normal period distribution can be excluded, as it under-predicts the number of spectroscopic binaries, even if the model binary fraction is 100\%.

% ============================================================================
% ============================================================================
% ============================================================================
% ============================================================================

\subsubsection{The ambiguity in deriving $f_a(a)/f_P(P)$ and $f_q(q)$} \label{section: fa_vs_fq}

\begin{figure}[!btp]
  \centering
  \includegraphics[width=1\textwidth,height=!]{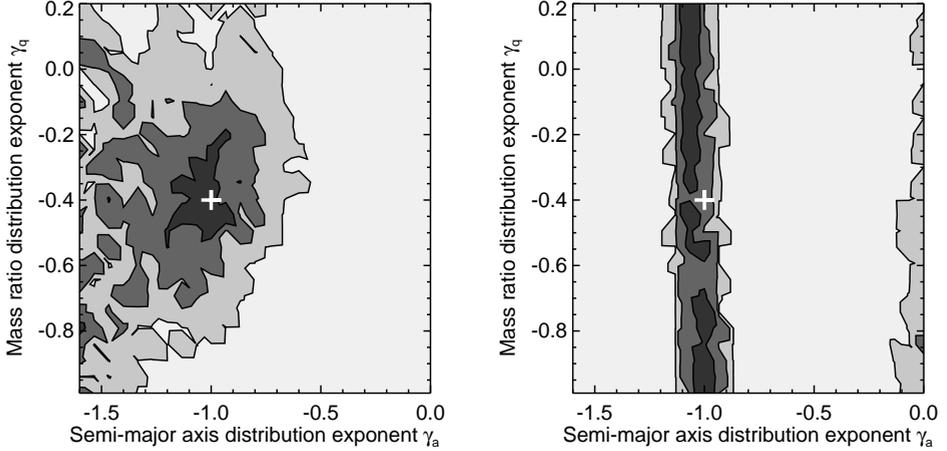}
  \caption{How well do models with a semi-major axis distribution $f_a(a) \propto a^{\gamma_a}$ ({\em horizontal axis}) and a mass ratio distribution $f_q(q) \propto q^{\gamma_q}$ ({\em vertical axis}) correspond to the observations of KO5 and SHT?
This figure is the two-dimensional equivalent of Figures~\ref{figure: bestfit_q} and~\ref{figure: bestfit_sma}. The left-hand panel shows the consistency between the observed distribution $\tilde{f}_{\rho,q}(\rho,q)$ and that of simulated observations, for models with different values of $\gamma_q$ and $\gamma_a$, respectively. The right-hand panel shows the consistency between the observed binary fraction and that of the simulated observations. In each model we adopt a binary fraction of 100\%. Models with a binary fraction significantly smaller than 100\% are inconsistent with the observations. In both panels the darkest colors indicate the best-fitting model. From dark to light, the three contours indicate the $1\sigma$, $2\sigma$, and $3\sigma$ confidence limits for rejection of the model, respectively. The white plus in each panel indicates the values of $\gamma_a$ and $\gamma_q$ of the models that are most consistent with the observations. \label{figure: bestfit_smaq} }
\end{figure}

In the sections above we have constrained $f_a(a)$ and $f_q(q)$ independently. A risk associated with this approach is that one may find a {\em local}, rather than a {\em global} best fit. In this section we demonstrate our choice has been appropriate, and that we have indeed found the best fitting model. 
Although the distributions $f_a(a)$ and $f_q(q)$ are assumed to be independent in our models, there could be a correlation between $a$ and $q$ among the detected binaries. The latter effect is usually important for the analysis of the results of a spectroscopic binary survey \citep[e.g.,][]{kobulnicky2006}, but is of less importance for visual binaries. In a visual binary survey, $a$ is related to the angular separation $\rho$, and $q$ is related to the brightness contrast between primary and companion. As the maximum detectable magnitude difference depends on the separation between the binary components, a correlation between $a$ and $q$ is introduced.

In order to check whether this effect is important for our analysis, we compare the two-dimensional distribution $\tilde{f}_{\rho,q}(\rho,q)$ of the simulated observations of the combined KO5 and SHT datasets with that of the observations. We compare the observed distribution $\tilde{f}_{\rho,q}(\rho,q)$ using the two-dimensional KS test \citep[e.g.,][]{numericalrecipies}, and we compare the observed binary fraction using the Pearson $\chi^2$ test. Figure~\ref{figure: bestfit_smaq} shows the two-dimensional equivalent of Figures~\ref{figure: bestfit_q} and~\ref{figure: bestfit_sma}. The best-fitting values for $\gamma_a$ and $\gamma_q$ in these two-dimensional diagrams is equal to that of the independent derivations in Sections~\ref{section: true_pairingfunction} and~\ref{section: true_smadistribution}, indicating that our independent (and more precise) derivation of $f_a(a)/f_P(P)$ and $f_q(q)$ is correct. Note that for the log-normal period distribution, the above-mentioned ambiguity is not present, as the distribution $f_P(P)$ is fixed. 
Note that in our models we adopt a binary fraction that is independent of semi-major axis and mass ratio. No correlation is introduced by the selection effects between $\tilde{F}_{\rm M}$ on the one hand, and $\tilde{f}_a(a)/\tilde{f}_P(P)$ and $\tilde{f}(q)$ on the other hand. Under the assumption that this independency holds, the binary fraction can be determined after $f_a(a)$ and $f_q(q)$ are derived; a three-dimensional search of the parameter space is not necessary.

% ============================================================================
% ============================================================================
% ============================================================================
% ============================================================================

\subsection{Recovering the eccentricity distribution} \label{section: true_eccentricitydistribution}

The LEV sample consists of 52~confirmed members of Sco~OB2. Among these targets there are 8~SB1s, 7~SB2s, 23~RVVs, and 14~targets with a constant radial velocity (within the errors). For the 23~RVVs it is unknown whether these are truly binary systems, as the radial velocity variations could also result from line profile variability. The observed binary fraction is thus $29\%-73\%$, depending on how many of the RVVs are truly binary systems.

In our modeling of spectroscopic binaries we only determine whether a binary is detected or not. We do not determine whether a binary is detected as SB1, SB2, or RVV. We therefore cannot compare the eccentricity distribution resulting from the simulated observations (SB1s, SB2s, and RVVs) with the observed eccentricity distribution (SB1s and SB2s only). However, it is possible to set several constraints on the eccentricity distribution, using the properties of the observed eccentricity distribution. 

By analyzing Figure~\ref{figure: observed_distributions}, we can rule out an eccentricity distribution of the form $f_{e_0}(e) = \delta(e-e_0)$, where, all binaries have the same eccentricity $e_0$. For associations with $f_{e_0}(e) = \delta(e-e_0)$ the distribution $\tilde{f}_e(e)$ is unbiased, as all binaries have $e_O$. However, an error is associated with each eccentricity measurement, so that $\tilde{f}_e(e)$ is broader than $f_e(e)$. In the spectroscopic binary sample, the error in the eccentricity is of order 0.05, ruling out the best-fitting distribution with $e_0=0.27$ with $\sim 3\sigma$ confidence. 

The distribution better resembles a thermal eccentricity distribution $f_{2e}(e)=2e$ or a flat eccentricity distribution $f_{\rm flat}(e)=1$. For these distributions, the relative fraction of binary systems with $e<0.5$ is:
\begin{equation}
E_{2e} = \frac{\mbox{\# binaries with $e<0.5$}}{\mbox{\# binaries}} = 25\%
\quad  \mbox{and} \quad
E_{\rm flat} = \frac{\mbox{\# binaries with $e<0.5$}}{\mbox{\# binaries}} = 50\% \,.
\end{equation}
In the LEV sample, 13~of the 15~targets with eccentricity measurements have $e<0.5$, and 2~have $e>0.5$. The total number of spectroscopic binaries is uncertain, as we do not know the true nature of the 23 RVVs in reported by LEV. The apparent overabundance of low-eccentricity ($e < 0.5$) systems (see Figure~\ref{figure: observed_distributions}) can partially be explained by selection effects. Highly eccentric systems spend a large fraction of their orbit near apastron, and are therefore more difficult to detect.

In Section~\ref{section: true_smadistribution} we have shown that the observed distribution $\tilde{f}_\rho(\rho)$ and binary fraction $\tilde{F}_{\rm M}$ of visual binaries are consistent with both \"{O}pik's law and the log-normal period distribution. For each of these distributions we predict the number of spectroscopic binaries in the LEV dataset. Our results indicate that LEV would be able to detect $\sim 38\%$ of the binary systems if \"{O}pik's law holds, and $\sim 23\%$ if the log-normal period distribution holds. These fractions are practically independent of the eccentricity distribution; the difference between a flat and a thermal eccentricity distribution is of order $1\%$. 

The detected number of binaries in the LEV dataset also depends on the binary fraction $F_{\rm M}$ of the association. The expected fraction of binary systems with $e < 0.5$ among the observed targets is thus given by:
\begin{eqnarray} \label{equation: e_values_levato1}
  \tilde{E}_{2e,{\rm D\&M}}    & = 23\% \times F_{\rm M} \times E_{2e}       &\leq 5.8\% \\
  \tilde{E}_{2e,{\rm Opik}}    & = 38\% \times F_{\rm M} \times E_{2e}       &\leq 9.5\% \\\label{equation: e_values_levato2}
  \tilde{E}_{{\rm flat, D\&M}} & = 23\% \times F_{\rm M} \times E_{\rm flat} &\leq 11.5\% \\\label{equation: e_values_levato3}
  \tilde{E}_{{\rm flat, Opik}} & = 38\% \times F_{\rm M} \times E_{\rm flat} &\leq 19.0\% \,,\label{equation: e_values_levato4}
\end{eqnarray}
where the limits on the right-hand side are obtained for models with a binary fraction of 100\%. The {\em observed} fraction of binary systems with $e < 0.5$ is not exactly known. For each of the 23~RVVs, we do not know whether it is truly a binary, and if so, we do not know its eccentricity. We therefore consider two extreme cases for $\tilde{E}_{\rm LEV}$: (1) none of the RVVs have $e<0.5$ (i.e., they are either spectroscopically single or are binaries $e>0.5$), and (2) all RVVs are binary systems with $e<0.5$. In these extreme cases, $ \tilde{E}_{\rm LEV} $ is constrained by
\begin{eqnarray}
  \tilde{E}_{\rm LEV} & = \frac{\mbox{number~of~binaries~with~$e<0.5$}}{\mbox{52~targets}} & =  25\%-73\% \,.
\end{eqnarray}
The observed eccentricity distribution is inconsistent with the predictions in Equations~\ref{equation: e_values_levato1}--\ref{equation: e_values_levato3}. The models with the log-normal period distribution or the thermal eccentricity distribution do not reproduce the observations; these under-predict the number of binaries in the LEV sample with $e<0.5$, even if {\em all} RVVs detected by LEV are spurious, and the model binary fraction is 100\%. The model with \"{O}pik's law, a flat eccentricity distribution, and a binary fraction of 100\% (Equation~\ref{equation: e_values_levato3}) is consistent with the observations, but only if most of the RVV candidate binaries are in reality single. 

Note that the discussion on the eccentricity distribution is constrained only by a small number of short-period binaries with a massive primary. The results presented above do not necessarily apply to the whole binary population in Sco~OB2. Further radial velocity and astrometric surveys among intermediate- and low-mass members of Sco~OB2 are necessary to further characterize the eccentricity distribution.

% ============================================================================
% ============================================================================
% ============================================================================
% ============================================================================

\subsection{The binary fraction in Sco~OB2} \label{section: true_binaryfraction}

In the previous sections we have constrained the pairing function, mass ratio distribution, semi-major axis distribution, and the eccentricity distribution. From the analysis of these properties we conclude that the binary fraction among intermediate mass stars in Sco~OB2 must be close to 100\%. 

Table~\ref{table: fm_inferred} shows for various models the expected fraction of binaries that is detected with the different observing techniques. For each main model we vary the lower and upper limits of the size of the orbit between the extremes derived in Sections~\ref{section: recover_amin} and~\ref{section: recover_amax}. Each of these models contains a binary fraction of 100\%. The predicted binary fraction for each technique therefore provides an upper limit, under our assumptions. Models that predict a significantly lower binary fraction than the observed binary fraction, can therefore be excluded.

The observed binary fraction based on the imaging surveys is consistent with \"{O}pik's law, as well as with  the log-normal period distribution. The observed binary fraction based on the radial velocity surveys, on the other hand, is inconsistent with the log-normal period distribution. Models with the log-normal period distribution under-predict the number of spectroscopic binaries, even if the model binary fraction is 100\%. 

If {\em all} RVV binaries in the LEV dataset are single, the predicted binary fraction of $23\pm 1\%$ is smaller than the observed binary fraction of $29\pm5\%$, although a difference between these fractions can only be stated with $1\sigma$ confidence. However, it is unrealistic that {\em all} RVV binaries are single stars. Many of these are binaries, for which too few radial velocity measurements are available to accurately fit a radial velocity curve to the data. In the more realistic case that 50\% or 100\% of the RVVs are binaries, the log-normal period distribution can be excluded with $4\sigma$ and $5\sigma$ confidence, respectively.

The {\em Hipparcos} selection effects for the (X), (O), and (G) binaries are not strict enough. This is because in our models we consider {\em all} binaries with the properties listed in Table~\ref{table: true_hipparcosbiases} as astrometric binaries. In reality, however, only a subset of these would have been marked as an astrometric binary by {\em Hipparcos}. The predicted number of (C) binaries on the other hand, is well-modeled. For models with a binary fraction of 100\% is in good agreement with the observations.

Using the observed binary fraction of each survey, and adopting \"{O}pik's law, we find that a binary fraction of 100\% best describes our observations. Models with a smaller binary fraction are less consistent with the visual binary fraction, i.e., KO5, SHT, and HIP-(C). An inspection of these binary fractions and their associated statistical errors indicates that the binary fraction among intermediate mass stars in Sco~OB2 must be larger than $85\%$ at the $2\sigma$ confidence level, and larger than $70\%$ at the $3\sigma$ confidence level. The spectroscopic binary fractions cannot be used to constrain the binary fraction further, as we do not know which of the RVV targets are truly binaries. The astrometric binary fraction merely provides a lower limit to the binary fraction, that is consistent with the derived binary fraction based on the other datasets.

In Section~\ref{section: true_pairingfunction} we concluded that the pairing function is either PCP-I, PCP-II, or PCP-III. Due to the lack of a large dataset of low-mass binaries in Sco~OB2, we are unable to discriminate between these three pairing functions. If pairing function PCP-II holds (e.g., if all planetary companions are {\em not} considered as companions), the binary fraction is $\sim 100\%$ among intermediate mass stars, $\sim 93\%$ among solar-type stars, and $\sim 75\%$ among low-mass stars. On the other hand, if either the pairing function PCP-I or PCP-III holds, the binary fraction in Sco~OB2 is for all spectral types equal to the binary fraction of intermediate mass stars, i.e., $\sim 100\%$.

\begin{table}
  \small
  \begin{tabular}{cc}
  \begin{tabular}{ll ccc}
    \hline
    \multicolumn{2}{l}{$f_a(a)\propto a^{\gamma_a}$, $F_{\rm M} = 100\%$} & $\gamma_a=-1.0$ & $-1.05$      & $-1.1$   \\
    \hline
    $a_{\rm min}$ & $a_{\rm max}$    & $\tilde{F}_{\rm M}$  & $\tilde{F}_{\rm M}$  & $\tilde{F}_{\rm M}$  \\
    \hline
    \multicolumn{5}{l}{{\bf (A1)} KO5 dataset} \\
    \hline
    2~R$_\odot$   & $10^6$~R$_\odot$ & 29\% &  22\% & 18\%  \\
    2~R$_\odot$   & $10^7$~R$_\odot$ & 22\% &  21\% & 16\%   \\
    10~R$_\odot$  & $10^6$~R$_\odot$ & 32\% &  26\% & 21\%  \\
    10~R$_\odot$  & $10^7$~R$_\odot$ & 25\% &  24\% & 22\%                \\
    \multicolumn{2}{l}{Observations} & \multicolumn{3}{c}{$\longleftarrow${\bf 33\%}$\longrightarrow$}                  \\
    \hline
    \multicolumn{5}{l}{{\bf (A2)} SHT dataset} \\
    \hline
    2~R$_\odot$   & $10^6$~R$_\odot$ & 27\%  &  22\% & 18\%               \\
    2~R$_\odot$   & $10^7$~R$_\odot$ & 24\%  &  21\% & 17\%                \\
    10~R$_\odot$  & $10^6$~R$_\odot$ & 29\%  &  26\% & 20\%              \\
    10~R$_\odot$  & $10^7$~R$_\odot$ & 27\%  &  23\% & 20\%                \\
    \multicolumn{2}{l}{Observations} &  \multicolumn{3}{c}{$\longleftarrow${\bf 29\%}$\longrightarrow$}                \\
    \hline
    \multicolumn{5}{l}{{\bf (A3)} LEV dataset} \\
    \hline
    2~R$_\odot$   & $10^6$~R$_\odot$ & 43\%    &  50\% & 57\%          \\
    2~R$_\odot$   & $10^7$~R$_\odot$ & 37\%    &  45\% & 52\%           \\
    10~R$_\odot$  & $10^6$~R$_\odot$ & 39\%    &  42\% & 51\%          \\
    10~R$_\odot$  & $10^7$~R$_\odot$ & 31\%    &  35\% & 42\%          \\
    \multicolumn{2}{l}{SB1/SB2 only}  &  \multicolumn{3}{c}{$\longleftarrow${\bf $>$29\%}$\longrightarrow$}                \\
    \multicolumn{2}{l}{SB1/SB2/RVV } &  \multicolumn{3}{c}{$\longleftarrow${\bf $<$73\%}$\longrightarrow$}                 \\
    \hline
    \multicolumn{5}{l}{{\bf (A4)} BRV dataset} \\
    \hline
    2~R$_\odot$   & $10^6$~R$_\odot$ & 45\%    &  52\% & 60\%            \\
    2~R$_\odot$   & $10^7$~R$_\odot$ & 39\%    &  45\% & 54\%            \\
    10~R$_\odot$  & $10^6$~R$_\odot$ & 41\%    &  44\% & 52\%         \\
    10~R$_\odot$  & $10^7$~R$_\odot$ & 35\%    &  37\% & 46\%            \\
    \multicolumn{2}{l}{BRV analysis -- RVV        } &  \multicolumn{3}{c}{$\longleftarrow${\bf $<$60\%}$\longrightarrow$}              \\
    \multicolumn{2}{l}{SB1/SB2/RVV       } &  \multicolumn{3}{c}{$\longleftarrow${\bf $<$65\%}$\longrightarrow$}             \\
    \hline
    \multicolumn{5}{l}{{\bf (A5)} HIP astrometric binaries} \\
    \hline
    2~R$_\odot$   & $10^6$~R$_\odot$ & $<$35\%     &  $<$41\% & $<$46\%           \\
    2~R$_\odot$   & $10^7$~R$_\odot$ & $<$31\%     &  $<$38\% & $<$44\%           \\
    10~R$_\odot$  & $10^6$~R$_\odot$ & $<$34\%     &  $<$37\% & $<$45\%           \\
    10~R$_\odot$  & $10^7$~R$_\odot$ & $<$28\%     &  $<$33\% & $<$38\%            \\
    \multicolumn{2}{l}{(X)/(O)/(G) binaries    }  &  \multicolumn{3}{c}{$\longleftarrow${\bf 4\%}$\longrightarrow$}                   \\
    \multicolumn{2}{l}{(X)/(O)/(G)/(S) binaries} &  \multicolumn{3}{c}{$\longleftarrow${\bf 9\%}$\longrightarrow$}                   \\
    \hline
    \multicolumn{5}{l}{{\bf (A6)} HIP photometric binaries} \\
    \hline
    2~R$_\odot$   & $10^6$~R$_\odot$ & 14\%    &  12\% & 9\%           \\
    2~R$_\odot$   & $10^7$~R$_\odot$ & 13\%    &  11\% & 9\%             \\
    10~R$_\odot$  & $10^6$~R$_\odot$ & 16\%    &  14\% & 11\%           \\
    10~R$_\odot$  & $10^7$~R$_\odot$ & 13\%    &  13\% & 10\%             \\
    \multicolumn{2}{l}{(C) binaries} &  \multicolumn{3}{c}{$\longleftarrow${\bf 15\%}$\longrightarrow$}                  \\
    \hline
  \end{tabular}
  &
  \begin{tabular}{ll c}
    \hline
    \multicolumn{3}{l}{$f_P(P)$ log-normal, $F_{\rm M} = 100\%$} \\
    \hline
    $P_{\rm min}$ & $P_{\rm max}$    & $\tilde{F}_{\rm M}$ \\
    \hline
    \multicolumn{3}{l}{{\bf (P1)} KO5 dataset} \\
    \hline
    12~hour& 0.15~Myr       & 40\%              \\
    12~hour& 4~Myr          & 37\%              \\
    1~day  & 0.15~Myr       & 40\%               \\
    1~day  & 4~Myr          & 34\%              \\
    \multicolumn{2}{l}{Observations} & {\bf 30\% }               \\
    \hline
    \multicolumn{3}{l}{{\bf (P2)} SHT dataset} \\
    \hline
    12~hour& 0.15~Myr       & 40\%              \\
    12~hour& 4~Myr          & 36\%             \\
    1~day  & 0.15~Myr       & 36\%              \\
    1~day  & 4~Myr          & 34\%               \\
    \multicolumn{2}{l}{Observations} & {\bf 29\% }                 \\
    \hline
    \multicolumn{3}{l}{{\bf (P3)} LEV dataset} \\
    \hline
    12~hour& 0.15~Myr       & 23\%              \\
    12~hour& 4~Myr          & 22\%              \\
    1~day  & 0.15~Myr       & 24\%              \\
    1~day  & 4~Myr          & 22\%             \\
    \multicolumn{2}{l}{SB1/SB2 only   }  & {\bf $>$29\% }                \\
    \multicolumn{2}{l}{SB1/SB2/RVV} & {\bf $<$73\% }                 \\
    \hline
    \multicolumn{3}{l}{{\bf (P4)} BRV dataset} \\
    \hline
    12~hour& 0.15~Myr       & 27\%             \\
    12~hour& 4~Myr          & 23\%              \\
    1~day  & 0.15~Myr       & 28\%               \\
    1~day  & 4~Myr          & 25\%             \\
    \multicolumn{2}{l}{BRV analysis -- RVV    } & {\bf $<$60\% }             \\
    \multicolumn{2}{l}{SB1/SB2/RVV   } & {\bf $<$65\% }             \\
    \hline
    \multicolumn{3}{l}{{\bf (P5)} HIP astrometric binaries} \\
    \hline
    12~hour& 0.15~Myr       & $<$27\%              \\
    12~hour& 4~Myr          & $<$24\%             \\
    1~day  & 0.15~Myr       & $<$27\%             \\
    1~day  & 4~Myr          & $<$26\%                \\
    \multicolumn{2}{l}{(X)/(O)/(G) binaries     }  & {\bf 4\% }                  \\
    \multicolumn{2}{l}{(X)/(O)/(G)/(S) binaries } & {\bf 9\% }                  \\
    \hline
    \multicolumn{3}{l}{{\bf (P6)} HIP photometric binaries} \\
    \hline
    12~hour& 0.15~Myr       & 19\%                \\
    12~hour& 4~Myr          & 20\%                \\
    1~day  & 0.15~Myr       & 20\%                \\
    1~day  & 4~Myr          & 18\%                \\
    \multicolumn{2}{l}{(C) binaries} & {\bf 15\%}                \\
    \hline
  \end{tabular}
  \\
  \end{tabular}
    \caption{The observed binary fraction $\tilde{F}_{M}$ depends on the observing technique, on $F_{\rm M}$, and on $f_a(a)$ or $f_P(P)$. In this table we list for various association models the value of $\tilde{F}_{M}$ in the simulated observations. Parts (A1)--(A6) of the table list the results for models $f_a(a) \propto a^{\gamma_a}$
and parts (P1)--(P6) for models with the log-normal period distribution. Each model has $F_{\rm M}=100\%$, so that the predicted values for $\tilde{F}_{M}$ are upper limits. The corresponding {\em observed} values of $\tilde{F}_{M}$ are indicated in boldface. The observed values of $\tilde{F}_{M}$ for visual binaries suggest a binary fraction of $\sim 100\%$ for the models in the left column, and $75-100\%$ for the models in the right column. However, the models with the log-normal period distribution under-predict the value of $\tilde{F}_{M}$ for spectroscopic binaries, even for models with $F_{\rm M}=100\%$. The binary fraction among intermediate mass stars in Sco~OB2 must therefore be $\sim 100\%$, and the semi-major axis distribution similar to \"{O}pik's law ($\gamma_a=-1$). 
 \label{table: fm_inferred} }
\end{table}

% ============================================================================
% ============================================================================
% ============================================================================
% ============================================================================

\section{Discussion} \label{section: true_discussion}

% ============================================================================
% ============================================================================
% ============================================================================
% ============================================================================

\subsection{A comparison between KO5 and KO6}

In their follow-up program, KO6 re-observed 22~targets of the 199~targets in the sample of KO5. Due to their different instrument and observing strategy, three new close companions are detected, while they remained unresolved in the survey of KO5. Five wide companions are detected with KO5 (in mosaic-mode), but they could not be detected by KO6, as they have a smaller field of view. 
Here we illustrate that the results for the simulated observations are similar, supporting our correct modeling of the selection effects.
The number of wide KO5 binaries that is expected to be missing in the KO6 dataset is of order $5\times 22/199 = 0.55$. This number is in agreement with the one ``missing'' component of HIP80142, which was detected by KO5, but is outside the field of view in KO6 (the other ``missing'' object is a faint background star near HIP81949). The number of newly detected close companions in the KO6 observations is 3, while the expected number is $5\times 22/199 = 0.33$. The observed number of newly found close KO6 binaries is statistically in agreement with the predictions, although somewhat higher.

% ============================================================================
% ============================================================================
% ============================================================================
% ============================================================================

\subsection{Comparison with \cite{heacox1995}}

\cite{heacox1995} derived the mass ratio distribution from the LEV dataset. He used the observations of all 22~binaries with spectroscopic elements. Using three different techniques the mass ratio distribution for the LEV sample is derived, resulting in a mean mass ratio $\langle q \rangle \approx 0.265$ with a standard derivation $\sigma_q \approx 0.037$. 
In our analysis we adopt a mass ratio distribution of the form $f_q(q) \propto q^{\gamma_q}$. If this distribution is adopted, the mean mass ratio is $\langle q \rangle = (\gamma_q+1)/(\gamma_q+2)$, and the standard deviation is $\sigma_q = (\gamma_q+1)/(\gamma_q+3) - \langle q \rangle^2$. Our best-fitting solution has $\gamma_q = -0.40$, so that $\langle q \rangle = 0.375$ and $\sigma_q=0.300$. Especially the latter value is obviously larger than the value derived by \cite{heacox1995}. However, for $\gamma_q=-0.5$, which is consistent with our observations at the $1\sigma$ level, one gets $\langle q \rangle = 0.33$ and $\sigma_q=0.08$. The latter values, based on our analysis of {\em visual} binaries, are in good agreement with the results of \cite{heacox1995}. Note that in the latter paper the mass ratio distribution is derived from a much smaller, but {\em independent} sample. \cite{heacox1995} derives the mass ratio distribution for {\em spectroscopic} binaries, while in our analysis we derive the distribution from observations of {\em visual} binaries.

% ============================================================================
% ============================================================================
% ============================================================================
% ============================================================================

\subsection{Background stars}

\begin{table}
  \small
%  \begin{tabular}{p{0.6cm} p{0.3cm}p{0.2cm} cc cc cc p{1.4cm}}
  \begin{tabular}{l cc cc cc cc l}
    \hline
    \hline
    Group            & $l$ & $b$                 & N$_{\rm KO5}$ & N$_{\rm KO5}$ & N$_{\rm KO6}$ & N$_{\rm KO6}$ & N$_{\rm SHT}$ & N$_{\rm SHT}$ & $C$ (arcsec$^{-2}$) \\
    \multicolumn{3}{l}{$K_S$ limit (mag)}        & $< 12$& $<18$ & $<12$ & $<18$ & $<12$ & $<18$& \\ 
    \hline
    US               & $352^\circ$ & $20^\circ$  & 0.002 & 0.19  & 0.001 & 0.10  & 0.002 & 0.15 &  $1.93 \times 10^{-9}$  \\
    UCL              & $328^\circ$ & $13^\circ$  & 0.010 & 0.87  & 0.006 & 0.47  & 0.008 & 0.67 &  $8.71 \times 10^{-9}$  \\
    LCC              & $299^\circ$ & $6^\circ$   & 0.016 & 1.35  & 0.009 & 0.73  & 0.013 & 1.05 &  $13.5 \times 10^{-9}$  \\
    GP               & $300^\circ$ & $0^\circ$   & 0.057 & 4.78  & 0.031 & 2.59  & 0.044 & 3.70 &  $48.0 \times 10^{-9}$  \\
    \hline
    \hline
  \end{tabular}
  \caption{The number of background stars expected {\em per field of view} for each of the three imaging surveys discussed in this paper, according our model. The field of view size is 361.2~arcsec$^2$ for KO5, 196.0~arcsec$^2$ for KO6, and 280.1~arcsec$^2$ for SHT. We list the results for three different pointings: to the centers of the three subgroups US, UCL, and LCC, and to the intersection of LCC with the Galactic plane. Columns~4, 6, and 8 list the number of background stars brighter than $K_S=12$~mag {\em per field of view}. Columns~5, 7, and 9 list the number of background stars brighter than $K_S=18$~mag {\em per field of view}. The last column lists for each group the value of the normalization constant $C$ in Equation~\ref{equation: backgroundstarsformula}. \label{table: backgroundstarcounts}}
\end{table}

In imaging surveys it is not always clear whether a secondary is a physical companion or a background star. In our models for the association this is obviously not a problem, but in practice it is. In this section we discuss briefly how background stars can affect the interpretation of the results of a visual binary survey.
We use the prescription derived in Chapter~3 for the number of background stars brighter than $K_S$ and with an angular separation smaller than $\rho$, as a function of $K_S$ and $\rho$:
\begin{equation} \label{equation: backgroundstarsformula}
  N(K_S,\rho) = C \cdot 10^{\gamma \cdot K_S} \cdot  A(\rho) \,,
\end{equation}
where $A(\rho)$ is the enclosed area in the field of view within a radius $\rho$, $\gamma=0.32\pm 0.01$~mag$^{-1}$, and $C$ is a constant. The $K_S$ dependency was derived by KO6 using the Besan\c{c}on model of the Galaxy \citep{besancon}, and the normalization constant $C$ is determined using the sky-observations of SHT. Table~\ref{table: backgroundstarcounts} lists for each of the three visual surveys (KO5, KO6, and SHT) the expected number of background stars {\em per field of view}. We list the number of background stars with $K_S < 12$~mag and with $K_S < 18$~mag for four different pointings in the Sco~OB2 region.

\begin{figure}[!btp]
  \centering
  \includegraphics[width=1\textwidth,height=!]{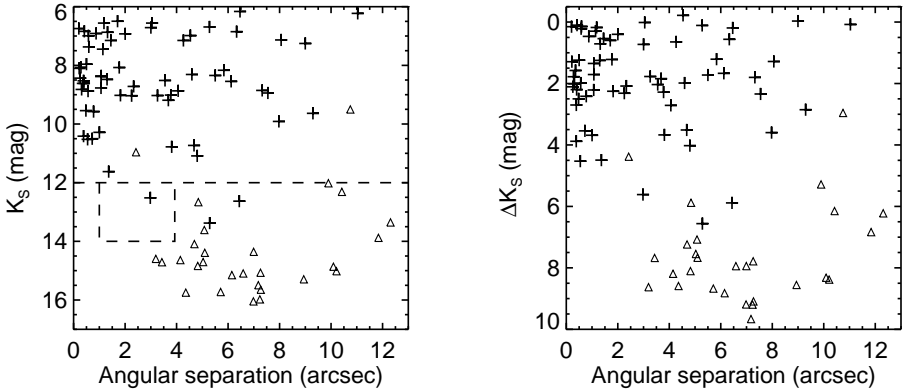}  \caption{The simulated KO5 data for a realization of the best-fitting model if Sco~OB2 (cf. Figure~2.3). The plots show physical companions (plusses) and background stars (triangles). The horizontal dashed line indicates the criterion used by KO5 (based on the analysis of SHT) to statistically separate companion stars and background stars. The simulated sample consists of 199~targets, next to which we detect 64~companion stars and 28~background stars. Although two companions have $K_S > 12$~mag, and two~background stars have $K_S <12$~mag in this example, the majority of the secondaries is correctly classified if $K_S=12$~mag is used to separate companions and background stars.
\label{figure: detlim_plot_adonis} }
\end{figure}

For the $N$ targets in our model, we assume that $N/3$ targets are in each of the regions US and UCL, and that $N/6$ targets are in each of the regions LCC and the Galactic plane (GP). After the number of background stars $N_{\rm bg}$ is determined, each background star is assigned a $K_S$ magnitude randomly drawn from the generating distribution corresponding to Equation~\ref{equation: backgroundstarsformula}. Each companion star is assigned a random position in the field of a random target star. Finally, the angular separation and magnitude difference are drawn. For each background star we then decide whether it would be detected in the simulated observations, i.e., whether it satisfies the separation constraint and the contrast constraint. Figure~\ref{figure: detlim_plot_adonis} shows the result for one of our models. Note how well the results in this figure resemble those of the KO5 observations in Figure~2.3.

SHT consider all new companions with $K_S > 12$~mag, $J > 13$~mag, or $J-K_S > 1$~mag as background stars. KO5 consider all new companions with $K_S>12$~mag as background stars. The expected number of background stars with $K_S < 12$~mag in the KO5 and SHT datasets can now be calculated. Among the 199~targets in the KO5 dataset, we expect $2-3$ background stars with $K_S < 12$~mag. Note that, of the background stars with $K_S < 12$~mag, most have a magnitude comparable to $K_S=12$~mag. In their follow-up study, KO6 have performed multi-color observations of several doubtful candidate companions, and identified six of these as possible background stars. Among the 80~confirmed Sco~OB2 members in the SHT dataset we expect $\sim 1$ background star with  $K_S < 12$~mag.
The expected number of bright background stars is small, and many background stars have been removed by magnitude criteria and the follow-up study of KO6. The possible presence of background stars among the candidate binaries has a negligible effect on the conclusions of our analysis.

%
% ============================================================================
% ============================================================================
% ============================================================================
% ============================================================================

\subsection{A brown dwarf desert in Sco~OB2?}

In KO6 we presented evidence for a possible brown dwarf desert in Sco~OB2. Our analysis was based on the companions among A and late-B stars in Sco~OB2, in the angular separation range $1''-4''$. We only considered the companions brighter than $K_S=14$~mag, as we are incomplete for fainter companions. 

\begin{figure}[!btp]
  \centering
  \includegraphics[width=0.5\textwidth,height=!]{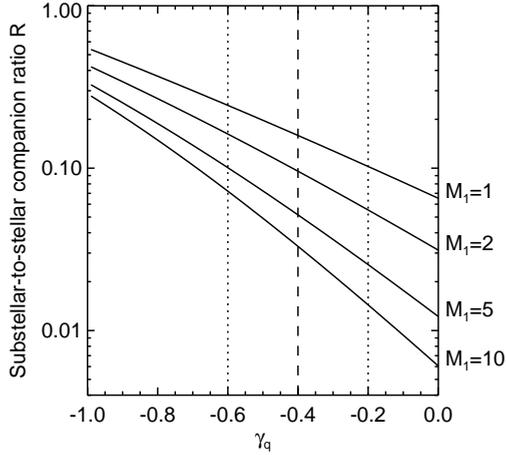}
  \caption{The substellar-to-stellar companion ratio $R$ (Equation~\ref{equation: browndwarfratio}) versus the exponent $\gamma_q$ of the mass ratio distribution. From top to bottom the solid curves represent the dependency $(\gamma_q,R)$ for a sample of binaries with primary mass $M_1=1$, 2, 5, and 10~M$_\odot$. The best-fitting value of $\gamma_q$ and its $1\sigma$ errors are indicated with the dashed line and the dotted lines, respectively. \label{figure: browndwarfs_r_ratio} }
\end{figure}

In our current analysis we find that the observed mass ratio distribution among early-type stars in Sco~OB2 is consistent with an intrinsic mass ratio distribution of the form $f_q(q)\propto q^{\gamma_q}$ with $\gamma_q = -0.4 \pm 0.2$. For a power-law mass ratio distribution, the substellar-to-stellar companion ratio for a sample of binary systems with primary mass $M_1$ is given by 
\begin{equation} \label{equation: browndwarfratio}
R_{M_1}(\gamma_q) = \frac{ \mbox{number of brown dwarf companions} }{ \mbox{number of stellar companions} }
  = \frac{ 1 - \left(\displaystyle\frac{0.02~\mbox{M}_\odot}{0.08~\mbox{M}_\odot}\right)^{\gamma_q+1} }{  \left(\displaystyle\frac{M_1}{0.08~\mbox{M}_\odot}\right)^{\gamma_q+1} - 1 }
  \,,
\end{equation}
where $\gamma_q \neq -1$. Figure~\ref{figure: browndwarfs_r_ratio} shows $R_{M_1}$ as a function of $\gamma_q$ for various values of $M_1$. The figure illustrates that for binaries with A and B-type primaries, $R_{\rm AB}(-0.4)$ is of order 0.06. This means that among these primaries, about 6\% of the companions are substellar, and 94\% are stellar. The interpretation of this small brown dwarf companion fraction and low substellar-to-stellar companion ratio was discussed in detail in Chapter~3. If star formation indeed results in a mass ratio distribution of the form $f_q(q)\propto q^{\gamma_q}$ with $\gamma_q = -0.4 \pm 0.2$, the brown dwarf desert (in terms of $R$) is a natural outcome the of star formation process. In the latter case, there is no need to invoke an embryo ejection scenario to explain the low number of brown dwarfs.

%

% ============================================================================
% ============================================================================
% ============================================================================
% ============================================================================

\subsection{The primordial binary population in Sco~OB2} \label{section: primordialbinarypopulation}

Sco~OB2 is a young OB~association (5--20~Myr) with a very low stellar density ($\sim 0.1~{\rm M}_\odot\,{\rm pc}^{-3}$);  one expects that stellar and dynamical evolution have only mildly altered properties of the binary population. In this section we study which binary systems in Sco~OB2 may have changed one or more of its parameters since the time of formation. We first consider whether stellar (and binary) evolution will have affected the binary population, and subsequently investigate the importance of dynamical evolution.

For stellar populations with an age less than $\la 20$~Myr, the fraction of binaries of which the properties have changed due to stellar evolution is small. The maximum age $\tau_{\rm MS}$ of a main sequence star as a function its mass $M$ can be expressed as
\begin{equation}
  \log \left( \frac{M}{\mbox{M}_\odot} \right) \approx 1.524 - 0.370 \log \left( \frac{\tau_{\rm MS}}{\mbox{Myr}} \right)
\end{equation}
\citep{maeder1988}. This means that stars more massive than $\sim 18$~M$_\odot$ (B0\,V) in US, and stars more massive than $\sim 11$~M$_\odot$ (B2\,V) in UCL and LCC have evolved away from the main sequence. Several of the binary systems with an initially more massive primary are disrupted; their original components will have obtained a space velocity different from the mean space velocity of the parent association (runaway stars); other remain bound.  The properties of compact binary systems including at least one star with a higher initial mass may have changed due to binary evolution. For $\ga 99.9\%$ of the binaries in Sco~OB2, however, both components have an initial mass (much) less than those mentioned above; stellar evolution will not have affected these binaries (see Table~\ref{figure: sos_massdistributions}). 
The change in the binary population due to stellar evolution can therefore be neglected. 
The change due to dynamical evolution is more complicated to quantify. Below we calculate which binary systems are affected by dynamical evolution, at the {\em current} stellar density of the association. After that we briefly discuss the consequences of the fact that the subgroups of Sco~OB2 may have been denser in the past.

The subgroups of Sco~OB2 are more or less spherical, with radii of order $R = 20-30$~pc (see Table~\ref{table: subgroups}). The structure and kinematics of Sco~OB2 have been studied by  \cite{debruijne1999}, who finds that the one-dimensional internal velocity dispersion for each of the subgroups  is $\sigma_v \la 1.0-1.5$~km\,s$^{-1}$. The crossing time $\tau_{\rm cross}$ of each subgroup is thus $R/\sigma_v \ga 16$~Myr. 
\cite{preibisch2002} find that the US subgroup contains approximately 2500~members more massive than 0.1~M$_\odot$. The total number of members $N$ in the substellar regime depends on the unknown value of $\alpha$ in Equation~\ref{equation: preibischimf}. For reasonable value of $\alpha$, about 20\% of the members have a mass smaller than 0.1~M$_\odot$, so that $N\approx 3200$. Adopting pairing function PCP-III and a binary fraction of 100\%, the total mass of US is of order 2000~M$_\odot$ (see Table~\ref{table: totalmass_differences}). We assume that the UCL and LCC subgroups contain a similar number of members. The mass density of each subgroup is therefore of order 0.03~M$_\odot\,\mbox{pc}^{-3}$, corresponding to 0.048~systems (or 0.098~individual stars) per cubic parsec.

An estimate for the relaxation time, i.e., the typical timescale at which a system of $N$ {\em single} stars has ``forgotten'' the initial conditions, is 
\begin{equation}
  \tau_{relax} \approx \frac{N}{8 \ln N} \times \tau_{\rm cross} \approx 1~\mbox{Gyr} 
\end{equation}
\citep{binneytremaine}. As the cross-section for binary systems is significantly larger than that of single stars, they interact more frequently, so that in reality the timescale of dynamical evolution is shorter than 1~Gyr. Whether a binary system experiences a strong encounter, depends not only on the properties of the association, but also on the size (more specifically, the binding energy) of the binary systems. A binary system is often classified as {\em hard} or {\em soft}, depending on whether it is likely to experience a strong encounter within a relaxation time. A binary system is called ``hard'' if its orbital energy $E_{\rm orb}$ is significantly larger than the mean kinetic energy $\langle K \rangle$ of the surrounding stars: $E_{\rm orb} > 3\langle K \rangle$ \citep{heggie1975,hills1975}; otherwise a binary system is called ``soft''. The orbital energy of a binary system is given by
\begin{equation}
E_{\rm orb} = \frac{qGM_1^2}{2a}
\end{equation}
where $q$ is the mass ratio, $G$ the gravitational constant, $M_1$ the primary mass, and $a$ the semi-major axis of the binary system. The mean kinetic energy is given by 
\begin{equation}
\langle K \rangle = \tfrac{1}{2} \langle M_T \rangle \sigma_v^2 \,,
\end{equation}
where $ \langle M_T \rangle $ is the average mass of a field object, and $\sigma_v$ its velocity dispersion. Assuming a binary fraction of 100\%, the average mass of a field object (i.e., another binary system in the association), is $ \langle M_T \rangle \approx 0.3$~M$_\odot$ (see Table~\ref{table: totalmass_differences}). The velocity dispersion is $\sigma_v \la 1.0-1.5$~km\,s$^{-1}$ \citep{debruijne1999}. The hard/soft boundary $a_{\rm hs}$ for a binary star with primary mass $M_1$ and mass ratio $q$ is thus given by $E_{\rm orb} = 3\langle K \rangle$, i.e.,
\begin{equation}
  a_{\rm hs} = \frac{G}{3\langle M_T \rangle \sigma_v^2}\, q M_1^2 = 1.0\times 10^3 \ q \left(\frac{M_1}{\mbox{M}_\odot} \right)^2~\mbox{AU} = 2.1\times 10^5 \ q \left(\frac{M_1}{\mbox{M}_\odot} \right)^2~\mbox{R}_\odot\,.
\end{equation}
The latter value is in good agreement with the widest known binaries in Sco~OB2 (see \S~\ref{section: recover_amax}), which all have a primary mass of order 3~M$_\odot$. The datasets studied in this paper contain mainly A and B type members of Sco~OB2, which have $M_1 \ga 1.5$~M$_\odot$. Among these binaries, about 90\% is hard, while the widest 10\% of the binaries is soft. The population of A and B stars must therefore be close to primordial, given the young age of the association. 
On the other hand, among the lower-mass binaries (0.02~M$_\odot \la M_1 \la 0.5$~M$_\odot$) about 50\% of the binaries is hard. Although about half of the binaries (i.e., the soft low-mass binaries) is expected to experience a close encounter in a relaxation time, this population is unlikely to have changed significantly, due to the young age of the association.
The majority of binaries, in particular the binaries studied in this paper, are unlikely to have experienced a strong encounter during the lifetime of the association, assuming that the density did not change over time. 

In the past the density of Sco~OB2 could have been much higher, but the simulations of \cite{vandenberk2006} suggest that it is unlikely that many binaries are affected. In the latter paper, simulations of small ($N =100$) and initially dense ($\sim 10^5$ systems per cubic parsec) expanding star clusters are presented. Their simulations include triple systems for which the period of the outer orbit reaches up to 1000~years. Although the initial stellar density in these simulations is more than a million times higher than the current density of Sco~OB2, the properties of the binary population do not change significantly within 20~Myr. As about 60\% of the binary systems in Sco~OB2 has an orbital period smaller than 1000~years, these binaries are expected not to have changed significantly since the birth of the association, even if the density of Sco~OB2 was orders of magnitude larger at the time of formation.

The statements above provide estimates, and have to be validated with large N-body simulations of expanding OB~associations. By varying the initial conditions, evolving each simulated association for 5--20~Myr, and comparing the outcome with the observations, the primordial binary population can be recovered. The latter technique is referred to as inverse dynamical population synthesis \citep[see][]{kroupa1995a,kroupa1995c}. If the primordial binary population is indeed very similar to the current binary population, it is likely that {\em all} stars are formed in a binary or multiple system.

% ====================================================================
% ====================================================================
% ====================================================================
% ==INTRODUCTION======================================================
% ====================================================================
% ====================================================================
% ====================================================================

\section{Summary and outlook} \label{section: summary}

We have recovered the intrinsic binary population in the nearby OB~association Sco~OB2, with the aim of finding the primordial binary population, that is present just after star formation. We have performed Monte Carlo simulations, and compared for each association model the {\em simulated observations} with the results of surveys for visual, spectroscopic, and astrometric binary systems in Sco~OB2. Our main results of our study are the following:
\begin{itemize}\addtolength{\itemsep}{-0.5\baselineskip}
  \item[--] The binary fraction among A and B stars in Sco~OB2 is {\em at least} 70\% ($3\sigma$ confidence). The best agreement with the observations is obtained for models with a binary fraction of 100\%.
  \item[--] The semi-major axis distribution, and the observed number of visual, spectroscopic, and astrometric binaries are consistent with \"{O}pik's law, as $f_a(a) \propto a^{\gamma_a}$ with $\gamma_a= -1.0 \pm 0.15$. The log-normal period distribution found by \cite{duquennoy1991} is excluded, as it significantly under-predicts the number of spectroscopic binaries (even with a binary fraction of 100\%), and the ratio between visual and spectroscopic binaries.
  \item[--] The pairing properties of the binaries in Sco~OB2 are consistent with pairing functions PCP-I, PCP-II, and PCP-III. Random pairing (RP) and primary-constrained random pairing (PCRP) from the mass distribution are excluded with high confidence.
  \item[--] The mass ratio distribution has the form $f_q(q) \propto q^{\gamma_q}$ , with $\gamma_q = -0.4 \pm 0.2$.
  \item[--] Sco~OB2 is a young association with a low stellar density. Stellar and binary evolution has only affected the binaries with components more massive than $\sim 11$~M$_\odot$ in US, and  $\sim 18$~M$_\odot$ in UCL and LCC. Dynamical evolution has only mildly affected the binary population. The current binary population of Sco~OB2, as described above, is expected to be very similar to the primordial binary population of Sco~OB2.
\end{itemize}
Practically all stars in Sco~OB2 are part of a binary (or multiple) system. Our results indicate that multiplicity is a fundamental parameter in the star forming process.

In our recovery of the current and primordial binary population in Sco~OB2 we have made several assumptions and simplifications. The modeling of the selection effects of the spectroscopic and astrometric binaries needs to be improved, so that the simulated observations and be directly compared to the observed binary parameter distributions. Improved models will include an accurate model for the background star population (which makes the $K_S=12$~mag criterion redundant), and for stellar variability, as both of these could result in spurious binaries in a catalog. 

In this chapter we have included the results of six major binarity surveys among Sco~OB2 members. We have not included the smaller surveys and individual discoveries, as each of these has its associated, often not well-documented selection effects. Inclusion of these will provide a better description of the binary population in Sco~OB2. 
Although the literature study and observations in Chapters~2 and~3 have shown that triple and higher-order systems are present in Sco~OB2, we have neglected these systems here, because of the non-trivial comparison with the observations, the complicated selection effects, and the small number of known higher-order multiples. 

The current binary population in Sco~OB2 may well 
be a fossil record of the primordial binary population, as 
the young age and low stellar density of Sco~OB2 suggest that stellar evolution
has affected only a handful of the most massive binaries, and that dynamical
evolution of the binary population has been modest. 
The latter statement needs to be
verified using numerical simulations of evolving OB~associations. Whether the effect of dynamical
evolution has been negligible over the lifetime of Sco~OB2, depends on
the initial conditions. If Sco~OB2 was born as a low-density
association, similar to its present state, the binary population is expected to have changed only modestly due to dynamical evolution. On the other
hand, if the association has expanded significantly over the last
5--20~Myr, dynamical evolution may have been more prominent.

The assumed independence of the binary parameters, as well as the properties of the low-mass binary population need to be addressed observationally. Due to selection effects a relatively small number of binaries is known among the low-mass members of Sco~OB2, making it difficult to derive the properties of these. This issue can be addressed using the results of the {\em Gaia} space mission \citep{perryman2001,turon2005}, which is a project of the European Space Agency, expected to be launched in 2011. {\em Gaia} will survey over a billion stars in our Galaxy and the Local Group, and will provide an enormous dataset of visual, eclipsing, spectroscopic, and astrometric binaries \citep{soderhjelm2005}. 
The membership and stellar content of nearby OB~associations can accurately determined using the results of {\em Gaia}, down into the brown dwarf regime. The {\em Gaia} dataset will be homogeneous, and its selection effects can therefore be modeled in detail. The available dataset of binaries will be larger and more complete than any other binarity survey in Galactic star clusters and OB~associations thus far.

% ====================================================================
% ====================================================================
% ====================================================================
% ==INTRODUCTION======================================================
% ====================================================================
% ====================================================================
% ====================================================================

\section*{Acknowledgments}

This research was supported by NWO under project number 614.041.006, the Royal Netherlands Academy of Sciences (KNAW) and  the Netherlands Research School for Astronomy (NOVA).

% ====================================================================
% ====================================================================
% ====================================================================
% ==INTRODUCTION======================================================
% ====================================================================
% ====================================================================
% ====================================================================

\clearpage

\markright{Appendix A: Datasets used}
\addcontentsline{toc}{section}{Appendix A: Datasets used}
\section*{Appendix A: Datasets used}

\setlength{\LTcapwidth}{1.0\textwidth}
\begin{longtable}{|lccc cr ccc|}
  \caption{Properties of the KO5 dataset that we use in our analysis. Note that not all candidate companions are listed here, as in our analysis we only consider single stars and binary systems. The members HIP68532 and HIP69113 (marked with a star) both have two companions with a similar separation, position angle, and brightness. In our analysis we consider these ``double companions'' as a single companion. \label{table: data_adonis}} \\
  \hline
  HIP & $K_{S,1}$ & $K_{S,2}$ & $\Delta K_S$ & $\rho$  & $\varphi$ & $M_1$     & $M_2$     & $q$ \\
      & mag       & mag       & mag          & arcsec  & deg       & M$_\odot$ & M$_\odot$ &     \\
  \endfirsthead
  \hline
  \multicolumn{9}{|l|}{\tablename\ \thetable{} -- continued from previous page} \\
  \hline
  HIP & $K_{S,1}$ & $K_{S,2}$ & $\Delta K_S$ & $\rho$  & $\varphi$ & $M_1$     & $M_2$     & $q$ \\
      & mag       & mag       & mag          & arcsec  & deg       & M$_\odot$ & M$_\odot$ &     \\
  \hline
  \endhead
  \hline
  \multicolumn{9}{|l|}{{Continued on next page}} \\ 
  \hline
  \endfoot
  \hline 
  \endlastfoot
  \hline
  HIP50520  &  6.23  &  6.39  &  0.16  &  2.51  &  313  &  2.12  &  1.98  &  0.93   \\
HIP52357  &  7.64  &  7.65  &  0.01  &  0.53  &  73  &  1.60  &  0.22  &   0.14  \\
HIP56993  &  7.38  &  11.88  &  4.50 &  1.68  &  23  &  1.90  &  0.180  &  0.09   \\
HIP58416  &  7.03  &  8.66  &  1.63  &  0.58  &  166  &  1.86  &  1.00  &  0.54   \\
HIP59413  &  7.46  &  8.18  &  0.72  &  3.18  &  100  &  1.62  &  1.34  &  0.83   \\
HIP59502  &  6.87  &  11.64  &  4.77  &  2.94  &  26  &  1.80  &  0.14  &  0.08   \\
HIP60084  &  7.65  &  10.10  &  2.45  &  0.46  &  330  &  1.66  &  0.62  & 0.37   \\
HIP61265  &  7.46  &  11.38  &  3.92  &  2.51  &  67  &  1.82  &  0.27  &  0.15   \\
HIP61639  &  6.94  &  7.06  &  0.12  &  1.87  &  182  &  1.82  &  1.74  &  0.96   \\
HIP61796  &  6.37  &  11.79  &  5.42  &  9.89  &  109  &  2.46  &  0.14  & 0.06   \\
HIP62002  &  7.09  &  7.65  &  0.56  &  0.38  &  69  &  1.68  &  1.20  &   0.71   \\
HIP62026  &  6.31  &  7.86  &  1.55  &  0.23  &  6  &  2.45  &  1.19  &    0.49   \\
HIP62179  &  7.20  &  7.57  &  0.37  &  0.23  &  283  &  1.84  &  1.56  &  0.85   \\
HIP64515  &  6.78  &  6.94  &  0.16  &  0.31  &  166  &  1.96  &  1.84  &  0.94   \\
HIP65822  &  6.68  &  11.08  &  4.40 &  1.82  &  304  &  2.91  &  0.38  &  0.13   \\
HIP67260  &  6.98  &  8.36  &  1.38  &  0.42  &  229  &  2.00  &  1.10  &  0.55   \\
HIP67919  &  6.59  &  9.10  &  2.51  &  0.69  &  297  &  1.97  &  0.75  &  0.38   \\
HIP68080  &  6.28  &  7.19  &  0.91  &  1.92  &  10  &  2.91  &  1.92  &   0.66   \\
HIP68532$^\star$  &  7.02  & 9.54   &  2.53  &  3.05  &  289  &  1.95  &  1.12  &   0.57   \\
HIP68867  &  7.17  &  11.61  &  4.44  &  2.16  &  285  &  2.18  &  0.24  & 0.11   \\
HIP69113$^\star$  &  6.37  & 10.29  &  3.92  &  5.34  &  65  &  3.87  &  1.49  &    0.39   \\
HIP69749  &  6.62  &  11.60  &  4.98  &  1.50  &  1  &  3.81  &  0.38  &   0.10   \\
HIP70998  &  7.06  &  10.83  &  3.77  &  1.17  &  355  &  2.54  &  0.48  & 0.19   \\
HIP71724  &  6.79  &  9.70  &  2.91  &  8.66  &  23  &  2.62  &  0.82  &   0.31   \\
HIP71727  &  6.89  &  7.80  &  0.91  &  9.14  &  245  &  2.46  &  1.64  &  0.67  \\
HIP72940  &  6.85  &  8.57  &  1.72  &  3.16  &  222  &  1.82  &  0.96  &  0.53   \\
HIP72984  &  7.05  &  8.50  &  1.45  &  4.71  &  260  &  1.90  &  1.06  &  0.56   \\
HIP74066  &  6.08  &  8.43  &  2.35  &  1.22  &  110  &  2.68  &  1.02  &  0.38   \\
HIP74479  &  6.31  &  10.83  &  4.52  &  4.65  &  154  &  3.03  &  0.38  & 0.13   \\
HIP75056  &  7.31  &  11.17  &  3.86  &  5.19  &  35  &  1.92  &  0.30  &  0.16   \\
HIP75151  &  6.65  &  8.09  &  1.44  &  5.70  &  121  &  3.19  &  1.64  &  0.51   \\
HIP75915  &  6.44  &  8.15  &  1.71  &  5.60  &  229  &  2.89  &  1.22  &  0.42   \\
HIP76001  &  7.60  &  7.80  &  0.20  &  0.25  &  3  &  1.54  &  1.36  &     0.88   \\
HIP76071  &  7.06  &  10.87  &  3.81  &  0.69  &  41  &  2.70  &  0.23  &  0.09   \\
HIP77315  &  7.24  &  7.92  &  0.68  &  0.68  &  67  &  2.08  &  1.56  &   0.75  \\
HIP77911  &  6.68  &  11.84  &  5.16  &  7.96  &  279  &  2.80  &  0.09  & 0.03   \\
HIP77939  &  6.56  &  8.09  &  1.53  &  0.52  &  119  &  3.85  &  1.82  &  0.47   \\
HIP78756  &  7.16  &  9.52  &  2.36  &  8.63  &  216  &  2.30  &  0.92  &  0.40 \\
HIP78809  &  7.51  &  10.26  &  2.75  &  1.18  &  26  &  2.03  &  0.30  &  0.15   \\
HIP78847  &  7.32  &  11.30  &  3.98  &  8.95  &  164  &  2.20  &  0.160  &0.07   \\
HIP78853  &  7.50  &  8.45  &  0.95  &  1.99  &  270  &  1.82  &  1.14  &  0.63  \\
HIP78956  &  7.57  &  9.04  &  1.47  &  1.02  &  49  &  2.40  &  1.16  &   0.48   \\
HIP79124  &  7.13  &  10.38  &  3.25  &  1.02  &  96  &  2.48  &  0.33  &  0.13   \\
HIP79156  &  7.61  &  10.77  &  3.16  &  0.89  &  59  &  2.09  &  0.27  &  0.13   \\
HIP79250  &  7.49  &  10.71  &  3.22  &  0.62  &  181  &  1.42  &  0.140  &0.10   \\
HIP79530  &  6.60  &  8.34  &  1.74  &  1.69  &  220  &  3.73  &  1.58  &  0.42   \\
HIP79631  &  7.17  &  7.61  &  0.44  &  2.94  &  128  &  1.90  &  1.58  &  0.83   \\
HIP79739  &  7.08  &  11.23  &  4.15  &  0.96  &  118  &  2.32  &  0.16  & 0.07   \\
HIP79771  &  7.10  &  10.89  &  3.79  &  3.67  &  313  &  2.14  &  0.19  & 0.09   \\
HIP80238  &  7.34  &  7.49  &  0.15  &  1.03  &  318  &  1.94  &  1.67  &  0.86   \\
HIP80324  &  7.33  &  7.52  &  0.19  &  6.23  &  152  &  1.70  &  1.54  &  0.91   \\
HIP80371  &  6.40  &  8.92  &  2.52  &  2.73  &  141  &  3.43  &  0.94  &  0.27   \\
HIP80425  &  7.40  &  8.63  &  1.23  &  0.60  &  156  &  2.08  &  1.16  &  0.56   \\
HIP80461  &  5.92  &  7.09  &  1.17  &  0.27  &  286  &  5.29  &  2.97  &  0.56   \\
HIP80799  &  7.45  &  9.80  &  2.35  &  2.94  &  205  &  1.86  &  0.34  &  0.18   \\
HIP80896  &  7.44  &  10.33  &  2.89  &  2.28  &  177  &  1.81  &  0.24  & 0.13   \\
HIP81624  &  5.80  &  7.95  &  2.15  &  1.13  &  224  &  6.53  &  2.30  &  0.35   \\
HIP81972  &  5.87  &  11.77  &  5.90 &  5.04  &  213  &  4.92  &  0.35  &  0.07   \\
HIP83542  &  5.38  &  9.90  &  4.52  &  8.86  &  196  &  1.10  &  0.91  &  0.83   \\
HIP83693  &  5.69  &  9.26  &  3.57  &  5.82  &  78  &  4.95  &  1.06  &   0.21   \\

\end{longtable}

\begin{longtable}{|lccc cr ccc|}
  \caption{Properties of the KO6 dataset that we use in our analysis. Note that not all candidate companions are listed here, as in our analysis we only consider single stars and binary systems. The members HIP68532 and HIP69113 (marked with a star) both have two companions with a similar separation, position angle, and brightness. In our analysis we consider these ``double companions'' as a single companion. \label{table: data_naco}} \\
  \hline
  HIP & $K_{S,1}$ & $K_{S,2}$ & $\Delta K_S$ & $\rho$  & $\varphi$ & $M_1$     & $M_2$     & $q$ \\
      & mag       & mag       & mag          & arcsec  & deg       & M$_\odot$ & M$_\odot$ &     \\
  \endfirsthead
  \hline
  \multicolumn{9}{|l|}{\tablename\ \thetable{} -- continued from previous page} \\
  \hline
  HIP & $K_{S,1}$ & $K_{S,2}$ & $\Delta K_S$ & $\rho$  & $\varphi$ & $M_1$     & $M_2$     & $q$ \\
      & mag       & mag       & mag          & arcsec  & deg       & M$_\odot$ & M$_\odot$ &     \\
  \hline
  \endhead
  \hline
  \multicolumn{9}{|l|}{{Continued on next page}} \\ 
  \hline
  \endfoot
  \hline 
  \endlastfoot
  \hline
  HIP59502 & 6.87 & 11.64 & 4.77 & 2.935 & 26 & 1.80 & 0.14 & 0.08   \\
HIP60851 & 6.06 & 13.69 & 7.63 & 8.159 & 231& 2.63 & 0.04 & 0.02   \\
HIP61265 & 7.46 & 11.38 & 3.93 & 2.505 & 67 & 1.82 & 0.27 & 0.15   \\
HIP62026 & 6.31 & 7.86  & 1.55 & 0.232 & 6  & 2.45 & 1.19 & 0.49   \\
HIP63204 & 6.78 & 8.40  & 1.62 & 0.153 & 237& 2.05 & 1.06 & 0.52   \\
HIP67260 & 6.98 & 8.36  & 1.38  & 0.423 & 229& 2.00 & 1.10 & 0.55  \\
HIP67919 & 6.59 & 9.10  & 2.51 & 0.685 & 297& 1.97 & 0.75 & 0.38  \\
HIP68532$^\star$ & 7.02 & 9.54  & 2.53 & 3.052 & 288& 1.95 & 1.12 & 0.57  \\
HIP69113$^\star$ & 6.37 & 10.29 & 3.92  & 5.344 & 65 & 3.87 & 1.49 & 0.39   \\
HIP73937 & 6.23 & 8.37  & 2.14 & 0.242 & 191& 2.94 & 1.11 & 0.38   \\
HIP78968 & 7.42 & 14.26 & 6.84 & 2.776 & 322& 2.33 & 0.02 & 0.01   \\
HIP79739 & 7.08 & 11.23 & 4.15 & 0.959 & 118& 2.32 & 0.16 & 0.07   \\
HIP79771 & 7.10 & 11.42 & 4.33 & 0.435 & 129& 2.14 & 0.19 & 0.09   \\
HIP80799 & 7.45 & 9.80  & 2.35 & 2.940 & 205& 1.86 & 0.34 & 0.18   \\
HIP80896 & 7.44 & 10.33 & 2.89 & 2.278 & 177& 1.81 & 0.24 & 0.13   \\
HIP81949 & 7.33 & 15.52 & 8.19 & 5.269 & 341& 2.26 & 0.02 & 0.01   \\
HIP81972 & 5.87 & 11.77 & 5.90 & 5.040 & 213& 4.92 & 0.35 & 0.07   \\
HIP83542 & 5.38 & 9.90  & 4.52 & 8.864 & 196& 1.10 & 0.91 & 0.83   \\

\end{longtable}

\begin{table}[!tbhp]
  \centering
  \caption{Properties of the SHT dataset that we use in our analysis. Note that not all candidate companions are listed here, as in our analysis we only consider single stars and binary systems. The six members at the bottom of the list were not explicitly observed by SHT. Due to the presence of (known) close companions these were not suitable for wavefront sensing. We have included these targets for our analysis to avoid a bias towards low binarity. \label{table: data_tokovinin}}
  \begin{tabular}{|lccc cr ccc|}
    \hline
    \hline
  HIP & $K_{S,1}$ & $K_{S,2}$ & $\Delta K_S$ & $\rho$  & $\varphi$ & $M_1$     & $M_2$     & $q$ \\
      & mag       & mag       & mag          & arcsec  & deg       & M$_\odot$ & M$_\odot$ &     \\
    \hline
    HIP55425  &  4.66  &  5.86  &  1.20  &  0.354  &  144  &  4.65  &  2.70  &  0.58   \\
HIP56561  &  3.17  &  6.81  &  3.64  &  0.734  &  135  &  8.30  &  2.22  &  0.27   \\
HIP58884  &  5.67  &  7.00  &  1.33  &  0.698  &  158  &  3.17  &  1.75  &  0.55   \\
HIP61585  &  3.41  &  10.94 &  7.53  &  4.853  &  198  &  6.30  &  0.19  &  0.03   \\
HIP63945  &  5.80  &  9.16  &  3.36  &  1.551  &  268  &  3.60  &  0.135 &  0.04   \\
HIP65271  &  5.12  &  7.03  &  1.91  &  0.164  &  135  &  4.25  &  1.80  &  0.42   \\
HIP67472  &  3.97  &  10.06 &  6.09  &  4.637  &  304  &  7.95  &  0.75  &  0.09   \\
HIP72683  &  5.27  &  6.84  &  1.57  &  0.099  &  86   &  4.52  &  2.17  &  0.48   \\
HIP72800  &  5.54  &  9.43  &  3.89  &  1.046  &  161  &  3.75  &  0.76  &  0.20   \\
HIP73334  &  4.09  &  5.46  &  1.37  &  0.128  &  156  &  7.83  &  5.33  &  0.68   \\
HIP75264  &  4.28  &  5.55  &  1.27  &  0.279  &  149  &  7.25  &  4.84  &  0.67  \\
HIP76945  &  5.79  &  9.47  &  3.68  &  0.507  &  133  &  3.50  &  0.80  &  0.23   \\
HIP77939  &  6.13  &  7.78  &  1.65  &  0.524  &  120  &  4.95  &  2.16  &  0.44   \\
HIP78820  &  3.86  &  6.80  &  2.94  &  0.292  &  171  &  11.20 &  2.98  &  0.27   \\
HIP79374  &  4.20  &  5.14  &  0.94  &  1.334  &  2    &  8.32  &  5.47  &  0.66   \\
HIP79530  &  6.31  &  8.07  &  1.76  &  1.693  &  220  &  3.23  &  1.35  &  0.42   \\
HIP80112  &  2.61  &  4.77  &  2.16  &  0.469  &  244  &  19.96 & 10.40  &  0.52   \\
\hline
HIP57851  &  ---    &  ---    &  ---     &  1.549  &  158  &  4.15  &  1.83  &  0.44   \\
HIP62322  &  ---    &  ---    &  ---     &  1.206  &  35   &  7.35  &  6.40  &  0.87  \\
HIP64425  &  ---    &  ---    &  ---     &  0.185  &  7    &  4.07  &  3.10  &  0.76   \\
HIP74117  &  ---    &  ---    &  ---     &  0.193  &  210  &  6.49  &  4.94  &  0.76   \\
HIP76371  &  ---    &  ---    &  ---     &  2.150  &  8    &  5.75  &  2.79  &  0.49   \\
HIP77840  &  ---    &  ---    &  ---     &  2.162  &  270  &  6.05  &  2.39  &  0.40   \\

    \hline
    \hline
  \end{tabular}
\end{table}

\begin{table}
  \centering
  \begin{tabular}{cc}
    \begin{tabular}{lc cr l}
      \hline
      \hline
      HIP & $P$  & $e$ & \multicolumn{1}{c}{$\omega$} & Group \\
          & days &     & \multicolumn{1}{c}{deg}      &       \\
      \hline
      HIP67464	& 2.6253 & 0.13 & 222 & UCL \\
HIP75647	& 3.8275 & 0.25 & 22 & UCL \\
HIP76297	& 2.8081 & 0.10 & 97 & ULC \\
HIP76503	& 5.2766 & 0.33 & 86 & US \\
HIP76600	& 3.2907 & 0.28 & 114 & UCL \\
HIP77858	& 1.9235 & 0.36 & 309 & US \\
HIP77911	& 1.264 & 0.61 & 330 & US \\
HIP78104	& 4.0031 & 0.27 & 231 & US \\
HIP78168	& 10.0535 & 0.58 & 340 & US \\
HIP78265	& 1.5701 & 0.15 & 25 & US \\
HIP78820	& 6.8281 & 0.28 & 38 & US \\
HIP79404	& 5.7805 & 0.19 & 115 & US \\
HIP79374	& 5.5521 & 0.11 & 267 & US \\
HIP80112	& 34.23 & 0.36 & 308 & US \\
HIP80569	& 138.8 & 0.44 & 325 & US \\

      \hline
      \hline
    \end{tabular}
    &
    \begin{tabular}{ll}
      \hline 
      \hline
      \multicolumn{2}{l}{Radial velocity variables}\\
      \\
      \hline
      HIP67472 & 
      HIP68245 \\ 
      HIP68862 & 
      HIP70300 \\ 
      HIP73334 & 
      HIP74100 \\ 
      HIP75141 & 
      HIP75304 \\ 
      HIP76633 & 
      HIP77635 \\ 
      HIP77900 & 
      HIP77939 \\ 
      HIP78246 & 
      HIP78384 \\ 
      HIP78530 & 
      HIP78549 \\ 
      HIP78655 & 
      HIP79031 \\ 
      HIP79374 & 
      HIP79739 \\ 
      HIP80024 & 
      HIP81266 \\ 
      HIP82545 &  \\ 
      \\
      \\
      \\
      \hline
      \hline
    \end{tabular}  
    \\
  \end{tabular}
  \caption{The LEV dataset used in this sample, consisting of 15~binaries with orbital elements (SB1 or SB2), and 23 radial velocity variables (RVV), for which no orbital elements are available. {\em Left}: the 15~spectroscopic binaries with orbital elements among the confirmed members of Sco~OB2, in the LEV dataset. LEV observed 52~confirmed members of Sco~OB2, of which 8~SB1s, 7~SB2s, 23~RVVs, and 14~targets with a constant radial velocity. {\em Right}: the 23 RVVs. \label{table: data_levato}}
\end{table}

\renewcommand{\figurename}{Figuur}
\hyphenpenalty=5000
\tolerance=1000

\chapter{Samenvatting in het Nederlands}

Onderzoek in de laatste decennia heeft aangetoond dat de meeste sterren zich in een dubbel- of meervoudig stersysteem bevinden. De vorming van sterren en de evolutie van verschillende sterpopulaties zijn twee van de belangrijkste vraagstukken in de hedendaagse sterrenkunde. Om deze te kunnen begrijpen, is het van groot belang de eigenschappen van dubbelsterren te onderzoeken.  Het doel van het onderzoek dat beschreven wordt in dit proefschrift is het vinden van ``de oorsponkelijke dubbelsterpopulatie''. Dit betekent dat onderzocht wordt wat de eigenschappen van dubbelsterren zijn, n\`{e}t nadat de sterren gevormd zijn. 

Al in de Oudheid ontdekten sterrenkundigen dat de sterren niet willekeurig verdeeld zijn  aan de hemel. De Griekse wetenschapper Ptolemeus merkte in zijn werk {\em Almagest} op dat sommige sterren zich in paren aan de hemel bevinden, en noemde deze ``dubbele sterren''. Pas veel later, nadat Isaac Newton in 1686 zijn ontdekkingen over de zwaartekracht gepubliceerd had, begon men zich te realiseren dat deze dubbele sterren door zwaartekracht gebonden zouden kunnen zijn, en dus om elkaar heen bewegen. Recente waarnemingen hebben aangetoond dat een groot deel van de sterren zich in een dubbel- of meervoudig stersysteem bevindt. Dit heeft zelfs geleid tot speculaties over een begeleider van onze Zon, {\em Nemesis}. In deze samenvatting beschrijf ik kort de methoden om dubbelsterren waar te nemen, de complicaties bij het interpreteren van de waarnemingen, en de noodzaak van het vinden van de oorspronkelijke dubbelsterpopulatie. Tenslotte geef ik een overzicht van de inhoud van dit proefschrift.

\begin{figure}[t]
  \centering
  \includegraphics[width=1\textwidth,height=!]{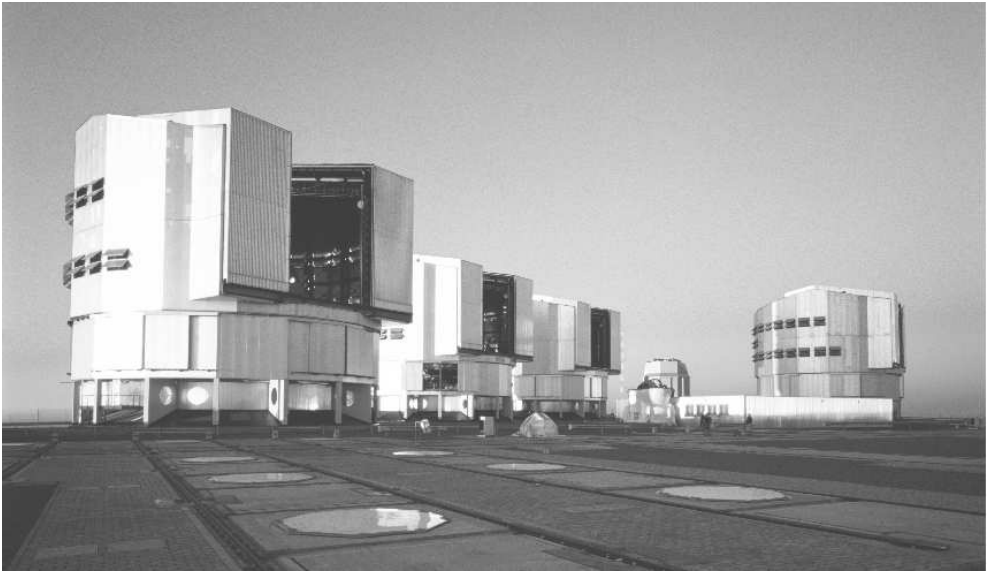}
  \caption{De ESO {\em Very Large Telescope} (VLT) bestaat uit vier optische telescopen: Antu, Kueyen, Melipal en Yepun. Elke telescoop heeft een spiegel met een doorsnede van 8.2~meter. De VLT is onderdeel van Paranal Observatory, 2600~meter boven zeeniveau in het noorden van Chili. De VLT behoort tot de grootste optische telescopen ter wereld, en wordt beheerd door de European Southern Observatory. De resultaten in Hoofdstuk~3 van dit proefschrift zijn gebaseerd op waarnemingen met het adaptieve optica instrument NAOS/CONICA, geinstalleerd op de Antu telescoop (links). Bron: European Southern Observatory (ESO). \label{figure: samen_vlt} }
\end{figure}

\begin{figure}[t]
  \centering
  \includegraphics[width=1\textwidth,height=!]{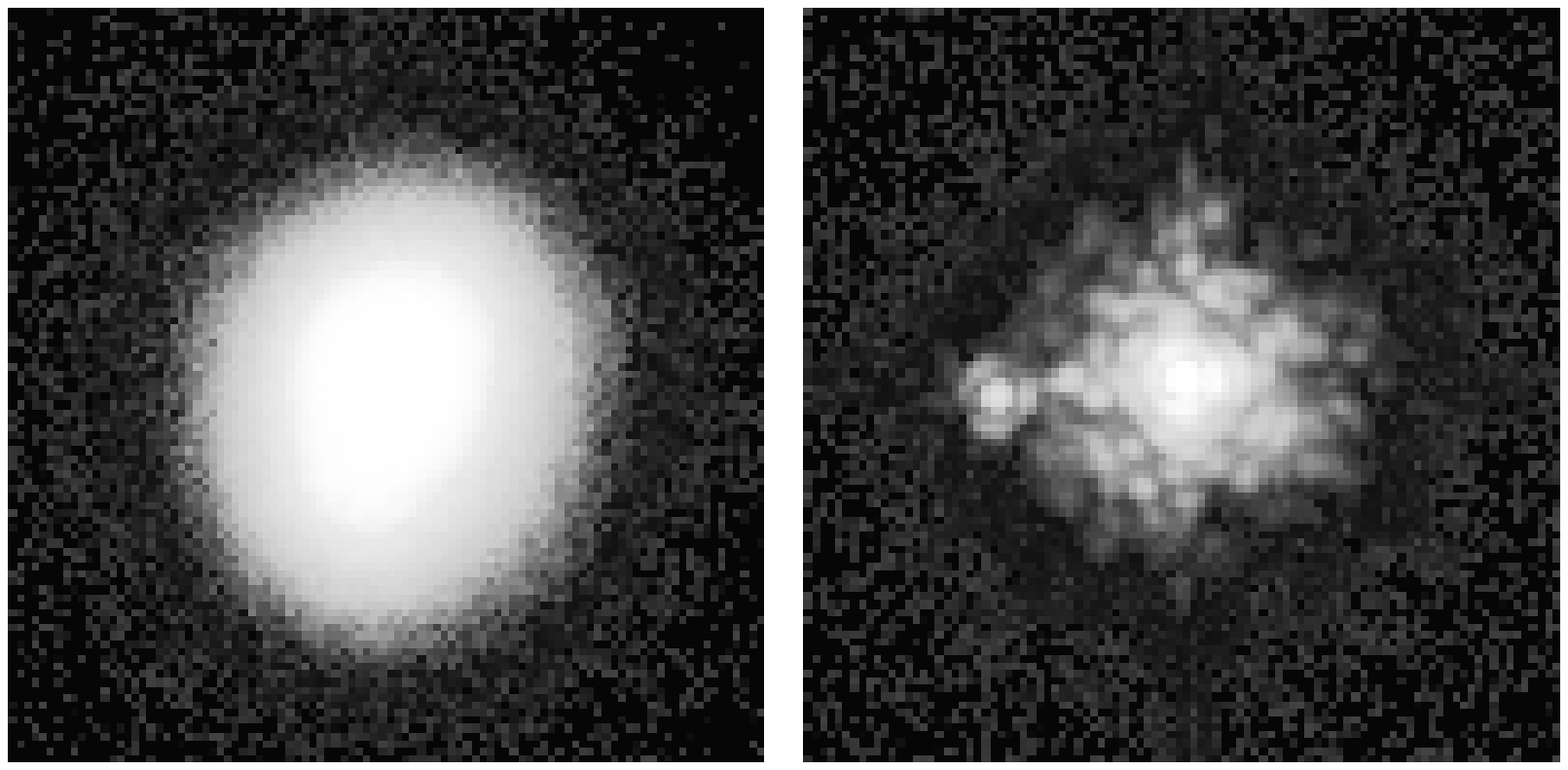}
  \caption{Vanwege turbulentie in de Aardatmosfeer was het tot voor kort onmogelijk om een ruimtelijke resolutie van meer dan ongeveer \'{e}\'{e}n boogseconde te behalen. Door middel van de recent ontwikkelde adaptive optica techniek is het mogelijk voor de effecten van turbulentie te corrigeren. Dit figuur laat twee opnamen van de ster HD101452 zien, waargenomen in het jaar 2000 met het instrument ADONIS op de ESO 3.6\,meter telescoop in La Silla, Chili. Links staat een afbeelding van de ster {\em z\'{o}nder}, en rechts {\em m\'{e}t} adaptive optica. Met behulp van adaptieve optica is ontdekt dat HD101452 een nauwe dubbelster is. De afmeting van elke afbeelding is $4.9\times 4.9$ boogseconden. Links naast de hoofdster ($K_S=6.8$~mag), op een afstand van 1.2 boogseconde, bevindt zich de bijna veertig keer zwakkere begeleider ($K_S=10.7$~mag).  Bron: Anthony Brown. \label{figure: adaptieve_optica} }
\end{figure}

\section{Het waarnemen van dubbelsterren}

De baan van een dubbelstersysteem in de ruimte wordt volledig vastgelegd door zeven grootheden: de massa $M_1$ van de hoofdster, de massa $M_2$ van de begeleider (of de massaverhouding $q=M_2/M_1$), de halve lange as $a$ (of omlooptijd $P$) van het dubbelstersysteem, de eccentriciteit $e$ van de baan ($e=0$ correspondeert met een cirkelbaan), en de orientatie van de baan (gegeven door de inclinatie $i$, de periastronhoek $\omega$, en de positiehoek van de klimmende knoop $\Omega$). Tenslotte wordt de fase van het dubbelstersysteem vastgelegd met het tijdstip $\tau$ waarop het systeem in periastron is.
Welke van deze baanparameters daadwerkelijk gemeten kunnen worden, hangt af van de eigenschappen van het dubbelstersysteem en van de waarneemtechniek. Er zijn verschillende manieren om dubbelsterren waar te nemen. Vanuit het oogpunt van de waarnemingen zijn de vier belangrijkste typen dubbelsterren de volgende:
\begin{itemize}\addtolength{\itemsep}{-0.5\baselineskip}
\item[--] {\bf Visuele dubbelsterren}. Dit zijn dubbelsterren waarvan de componenten vlak bij elkaar aan de hemel staan op een fotografische opname (of met het blote oog). Deze dubbelsterren kunnen een dusdanig lange omlooptijd hebben dat ze niet zichtbaar bewegen. Soms staan twee sterren ``toevallig'' dicht bij elkaar aan de hemel, terwijl ze niet gravitationeel gebonden zijn; dit zijn zogenaamde ``optische dubbelsterren''. Het verschil tussen optische dubbelsterren en echte dubbelsterren is vaak moeilijk te bepalen.
\item[--] {\bf Astrometrische dubbelsterren}. Een astrometrische dubbelster lijkt op een visuele dubbelster, waarvan \'{e}\'{e}n van de sterren onzichtbaar is. Men constateert dat dit een dubbelster is omdat men de zichtbare ster aan de hemel in de loop der tijd een baan ziet beschrijven. Waarnemingen van astrometrische dubbelsterren kunnen veel informatie over het systeem opleveren, maar astrometrische dubbelsterren zijn vaak moeilijk waarneembaar.
\item[--] {\bf Bedekkingsveranderlijken} (eclipserende dubbelsterren). Dit zijn dubbelsterren die periodiek voor elkaar langs bewegen. Hierdoor treedt er een (gedeeltelijke) ster\-verduistering op. Alhoewel de twee sterren vaak niet visueel van elkaar te onderscheiden zijn, kan er uit de lichtcurve toch informatie over het systeem afgeleid worden, met name over de afmetingen van de twee sterren en de omlooptijd. De meeste bedekkingsveranderlijken hebben een korte omlooptijd, in de orde van dagen tot weken.
\item[--] {\bf Spectroscopische dubbelsterren}. Deze dubbelsterren hebben een korte omlooptijd (van enkele dagen tot jaren), en bewegen vaak met meer dan een kilometer per seconde om elkaar heen. Vanwege de hoge snelheid van de sterren is de beweging te meten in de spectraallijnen van de beide sterren (vanwege het Doppler effect). Uit deze beweging kan worden afgeleid dat een ster dubbel is, en soms ook enkele andere eigenschappen van het systeem. Doordat deze dubbelsterren erg nauw zijn, is het meestal niet mogelijk de componenten ruimtelijk van elkaar te onderscheiden.
\end{itemize}
Een dubbelster kan ook op meerdere manieren waargenomen zijn; het kan bijvoorbeeld zowel een spectroscopisch dubbelster als een bedekkingsveranderlijke zijn. Elk van de waarneemtechnieken is gevoelig voor dubbelsterren met specifieke eigenschappen. Door de resultaten van verschillende waarneemtechnieken te combineren, kan de dubbelsterpopulatie optimaal bestudeerd worden.
Alhoewel van veel sterren bekend is dat ze een begeleider hebben, is het niet eenvoudig de dubbelsterfractie te bepalen. Sommige begeleiders bevinden zich erg dichtbij of erg ver van de hoofdster, of zijn te lichtzwak om waargenomen te kunnen worden; onze huidige kennis van statistiek van dubbelsterren is daarom beperkt.

Door turbulentie in de Aardatmosfeer was het tot de jaren negentig vanaf de Aarde vrijwel onmogelijk om met een hogere resolutie dan ongeveer \'{e}\'{e}n boogeseconde waar te nemen. Hierin kwam verandering toen men de techniek van ``adaptieve optica'' onder controle kreeg. Met adaptieve optica is het mogelijk te corrigeren voor het versmerende effect dat de atmosfeer heeft op de afbeelding van een ster; zie Figuur~\ref{figure: adaptieve_optica}. Deze correctie wordt gedaan met behulp van een vervormbare spiegel die zich elke fractie van een seconde aanpast. Hierdoor werd het mogelijk om met tien maal hogere ruimtelijke resolutie waar te nemen. Met behulp van de adaptieve optica techniek is het mogelijk het gat tussen de wijdste spectroscopische dubbelsterren en de nauwste visuele dubbelsterren (deels) te overbruggen.

\section{De interpretatie van dubbelsterwaarnemingen}

\begin{figure}[!tbp]
  \centering
  \includegraphics[width=1\textwidth,height=!]{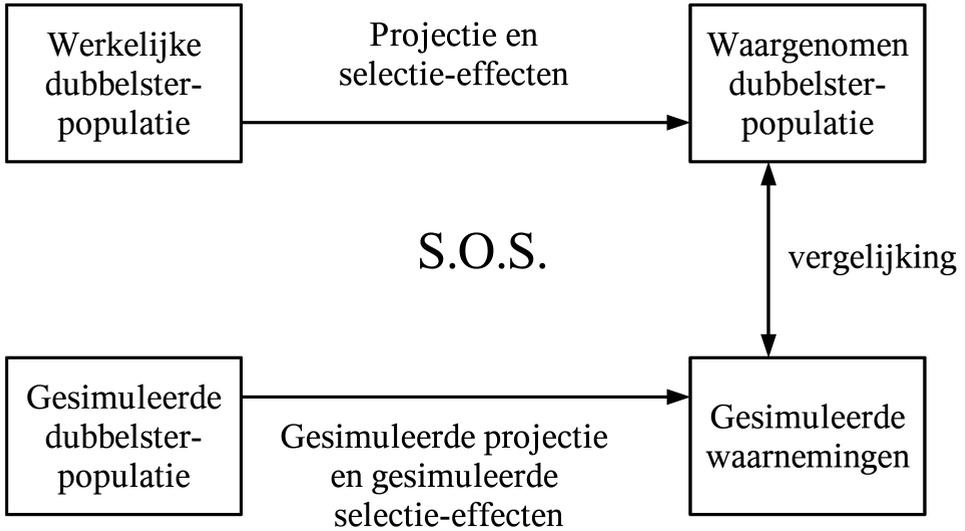}
  \caption{Een schematische weergave van de methode van gesimuleerde waarnemingen 
    ({\em Simulating Observations of Simulations; S.O.S.}). We gebruiken deze methode
    om de {\em werkelijke} dubbelsterpopulatie te reconstrueren uit de
    {\em waargenomen} dubbelsterpopulatie. In deze methode worden de eigenschappen
    van dubbelsterren in een gesimuleerde sterrengroep omgezet naar nagebootste
    waarnemingen. Door de eigenschappen van de gesimuleerde sterrengroep te vari\"{e}ren,
    en elke keer de gesimuleerde waarnemingen met de \'{e}chte waarnemingen te vergelijken,
    kan de in principe werkelijke dubbelsterpopulatie herleid worden.
  \label{figure: NL_sosmethod} }
\end{figure}

De interpretatie van de resultaten van een waarneemcampagne voor dubbelsterren is niet eenvoudig. Door selectie-effecten wordt in het algemeen slechts een klein deel van de dubbelsterren in een sterrengroep daadwerkelijk gevonden. Deze selectie-effecten zijn specifiek voor elk van de hierboven besproken waarneemtechnieken, en hangen af van de waarneemstrategie, de eigenschappen van de telescoop en het instrument, en van de weersomstandigheden. Voor een correcte interpretatie van de waarneemresultaten is een goed begrip van de selectie-effecten van groot belang.
In dit proefschrift gebruiken we de methode van gesimuleerde waarnemingen ({\em Simulating Observations of Simulations; S.O.S.}) om de dubbelsterpopulatie te herleiden. Een schematisch overzicht van deze techniek is gegeven in Figuur~\ref{figure: NL_sosmethod}. Kort samengevat, houdt deze methode in dat de waarneemresultaten vergeleken worden met {\em nagebootste} waarnemingen van een computersimulatie. Het grote voordeel van het gebruik van deze methode is dat de eigenschappen van een dubbelsterpopulatie herleid kunnen worden, in principe zonder dat er extrapolaties gemaakt hoeven te worden. Voorwaarde is dan wel dat de selectie-effecten goed gemodelleerd worden, dat de parameterruimte goed afgezocht wordt, en dat er voldoende waarneemresultaten beschikbaar zijn.

\section{De oorsponkelijke dubbelsterpopulatie}

De oorspronkelijke dubbelsterpopulatie ({\em primordial binary population}) van een sterrengroep is de dubbelsterpopulatie die aanwezig is n\`{e}t nadat het gas dat is overgebleven na de stervorming, verdwenen is (bijvoorbeeld weggeblazen is door sterwinden). Aangezien waarschijnlijk de meeste sterren in een dubbelstersysteem vormen, kan kennis over de oorspronkelijke dubbelsterpopulatie een belangrijke bijdrage leveren aan ons begrip van het stervormingsproces.
Naarmate de tijd verstrijkt, veranderen de eigenschappen van de dubbelsterpopulatie in een sterrengroep, als gevolg van sterevolutie en dynamische evolutie. Door sterk massaverlies van een ster (bijvoorbeeld door een supernova) veranderen de eigenschappen van de dubbelsterbaan. In enkele gevallen resulteert dit in de ontbinding van het dubbelstersysteem. 
Daar kortlevende zware sterren zeldzaam zijn, zal het overgrote deel van de dubbelsterren in jonge sterrenhopen nog geen effecten van sterevolutie hebben ondergaan.
Dynamische evolutie speelt voor deze sterrenhopen in het algemeen een belangrijkere rol dan sterevolutie. Doordat dubbelsterren in de loop van de tijd dicht in de buurt komen van andere systemen, kunnen de dubbelstereigenschappen veranderen, of kan door dynamische effecten het systeem zelfs ontbonden worden. Dynamische evolutie is vooral van belang voor de wijdste dubbelsterren, omdat deze de grootste kans hebben een interactie te ondergaan. Nauwe dubbelsterren worden vrijwel niet be\"{i}nvloed door dynamische evolutie.
Het herleiden van de dubbelsterpopulatie kan gedaan worden met behulp van numerieke simulaties, op een manier vergelijkbaar aan die in Figuur~\ref{figure: NL_sosmethod}. Door associaties met verschillende initi\"{e}le eigenschappen te laten evolueren, en de gesimuleerde waarnemingen voor iedere associatie te vergelijken met de werkelijke waarnemingen, kan het best passende model gevonden worden. 

\section{In dit proefschrift\dots}

\subsection*{Dubbelsterren in de OB~associatie Sco~OB2}

OB~associaties zijn jonge stergroepen (5--50 miljoen jaar) met een lage sterdichtheid (ongeveer \'{e}\'{e}n ster per kubieke parsec) en bevatten zowel zware als lichte sterren. Vanwege deze eigenschappen zijn OB~associaties uitermate geschikt om de oorspronkelijke dubbelsterpopulatie af te leiden. Sterevolutie heeft alleen de zwaarste dubbelstersystemen be\"{i}nvloed, terwijl dynamische evolutie de dubbelsterpopulatie nauwelijks heeft kunnen veranderen vanwege de jonge leeftijd en lage sterdichtheid. Verder bevatten OB~associaties vrijwel het volledige spectrum aan stermassa's, van B~sterren tot bruine dwergen, in tegenstelling tot de oude populatie in de buurt van de Zon. Bovendien zijn de meeste OB~associaties vrijwel ontdaan van het gas waaruit de sterren gevormd zijn, hetgeen het waarnemen aanzienlijk vereenvoudigt in vergelijking met het waarnemen van stervormingsgebieden.
Scorpius-OB2 (Sco~OB2) is de meest nabije jonge OB~associatie, en daarmee de meest geschikte kandidaat voor het onderzoek van de oorsponkelijke dubbelsterpopulatie. Deze sterrengroep bevindt zich in de sterrenbeelden Schorpioen, Centaur, Wolf en Zuiderkruis. Sco~OB2 bestaat uit drie subgroepen: Upper Scorpius (US), Upper Centaurus Lupus (UCL), en Lower Centaurus Crux (LCC), die op een afstand van respectievelijk 145, 140 en 118~parsec staan. De leeftijd van US is 5~miljoen jaar; die van UCL en LCC ongeveer 20~miljoen jaar. Figuur~\ref{figure: scocen_subgs} in de inleiding laat deze drie subgroepen zien, met het $\rho$~Ophiuchus stervormingsgebied ernaast. 
De spreiding van de leeftijd van de subgroepen suggereert dat zich een stervormingsgolf in \'{e}\'{e}n richting door de oorspronkelijke moleculaire wolk heeft voortgeplant.

\subsection*{Inhoud van dit proefschrift}

Het doel van het onderzoek in dit proefschrift is het vinden van de oorspronkelijke dubbelsterpopulatie in Sco~OB2. De belangrijkste vragen die in dit proefschrift aan de orde komen zijn: ``Wat zijn de eigenschappen van de dubbelsterpopulatie in Sco~OB2?'' en ``Welke fractie van de sterren wordt gevormd in een dubbel- of meervoudig stersysteem?''
Om deze vragen te kunnen beantwoorden verzamelen we alle informatie die bekend is over dubbelsterren in Sco~OB2, en vullen deze aan met nieuwe waarnemingen, gebruik makend van adaptieve optica in het nabij-infrarood (Hoofdstuk~2 en~3). Vervolgens identificeren we in Hoofdstuk~4 de selectie-effecten voor verschillende waarnemingstechnieken, en beschrijven en demonstreren de methode van gesimuleerde waarnemingen, waarmee de meeste intrinsieke eigenschappen van een associatie herleid kunnen worden. Vervolgens gebruiken we de waarnemingen en technieken beschreven in deze hoofdstukken om de eigenschappen van de huidige en oorspronkelijke dubbelsterpopulatie in Sco~OB2 af te leiden. De belangrijkste resultaten van deze vier hoofdstukken zijn hieronder samengevat.
\vskip0.3cm
\noindent
{\bf Hoofdstuk 2} 
\vskip0.3cm
\noindent
In dit hoofdstuk presenteer ik samen met mijn mede-auteurs de resultaten van een dubbelsterstudie onder sterren in de Sco~OB2 associatie. In de waarneemcampagne zijn 199~sterren van spectraaltype A en B waargenomen in de $K_S$ band met het instrument ADONIS op de ESO 3.6\,meter telescoop in La Silla, Chili. Door gebruik te maken van adaptieve optica, overbruggen we het onbekende terrein tussen de nauwe spectroscopische dubbelsterren en de wijde visuele dubbelsterren. We vinden 151~sterren rond de 199~waargenomen leden van de associatie. Onze analyse wijst uit dat er hiervan 74~met grote waarschijnlijkheid begeleiders zijn, terwijl er 77~achtergrondsterren zijn. Van deze 74 kandidaat-begeleiders zijn er 41~nieuw, en 33~waren reeds bekend. 
De scheiding tussen begeleiders en achtergrondsterren is uitgevoerd aan de hand van een statistisch onderbouwd magnitudecriterium. Sterren helderder dan $K_S=12$~mag zijn geclassificeerd als begeleiders; sterren zwakker dan  $K_S=12$~mag als achtergrondsterren.  
Met behulp van de afstand en leeftijd bepalen we de massa van elke hoofdster en zijn begeleider(s). De analyse van de volledige set van waarnemingen laat zien dat de verdeling van de massaverhoudingen de vorm $f_q(q) \propto q^{-0.33}$ heeft. Met behulp van deze verdeling kan worden uitgesloten dat de hoofdsterren en hun begeleiders willekeurig gepaard zijn. Een ander opmerkelijk resultaat van deze studie is het kleine aantal bruine-dwergbeleiders met een korte omlooptijd. 
\vskip0.3cm
\noindent
{\bf Hoofdstuk 3}
\vskip0.3cm
\noindent
In Hoofdstuk~2 zijn verschillende nieuwe kandidaat-begeleiders gevonden, en er is gevonden dat er zeer weinig nauwe bruine-dwergbegeleiders zijn. Om deze twee eigenschappen verder te onderzoeken, en om de juistheid van het $K_S=12$~mag criterium (voor de scheiding tussen begeleiders en achtergrondsterren) te controleren, hebben we een vervolgstudie gedaan. We hebben 22~sterren opnieuw waargenomen met het instrument NAOS/CONICA op de {\em Very Large Telescope} in Paranal, Chili. De waarnemingen zijn gedaan in het nabij-infrarood met de filters $J$, $H$, en $K_S$. Door de locatie van elke kandidaatbegeleider en kandidaat-achtergrondster te vergelijken met die van de theoretische isochroon in het kleur-magnitude diagram, kan de status van de ster met grotere waarschijnlijkheid vastgesteld worden. Onze analyse laat zien dat het $K_S=12$~mag criterium nauwkeurig is, en dus de meeste begeleiders en achtergrondsterren correct classificeert. Het kleine aantal nauwe bruine-dwergbegeleiders dat gevonden is, is echter in overeenstemming met een extrapolatie van de massaverhoudingsverdeling $f_q(q) \propto q^{-0.33}$ in het bruine-dwergregime. Het kleine aantal bruine-dwergbeleiders ten opzichte van het aantal stellaire begeleiders wordt verwezen als de {\em brown dwarf desert}, kan daarom verklaard worden door deze extrapolatie. De {\em embryo ejection} theorie, die ooit in het leven geroepen is om het kleine aantal bruine-dwergbeleiders te verklaren, is dus niet nodig om onze waarnemingen te verklaren. Onze resultaten suggereren dat bruine-dwergbegeleiders net als stellaire begeleiders gevormd worden, en niet zoals planeten.
\vskip0.3cm
\noindent
{\bf Hoofdstuk 4}
\vskip0.3cm
\noindent
Het waarnemen van dubbelsterren en het afleiden van de dubbelstereigenschappen van een OB~associatie is niet triviaal. Vanwege selectie-effecten (door de keuze van de waarnemer, door de waarneemtechniek en instrument, en door weersomstandigheden) worden veel dubbelsterren niet waargenomen. In dit hoofdstuk beschrijven we in detail de selectie-effecten die van belang zijn voor de interpretatie van waarnemingen van visuele, spectroscopische, en astrometrische dubbelsterren. We beschrijven vijf manieren om de begeleiders aan de hoofdsterren te koppelen (de zogenaamde {\em pairing functions}), en bediscussi\"{e}ren de consequenties van deze keuzes. In dit hoofdstuk demonstreren we de methode van gesimuleerde waarnemingen, waarmee de eigenschappen van een dubbelsterpopulatie uit de waarnemingen gereconstrueerd kunnen worden.
\vskip0.3cm
\noindent
{\bf Hoofdstuk 5}
\vskip0.3cm
\noindent
In het laatste hoofdstuk leiden we de dubbelsterpopulatie van Sco~OB2 af, gebruik makend van de techniek van gesimuleerde waarnemingen. We leiden af dat de dubbelsterfractie in Sco~OB2 erg hoog is; de modellen die het best passen bij de waarnemingen hebben een dubbelsterfractie van 100\%. De massaverhoudingsverdeling heeft de vorm $f_q(q) \propto q^{\gamma_q}$, met $\gamma_q = -0.4 \pm 0.2$, waarmee we kunnen uitsluiten dat de hoofdsterren en begeleiders willekeurig gepaard zijn. De waarnemingen duiden aan dat de Wet van \"{O}pik, $f_a(a)\propto a^{-1}$ geldt voor Sco~OB2. De eccentriciteitsverdeling in de associatie is consistent met een vlakke verdeling, alhoewel er niet genoeg waarnemingen zijn om deze verdeling nauwkeurig vast te stellen. Vanwege de lage dichtheid en de jonge leeftijd van Sco~OB2 lijkt de huidige dubbelsterpopulatie naar verwachting erg op de oorspronkelijke dubbelsterpopulatie. Sterevolutie heeft alleen enkele van de zwaarste dubbelsterren be\"{i}nvloed, en dynamische evolutie alleen de wijdste dubbelsterren. Onze analyse duidt aan dat vrijwel alle sterren in een dubbelstersysteem geboren worden; een fundamenteel gegeven voor ons begrip van de vorming van sterren.

\subsection*{Wat zal de toekomst brengen?}

Het doel van het onderzoek in dit proefschrift is het afleiden van de oorspronkelijke dubbelsterpopulatie. Hierbij zijn een aantal aannames gemaakt, die door waarnemingen bevestigd zullen moeten worden. 
Tijdens het afleiden van de dubbelsterpopulatie hebben we bijvoorbeeld de aanname gemaakt dat er zich in Sco~OB2 alleen enkelvoudige sterren en dubbelstersystemen bevinden. Hoewel waarschijnlijk slechts een klein deel van de sterren zich in drie- of meervoudige systemen bevindt, zullen deze systemen in de verdere analyse meegenomen moeten worden. In dit proefschrift hebben we het vaak over {\em de} oorspronkelijke dubbelsterpopulatie, waarmee we impliciet verwijzen naar die van Sco~OB2. Het is erg interessant om de oorspronkelijke dubbelsterpopulatie van andere stergroepen te analyseren, en de eigenschappen van deze als functie van omgevingsfactoren (zoals de temperatuur, de dichtheid, en het metaalgehalte van de oorspronkelijke gaswolk) te bestuderen. De resultaten van een dergelijk onderzoek zullen een grote bijdrage leveren aan ons begrip van stervorming.
In het laatste hoofdstuk van dit proefschrift bespreken we kort hoe de dubbelsterpopulatie in Sco~OB2 is veranderd sinds de associatie geboren werd. Voor de mate van verandering kunnen schattingen gemaakt worden, maar de resultaten hangen af van de oorspronkelijke sterdichtheid van de expanderende associatie. Het is daarom van belang numerieke simulaties van evoluerende associaties met verschillende initi\"{e}le eigenschappen uit te voeren, en de resulaten te vergelijken met de waarnemingen. 
Verder hebben we onze analyse gebaseerd op waarnemingen van hoofdzakelijk A en B sterren in Sco~OB2, omdat voor deze heldere sterren de grootste hoeveelheid waarneemgegevens beschikbaar is. Echter, omdat een groot deel van de populatie uit sterren van lagere massa bestaat, is het van belang ook deze groep in te detail te bestuderen. De {\em Gaia} satelliet, die rond 2011 gelanceerd zal worden door de {\em European Space Agency}, biedt hiervoor de oplossing. {\em Gaia} zal meer dan een miljard sterren bestuderen en met ongekende nauwkeurigheid hun lokatie en snelheid in de ruimte bepalen. {\em Gaia} zal een enorme hoeveelheid nieuwe visuele, astrometrische en spectroscopische dubbelsterren opleveren, alsmede miljoenen bedekkingsveranderlijken en duizenden exoplaneten en dubbele bruine dwergen. De missie zal een schat aan informatie opleveren over de dubbelsterpopulatie in nabije veldsterpopulatie, sterclusters en OB~associaties. {\em Gaia} zal hiermee een essenti\"{e}le bijdrage leveren aan het ontraadselen van het stervormingsproces.

\addcontentsline{toc}{chapter}{Bibliography}
\bibliographystyle{aatitle}
\bibliography{references/bibliography}

\begin{thebibliography}{207}
\expandafter\ifx\csname natexlab\endcsname\relax\def\natexlab#1{#1}\fi

\bibitem[{{Abt}(1983)}]{abt1983}
{Abt}, H.~A. 1983, \emph{{Normal and abnormal binary frequencies}}.
\newblock \araa, 21, 343

\bibitem[{{Abt} {et~al.}(1990){Abt}, {Gomez}, \& {Levy}}]{abt1990}
{Abt}, H.~A., {Gomez}, A.~E., \& {Levy}, S.~G. 1990, \emph{{The frequency and
  formation mechanism of B2-B5 main-sequence binaries}}.
\newblock \apjs, 74, 551

\bibitem[{{Abt} \& {Levy}(1976)}]{abtlevy1976}
{Abt}, H.~A. \& {Levy}, S.~G. 1976, \emph{{Multiplicity among solar-type
  stars}}.
\newblock \apjs, 30, 273

\bibitem[{{Aitken}(1918)}]{aitken1918}
{Aitken}, R.~G. 1918, {The binary stars} (D.~C.~McMurtrie, New York)

\bibitem[{{Aitken}(1932)}]{aitken1932}
{Aitken}, R.~G. 1932, {New General Catalogue of Double Stars} (Carnegie
  Institution, Washington)

\bibitem[{{Alencar} {et~al.}(2003){Alencar}, {Melo}, {Dullemond}, {Andersen},
  {Batalha}, {Vaz}, \& {Mathieu}}]{alencar2003}
{Alencar}, S.~H.~P., {Melo}, C.~H.~F., {Dullemond}, C.~P., {et~al.} 2003,
  \emph{{The pre-main sequence spectroscopic binary AK Scorpii revisited}}.
\newblock \aap, 409, 1037

\bibitem[{{Andersen} {et~al.}(1993){Andersen}, {Clausen}, \&
  {Gimenez}}]{andersen1993}
{Andersen}, J., {Clausen}, J.~V., \& {Gimenez}, A. 1993, \emph{{Absolute
  Dimensions of Eclipsing Binaries - Part Twenty - Gg-Lupi - Young Metal
  Deficient B-Stars}}.
\newblock \aap, 277, 439

\bibitem[{{Armitage} \& {Bonnell}(2002)}]{armitage2002}
{Armitage}, P.~J. \& {Bonnell}, I.~A. 2002, \emph{{The brown dwarf desert as a
  consequence of orbital migration}}.
\newblock \mnras, 330, L11

\bibitem[{{Bahcall} {et~al.}(1985){Bahcall}, {Hut}, \&
  {Tremaine}}]{bahcall1985}
{Bahcall}, J.~N., {Hut}, P., \& {Tremaine}, S. 1985, \emph{{Maximum mass of
  objects that constitute unseen disk material}}.
\newblock \apj, 290, 15

\bibitem[{{Balega} {et~al.}(1994){Balega}, {Balega}, {Belkin}, {Maximov},
  {Orlov}, {Pluzhnik}, {Shkhagosheva}, \& {Vasyuk}}]{balega1994}
{Balega}, I.~I., {Balega}, Y.~Y., {Belkin}, I.~N., {et~al.} 1994, \emph{{Binary
  star speckle measurements during 1989-1993 from the SAO 6 M and 1 M
  telescopes in Zelenchuk}}.
\newblock \aaps, 105, 503

\bibitem[{{Baraffe} {et~al.}(1998){Baraffe}, {Chabrier}, {Allard}, \&
  {Hauschildt}}]{baraffe1998}
{Baraffe}, I., {Chabrier}, G., {Allard}, F., \& {Hauschildt}, P.~H. 1998,
  \emph{{Evolutionary models for solar metallicity low-mass stars:
  mass-magnitude relationships and color-magnitude diagrams}}.
\newblock \aap, 337, 403

\bibitem[{{Barbier-Brossat} {et~al.}(1994){Barbier-Brossat}, {Petit}, \&
  {Figon}}]{barbier1994}
{Barbier-Brossat}, M., {Petit}, M., \& {Figon}, P. 1994, \emph{{Third
  bibliographic catalogue of stellar radial velocities}}.
\newblock \aaps, 108, 603

\bibitem[{{Barnard}(1945)}]{barnard1945}
{Barnard}, G.~A. 1945, \emph{{A new test for $2\times 2$ tables}}.
\newblock Nature, 156, 177

\bibitem[{{Bate} {et~al.}(2003){Bate}, {Bonnell}, \& {Bromm}}]{bate2003}
{Bate}, M.~R., {Bonnell}, I.~A., \& {Bromm}, V. 2003, \emph{{The formation of a
  star cluster: predicting the properties of stars and brown dwarfs}}.
\newblock \mnras, 339, 577

\bibitem[{{Batten}(1967)}]{batten1967}
{Batten}, A.~H. 1967, \emph{{The Orientation of the Orbital Planes of Fifty-Two
  Visual Binary Systems}}.
\newblock in IAU Symp. 30: Determination of Radial Velocities and their
  Applications, 199

\bibitem[{{Batten}(1973)}]{batten1973}
{Batten}, A.~H. 1973, {Binary and Multiple Systems of Stars} (Pergamon Press,
  Oxford)

\bibitem[{{Batten} {et~al.}(1997){Batten}, {Fletcher}, \&
  {MacCarthy}}]{batten1997}
{Batten}, A.~H., {Fletcher}, J.~M., \& {MacCarthy}, D.~G. 1997, \emph{{Eighth
  Orbital Elements of Spectroscopic Binaries}}.
\newblock VizieR Online Data Catalog, 5064, 0

\bibitem[{{Bertelli} {et~al.}(1994){Bertelli}, {Bressan}, {Chiosi}, {Fagotto},
  \& {Nasi}}]{bertelli1994}
{Bertelli}, G., {Bressan}, A., {Chiosi}, C., {Fagotto}, F., \& {Nasi}, E. 1994,
  \emph{{Theoretical isochrones from models with new radiative opacities}}.
\newblock \aaps, 106, 275

\bibitem[{{Bertiau}(1958)}]{bertiau1958}
{Bertiau}, F.~C. 1958, \emph{{Ansolute Magnitudes of Stars in the
  Scorpio-Centaurus Association.}}
\newblock \apj, 128, 533

\bibitem[{{Bessel}(1844)}]{bessel1844}
{Bessel}, F.~W. 1844, \emph{{Extract of a letter from on the proper motions of
  Procyon and Sirius}}.
\newblock \mnras, 6, 136

\bibitem[{{Beuzit} {et~al.}(1997){Beuzit}, {Demailly}, {Gendron}, {Gigan},
  {Lacombe}, {Rouan}, {Hubin}, {Bonaccini}, {Prieto}, {Chazallet}, {Rabaud},
  {Madec}, {Rousset}, {Hofmann}, \& {Eisenhauer}}]{beuzit1997}
{Beuzit}, J.-L., {Demailly}, L., {Gendron}, E., {et~al.} 1997, \emph{{Adaptive
  Optics on a 3.6-Meter Telescope. The ADONIS System.}}
\newblock Experimental Astronomy, 7, 285

\bibitem[{{Binney} \& {Tremaine}(1987)}]{binneytremaine}
{Binney}, J. \& {Tremaine}, S. 1987, {Galactic dynamics} (Princeton, NJ,
  Princeton University Press)

\bibitem[{{Blaauw}(1961)}]{blaauw1961}
{Blaauw}, A. 1961, \emph{{On the origin of the O- and B-type stars with high
  velocities (the ''run-away'' stars), and some related problems}}.
\newblock \bain, 15, 265

\bibitem[{{Blaauw}(1964)}]{blaauw1964A}
{Blaauw}, A. 1964, \emph{{The O Associations in the Solar Neighborhood}}.
\newblock \araa, 2, 213

\bibitem[{{Blaauw}(1978)}]{blaauw1978}
{Blaauw}, A. 1978, {Internal Motions and Age of the Sub-Association Upper
  Scorpio} (Problems of Physics and Evolution of the Universe), 101

\bibitem[{{Blaauw}(1991)}]{blaauw1991}
{Blaauw}, A. 1991, \emph{{OB Associations and the Fossil Record of Star
  Formation}}.
\newblock in NATO ASIC Proc. 342: The Physics of Star Formation and Early
  Stellar Evolution, 125

\bibitem[{{Bouy} {et~al.}(2006){Bouy}, {Mart{\'{\i}}n}, {Brandner},
  {Zapatero-Osorio}, {B{\'e}jar}, {Schirmer}, {Hu{\'e}lamo}, \&
  {Ghez}}]{bouy2006}
{Bouy}, H., {Mart{\'{\i}}n}, E.~L., {Brandner}, W., {et~al.} 2006,
  \emph{{Multiplicity of very low-mass objects in the Upper Scorpius OB
  association: a possible wide binary population}}.
\newblock \aap, 451, 177

\bibitem[{{Brice{\~n}o} {et~al.}(2005){Brice{\~n}o}, {Calvet}, {Hern{\'a}ndez},
  {Vivas}, {Hartmann}, {Downes}, \& {Berlind}}]{briceno2005}
{Brice{\~n}o}, C., {Calvet}, N., {Hern{\'a}ndez}, J., {et~al.} 2005, \emph{{The
  CIDA Variability Survey of Orion OB1. I. The Low-Mass Population of Ori OB1a
  and 1b}}.
\newblock \aj, 129, 907

\bibitem[{{Brown}(2001)}]{brown2001}
{Brown}, A. 2001, \emph{{The binary population in OB associations}}.
\newblock Astronomische Nachrichten, 322, 43

\bibitem[{{Brown} {et~al.}(1997){Brown}, {Arenou}, {van Leeuwen}, {Lindegren},
  \& {Luri}}]{brown1997}
{Brown}, A.~G.~A., {Arenou}, F., {van Leeuwen}, F., {Lindegren}, L., \& {Luri},
  X. 1997, \emph{{Some Considerations in Making Full Use of the HIPPARCOS
  Catalogue}}.
\newblock The First Results of Hipparcos and Tycho, 23rd meeting of the IAU,
  Joint Discussion 14, 25 August 1997, Kyoto, Japan, meeting abstract., 14

\bibitem[{{Brown} {et~al.}(1999){Brown}, {Blaauw}, {Hoogerwerf}, {de Bruijne},
  \& {de Zeeuw}}]{brown1999}
{Brown}, A.~G.~A., {Blaauw}, A., {Hoogerwerf}, R., {de Bruijne}, J.~H.~J., \&
  {de Zeeuw}, P.~T. 1999, \emph{{OB Associations}}.
\newblock in NATO ASIC Proc. 540: The Origin of Stars and Planetary Systems,
  ed. C.~J. {Lada} \& N.~D. {Kylafis}, 411

\bibitem[{{Brown} \& {Verschueren}(1997)}]{brownverschueren}
{Brown}, A.~G.~A. \& {Verschueren}, W. 1997, \emph{{High S/N Echelle
  spectroscopy in young stellar groups. II. Rotational velocities of early-type
  stars in SCO OB2.}}
\newblock \aap, 319, 811

\bibitem[{{Burgasser} {et~al.}(2003){Burgasser}, {Kirkpatrick}, {Reid},
  {Brown}, {Miskey}, \& {Gizis}}]{burgasser2003}
{Burgasser}, A.~J., {Kirkpatrick}, J.~D., {Reid}, I.~N., {et~al.} 2003,
  \emph{{Binarity in Brown Dwarfs: T Dwarf Binaries Discovered with the Hubble
  Space Telescope Wide Field Planetary Camera 2}}.
\newblock \apj, 586, 512

\bibitem[{{Buscombe} \& {Kennedy}(1962)}]{buscombe1962}
{Buscombe}, W. \& {Kennedy}, P.~M. 1962, \emph{{Two B-Type Spectroscopic
  Binaries}}.
\newblock \pasp, 74, 323

\bibitem[{{Carpenter}(2001)}]{carpenter2001}
{Carpenter}, J.~M. 2001, \emph{{Color Transformations for the 2MASS Second
  Incremental Data Release}}.
\newblock \aj, 121, 2851

\bibitem[{{Carter} \& {Meadows}(1995)}]{carter1995}
{Carter}, B.~S. \& {Meadows}, V.~S. 1995, \emph{{Fainter Southern JHK Standards
  Suitable for Infrared Arrays}}.
\newblock \mnras, 276, 734

\bibitem[{{Chabrier} {et~al.}(2000){Chabrier}, {Baraffe}, {Allard}, \&
  {Hauschildt}}]{chabrier2000}
{Chabrier}, G., {Baraffe}, I., {Allard}, F., \& {Hauschildt}, P. 2000,
  \emph{{Evolutionary Models for Very Low-Mass Stars and Brown Dwarfs with
  Dusty Atmospheres}}.
\newblock \apj, 542, 464

\bibitem[{{Chabrier} {et~al.}(2005){Chabrier}, {Baraffe}, {Allard}, \&
  {Hauschildt}}]{chabrier2005}
{Chabrier}, G., {Baraffe}, I., {Allard}, F., \& {Hauschildt}, P.~H. 2005,
  \emph{{Review on low-mass stars and brown dwarfs}}.
\newblock in Resolved Stellar Populations, ASP Conference Series, eds.
  Valls-Gabaud \& Chavez

\bibitem[{{Chanam{\'e}} \& {Gould}(2004)}]{chaname2004}
{Chanam{\'e}}, J. \& {Gould}, A. 2004, \emph{{Disk and Halo Wide Binaries from
  the Revised Luyten Catalog: Probes of Star Formation and MACHO Dark Matter}}.
\newblock \apj, 601, 289

\bibitem[{{Chen} {et~al.}(2006){Chen}, {Henning}, {van Boekel}, \&
  {Grady}}]{chen2006}
{Chen}, X.~P., {Henning}, T., {van Boekel}, R., \& {Grady}, C.~A. 2006,
  \emph{{VLT/NACO adaptive optics imaging of the Herbig Ae star HD 100453}}.
\newblock \aap, 445, 331

\bibitem[{{Close} {et~al.}(1990){Close}, {Richer}, \& {Crabtree}}]{close1990}
{Close}, L.~M., {Richer}, H.~B., \& {Crabtree}, D.~R. 1990, \emph{{A complete
  sample of wide binaries in the solar neighborhood}}.
\newblock \aj, 100, 1968

\bibitem[{{Couteau}(1995)}]{couteau1995}
{Couteau}, P. 1995, \emph{{Catalogue of 2700 double stars}}.
\newblock VizieR Online Data Catalog, 1209, 0

\bibitem[{{Cox}(2000)}]{allen2000}
{Cox}, A.~N. 2000, {Allen's astrophysical quantities} (Springer, New York,
  2000.~Edited by Arthur N.~Cox.)

\bibitem[{{Cruzalebes} {et~al.}(1998){Cruzalebes}, {Lopez}, {Bester},
  {Gendron}, \& {Sams}}]{cruzalebes1998}
{Cruzalebes}, P., {Lopez}, B., {Bester}, M., {Gendron}, E., \& {Sams}, B. 1998,
  \emph{{Near-infrared adaptive optics imaging of dust shells around five
  late-type stars with COME-ON+}}.
\newblock \aap, 338, 132

\bibitem[{{Cunha} \& {Lambert}(1994)}]{cunha1994}
{Cunha}, K. \& {Lambert}, D.~L. 1994, \emph{{Chemical evolution of the Orion
  association. 2: The carbon, nitrogen, oxygen, silicon, and iron abundances of
  main-sequence B stars}}.
\newblock \apj, 426, 170

\bibitem[{{Cutri} {et~al.}(2003){Cutri}, {Skrutskie}, {van Dyk}, {Beichman},
  {Carpenter}, {Chester}, {Cambresy}, {Evans}, {Fowler}, {Gizis}, {Howard},
  {Huchra}, {Jarrett}, {Kopan}, {Kirkpatrick}, {Light}, {Marsh}, {McCallon},
  {Schneider}, {Stiening}, {Sykes}, {Weinberg}, {Wheaton}, {Wheelock}, \&
  {Zacarias}}]{2mass}
{Cutri}, R.~M., {Skrutskie}, M.~F., {van Dyk}, S., {et~al.} 2003, \emph{{2MASS
  All Sky Catalog of point sources.}}
\newblock VizieR Online Data Catalog, 2246

\bibitem[{{Davis} {et~al.}(1984){Davis}, {Hut}, \& {Muller}}]{davis1984}
{Davis}, M., {Hut}, P., \& {Muller}, R.~A. 1984, \emph{{Extinction of species
  by periodic comet showers}}.
\newblock \nat, 308, 715

\bibitem[{{de Bruijne}(1999)}]{debruijne1999}
{de Bruijne}, J.~H.~J. 1999, \emph{{Structure and colour-magnitude diagrams of
  Scorpius OB2 based on kinematic modelling of Hipparcos data}}.
\newblock \mnras, 310, 585

\bibitem[{{de Geus}(1992)}]{degeus1992}
{de Geus}, E.~J. 1992, \emph{{Interactions of stars and interstellar matter in
  Scorpio Centaurus}}.
\newblock \aap, 262, 258

\bibitem[{{de Geus} {et~al.}(1989){de Geus}, {de Zeeuw}, \& {Lub}}]{degeus1989}
{de Geus}, E.~J., {de Zeeuw}, P.~T., \& {Lub}, J. 1989, \emph{{Physical
  parameters of stars in the Scorpio-Centaurus OB association}}.
\newblock \aap, 216, 44

\bibitem[{{De Marco} {et~al.}(2004){De Marco}, {Bond}, {Harmer}, \&
  {Fleming}}]{demarco2004}
{De Marco}, O., {Bond}, H.~E., {Harmer}, D., \& {Fleming}, A.~J. 2004,
  \emph{{Indications of a Large Fraction of Spectroscopic Binaries among Nuclei
  of Planetary Nebulae}}.
\newblock \apjl, 602, L93

\bibitem[{{de Zeeuw} {et~al.}(1999){de Zeeuw}, {Hoogerwerf}, {de Bruijne},
  {Brown}, \& {Blaauw}}]{dezeeuw1999}
{de Zeeuw}, P.~T., {Hoogerwerf}, R., {de Bruijne}, J.~H.~J., {Brown}, A.~G.~A.,
  \& {Blaauw}, A. 1999, \emph{{A HIPPARCOS Census of the Nearby OB
  Associations}}.
\newblock \aj, 117, 354

\bibitem[{{de Zeeuw} \& {Brand}(1985)}]{dezeeuw1985}
{de Zeeuw}, T. \& {Brand}, J. 1985, \emph{{Photometric age determination of
  OB-association}}.
\newblock in ASSL Vol. 120: Birth and Evolution of Massive Stars and Stellar
  Groups, ed. W.~{Boland} \& H.~{van Woerden}, 95--101

\bibitem[{{Devillard}(1997)}]{devillard1997}
{Devillard}, N. 1997, \emph{{The eclipse software}}.
\newblock The Messenger, 87, 19

\bibitem[{{Diolaiti} {et~al.}(2000){Diolaiti}, {Bendinelli}, {Bonaccini},
  {Close}, {Currie}, \& {Parmeggiani}}]{diolaiti2000}
{Diolaiti}, E., {Bendinelli}, O., {Bonaccini}, D., {et~al.} 2000,
  \emph{{Analysis of isoplanatic high resolution stellar fields by the
  StarFinder code}}.
\newblock \aaps, 147, 335

\bibitem[{{Duflot} {et~al.}(1995){Duflot}, {Figon}, \&
  {Meyssonnier}}]{duflot1995}
{Duflot}, M., {Figon}, P., \& {Meyssonnier}, N. 1995, \emph{{Vitesses radiales.
  Catalogue WEB: Wilson Evans Batten. Subtittle: Radial velocities: The
  Wilson-Evans-Batten catalogue.}}
\newblock \aaps, 114, 269

\bibitem[{{Duquennoy} \& {Mayor}(1991)}]{duquennoy1991}
{Duquennoy}, A. \& {Mayor}, M. 1991, \emph{{Multiplicity among solar-type stars
  in the solar neighbourhood. II - Distribution of the orbital elements in an
  unbiased sample}}.
\newblock \aap, 248, 485

\bibitem[{{ESA}(1997)}]{esa1997}
{ESA}. 1997, \emph{{The Hipparcos and Tycho Catalogues (ESA 1997)}}.
\newblock VizieR Online Data Catalog, 1239, 0

\bibitem[{{Fischer} \& {Marcy}(1992)}]{fischer1992}
{Fischer}, D.~A. \& {Marcy}, G.~W. 1992, \emph{{Multiplicity among M dwarfs}}.
\newblock \apj, 396, 178

\bibitem[{{Fisher}(1925)}]{fisher1925}
{Fisher}, R.~A. 1925, {Statistical Methods for Research Workers} (Oliver and
  Boyd, Edinburgh)

\bibitem[{{Fryer} \& {Kalogera}(1997)}]{fryer1997}
{Fryer}, C. \& {Kalogera}, V. 1997, \emph{{Double Neutron Star Systems and
  Natal Neutron Star Kicks}}.
\newblock \apj, 489, 244

\bibitem[{{Fryer} {et~al.}(1999){Fryer}, {Woosley}, {Herant}, \&
  {Davies}}]{fryer1999}
{Fryer}, C.~L., {Woosley}, S.~E., {Herant}, M., \& {Davies}, M.~B. 1999,
  \emph{{Merging White Dwarf/Black Hole Binaries and Gamma-Ray Bursts}}.
\newblock \apj, 520, 650

\bibitem[{{Garmany} {et~al.}(1980){Garmany}, {Conti}, \&
  {Massey}}]{garmany1980}
{Garmany}, C.~D., {Conti}, P.~S., \& {Massey}, P. 1980, \emph{{Spectroscopic
  studies of O type stars. IX - Binary frequency}}.
\newblock \apj, 242, 1063

\bibitem[{{Ghez} {et~al.}(1997){Ghez}, {McCarthy}, {Patience}, \&
  {Beck}}]{ghez1997}
{Ghez}, A.~M., {McCarthy}, D.~W., {Patience}, J.~L., \& {Beck}, T.~L. 1997,
  \emph{{The Multiplicity of Pre-Main-Sequence Stars in Southern Star-forming
  Regions}}.
\newblock \apj, 481, 378

\bibitem[{{Ghez} {et~al.}(1993){Ghez}, {Neugebauer}, \& {Matthews}}]{ghez1993}
{Ghez}, A.~M., {Neugebauer}, G., \& {Matthews}, K. 1993, \emph{{The
  multiplicity of T Tauri stars in the star forming regions Taurus-Auriga and
  Ophiuchus-Scorpius: A 2.2 micron speckle imaging survey}}.
\newblock \aj, 106, 2005

\bibitem[{{Gies}(1987)}]{gies1987}
{Gies}, D.~R. 1987, \emph{{The Kinematical and Binary Properties of Association
  and Field O stars}}.
\newblock \baas, 19, 715

\bibitem[{{Gillett}(1988)}]{gillett1988}
{Gillett}, S.~L. 1988, \emph{{Orbital planes of visual binary stars are
  randomly oriented - A statistical demonstration}}.
\newblock \aj, 96, 1967

\bibitem[{{Girardi} {et~al.}(2002){Girardi}, {Bertelli}, {Bressan}, {Chiosi},
  {Groenewegen}, {Marigo}, {Salasnich}, \& {Weiss}}]{girardi2002}
{Girardi}, L., {Bertelli}, G., {Bressan}, A., {et~al.} 2002, \emph{{Theoretical
  isochrones in several photometric systems. I. Johnson-Cousins-Glass,
  HST/WFPC2, HST/NICMOS, Washington, and ESO Imaging Survey filter sets}}.
\newblock \aap, 391, 195

\bibitem[{{Gizis} {et~al.}(2001){Gizis}, {Kirkpatrick}, {Burgasser}, {Reid},
  {Monet}, {Liebert}, \& {Wilson}}]{gizis2001}
{Gizis}, J.~E., {Kirkpatrick}, J.~D., {Burgasser}, A., {et~al.} 2001,
  \emph{{Substellar Companions to Main-Sequence Stars: No Brown Dwarf Desert at
  Wide Separations}}.
\newblock \apjl, 551, L163

\bibitem[{{Glebocki}(2000)}]{glebocki2000}
{Glebocki}, R. 2000, \emph{{Orientation of the Orbital Planes of Visual Binary
  Systems}}.
\newblock Acta Astronomica, 50, 211

\bibitem[{{Greaves}(2004)}]{greaves2004}
{Greaves}, J. 2004, \emph{{New Northern hemisphere common proper-motion
  pairs}}.
\newblock \mnras, 355, 585

\bibitem[{{Halbwachs}(1981)}]{halbwachs1981}
{Halbwachs}, J.~L. 1981, \emph{{Statistical models for spectroscopic and for
  eclipsing binary stars}}.
\newblock \aap, 102, 191

\bibitem[{{Halbwachs}(1983)}]{halbwachs1983}
{Halbwachs}, J.~L. 1983, \emph{{Binaries among the bright stars - Estimation of
  the bias and study of the main-sequence stars}}.
\newblock \aap, 128, 399

\bibitem[{{Halbwachs}(1986)}]{halbwachs1986}
{Halbwachs}, J.~L. 1986, \emph{{Binaries among bright stars - Systems with
  evolved primary components and their relation to the properties of
  main-sequence binaries}}.
\newblock \aap, 168, 161

\bibitem[{{Halbwachs} {et~al.}(2003){Halbwachs}, {Mayor}, {Udry}, \&
  {Arenou}}]{halbwachs2003}
{Halbwachs}, J.~L., {Mayor}, M., {Udry}, S., \& {Arenou}, F. 2003,
  \emph{{Multiplicity among solar-type stars. III. Statistical properties of
  the F7-K binaries with periods up to 10 years}}.
\newblock \aap, 397, 159

\bibitem[{{Han} \& {Chang}(2006)}]{han2006}
{Han}, C. \& {Chang}, H.-Y. 2006, \emph{{Determination of Stellar Ellipticities
  in Future Microlensing Surveys}}.
\newblock \apj, 645, 271

\bibitem[{{Hartkopf} {et~al.}(2001){Hartkopf}, {McAlister}, \&
  {Mason}}]{hartkopf2001}
{Hartkopf}, W.~I., {McAlister}, H.~A., \& {Mason}, B.~D. 2001, \emph{{The 2001
  US Naval Observatory Double Star CD-ROM. III. The Third Catalog of
  Interferometric Measurements of Binary Stars}}.
\newblock \aj, 122, 3480

\bibitem[{{Heacox}(1997)}]{heacox1997}
{Heacox}, W. 1997, \emph{{Statistical Distributions of Binary Orbital Dynamic
  Characteristics}}.
\newblock in ASP Conf. Ser. 130: The Third Pacific Rim Conference on Recent
  Development on Binary Star Research, ed. K.-C. {Leung}, 13

\bibitem[{{Heacox}(1995)}]{heacox1995}
{Heacox}, W.~D. 1995, \emph{{On the Mass Ratio Distribution of Single-Lined
  Spectroscopic Binaries}}.
\newblock \aj, 109, 2670

\bibitem[{{Heggie}(1975)}]{heggie1975}
{Heggie}, D.~C. 1975, \emph{{Binary evolution in stellar dynamics}}.
\newblock \mnras, 173, 729

\bibitem[{{Heintz}(1969)}]{heintz1969}
{Heintz}, W.~D. 1969, \emph{{A Statistical Study of Binary Stars}}.
\newblock \jrasc, 63, 275

\bibitem[{{Heintz}(1978)}]{heintz1978}
{Heintz}, W.~D. 1978, \emph{{Double stars, Revised edition}}.
\newblock Geophysics and Astrophysics Monographs, 15

\bibitem[{{Herschel}(1802)}]{herschel1802}
{Herschel}, W. 1802, \emph{{Catalogue of 500 New Nebulae, Nebulous Stars,
  Planetary Nebulae, and Clusters of Stars; With Remarks on the Construction of
  the Heavens}}.
\newblock Philosophical Transactions Series I, 92, 477

\bibitem[{{Herschel}(1804)}]{herschel1804}
{Herschel}, W. 1804, \emph{{Continuation of an Account of the Changes That Have
  Happened in the Relative Situation of Double Stars}}.
\newblock Philosophical Transactions Series I, 94, 353

\bibitem[{{Hilditch}(2001)}]{hilditch2001}
{Hilditch}, R.~W. 2001, {An Introduction to Close Binary Stars} (An
  Introduction to Close Binary Stars, by R.~W.~Hilditch, pp.~392.~ISBN
  0521241065.~Cambridge, UK: Cambridge University Press, March 2001.)

\bibitem[{{Hillebrandt} \& {Niemeyer}(2000)}]{hillebrandt2000}
{Hillebrandt}, W. \& {Niemeyer}, J.~C. 2000, \emph{{Type IA Supernova Explosion
  Models}}.
\newblock \araa, 38, 191

\bibitem[{{Hills}(1975)}]{hills1975}
{Hills}, J.~G. 1975, \emph{{Encounters between binary and single stars and
  their effect on the dynamical evolution of stellar systems}}.
\newblock \aj, 80, 809

\bibitem[{{Hoffleit} \& {Jaschek}(1982)}]{hoffleit1982}
{Hoffleit}, D. \& {Jaschek}, C. 1982, {The Bright Star Catalogue} (New Haven:
  Yale University Observatory (4th edition), 1982)

\bibitem[{{Hoffleit} {et~al.}(1983){Hoffleit}, {Saladyga}, \&
  {Wlasuk}}]{hoffleit1983}
{Hoffleit}, D., {Saladyga}, M., \& {Wlasuk}, P. 1983, {Bright star catalogue.
  Supplement} (New Haven: Yale University Observatory, 1983)

\bibitem[{{Hogeveen}(1990)}]{hogeveen1990}
{Hogeveen}, S.~J. 1990, \emph{{The mass-ratio distribution of visual binary
  stars}}.
\newblock \apss, 173, 315

\bibitem[{{Hogeveen}(1992{\natexlab{a}})}]{hogeveen1992a}
{Hogeveen}, S.~J. 1992{\natexlab{a}}, \emph{{Statistical properties of
  spectroscopic binary stars - As derived from The Eight Catalogue of the
  Orbital Elements of Spectroscopic Binary Stars}}.
\newblock \apss, 193, 29

\bibitem[{{Hogeveen}(1992{\natexlab{b}})}]{hogeveen1992b}
{Hogeveen}, S.~J. 1992{\natexlab{b}}, \emph{{The mass-ratio distribution of
  spectroscopic binary stars}}.
\newblock \apss, 196, 299

\bibitem[{{Hoogerwerf}(2000)}]{hoogerwerf2000}
{Hoogerwerf}, R. 2000, \emph{{OB association members in the ACT and TRC
  catalogues}}.
\newblock \mnras, 313, 43

\bibitem[{{Hoogerwerf} {et~al.}(2001){Hoogerwerf}, {de Bruijne}, \& {de
  Zeeuw}}]{hoogerwerf2001}
{Hoogerwerf}, R., {de Bruijne}, J.~H.~J., \& {de Zeeuw}, P.~T. 2001, \emph{{On
  the origin of the O and B-type stars with high velocities. II. Runaway stars
  and pulsars ejected from the nearby young stellar groups}}.
\newblock \aap, 365, 49

\bibitem[{{Hu{\'e}lamo} {et~al.}(2001){Hu{\'e}lamo}, {Brandner}, {Brown},
  {Neuh{\"a}user}, \& {Zinnecker}}]{huelamo2001}
{Hu{\'e}lamo}, N., {Brandner}, W., {Brown}, A.~G.~A., {Neuh{\"a}user}, R., \&
  {Zinnecker}, H. 2001, \emph{{ADONIS observations of hard X-ray emitting late
  B-type stars in Lindroos systems}}.
\newblock \aap, 373, 657

\bibitem[{{Hut} {et~al.}(1992){Hut}, {McMillan}, {Goodman}, {Mateo}, {Phinney},
  {Pryor}, {Richer}, {Verbunt}, \& {Weinberg}}]{hut1992}
{Hut}, P., {McMillan}, S., {Goodman}, J., {et~al.} 1992, \emph{{Binaries in
  globular clusters}}.
\newblock \pasp, 104, 981

\bibitem[{{Hut} {et~al.}(2003){Hut}, {Shara}, {Aarseth}, {Klessen}, {Lombardi},
  {Makino}, {McMillan}, {Pols}, {Teuben}, \& {Webbink}}]{modest1}
{Hut}, P., {Shara}, M.~M., {Aarseth}, S.~J., {et~al.} 2003, \emph{{MODEST-1:
  Integrating stellar evolution and stellar dynamics}}.
\newblock New Astronomy, 8, 337

\bibitem[{{Jilinski} {et~al.}(2006){Jilinski}, {Daflon}, {Cunha}, \& {de La
  Reza}}]{jilinski2006}
{Jilinski}, E., {Daflon}, S., {Cunha}, K., \& {de La Reza}, R. 2006,
  \emph{{Radial velocity measurements of B stars in the Scorpius-Centaurus
  association}}.
\newblock \aap, 448, 1001

\bibitem[{{Jordi} {et~al.}(1997){Jordi}, {Ribas}, {Torra}, \&
  {Gimenez}}]{jordi1997}
{Jordi}, C., {Ribas}, I., {Torra}, J., \& {Gimenez}, A. 1997,
  \emph{{Photometric versus empirical surface gravities of eclipsing
  binaries.}}
\newblock \aap, 326, 1044

\bibitem[{{Kenyon} \& {Hartmann}(1995)}]{kenyon1995}
{Kenyon}, S.~J. \& {Hartmann}, L. 1995, \emph{{Pre-Main-Sequence Evolution in
  the Taurus-Auriga Molecular Cloud}}.
\newblock \apjs, 101, 117

\bibitem[{{Klessen} {et~al.}(2000){Klessen}, {Heitsch}, \& {Mac
  Low}}]{klessen2000}
{Klessen}, R.~S., {Heitsch}, F., \& {Mac Low}, M.-M. 2000, \emph{{Gravitational
  Collapse in Turbulent Molecular Clouds. I. Gasdynamical Turbulence}}.
\newblock \apj, 535, 887

\bibitem[{{Kobulnicky} {et~al.}(2006){Kobulnicky}, {Fryer}, \&
  {Kiminki}}]{kobulnicky2006}
{Kobulnicky}, H.~A., {Fryer}, C.~L., \& {Kiminki}, D.~C. 2006, \emph{{A Fresh
  Look at the Binary Characteristics Among Massive Stars with Implications for
  Supernova and X-Ray Binary Rates}}.
\newblock astro-ph/0605069

\bibitem[{{Koehler} {et~al.}(2006){Koehler}, {Petr-Gotzens}, {McCaughrean},
  {Bouvier}, {Duchene}, {Quirrenbach}, \& {Zinnecker}}]{koehler2006}
{Koehler}, R., {Petr-Gotzens}, M.~G., {McCaughrean}, M.~J., {et~al.} 2006,
  \emph{{Binary Stars in the Orion Nebula Cluster}}.
\newblock astro-ph/0607670

\bibitem[{{K{\"o}hler} {et~al.}(2000){K{\"o}hler}, {Kunkel}, {Leinert}, \&
  {Zinnecker}}]{koehler2000}
{K{\"o}hler}, R., {Kunkel}, M., {Leinert}, C., \& {Zinnecker}, H. 2000,
  \emph{{Multiplicity of X-ray selected T Tauri stars in the Scorpius-Centaurus
  OB association}}.
\newblock \aap, 356, 541

\bibitem[{{Kouwenhoven} {et~al.}(2006{\natexlab{a}}){Kouwenhoven}, {Brown}, \&
  {Kaper}}]{kouwenhoven2006a}
{Kouwenhoven}, M.~B.~N., {Brown}, A., \& {Kaper}, L. 2006{\natexlab{a}},
  \emph{{A brown dwarf desert for intermediate mass stars in Sco~OB2?}}
\newblock Submitted to \aap

\bibitem[{{Kouwenhoven} {et~al.}(2006{\natexlab{b}}){Kouwenhoven}, {Brown},
  {Portegies Zwart}, \& {Kaper}}]{kouwenhoven2006b}
{Kouwenhoven}, M.~B.~N., {Brown}, A., {Portegies Zwart}, S., \& {Kaper}, L.
  2006{\natexlab{b}}, \emph{{Recovering the true binary population from
  observations}}.
\newblock To be submitted to \aap

\bibitem[{{Kouwenhoven} {et~al.}(2004){Kouwenhoven}, {Brown}, {Gualandris},
  {Kaper}, {Portegies Zwart}, \& {Zinnecker}}]{kouwenhoven2004a}
{Kouwenhoven}, M.~B.~N., {Brown}, A.~G.~A., {Gualandris}, A., {et~al.} 2004,
  \emph{{The Primordial Binary Population in OB Associations}}.
\newblock in Revista Mexicana de Astronom\'{i}a y Astrof\'{i}sica Conference
  Series, ed. C.~{Allen} \& C.~{Scarfe}, 139--140

\bibitem[{{Kouwenhoven} {et~al.}(2005{\natexlab{a}}){Kouwenhoven}, {Brown},
  {Zinnecker}, {Kaper}, \& {Portegies Zwart}}]{kouwenhoven2005}
{Kouwenhoven}, M.~B.~N., {Brown}, A.~G.~A., {Zinnecker}, H., {Kaper}, L., \&
  {Portegies Zwart}, S.~F. 2005{\natexlab{a}}, \emph{{The primordial binary
  population. I. A near-infrared adaptive optics search for close visual
  companions to A star members of Scorpius OB2}}.
\newblock \aap, 430, 137

\bibitem[{{Kouwenhoven} {et~al.}(2005{\natexlab{b}}){Kouwenhoven}, {Brown},
  {Zinnecker}, {Kaper}, {Portegies Zwart}, \& {Gualandris}}]{kouwenhoven2005ao}
{Kouwenhoven}, T., {Brown}, A., {Zinnecker}, H., {et~al.} 2005{\natexlab{b}},
  \emph{{A Search for Close Companions in Sco OB2}}.
\newblock in Science with Adaptive Optics, ed. W.~{Brandner} \& M.~E. {Kasper},
  203

\bibitem[{{Kraicheva} {et~al.}(1989){Kraicheva}, {Popova}, {Tutukov}, \&
  {Yungelson}}]{kraicheva1989}
{Kraicheva}, Z.~T., {Popova}, E.~I., {Tutukov}, A.~V., \& {Yungelson}, L.~R.
  1989, \emph{{Statistical properties of the stars in the third edition of the
  catalogue of the physical parameters of the spectroscopic binary stars}}.
\newblock Nauchnye Informatsii, 67, 3

\bibitem[{{Kroupa}(1995{\natexlab{a}})}]{kroupa1995a}
{Kroupa}, P. 1995{\natexlab{a}}, \emph{{Inverse dynamical population synthesis
  and star formation}}.
\newblock \mnras, 277, 1491

\bibitem[{{Kroupa}(1995{\natexlab{b}})}]{kroupa1995c}
{Kroupa}, P. 1995{\natexlab{b}}, \emph{{Star cluster evolution, dynamical age
  estimation and the kinematical signature of star formation}}.
\newblock \mnras, 277, 1522

\bibitem[{{Kroupa}(1995{\natexlab{c}})}]{kroupa1995b}
{Kroupa}, P. 1995{\natexlab{c}}, \emph{{The dynamical properties of stellar
  systems in the Galactic disc}}.
\newblock \mnras, 277, 1507

\bibitem[{{Kroupa}(2001)}]{kroupa2001}
{Kroupa}, P. 2001, \emph{{On the variation of the initial mass function}}.
\newblock \mnras, 322, 231

\bibitem[{{Kroupa}(2002)}]{kroupa2002}
{Kroupa}, P. 2002, \emph{{The Initial Mass Function of Stars: Evidence for
  Uniformity in Variable Systems}}.
\newblock Science, 295, 82

\bibitem[{{Kroupa} {et~al.}(2001){Kroupa}, {Aarseth}, \&
  {Hurley}}]{kroupaaarseth2001}
{Kroupa}, P., {Aarseth}, S., \& {Hurley}, J. 2001, \emph{{The formation of a
  bound star cluster: from the Orion nebula cluster to the Pleiades}}.
\newblock \mnras, 321, 699

\bibitem[{{Kroupa} \& {Boily}(2002)}]{kroupaboily2002}
{Kroupa}, P. \& {Boily}, C.~M. 2002, \emph{{On the mass function of star
  clusters}}.
\newblock \mnras, 336, 1188

\bibitem[{{Kroupa} \& {Bouvier}(2003)}]{kroupabouvier2003}
{Kroupa}, P. \& {Bouvier}, J. 2003, \emph{{On the origin of brown dwarfs and
  free-floating planetary-mass objects}}.
\newblock \mnras, 346, 369

\bibitem[{{Kroupa} {et~al.}(1991){Kroupa}, {Gilmore}, \& {Tout}}]{kroupa1991}
{Kroupa}, P., {Gilmore}, G., \& {Tout}, C.~A. 1991, \emph{{The effects of
  unresolved binary stars on the determination of the stellar mass function}}.
\newblock \mnras, 251, 293

\bibitem[{{Kuiper}(1935{\natexlab{a}})}]{kuiper1935a}
{Kuiper}, G.~P. 1935{\natexlab{a}}, \emph{{Problems of Double-Star Astronomy.
  I}}.
\newblock \pasp, 47, 15

\bibitem[{{Kuiper}(1935{\natexlab{b}})}]{kuiper1935b}
{Kuiper}, G.~P. 1935{\natexlab{b}}, \emph{{Problems of Double-Star Astronomy.
  II}}.
\newblock \pasp, 47, 121

\bibitem[{{Lada}(2006)}]{lada2006}
{Lada}, C.~J. 2006, \emph{{Stellar Multiplicity and the Initial Mass Function:
  Most Stars Are Single}}.
\newblock \apjl, 640, L63

\bibitem[{{Lada} \& {Lada}(2003)}]{LL2003}
{Lada}, C.~J. \& {Lada}, E.~A. 2003, \emph{{Embedded Clusters in Molecular
  Clouds}}.
\newblock \araa, 41, 57

\bibitem[{{Larson}(2001)}]{larson2001}
{Larson}, R.~B. 2001, \emph{{Implications of Binary Properties for Theories of
  Star Formation}}.
\newblock in IAU Symposium, ed. H.~{Zinnecker} \& R.~{Mathieu}, 93

\bibitem[{{Leinert} {et~al.}(1993){Leinert}, {Zinnecker}, {Weitzel},
  {Christou}, {Ridgway}, {Jameson}, {Haas}, \& {Lenzen}}]{leinert1993}
{Leinert}, C., {Zinnecker}, H., {Weitzel}, N., {et~al.} 1993, \emph{{A
  systematic approach for young binaries in Taurus}}.
\newblock \aap, 278, 129

\bibitem[{{Lenzen} {et~al.}(1998){Lenzen}, {Hofmann}, {Bizenberger}, \&
  {Tusche}}]{lenzen1998}
{Lenzen}, R., {Hofmann}, R., {Bizenberger}, P., \& {Tusche}, A. 1998,
  \emph{{CONICA: the high-resolution near-infrared camera for the ESO VLT}}.
\newblock in Proc. SPIE Vol. 3354, p. 606-614, Infrared Astronomical
  Instrumentation, Albert M. Fowler; Ed., 606--614

\bibitem[{{Levato} {et~al.}(1987){Levato}, {Malaroda}, {Morrell}, \&
  {Solivella}}]{levato1987}
{Levato}, H., {Malaroda}, S., {Morrell}, N., \& {Solivella}, G. 1987,
  \emph{{Stellar multiplicity in the Scorpius-Centaurus association}}.
\newblock \apjs, 64, 487

\bibitem[{{Lindegren} {et~al.}(1997){Lindegren}, {Mignard}, {S{\"o}derhjelm},
  {Badiali}, {Bernstein}, {Lampens}, {Pannunzio}, {Arenou}, {Bernacca},
  {Falin}, {Froeschl{\'e}}, {Kovalevsky}, {Martin}, {Perryman}, \&
  {Wielen}}]{lindgren1997}
{Lindegren}, L., {Mignard}, F., {S{\"o}derhjelm}, S., {et~al.} 1997,
  \emph{{Double star data in the HIPPARCOS Catalogue}}.
\newblock \aap, 323, L53

\bibitem[{{Lindroos}(1985)}]{lindroos1985}
{Lindroos}, K.~P. 1985, \emph{{A study of visual double stars with early type
  primaries. IV Astrophysical data}}.
\newblock \aaps, 60, 183

\bibitem[{{Maeder} \& {Meynet}(1988)}]{maeder1988}
{Maeder}, A. \& {Meynet}, G. 1988, \emph{{Tables of evolutionary star models
  from 0.85 to 120 solar masses with overshooting and mass loss}}.
\newblock \aaps, 76, 411

\bibitem[{{Malkov} \& {Zinnecker}(2001)}]{malkovzinnecker2001}
{Malkov}, O. \& {Zinnecker}, H. 2001, \emph{{Binary stars and the fundamental
  initial mass function}}.
\newblock \mnras, 321, 149

\bibitem[{{Malkov}(1993)}]{malkov1993}
{Malkov}, O.~Y. 1993, \emph{{Catalogue of astrophysical parameters of binary
  systems}}.
\newblock Bulletin d'Information du Centre de Donnees Stellaires, 42, 27

\bibitem[{{Mamajek} {et~al.}(2002){Mamajek}, {Meyer}, \&
  {Liebert}}]{mamajek2002}
{Mamajek}, E.~E., {Meyer}, M.~R., \& {Liebert}, J. 2002, \emph{{Post-T Tauri
  Stars in the Nearest OB Association}}.
\newblock \aj, 124, 1670

\bibitem[{{Marcy} \& {Butler}(2000)}]{marcy2000}
{Marcy}, G.~W. \& {Butler}, R.~P. 2000, \emph{{Planets Orbiting Other Suns}}.
\newblock \pasp, 112, 137

\bibitem[{{Marcy} {et~al.}(2003){Marcy}, {Butler}, {Fischer}, \&
  {Vogt}}]{marcy2003}
{Marcy}, G.~W., {Butler}, R.~P., {Fischer}, D.~A., \& {Vogt}, S.~S. 2003,
  \emph{{Properties of Extrasolar Planets}}.
\newblock in ASP Conf. Ser. 294: Scientific Frontiers in Research on Extrasolar
  Planets, ed. D.~{Deming} \& S.~{Seager}, 1--58381

\bibitem[{{Mart{\'{\i}}n} {et~al.}(2004){Mart{\'{\i}}n}, {Delfosse}, \&
  {Guieu}}]{martin2004}
{Mart{\'{\i}}n}, E.~L., {Delfosse}, X., \& {Guieu}, S. 2004,
  \emph{{Spectroscopic Identification of DENIS-selected Brown Dwarf Candidates
  in the Upper Scorpius OB Association}}.
\newblock \aj, 127, 449

\bibitem[{{Mason}(1995)}]{mason1995}
{Mason}, B.~D. 1995, \emph{{Speckles and Shadow Bands}}.
\newblock \pasp, 107, 299

\bibitem[{{Mathieu}(1994)}]{mathieu1994}
{Mathieu}, R.~D. 1994, \emph{{Pre-Main-Sequence Binary Stars}}.
\newblock \araa, 32, 465

\bibitem[{{Mathis}(1990)}]{mathis1990}
{Mathis}, J.~S. 1990, \emph{{Interstellar dust and extinction}}.
\newblock \araa, 28, 37

\bibitem[{{Mazeh} \& {Goldberg}(1992)}]{mazeh1992}
{Mazeh}, T. \& {Goldberg}, D. 1992, \emph{{On the study of the mass ratio of
  spectroscopic binaries}}.
\newblock \apj, 394, 592

\bibitem[{{McAlister} {et~al.}(1993){McAlister}, {Mason}, {Hartkopf}, \&
  {Shara}}]{mcalister1993}
{McAlister}, H.~A., {Mason}, B.~D., {Hartkopf}, W.~I., \& {Shara}, M.~M. 1993,
  \emph{{ICCD speckle observations of binary stars. X - A further survey for
  duplicity among the bright stars}}.
\newblock \aj, 106, 1639

\bibitem[{{McCarthy} \& {Zuckerman}(2004)}]{mccarthy2004}
{McCarthy}, C. \& {Zuckerman}, B. 2004, \emph{{The Brown Dwarf Desert at
  75-1200 AU}}.
\newblock \aj, 127, 2871

\bibitem[{{Miller} {et~al.}(1990){Miller}, {Freund}, \& {Johnson}}]{miller1990}
{Miller}, I., {Freund}, J.~E., \& {Johnson}, R.~A. 1990, { Probability and
  Statistics for Engineers} (Prentice-Hall, USA)

\bibitem[{{Miura} {et~al.}(1992){Miura}, {Baba}, {Ni-Ino}, {Ohtsubo},
  {Noguchi}, \& {Isobe}}]{miura1992}
{Miura}, N., {Baba}, N., {Ni-Ino}, M., {et~al.} 1992, \emph{{Speckle
  observations of visual and spectroscopic binaries. IV}}.
\newblock Publications of the National Astronomical Observatory of Japan, 2,
  561

\bibitem[{{Mokiem} {et~al.}(2006){Mokiem}, {de Koter}, {Evans}, {Puls},
  {Smartt}, {Crowther}, {Herrero}, {Langer}, {Lennon}, {Najarro}, {Villamariz},
  \& {Yoon}}]{mokiem2006}
{Mokiem}, M.~R., {de Koter}, A., {Evans}, C.~J., {et~al.} 2006, \emph{{The
  VLT-FLAMES survey of massive stars: Mass loss and rotation of early-type
  stars in the SMC}}.
\newblock astro-ph/0606403

\bibitem[{{Moraux} {et~al.}(2003){Moraux}, {Bouvier}, {Stauffer}, \&
  {Cuillandre}}]{moraux2003}
{Moraux}, E., {Bouvier}, J., {Stauffer}, J.~R., \& {Cuillandre}, J.-C. 2003,
  \emph{{Brown dwarfs in the Pleiades cluster: Clues to the substellar mass
  function}}.
\newblock \aap, 400, 891

\bibitem[{{Nitschelm}(2004)}]{nitschelm2004}
{Nitschelm}, C. 2004, \emph{{Discovery and confirmation of some double-lined
  spectroscopic binaries in the Sco-Cen Complex}}.
\newblock in ASP Conf. Ser. 318: Spectroscopically and Spatially Resolving the
  Components of the Close Binary Stars, ed. R.~W. {Hidlitch}, H.~{Hensberge},
  \& K.~{Pavlovski}, 291--293

\bibitem[{{Oblak}(1978)}]{oblak1978}
{Oblak}, E. 1978, \emph{{UVBY photometry of wide visual double stars with B, A
  and F spectral type- I.}}
\newblock \aaps, 34, 453

\bibitem[{{Palla} \& {Stahler}(1999)}]{palla1999}
{Palla}, F. \& {Stahler}, S.~W. 1999, \emph{{Star Formation in the Orion Nebula
  Cluster}}.
\newblock \apj, 525, 772

\bibitem[{{Pedoussaut} {et~al.}(1996){Pedoussaut}, {Capdeville}, {Ginestet}, \&
  {Carquillat}}]{pedoussaut1996}
{Pedoussaut}, A., {Capdeville}, A., {Ginestet}, N., \& {Carquillat}, J.~M.
  1996, \emph{{List of Spectroscopic Binaries (Pedoussaut+ 1985)}}.
\newblock VizieR Online Data Catalog, 4016, 0

\bibitem[{{Perryman} {et~al.}(2001){Perryman}, {de Boer}, {Gilmore}, {H{\o}g},
  {Lattanzi}, {Lindegren}, {Luri}, {Mignard}, {Pace}, \& {de
  Zeeuw}}]{perryman2001}
{Perryman}, M.~A.~C., {de Boer}, K.~S., {Gilmore}, G., {et~al.} 2001,
  \emph{{GAIA: Composition, formation and evolution of the Galaxy}}.
\newblock \aap, 369, 339

\bibitem[{{Perryman} {et~al.}(1997){Perryman}, {Lindegren}, {Kovalevsky},
  {Hoeg}, {Bastian}, {Bernacca}, {Cr{\'e}z{\'e}}, {Donati}, {Grenon}, {van
  Leeuwen}, {van der Marel}, {Mignard}, {Murray}, {Le Poole}, {Schrijver},
  {Turon}, {Arenou}, {Froeschl{\'e}}, \& {Petersen}}]{hipparcos}
{Perryman}, M.~A.~C., {Lindegren}, L., {Kovalevsky}, J., {et~al.} 1997,
  \emph{{The HIPPARCOS Catalogue}}.
\newblock \aap, 323, L49

\bibitem[{{Persson} {et~al.}(1998){Persson}, {Murphy}, {Krzeminski}, {Roth}, \&
  {Rieke}}]{persson1998}
{Persson}, S.~E., {Murphy}, D.~C., {Krzeminski}, W., {Roth}, M., \& {Rieke},
  M.~J. 1998, \emph{{A New System of Faint Near-Infrared Standard Stars}}.
\newblock \aj, 116, 2475

\bibitem[{{Piskunov} \& {Mal'Kov}(1991)}]{piskunov1991}
{Piskunov}, A.~E. \& {Mal'Kov}, O.~I. 1991, \emph{{Unresolved binaries and the
  stellar luminosity function}}.
\newblock \aap, 247, 87

\bibitem[{{Podsiadlowski} {et~al.}(1992){Podsiadlowski}, {Joss}, \&
  {Hsu}}]{podsialowski1992}
{Podsiadlowski}, P., {Joss}, P.~C., \& {Hsu}, J.~J.~L. 1992,
  \emph{{Presupernova evolution in massive interacting binaries}}.
\newblock \apj, 391, 246

\bibitem[{{Portegies Zwart} {et~al.}(2001){Portegies Zwart}, {McMillan}, {Hut},
  \& {Makino}}]{ecology4}
{Portegies Zwart}, S.~F., {McMillan}, S.~L.~W., {Hut}, P., \& {Makino}, J.
  2001, \emph{{Star cluster ecology - IV. Dissection of an open star cluster:
  photometry}}.
\newblock \mnras, 321, 199

\bibitem[{{Portegies Zwart} \& {Verbunt}(1996)}]{spzverbunt1996}
{Portegies Zwart}, S.~F. \& {Verbunt}, F. 1996, \emph{{Population synthesis of
  high-mass binaries.}}
\newblock \aap, 309, 179

\bibitem[{{Pourbaix} {et~al.}(2003){Pourbaix}, {Platais}, {Detournay},
  {Jorissen}, {Knapp}, \& {Makarov}}]{pourbaix2003}
{Pourbaix}, D., {Platais}, I., {Detournay}, S., {et~al.} 2003, \emph{{How many
  Hipparcos Variability-Induced Movers are genuine binaries?}}
\newblock \aap, 399, 1167

\bibitem[{{Poveda} \& {Allen}(2004)}]{poveda2004}
{Poveda}, A. \& {Allen}, C. 2004, \emph{{The distribution of separations of
  wide binaries of different ages}}.
\newblock in Revista Mexicana de Astronomia y Astrofisica Conference Series,
  ed. C.~{Allen} \& C.~{Scarfe}, 49--57

\bibitem[{{Poveda} {et~al.}(1982){Poveda}, {Allen}, \& {Parrao}}]{poveda1982}
{Poveda}, A., {Allen}, C., \& {Parrao}, L. 1982, \emph{{Statistical studies of
  visual double and multiple stars. I - Incompleteness of the IDS, intrinsic
  fraction of visual doubles and multiples, and number of optical systems}}.
\newblock \apj, 258, 589

\bibitem[{{Preibisch} {et~al.}(2002){Preibisch}, {Brown}, {Bridges},
  {Guenther}, \& {Zinnecker}}]{preibisch2002}
{Preibisch}, T., {Brown}, A.~G.~A., {Bridges}, T., {Guenther}, E., \&
  {Zinnecker}, H. 2002, \emph{{Exploring the Full Stellar Population of the
  Upper Scorpius OB Association}}.
\newblock \aj, 124, 404

\bibitem[{{Preibisch} {et~al.}(2003){Preibisch}, {Stanke}, \&
  {Zinnecker}}]{preibisch2003}
{Preibisch}, T., {Stanke}, T., \& {Zinnecker}, H. 2003, \emph{{Constraints on
  the IMF and the brown dwarf population of the young cluster IC 348}}.
\newblock \aap, 409, 147

\bibitem[{{Preibisch} \& {Zinnecker}(1999)}]{preibisch1999}
{Preibisch}, T. \& {Zinnecker}, H. 1999, \emph{{The History of Low-Mass Star
  Formation in the Upper Scorpius OB Association}}.
\newblock \aj, 117, 2381

\bibitem[{{Press} {et~al.}(1992){Press}, {Teukolsky}, {Vetterling}, \&
  {Flannery}}]{numericalrecipies}
{Press}, W.~H., {Teukolsky}, S.~A., {Vetterling}, W.~T., \& {Flannery}, B.~P.
  1992, {Numerical recipes in FORTRAN. The art of scientific computing}
  (Cambridge: University Press, |c1992, 2nd ed.)

\bibitem[{{Quist} \& {Lindegren}(2000)}]{quist2000}
{Quist}, C.~F. \& {Lindegren}, L. 2000, \emph{{Statistics of Hipparcos
  binaries: probing the 1-10 AU separation range}}.
\newblock \aap, 361, 770

\bibitem[{{Reipurth} \& {Clarke}(2001)}]{reipurth2001}
{Reipurth}, B. \& {Clarke}, C. 2001, \emph{{The Formation of Brown Dwarfs as
  Ejected Stellar Embryos}}.
\newblock \aj, 122, 432

\bibitem[{{Reipurth} \& {Zinnecker}(1993)}]{reipurth1993}
{Reipurth}, B. \& {Zinnecker}, H. 1993, \emph{{Visual binaries among pre-main
  sequence stars}}.
\newblock \aap, 278, 81

\bibitem[{{Robin} {et~al.}(2003){Robin}, {Reyl{\' e}}, {Derri{\` e}re}, \&
  {Picaud}}]{besancon}
{Robin}, A.~C., {Reyl{\' e}}, C., {Derri{\` e}re}, S., \& {Picaud}, S. 2003,
  \emph{{A synthetic view on structure and evolution of the Milky Way}}.
\newblock \aap, 409, 523

\bibitem[{{Rousset} {et~al.}(2000){Rousset}, {Lacombe}, {Puget}, {Gendron},
  {Arsenault}, {Kern}, {Rabaud}, {Madec}, {Hubin}, {Zins}, {Stadler},
  {Charton}, {Gigan}, \& {Feautrier}}]{rousset2000}
{Rousset}, G., {Lacombe}, F., {Puget}, P., {et~al.} 2000, \emph{{Status of the
  VLT Nasmyth adaptive optics system (NAOS)}}.
\newblock in Proc. SPIE Vol. 4007, p. 72-81, Adaptive Optical Systems
  Technology, Peter L. Wizinowich; Ed., 72--81

\bibitem[{{Salpeter}(1955)}]{salpeter1955}
{Salpeter}, E.~E. 1955, \emph{{The Luminosity Function and Stellar Evolution.}}
\newblock \apj, 121, 161

\bibitem[{{Sartori} {et~al.}(2003){Sartori}, {L{\'e}pine}, \&
  {Dias}}]{sartori2003}
{Sartori}, M.~J., {L{\'e}pine}, J.~R.~D., \& {Dias}, W.~S. 2003,
  \emph{{Formation scenarios for the young stellar associations between
  galactic longitudes l = 280degr - 360degr}}.
\newblock \aap, 404, 913

\bibitem[{{Saunders} {et~al.}(2006){Saunders}, {Naylor}, \&
  {Allan}}]{saunders2006}
{Saunders}, E.~S., {Naylor}, T., \& {Allan}, A. 2006, \emph{{Optimal placement
  of a limited number of observations for period searches}}.
\newblock astro-ph/0605421

\bibitem[{{Savage} \& {Mathis}(1979)}]{savage1979}
{Savage}, B.~D. \& {Mathis}, J.~S. 1979, \emph{{Observed properties of
  interstellar dust}}.
\newblock \araa, 17, 73

\bibitem[{{Shatsky} \& {Tokovinin}(2002)}]{shatsky2002}
{Shatsky}, N. \& {Tokovinin}, A. 2002, \emph{{The mass ratio distribution of
  B-type visual binaries in the Sco OB2 association}}.
\newblock \aap, 382, 92

\bibitem[{{Siess} {et~al.}(2000){Siess}, {Dufour}, \& {Forestini}}]{siess2000}
{Siess}, L., {Dufour}, E., \& {Forestini}, M. 2000, \emph{{An internet server
  for pre-main sequence tracks of low- and intermediate-mass stars}}.
\newblock \aap, 358, 593

\bibitem[{{Sills} {et~al.}(2002){Sills}, {Adams}, {Davies}, \&
  {Bate}}]{sills2002}
{Sills}, A., {Adams}, T., {Davies}, M.~B., \& {Bate}, M.~R. 2002,
  \emph{{High-resolution simulations of stellar collisions between equal-mass
  main-sequence stars in globular clusters}}.
\newblock \mnras, 332, 49

\bibitem[{{Sills} {et~al.}(2003){Sills}, {Deiters}, {Eggleton}, {Freitag},
  {Giersz}, {Heggie}, {Hurley}, {Hut}, {Ivanova}, {Klessen}, {Kroupa},
  {Lombardi}, {McMillan}, {Portegies Zwart}, \& {Zinnecker}}]{modest2}
{Sills}, A., {Deiters}, S., {Eggleton}, P., {et~al.} 2003, \emph{{MODEST-2: a
  summary}}.
\newblock New Astronomy, 8, 605

\bibitem[{{Simon} {et~al.}(1995){Simon}, {Ghez}, {Leinert}, {Cassar}, {Chen},
  {Howell}, {Jameson}, {Matthews}, {Neugebauer}, \& {Richichi}}]{simon1995}
{Simon}, M., {Ghez}, A.~M., {Leinert}, C., {et~al.} 1995, \emph{{A lunar
  occultation and direct imaging survey of multiplicity in the Ophiuchus and
  Taurus star-forming regions}}.
\newblock \apj, 443, 625

\bibitem[{{Slesnick} {et~al.}(2006){Slesnick}, {Carpenter}, \&
  {Hillenbrand}}]{slesnick2006}
{Slesnick}, C.~L., {Carpenter}, J.~M., \& {Hillenbrand}, L.~A. 2006, \emph{{A
  Large-Area Search for Low-Mass Objects in Upper Scorpius. I. The Photometric
  Campaign and New Brown Dwarfs}}.
\newblock \aj, 131, 3016

\bibitem[{{Slettebak}(1968)}]{slettebak1968}
{Slettebak}, A. 1968, \emph{{Stellar Rotation in the Scorpio-Centaurus
  Association}}.
\newblock \apj, 151, 1043

\bibitem[{{S{\"o}derhjelm}(2000)}]{soderhjelm2000}
{S{\"o}derhjelm}, S. 2000, \emph{{Binary statistics from Hipparcos data - a
  progress report}}.
\newblock Astronomische Nachrichten, 321, 165

\bibitem[{{S{\"o}derhjelm}(2005)}]{soderhjelm2005}
{S{\"o}derhjelm}, S. 2005, \emph{{Census of Binaries the Big Picture}}.
\newblock in ESA SP-576: The Three-Dimensional Universe with Gaia, ed.
  C.~{Turon}, K.~S. {O'Flaherty}, \& M.~A.~C. {Perryman}, 97

\bibitem[{{Sowell} \& {Wilson}(1993)}]{sowell1993}
{Sowell}, J.~R. \& {Wilson}, J.~W. 1993, \emph{{All-sky Stromgren photometry of
  speckle binary stars}}.
\newblock \pasp, 105, 36

\bibitem[{{Svechnikov} \& {Bessonova}(1984)}]{svechnikov1984}
{Svechnikov}, M.~A. \& {Bessonova}, L.~A. 1984, \emph{{A Catalogue of Orbital
  Elements Masses and Luminosities of Close Double Stars}}.
\newblock Bulletin d'Information du Centre de Donnees Stellaires, 26, 99

\bibitem[{{ten Brummelaar} {et~al.}(2000){ten Brummelaar}, {Mason},
  {McAlister}, {Roberts}, {Turner}, {Hartkopf}, \&
  {Bagnuolo}}]{tenbrummelaar2000}
{ten Brummelaar}, T., {Mason}, B.~D., {McAlister}, H.~A., {et~al.} 2000,
  \emph{{Binary Star Differential Photometry Using the Adaptive Optics System
  at Mount Wilson Observatory}}.
\newblock \aj, 119, 2403

\bibitem[{{Terman} {et~al.}(1998){Terman}, {Taam}, \& {Savage}}]{terman1998}
{Terman}, J.~L., {Taam}, R.~E., \& {Savage}, C.~O. 1998, \emph{{A population
  synthesis study of high-mass X-ray binaries}}.
\newblock \mnras, 293, 113

\bibitem[{{Testi} {et~al.}(1998){Testi}, {Palla}, \& {Natta}}]{testi1998}
{Testi}, L., {Palla}, F., \& {Natta}, A. 1998, \emph{{A search for clustering
  around Herbig Ae/Be stars. II. Atlas of the observed sources}}.
\newblock \aaps, 133, 81

\bibitem[{{Thackeray} \& {Hutchings}(1965)}]{thackeray1965}
{Thackeray}, A.~D. \& {Hutchings}, F.~B. 1965, \emph{{Orbits of two
  double-lined binaries HD 140008 and 178322}}.
\newblock \mnras, 129, 191

\bibitem[{{Todd} {et~al.}(2005){Todd}, {Pollacco}, {Skillen}, {Bramich},
  {Bell}, \& {Augusteijn}}]{todd2005}
{Todd}, I., {Pollacco}, D., {Skillen}, I., {et~al.} 2005, \emph{{A survey of
  eclipsing binary stars in the eastern spiral arm of M31}}.
\newblock \mnras, 362, 1006

\bibitem[{{Tohline}(2002)}]{tohline2002}
{Tohline}, J.~E. 2002, \emph{{The Origin of Binary Stars}}.
\newblock \araa, 40, 349

\bibitem[{{Tokovinin}(1997)}]{tokovininmsc}
{Tokovinin}, A.~A. 1997, \emph{{MSC - a catalogue of physical multiple stars}}.
\newblock \aaps, 124, 75

\bibitem[{{Tout}(1991)}]{tout1991}
{Tout}, C.~A. 1991, \emph{{On the relation between the mass-ratio distribution
  in binary stars and the mass function for single stars}}.
\newblock \mnras, 250, 701

\bibitem[{{Trimble}(1990)}]{trimble1990}
{Trimble}, V. 1990, \emph{{The distributions of binary system mass ratios - A
  less biased sample}}.
\newblock \mnras, 242, 79

\bibitem[{{Trumpler} \& {Weaver}(1953)}]{trumpler1953}
{Trumpler}, R.~J. \& {Weaver}, H.~F. 1953, {Statistical astronomy} (Dover Books
  on Astronomy and Space Topics, New York: Dover Publications)

\bibitem[{{Turon} {et~al.}(2005){Turon}, {O'Flaherty}, \&
  {Perryman}}]{turon2005}
{Turon}, C., {O'Flaherty}, K.~S., \& {Perryman}, M.~A.~C., eds. 2005, {The
  Three-Dimensional Universe with Gaia}

\bibitem[{{van Albada}(1968)}]{vanalbada1968}
{van Albada}, T.~S. 1968, \emph{{Statistical properties of early-type double
  and multiple stars}}.
\newblock \bain, 20, 47

\bibitem[{{van den Berk} {et~al.}(2006){van den Berk}, {Portegies Zwart}, \&
  {McMillan}}]{vandenberk2006}
{van den Berk}, J., {Portegies Zwart}, S., \& {McMillan}, S. 2006, \emph{{The
  formation of higher-order hierarchical systems in star clusters}}.
\newblock astro-ph/0607456

\bibitem[{{van den Heuvel}(1994)}]{vandenheuvel1994}
{van den Heuvel}, E.~P.~J. 1994, \emph{{The binary pulsar PSRJ 2145-0750: A
  system originating from a low or intermediate mass X-ray binary with a donor
  star on the asymptotic giant branch?}}
\newblock \aap, 291, L39

\bibitem[{{van der Bliek} {et~al.}(1996){van der Bliek}, {Manfroid}, \&
  {Bouchet}}]{vanderbliek1996}
{van der Bliek}, N.~S., {Manfroid}, J., \& {Bouchet}, P. 1996, \emph{{Infrared
  aperture photometry at ESO (1983-1994) and its future use.}}
\newblock \aaps, 119, 547

\bibitem[{{Vereshchagin} {et~al.}(1988){Vereshchagin}, {Tutukov}, {Iungelson},
  {Kraicheva}, \& {Popova}}]{vereshchagin1988}
{Vereshchagin}, S., {Tutukov}, A., {Iungelson}, L., {Kraicheva}, Z., \&
  {Popova}, E. 1988, \emph{{Statistical study of visual binaries}}.
\newblock \apss, 142, 245

\bibitem[{{Vereshchagin} {et~al.}(1987){Vereshchagin}, {Kraicheva}, {Popova},
  {Tutukov}, \& {Yungelson}}]{vereshchagin1987}
{Vereshchagin}, S.~V., {Kraicheva}, Z.~T., {Popova}, E.~I., {Tutukov}, A.~V.,
  \& {Yungelson}, L.~R. 1987, \emph{{Physical Parameters of Visual Binaries - a
  Statistical Analysis}}.
\newblock Soviet Astronomy Letters, 13, 26

\bibitem[{{Verschueren} {et~al.}(1997){Verschueren}, {Brown}, {Hensberge},
  {David}, {Le Poole}, {de Geus}, \& {de Zeeuw}}]{verschueren1997}
{Verschueren}, W., {Brown}, A.~G.~A., {Hensberge}, H., {et~al.} 1997,
  \emph{{High S/N Echelle Spectroscopy in Young Stellar Groups. I. Observations
  and Data Reduction}}.
\newblock \pasp, 109, 868

\bibitem[{{Verschueren} {et~al.}(1996){Verschueren}, {David}, \&
  {Brown}}]{verschueren1996}
{Verschueren}, W., {David}, M., \& {Brown}, A.~G.~A. 1996, \emph{{A Search for
  Early-Type Binaries in the SCO OB2 Association}}.
\newblock in ASP Conf. Ser. 90: The Origins, Evolution, and Destinies of Binary
  Stars in Clusters, ed. E.~F. {Milone} \& J.-C. {Mermilliod}, 131

\bibitem[{{Whitmire} \& {Jackson}(1984)}]{whitmire1984}
{Whitmire}, D.~P. \& {Jackson}, A.~A. 1984, \emph{{Are periodic mass
  extinctions driven by a distant solar companion?}}
\newblock \nat, 308, 713

\bibitem[{{Worley}(1978)}]{worley1978}
{Worley}, C.~E. 1978, \emph{{Micrometer measures of 1,980 double stars}}.
\newblock Publications of the U.S.~Naval Observatory Second Series, 24, 1

\bibitem[{{Worley} \& {Douglass}(1997)}]{wds1997}
{Worley}, C.~E. \& {Douglass}, G.~G. 1997, \emph{{The Washington Double Star
  Catalog (WDS, 1996.0)}}.
\newblock \aaps, 125, 523

\bibitem[{{Yungelson} \& {Livio}(1998)}]{yungelson1998}
{Yungelson}, L. \& {Livio}, M. 1998, \emph{{Type IA Supernovae: an Examination
  of Potential Progenitors and the Redshift Distribution}}.
\newblock \apj, 497, 168

\end{thebibliography}

\addcontentsline{toc}{chapter}{Curriculum Vitae}
\chapter*{Curriculum Vitae}

Op 25~februari 1978 werd ik geboren in Kwintsheul. In juli 1996 voltooide ik mijn VWO opleiding aan het Interconfessioneel Westland College in Naaldwijk. Daarna ben ik sterrenkunde gaan studeren aan de Universiteit van Leiden. Ik deed daar twee afstudeer\-onderzoeken, getiteld ``{\em The large-scale structure and kinematics of the magelanic dwarf galaxy ESO\,364-G029}'' bij Martin Bureau, en ``{\em The dynamical distance and internal structure of the globular cluster $\omega$\,Centauri}, bij Ellen Verolme en Tim de Zeeuw. Tegen het einde van mijn studie bezocht ik een colloquium getiteld ``{\em The binary population in OB~associations}'' in \mbox{Leiden}, waarin Anthony Brown het onderzoeksvoorstel presenteerde, waar dit proefschrift het resultaat van is. Op 27~augustus 2002 behaalde ik mijn bul in Leiden, en drie dagen later begon ik als onderzoeker-in-opleiding op het Sterrenkundig Instituut Anton Pannekoek aan de Universiteit van Amsterdam.

Dit proefschrift is gebaseerd op het onderzoek dat ik in Amsterdam gedaan heb, onder begeleiding van Lex Kaper, Anthony Brown, Simon Portegies Zwart, en Ed van den Heuvel. Tijdens mijn onderzoek heb ik gebruik gemaakt van de ESO {\em Very Large Telescope} te Paranal, Chili. Vanwege het multi-disciplinaire karakter van mijn onderzoek (zowel observationele als computationele sterrenkunde) heb ik veel astronomen in binnen- en buitenland ontmoet. Voor mijn onderzoek ben ik onder andere afgereisd naar Itali\"{e}, Australi\"{e}, Duitsland, Frankrijk, Tsjechi\"{e}, het Verenigd Koninkrijk en Mexico. Behalve het doen van onderzoek behoorde ook onderwijs tot mijn taken; ik heb verschillende werkcolleges en practica gegeven, waaronder het waarneempracticum op La~Palma (Canarische Eilanden). 

Tijdens mijn aanstelling in Amsterdam deed ik aan outreach door middel van publieke lezingen, interviews op de radio, de Intel Science and Engineering Fair, en de Weekendschool voor basisscholieren. Ik contact gelegd met een aantal overheidsorganisaties en universiteiten in de Filippijnen, welke hopelijk binnenkort overtuigd zullen zijn om aldaar op kleine schaal aan sterrenkunde te gaan doen. Daarnaast heb ik me ook buiten de sterrenkunde ingezet voor Rotaract, de Nederlandse Coeliakievereniging, en de Association of European Coeliac Societies.
Op 1 oktober 2006 zal ik beginnen aan mijn eerste postdoctorale positie aan de Universiteit van Sheffield (Verenigd Koninkrijk), alwaar ik het werk uit dit proefschrift voort zal zetten, en zal werken aan de evolutie van jonge sterclusters.

\addcontentsline{toc}{chapter}{List of publications}
\chapter*{List of publications}

\section*{Refereed publications}

\begin{itemize}\addtolength{\itemsep}{-0.2\baselineskip}
\item[--] Kouwenhoven, M.B.N., Brown, A.G.A., \& Kaper, L. 2006, {\em A brown dwarf desert for intermediate mass stars in Sco OB2?}, submitted to \aap
\item[--] Kouwenhoven, M.B.N., Brown, A.G.A., Zinnecker, H., Kaper, L., \& Portegies Zwart, S.F. 2005, {\em The primordial binary population. I. A near-infrared adaptive optics search for close visual companions to A star members of Scorpius OB2}, \aap, 430, 137
\end{itemize}

\section*{Non-refereed publications}

\begin{itemize}\addtolength{\itemsep}{-0.2\baselineskip}
\item[--] Kouwenhoven, M.B.N. 2006, {\em The Primordial Binary Population in OB Associations}, to appear in the proceedings of the ESO Workshop on Multiple Stars across the H-R Diagram, held in Garching, 12-15 July 2005 (astroph/0509393)
\item[--] Kouwenhoven, M.B.N., Brown, A.G.A., Zinnecker, H., Kaper, L., Portegies Zwart, S.F., \& Gualandris, A. 2005, {\em A Search for Close Companions in Sco OB2} in Science with Adaptive Optics, eds. W. Brandner \& M.E. Kasper, 203
\item[--] Kouwenhoven, M.B.N., Brown, A.G.A., Gualandris, A., Kaper, L., Portegies Zwart, S.F., \& Zinnecker, H.  2004, {\em The Primordial Binary Population in OB Associations} in Revista Mexicana de Astronom\'{i}a y Astrof\'{i}sica Conference Series, eds. C. Allen \& C. Scarfe, 139140
\item[--] Kouwenhoven, M.B.N., Portegies Zwart, S.F., Gualandris, A., Brown, A.G.A., Kaper, L., \& Zinnecker, H. 2004, {\em Recovering the Primordial Binary Population using Simulating Observations of Simulations}. Astronomische Nachrichten Supplement, 325, 101
\item[--] Kouwenhoven, M.B.N., Brown, A.G.A., Gualandris, A., Kaper, L., Porgeties Zwart, S.F., \& Zinnecker, H. 2003, {\em The Primordial Binary Population in OB Associations} in Star Formation at High Angular Resolution, IAU Symposium 221, held 22-25 July, 2003 in Sydney, Australia, 49
\end{itemize}

\addcontentsline{toc}{chapter}{Nawoord}
\chapter*{Nawoord}

Als eerste wil ik mijn begeleiders (tevens co-auteurs) bedanken. Anthony, Lex, Simon en Ed, jullie hebben me de sterrenkunde van verschillende kanten laten zien. De combinatie van jullie verschillende manieren van aanpak is erg leerzaam geweest en is onmisbaar geweest voor de totstandkoming van dit proefschrift.

Ik heb een geweldige tijd gehad op het API en ben iedereen daar erg dankbaar voor: mijn kamergenoten (Dominik, Marc en Evert), het N-body team (Alessia en Evghenii) en de \mbox{cola-,} Quake-, East-, and keuken-break API's (Rien, Nick, Rohied, Annique, Rhaana, Wing-Fai, Simone, Elena, alle Arjan's en Arjen's, Joke, Klaas, Kazi en vele anderen). Martin, je hebt vele pogingen gedaan om me achter de computer vandaan te halen op vrijdagavond. Alhoewel het niet altijd lukte, was ik erg blij met elke uitnodiging voor de East.
Al dit werk zou niet mogelijk geweest zijn zonder de support van Minou (ook bedankt voor de extra hulp), Lidewijde (in het bijzonder voor de huisvesting), Annemieke en Fieke. 

Dit onderzoek is mogelijk gemaakt door steun van NWO (projectnummer 614.041.006), NOVA, de Universteit van Amsterdam, het LKBF en de IAU. I wish to thank ESO and the Paranal Observatory for efficiently conducting the Service-Mode observations and for their support. Ook de on-line services van 2MASS, Simbad, Vizier, ADS en Google waren onmisbaar voor mijn onderzoek.

During my PhD I have been a member of J\&G, AOECS-CYE, HC, and Rotaract; participation in these organizations has resulted in a lot of fun, many new friends, and has indirectly been of great importance for this thesis and my career. Nais kong ipahatid ang taos puso kong pasasalamat sa aking mga kaibigang pinoy sa pagpapakilala sa akin ng bagong mundo at sa pagbibigay sa akin ng pangalawang tahanan. Maraming maraming salamat sa inyo!  Chris, Barry, Jan, Mark, Robert en Ingrid, volgende Koninginnenach ben ik er weer bij! Swen, zet de apfelkorn alvast maar klaar. 
Mijn interesse voor de astronomie begon toen mijn vader me in 1985 voor het eerst wegwijs maakte in de sterrenhemel. Ik ben mijn vader en moeder enorm dankbaar voor de inspiratie en de keuzevrijheid die ze me gegeven hebben. Don, Els, Kees en Marieke, bedankt voor jullie geduld en voor de leuke tijd in De Heul.

Finally, I wish to finish this thesis with a message to the indirect (financial) supporters of this project: I am very thankful to the Dutch society and European community for strongly supporting education and research in the wonderful field of astronomy!

\cleardoublepage

\thispagestyle{empty}
\includegraphics[width=0.95\textwidth,height=!]{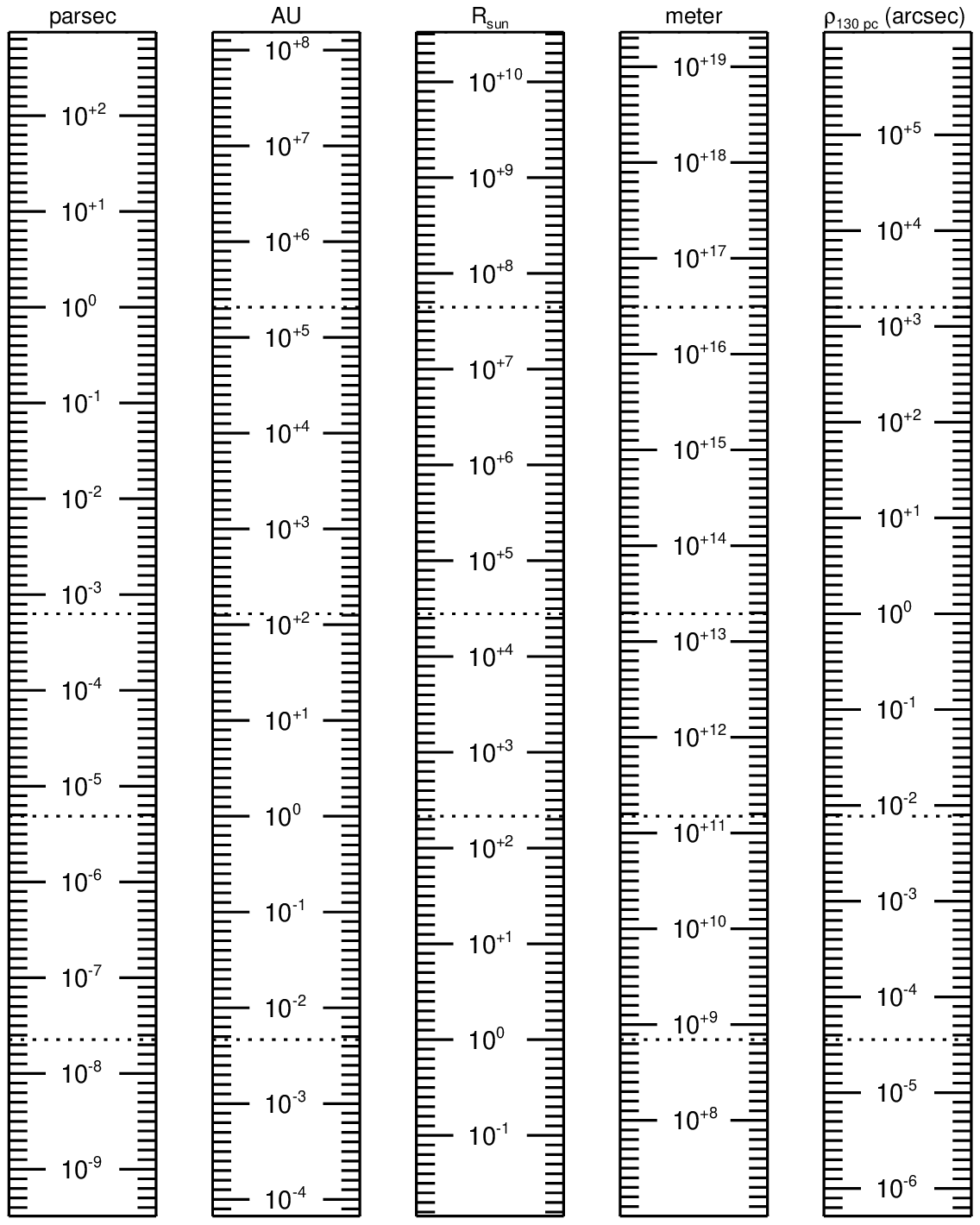}

\vskip0.5cm

\begin{tabular}{llp{0.75cm}ll}
1 parsec    &=& 3.08568 &$\times 10^{16}$ &meter \\
1 AU        &=& 1.49598 &$\times 10^{11}$ &meter \\
1 R$_\odot$ &=& 6.9599  &$\times 10^8$    &meter \\
\end{tabular}

\vfill

\begin{flushright}
{\small Thijs Kouwenhoven, 2006}
\end{flushright}

\cleardoublepage

\end{document}